\documentstyle[12pt,epsf]{article}
\newcommand{\sba}{{\mbox{\scriptsize            \bf             {A}}}}
\newcommand{\sbz}{{\mbox{\scriptsize            \bf             {Z}}}}
\newcommand{\tbz}{{\mbox{\tiny            \bf             {Z}}}}

\newcommand{\sbc}{_{\mbox{\scriptsize            \bf            {C}}}}
\newcommand{\sbp}{_{\mbox{\scriptsize            \bf            {P}}}}
\newcommand{\sbnul}{_{\mbox{\scriptsize           \bf           {0}}}}
\newcommand{\sbunit}{{\mbox{\scriptsize          \bf           {1}}}}
\newcommand{\bunit}{\mbox{\bf{1}}}
\newcommand{\bpone}{\mbox{\bf{P}}_{1}}
\newcommand{\bptwo}{\mbox{\bf            {P}}_{2}}
\newcommand{\bpprime}{\mbox{\bf  {P'}}}
\newcommand{\baone}{\mbox{\bf{A}}_{1}}
\newcommand{\batwo}{\mbox{\bf            {A}}_{2}}
\newcommand{\sbtwo}{\mbox{\scriptsize            \bf            {2}}}
\newcommand{\sbthree}{\mbox{\scriptsize            \bf            {3}}}
\newcommand{\cartprod}{\mbox{{\bf $\times$}}^{\mbox{{\tiny cart. prod.}}}}
\newcommand{\link}{\mbox{\begin{picture}(4.15,10)
\put(0,3){\circle*{2}}                     \put(0,2.75){\line(1,0){7}}
\put(7.3,3){\circle*{2}}\mbox{  }  \end{picture}  }  }

\newcommand{\isomorph}{\stackrel{-}{\simeq}}

\newcommand{\fsps}{fundamental spacetime points}
\newcommand{\fspx}{fundamental spacetime point }
\newcommand{\fspsx}{fundamental spacetime points }
\newcommand{\site}{\mbox{\begin{picture}(4.15,4)
\put(2,2){\circle*{2}} \end{picture}  }  }
\newcommand{\on}{U_{sign\;U(\Box)}(\Box)}
\newcommand{\B}{U_{BIO}(\Box)}
\newcommand{\sitex}{\site^{\!\!\!x^{\mu}}}
\newcommand{\sitey}{\site^{\!\!\!\!\!y^{\mu}}}
\newcommand{\sites}{\site^{\!\!\!\!\!s^{\mu}}}
\newcommand{\sitespa}{\site^{\!\!\!\!\!s^{\mu}+a\delta^{\mu}_{\nu}}}
\newcommand{\linklo}{\link^{\!\!\!\!l_0}}
\newcommand{\linkxy}{\;\link^{\!\!\!\!\!\!\!x^{\mu}\;\;y^{\mu}}}
\newcommand{\linksspa}{\;\link^{\!\!\!\!\!\!\!s^{\mu}\;\;s^{\mu}+
a\delta^{\mu}_{\nu}} }
\newcommand{\linkx}{\;\link^{\!\!\!\!\!\!\!x^{\mu}\;\;}}
\newcommand{\p}{\partial}
\newcommand{\mi}{{\cal M}_i}
\newcommand{\beq}{\begin{equation}}
\newcommand{\eeq}{\end{equation}}
\newcommand{\nin}{\noindent}
\newcommand{\mc}{\multicolumn}
\newcommand{\tla}{\mbox{{\bf{\sf                          T}}}_{L(A)}}
\newcommand{\ba}{\mbox{\bf {A}}}
\newcommand{\dof}{degrees of freedom}
\newcommand{\dofx}{degrees of freedom }
\newcommand{\pcp}{partially       confining       phase }
\newcommand{\pcps}{partially        confining         phases }
\newcommand{\ppm}{$A^b_{\mu,\;Peter}, A^b_{\mu,\;Paul}, \cdots ,
A^b_{\mu,\;N_{gen.}}$ }

\newcommand{\bb}{\mbox{\bf{B}}}
\newcommand{\bc}{\mbox{\bf    {C}}}
\newcommand{\bp}{\mbox{\bf{P}}}
\newcommand{\bx}{\mbox{\bf    {X}}}
\newcommand{\bz}{\mbox{\bf{Z}}}
\newcommand{\br}{\mbox{\bf{R}}}
\newcommand{\bu}{\mbox{\bf{U}}}

\newcommand{\bnul}{\mbox{\bf{0}}}
\newcommand{\gena}{\frac{\mbox{\boldmath$\lambda^{a}$}}{2}}

\newcommand{\genb}{\frac{\mbox{\boldmath$\lambda^{b}$}}{2}}
\newcommand{\genai}{\frac{\mbox{\boldmath$\lambda_{a_i}$}}{2}}
\newcommand{\genbi}{\frac{\mbox{\boldmath$\lambda_{b_i}$}}{2}}

\newcommand{\orb}{ORB_{large\;fluc\;in \cartprod_{i\in B(s^{\mu})}
{\cal M}_i}}

\begin{document}
\bibliographystyle{unsrt}
\pagenumbering{roman}

\begin{flushright}
\begin{tiny}
31 May 96 final version 
\end{tiny}
\end{flushright}

\addtolength{\textwidth}{1in}

\begin{center}

{\Large \bf Multiple Point Criticality, Nonlocality}\vspace{.3cm}\\{\bf and} \\
\vspace{.3cm}  {\Large  \bf
Finetuning in Fundamental Physics:}\\
\vspace{.3cm}  {\Large  \bf Predictions for Gauge Coupling Constants}\\
\vspace{.3cm} {\Large \bf gives $\alpha^{-1}=136._8 \pm 9$}
\vspace{40pt}

{\large \sl D.L. Bennett}

{\footnotesize  Ph.D. Thesis}\\
\vspace{.6cm}

{\small The Niels Bohr Institute \\ Blegdamsvej 17,\\ DK-2100 Copenhagen {\O},
Denmark}

\end{center}

\tableofcontents
\addtolength{\textwidth}{-1in}

\newpage
\pagenumbering{arabic}
\setcounter{page}{1}
\vspace{1.1in}

\begin{flushright}
\begin{tiny}
the1.tex 31 May 96 alf 
\end{tiny}
\end{flushright}

\section{Introduction}

The so-called principle of multiple point
criticality\cite{abel,bomb,nonabel,albu}
states that Nature - in for example a field theory -
seeks out values of action parameters that are located at the junction
of a maximum number of
phases in a phase diagram of a system that undergoes phase transitions. The
phases to which we here ascribe physical importance are normally regarded as
artifacts of a calculational regulation procedure.
This latter often takes the form of a lattice.
Contrary to the notion that a regulator is just a calculational device,
we claim that the consistency of any field theory in the ultraviolet limit
requires an ontological fundamental scale regulator. In light of this claim,
the ``lattice artifact'' phases of, for example, a lattice gauge theory
acquire the status of physically distinguishable fluctuation patterns at the
scale of the fundamental regulator that can have important consequences for
fundamental physics.

We have applied the principle of multiple point
criticality\cite{abel,nonabel,albu}
to the system of different (Planck scale) lattice
phases that can be provoked using a suitably generalised action in a
lattice gauge theory with a gauge group that is taken to be a Planck scale
predecessor  to the {\bf S}tandard {\bf M}odel {\bf G}roup (SMG) - namely
the $N_{gen}$-fold
Cartesian product of the SMG (here $N_{gen}$ denotes the
number of fermion generations). This gauge group, denoted as $SMG^{N_{gen}}$,
is referred to as the {\bf A}nti {\bf G}rand {\bf U}nified {\bf T}heory
(AGUT) gauge group. The number of generations $N_{gen}$ is taken to be three
in accord with experimental evidence; the AGUT gauge group $SMG^3$
has one $SMG$ factor for each family of quarks and leptons.
Ambiguities that arise under mappings
of the gauge group $SGM^3$ onto itself result in the
Planck scale breakdown
to the diagonal subgroup of $SMG^3$. The diagonal subgroup is isomorphic
to the usual standard model group.

In the context of a Yang-Mills lattice gauge theory, the principle of
multiple point
criticality states that Nature seeks out the point in the phase diagram
at which a maximum number of phases convene. This is the multiple point.
The physical values of the three SMG  gauge  couplings  at  the
Planck scale are predicted  to  be  equal  to  the  diagonal  subgroup
couplings corresponding to the multiple point action parameters of the
Yang Mills  lattice  gauge  theory  having  as  the  gauge  group  the
AGUT gauge group $SMG^3$.
It is indeed truly
remarkable that this prediction leads to agreement with experiment
to within  10 \% for the non-Abelian couplings and 5\% for
the U(1) gauge coupling. For the Abelian as well as the non-Abelian cases,
the deviation is of the order of the calculational uncertainty.

In order to  compare  Planck  scale  predictions  for  gauge  coupling
constants with experiment, it is of course  necessary  to  extrapolate
experimental  values  to  the  Planck  scale.  This  is  done  with  a
renormalization group extrapolation in which a ``desert'' scenario  is
assumed.
It should be emphasised that the prediction put forth here for the
gauge couplings using our model with the AGUT gauge  group  $SMG^3$
is  incompatible  with  the  currently  popular  $SU(5)$  or   $SU(5)$
super-symmetric grand unified models and is therefore to be regarded as
a rival to these.

In a more general context, the
{\bf M}ultiple {\bf P}oint {\bf C}riticality {\bf P}rinciple (MPCP)
is proposed as a fundamental
principle of Nature that also may be able to explain
essentially all of the well known
fine-tuning enigma in high energy physics\cite{nl1,nl2,bomb}.
Indeed, the conspicuous
values assumed by many physical
constants (e.g., fine-structure constants, the vanishing effective cosmological
constant, the smallness
of Higgs mass compared to Planck scale, $\Theta_{QCD}$) seem to coincide with
values that are obtained if one assumes that Nature in general seeks out
multiple
point values for intensive parameters.

Multiple point values of intensive parameters could be explained
- indeed, would be expected -
by having the presence of many coexisting phases
separated by first order transitions.
Phase coexistence would be
enforced for many combinations of universally fixed but not fine-tuned
amounts of extensive quantities. The intensive parameters  conjugate to
such extensive quantities would then have fine-tuned values.
And the higher the degree of first-orderness of the phase transition between
the coexisting phases, the greater the number of combinations of the extensive
quantities that could only be realized by having coexisting phases.
As a useful illustrative  prototype,  one  can
think of an equilibrium system consisting of a container within  which
there is water in all three phases: solid, liquid,  and  ice.  If  the
container  is  rigid  and  also  impenetrable  for  heat  and  water
molecules, we have  accordingly  the fixed amounts of  the  extensive
quantities energy, mole number of water, and volume. If these quantities are
fixed at values within some rather wide ranges, the fact that the heats of
melting, vaporisation, and sublimation are finite
forces the system to maintain the presence of all three phases
of water.
The  permanent   coexistence   of   all   three   phases   accordingly
``fine-tunes'' the values of the intensive parameters  temperature  and
pressure to those of the triple point of water.

However, having fixed amounts of such extensive quantities is tantamount to
having long range non-local interactions of a special type: these interactions
are identical between fields at all pairs of space-time points regardless
of the space-time distance between them.
Such omnipresent nonlocal interactions, which are present in
a very general form of a reparameterization invariant action\cite{frogniel},
would not be perceived as ``action at a distance'' but rather most likely
incorporated into our theory as constants of Nature. Hence one can
speculate\cite{nl1,nl2} that this mild form
of non-locality is the underlying explanation of Nature's
affinity for the multiple point.
We also speculate that nonlocal effects, described by
fields depending on two space-time points, may be responsible for the
replication of the fields in three generations\cite{nl2}. Such a nonlocal
mechanism would also triple the number of boson fields. As this feature is
inherent to the AGUT model, a tripling of boson fields is a welcome
prediction.


Originally, the MPCP was
suggested on phenomenological grounds in conjunction with the development
of methods for constructing phase diagrams for lattice gauge theories with
non-simple gauge groups. These methods have been used to implement the
MPCP in the most recent of a series of models that have been developed
with the aim of calculating the standard model gauge coupling constants.

Indeed, the theoretical calculation of the fine structure constant $\alpha =
1/137.0360...$ and  the  other  Yang  Mills  coupling  constants,  the
information content of which is identical  to  that  of  the  Weinberg
angle $\theta_{W}$ and the scale parameter  $\Lambda_{\overline{MS}}$,
continues to pose a challenge to be surmounted by theories at  a  more
fundamental level.

The Multiple Point Criticality
Principle (MPCP), developed by Holger Bech Nielsen and me,
is but one of the more  recent  results  in  a
long series of interwoven projects involving many people in which the
undisputed central figure and prime instigator has been my
teacher  and  colleague  Holger  Bech
Nielsen. In the course of the last 15 years or so, I have  also
had the privilege of working together with H.B. Nielsen and others on
several other projects some of which have been predecessors to the
multiple point criticality idea. Also included among the projects
to which I have contributed is some of the work on Random Dynamics. The
philosophy of Random Dynamics is the creation of  H.  B.  Nielsen\cite{roysoc}
and came to my attention over 15 years ago\cite{scot,gamma36,gamma37}.

\vspace{.3cm}

{\bf Section~\ref{hist}} reviews some earlier work belonging more or less
directly to the convoluted ancestry of the multiple point criticality idea.
Being somewhat historical, the presentation in this Section
is not logically streamlined but rather emphasises the inspirational
role that Random Dynamics has played in a number of contexts that
somehow are part of the lineage of the multiple point criticality
principle. {\bf Section~\ref{plscalineq}} describes a sort of forerunner to the
MPCP that has emerged
in a number of models considered  - namely an inequality relating gauge
couplings at the Planck scale,
the number  $N_{gen}$ of quark and lepton generations and the
critical couplings in a lattice gauge theory. {\bf Section~\ref{agut}}
describes the AGUT gauge group $SMG^{N_{gen}}$ which is inextricably
interwoven with the development of the multiple point criticality principle
and various Random Dynamics-inspired models. The AGUT gauge group

\beq SMG^{N_{gen}}\stackrel{def}{=}  \underbrace{SMG\times  SMG\times
\cdots \times SMG}_{N_{gen.}\mbox{ \footnotesize Cartesian product factors}}.
\eeq

\nin is assumed to be a more fundamental predecessor to the
phenomenologically established standard model group. The latter
arises as the diagonal subgroup of $SMG^{N_{gen}}$ that survives
the Planck scale breakdown of $SMG^{N_{gen}}$. In particular,
this Section describes the way that inverse squared gauge couplings are
enhanced by a $N_{gen}$-related factor in conjunction with the
breakdown to this diagonal subgroup. By definition, the
diagonal subgroup of $SMG$ has excitations of the (group valued) lattice link
variables that are identical for each of the $N_{gen}$ Standard Model Group
factors of $SMG^3$.
The philosophy  of Random Dynamics is sketched in {\bf Section~\ref{phil}}.
The Random Dynamics approach is used in the rather lengthly
{\bf Section~\ref{ftg}}
to ``derive'' gauge symmetry in the context of a field theory glass. This
Section develops and relates many ideas the rudiments of which have been
presented
by H.B. Nielsen in lectures given at his inspiring
course series ``Q.C.D. etc''. This legendary course is a veritable
forum of new
ideas in physics.
The so-called ``confusion'' mechanism by which the AGUT gauge group $SMG^3$
breaks down at the
Planck scale to its diagonal subgroup is
reviewed in {\bf Section~\ref{confbrkdn}}. This happens as a result of
ambiguities that
arise under group automorphic symmetry operations.
The AGUT gauge group is viewed as the final link in a
chain of increasingly robust SMG predecessors selected by Random Dynamics
in going to lower and lower energies at roughly the Planck scale.
In {\bf Section~\ref{kaluza}}, a model with a string-like regularization
in a Kaluza-Klein space-time at the fundamental scale is used to derive the
inequality of Section~\ref{plscalineq}.

{\bf Section~\ref{pmpc}} is devoted to the {\bf M}ultiple {\bf P}oint
{\bf C}riticality {\bf P}rinciple (MPCP).
Though the MPCP was from the start formulated in the context of a lattice
gauge theory in order to predict standard model gauge coupling constants,
generalisations\cite{albu,glas94,nl1,nl2,bomb} in the formulation and
applicability of this principle have been considered.
Multiple point criticality as a way of explaining  ``fine-tuned''
quantities in Nature is  discussed in {\bf Section~\ref{mpcfintn}}. This
fine-tuning mechanism entails having universally fixed amounts of
extensive quantities conjugate to intensive quantities
(e.g., fine-structure constants); these latter have a finite probability for
being fine-tuned that increases as the number of
combinations
of extensive variables that cannot be realized as a single phase
becomes larger.  However,
having universally fixed amounts  of
extensive quantities actually implies having a mild form of non-locality
that is analogous to that inherent to a micro-canonical ensemble  in
statistical mechanics: an inherent feature of a micro-canonical ensemble is the
introduction of long range correlations that strictly speaking breaks
locality. However  it  has
been shown \cite{frogniel} that  non-locality of the type analogous to that
introduced  by  the
assumption of a micro-canonical ensemble is harmless insofar as it does
not lead to experimentally observable violations of locality. Multiple
point criticality as related to the problem of fine-tuning and the
presence of non-locality is discussed at some length in
{\bf Section~\ref{fintnnonloc}}.





{\bf Section~\ref{phasecon}} deals with the phases distinguishable
at the scale of the lattice in a lattice gauge theory
implementation of the
MPCP. The junction in the phase diagram to which the maximum
number of such phases are adjoined is sought
out for the purpose of determining gauge coupling constants.
Such phases   are really what
would normally be regarded as
``lattice artifacts''. However, in light of our philosophy that a lattice
is one of perhaps many ways of implementing what is assumed to be the
necessity of a ontological Planck scale regulator,
these ``artifact phases'' acquire a physical meaning.
These phases  are  governed  by  which  {\em  micro}
physical fluctuation patterns yield the maximum  value  of  $\log
Z$ where $Z$ is the partition function.
Qualitatively  different  short  distance  physics
could consist  of
different distributions of group elements along various
subgroups  for different regions of
(bare)
plaquette action parameter space. For example, if in going from one
region of parameter space to another, the
correlation length goes from being shorter than the lattice constant to
being of the order of several lattice constants, we would
see a transition from a confining to a ``Coulomb phase'' if the determination
of phase was done at the scale of the lattice. Even if such different
lattice scale phases become indistinguishable in going to long
wavelengths\footnote{E.g., non-Abelian groups for which (if matter fields are
ignored as is the case here) there are no
long range
correlations (corresponding to finite glue-ball masses)
in going to sufficiently large distances.}
because such phases turn out to be confining,
this in no way precludes {\em physical} significance for lattice-scale  phases
that come about as a result of a {\em physically existing} fundamental
regulator.
At the scale of what is
assumed to be the fundamental
regulator, there is a distinguishable ``phase'' for each invariant subgroup
$H$ of a non-simple gauge group $G$ including {\em discrete} (invariant)
subgroups.
Any invariant subgroup $H$ of the gauge group $G$ labels
a confinement-like phase 
the defining feature of which
is that Bianchi variables
(i.e., the variables that must coincide with the unit element in
order that the Bianchi identity be fulfilled) have a distribution
on the group such that essentially all  elements of $H$ would be
accessed by quantum fluctuations if the fluctuations in the
plaquette variables did not have to be correlated in such a way
as the fulfil the Bianchi identities. So a phase confined w.r.t.
$H$ has appreciable quantum fluctuations along $H$ while the
degrees of freedom corresponding to the cosets of the factor
group $G/H$  have a distribution peaked at the coset
for which the identity of $G$ is a representative. These latter
degrees of freedom are said to have Coulomb-like behaviour. Such a phase
is often referred to as a {\em \pcp}.
The classification of these \pcps is the subject of
{\bf Section~\ref{phclass}}.


{\bf Section~\ref{secgenact}} addresses the problem of distinguishing
all (or some chosen set)  of  the  possible  \pcps
(with each phase  corresponding  to  an  invariant  subgroup $H$). This
requires using
a class  of  plaquette  actions  general
enough to provoke  these phases. Some general features of such actions
are examined in {\bf Section~\ref{genidea}}.
Phase diagrams for non-simple gauge groups
suitable for seeking the multiple point are considered in
{\bf Sections~\ref{iftyofph}} and {\bf \ref{practical}}.
{\bf Section~\ref{seekingthemp}} outlines problems encountered in implementing
the principle of
multiple point criticality in the case of $U(1)$ (or $U(1)^3$)
as compared to the simpler case of the non-Abelian subgroups of the
standard model. One of these problems,
related to the ``Abelian-ness'' of $U(1)$, is a result of the interactions
between the $N_{gen}$ replicas of $U(1)$ in the AGUT gauge group $SMG^3$. In
the roughest approximation, these interactions result in a weakening of the
diagonal subgroup coupling of $U(1)\in SMG^3$ by a factor of
$\frac{1}{2}N_{gen}(N_{gen}+1)=6$
instead of the weakening factor $N_{gen}=3$ that applies to the non-Abelian
subgroups (for which such interaction are not gauge invariant).

In constructing actions suitable for implementing the MPCP for the purpose of
determining gauge couplings, we reach in
Section~\ref{mpcnonabel}  a point in the development where it is
expedient to consider separately the Abelian and non-Abelian
couplings. The  simpler case of the non-Abelian couplings is treated
first in Section~\ref{mpcnonabel} followed by the Abelian case in
Sections~\ref{mpcabelmeth} and \ref{calculation}.
These latter two sections contain the essence of very recent work\cite{abel}.

{\bf Section~\ref{mpcnonabel}} deals with the determination of the multiple
point
couplings (with the AGUT gauge group
$SMG^3$) for the non-Abelian subgroups of the $SMG$.
After writing down some formalism in {\bf Section~\ref{secgenrem}}, the {\em a
priori}
lack of universality of the model is discussed in
{\bf Section~\ref{universality}}
inasmuch as the phase transitions at the multiple point are typically  at
first  order.  However,   our restriction on the form of the plaquette action
nurtures the  hope  of at least an approximate universality.
{\bf Section~\ref{modmantonact}} develops a modified Manton action
that  leads  to  distributions
$e^{S_{\Box}}$  of  group-valued  plaquette  variables  consisting  of
narrow maxima centred at elements $p$ belonging to  certain  discrete
subgroups of the centre of the gauge group. The action at these  peaks
is then expressed as truncated Taylor expansions around  the  elements
$p$.  With  this  action   ansatz,   it   is   possible   to   provoke
confinement-like or  Coulomb-like  behaviour  {\em independently}
(approximately at least) for the 5 ``constituent'' invariant subgroups
$\bz_2,\;\bz_3,\;U(1),\;  SU(2),$  and  $SU(3)$  of  the  SMG  which
``span'' the set of ``all'' invariant subgroups\footnote{Here we for the
most part do not
consider the infinity of invariant subgroups  $\bz_N\subset
U(1)$ for $N>3$.} of the SMG. In our approximation, the mutually
un-coupled  variation  in
the distributions along these 5 ``constituent'' invariant subgroups is
accomplished    using    5    action    parameters:    3    parameters
$\beta_1,\;\beta_2,$ and  $\beta_3$  that  allow  adjustment  of  peak
widths in the $U(1),\;SU(2),$ and $SU(3)$ directions on the group
manifold {\em and} 2
parameters $\xi_2$ and $\xi_3$ that  make  possible  the
adjustment of the relative heights of the peaks centred  at  elements
$p\in \mbox{span}\{ \bz_2,\bz_3 \}$.
A lengthly digression in {\bf Section~\ref{digression}} develops techniques for
constructing phase diagrams for non-simple gauge group in the simpler
approximation in which phases
solely confined w.r.t discrete subgroups are not included. Methods for
constructing phase diagrams also
having this latter type of phase (which are necessary for having a
multiple point) are then considered in {\bf Section~\ref{usingdiscrete}}.
Correction due to quantum fluctuations are considered in
{\bf Section~\ref{quantumcorr}}.
The first four {\bf Appendices~\ref{apphaar}, \ref{appcorr}, \ref{appnoncom}}
and {\bf \ref{appint}}, which are not
essential to the continuity of the thesis, deal with various
improvements to the methods of constructing approximate phase diagrams
considered in Section~\ref{mpcnonabel}. In Appendix~\ref{appint} for example,
interactions between the ``constituent'' invariant subgroups mentioned
above are considered.

{\bf Section~\ref{mpcabelmeth}} considers the gauge group
$U(1)^3$ that is used as an
approximation to the AGUT $SMG^3$ for the purpose of determining the
$U(1)$ gauge coupling. The normalisation
problems with $U(1)$ are considered in {\bf Sections~\ref{onlyz6}, \ref{nor2}}
and {\bf \ref{resolve}}.
Phase diagrams for the
gauge group $U(1)^3$
in which we can seek out multiple point parameter values are needed.
In {\bf Section~\ref{sec-portray}}, a
formalism is developed that allows us to seek multiple point parameter
values by
adjusting the metric (which amounts to adjusting the parameters of
a Manton action) in a $N_{gen}$-dimensional space upon which
is superimposed an hexagonally symmetric lattice of points identified
with the identity
of $U(1)^3$. The hexagonal symmetry takes into account the allowed
interactions between the $N_{gen}=3$ $U(1)$ factors of $U(1)^3$. Using this
formalism, two approximative methods of determining
phase boundaries are developed: the independent monopole approximation
({\bf Section~\ref{moncon}}) and the group
volume approximation ({\bf Section~\ref{grvolapprox}}).
These describe respectively phase transitions that are purely
second order and strongly first order.

{\bf Section~\ref{calculation}} is devoted to calculations
where we interpolate between the
extreme situations described by the group volume and independent monopole
approximations. This interpolation is done by calculating the discontinuity
$\Delta \gamma_{eff}$ at the multiple point in an effective coupling
$\gamma_{eff}$ (introduced in {\bf Section~\ref{secgamef}}).
In {\bf Section~\ref{deltagam}} it is seen that the dominant contributions to
$\Delta \gamma_{eff}$ are due to
multiple point transitions between phases that differ by the confinement
of discrete subgroups (rather than continuous subgroups). The
calculated $\Delta \gamma_{eff}$ reflects the degree of first-orderness
of these transitions. As a result of including this effect,
 the weakening factor $N_{gen}(N_{gen}+1)/2=6$ is seen in
{\bf Section~\ref{enhance}}
to increase to about 6.5. The quantity $\Delta \gamma_{eff}$ is also used
(together with $\gamma_{eff}$) to calculate the continuum $U(1)$ coupling
corresponding to the multiple point of a single $U(1)$ in
{\bf Section~\ref{contcoup}}. In the tables at the end of
{\bf Section~\ref{contcoup}}, this value of the continuum $U(1)$
coupling
is multiplied by the weakening factor of about 6.5
(calculated in Section~\ref{enhance})
to get our prediction for
the value of the running $U(1)$ coupling at the Planck scale.

{\bf Section~\ref{conclusion}} presents the results from multiple point
criticality for all
three gauge couplings. Values are given at the Planck scale as well as
at the scale of $M_Z$. The latter are obtained
using the
assumption of the minimal standard model in doing the renormalization group
extrapolation.
In the case of $U(1)$, a  number of slightly different values are
presented that
reflect the differences that arise due to approximations
that differ in how some details are treated.
In presenting what we take to be the ``most correct'' result, we compute the
uncertainty from the deviations arising from  plausibly correct
ways of making distinctions in how different discrete subgroups
enter into the calculation of $\Delta \gamma_{eff}$.
The value of $\alpha^{-1}$ predicted from
multiple point criticality is calculated to be $136.8\pm9$. This is to be
compared with the experimental value of $137.036\cdots$. The thesis  ends with
some concluding remarks.

Although the presentation of  the  current  state  of  the MPC  model  for
predicting gauge coupling constants and  the  techniques  devised  for
implementing it will constitute the major part of this thesis, earlier
work will be reviewed and the most recent developments in ongoing work
will be included.

\newpage

\section{History of the project including the inspirational role of Random
Dynamics}\label{hist}

\subsection{A Planck scale inequality relating gauge coupling to number of
fermion generations}\label{plscalineq}

The evolution of the MPC model for predicting
the Standard Model gauge coupling constants is inseparably tied together with
the ideas of Random Dynamics as well as
speculations as to the origins of the SMG and the number $N_{gen}$ of
generations of fermions.
The work preceding the MPC principle has involved
a number of models all of which lead to or at least suggest features of
an  {\em inequality} relating  the gauge coupling constants at the Planck
scale,
the number $N_{gen}$ of quark and lepton generations, and the critical
values of inverse squared gauge couplings $\beta_{crit}$
in a regularised (i.e., latticised in most cases) gauge theory:

\beq \beta_{experimental}(\mu_{Planck})\geq N_{gen}\beta_{crit}\label{ineq}\eeq

This inequality, originally suggested on phenomenological
grounds\cite{ngen2,kk1,kk2,ngen1,kk3,prot}, is, when supplemented with
arguments for why it is
realised in Nature as an {\em equality},
the forerunner of the multiple point criticality idea. Various
``derivations'' of this inequality
share some common and interrelated features (that, depending on
which model is considered, are used as assumptions or show up as
consequences):

\begin{enumerate}

\item The phenomenologically observed Standard Model Group (SMG)
is,  at the fundamental
($\approx$ Planck) scale, replicated a number of times. The Cartesian product
of these replicas of the SMG, assumed to be a predecessor to
the usual SMG, breaks down at the Planck scale to the diagonal subgroup
of the Cartesian product.

\item In order to be phenomenologically  relevant, the replicas of the
Standard Model Group at the fundamental scale must have coupling constants
that are on the weak coupling side of the ``critical'' value in order to
avoid a confinement-like phase already at the fundamental scale. This amounts
to an upper bound on allowed Planck scale couplings.

\item The Planck scale criticality referred to in {\bf 2.}
pertains to transitions between  ``phases''
that conventionally would be regarded as artifacts of the regularization
procedure used (which, for almost all the models considered up to now,
means a lattice).
Ascribing physical significance to such ``phases'' is tantamount to
assuming  the existence of  a regulator as an intrinsic property
of fundamental scale physics.

\item The upper bound on Planck couplings for the SMG replicas
appears to be saturated; i.e., couplings assume the largest
possible (i.e., critical) values that are
consistent with avoiding confinement. This feature, which may be necessary
in order to avoid a Higgsed phase at the Planck scale, is intrinsic  to the
idea of MPC.

\item It is assumed that the more fundamental Cartesian product gauge
group assumed in {\bf 1.} above contains (at least) one SMG factor
for each generation of quarks and leptons.
Collaboration of the $N_{gen}$ replicated  SMG factors near the Planck scale
reduces the gauge symmetry from that of the
Cartesian product group to that of the
diagonal subgroup (which of course is isomorphic to the usual SMG).
Assuming the validity of the saturation
property in {\bf 4.}  above for each SMG factor in the Cartesian product
gauge group, this spontaneous reduction in the gauge
symmetry is accompanied by an enhancement in the values of
the three SMG inverse squared gauge couplings of the diagonal subgroup
by a factor equal to the number $N_{gen}$ of quark and lepton generations.
This was originally suggested on  phenomenological
grounds: early on the observation was made that the
magnitude of the non-Abelian gauge coupling constants is of the order one
divided by the square root of the number of generations $N_{gen}$ provided
that unit coupling strength is taken to be that at the transition between
the confined and Coulomb phases in the mean field approximation.

\end{enumerate}

\subsection{The \underline{A}nti \underline{G}rand \underline{U}nified
\underline{T}heory Gauge Group (AGUT Gauge Group) SMG$^3$}\label{agut}

As mentioned several times already, a central feature that emerges or
that at least is suggested in the context of
various different models is that
the phenomenologically
well-established Standard Model Group SMG stems from
a more fundamental predecessor  referred to as the
``anti-grandunified theory'' (AGUT)
gauge group and denoted by $SMG^{N_{gen}}$. This group is the $N_{gen}$-fold
Cartesian product of essentially SMG factors with one SMG factor for each
of the $N_{gen}$ generations of quarks and leptons.
In terms of the Lie algebra

\begin{small}

\begin{equation}
SMG^{N_{gen}}\stackrel{def}{=} \end{equation}

\[ =\underbrace{U(1)\times U(1)\times \cdots \times  U(1)}_{N_{gen}}\times
\underbrace{SU(2)\times               SU(2)\times               \cdots
\times SU(2)}_{N_{gen}}\times
\underbrace{SU(3)\times        SU(3)\times        \cdots        \times
SU(3)}_{N_{gen}}. \]

\end{small}

\nin The
identification of the number of SMG factors in the Cartesian product
$SMG^{N_{gen}}$ with the number of families $N_{gen}$ allows the possibility of
having
different gauge quantum numbers for the $N_{gen}$ different families.
The integer  $N_{gen}$
designates the number of  generations and is taken  to
have the value $N_{gen}=3$  in  accord  with  experimental  results.

In this work, an alternative to the usual Standard Model {\em Group}
$U(1)\times SU(2)\times SU(3)$ will be used; here the
Standard Model Group (SMG) is defined as

\begin{footnotesize}
\beq SMG\stackrel{def.}{=}S(U(2)\times            U(3))\stackrel{def.}{=}
\left\{\left.\left( \begin{array}{cc} \bu_2 & \begin{array}{ccc} 0 & 0
& 0 \\ 0 & 0 & 0 \end{array} \\ \begin{array}{cc} 0 & 0 \\ 0 & 0 \\  0
& 0 \end{array} & \bu_3 \end{array} \right ) \right|  \begin{array}{l}
\bu_2\in   U(2),   \\   \bu_3\in   U(3),    \\    \mbox{det}\bu_2\cdot
\mbox{det}\bu_3=1 \end{array} \right\} \eeq  \end{footnotesize}

\nin
This {\em group} is suggested by the representation spectrum  of  the
standard  model\cite{michel}; it has of course the same Lie algebra as
the  more  commonly  used group $U(1)\times SU(2)\times SU(3)$.

The usual  standard model  description  of
high energy physics comes about as  the  diagonal  subgroup of $SMG^3$:

\begin{equation}      SMG^3      \stackrel{breakdown}{\longrightarrow}
(SMG^3)_{diag.\;subgr.}\stackrel{def.}{=}\{ (g,g,g)\mid g  \in  SMG  \}
\end{equation}

\nin  resulting
from the Planck scale \footnote{The choice of the Planck scale for the
breaking of the (grand) ``anti-unified'' gauge group  $SMG^3$  to  its
diagonal subgroup is not completely arbitrary insofar as  gravity  may
in some sense be critical at the Planck scale. Also,  our  predictions
are rather insensitive to  variations  of  up  to  several  orders  of
magnitude in the choice of energy at which the Planck scale is fixed.}
breakdown of the gauge group $SMG^3$. The diagonal subgroup of $SMG^3$
is of course isomorphic to the SMG.  The  breakdown to the diagonal
subgroup
can come about due to ambiguities  that  arise  under  group  automorphic
symmetry operations. This is usually referred to  as  the  ``confusion''
mechanism\cite{conf1,conf2,gaeta}.

The breakdown of the  group  $SMG^3$  to  the  diagonal  subgroup  has
consequences\cite{rajpoot} for the SMG gauge couplings that we now briefly
describe.
Recalling that the diagonal  subgroup of $SMG^3$ corresponds  by  definition
to identical excitations of the $N_{gen}=3$ isomorphic gauge
fields (with the gauge couplings absorbed)
and using the names $Peter$, $Paul,\cdots $  as  indices  that
label the $N_{gen.}$ different isomorphic Cartesian product factors of
$(SMG)^{N_{gen.}}$, one has\footnote{As it is $gA_{\mu}$ rather
than $A_{\mu}$ that appears in
the (group valued) link variables $u\propto e^{iagA_{\mu}}$, it is the
quantities $(gA_{\mu})_{Peter}$, $(gA_{\mu})_{Paul}$, etc. which are equal in
the diagonal subgroup.}

\begin{equation} (g{\cal A}_{\mu}(x))_{Peter} =(g{\cal A}_{\mu}(x))_{Paul} =
\cdots  =   (g{\cal   A}_{\mu}(x))_{N_{gen.}}   \stackrel{def.}{=}(   g{\cal
A}_{\mu}(x))_{diag.};\label{diagconfig} \end{equation}

\nin this has the consequence that the common $(gF_{\mu \nu})^2_{diag}$
in each term of the Lagrangian density
for $(SMG)^{N_{gen.}}$ can be factored out:

\begin{equation}  {  \cal
L}=-1/(4g_{Peter}^2)(gF_{\mu                      \nu}^a(x))^2_{Peter}
-1/(4g_{Paul}^2)(gF_{\mu      \nu}^a(x))^2_{Paul}       -       \cdots
-1/(4g_{N_{gen.}}^2)(gF_{\mu \nu}^a(x))^2_{N_{gen.}} \end{equation}

\begin{equation} =(-1/(4g_{Peter}^2)  -1/(4g_{Paul}^2)-\cdots
-1/(4g_{N_{gen.}}^2)) \cdot (F_{\mu \nu}^a(x))^2_{diag}  =-1/(4g_{diag}^2)
\cdot (gF_{\mu\nu}^a(x))^2_{diag}. \end{equation}

The inverse squared couplings for the
diagonal subgroup is  the sum of the inverse  squared  couplings
for each of the $N_{gen.}$ isomorphic  Cartesian  product  factors  of
$(SMG)^3$. Additivity in the inverse squared couplings in going to the
diagonal subgroup  applies  separately  for  each  of  the invariant
Lie
subgroups\footnote{For $U(1)$, a modification is  required.}
$i \in \{SU(2), SU(3)\}\subset SMG$. However, for $U(1)$ it  is  possible
to    have    terms    in    the    Lagrangian     of     the     type
$F_{\mu\nu}^{Peter}F^{\mu\nu\;Paul}$  in  a   gauge   invariant   way.
Therefore it becomes more complicated as to how one should  generalise
this notion of additivity. Terms of this type can  directly  influence
the $U(1)^3$  continuum  couplings\footnote{In  seeking  the  multiple
point for $SMG^3$, one is lead to seek criticality separately for  the
Cartesian  product  factors  as  far  as  the  non-Abelian  groups  are
concerned. For $U(1)$, one  should  seek
the multiple point for the whole group $U(1)^3$ rather than  for  each
of the $N_{gen.}=3$ factors $U(1)$ separately.  The  reason  for  this
complication concerning Abelian groups  (continuous  or  discrete)  is
that these have subgroups and thereby invariant subgroups  (infinitely
many for continuous Abelian groups) that cannot be regarded as being a
subgroup of one of the $N_{gen.}$ factors of $SMG^3$  or  a  Cartesian
product of such subgroups. A phenomenologically desirable  factor of
approximately ``6'' is indicated for the ratio
$\frac{\alpha_{crit.,\;U(1)}}{\alpha_1(\mu_{_{Pl.}})}$          (where
$\alpha_{crit.,\;U(1)}$ is the critical coupling for the  gauge  group
$U(1)$) instead of the  factor  ``3''  (from  $N_{gen.}$)  that  would
naively be expected for this ratio by analogy to the  predictions  for
the non-Abelian couplings.}. But for the non-Abelian couplings we simply
get

\begin{equation}
\frac{1}{g_{i,diag}^2}=\frac{1}{g_{i,Peter}^2}+\frac{1}{g_{i,Paul}^2}+\cdots
+\frac{1}{g_{i,N_{gen}}^2}     \;\;(i\in     \{SU(2),\;SU(3)\})      .
\label{confadd} \end{equation}

Assuming that the inverse squared
couplings for a given $i$ but different labels  $\{
Peter, Paul, \cdots ,N_{gen.} \}$ are all driven to the  multiple
point in accord with the principle of multiple point criticality (discussed
at length in a later section),
these $\{ Peter, Paul, \cdots,N_{gen} \}$ couplings
all become equal to the multiple point value $g_{i,multi.\; point}$; i.e.,:

\begin{equation}
\frac{1}{g_{i,Peter}^2}=\frac{1}{g_{i,Paul}^2}=\cdots
=\frac{1}{g_{i,N_{gen}}^2}=\frac{1}{g_{i,\;multi.\; point }^2}.
\end{equation}

It is seen that the inverse squared coupling $1/g_{i, \;  diag}^2$  for  the
$i$th subgroup of the diagonal subgroup is enhanced  by  a
factor $N_{gen}$ relative to the corresponding subgroup $i$ of each of
the $N_{gen}$ Cartesian product  factors  $Peter$,  $Paul,
\cdots, N_{gen.}$ of $(SMG)^{N_{gen.}}$:

\begin{equation}
\frac{1}{g_{i,diag}^2}=\frac{N_{gen}}{g_{i,\;multi.\; point }^2}.
\end{equation}

It is this weakening of the coupling for each of the  subgroups  $i\in
\{SU(2),SU(3)\}$ of the  diagonal  subgroup  (i.e.,  the  SMG)  that
constitutes the main role of the anti-unification scheme in our model.
Anticipating the later discussion of the role of the multiple point,
we  point  out  prematurely  that  while  it  is   the
$g_{i,\;multi.\; point }$ (i.e., $i=SU(2)$ or $SU(3)$) which are to be
identified with  the  critical  values  (at  the  multiple  point)  of
coupling  constants  for  the  bulk  phase  transition  of  a  lattice
Yang-Mills theory with gauge group $i$, it is the
$g_{i,\;diag.}= g_{i,\;multi.\; point }/\sqrt{N_{gen.}}$ that, in  the
continuum  limit,  are  to  be  identified  with   the   corresponding
experimentally  observed  couplings   extrapolated   to   the   Planck
scale\cite {kim,amaldi}.

The validity of the principle of multiple  point  criticality
together with  the assumption of the AGUT gauge group $SMG^{N_{gen}}$
as the immediate predecessor to the usual SMG at the Planck scale
can be claimed to be justifiable {\em a  posteori}  alone  on
the grounds of phenomenological success in predicting gauge coupling
constants. Even though the idea of multiple point criticality can stand
alone as a model that predicts gauge coupling constants and as a
plausible candidate for explaining fine-tuned quantities in Nature,
I think it is important to emphasise the important
inspirational role that Random Dynamics has played in  developing the various
models that culminated in the principle of multiple point criticality.


\subsection{The philosophy of Random Dynamics}\label{phil}

The idea behind Random Dynamics is outlined in this section.
In the following three Sections~\ref{ftg}, \ref{confbrkdn} and \ref{kaluza}
three representative applications of Random Dynamics have been chosen from
various models that motivate various aspects that have been important in
arriving at what today refer to as the principle of multiple point
criticality with AGUT gauge group.
In Section~\ref{ftg} we present
a field theory glass model for Planck scale physics that, in keeping with the
idea
of Random Dynamics, is proposed as being sufficiently general and unrestricted
so as to have a good chance of yielding low energy physics (LEP) in the long
wavelength limit. This Section ends with arguments that indicate that the
phenomenologically suggested
realisation of the inequality (\ref{ineq}) as an equality can be understood
in the context of a field theory glass.
In Section~{\ref{confbrkdn}, a brief review
is given of a possible
mechanism - the {\em confusion mechanism} - for the Planck scale breakdown
of the AGUT gauge  group $SMG^3$ to its diagonal subgroup.
In Section~\ref{kaluza} we discuss a model in which
experimental couplings extrapolated to
the Planck scale suggest constraints on the volume of the compactification
space in a model with
a Kaluza-Klein space-time. This leads to a constraint on the value of $N_{gen}$
as well as the scale at which grand-unification, if realised in Nature,
could take place.
All three models
involve arguments suggested by the ideas of Random Dynamics.

\vspace{.3cm}

The idea behind the Random Dynamics principle is  that  at  very  high
energies  (i.e.,  $\geq$  Planck  energies),  almost  any  model   for
fundamental  physics  that   possesses   sufficient   complexity   and
generality will in the low energy limit (i.e., at energies  accessible
to experiment) yield physics as we know it. In other words, it is  the
constraints dictated by the process of taking the low energy limit (of
a fundamental model for supra-Planck scale physics) that are  decisive
in determining the form of low energy physics. Taking  this  viewpoint
means  that  essentially  any  (e.g.,  randomly  chosen)   model   for
supra-Planck scale physics will be shaped into low energy  physics  as
we know it because the process of  going  to  low  energies  ``filters
away'' all  features  of  any  supra-Planck  scale  model  except  the
features that characterise low energy physics. These  latter  features
``survive'' a sort of selection process that is
                                                   assumed to be inherent
in taking the long wavelength limit of  any  fundamental  theory.
The assertion that we get the same low energy physics (LEP) for almost
{\em }any sufficiently general set of  assumptions for
fundamental scale physics suffices as a starting point for taking a long
wavelength limit is  equivalent to deriving LEP from almost no
assumptions about fundamental scale physics. This is because
few if any assumptions are so important that they couldn't be excluded from
some set of sufficiently general assumptions that would also yield LEP
in the long wavelength limit.

\subsection{Gauge symmetry from a field theory glass}\label{ftg}

Here a {\em field theory glass}\cite{randyn2,pc} is taken as the
starting point for a Random Dynamics ``derivation'' of the gauge symmetry
of the Standard Model description of LEP physics.
By examining a field
theory glass, one hopes to find as a generic possibility the approximate
gauge symmetry needed as the starting point for the
the FNNS
\footnote{F\"{o}rster, Nielsen, Ninomia and Schenker;
the remarkable result of the FNNS mechanism is illustrated by a simple
example using an approximate $U(1)$ lattice gauge theory: even for a action
having an explicit gauge breaking term $S_{g.b.}$ (in addition to
a gauge invariant term $S_{g.i.}$):
for an action of the
form

\beq S_{fund.}=S_{g.\;i.}+S_{g.\;b.}=\beta \sum_{\Box} Re U(\Box)  +\kappa
\sum_{\link} Re U(\link)\eeq

\nin there is a whole range of values for $\beta$ and $\kappa$ for which
$\beta$ is large enough to avoid confinement and $\kappa$ is small enough
so as not to bring about a global breakdown of gauge symmetry due to
Higgsing.}
gauge symmetry exactification mechanism\cite{fnns}. Subsequently, we shall use
the
formal technique used in demonstrating the FNNS mechanism  to argue
that in a statistical sense MPC offers the  best chance for reconciling a
conflict between
on one hand avoiding confinement and the other hand avoiding Higgsing.

The field theory  glass is envisioned as
residing in a discretized space-time. In other words,  each  physically
realisable space-time event corresponds to a site in a  very  irregular
lattice. As regards  Lorentz invariance, which is obviously absent in
this discretized space-time, the hope is that it can be recovered in the
long wavelength limit.

Denoting the fundamental set of such space-time points
as  $\{j\}$,  we
define a generalised field $\phi$ that for each site $j$ takes  values
on  a  site-associated  manifold  ${\cal  M}_j$.  The  site-associated
manifolds  ${\cal  M}_j$,  each  of  which  is  presumed  to  be
individually very
complicated, depend on $j$ in a quenched random way.  The  generalised
field $\phi$ is described by the mapping \cite{randyn1,randyn2}

\beq \phi: \{ j \} \rightarrow \cup_j \{{\cal M}_j  \}\;\mbox{  with  the
restriction}\; \phi(j) \in {\cal M}_j. \eeq

Having the field theory glass \dof, we want now to define  a  very  general
action subject to the
constraint  that  (semi)locality  is  to  be   retained.   This   is
accomplished by defining the action to be  additive  in  contributions
from small quenched randomly chosen  space-time regions
(generally overlapping) distinguished here
by  the  index
``$r$'': $S[\phi]=\sum_r S_r(\phi \mid_r)$ where $\phi \mid_r$ denotes
the restriction of the mapping  $\phi$  to  the  sites $i\in r$.
Each regional contribution $S_r$ to the action is a mapping $S_r$:
$\times^{cart\;prod}_{j\in r} {\cal M}_j \rightarrow \br$ that depends
on the region
$r$ in a quenched random way. This could be accomplished by assigning
to each region $r$  a random set of expansion coefficients for the
action $S_r$  expressed in terms of a (complete) system of orthogonal
functions.
The quantum field theory based on this quenched random structure
is  what  we  refer  to  as a  field theory glass.

It is instructive to think of how one might in principle use a computer
simulation procedure to study the way in which gauge symmetry at LEP might
evolve from a field theory glass model for fundamental scale physics. To
begin such a computer study, one could proceed in the following manner.

\begin{enumerate}

\item Set up a random set $\{j\}$ of points $j$ in 4 dimensions that are the
space-time points that exist in the theory.

\item Set up a field $\phi$ on these space-time points such that the values
that the field $\phi$ can take at the space-time point $j$ lie on a randomly
chosen manifold ${\cal M}_j$ that is assigned to the space-time point $j$.

\item Choose in a random way overlapping regions $r$ of space-time points
$i\in r$. The overlap is necessary in order to have correlations between
space-time regions.

\item Assign in a random way an action $S_r$ (i.e. a set $\{\beta_r\}$
of action parameters) to each region $r$ such that $S_r$ depends only on
the values $\phi(i)$ of the field $\phi$ that correspond to $i\in r$.
This means that the action $S=\sum_r S_r$ is
                                    semi-local in the sense that
the total action $S$ is a sum of possibly non-local action contributions $S_r$
defined on small localised regions $r$; it is only within the small localised
regions $r$ that there can be non-locality.

Such a very random action $S=\sum_rS_r$ could a priori be taken as an
expansion in some system of orthogonal functions with a set $\{\beta_r\}$ of
expansion coefficients that, for each region $r$, is chosen as a
quenched random set.

\end{enumerate}

It is important to emphasise that the above features of the model
(i.e., sites $\{i\}$,  field  target
spaces $\{{\cal M}_i \}$,  local action regions $r$, and the parameters
$\{\beta_r\}$
associated with each region $r$ that define the local action  contribution
$S_r$) are quenched
random; that is, they are beforehand randomly fixed once and for all.
Accordingly they are held constant  under  the  functional  integration
used to get the partition function.

\begin{itemize}

\item Parallel to the discussion of the general case of the field theory
glass, a very restricted form of a field theory glass will also be considered
as a concrete example.
In this very special case, let the set of \fspsx $\{i\}$ coincide with
the middle of the links of a hyper-cubic lattice; let $ {\cal M}_i=U(1)
\stackrel{-}{\simeq}S^1\;\; \forall\;\; i\in \{j\}$. Let each
of the (non)local action regions $r$ include just the four link-centred $i$
of a simple plaquette. Finally, let us assume that the (semi)local action
contribution  $S_r$  defined on each
(non)region $r$ is of the simple identical form

\beq S_r=\beta Re U(\Box)+\kappa\sum_{\link\;\in\Box} Re U(\link);
\label{actgg}\eeq

\nin i.e., we assign the {\em same}  quenched random set
$\{\beta,\kappa \}$ to each (non)local action region $r$.

\end{itemize}

\begin{figure}
\centerline{\epsfxsize=\textwidth \epsfbox{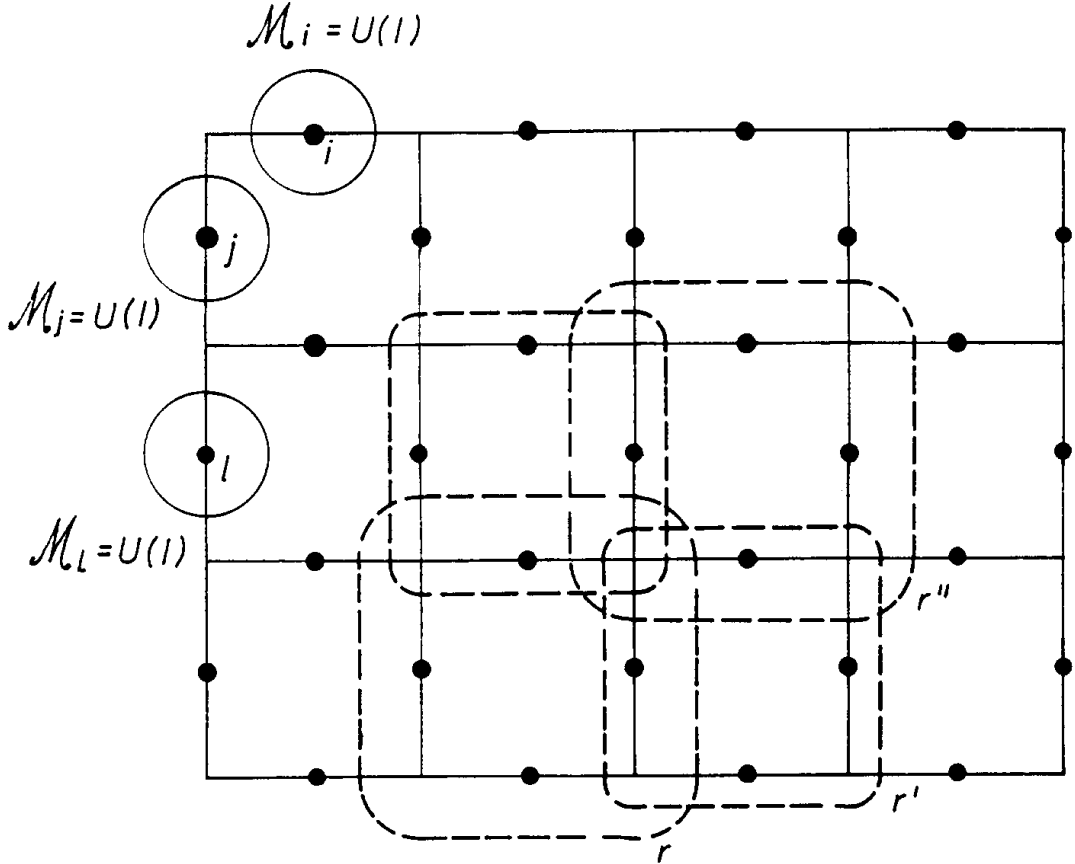}}
\caption[Special case field theory glass]{\label{u1ftg}In this very special
case field theory glass the fundamental space-time points $\{j\}$ (indicated
by the ``$\bullet$'') lie at the centres of the links
of a hyper-cubic lattice. The target space ${\cal M}_i$ of each
$\phi(i)$ is assumed to be a $U(1)$ group manifold. The ``quenched random''
regions $r,r^{\prime},r^{\prime\prime},\cdots$ on which the semi-local action
contributions $S_r$ are defined are the
overlapping, round-cornered ``squares'' drawn with broken lines each of which
contains just four of the fundamental space-time points $\bullet$ of the
set $\{j\}$. The
``quenched random'' semi-local action contribution $S_r$ is taken to be of
the identical form
$\beta Re U(\Box)+\kappa \sum_{\link\;\;\;\;\in r}Re U(\link)$
for each non-local
action region $r$. In other words, the quenched random set
$\{\beta_r\}=\{\beta,\kappa\}$
associated with a non-local action region $r$ is the same
for all regions in this very special
case.}
\end{figure}

Roughly speaking, the hope is by some means to discover degrees of
freedom that have patterns of quantum fluctuations that are
independent of the manner in which distant boundary conditions are chosen.
This behaviour is assumed to be the characteristic feature of (physical) gauge
degrees of freedom because even by a very ingenious choice of boundary
conditions (e.g., a fixing of boundary conditions in a gauge variant way) it
is not possible to influence the fluctuation pattern of gauge \dofx inasmuch
as these are not in any way coupled to anything on a distant boundary.

One can think of doing a search for such (would be) gauge field directions
(i.e., patterns of quantum fluctuations independent of distant boundary
conditions) in
configuration space - that is, the space that is the Cartesian product
of all the (fundamental site associated) target spaces ${\cal M}_i$. This
Cartesian product space is denoted as
$\cartprod_{i\in \{j\}}{\cal M}_i$. A point in this space - a microstate - is
sometimes denoted as $\Phi_{\{j\}}$.

By trial and error one could imagine  using a computer routine
(for illustrative purposes one might also envision
enlisting the assistance of a small ``demon'')
to find approximate local gauge symmetries (in practice, it is uncertain
whether large enough computers are available). By this imagined procedure
is meant the discovery
of space-time neighbourhoods $B(s^{\mu})$ labelled by a space-time point
$s^{\mu}$ such
that the $\phi(i)$ for $i\in B(s^{\mu}))$ undergo large
fluctuations along some orbit in the configuration space
$\cartprod_{i\in B(s^{\mu})}{\cal M}_i$.
Here $s^{\mu}$ denotes a space-time point (generally not
coinciding with an $i$ of the set $\{j\}$ of \fsps) that the demon uses to
label such a neighbourhood $B(s^{\mu})$. Such large fluctuation directions
or orbits -
designated as $\orb$ - are subsets of $\cartprod_{i\in B(s^{\mu})}{\cal M}_i$:

\beq \orb \subset \cartprod_{i\in B(s^{\mu})}{\cal M}_i.\eeq

These orbits can be considered as possible
candidates for what can turn out to be a gauge transformation
direction along which there is approximate invariance of the
action contributions $S_r$ corresponding to region(s)
$r$ having a non-vanishing
intersection with $B(s^{\mu})$. Guided by the $\orb$, let us assume
that the demon
can by trial and
error put together transformations $\Lambda(s^{\mu})$ of the
$\phi(i)$ for $i \in B(s^{\mu})$
that leave  action contributions $S_r$ invariant
(when $r\cap B(s^{\mu})\neq \emptyset$
for the region $r$ associated with $S_r$). In general, such  transformation
$\Lambda(s^{\mu})$ transforms {\em all}
the $\phi(i)|_{i\in B(s^{\mu})}$ - perhaps differently but in a coordinated
way -
so as to leave the  $S_r$ with $r\cap B(s^{\mu})\neq \emptyset$ approximately
invariant.

It should be pointed out that it is generically unlikely a priori that
local fluctuations would be coupled to distant boundary conditions.
Fluctuation patterns sensitive to distant boundary conditions need
long range correlations the presence of which would imply massless
(or light) particles. In the absence of strictly imposed symmetries (we
assume no symmetries in a field theory glass model), having such particles
would, naively at least, seem very unlikely as a generic possibility.
However, the essence of the FNNS mechanism is precisely that the
emergence of exact gauge symmetry is a generic possibility when approximate
symmetries are present in a model such as a field theory glass.

Having in some way exact gauge symmetry, we would expect to
see massless gauge particles that would survive down to low energies. Such
degrees of freedom could couple to distant boundary conditions. Potential
gauge symmetry directions in configuration space - coinciding with closed
orbits
along which there are large fluctuations - should not be affected by
changes in distant boundary conditions in the sense that such
changes either should not change gauge symmetry orbits at all or at most
``parallel translate'' such orbits in a direction in configuration
space corresponding to massless \dofx (having long range correlations)
that can couple to distant boundary conditions\footnote{There is a problem
here. If the photon field is set up in a configuration space direction
orthogonal to the direction corresponding to gauge transformations, how do
we get the spontaneous breakdown of gauge symmetry under gauge
transformations $\Lambda_{Linear}$ having linear gauge functions as in
Section~\ref{lingf} will be espoused as the defining feature of a Coulomb
phase? Presumably this problem is a statement of Elitzur's Theorem
in disguise. It is well-known that various tricks must be used to put this
theorem out of commission if spontaneous symmetry breaking is to be achieved.}.


Let us make a few remarks about the transformations $\Lambda(s^{\mu})$
associated
with the gauge ball $B(s^{\mu})$. First, we point out that the requirement
that these transformations leave the relevant $S_r$ invariant (together with
the requirement that the Cartesian product structure
$\cartprod_{i\in B(s^{\mu})}
{\cal M}_i$
should remain intact) essentially insures that the
$\Lambda(s^{\mu})$  have the structure of a \label{grpfn}group\footnote{The
subset
$\orb\subset \cartprod_{i\in B(s^{\mu})}{\cal M}_i$ is defined (or discovered)
as that
corresponding to combinations of values of fields variables $\phi(i)|_{i\in
B(s^{\mu})}$
for which the $S_r$ for which $r\cap B(s^{\mu})\neq \emptyset$ are roughly
constant. The
transformations
$\Lambda(s^{\mu})$ are just bijective mappings of such a subset $\orb$
onto itself. It turns out that the invariance requirement defines a subset
$\orb$ of points in $\cartprod_{i\in B(s^{\mu})}{\cal M}_i$ within which
certain
permutations are allowed. These correspond to subgroups of the
group of all permutations.
The composition of elements of such a permutation subgroup has of course
the structure of a group.}. Let us denote this group of transformations
associated with $B(s^{\mu})$ as $G(s^{\mu})$: i.e., $\Lambda(s^{\mu})\in
G(s^{\mu})$.
It should be understood that, by definition of a gauge ball $B(s^{\mu})$, the
field variables $\phi(k)$ corresponding to $k\not\in B(s^{\mu})$ are
transformed
trivially under the $\Lambda\in G(s^{\mu})$.

If the demon were, in the manner outlined above,
 to discover another
neighbourhood $B(s^{\prime})$ of another site $s^{\prime}$ such that
$B(s^{\mu}) \cap B(s^{\prime})\neq \emptyset$, then he would hope to find
among the (hopefully large) set of variables $\{\phi(i)\}$
for which $i\in B(s^{\mu}) \cap B(s^{\prime})\neq \emptyset$ some
that behave as link variables in the sense that they are
transformed {\em both} by transformations (necessarily corresponding to the
same
representation) associated with the ``site''
$s^{\mu}$ and with transformations associated with the ``site'' $s^{\mu
\prime}$
as indeed is characteristic of a link variable. Such a ``link-like''
configuration is illustrated in Figure \ref{linkvar}.

\begin{figure}
\centerline{\epsfxsize=\textwidth \epsfbox{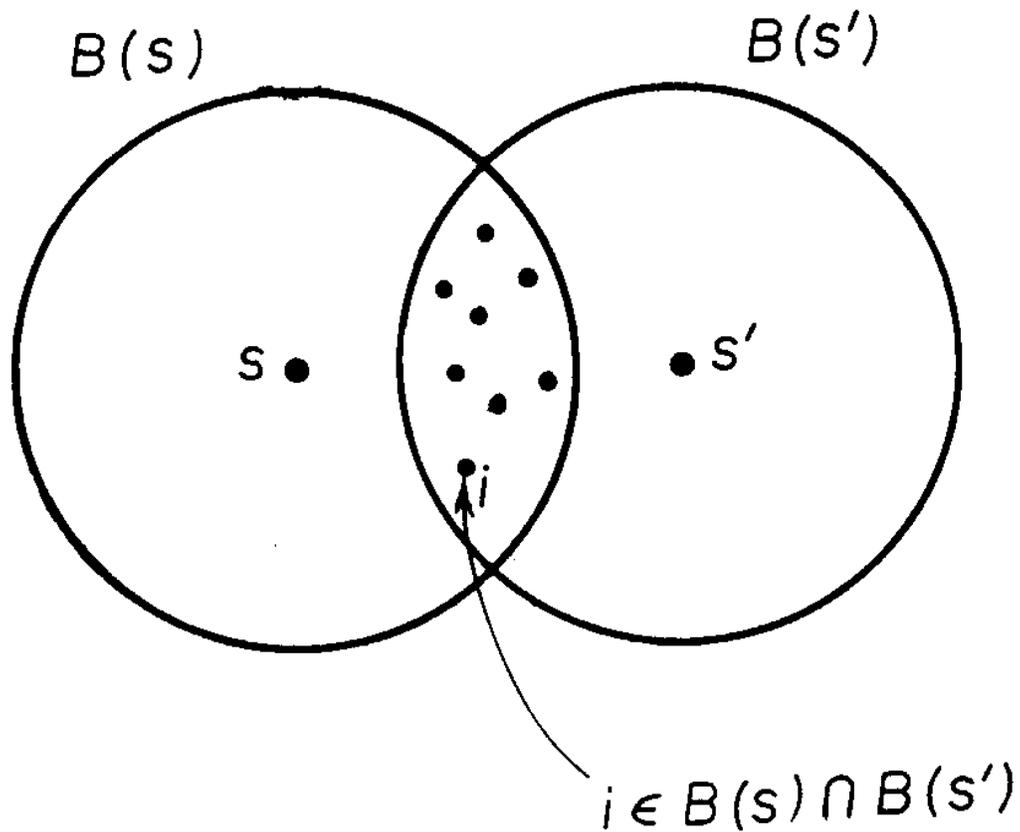}}
\caption[Link-like variables]{\label{linkvar}If the number of fundamental
space-time
points $i\in\{j\}$ lying within $B(s^{\mu})\cap B(s^{\prime\mu})$ is large
enough,
then it becomes likely that there will be at least a single
fundamental space-time point $i$ for which the corresponding field $\phi(i)$
transforms under the same representation
of a common subgroup of $G(s^{\mu})$ and $G(s^{\prime\mu})$. Such a field
variable
will be said to have link-like behaviour.}  \end{figure}

In order to have a microstate configuration that has a chance of giving
rise to invariant terms of a (semi)local action contribution $S_r$, it is
necessary that ``link-like'' variables are sufficiently profuse so as to ensure
what we can call ``plaquette-like'' variables as a generic possibility:
if ``link-like'' variables are sufficiently copious,
one could hope to have
part of the overlap of each of  (at least) three ``link-like'' pairs of three
gauge balls $B(s_n^{\mu})$ ($n\in \{1,2,3\}$) that intersect a
(non)local action region $r$.
Field variables
from each of three such overlap regions (``link-like''
variables - see last paragraph) of  three gauge balls can be combined so as
to simulate a ``plaquette-like'' variable (see Figure \ref{plaqvar}). If the
coefficient of such a ``plaquette
variable'' term in the action turns out to be large compared say to the
coefficients of
non-invariant contributions to $S_r$ (e.g., a ``link-like'' term), $S_r$ would
be approximately invariant
under the groups
of (local) transformations $G(s_n)$ ($n\in\{1,2,3\}$) associated with the three
gauge balls $B(s_n)$.

\begin{figure}
\centerline{\epsfxsize=\textwidth \epsfbox{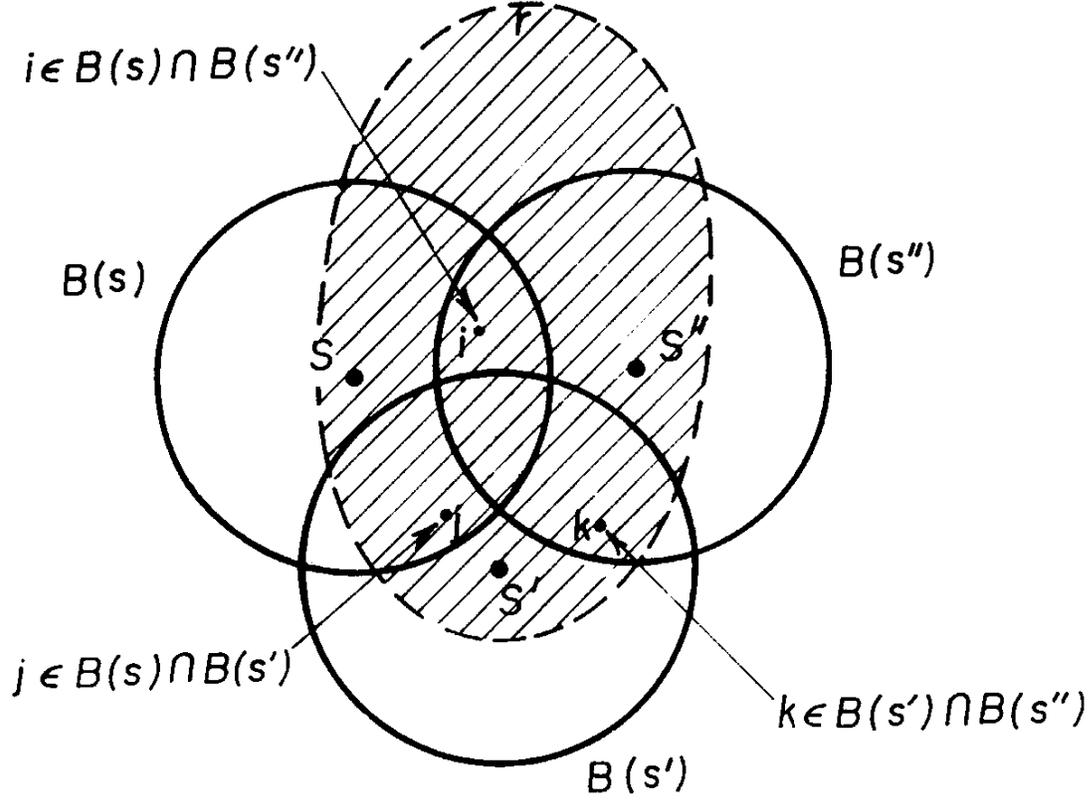}}
\caption[Plaquette variables in a field theory glass]{\label{plaqvar}
Consider three gauge balls
$B(s^{\mu})$, $B(s^{\prime\mu})$
and $B(s^{\prime\prime\mu})$ for which the pair-wise overlap
of each of three  gauge ball pairs has a non-vanishing
intersection with a (non)local action region $r$. Assume that the
overlap of each such pair of gauge balls is populated by a number
of fundamental space-time points $i$.
Then it may be possible, for example, to find three fields
$\phi(j)$, $\phi(k)$ and $\phi(i)$ - with let us say
$j\in B(s^{\mu})\cap B(s^{\prime\mu})\cap r \neq \emptyset$,
$k\in B(s^{\prime\mu})\cap B(s^{\prime\prime\mu})\cap r \neq \emptyset$ and
$i\in B(s^{\prime\prime\mu})\cap B(s^{\mu})\cap r \neq \emptyset$ - such that
all three
fields transform as elements (in a common representation)  of a subgroup
common to $G(s^{\mu})$, $G(s^{\prime\mu})$ and $G(s^{\prime\prime\mu})$.
This being the
case, the composition $\phi(i)\cdot\phi(j)\cdot\phi(k)$ would be
meaningful and would behave in a plaquette-like way in that a
term of $S_r$ containing the trace of $\phi(i)\cdot\phi(j)\cdot\phi(k)$ would
be approximately invariant under transformations corresponding to the
above-mentioned common subgroup of
$G(s^{\mu})$, $G(s^{\prime\mu})$ and
$G(s^{\prime\prime\mu})$ associated with the three gauge balls
$B(s^{\mu})$, $B(s^{\prime\mu})$ and
$B(s^{\prime\prime\mu})$.}
\end{figure}

\begin{figure}
\centerline{\epsfxsize=\textwidth \epsfbox{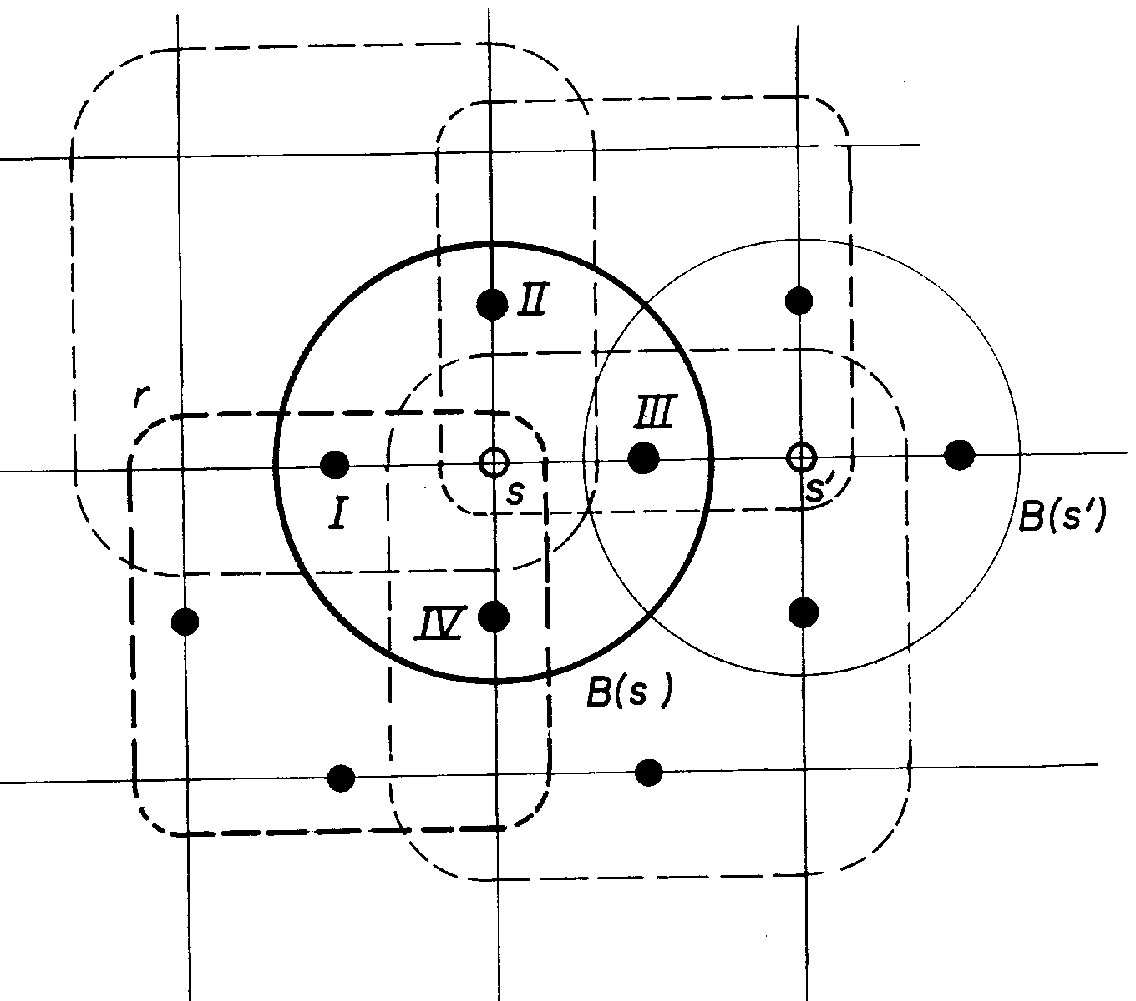}}
\caption[Gauge balls in the field theory glass]{\label{figu1ftg2}
The 2-dimensional special case
field theory glass is readily revealed by the demon as a priori being
very nearly
a gauge glass. Moving a point in
the configuration space (\ref{ggconfig}) along the orbit corresponding
to the application
of the elements of (\ref{ggorb}) to the fields $\phi$ at the four \fspsx
$i\in \{I,II,III,IV\}$
lying within the gauge ball $B(s^{\mu})$ leave the action
contributions $S_r=\beta ReU(\Box)+\kappa\sum_{\link\;\;\in r}ReU(\link)$
(with $r\cap B(s^{\mu})\neq\emptyset$)
approximately invariant if $\kappa$ is small. The index $r$ labels the
non-local action regions (on which the $S_r$ are defined); these are drawn
as broken-line rounded squares.}
\end{figure}

\begin{itemize}

\item In the ``special case'' field theory glass (really it turns out
a priori to be close to being a gauge glass, but it is still illustrative
to see how
a demon might reveal this),  a demon would eventually
discover a gauge ball $B(s^{\mu})$ centred at an $s^{\mu}$ that coincides with
a
site of the hyper-cubic lattice (such a site is \underline{not} one of the
\fspx $i\in \{j\}$ located at the centres of the links of the
lattice). Let such a gauge ball $B(s^{\mu})$ contain the $i$'s located at the
centres of the $2d$ ($d$ denotes dimension) links of the lattice emanating
from the centre $s^{\mu}$ of $B(s^{\mu})$. In Figure \ref{figu1ftg2} (with
$d$=2), $B(s^{\mu})$ is
indicated by the circle that contains four ``fundamental space-time points''
$i$ (labelled by the set
$\{I,II,III,IV\}$) and has a non-vanishing
intersection with four (overlapping) non-local action regions $r$
(corresponding to plaquettes containing four $i$'s). Each region $r$
such that $r\cap B(s^{\mu})\neq\emptyset$ contains
two (adjacent) $i$'s of the four $i\in B(s^{\mu})=\{I,II,III,IV\}$.
Let us make
the (fortuitous) assumption that the action parameters $\{\beta,\kappa\}$
assigned to each non-local action region $r$ are (in addition to being
identical for each region $r$) such that $\beta$ is large
and $\kappa$ is small. Then the demon would, for example,
observe large fluctuations on

\beq \times^{cart.\;prod}_{i\in B(s^{\mu})}{\cal M}_i=
\label{ggconfig}\eeq

\[{\cal M}_{i=I}\times {\cal M}_{i=II}\times {\cal M}_{i=III}\times {\cal
M}_{i=IV}=
U(1)_{i=I}\times U(1)_{i=II}\times U(1)_{i=III}\times U(1)_{i=IV}
\]
along an orbit coinciding with the elements of the group $G(s^{\mu})$ of
symmetry
operations

\beq G(s^{\mu})=\{ (\exp(i\theta_I),\exp(i\theta_{II}),\exp(i\theta_{III}),
\exp(i\theta_{IV}))| \label{ggorb}\eeq

\[ |\theta_I=-\theta_{II}=\theta_{II}=-\theta_{IV}\;[
e^{i\theta_I}\in U(1)_I \wedge e^{i\theta_{II}}\in U(1)_{II}\wedge
e^{i\theta_{III}}\in U(1)_{III}\wedge e^{i\theta_{IV}}\in U(1)_{IV}] \}; \]

\nin i.e., $G(s^{\mu})$ is the group of (approximate) symmetry operations
associated with the gauge ball $B(s^{\mu})$.

For small $\kappa$ and large $\beta$, this group of transformations leaves
the four non-local action contributions corresponding to the four regions
$r$ that overlap $B(s^{\mu})$ approximately invariant.

\end{itemize}

\subsubsection{The FNNS mechanism of exactification of an approximate gauge
symmetry}

The essential result of the FNNS mechanism is that the emergence of exact
gauge symmetry in the long wavelength limit is, without fine-tuning, a
generic possibility for a very broad class of field theories.

A prerequisite needed in order for the FNNS mechanism to work is
an approximate gauge symmetry at say the fundamental scale. Then FNNS promises
exact gauge symmetry (i.e. massless gauge bosons) in going to long
wavelengths.
Let us assume that such an approximate gauge symmetry has, in the manner
sketched above, been found on a field theory glass - presumably
from observing
directions in configuration space along which there are large quantum
fluctuations. Large fluctuations
are expected in
directions corresponding to orbits in configuration space along which the
action $S$ is almost independent of
the combinations of the  $\phi(i)$ lying on such an orbit.

The validity of the FNNS statement
is hard to see unless one uses the technique that the founders of FNNS used
to construct the argument leading to the conclusion
that the emergence of massless
gauge bosons is a
generic possibility for a broad class of fundamental scale field theories.

The technique consists in the (formal) rewriting of
the (single) ``God given'' field $\phi$ in terms of new
fictive fields\footnote{This procedure was first described
by H.B. Nielsen {\em et al} in \cite{fnns}; since then, developments in and
reviews
of this idea has appeared in many works; e.g.,
\cite{roysoc,randyn2,frogniel}.}
$\phi_h(i)$  and $H(s^{\mu})$ defined respectively on fundamental
space-time points $i$ and gauge ball centres $s^{\mu}$. These new variables are
defined
by

\beq \phi(i)\stackrel{def}{=}\phi_h^{H(s^{\mu})}(i)  \eeq

With this formal replacement, we trivially acquire a formal symmetry
under the  transformations

\beq \phi_h\rightarrow \phi_h^{\Omega} \eeq

\beq H \rightarrow \Omega^{-1}H \eeq

\nin inasmuch as transforming back and forth
between $\phi_h$ and $H$ in this way that doesn't change the $\phi$ field
containing the physics; i.e.,

\beq \phi=\phi_h^H\rightarrow (\phi_h^{\Omega})^{\Omega^{-1}H}=\phi_h^H=\phi
\label{formsym}\eeq

\nin leaves $S[\phi_h,H]\stackrel{def}{=}S[\phi]$ invariant.

Having this formal symmetry also
allows the freedom of choosing a  gauge
condition for the formal symmetry. The fact that these just formal
manipulations
will be done in a special way so as to make possible the analysis leading to
the FNNS
result in no way limits the (completely general) validity of the
FNNS mechanism conclusion (i.e., photons without fine-tuning)
because these formal manipulations are completely decoupled
from the physics. In fact it is precisely because the formal
manipulations of the $\phi_h$
and $H$ fields do not affect the physics that we can conclude that
a physical result obtained using a very special manipulation of these fictive
variables will remain valid in general (also when the fictive variables are
manipulated away). The formal manipulations are however important in the sense
that they reveal ``hidden'' physics that is otherwise not easy to see.

In the FNNS mechanism,
the freedom to choose a gauge is used to rewrite the field $\phi(i)$
(with $i\in B(s^{\mu})$ say) as a site ``$s$'' associated part $H(s^{\mu})$
that is somehow common to the field variables $\phi(i)$  (with $i\in
B(s^{\mu})$
and a part $\phi_h(i)$ that is the part of $\phi(i)$ that cannot
be described by $H(s^{\mu})$. So the gauge choice
that is made fixes $H$ to be the part of the $\phi(i)$
fluctuation pattern that for all $i\in B(s^{\mu})$ is common to the $\phi(i)$
field. Even though each field $\phi(i)$ takes values on a different target
space ${\cal M}_i$, the pattern of fluctuation along different ${\cal M}_i$
can be correlated in the sense that in moving a point in configuration space
along an orbit of large fluctuations results in changes in the  various fields
$\phi(i)$ that are correlated. For example, a
common phase factor of fixed norm could be ``factored out'' of the
fluctuation pattern of each $\phi(i)$ (with $i\in B(s^{\mu})$) and absorbed
into
the $H(s^{\mu})$ field defined on the centre of the gauge ball $B(s^{\mu})$.
This renders the $H$ field
a sort of non-linear Higgs field that fluctuates wildly
in the target (configuration) space $\cartprod_{i\in B(s^{\mu})}{\cal M}_i$.
Since the $H(s^{\mu})$ field has the same value
at all \fspsx $i\in B(s^{\mu})$ as a result of the choice of gauge, the action
$S$ is
roughly independent of $H$. Hence $H(s^{\mu})$ can have large fluctuations
that can prevent the theory from Higgsing.

The remarkable result of the FNNS mechanism is that if it is possible to
formally choose the gauge
so that to a large extent the $\phi(i)$ field fluctuations come to reside
in the site associated $H(s^{\mu})$ field, there is a generically good chance
for having the $\phi_h$
field in a Coulomb phase {\em without} fine-tuning.

So far the $H$ field is not a proper Higgs field in that in general it
will be a non-linear field that fluctuates on a non-convex (e.g. group)
target space manifold.
However, by block spinning the $H$ field, one can effectively introduce
a new variable (not present on the group manifold) that allows the
$H$ to effectively become linear. Block spinning essentially re-expresses the
fluctuations  in (the non-linear) $H$ on the target manifold in terms of this
new variable that
in effect fills out the non-convex target space  manifold so as to form
the convex
closure of the latter. Of course the space in which the convex closure of the
target space comes to reside must be postulated as being a reality.
Such a constructed extension of the target space could be taken as the
simplest possible space in which $H$ can be embedded linearly. For example,
if a non-linear $H$ field takes values in a $U(1)$ target space, it could,
by block spinning, become a normal linear Higgs field if $U(1)$ is first
embedded
in $\bc$.
For large $H$ fluctuations, $H$ can even come to lie in the symmetric point of
the convex envelope of the group manifold thereby attaining a vanishing
value as indeed is also possible for a proper
Higgs field. The founders of the FNNS mechanism\cite{fnns}  have
demonstrated in
a number of field theory models that there is a whole range of action
parameters for
which the fluctuations in $H$ are large enough to prevent the theory
from being Higgsed and for which the correlations in the $\phi_h$ field are
of sufficiently long range to yield a Coulomb phase.

\begin{itemize}

\item In the special case ``field theory glass'' for which there a priori
is approximate $U(1)$ gauge symmetry, the derivation of the FNNS mechanism
would first involve the introduction of the formal variables\cite{pc}
$\phi_h(i)$ and $H(s^{\mu})$. The first is

\beq \phi_h(i)=U_h(\linksspa) \in U(1)
\mbox{ defined on``links'' $i$ of lattice} \eeq

\nin ($a$ is the distance between adjacent gauge ball centres; e.g.,
the centre adjacent to that with coordinate $s^{\mu}$ in the the direction
$\nu$ has coordinates $s^{\mu}+a\delta^{\mu}_{\nu}$).

\nin The second new variable is

\beq H(s^{\mu})= H(\sites) \in U(1)
\mbox{ defined on ``sites'' $s^{\mu}$ of lattice}. \eeq

\nin These two new variable are defined by

\beq  \phi(i)=H^{-1}(\sites)U_h(\linksspa)H(\sitespa).  \eeq

\nin The action

\beq S[U_h,H]\stackrel{def}{=}S[\phi(i)]=
S[H^{-1}(\sites)U_h(\linksspa)H(\sitespa)] \eeq

\nin is readily shown to be invariant under the following (formal) gauge
transformations of the formal variables $U_h$ and $H$:

\beq
U_h(\linksspa)\rightarrow\Lambda^{-1}(\sites)U_h(\linksspa)\Lambda(\sitespa)
\eeq

\beq H(\sites)\rightarrow\Lambda^{-1}(\sites)H(\sites). \eeq

\nin In terms of the new formal variables $U_h$ and $H$ the ``semi-local''
action contributions (\ref{actgg}) (identical for each region $r$ in this very
special case) each become

\beq S_r=S[U_h,H]=\beta Re U_h(\Box)+\kappa\sum_{\linksspa\in\Box}
Re (H^{-1}(\sites) U_h(\linksspa) H(\sitespa). \label{actggfnns}\eeq

Let us now choose a gauge: e.g., $\prod_{\linkx \mbox{\tiny  meeting at }
\sitex} U_h(\link)=1$ (the lattice equivalent of the Lorentz gauge) or
$U_h(\link)=1$ for links in the direction $x^0$ (temporal gauge).
Now, if $\beta$ is sufficiently large
we will have

\beq U_h(\link)\approx 1 \mbox{  in the measure }
{\cal  D}U_h(\link) {\cal D}H(\site)e^{S[U_h,H]}/Z \eeq

and the action (\ref{actggfnns}) becomes

\beq S_r=S[U_h=1\mbox{ $\forall$ $\link$}  ,H]=\sum \beta \bunit+
\kappa\sum_{\linksspa\in\Box} Re (H^{-1}(\sites) H(\sitespa)\eeq

\[ \propto  \kappa\sum_{\linksspa} Re (H^{-1}(\sites) H(\sitespa). \]

\nin This is recognised as the ``$x-y$''-model in 4-dimensions;   from the
decay correlation theorem it is known that there are no long range
correlations for  sufficiently small $\kappa$. Hence the Higgsed phase is
avoided and for sufficiently large $\beta$ a Coulomb phase emerges complete
with photons!

\end{itemize}

\subsubsection{Relating microstates to macroscopic gauge fields}

Starting with the microstate vacuum, we shall now demonstrate a procedure
for setting up a macroscopic
gauge potential $A^a_{\mu}(y_I)$. This will be done by transforming
the field variables $\phi(i)$ at the fundamental space-time points
$i\in\{j\}$ using the local microlevel
transformations
$\Lambda (s^{\mu})\in G(s^{\mu})$ that are associated with the gauge balls
$B(s^{\mu})$
within which a fundamental space-time point $i$ (corresponding to $\phi(i)$)
lies. We shall also demonstrate that there is a microstate transformation
that corresponds to a pure gauge transformation of a macroscopic
$A^a_{\mu}$ field.
Recall that a microstate is specified by a point in $\cartprod_{i\in \{j\}}
{\cal M}_i$ (i.e., configuration space). Such a point corresponds to a
value of $\phi(i)$ for each $i\in \{j\}$ where $\{j\}$ denotes the set of
fundamental space-time sites.

We begin by choosing an (arbitrary) partition
of the set of fundamental space-time sites
$\{j\}$ in the field theory glass into a set $\{{\cal P}_{y^{\mu}_J}\}$
of non-overlapping cells ${\cal P}_{y^{\mu}_I}\in\{{\cal P}_{y^{\mu}_J}\}$
in such a way that
every fundamental space-time point $i$ lies in one and only one cell
${\cal P}_{y^{\mu}_I}$. Cells are labelled by the coordinate
$y^{\mu}_I$ that by definition lies within the cell ${\cal P}_{y^{\mu}_I}$.
We require that a cell is
small to a degree sufficient to validate the assumption that the variation
of $A^a_{\mu}$ within any cell is negligible.

Any
fundamental site $i\in\{j\}$ will always fall within
a unique cell  of the partition - let us say that $i$ falls within the cell
${\cal P}_{y^{\mu}_J}$:
\beq i\in {\cal P}_{y^{\mu}_J}\in\{{\cal P}_{y^{\mu}_I}\}.\eeq In general,
$i$ will also  belong to a set of gauge balls:

\beq \{B(s^{\prime\mu})|i\in B(s^{\prime\mu})\}. \label{olapset} \eeq

\nin The set (\ref{olapset}) can be empty or contain a number of gauge
balls depending on the density of gauge balls.

Choose now some cell ${\cal P}_{y^{\mu}_J}$ and consider the following
subset of the set $\{j\}$ of fundamental space-time points:

\beq \{k|k\in\{j\}\wedge k\in {\cal P}_{y^{\mu}_J}\wedge k\in
B(s^{\prime\mu})\}.
\label{setyj} \eeq

\nin Now let each fundamental space-time point $k$ belonging to the set
(\ref{setyj}) be transformed according to

\beq \lambda^a(s^{\prime\mu})= A^a_{\mu}(y_J)(s^{\prime\mu}-y^{\mu}_J)
+ c(y_J).
\label{liealg} \eeq

\nin where $A^a_{\mu}$ is the macroscopic field that we want to set up
at $y^{\mu}_J$ and $s^{\prime\mu}$ labels the gauge balls $B(s^{\prime\mu})$
for which $B(s^{\prime\mu})\cap {\cal P}_{y^{\mu}_J}\neq\emptyset$.
The $c(y^{\mu}_J)$ is a quantity depending only on $y^{\mu}_J$
that without loss of generality can be set to zero (because we assume the
theory is not Higgsed\footnote{An un-Higgsed system is invariant under any
global gauge transformation. In particular, this is true of a global
transformation generated by $c(y^{\mu}_J)$.}).
This $\lambda^a(s^{\prime\mu})$ is the generator of an element
$\Lambda(s^{\prime\mu})\in G(s^{\prime\mu})$

\beq  \Lambda(s^{\prime\mu})=\exp(i\lambda^a(s^{\prime\mu})\tau^a/2) \eeq

\nin that transforms all $\phi(i)$ with $i\in B(s^{\prime\mu})$.
Recall that $G(s^{\prime\mu})$ is the
set of symmetry operations associated with the gauge ball $B(s^{\prime\mu})$
such that each  element of $\Lambda(s^{\mu})\in G(s^{\mu})$ transforms all the
$\phi(i)$ corresponding to \fspsx $i$
within the gauge-ball $B(s^{\mu})$. These transformations are such that
action contributions $S_r$ associated with regions $r\in \{r^{\prime}|
r^{\prime}\cap B(s^{\mu})\neq \emptyset\}$ remain approximately invariant
{\bf {\em if and only if}}
all $k\in B(s^{\mu})$ are transformed by the {\bf{\em same}} element
$\Lambda(s^{\mu})\in G(s^{\mu})$.

In general this is not the case. From (\ref{liealg}) it is seen that
\fspsx get transformed by transformations that depend on the cell
${\cal P}_{y^{\mu}_J}$ (labelled by $y^{\mu}_J$) within which these
points lie. The important point is that if a gauge ball $B(s^{\mu})$ falls
within
more than one cell, then $k\in B(s^{\mu})$ get transformed by (different)
$\Lambda(s^{\mu})\in G(s^{\mu})$ according to which cell $k$ belongs.
If an action region
$r$ (corresponding to an $S_r$ that depends only on the field variables
$\phi(i)$ with $i\in r$) contains \fspsx lying within the same gauge ball
$B(s^{\mu})$ but different cells of the partition, then fields $\phi$ at
the \fspsx in
different cells get transformed  by {\bf{\em different}} elements
$\Lambda(s^{\mu})\in G(s^{\mu})$.
The result is an operation under which $S_r$ is not invariant.
Such operations can be used to alter the microstate vacuum so as to set up
any prescribed macroscopic $A_a^{\mu}(y_J)$ corresponding to
a field configuration with non-vanishing $F^{\mu\nu}$.

By way of example, consider the case where the two
field variables $\phi(l)$ and $\phi(m)$ with $l,m \in B(s^{\mu})$ fall
respectively
into two different cells ${\cal P}_{y^{\mu}_I}$ and ${\cal P}_{y^{\mu}_K}$ of
the partition $\{{\cal P}_{y^{\mu}_J}\}$. Then
$\phi_l$ and $\phi_m$ are  transformed by {\em different} group elements
of $G(s^{\mu})$: the field variable $\phi_l$ is transformed by the element
$\Lambda(s^{\mu})\in G(s^{\mu})$
generated by the Lie algebra element
$\lambda(s^{\mu})=A(y_I)(s^{\mu}-y^{\mu}_I)$ while $\phi_m$ is
transformed by the element of $G(s^{\mu})$ generated by
$\lambda(s^{\mu})=A(y_K)(s^{\mu}-y^{\mu}_K)$. Were the
field variables $\phi_l$ and $\phi_m$ ``links'' of a plaquette lying in
in some local action contribution $S_r$, the corresponding plaquette
term of $S_r$ would not be invariant under the modification of the microstate
vacuum outlined above. Indeed, such a modification would in general lead
to a non-vanishing curvature for such a plaquette.

In the special case that a gauge ball $B(s^{\mu})$ lies entirely within a
single
cell ${\cal P}_{y^{\mu}_I}$ of the partition $\{{\cal P}_{y^{\mu}_J}\}$,
the \fspsx in this gauge ball are all transformed by the {\bf{\em same}}
element

\beq \Lambda(s^{\mu})= \exp(iA^a_{\mu}(y_J)(s^{\mu}-y^{\mu}_J)\tau^a/2)\in
G(s^{\mu})
\eeq

\nin which just leaves us in the vacuum. This will be seen to correspond simply
to a gauge transformation of the microstate vacuum.

\begin{itemize}

\item In the special case ``field theory glass'' (i.e., in the case where
we almost have a gauge glass from the start), let the partition
${\cal P}_{\{y^{\mu}_J\}}$ be the squares  formed by the dashed diagonal lines
in Figure~\ref{amu}. Assume that the points $y^{\mu}_K$ and $y^{\mu}_J$ are
arbitrarily situated in respectively the cells ${\cal P}_{y^{\mu}_K}$ and
${\cal P}_{y^{\mu}_J}$. Then the field $\phi(I)$ at the \fspx I is transformed
by the element
$\Lambda(s^{\mu})=\exp(iA^{\mu}(y_K)(s^{\mu}-y^{\mu}_K))\in G(s^{\mu})$
corresponding to the setup of the field value $A^{\mu}(y_K)$ at $y^{\mu}_K$
in ${\cal P}_{y^{\mu}_K}$. In an
analogous fashion, the field $\phi(II)$ at the \fspx II is transformed
by the element
$\Lambda^{\prime}(s^{\mu})=\exp(iA^{\mu}(y_{II})(s^{\mu}-y^{\mu}_{II}))\in
G(s^{\mu})$
corresponding to the
set-up of the field value $A^{\mu}(y_{II})$ at $y^{\mu}_{II}\in {\cal
P}_{y^{\mu}_{II}}$.
The essential point is that $\phi(I)$ and $\phi(II)$ are transformed by
{\em different} elements $\Lambda(s^{\mu})$ and $\Lambda^{\prime}(s^{\mu})$ of
the symmetry
group $G(s^{\mu})$ associated with the same gauge ball (namely the gauge ball
containing both the \fspsx $I$ and $II$. The (semi)local action contribution
$S_r$ defined on the (non)local region $r$ within which $I$ and $II$ lie is
approximately invariant when the fields at the \fspsx (within any gauge ball
$B(s^{\mu})$ with $B(s^{\mu})\cap r\neq\emptyset$) are all transformed by the
{\em same}
element of the gauge group $G(s^{\mu})$. But this is in general not true when
such
fields $\phi(I)$ and $\phi(II)$ are transformed by {\em different} elements
of $G(s^{\mu})$. It is precisely this situation - i.e., different fields
$\phi(I)$ and $\phi(II)$ corresponding to \fspsx within the same gauge ball
that are transformed by different elements of $G(s^{\mu})$ - that is needed in
order
to set up macroscopic fields $A_{\mu}$ of any desired curvature.
\end{itemize}

\begin{figure}
\centerline{\epsfxsize=\textwidth \epsfbox{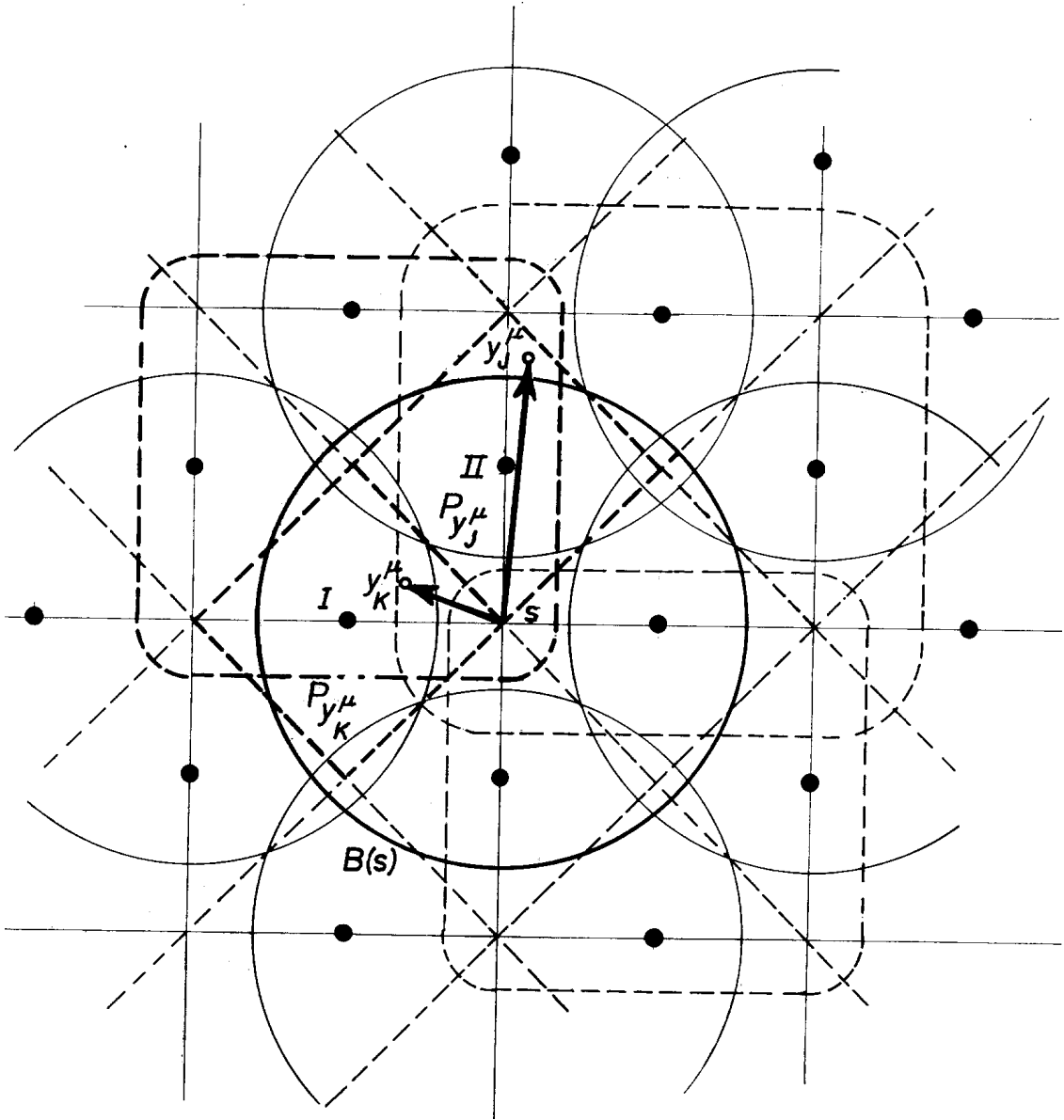}}
\caption[Setting up an $A^{\mu}$ field on special case field theory glass]
{\label{amu} In the figure, the chosen partition of the special field theory
glass is shown as the diagonal broken-line grid; ${\cal P}_{y^{\mu}_K}$ and
${\cal P}_{y^{\mu}_J}$ denote cells of this partition containing respectively
the (arbitrarily placed) points $y^{\mu}_K$ and $y^{\mu}_J$. Consider by way
of example the fields $\phi(I)$ and $\phi(II)$ at the \fspsx $I$ and $II$ that
both fall within the same gauge ball $B(s^{\mu})$. As $\phi(I)$ (with $I\in
{\cal P}_{y^{\mu}_K}$) and $\phi(II)$ (with $I\in
{\cal P}_{y^{\mu}_J}$) are transformed by the {\em different} elements
$\exp(A_{\mu}(y_K)(s^{\mu}-y^{\mu}_K))$ and $\exp(A_{\mu}(y_J)
(s^{\mu}-y^{\mu}_J))$,
of $G(s^{\mu})$, we can choose these elements so as to correspond to mutually
independent fields $A_{\mu}(y_K)(s^{\mu}-y^{\mu}_K)$ and
$A_{\mu}(y_J)(s^{\mu}-y^{\mu}_J)$ of any desired values.}
\end{figure}

It has been demonstrated that any  macroscopic $A_{\mu}$ field can be set
up by a modification of the fields $\phi_k$ of the (quenched)
microstate vacuum using
(local) transformations from (approximate) symmetry groups $G(s^{\mu})$
associated
with the
different gauge balls $B(s^{\mu})$ of the overlapping system of gauge balls
that
contain the
fundamental space-time sites $k$ at which the fields $\phi(k)$ are defined.

It is important to see that there is also a correspondence between a usual
gauge transformation of a macroscopic gauge field $A_{\mu}$ and a
modification of the microstate vacuum that corresponds to a (pure) gauge
transformation. To see this, consider a usual gauge transformation of a
macroscopic gauge field $A_{\mu}\stackrel{def}{=} A_{\mu}^a\tau^a/2$:


\beq A_{\mu}(x)\rightarrow A_{\mu}(x)^{\Omega(x)}\stackrel{def}{=}
\Omega^{-1}(x)A_{\mu}(x)\Omega(x)+
i\Omega^{-1}(x)\partial_{\mu} \Omega(x) \label{reggt} \eeq

\nin We want to see the relation between two microstate
transformations leading to two macroscopic gauge fields
that are related by (\ref{reggt}).
In doing this, it is easier to work with the {\em group elements}
$\Lambda(s^{\mu})\in G(s^{\mu})$ that transform the microstate in the desired
way
rather than the Lie algebra elements $\lambda(s^{\mu})$ that generates this
transformation.

In setting up some macroscopic $A_{\mu}$ field by performing
transformations of microstates, we can deal with one cell at a time.
Consider therefore some cell ${\cal P}_{y^{\mu}_J}$; this cell
${\cal P}_{y^{\mu}_J}$
generally contains some subset of the set $\{j\}$ of
\fsps :
\beq \{k|k\in\{j\}\wedge k\in  {\cal P}_{y^{\mu}_J}\}\label{spsinp}\eeq
and intersects some set of gauge balls
\beq \{B(s^{\prime\mu})|B(s^{\prime\mu})\cap{\cal P}_{y^{\mu}_J} \neq
\emptyset\}.
\label{ballsinp}\eeq
For each of the gauge balls $B(s^{\mu})$ in the set (\ref{ballsinp}) we perform
the microstate transformation

\beq \Lambda^{A_{\mu}}_{y^{\nu}_J}(s^{\nu})\stackrel{def}{=}
P\left(\exp\left(i\int_{y^{\nu}_J}^{s^{\nu}}A_{\mu}(x^{\nu})dx^{\mu}\right)
\right)\in G(s^{\nu})
\label{lamj}\eeq

\nin that is determined by the $A_{\mu}$ field that we want to set up
(hence the superscript ``$A_{\mu}$'' on
$\Lambda^{A_{\mu}}_{y^{\nu}_J}(s^{\nu})$). The subscript indicates that this is
a
(microstate) transformation of the fields $\phi(k)$ with
$k\in {\cal P}_{y^{\nu}_J}$. The argument $s^{\nu}$ indicates that the fields
$\phi(i)$ that get transformed are those associated with $i$ lying in
$B(s^{\mu})\cap{\cal P}_{y^{\mu}_J}$. The ``$P$'' preceding the integral
indicates
that a path ordered product is to be taken.

The number of such transformations performed in each cell in setting up a
given macroscopic $A_{\mu}$ field is just the number of elements in the set
(\ref{ballsinp}); a field $\phi(j)$ at a \fspx $j$ contained in the set
(\ref{spsinp}) gets transformed once for each gauge ball of the set
(\ref{ballsinp}) within which  $j$ lies.

Now if we want to set up the field that has been gauge transformed according to
(\ref{reggt})), then we want to use the microstate
transformations (\ref{lamj}) after these have been transformed according
to

\beq \Lambda^{A_{\mu}}_{y^{\nu}_J}(s^{\nu})\rightarrow
\Lambda^{A_{\mu}^{\Omega}}{y^{\nu}_J}(s^{\nu})=
\Omega^{-1}(s^{\nu})\Lambda^{A_{\mu}}_{y^{\nu}_J}(s^{\nu})\Omega(y^{\nu}_J)
\;\;(\Omega\in G(s^{\nu})). \label{lamjtrans} \eeq

In order to establish that this corresponds to a pure gauge
transformation of the microstate vacuum, we need to show that {\em all}
the $\phi(k)$ corresponding to $k$ lying within a given gauge ball
$B(s^{\mu})$ get transformed by only {\em one} element $\Lambda(s^{\mu})\in
G(s^{\mu})$ of the
group $G(s^{\mu})$ associated with this gauge ball. This is the opposite
of the situation needed to set up an $A_{\mu}$ in general (with
non-vanishing curvature):
recall from above that
in setting up an  $A_{\mu}$ field in general, it was essential
that  the fields $\phi(k)$ corresponding to $k\in B(s^{\mu})$  transform
in a cell dependent way. This being the case, a gauge ball intersected by
more than one cell could have  fields $\phi(k)$ and $\phi(l)$ (corresponding
to  \fspsx in different cells)  that would be transformed
by {\em different} elements of $G(s^{\mu})$ with the consequence that
$A_{\mu}$ fields with non-vanishing (or modified) $F_{\mu\nu}$ could be set
up.
In order to show that $\Lambda^{A_{\mu}}_{y^{\nu}_J}$ and
$\Lambda^{A^{\Omega(x)}_{\mu}}_{y^{\nu}_J}$ both set up macroscopic $A_{\mu}$
fields having the same $F_{\mu\nu}$ (i.e., macroscopic fields related by a
pure gauge transformation), we need to show that the
transformation~(\ref{lamjtrans}) takes place in a
cell independent way. Looking at (\ref{lamjtrans}), this would at first glance
seem difficult because (\ref{lamjtrans}) involves a transformation
$\Omega(y^{\mu}_J)$ with a cell dependent argument $y_J^{\mu}$:
while it is true that  in (\ref{lamjtrans}) both $\Omega(y_J^{\nu})$ and
$\Omega(s^{\nu})$ are elements of
$G(s^{\nu})$, they are {\em not} the the same element of $G(s^{\nu})$.
The element $\Omega(y_J^{\nu})\in G(s^{\nu})$ is obtained by the parallel
transport of $G(s^{\nu})$ from $s^{\nu}$ to $y_J^{\nu}$ along
$y_J^{\nu}-s^{\nu}$ using (\ref{lamj}).

 But now
we make use of the fact that we are assuming that our field theory glass
is un-Higgsed. This means that two vacua that are related by a global gauge
transformation are really exactly the same vacuum. In particular, we can
do global transformations for each cell; when we get to the cell
${\cal P}_{y_J^{\mu}}$, we perform the global gauge transformation
$\Omega(y^{\mu}_J)^{-1}$ on {\em all} gauge balls.
Letting $\Omega^{-1}(y_J^{\mu})$ act on~(\ref{lamjtrans}) from the right
yields a  transformation

\beq \Lambda^{A^{\Omega(x)}_{\mu}}_{y^{\nu}_J}\Omega(y^{\nu}_J)^{-1}=
\Omega^{-1}(s^{\nu})\Lambda^{A_{\mu}}_{y^{\nu}_J}(s^{\nu})
\label{lamj2trans}\eeq

\nin of the {\em same} vacuum
that is completely equivalent to (\ref{lamjtrans}).
The right-hand side of (\ref{lamj2trans}) is a single element of $G(s^{\nu})$
namely that obtained as the group product of $\Omega^{-1}(s^{\nu})\in
G(s^{\nu})$
and the transformation $\Lambda^{A_{\mu}}_{y^{\nu}_J}(s^{\nu})\in G(s^{\nu})$
that sets up the
macroscopic field $A_{\mu}$ before it is subjected to the gauge
transformation (\ref{reggt}). The important point is that the
transformation (\ref{lamj2trans}) depends only on $s^{\nu}$ and {\em} not on
the cell ${\cal P}_{y^{\nu}_J}$.

Repeating this procedure
for each cell ${\cal P}_{y_J^{\mu}}$ of the partition, it is seen
that the net result is that the fields $\phi(k)$ for $k\in B(s^{\mu})$
always get transformed by the {\em same} element $\Lambda(s^{\mu})\in
G(s^{\mu})$ even if
such a gauge ball lies in more than one cell of the partition.
Accordingly, we can conclude that the application of the microstate
transformations
$\Lambda^{A_{\mu}}_J$ and
$\Lambda^{A^{\Omega(x)}_{\mu}}_J$ to the microstate vacuum sets up
respectively macroscopic fields $A_{\mu}(x)$ and $A_{\mu}^{\Omega(x)}(x)$
that are related to each other by a pure gauge transformation.
This was what we set out to show.

\subsubsection{Multiple point criticality from a field theory glass}

We have demonstrated a procedure for setting up a macroscopic $A_{\mu}$ field
locally in space-time regions delineated by gauge balls $B(s^{\mu})$ using the
gauge
ball-associated group $G(s^{\mu})$
of (approximate) symmetry transformations $\Lambda(s^{\mu})\in G(s^{\mu})$ to
modify the microstate vacuum $\Phi_{\{i\}}$ at space-time points $i$ lying
within the gauge ball $B(s^{\mu})$.
More specifically, it was seen that in order to set up a gauge
field having non-vanishing curvature, it is necessary
that a gauge-ball $B(s^{\mu})$ be intersected by more than one cell of an
(arbitrary)
partition $\{{\cal P}_{y^{\mu}_J}\}$ of the fundamental set of space-time sites
$\{i\}$. This being the case, it is generically possible
to find two fundamental space-time sites  $j$ and $k$ such that even when
$j,k\in B(s^{\mu})$, the
associated $\phi_j$ and $\phi_k$ get transformed by different group elements
of the set of gauge transformations $G(s^{\mu})$ that are approximate
symmetries
of (non)local action contributions $S_r$ for which $r\cap B(s^{\mu})\neq
\emptyset$.  This will generally be the case when $j$ and $k$ belong to
different cells ${\cal P}_{y^{\mu}_I}$
and ${\cal P}_{y^{\mu}_L}$ of the partition $\{{\cal P}_{y^{\mu}_J}\}$
in which case
the fields $\phi_j$ and $\phi_k$ transform according to
$\phi_j\rightarrow \exp(iA_{\mu}(y_I)(s^{\mu}-y^{\mu}_I))\phi_j$ and
$\phi_k\rightarrow \exp(iA_{\mu}(y_L)(s^{\mu}-y_L^{\mu}))\phi_k$.
In general,
such a combination of transformations
does not coincide with just a single element $\Lambda(s^{\mu})$ of the set
$G(s^{\mu})$ of
(approximate) symmetries of the $S_r$ for which
$r\cap B(s^{\mu})\neq \emptyset$.

An implicit assumption in this procedure is that there are
microstate field variable degrees of freedom $\phi_i$ that {\em can} be
modified non-trivially
under the transformations of the various $G(s^{\mu})$ associated with the
various
gauge balls $B(s^{\mu})$; otherwise the action can only remain constant. The
point
is that a continuum limit $\frac{1}{g^2}$ must stem from a sum of
contributions coming from microstate configurations that can represent a
$gA^{\mu}$ field. An essential prerequisite for setting up such macroscopic
$gA_{\mu}$ fields in
the manner outlined above is that there is a sufficient density of sites
among the set $\{i\}$ of
fundamental space-time sites at which the associated field variables $\phi(i)$
transform
non-trivially under the approximate symmetry group $G(s^{\mu})$ associated with
some gauge ball $B(s^{\mu})$.

When the demon succeeds in finding a set of transformations $\Lambda(s^{\mu})
\in G(s^{\mu})$
where $G(s^{\mu})$ is a group
associated with a gauge ball $B(s^{\mu})$, he was presumably helped by the
observation of large quantum fluctuations along a (closed)  orbit
$\orb\subset \cartprod_{i\in B(s^{\mu})} {\cal M}_i$ (or a set of
``parallel'' orbits corresponding to different choices of (distant) physical
boundary conditions) on the manifold
$\cartprod_{i\in B(s^{\mu})}{\cal M}_i$. That large fluctuations are allowed
along
these orbits is a indication that the action is almost
constant along such orbits. For a given set of distant
boundary
conditions, the different points on $\orb \mi$
are related by the transformations $\Lambda(s^{\mu})\in G(s^{\mu})$.

Along such orbits, the distributions of target space values taken by
the fields $\phi(i)$ are such that there are correlations in the way
that the values assumed  by these $\phi$ change. In other words, in moving
the configuration space point along such an orbit, we expect that the
different fields $\phi(i)$ (with $i\in B(s^{\mu})$) will change in a correlated
way.
This behaviour would also follow from the properties that we expect
to be characteristic of such an orbit. Recall  that having
such an orbit is presumably tantamount to having found
a subset of the set of possible field variable combinations
$\phi:\{i\}\cap B(s^{\mu})\rightarrow \cartprod_{i\in B(s^{\mu})}{\cal M}_i$
                                         for which the $S_r$
(with $r\cap B(s^{\mu})\neq \emptyset$) are invariant and for which the
Cartesian
product structure
of $\cartprod_{i\in B(s^{\mu})} {\cal M}_i$ is intact. If these properties are
fulfilled, there will be  points along such a configuration space orbit that
can be transformed into each other under the action of a group (see footnote
on page~\pageref{grpfn}). The effect of such group operations is to permute
points in configuration space (on the orbit) that correspond to whole sets of
values of the $\phi(i)$ on such an orbit.
This permutation symmetry is in itself an expression of the correlated way in
which the $\phi(i)$ change when a point in configuration space is moved along
such an orbit $\orb \subset \cartprod_{i\in B(s^{\mu})} {\cal M}_i$.

We seek now to extract the common variation in the various $\phi(i)$ field
combinations corresponding to permutations (i.e., displacements)
of configuration points on such orbits. The idea is to incorporate this
common movement of the $\phi(i)$'s into a fictive (formal) field variable
$H(s^{\mu})$ that takes values in $G(s^{\mu})$. The fictive variable
$H(s^{\mu})$ (that maps
sites $s$ into configuration space) is defined
together with another fictive variable $\phi_h(i)$ by

\beq  \phi(i)=\phi_h^{H(s^{\mu})}(i). \label{newhum}
\eeq

\nin This completely formal replacement of a ``God-given'' variable by
a combination of formal variables is reminiscent of the technique used
in establishing the FNNS Theorem. Recall that the physical content of the
FNNS Theorem is revealed by formal manipulations un-coupled to the
physics of a field theory but which are extremely useful in exposing
the validity of the physical content of the FNNS mechanism. That real
physics can be uncovered using an analysis with fictive variables relies
on the fact that such formal operations cannot modify the physical content
of a theory. However if such formal manipulations help to reveal real
physics, such real physics is still there even when such fictive variables
are manipulated in some other way (and in particular when such fictive
variables are completely manipulated away).

The argumentation to be given below suggests that MPC actually results from a
rather precise compromise between competing behaviour the one
extreme of which favours the avoidance of a Higgs phase by having
confinement while the other extreme favours the avoidance of confinement
by having a Higgsed phase. It will be argued that at the multiple
point, the chances of avoiding confinement and Higgsing are best.

In the spirit of the FNNS fictive variable technique, the second new
field variable $\phi_h(i)$
corresponds to the part of the fluctuation pattern of the $\phi(i)$
(with $i\in B(s^{\mu})$) that remains after ``correlated variations''
in the values assumed by the
variables $\phi(i)$
have been absorbed\footnote{Even though fluctuations of the $\phi(i)$ occur
on different target spaces ${\cal M}_i$, it is still meaningful to consider
correlations in the pattern of fluctuation.} into the new field
$H(s^{\mu})$ defined
at the gauge ball centre~$s$. This amounts to choosing a gauge for the
formal symmetry that comes from introducing fictive variables in such a
way that the $\phi(i)_h$ have smaller fluctuations than the original fields
$\phi(i)$.

Now recall that the orbit $\orb\subset \cartprod_{i\in B(s^{\mu})} {\cal M}_i$
corresponds
to transformations that are only {\em approximately} symmetries of the
action contributions $S_r$ corresponding to the (non)local regions $r$
that overlap the gauge ball $B(s^{\mu})$. There can be small imperfections -
i.e.,
points on the orbit
$\orb\subset \cartprod_{i\in B(s^{\mu})} {\cal M}_i$ corresponding to (shallow)
relative minima in one or more of the (coupled)  $S_r$ corresponding to
(non)local regions $r$ that overlap $B(s^{\mu})$.

Having a shallow minimum in an $S_r$ - coupled to other (semi)local action
contributions $S_{r^{\prime}}$, $S_{r^{\prime\prime}}$, $\dots$ due to the
overlap
of (non)local regions $r$, $r^{\prime}$, $r^{\prime\prime}$, $\dots$ with
$B(s^{\mu})$ -
makes for the risk of an alignment of the field $H$ at $s$ that can
become correlated with $H$ at other points $s^{\prime}$, $s^{\prime\prime}$,
$\dots$ separated from $s$ by distances large enough to lead to Higgsing.

However, Higgsing can be rendered less likely if the fluctuations in
$\phi_h(i)$ are
large enough (corresponding to not having a coupling for the
field $\phi_h(i)$ that is too weak)
to inhibit such correlations in $H$ over large distances. Presumably, the
weaker the coupling of the $\phi(i)$, the more of the original fluctuation
pattern is common to the $\phi(i)$ (with $i\in B(s^{\mu})$) and therefore
incorporated into the $H(s^{\mu})$ field. The remainder of the fluctuation
pattern of the $\phi(i)$ fields - the incoherent part that cannot be put
into the $H(s^{\mu})$ field - resides in the $\phi_h(i)$ and in a statistical
sense at least can help to drown out imperfections in the approximate
symmetries under the group $G(s^{\mu})$ that could lead to Higgsing.

But if the coupling is too strong, the fluctuations in the (new) variables
$\phi_h(i)$ are so large that we get confinement of these degrees of
freedom (and  at the same time more effectively reduce the risk of
Higgsing of the
$H$-fields).

What we want is long range correlations for the \dofx corresponding to the new
variables $\phi_h(i)$ while
at the same avoiding a Higgsing of the new variable $H(s^{\mu})$. This is the
compromise that we claim is sought out by the MPCP.

The weaker the coupling for the variables
$\phi_h(i)$ - (corresponding to smaller fluctuations in $\phi_h(i)$ that
accordingly are less effective in preventing correlations in the $H$
field over long distances) -  the more near perfect must be the
``approximate'' gauge symmetries found by the demon if the small
uncorrelated fluctuations in $\phi_h(i)$ (with $i\in B(s^{\mu})$) are
- at least statistically speaking - to be
effective in reducing correlations in the
$H$-field over distances that can lead to Higgsing.

Consider a gauge ball $B(s^{\mu})$.
We want to define a quantity that expresses the amount by which a group
of transformations $G(s^{\mu})$ associated with this gauge ball $B(s^{\mu})$
deviates from
being a perfect symmetry. Such a quantity, denoted by $\Delta_{dev}$, is
considered for each local action contribution $S_r$ for which the
corresponding (non)local
region $r$ (containing all the field variables $\phi_i$ on which
$S_r$ depends non-trivially) is such that $r\cap B(s^{\mu})\neq \emptyset$.
This quantity $\Delta_{dev}$ is defined by

\beq \Delta_{dev}(G(s^{\mu}))=
\stackrel{max}{\tiny over\;r}||S_r[\phi^{\Lambda}]-
\langle S_r[\phi^{\Lambda}]\rangle_{\Lambda\in G(s^{\mu})}||
\mbox{ for }r\cap B(s^{\mu})\neq \emptyset \mbox{ and }
\Lambda(s^{\mu})\in G(s^{\mu}). \label{crit} \eeq

\nin According to the argumentation above, the quantity
$\Delta_{dev}(G(s^{\mu}))$ must be
smaller the larger the inverse squared coupling if the risk for Higgsing
due to deviations from perfect symmetry is not to increase.
We can express this requirement by an inequality that must be satisfied:

\beq \Delta_{dev}(G(s^{\mu})) \leq f(\frac{1}{g^2}) \label{useful} \eeq

\nin where $f(\frac{1}{g^2})$ is a monotonically decreasing function of
$\frac{1}{g^2}$ and $\Delta_{dev}(G(s^{\mu}))$ is given by (\ref{crit}).

So the gauge ball $B(s^{\mu})$ is useful only if the associated group of
transformations $G(s^{\mu})$ satisfy the criterion~(\ref{useful})
above. The weaker the coupling (i.e., the larger the value of
$\frac{1}{g^2}$) the smaller the allowed deviation from perfect symmetry
($\Delta_{dev}$) and the less likely it will be that a gauge ball is useful
in the sense that
(\ref{useful}) is satisfied. The density of such useful gauge balls decreases
as the coupling for the $\phi_h(i)$ variables decreases; concurrently,
the fundamental space-time points $i$  and associated field variables $\phi(i)$
lying within  the gauge balls ``rejected'' according to the criterion
(\ref{useful}) are no longer available for use in setting up
a macroscopic $gA^{\mu}$ field. But it is necessary
that such
$gA^{\mu}$ can be set up if there are to be contributions to $\frac{1}{g^2}$ in
the
continuum limit.

Let us denote the number of gauge balls to which are associated sufficiently
accurate symmetry $G(s^{\mu})$ groups (i.e., useful gauge balls) as
\beq \#\{B(s^{\mu})| \Delta_{inv}(G(s^{\mu}))\leq f(\frac{1}{g^2})\}.\eeq

There are  two competing relationships  between $\frac{1}{g^2}$ and
$\#\{B(s^{\mu})| \Delta_{inv}(G(s^{\mu}))\leq f(\frac{1}{g^2})\}$ (see
Figure~\ref{2rel})
that can be stated
as follows:

\begin{enumerate}

\item the larger the number of ``useful'' gauge balls
$\#\{B(s^{\mu})| \Delta_{inv}(G(s^{\mu}))\leq f(\frac{1}{g^2})\}$, the larger
the number
of microstate degrees of freedom $\phi$ that are connected to the macroscopic
field
$gA^{\mu}$ and hence the larger is $\frac{1}{g^2}$.

\item The weaker the coupling the more readily will there be long distance
correlations in the field $H(s^{\mu})$ with the danger of Higgsing as a
consequence;
avoiding such correlations necessitates a smaller allowed
deviation  (\ref{useful}) from perfect symmetry for the groups $G(s^{\mu})$
associated with gauge balls $B(s^{\mu})$ and consequently a reduction in the
number
of ``active'' gauge balls
$\#\{B(s^{\mu})| \Delta_{inv}(G(s^{\mu}))\leq f(\frac{1}{g^2})\}$.

\end{enumerate}

\begin{figure}
\centerline{\epsfxsize=\textwidth \epsfbox{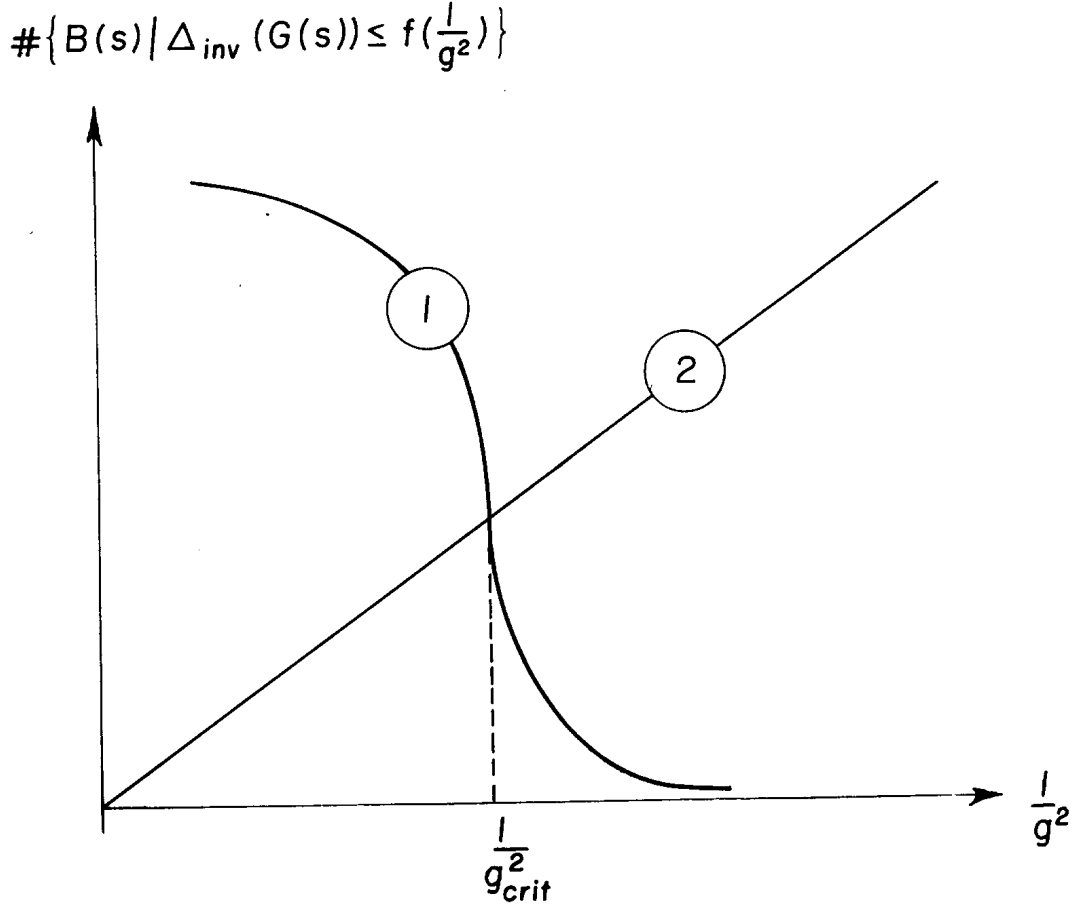}}
\caption[fig2rel]{\label{2rel} There are two competing relationships between
$\#\{B(s^{\mu})| \Delta_{inv}(G(s^{\mu}))\leq f(\frac{1}{g^2})\}$ and
$\frac{1}{g^2}$
shown as curves 1 and 2. Curve 1 suggests the manner in which more and more
gauge  balls fail to satisfy (\ref{useful}) as $\frac{1}{g^2}$ increases.
The suddenness with which curve 1 is depicted as dropping off with increasing
$\frac{1}{g^2}$ is intended to suggest that
$\#\{B(s^{\mu})| \Delta_{inv}(G(s^{\mu}))\leq f(\frac{1}{g^2})\}$ varies
rapidly at
the confinement to Coulomb phase transition.
Curve 2 shows that $\frac{1}{g^2}$ increases
with the number of useful gauge balls. Because of the rapid variation of
curve 1 near $\frac{1}{g^2_{crit}}$, the intersection of the two curves
can be expected to be at least very close to critical coupling values.
Applying this argumentation to each possible partially confining phase,
the analogue of the
intersection of the two curves becomes the multiple point where all possible
combinations of \dofx can coexist in confinement  and in configurations
with long range correlations}
\end{figure}

The point to be made is that the field theory glass model for
fundamental scale physics is a Random Dynamics scenario that, apart from
yielding exact LEP gauge symmetry by the FNNS mechanism if there is an
approximate symmetry at the fundamental scale,
suggests that the inequality (\ref{ineq}) is obeyed in Nature as an
{\em equality}.

Point {\bf 1.} above implies that having a value of $\frac{1}{g^2}$ that at
the Planck scale is large enough to avoid confinement (i.e., the
fulfilment of the {\em inequality} (\ref{ineq}))
is really a question
of having sufficiently many gauge balls that can be connected to a
macroscopic $gA^{\mu}$ field. That the inequality (\ref{ineq}) must be
realised as an {\em equality} is suggested by point 2) above inasmuch as
weaker than necessary couplings increase the risk of correlations over
distances large enough to lead to Higgsing.

The two relations between $\frac{1}{g^2}$ and
$\#\{B(s^{\mu})| \Delta_{inv}(G(s^{\mu}))\leq f(\frac{1}{g^2})\}$ (points {\bf
1.} and
{\bf 2.} above) are depicted schematically in Figure (\ref{2rel}).
The suggestion that
the inequality (\ref{ineq}) is realised as a equality - if understood as
applying to all possible \pcps - is tantamount to suggesting the validity
of the principle of multiple point criticality.

\subsection{Breakdown of the AGUT $SMG^3$ by confusion}\label{confbrkdn}

As an inequality, (\ref{ineq}) expresses the important requirement that
Yang-Mills degrees of freedom at the  Planck  scale
that give rise to the observed Yang-Mills fields of the Standard Model
cannot, already at the Planck scale, have developed a strong
coupling/high temperature/confinement-like physics. We make the important
assumption
that only Yang-Mills \dofx that are {\em Coulomb-like} in behaviour at the
fundamental scale
have a chance of surviving down to experimentally accessible energies.
This is what is insured by the {\em in}equality (\ref{ineq}). In a simple
lattice gauge theory with a gauge invariant action $S_{g.i.}$ given by
$S_{g.i.}=\beta\sum_{\Box}ReU(\Box)$, confinement is avoided by having a large
enough $\beta$.

However, a direction in the configuration space of a field theory glass
along which there is only approximate gauge symmetry has accordingly
at least small gauge breaking action contributions $S_{g.b.}$
to the quenched random action. We call this latter the fundamental action
$S_{fund}$. Let us take as a prototype for $S_{g.b.}$ a term
$S_{g.b.}=\kappa\sum_{\link}U(\link)$ that explicitly breaks gauge symmetry.

The random dynamics philosophy for fundamental physics has
played a decisive role in motivating the theoretical
picture we have for the origin of the SMG via the $SMG^{N_{gen}}$ gauge
group.


A possible theoretical motivation for the AGUT gauge  group  $SMG^3$  and
its subsequent breakdown to the diagonal subgroup could start with a scenario
from ``random dynamics''\cite{randyn1,randyn2}: at an energy a little above the
Planck
scale, one has a multitude of gauge symmetries resulting from the
FNNS\footnote{F\"{o}rster,  Nielsen,  Ninomiya, Shenker}
exactification\cite{fnns}
of chance occurrences of approximate gauge symmetries.  This  collection   of
symmetries can be expected to be  dominated  by low-dimensional groups
as such symmetries are most
likely to occur by chance. We envision that the symmetry embodied by
this collection of groups is broken down by a succession of  steps  the
last of which, before the Weinberg-Salam breakdown, is  the  breakdown
of the AGUT group $SMG^3$ to its diagonal  subgroup. This
succession  of  symmetry  breakdowns  is  pictured  as  occurring   for
decreasing energies within a range of a few orders of magnitude at the
Planck scale.

The succession of breakdowns envisioned coincides  with  gauge  groups
that are more and  more  depleted  of  group  automorphisms\footnote{A
group automorphism is defined as a bijective map  of  the  group  onto
itself that preserves the group composition law. The set of all  group
automorphisms is itself a group some of  the  elements  of  which  are
inner automorphisms (i.e., just similarity transformations within the
group).  There  can  also  be  outer  automorphisms   (essential   for
confusion) which  are  defined  as  factor  groups  of  the  group  of
automorphisms modulo inner automorphisms. }. We
(and  others)  have  proposed  a
breakdown  mechanism\cite{conf1,conf2,gaeta}  called  ``confusion''  that   is
active when gauge groups possess automorphisms.

We speculate that confusion breaking - that can be called into play by
different types of automorphisms - can successively break very general
groups with many Cartesian product factors down  to  a  collection  of
groups with especially few automorphisms as is characteristic  of  the
SMG  itself.  It   is   noteworthy   that   the   SMG   has   been
shown in a certain sense \cite{skew1,skew2} to be the  group  of  rank  4
(and  dimension  less  than  19)  that  is  maximally   deficient   in
automorphisms. We propose the  group  $SMG^3$  as  the  last
intermediate step on the way to the SMG.

We now briefly explain
how the confusion breakdown mechanism functions for gauge groups with
outer automorphisms. First it is argued that,  in  the  spirit  of
assuming a fundamental physics that can be taken as  random,
one is forced to allow for the possibility of having  quenched  random
``confusion surfaces'' in space-time. The defining  property  of  these
surfaces  is  that  (e.g.  gauge)  fields  obey  modified   continuity
conditions at such surfaces; for example, the permutation of  a  gauge
field with an automorphic image of the field can  occur.  A  non-simply
connected  space-time  topology  is  essential  for  the  presence   of
nontrivial confusion surfaces; a discrete space-time structure such  as
a lattice is inherently non-simply connected because of  the  ``holes''
in the structure.

The essential feature of  the  ``confusion''  breakdown  mechanism  is
that,  in  the  presence  of  ``confusion  surfaces'',  the   distinct
identities of a field and its automorphic image can be maintained {\em
locally} but not {\em globally}. To see  how  this  ambiguity  arises,
imagine taking a journey along a  closed  path  on  the  lattice  that
crosses a confusion surface at which  the  labels  of  a  gauge  group
element and its automorphic image are permuted. Even if one could,  at
the onset of the journey, unambiguously assign say the names ``$Peter$''
and ``$Paul$'' to two gauge  fields  related  by  an  automorphism,  our
careful accounting of the field identities as  we  travel  around  the
loop would not, upon arriving back at the starting point,  necessarily
be in agreement with the  names  assigned  when  we  departed  on  our
journey. So an attempt to make independent global gauge transformations of
$Peter$ and $Paul$ (sub)groups would not succeed.
Therefore, for the action at confusion surfaces, there is not invariance
under global gauge transformations of the whole gauge group but only under
transformations of the subgroup left invariant by the automorphism.

The ambiguity under the automorphism caused by confusion (surfaces)
is removed by the breakdown of the gauge group
to the maximal subgroup which is left invariant under the  automorphism.
The diagonal  subgroup of the Cartesian
product of isomorphic groups is
the maximal invariant subgroup  of  the  permutation  automorphism(s);
i.e., because the diagonal subgroup is the subgroup left invariant  by
the automorphism, it has  the  symmetry  under  gauge  transformations
generated by constant gauge functions  (corresponding  to  the  global
part  of  a  local  gauge  transformation)  that  survives  after  the
ambiguity caused by the automorphism is removed by  breakdown  to  the
diagonal subgroup.

For the purpose of illustrating a possible origin of the
``anti-unified'' gauge group  and  its  subsequent  breakdown  to  the
standard model group, we describe  two  important  examples  of  group
automorphisms - {\em examples 1 and 2} below - that call the confusion
mechanism into play:

\begin{description}

\item[{\em Example 1.}]Many groups have a charge conjugation-like
automorphism corresponding in the $SU(N)$ case to complex  conjugation
of the matrices element by element. While for $SU(2)$ this is an inner
automorphism, it is for higher $SU(N)$ groups an  outer  automorphism.
According to the speculated confusion mechanism, such a  group  should
break down to the  subgroup  consisting  of  only  the  real  matrices
which  is  the  largest  subgroup  that   is   invariant   under   the
automorphism.  If  the  group  is  provided  with  C-breaking   chiral
fermions, the automorphism can be broken in this way thereby thwarting
the ``attack'' from the confusion mechanism.

\item[{\em Example 2.}] There can be automorphisms under the permutation
of identical group factors in a Cartesian product group: we argue that
the  symmetry   reduction   (at   the   Planck   scale)   from
$(SMG)^{N_{gen.}}$ to $SMG=S(U(2) \times U(3))$ is  triggered  by  the
symmetry under the automorphism  that  permutes  the  $N_{gen}$  SMG
factors in $(SMG)^{N_{gen.}}$.

\end{description}

Elaborating briefly on {\em example 1} above, we point  out  that  with
the exception of the semi-simple groups such as $SU(2)$, $SO(3)$ ,  the
odd $N$ spin or $SO(N)$-groups and the symplectic groups,  all  groups
have  outer  automorphisms  of  the  complex  conjugation-  or  charge
conjugation-like type. Following  a  series  of  confusion  breakdowns
activated by charge conjugation-like automorphisms, we expect that the
(intermediate)   surviving   gauge    symmetry    (i.e.,    that    of
$(SMG)^{N_{gen.}}$)  must  have  matter  fields  that   break   charge
conjugation-like symmetries. In other  words,  the  presence  of  such
matter fields serves to protect the surviving  symmetry  from  further
breakdown by eliminating the possibility for further confusion  of  the
surviving group with its automorphic image under charge conjugation.

In particular, we expect that a necessary condition for  the  survival
of gauge groups like $U(1)$ and $SU(3)$ is the presence of
some  matter  fields  not  invariant  under  charge  conjugation.
Protection against this sort of breakdown can be  provided  by  chiral
fields that break the charge conjugation symmetry of the gauge fields.
In the case of the Standard Model,  left-  and  right-handed  fermions
always appear in different representations so that confusion breakdown
by way of a charge conjugation automorphism is not possible. In  fact,
the number of particles in a single generation in combination with the
rather intricate way these are represented in the Standard  Model  can
be shown to be the simplest possible manner in which  gauge  anomalies
can be avoided\cite{anomalies1,anomalies2,anomalies3,korfu}.

As mentioned in  {\em  example  2}  above,  it  is  assumed  that  the
confusion breakdown of the intermediate  gauge  group  $SMG^3$  to  the
standard model group SMG (at the Planck scale) is activated  by  the
automorphism that permutes the $N_{gen}$ isomorphic Cartesian  product
factors in  $(SMG)^{N_{gen.}}$.   The   elimination   of   the
ambiguities that can arise in trying to keep track of  the  identities
of a group element and its automorphic image under  such  permutations
coincides with the breakdown to the standard model  group  $SMG=S(U(2)
\times   U(3))$   which,    being    the    diagonal    subgroup    of
$(SMG)^{N_{gen.}}$, is invariant under the automorphism that  permutes
the SMG group factors in $(SMG)^{N_{gen.}}$. In order for this final
confusion breakdown to work effectively, the Cartesian product factors
of $(SMG)^{N_{gen.}}$ must presumably be truly isomorphic - i.e.,  the
matter field content of each factor must  essentially  have  the  same
structure.  This  combined  with  the  fact  that  one  usual  fermion
generation is known to provide the least  complicated  arrangement  of
particles that avoids gauge anomalies would strongly suggest that  the
$N_{gen}$ factors of $(SMG)^{N_{gen.}}$ are simply dull repetitions of
the standard model group each having a comparable matter field content
but with the possibility of having different gauge quantum numbers  as
a distinguishing feature  of  the  different  families.  Each  of  the
$N_{gen}$  factors  is  the  ``ancestor''  to  one  of  the  $N_{gen}$
generations  of  the  diagonal  subgroup  identified  with  the  usual
Standard Model Group.

It should be emphasised that all  the  confusion  breakdowns  -  those
utilising   a   series    of    charge    conjugation    automorphisms
leading to $(SMG)^{N_{gen.}}$ as well as the final confusion breakdown
of the $SMG^3$  to  the  diagonal  subgroup  that  is  caused  by  the
permutation  automorphism
                  - are assumed to take place within a  rather  narrow
range of energies at the Planck scale.

Before leaving the confusion breakdown mechanism, we should point  out
that any mechanism that  breaks  the  $SMG^3$  down  to  the  diagonal
subgroup would suffice for our model. A Higgs  field  mechanism  could
for example provide an  alternative  to  the  confusion  mechanism  of
breakdown.

\subsection{A string-regulated model using a Kaluza-Klein
space-time}\label{kaluza}

In this model, the Random Dynamics-inspired input is that the (quenched random)
values of  gauge couplings are
given at the fundamental scale {\em independent of low energy physics}.
This assumption can be implemented by insisting that, from the point of
view of low energy physics, the value of gauge couplings at the fundamental
scale must appear to be random. In the context of this model which will
be briefly
described now, it is argued that there is a range of fundamental scale
coupling values that
must be avoided if the suspicion of fine-tuning is not to be aroused.
The weakest coupling in this range provides an upper limit on how strong
a coupling should be observed experimentally.

Using a regularised Kaluza-Klein  space-time at the fundamental
scale with Yang-Mills fields in $D-4$ compactified dimensions, we examine
the $\beta$-function for a dimensionless expression $\beta(1/a)=
(2/g^2)a^{D-4}$ for the coupling
constants $g$ in $D$-dimensions. Here $a$ is the lattice constant and defines
the renormalization point $\mu=1/a$. The ``running'' $\beta(\mu)$ is defined
so as to describe observed continuum physics as a lattice with lattice
constant $a=1/\mu$. The $\beta$-function for this $\beta(1/\mu)$ is shown to
have an ultraviolet stable fixed point $\beta_{crit}$.
The argument is that in order to avoid the suspicion
that the values at the fundamental scale are fine-tuned to experimentally
observed values (recall that couplings are assumed to be given at the
fundamental scale independent of low energy physics), a small range of
fundamental scale ($\approx$ Planck scale) coupling values centred
at $\beta_{crit}$
must be avoided. Due to a scale dependent effective dimensionality $D$, this
``forbidden'' interval of values expands due to
renormalization group effects into a large range of values in going towards
the infrared that should not be observed at low (i.e., experimentally
accessible) energies (see Figure~\ref{figkk1}. It is the larger $\beta$
boundary of the ``forbidden''
interval (at low energies) that provides an upper bound on ``allowed'' values
for experimental couplings inasmuch as this boundary corresponds to being on
the ``Coulomb-like'' side of $\beta_{crit}$ at the fundamental scale. The
small $\beta$ boundary of the ``forbidden'' interval at low energies would
correspond to a fundamental scale $\beta$ for which there was confinement
already at the fundamental scale and accordingly would not lead to physics
observable at low energies.

\begin{figure}
\centerline{\epsfxsize=\textwidth \epsfbox{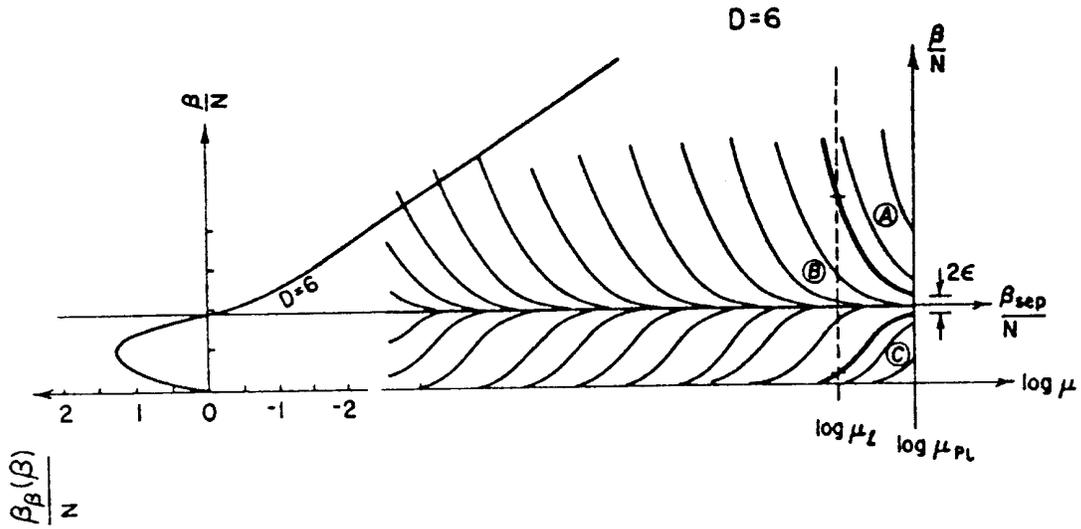}}
\caption[figurekk1]{\label{figkk1} In the string regulated model using a
Kaluza-Klein space-time
with $D-4$ compactified dimensions at the fundamental scale, the idea is that
in order not to arouse the suspicion of fine-tuning, a small interval
centred at $\beta_{crit}$ at the fundamental scale must be avoided. This
small ``forbidden'' interval $\epsilon$ expands due to renormalization group
effects into a
whole range of ``forbidden'' values of $\beta_{crit}$ that should not be
observed in going to low energies. It is the upper limit of this range that
provides an upper bound on allowed experimental couplings.}
\end{figure}

In this model, the scale dependence of the space-time dimensionality $D$
has several consequences. At the fundamental scale, the Kaluza-Klein
space $\br^4\times K$ is of dimension $D>4$ but in going to lower energies,
one at some point encounters  the energy corresponding to the linear dimension
$\rho$ of the compactification space $K$. At energies corresponding to
distances longer than $\rho$, the dimensionality of space-time is reduced to
$D=4$ and renormalization group effects are just the normal quantum
mechanical ones. At the energy $1/\mu\approx \rho$ coinciding with the
transition $D>4$ $\longrightarrow$  $D=4$, it is argued that couplings are
weakened by a factor roughly equal to the number of fundamental string
regulation volumes $reg.\;vol.$  that can be accommodated in the volume
$vol(K)$ of the compactification space. Subsequently it is proposed
that this factor cannot be
less than the number of generations $N_{gen}$  by invoking an argument
reminiscent of that sometimes encountered in string model ``$T. O. E.$'' the
essence of which is that the quark and lepton generations correspond to
various zero modes of a Weyl operator in the compactifying space.

Using string physics as a way of achieving a
regularization of the Yang-Mills fields in the limit of zero slope is
an alternative to a fundamental scale lattice as a way of implementing what
we regard as the
necessity of an ontological fundamental scale regularization in order that
field theories be consistent. In a string theory, the compactification space
$K$ is a continuum and what is needed, roughly speaking,  is an
argument for $vol(K)$ being large enough to accommodate a string from each
generation without an overlapping of the domains of the different strings.
A plausible line of reasoning would be that if the compactifying space $K$
becomes too small, the fields in $K$ corresponding to the zero modes could
become so compressed and thereby so strong that the fluctuations giving rise
to the extension of a string (corresponding for example to a gauge particle)
become limited because the maximal string extension is constrained to be of
the order of a typical length related to the field.
For instance, it is quite possible
that a string in a very curled-up Riemannian space with many niches and
corridors will be constrained to be inside one (or several) of these niches
even if this means that the string cannot have the usual
$\sqrt{\alpha^{\prime}}$
extension.

As an example, we can think of a string as a rigid rotor on an $S_2$-sphere.
If the sphere has a very small radius, even a string of small angular
momentum (e.g. $l=1$) about a pole of the sphere can readily attain maximal
extension - namely that corresponding to the ends of the string rotating in
the equatorial plane. So in this case, the length of the string state (leading
to a gauge particle) can be no more than $\pi r$ where  $r$ is the radius of
the
$S_2$-sphere. This means that it is the scale of the curvature of the
Riemannian space-time rather than the $\sqrt{\alpha^{\prime}}$ string extension
that determines the string size in a very strongly curved space-time. In
summary, the conclusion we want to make is that when a compactifying space of
given topology (Euler number) is diminished in linear dimension, at some
stage of this squeezing process, the property dictating the effective string
regularization switches from being a characteristic of the string to being a
characteristic of the curvature of the compactifying space. This in turn
has the effect of ``squeezing down'' the otherwise string-determined
regularization scale.

A variation on this scenario is the suggestion\cite{venez} that a string
cannot be affected by the components in the background (e.g. gravitational)
field having frequencies above the $\frac{1}{\sqrt{\alpha^{\prime}}}$ scale.
Effectively this means that such components do not exist in which case it
would be impossible to have the compactifying space-time of extent less than
that necessary in order to have a curvature that is of the order
$\alpha^{\prime}$. This would in turn mean that $\frac{vol(K)}{reg.\;vol.}$
would be at least of  the same order of magnitude as the Euler number of
$ K$.

Arguments of the type just presented aid in justifying the assertion that,
at least to order of magnitude validity, the effective number of
regularization volumes in $ K$ (namely $\frac{vol(K)}{reg.\;vol.}$)
is larger than  the number of zero modes $N_{gen}$.
This form of argumentation together with the requirement that suspicions
of fine-tuning be avoided leads to the inequality (\ref{ineq}) in the slightly
modified
form

\beq
\beta_{experimental}(\mu_{Planck})\geq \frac{vol(K)}{{\cal N}_0}\geq N_{gen}
\beta_{crit}. \label{ineq1} \eeq

\nin Here $\beta_{experimental}(\mu_{Planck})$ denotes the experimental
values of the inverse squared couplings (one for each of the SMG subgroups
$SU(3)$, $SU(2)$, and $U(1)$ that have been extrapolated to the Planck scale
using the assumption of a ``desert'' in the renormalization group
extrapolation. The lower bound
$N_{gen}\beta_{crit}$ of the inequality (\ref{ineq1}) depends obviously
on $N_{gen}$ as well as $\beta_{crit}$. The latter tends to be larger the
larger the group. This is corroborated for $SU(N)$ groups by the
approximate relationship
\beq \frac{\beta_{crit,\;M.F.A.}}{N}\approx 0.8 \eeq
known from studies using the Mean Field Approximation ($MFA$).
It is shown that for $N_{gen}=3$ and for a $SU(N)$ group with $N=3$, the
lower limit $N_{gen}\beta_{crit}$ is pushed so high up in value that the
experimentally observed inverse squared couplings are only just barely large
enough to escape the ``forbidden gap''. In fact, a gauge group no larger
than $SU(5)$ would be on the verge of predicting less than three generations
of fermions in the ``forbidden gap'' is to be avoided. This allows several
tentative conclusions:

\begin{enumerate}

\item There is not ``room'' for many more than the three experimentally
known generations of quarks and leptons.

\item Grand unification cannot be accommodated in the 4-dimensional scale
region as this would increase $\beta_{crit}$ prohibitively; in fact,
grand unification could only be tolerated at a scale close to the Planck scale.
This implies a large desert.
Already for $SU(2)$ and $SU(3)$ the inequality (\ref{ineq1}) is close
to being  saturated insofar as $N_{gen}$ is experimentally established as
not being less than three.

\item A ``blooming'' of the desert in excess of that already ``known'' to
couple to the SMG Yang-Mills fields is not allowed inasmuch as additional
scalars or fermions at low energies would make unwelcome contributions
to the 4-dimensional Callan-Symanzik $\beta$-function.

\item The compactifying space $K$ should be no larger ``than necessary'';
i.e.,., $vol(K)$ should not exceed the number of generations times the
regulation volume of a fundamental string region.

\item At the string or fundamental scale, the couplings are expected only
to be just enough larger so as not to arouse the suspicion of fine-tuning.

\end{enumerate}

\newpage

\section{The principle of multiple point criticality}\label{pmpc}


A central theme of this work can be stated as a proposed  fundamental
principle - the principle of multiple point criticality. This principle
has evolved from being a specific assertion about the values assumed by
the gauge fine-structure constants in the context of a Yang-Mills lattice
gauge theory to a general statement about Nature that essentially aspires
to be a solution to all
the fine-tuning paradoxes in fundamental physics. In the specific
context in which this principle originated, it can be stated as follows:

\hspace{.2cm}

\nin {\em At the fundamental scale (taken to be the Planck scale), the
actual running gauge coupling constants correspond to the  multiple point
critical values in the phase diagram of a lattice gauge theory.}

\hspace{.2cm}

\nin In this context, the multiple point is a point in the phase  diagram
of  the
lattice gauge theory at which all - or at least many - ``phases''\footnote
{These ``phases'' are explained in Section~\ref{phasecon}.}  convene.  This
point corresponds to  critical  values  for  the  parameters  used  to
describe the form of the  action.  In  the  rather  crude  mean  field
approximation considered here, there is one ``phase''\footnote{In  reality,
such ``phases'' are not necessarily  separated  by  a  phase  boundary
everywhere in the action parameter space; e.g., phase boundaries  that
end at a critical point can be circumvented in going from one phase to
another.} for each combination $(K,H)$ where $K$ is  a  subgroup
of the gauge group $G$ and $H$ is an invariant  subgroup  of  $K$;
i.e., there is a ``phase''  for  each  combination  $(K\subseteq  G,
H\lhd K)$ where $H$ can also be a discrete invariant subgroup.

If one adopts the viewpoint that the
actual existence of a fundamental regulator is a prerequisite for the
consistency of any field theory at sufficiently short distances,
then one must accept phases distinguishable at the scale of the regulator
as also being real and physically existing. In this context, the MPCP
would assert that there is an affinity for parameters corresponding
to the junction of a maximum number of ``regulator-scale'' distinguishable
phases regardless of how the
ontological regulator of field theories at the fundamental scale is
implemented. We have considered the MPCP in the context of a lattice
regulator; however, assuming that such a fundamental regulator can be
formulated or implemented in ways alternative to a lattice (e.g., as strings),
a credible MPCP would need to  give the same values of, for example,
fine-structure constants upon seeking out the junction of a maximum
number of phases distinguishable at the scale of the alternative regulator.


\subsection{Multiple Point Criticality: a prototype for
fine-tuning}\label{mpcfintn}

\subsubsection{What are fine-tuning problems?}

One has a fine-tuning problem when the experimental values of physical
constants are found to have very special values relative to an a priori
expectation.
An explanation of
why constants of Nature have seemingly non-generic values cries out for
a theoretical explanation.
Why for instance is the cosmological
constant so exceedingly small in terms of Planck scale units, which one
would naturally suspect were the fundamental units in Nature?
Why is the Higgs expectation value, which determines the
weak interaction scale, so small compared to
the Planck mass or,
if one believes in Grand Unification,
to the unification scale? Addressing the fine-tuning problems offers
the hope of being able to use hints coming directly from Nature -
rather than from
pure speculation - to learn about what the physics should be like at much
shorter distances than those presently accessible and known.

The values assumed by the fine-structure constants of the Standard Model
also constitute a fine-tuning problem in that these rather remarkably
take the values at ``the'' multiple point. In the formulation that we have
used so far, the multiple point is the point in the
phase diagram of a lattice gauge theory (having a sufficiently
general plaquette action)  at which all - or at least many -
phases convene.
Actually the experimental values of the fine-structure constants only
coincide with the multiple point values if we make the
assumption of
an AGUT gauge group\cite{nonabel,van,lap1,lap2,frogniel,conf1,conf2,gaeta}
with the gauge group $SMG^3$ (which is the 3-fold Cartesian product of the
``usual'' Standard Model Group (SMG): $SMG^3\stackrel{def}{=}
SMG\times SMG\times SMG$).
The ``usual'' Standard Model Group
$SMG=S(U(2)\times U(3))$ arises as the group surviving the Planck scale
breakdown of $SMG^3$ to its diagonal subgroup.
With the anti-grand unified gauge group $SMG^3$,
each generation acquires its own 12 gauge fields
just as in the Standard Model).
Hence, if the assumption of the gauge group $SMG^3$ etc. is accepted as the
immediate (Planck scale) predecessor to the $SMG$, it is indeed
a fine-tuning problem
that is addressed in explaining why the fine-structure constants
should take just the multiple point
values, on a  par with explaining the smallness of, for example, the
cosmological constant.

We propose
that all the fine-tuning problems, including the fine-structure
constant one, are unified (or at least reformulated)
if it is assumed that Nature in general has an affinity for
the multiple point, where
a lot of phases meet for a single combination of the
``intensive parameters''. The latter are really just parameters of the action.
Included among such parameters -
generalised ``coupling constants'' - are lattice artifact parameters.
This is because we take the lattice as really existing, in the sense that
a lattice is one of many ways of implementing what we assume
to be the actual existence of a
fundamental regularization at roughly the Planck scale.
This assumption is inspired by
the fact that the consistency of any field theory seems to
require a cut-off.

\subsubsection{How does multiple point criticality solve fine-tuning problems?}

So our basic explanation for the fine-tuning problems
is that, for some reason, the coupling constants etc. in Nature
take values that correspond to the multiple point where
``all'' (or as many as possible)  phases convene.

An analogous phenomenon is known from other fields of physics:
e.g., a mixture of ice and water (and vapour)
chooses its temperature and pressure to be that of the
melting point (the triple point). By mixtures of well chosen
but not fine-tuned amounts of various different molecules, it would be possible
to realise a multiple point with more than just three phases that
convene.
Here it is the enforced coexistence (insured by choosing combinations of
extensive
quantities like mole number, energy and volume such that the universe is
not realizable as a single phase\footnote{Such a choice is not in itself
a fine-tuned choice but can  be a generic possibility
if the analogies to the heat of melting are large.})
of the phases that consequently enforces the
multiple point values for the chemical potentials, temperature
and pressure so that there is a balance w.r.t. exchange
of molecules between the phases.

It is very tempting to speculate that an analogous scenario, in
which there are (e.g., primordially) fixed but not fine-tuned amounts
of perhaps a
great many extensive quantities (analogous to number of molecules, energy and
volume in the above example),  can provide
an explanation for all fine-tuned quantities in Nature. We shall see in the
sequel that having fixed amounts of extensive quantities in, say, the universe
implies a mild form of non-locality (or vice versa)
that, in turn, implies multiple point
criticality and thereby universally fixed physical constants.


\subsection{Fine-tuning demands non-locality}\label{fintnnonloc}

In this section it is argued that at least the cosmological
constant fine-tuning problem really calls for the breakdown
of the principle of locality in the mild sense referred to above.
Any fine-tuning problem
concerning coupling
constants - among which we may also include the cosmological constant -
calls for some way by which
                   these coupling constants are rendered ``dynamical'', in
the sense that their values are not simply fixed a priori
but can in some way take on values that must (for the sake of
translational invariance) be maintained at a constant
value. That a physical constant (e.g., coupling constant) can depend
on something (i.e., in spite of being constant
as a function of space-time, is not simply fixed a priori) is
the most important content of the
baby-universe\cite{baby,baby1,baby2,baby3,baby4} theory. The latter
theory
indeed aspires with some success to solve the cosmological constant problem.
The baby-universe theory also makes use of an effective
breakdown of the principle of locality and renders the coupling
constants dynamical. Hereby this theory has the right
ingredients needed
if the goal of explaining why the cosmological
constant is small is to be achieved \footnote{Tsamis and Woodard\cite{woodard}
may have a way around this.}.

The problem in local theories - i.e., healthy theories inasmuch as
locality is seemingly well confirmed - is that, if the coupling
constants and in particular the (bare) cosmological constant are
``dynamical'', the strict validity of a  principle of
locality in
the theory would imply that the bare dynamical cosmological
constant could only depend on the situation at the space time point
in question and, indirectly, also on previous times but certainly not
on the future!  However, a bare cosmological constant that is constant
in space-time should already in the first moment after the
Big Bang have had its value fine-tuned once and for all - up to, say, 120
decimal
places -
to the value which makes the {\em dressed} (renormalised) cosmological
constant so exceedingly small (as only can be seen in a background
so depleted of matter as is the case today). That means that the bare constant
had to ``know''
about the details of a vacuum that did not exist at the time when
the bare cosmological constant was already tuned in to the vacuum that would
eventually evolve! Such a tuning of the bare cosmological constant
seems to need some form of
pre-cognition! But this is precisely what is achieved by
breaking the principle of locality.
So we are forced to accept that at least a strict principle of locality
is not allowed,
if we are to explain  the cosmological constant problem in a way
commensurate with having
dynamical (bare) couplings and the renormalization
corrections of quantum field theory with a well-defined vacuum.

\subsubsection{The possibility for having non-locality commensurate with
phenomenology}\label{retrieve}

In~\cite{frogniel} it is argued that, even
if the principle of locality were indeed broken at the
fundamental level, it might be regained
effectively by restricting the breakdown of locality to a form that does not
violate the principle of general reparameterization invariance
of general relativity.

A theory having non-localities extending only over
fundamental scale distances may usually be considered local
when viewed at distances long compared to the fundamental scale.
So the form of non-locality that potentially could be in conflict with
the phenomenologically  obeyed
principle of locality must involve distances much longer than
the fundamental scale (the Planck scale say). We want to argue that
even non-locality over extremely large distances
is not in conflict with what we
regard as the phenomenological validity of the principle of locality,
{\bf \em if} the (long distance) non-locality is restricted to
being invariant under diffeomorphisms or reparameterization.
This class of non-locality includes that of interest to us - namely
nonlocal interactions surviving
at distances much longer than the fundamental scale {\bf and}
that {\em are the same between the fields at any
pair of points in space-time independent of the distance between
these points}.
\begin{figure}
\centerline{\epsfxsize=\textwidth \epsfbox{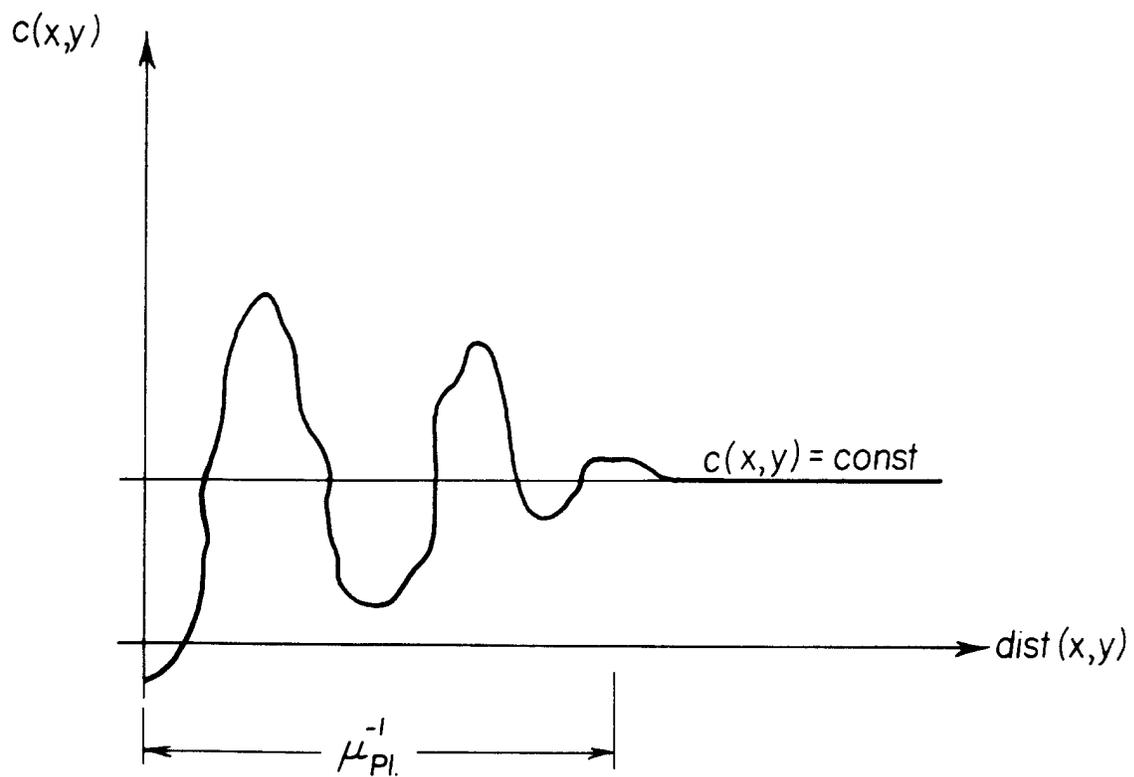}}
\caption[decay]{\label{fig7a} Exponential decay of interaction coefficients
$c(x,y)$ as a result of quantum gravity fluctuations\cite{frogniel} (Figure
from
a lecture by H.B. Nielsen
in his course ``Q.C.D. etc.'').}
\end{figure}

It can be argued\cite{frogniel} that quantum gravity fluctuations
will at large distances $\int_x^yds$ smooth out  the effective interaction
between a pair of fields $\phi(x)$ and $\phi(y)$
in such a way that interaction coefficients $c(x,y)$ decay exponentially
as a function of distance to values independent of
the distance $\int_x^yds$: i.e., $c(x,y)=const.$ Here the $c(x,y)$ are
defined by there being an action term

\begin{equation}
\int\int d^4xd^4y\sqrt{g(x)g(y)}c(x,y){\cal L}_i(x){\cal L}_j(y) \label{eq1}.
\end{equation}

The expected exponential decay of $c(x,y)$ to the long distance
constant value $const.$ has decay rates not
differing by more than a few orders of magnitude from the fundamental scale.
Hence, for the purposes of very long distances, (\ref{eq1}) becomes
$$ const.\cdot\int\int d^4xd^4y \sqrt{g(x)}\sqrt{g(y)}
{\cal L}_i(x){\cal L}_j(y)\stackrel{def}{=}const.\cdot I_iI_j$$
The interaction between a number of fields can similarly be taken into
consideration, in such a way that the long distance physics takes the form
of non-linear functions of integrals $I_j = \int d^4x\sqrt{g(x)} {\cal
L}_j(x)$.
Here the ${\cal  L}_i(x)$'s denote expressions of the type that could be
usual Lagrangian density terms.
The reparameterization invariance
of general relativity is in essence assumed in this argumentation.

Indeed  a
principle like reparameterization invariance
is needed, in order to have a symmetry between all
pairs of space-time points that implies
the same interaction between all such pairs regardless of the distance
separating them.

The important point is that an interaction that has the character of
being the same between the fields located at {\em any} pair of points
(regardless of separation)
is really hardly perceivable as a nonlocal interaction.
Rather we would tend to interpret such  effects as being a  part of the
laws of Nature, since such effects are forever everywhere the same. Such
an omnipresent effect is therefore effectively unobservable and we
would not in practice see any deviation from locality.

Finally, having once renounced a strict principle of locality,
it is natural to go a step further and enquire as to whether it might be
possible for fields to depend on
more than one space-time point (separated by large distances).
We propose that such fields might cause a spontaneous breakdown of
reparameterization  invariance, so that distant points in space-time become
related.
Degrees of freedom at distant points, related
by this breakdown,
would be interpreted as several degrees of freedom at the {\em same point}.
Such a field replication mechanism that comes from ``explaining away''
ontological non-locality would be welcome, as a possible explanation
for the 3-fold replication seen in the three generations of quarks and
leptons. This possibility is discussed more extensively in
Section~\ref{twopos}. That a 3-fold replication mechanism for fermions would,
probably unavoidably, also provide a 3-fold replication of bosons is
also  a very welcome prediction in the context of the
AGUT gauge group model that uses the gauge group $SMG^3$ (i.e., the
3-fold Cartesian product of the Standard Model Group). This gauge group,
the Planck scale breakdown of which yields the normal SMG in our model,
is an important ingredient in our predictions of gauge couplings using
multiple point criticality.

\subsubsection{Non-locality as the underlying explanation of the affinity of
Nature for a multiple point}\label{nonlocmpc}

We shall now argue that the assumption of non-locality implies having the
the  principle of multiple point criticality. For the purpose of
explaining why Nature seeks out the multiple point, we assume in accord with
the
argumentation of Section~\ref{retrieve} that we have
fields $\phi$ depending on a single space-time
point that interact non-locally, in such a way that
the long distance remnants of the
nonlocal interactions between
fields $\phi(x)$ and $\phi(y)$  are the {\em same} for all pairs of
space-time points
$x$ and $y$. As the reparameterization invariance of general relativity
implies this symmetry between space-time points, we write our nonlocal
action as a non-linear function of
reparameterization invariant integrals of the form

\begin{equation}
I_j \stackrel{def}{=} \int d^4x\sqrt{g(x)} {\cal L}_j(x)
\end{equation}
\noindent where the ${\cal L}_j$ denote the usual sort of terms in a local
Lagrangian density. An  ${\cal L}_j$ could, for example,  be a polynomial
of degree $n$ in
the (scalar) field $\phi(x)$: ${\cal L}_j=\phi^n(x)$ or the $k$th partial
derivative of such a field: ${\cal L}_j=\partial^k\phi^n(x)$ (somehow
made rotationally invariant).

We achieve non-locality by considering actions
$S_{nl}( I_1,I_2,...,I_N)$
that are {\em non-linear}
functions of the integrals $I_j$. Note that nonlinearity is tantamount to
non-locality, because
nonlinearity in the quantities $I_j$ implies having integrals with more
than one integration variable; e.g., an action term $\propto I_iI_j$
is indeed nonlocal because

\[ I_iI_j=\int\int d^4xd^4y \sqrt{g(x)}\sqrt{g(y)}
{\cal L}(x)_i{\cal L}(y)_j\]

\noindent contains contributions from fields at independent
(and therefore in general different) space-time points $x$ and $y$.
Note that  had we taken a linear function of the integrals $I_j$:
$S= \sum g_j I_j$,
we would get an ordinary local action.

An important property of the reparameterization invariant integrals $I_j$
is that any function of such integrals - even a non-linear and thereby
nonlocal one - is also reparameterization invariant.
So we can say that we restrict the non-locality allowed in our model
to the non-locality that comes about, due to having
an action that is a non-linear function of a
lot of integrals $I_j$ having integrands corresponding to the
various Lagrangian densities ${\cal L}_j$ being used.
Our speculation is that this form of non-locality (formulated with the $I_i$'s)
is really the only form that can survive at long distances, when
reparameterization invariance is insisted upon (see however
Section~\ref{twopos} for
a generalisation).

However, we now  want to argue that this restricted form of non-locality
would not be easily observable and could therefore really exist
in Nature without ever having been observed as, for example, an ``action at
a distance'' sort of  non-locality.
Rather we would say that the only traces of the restricted form of
non-locality that we consider are
(some) solutions of fine-tuning problems.

Formally we can think of having the functional integral of Nature with
a nonlocal action $\hat{S}_{nl}$ that is a functional of the fields $\phi$
of the theory:

\begin{equation}
\int {\cal D}\phi e^{i\hat{S}_{nl}[\phi]}.
\end{equation}
where

\begin{equation}
\hat{S}_{nl}[\phi]\stackrel{def}{=}S_{nl}(I_1[\phi],I_2[\phi],\dots I_N[\phi])
\end{equation}
and $\phi$ is used as a symbol for all the fields of the theory.
As with any classical approximation for a field theory, it can be argued that
this functional integral is dominated by field values in the
neighbourhood of the field values $\phi_0$ for which the action is stationary:
\begin{equation}
\frac{\delta \hat{S}_{nl}}{\delta \phi}|_{\phi_0}=0 \label{eq2}. \end{equation}

\noindent Were the quantities $I_i$ effectively independent, we would deduce
from Eq.~(\ref{eq2})  that

\begin{equation}
\frac{\partial S_{nl}(\{I_i\})}{\partial I_j}=0\label{eq3}.
\end{equation}

If there are some necessary
relations between the $I_i$'s, because of their functional forms as
 functionals  of the same fields $\phi$, the $I_i$'s may be constrained to
take values in only some allowed region of the space spanned by the
$\{I_i\}$ (see Figure~\ref{nl1} for an example).
In the event that $S_{nl}$ has an extremum on the border of the
allowed region, we should only require that the variation of $S_{nl}$
vanish along this border. In this event,

\begin{equation}
\frac{\partial S_{nl}(\{I_i\})}{\partial I_j}=\lambda a_j. \label{eq7}
\end{equation}
where the variation along the border obeys the restriction
$\sum a_jdI_j=0$ and $\lambda$ is a Lagrange multiplier. If the border is of
co-dimension greater than one, there will be a Lagrange multiplier for each
co-dimension and a corresponding contribution in Eq.~(\ref{eq7}).

We illustrate the idea of how a nonlocal reparameterization invariant action
can lead to fine-tuning by an example in  which we ignore
derivative terms in the action. Thus we consider only a nonlocal pure
scalar field potential type action, in which the
potential term is nonlocal at  very long distances in such a way that
the interaction is independent of the
distance between space-time points. This is insured by taking a nonlocal
potential $\hat{V}_{nl}[\phi]=V_{nl}(\{I_j[\phi]\})$ that is a (non-linear)
function of the quantities $\{I_j\}$.

We now seek the minimum
for a nonlocal potential $V_{nl}(\{I_j\})$ in a space spanned by
quantities $I_j/V$ - i.e., the volume densities of the quantities $I_j$.
For illustrative purposes we consider the simple situation in which
there are just two quantities $I_1$ and $I_2$ where

\begin{equation}
I_1=\int d^4x\sqrt{g(x)}\phi(x)\stackrel{def}{=}\int d^4x
\sqrt{g(x)}v_1(\phi(x))
\end{equation}
and
\begin{equation}
I_2=\int d^4x\sqrt{g(x)}(\phi^4(x)-5\phi^2(x)+\phi(x))\stackrel{def}{=}
\int d^4x\sqrt{g(x)}v_2(\phi(x))
\end{equation}
where $v_2(\phi(x))$ is some polynomial that, for illustrative purposes,
is taken as being 4$th$
order; e.g., $v_2(\phi(x))=\phi^4(x)-5\phi^2(x)+\phi(x)$.

One should bear in mind that the integrals $I_1$ and $I_2$
of field polynomials over space-time cannot take values that are
completely independent of each other. If, for instance,
the integral $I_2$ of $v_2(\phi(x))$ over space-time is required
to be rather
small, the value of $\phi$ cannot be too large over most of space-time. This
in turn would limit the possible values of the
integral over space-time of $\phi$ itself. Taking such relationships into
account leads to an allowed region of values for the $\{I_j\}$.
Including many polynomials in the fields $\phi$ can lead
to allowed regions that can be somewhat complicated. We shall continue to
restrict our example to the two quantities $I_1$ and $I_2$ defined above.
Figure~\ref{nl1} depicts schematically the allowed  region of $I_1$ and $I_2$
values,
with values of $I_1/V=\langle \phi \rangle$ plotted along the abscissa
and $I_2/V=\langle v_2(\phi) \rangle$ along the
ordinate. The average $\langle \rangle$ denotes an average over space-time.
The part of the boundary of the convex envelope of allowed values
drawn as the heavy solid
curve corresponds to having a constant $\phi(x)$ in space-time: $\phi(x)=
I_1/V$.

\begin{figure}
\centerline{\epsfxsize=\textwidth \epsfbox{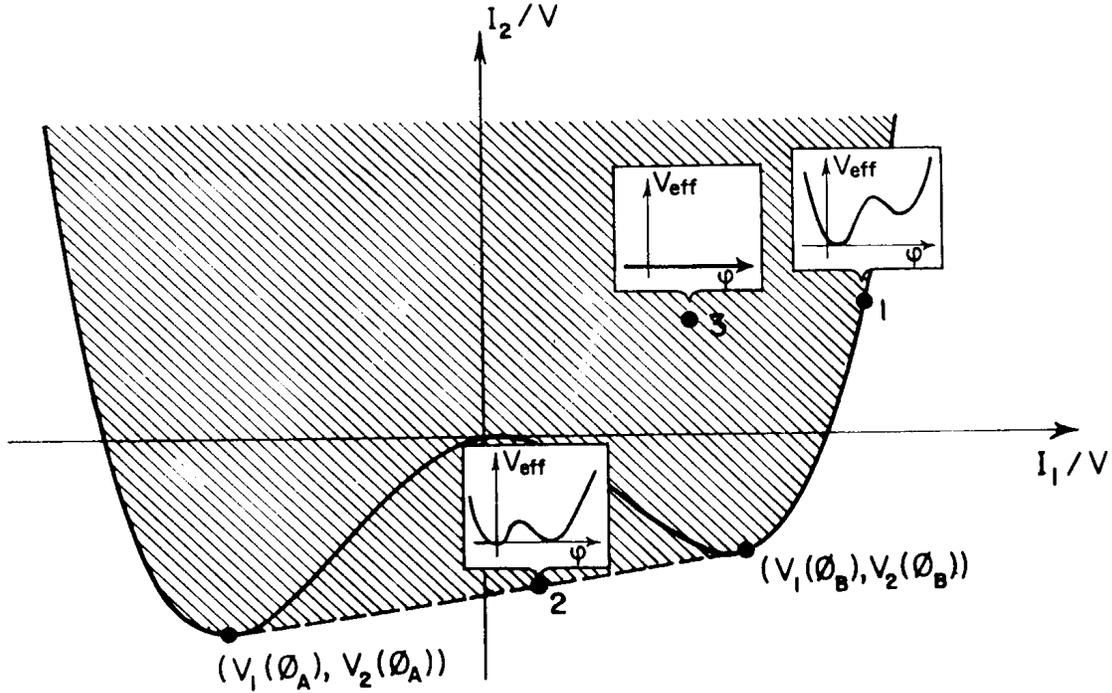}}
\caption[A prototype nonlocal potential]
{\label{nl1}The nonlocal potential $V_{nl}(I_1,I_2)$ can have its minimum at
any
point in the interior (e.g., point 3) or on the boundary (e.g., points 1 or 2)
of the convex closure
of allowed $(I_1/V,I_2/V)$ combinations (the cross-hatched area). The three
inserts show the characteristic form of the effective local potential
$V_{eff}$ at the three generic possibilities for the minima of $V_{nl}$. For
minima of $V_{nl}$ at interior points, $V_{eff}$ is just flat (see insert at
point 3). At minima of $V_{nl}$  on the heavy solid curve portion of the
boundary, the characteristic feature of $V_{eff}$ is one absolute minimum
(see insert at point 1) corresponding to $(I_1/V,I_2/V)$ combinations
realisable in a universe  with just one
(dominant) value of $\phi(x)$ in the vacuum. At minima of $V_{nl}$ located at
boundary points indicated by the heavy broken
line, the characteristic feature of $V_{eff}$ is {\em two} equally deep minima
(see insert at point 2), corresponding to $(I_1/V,I_2/V)$ combinations that
can be realised as the vacuum of a universe having different dominant
constant values of $\phi(x)$ in different space-time subregions.}
\end{figure}

A priori, the nonlocal potential $V_{nl}(I_1,I_2)$ can
have its minimum at
any point in the interior (the
cross-hatched region of Figure~\ref{nl1}) or on the boundary of the allowed
region
(convex envelope). The  heavy solid curve of Figure~\ref{nl1}
corresponds to the
$(I_1/V,I_2/V)$ combinations that can be realised in a universe
having just one dominant value of
(i.e., almost everywhere in space-time constant)
$\phi(x)$ in the vacuum. Here the symbol $V$ denotes the
volume of the universe.
That is, $\phi(x)=\langle \phi \rangle$ for
almost all of space(time).
Allowed $(I_1/V,I_2/V)$ combinations, not lying on the
heavy solid curve portion of the boundary of the convex envelope,
cannot be realised in a universe
having a single dominant (for all space-time)  constant value of $\phi(x)$.
However, such points can
be realised by means of a positively weighted linear combination
of points on the heavy solid curve.
Such points would correspond to a
universe the vacuum of which has different dominant constant values
of $\phi(x)$ in
different space-time subregions, where the extent of these subregions
in space-time is
proportional to the positive weights needed, in the combination
of the several constant values of $\phi(x)$, to
get a universe having the average values $\langle \phi \rangle = I_1/V$
and $\langle v_2(\phi) \rangle=I_2/V$.

In Figure~\ref{nl1}, we also indicate with the points 1, 2 and 3
representatives for
the three generic classes of points, in the convex envelope of allowed
$(I_1,I_2)$
combinations, at which $V_{nl}(I_1,I_2)$ can have its minimum:
point 3 represents
the interior, point 1 represents the class of points on the heavy solid curve
coinciding with the boundary of the convex envelope,
and point 2 is a prototype for the remainder of the boundary of the convex
envelope.
It is reasonable
to claim that all of these 3 prototypes represent generic possibilities
- even though one might
a priori think that a minimum on the border would require some degree of
fine-tuning\footnote{Actually, a point seeking a minimum in the allowed
region would statistically often tend to accumulate somewhere along the
border.}.

A moment's
reflection can perhaps convince the reader that a point such as 3
can be obtained as a suitably (positively) weighted combination of infinitely
many points on the heavy solid curve in Figure~\ref{nl1}. Points on this
heavy solid line in Figure~\ref{nl1} (i.e., points of type 1)
correspond to universes that can only be realised
with fields that are almost everywhere equal to the average values of
these fields
(i.e., essentially the same constant value
for $\phi(x)$ at almost all space-time points $x$):
$\forall x,\; \phi(x)=\langle \phi \rangle$.
Indeed a point such as 1 on the convex closure of the convex envelope of
allowed $(I_1/V, I_2/V)$
combinations
can only be obtained as a single-term
``combination'' of different
constant values of
$\phi(x)$ - namely the constant value of $\phi(x)$ at the point 1.

The final prototype point at which
$V_{nl}(I_1,I_2)$ can have its minimum - the interesting case as it
turns out - is  point 2, located on the closure of the
convex envelope that is not on the heavy solid curve.
Such  a point corresponds to a universe {\em not} realisable with a
{\em single} constant (i.e.,
everywhere in space-time  constant)
value of $\phi(x)$.

At such a point, there are only
two constant values of $\phi(x)$ (having one constant value
at  points in some space-time subregion and the other constant value
at all other points in space-time)
that together can participate in a weighted combination
that can realise the prototype  point 2.
These are the constant values, $\phi=\phi_A$ and
$\phi=\phi_B$,
at the points on the convex closure at which the heavy
broken line of universes, un-realisable with single constant values
of $\phi(x)$, is tangent
to the heavy solid curve (corresponding to all universes that
{\em are} realisable with a single
value of $\phi(x)=\langle \phi \rangle$):

\begin{equation}
\frac{I_1}{V}=\langle \phi_{point\;2}\rangle =
w_A\phi_A+w_B\phi_B\;\;\;(w_A+w_B=1)
\end{equation}
where $w_A$ is proportional to the  extent of the space-time region having the
constant value
$\phi_A$ and
$w_B$ to that of the space-time region having the constant value $\phi_B$.

We want to examine the effective local potential in the three cases, in which
the nonlocal potential is located at the three types of points 1, 2 and 3.

The effective local potential $V_{eff}$ is defined
as that function of $\phi$ for which the
derivatives are equal to the corresponding (functional) derivatives of
the nonlocal potential $V_{nl}$.
We can think of $V_{eff}$ as the potential observed in a
laboratory  very small compared to the volume
of the universe and arbitrarily placed at some space-time point.
The derivative of $V_{eff}$ is the change in $V_{eff}$
observed in the laboratory, when the value of the field $\phi$
is changed only in the laboratory and kept constant at all other points of
space.
If $\phi$ is changed by a finite amount in the laboratory,
the effective local potential can be integrated
up: $V_{eff}(\phi_A)-V_{eff}(\phi_B)=
\int^{\phi_B}_{\phi_A} V_{eff}^{\prime}(\phi)d\phi$.

Formally we make the definition

\begin{equation}
\frac{\partial V_{eff}(\phi(x))}{\partial \phi(x)} \stackrel{def}{=}
\frac{\delta V_{nl}(\{I_j[\phi]\})}{\delta \phi(x)}
\left|_{near\;\;min.} \right.=
\sum_i\left(\frac{\partial V_{nl}(\{I_j\})}{\partial I_i}\frac{\delta
I_i[\phi]}
{\delta \phi(x)}\right)\left|_{near\;\;min.} \right. \label{eq11}
\end{equation}
\[ =\sum_i\frac{\partial V_{nl}(\{I_j\})}{\partial I_i}
\left|_{near\;\;min.} \right. v_i^{\prime}(\phi(x))  \]
This definition implicitly assumes that, to a very good approximation,
$V_{nl}$ takes on its lowest
possible value. But this does not preclude small regions
of space-time from having $\phi$ values that deviate, by essentially any
desired amount, from the average value(s) in the vacuum or vacua\footnote
{i.e., more than one vacuum in the,
for us, interesting case of competing vacua corresponding to different
phases in different regions of space-time.}.
The subscript ``near min'' in this formula
denotes  the approximate ground state of the whole universe, up to deviations
of $\phi(x)$ from  its vacuum value (or vacuum values for a multi-phase vacuum)
in relatively small regions.

As a solution to Eq.~(\ref{eq11}) we have

\begin{equation}
V_{eff}(\phi)=\sum_i \frac{\partial V_{nl}(\{I_j\})}{\partial I_i}v_i(\phi)
\label{eq12} \end{equation}
where the $v_i(\phi)$ are the (field polynomial) integrands
of the
``extensive'' (reparameterization invariant) quantities
$I_j=\int d^4x\sqrt{g(x)}v_j(\phi(x))$. We can identify the
$\frac{\partial V_{nl}(\{I_j\})}{\partial I_i}$ as intensive quantities
conjugate to the $I_i$.

That Eq.~(\ref{eq12})
solves Eq.~(\ref{eq11})  is easily seen by differentiating
Eq.~(\ref{eq12}) and using
that the right hand side of Eq.~(\ref{eq11}) is
$\sum_j \frac{\partial V_{nl}}{\partial I_j}\frac{\delta I_j[\phi]}{\delta
\phi}
=\sum_j \frac{\partial V_{nl}}{\partial I_j}v_j^{\prime}(\phi(x))$. The seeming
$x$-dependence of this right-hand side of  (\ref{eq11}) for  prescribed
values of $\phi(x)$  is effectively absent due, at the end,  to
the reparameterization invariance hidden in the form of the $I_j$'s.

We now proceed with a study of the effective potential $V_{eff}$
for the field configurations $\phi$ near
the minimum of $V_{nl}$, when this minimum is near one of the three types
of points 1, 2 and 3.

At an interior point of type 3, the absolute minimum of $V_{nl}$ is also
a local minimum and $\frac{\partial V_{nl}(\{I_j\})}{\partial I_i}=0$
for all $i$. Accordingly,

\begin{equation}
V_{eff}=\sum_i\frac{\partial V_{nl}}{\partial I_i}\left|_{near\;\; min.\;\;
at\;\;``3"} v_i(\phi) =0 \right. \end{equation}
So when $V_{nl}$ has its minimum in the interior, the effective potential
$V_{eff}$ is flat. Recall that an interior point such as 3
can be obtained as a suitably (positively) weighted combination of infinitely
many points on the boundary of the allowed region. This is related
to the fact that $V_{eff}$ has infinitely many
minima (because it is flat) at an interior point at which $V_{nl}$ has its
minimum.

If $V_{nl}$ has its minimum at a point of the type 1 or 2 (i.e., on the border
of the convex envelope), we have in general that

\begin{equation}
V_{eff}=\sum_i\frac{\partial V_{nl}}{\partial I_i}\left|_{near\;\; min.\;\;
at \;\;``1"\;\;or\;\;``2"} \;v_i(\phi) \neq0 \right. \end{equation}
because in general
$\frac{\partial V_{nl}(\{I_j\})}{\partial I_i}\neq 0$ at an
absolute minimum of $V_{nl}$ located on the boundary of the
convex envelope.

For the minimum of $V_{nl}$ located at a point of the type 1, there is
only one value of $\phi(x)$ realised in the vacuum (i.e., in extended regions
of space-time) - namely the value $\langle \phi \rangle=I_1/V$ at which the
minimum
of $V_{nl}$ is located.
Accordingly, $V_{eff}$ has a {\em single} deepest minimum
 - namely that at
$ \phi_1$ where the latter denotes the value of $I_1/V$ at the
point 1 on the convex closure where $V_{nl}$ has its minimum.

That there is only one deepest minimum at a type 1 point
is readily seen, by showing that the assumption of a second equally
deep minimum at some other value $\phi_C$ would lead to a contradiction.
First we make the observation that the gradient of $V_{nl}$, which
cannot be zero for a generic point of type 1, is perpendicular
to the tangent to the convex envelope at point 1. Secondly, note that the
line connecting $\phi_1$ with $\phi_C$ determines a chord of the
convex envelope that necessarily lies in the interior of the convex envelope.
A displacement away from point 1, along such a chord, has therefore always a
component along the gradient of $V_{nl}$. But moving along this chord,
defined by the  two equally deep minima in
$V_{eff}$ at respectively $\phi_1$ and $\phi_C$, corresponds to replacing
$\phi_1$ by $\phi_C$ (or vice versa) in a small space-time region
at no cost in energy. This is inconsistent with the observation
that a displacement along this chord necessarily has a component
along the
gradient of $V_{nl}$. We conclude that $V_{eff}$ cannot have two
(or more) equally deep minima.

The most interesting case is that for which the minimum of $V_{nl}$ is located
at a point of type 2 (with coordinates denoted as
$(I_1,I_2)_{type\;2}$, see Figure~\ref{nl1}) on the convex closure of the
convex envelope of the
allowed region. Such a universe can\underline{{\bf not}} be realised with
$(\phi(x),v_2(\phi(x)))=(I_1/V,I_2/V)_{type\;2}$.
It can be shown that, in order to realise
$(I_1/V,I_2/V)_{type\;2}$, only
the two constant contributions $(\phi_A,v_2(\phi_A))$ and
$(\phi_B,v_2(\phi_B))$  can participate
in the (unique) weighted combination.
A universe corresponding to the point  $(I_1/V,I_2/V)_{type\;2}$
could be realised with the field
\begin{equation}
\phi(x)=\left\{ \begin{array}{c} \phi_A \mbox{\footnotesize
{  for $x$ $\in R_A$ }}  \\
\phi_B \mbox{\footnotesize{ for $x$ $\in R_B$}}
                               \end{array} \right.. \end{equation}
where $R_A$ and $R_B$ are large regions of space-time.

It is interesting that two minima of $V_{eff}$ will be seen to have the
{\bf same} depth.
This is tantamount to fine-tuning, in
that the relation $V_{eff}(\phi_A)=V_{eff}(\phi_B)$ can be used to
eliminate a bare parameter (for example, the bare Higgs mass $m_H$).
Having two equally deep minima of $V_{eff}$, for $\phi_A$ and $\phi_B$,
is characteristic of a vacuum with two coexisting phases. This is
tantamount to being at the multiple point.

That we in fact have $V_{eff}(\phi_A)=V_{eff}(\phi_B)$, when the minimum
of $V_{nl}$ is at a type 2 point, can be seen by considering the directional
derivative of $V_{nl}$ along the line connecting the points
$(\phi_A,v_2(\phi_A))$ and $(\phi_B,v_2(\phi_B))$. This line is
parameterised by
\begin{equation}
(I_1/V , I_2/V ) = \xi (v_1(\phi_A),v_2(\phi_A)) +
(1-\xi) (v_1(\phi_B),v_2(\phi_B))
\end{equation}
with $\xi$ as the parameter. Along this line we have
\begin{equation}
\frac{dI_j/V}{d\xi} = v_j(\phi_A)-v_j(\phi_B)
\end{equation}
for $j=1,2$. The directional derivative is
\begin{equation}
\frac{dV_{nl}}{d\xi}= \sum_j \frac{\partial V_{nl}}{\partial I_j}
\frac{dI_j}{d\xi}
= \sum_j  \frac{\partial V_{nl}}{\partial I_j} ( v_j(\phi_A)-v_j(\phi_B))V
=V_{eff}(\phi_A) -V_{eff}(\phi_B) \label{AB}
\end{equation}
which means that, if Eq.~(\ref{AB})  is zero as must be the case at an
(absolute) minimum at a point of type 2, the effective
potential $V_{eff}$ will take the same value in $\phi_A$ and
$\phi_B$. Let us emphasise that having demonstrated
$V_{eff}(\phi_A)=V_{eff}(\phi_B)$
amounts to having derived multiple point criticality
at least with finite
probability, i.e. in one generic situation.

\subsubsection{Multiple Point Criticality as the solution of non-locality
paradoxes}\label{paradoxx}

This section begins with the familiar example (in three space dimensions)
of temperature
fine-tuning that can be accomplished by enforcing the coexistence of ice and
water phases by fixing the values of extensive
quantities. The generalisation of phase coexistence enforced by having fixed
values of extensive quantities is subsequently examined for
(4-dimensional) space-time. It will be
seen that in general, intensive quantities (conjugate to fixed extensive
quantities) - we sometimes refer to such quantities as generalised
``coupling constants'' - depend on the {\em future} (as well as on the past).
Such behaviour is of course blatantly non-local and also can lead to
paradoxes. It will be seen however that such paradoxes can  be
resolved\cite{bomb}. The interesting point is that the resolution of paradoxes
that would arise by having non-locality is a compromise that Nature can only
realise by obeying the principle of multiple point criticality.

Consider first an ice-water system at 1 atm. pressure. The system -  let it be
enclosed in  a soft, insulated plastic bag - has a fixed total energy $E_{TOT}$
and number of moles $n_{H_2O}$. Under what conditions is the system forced to
exist in two phases? The coexistence of both phases is insured (and the
temperature fine-tuned to $0^o$ $C$ ($273^o$ $K$)) if the total energy
$E_{TOT}$ lies in the interval

\beq n_{H_2O}\int^{273^o\;K}_{0^o\;K}C_{p,ice}(T)dT < E_{TOT} <
n_{H_2O}\int^{273^o\;K}_{0^o\;K}C_{p,ice}(T)dT+
n_{H_2O}\cdot (molar\;heat\; of\;fusion) \label{interval} \eeq
\[ \mbox{($C_{p,ice}$ is the molar heat capacity of ice at constant
pressure)}\]

\nin because a total energy $E_{TOT}$ in this interval cannot be realised as
$n_{H^2O}$ moles of ice alone or $n_{H^2O}$ moles of water alone. Rather
an $E_{TOT}$ in the interval (\ref{interval}) (or equivalently,
if $E_{TOT}$ is such that $0<
\frac{E_{TOT}}{n_{H_2O}}-\int^{273^o\;K}_{0^o\;K}C_{p,ice}(T)dT <
molar\;heat\;of\;fusion$) can only be realised as a mixture
of a well defined (equilibrated) {\em mixture} of ice and water because the
heat of fusion is a finite quantity (because the ice-water transition is first
order). In fact the larger the heat of
fusion, the better is a system as a ``fine-tuner'' because there is a better
chance that a randomly chosen $E_{TOT}$ will fall in the interval
(\ref{interval}).

By also fixing the extensive quantity volume $V$ (in addition to $E_{TOT}$
and $n_{H_2O}$) using say a thermos flask (instead of a soft, insulated
plastic bag), there are many combinations of the extensive quantities
$V$, $E_{TOT}$ and $n_{H_2O}$ that can only be realised by maintaining the
presence of the three phases ice, liquid water, and water vapour at the
triple point of water. Here the temperature and pressure are fine-tuned to
the triple point values. The triple point is a multiple point where three
phases coexist instead of just two as is the case for  values
of $E_{TOT}$ and $n_{H_2O}$ satisfying (\ref{interval}) (with unconstrained
volume). As a consequence of also constraining the volume $V$, we can
have two fine-tuned intensive parameters instead of one.

In suggesting how we humans might explain the presence of fine-tuned quantities
in Nature, we can compare ourselves to an intelligent, well-educated fish
that makes the observation that the temperature of the water in which it
lives is suspiciously fine-tuned to the value of $0^{\circ}\;C$. Despite the
fact that the clever fish had never observed other than the liquid phase of
water, it might deduce that in some remote region of its large tank there
must be an ice phase of water that coexists in equilibrium (or close to
equilibrium) with the water phase in which the fish lives.

It should be emphasised that the mechanism of fine-tuning by enforced
coexistence of phases is, to be effective, contingent upon having phase
transitions
that are strongly first order (corresponding to the largest possible ``heats
of fusion'',  ``heats of vaporisation'' and ``heats of sublimation''
in the water system analogy). The more strongly first order the phase
transitions are, the more
combinations of fixed amounts of extensive variables - even random
combinations - that can only be realised as a mixture of phases.

Analogies to the ice-water system just described can be sought in
treating the universe having a (4-dimensional) space-time volume $V$.
What does it mean to have coexisting phases in (4-dimensional) space-time?
In addressing this question, consider first the effective potential
(\ref{eq12}) in the special case that

\beq V_{nl}(\{I_j\})=V_{nl}(I_2,I_4)\eeq
where $I_2$ and $I_4$ (the extensive quantities) are the
spacetime integrals over respectively the polynomials $|\phi^2(x)|$ and
$|\phi^4(x)$:

\beq I_2=\int_Vd^4x\sqrt{g(x)}|\phi^2(x)| \eeq

\nin and

\beq I_4=\int_Vd^4x\sqrt{g(x)}|\phi^4(x)|. \eeq

\nin In this case, (\ref{eq12}) becomes

\beq V_{eff}=\frac{\partial V_{nl}(I_2,I_4)}{\partial I_2}|\phi^2(x)|+
\frac{\partial V_{nl}(I_2,I_4)}{\partial I_4}|\phi^4(x)|\stackrel{def}{=}
\frac{1}{2}m^2_{Higgs}|\phi^2(x)|+
\frac{1}{4}\lambda|\phi^4(x)| \label{msqh}\eeq

\nin where the right hand side of this equation defines $m^2_{Higgs}$ and
$\lambda$; the right hand
side of (\ref{msqh}) is recognised as the prototype SMG scalar potential
at the tree level.
Here it serves to illustrate how ``coupling
constants'' (think of $m^2_{Higgs}$ as a prototype for a ``coupling
constant'') are related to the nonlocal potential $V_{nl}$:

\beq \frac{1}{2}m^2_{Higgs}=\frac{\partial V_{nl}(I_2,I_4)}{\partial I_2}
\label{cc}\eeq

\beq \frac{1}{4}\lambda=\frac{\partial V_{nl}(I_2,I_4)}{\partial I_4}. \eeq

\nin Of course the
form of $V_{nl}$ is, at least {\em a priori}, completely unknown to us
so - for example - (\ref{cc}) cannot be used to calculate the coupling
constant $m^2_{Higgs}$.

The potential

\beq V_{eff}=\frac{1}{2}m_{Higgs}^2\phi^2+\frac{1}{4}\lambda \phi^4
\label{tree}\eeq
is for $m_{Higgs}^2<0$
characteristically different from that for $m_{Higgs}^2>0$: in the usual
scenario, the case $m_{Higgs}^2<0$ leads to spontaneous symmetry breakdown
and thereby masses
for the gauge bosons and fermions without compromising renormalizability
if the Lagrangian has gauge symmetry.
This is just standard physics (without non-locality).
Actually we want to consider a potential having two minima; such a potential
comes about when radiative corrections to (\ref{tree}) are taken into account.
The next order approximation to (\ref{tree}) is obtained\cite{sher} by
considering  one particle irreducible (1PI)
diagrams with a single loop and with zero external momenta. Assuming that
scalar loops are negligible (justifiable if the Higgs scalar mass and thereby
$\lambda$ is small), the effective
potential $V_{eff\;1-loop}$ so obtained is of the form

\beq V_{eff\;1-loop}=\frac{1}{2}m^2_{Higgs}\phi^2+\phi^4(\frac{\lambda}{4}+
B\log\frac{\phi^2}{\Lambda}) \label{1loop} \eeq

\nin where $\Lambda$ is an arbitrary (non-vanishing) mass scale and
$B$ depends on the Yakawa coupling $g_Y$ and gauge couplings $g$ and
$g^{\prime}$:

\beq B=\frac{3}{16}(3g^4+2g^2g^{\prime 2}+g^{\prime 4})-g_Y^4 \eeq

Using a renormalization group
improved effective potential of the type (\ref{1loop}), it can be shown
that if the top quark is very heavy, the potential will be unbounded:
an asymmetric  minimum will still exist at the weak scale, but at some
large value of $\phi$, the potential will turn over and fall to negative
infinity\cite{sher}. If the top quark mass is not too heavy, the Yukawa
coupling will eventually be small enough so that $B$ will be positive; the
potential will then turn back around. There will be two relative
asymmetric minima (i.e., both minima at non-vanishing values of $\phi$).
The two relative minima of such a potential would  correspond to two
possible vacua of the universe.
Which of these vacua would dominate depends on
the value of, for example, the parameter $m^2_{Higgs}$. The case of interest
here will be that for which  the vacua are essentially degenerate; this is
tantamount to the coexistence of more than one phase.

\vspace{.4cm}
\nin{\bf Coexistence of more than one phase in a simple model of the
universe }
\vspace{.3cm}

Thinking now of the double well potential, we want to use a simple
cosmological model
for the universe in order to illustrate
that one of the ways of achieving
the coexistence of several
phases in space{\em time} is to have different vacua in different
epochs of the history of the universe.

\vspace{.4cm}
\nin{\bf Digression}
\vspace{.3cm}

Non-locality can be introduced
using the assumption that the values of (perhaps many)
space-time integrals $I_i$  - extensive quantities - are fixed
at  values $I_{i\;fixed}$. This amounts to choosing
the non-local action $S_{nl}(\{I_j\})$ such that

\beq \exp (S_{nl}(\{I_j\}))=\prod_i \delta(I_i-I_{i\;fixed}).\label{delact}
\eeq

\nin This highly non-linear function of the space-time integrals $I_i$
is just one of many possible ways of having non-locality. It turns out, as
will be discussed in the sequel, that the essential feature - namely that
the presence of non-local (but reparameterization invariant) Lagrangian
contributions renders multiple point
values of intensive parameters generically very probable - is rather
insensitive to which non-local function of the $I_i$ is considered.
The partition function corresponding to (\ref{delact}) is then

\beq Z=\int{\cal D}\phi\prod_i\delta(I_i-I_{i\;fixed}). \eeq

\nin Rewriting the $\delta$-function we have

\beq
Z=\int{\cal D}\phi\int\prod_j d\beta_j e^{i\beta_j(I_j-I_{j\;fixed})}
\eeq

\nin and then

\beq \langle I_i-I_{i\;fixed} \rangle=
\frac{1}{Z}
\int{\cal D}\phi\int \frac{\partial}{\partial\beta_i}
\prod_j d\beta_j e^{i\beta_j(I_j-I_{j\;fixed})}.
\eeq

\nin Requiring $\langle I_i\rangle -I_{i\;fixed}=0$ determines the
value of the dominant $\beta_i$
($\langle I_i \rangle$ is function of the set $\{\beta_j\}$).
The $\beta_i$ are the intensive variables conjugate to the $I_i$.
The important
point is that for a first order phase transition, there is a finite
probability that the set $\{I_{i\;fixed}\}$ will fall within some range
of values
that can only be realized as two or more coexisting phases.

\begin{figure}
\centerline{\epsfxsize=\textwidth \epsfbox{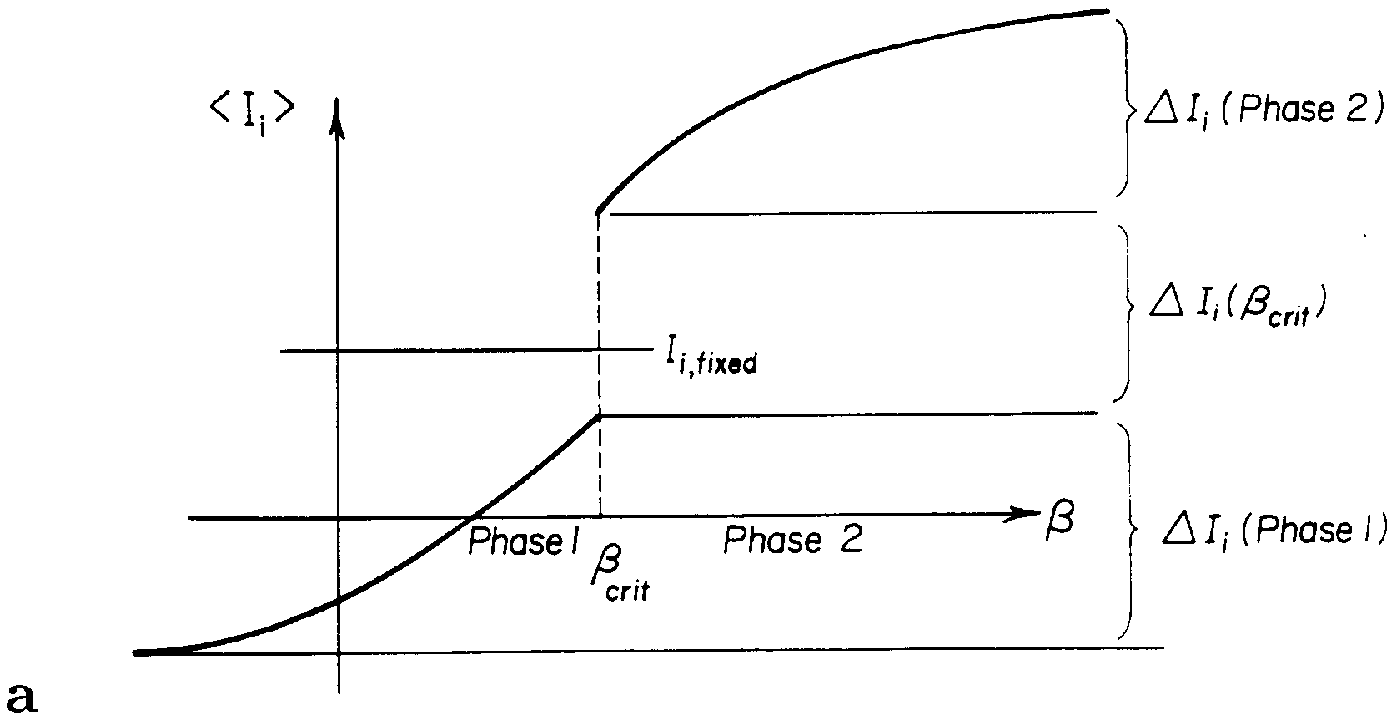}}
\centerline{\epsfxsize=\textwidth \epsfbox{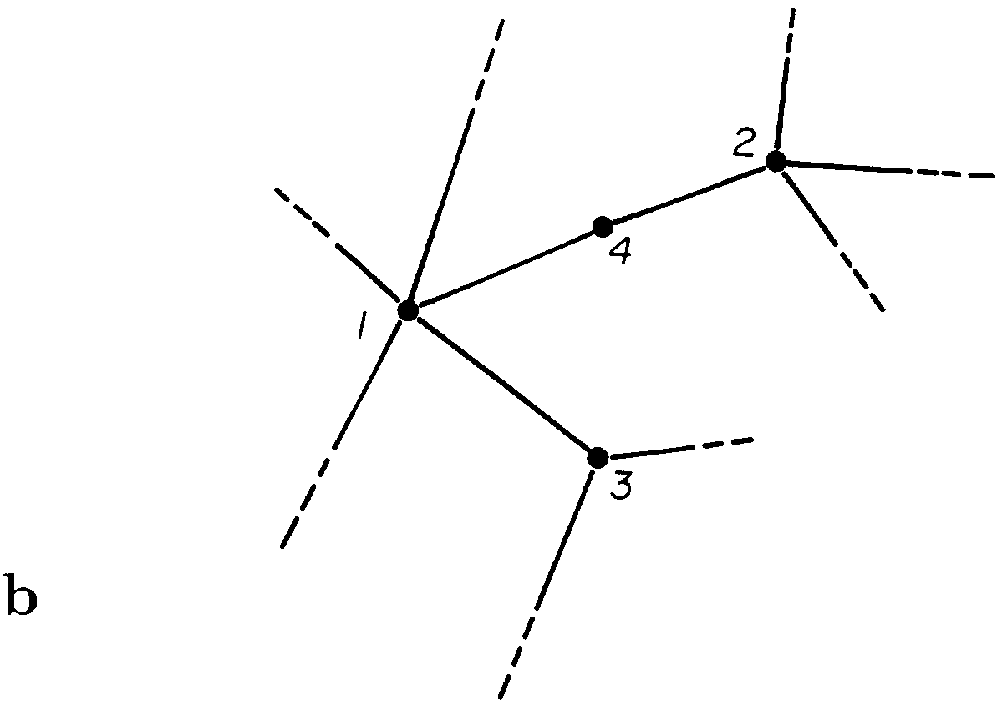}}
\caption[labyrinth]{\label{fig8a} The essential feature of the MPCP fine-tuning
mechanism is that phase transitions are {\em first order}. This being the
case, there will be finite volume in the space of possible combinations
of extensive variables associated with {\em single points} the in action
parameter. In part a) of the figure, it is seen that the volume
$\Delta I_i(\beta_{crit})$ that get mapped to the single point $\beta_{crit}$
is comparable in size the the volumes $\Delta I_i(Phase\;\; 2)$ and
$\Delta I_i(Phase\;\;1)$ that get mapped respectively to finite intervals
$\beta > \beta_{crit}$ and $\beta < \beta_{crit}$ in parameter
space that correspond to entire phases. Thinking of a complicated system of
phase boundaries (as suggested in part b) of the figure), one could assign
probabilities that random choices of extensive quantities would result in
fine-tuned values corresponding to the various multiple points
$1,2,3,4, \cdots$. These probabilities are proportional to the volumes in
the space of possible combinations of extensive variables that are mapped into
the points $1,2,3,4, \cdots $ (i.e., analogous to $\delta I_i(\beta_{crit})$
in part a) of the figure). The important point is that fine-tuning
by multiple point criticality work because a finite probability is
associated with
a set of points in parameter space of {\em measure zero}.}
\end{figure}

In general, a volume in action parameter space $\prod_i\Delta
\beta_i(\vec{\beta_0})$
(i.e. in phase-diagram space) at the point $\vec{\beta_0}$  can be mapped into
a corresponding volume
in the space of extensive variables (spanned by the $I_j$):

\beq  \prod_i\Delta
\beta_i(\vec{\beta_0})\longrightarrow \prod_j\Delta I_j(\prod_i\Delta
\beta_i(\vec{\beta_0})) \eeq

\nin This volume in $I_j$-space contains the combinations of values of
extensive quantities that lead to parameter space values within the volume
$\prod_i\Delta \beta_i(\vec{\beta_0})$. Generally, even an entire (single)
phase corresponds to some finite volume in the space of extensive quantities
spanned by the $\{I_j\}$. The useful feature for the purpose of fine-tuning
is that, for a first order phase transition,  the {\em single} points
(belonging to a
measure zero set) along phase borders (including the system of multiple
points) correspond to a finite volume in the space of extensive
quantities of size comparable to the volume associated
with the dense set of points in action parameter space corresponding to the
values of an entire (single) phase.

Thinking in terms of a random dynamics scenario,
the values of
extensive quantities $I_j$ are envisioned as having been randomly
fixed at some set of values corresponding to a point $\vec{I_0}$ in the
space of extensive quantities. The probability that the ``coupling
constants'' realized in Nature fall within a small volume
$\prod_j\Delta\beta_j(\vec{\beta})$ at
some set $\{\beta_{0\;i}\}=\vec{\beta_0}$ of values
is proportional to the volume
$\prod_i\Delta I_i$ in the space spanned by
the extensive quantities corresponding to this small volume
$\prod_j\Delta\beta_j(\vec{\beta})|_{\vec{\beta}=\vec{\beta_0}}$.

In the case that the values in a finite volume in the space of extensive
quantities all result in the action parameter values at a {\em single
point} (e.g., a multiple point) greatly increases the probability that a
random choice $\vec{I_0}$ of values in the space of extensive quantities
will result in action parameter values at a phase transition. At such
points (which must be first order phase transitions) the ``coupling
constants are ``infinitely well'' fine-tuned. The probability for the
realization of such ``infinitely well'' fine-tuned values for
``coupling constants'' is greater
the greater the ``heat of fusion'' at the transition.

If we think of an abstract phase diagram with a labyrinth of phase
boundaries having a system of multiple points at which some
number of the  various phase boundaries convene, we could ascribe
probabilities that one or another of the corresponding sets of values of
``coupling constants'' were realized in Nature
according to the volume in extensive parameter space associated with
these various multiple points.

The non-locality resulting from having ``fixed'' amounts
$I_i=I_{i\;fixed}$ of extensive quantities in
the space-time manifold is of the same type as
that introduced in the water example above
where the coexistence of three phases was enforced by appropriately
fixing the values of $V$, $E_{TOT}$ and $n_{H_2O}$ so as to fine-tune the
temperature and pressure to triple point values. However, non-locality
that comes about in this way can essentially be ``approximated away'' by the
standard technique of  approximating a microcanonical ensemble by
a canonical ensemble
\footnote{For a statistical mechanical system with Hamiltonian $H$, it is a
standard procedure to approximate a microcanonical ensemble with fixed energy
$E_{fixed}$  by a canonical ensemble:
$\delta(H-E_{fixed}){\cal D}\phi \approx e^{-\beta(H-E_{fixed})}$ when the
``Lagrange multiplier'' $\beta$ $(=1/kT)$ is determined so that

\beq \langle H \rangle \stackrel{def}{=}
\frac{\int H e^{\beta(H-E_{fixed})}{\cal D}\phi}
{\int e^{\beta(H-E_{fixed})}{\cal D}\phi}=E_{fixed}. \label{can} \eeq

For a large number of degrees of freedom, this becomes a very good
approximation because phase space volume (or functional integration measure)
${\cal D}\phi$ is a very rapidly varying function of energy. However, it
can happen that there is no $\beta$ that can fulfil (\ref{can}) because
$E_{fixed}$ falls within a range of energies corresponding to the ``jump''
(e.g., heat of fusion)  at a (first order) phase transition. In this
``forbidden interval'', the energy $E_{fixed}$ can only be realised as a
mixture of two phases (e.g., ice + water) at the fine-tuned parameter
value $\beta=\beta_{crit}$.}.

The long range correlations
that strictly speaking are present in microcanonical ensembles because
the total energy is ``known''(thinking classically)
will go unnoticed because they can be
absorbed into coupling constants. In other words, this  form  of  non-locality
is acceptable because it is manifested in all space-time in the same way -
namely  as contributions to constants of Nature.

In implementing the 4-dimensional space-time analogy to using a canonical
ensemble as
an approximation to a microcanonical one, the idea\cite{albu} is to assume that
the
Feynman  path  integral  of the ``quantum field theory of
Nature'' is constrained by  the requirement  of fixed amounts of some
(perhaps a multitude of)  extensive quantities $I_i$ by using an action
that satisfies (\ref{delact}).
This procedure results in ``coupling
constants'' that are maintained at critical values  in a
dynamical way. The mild form of non-locality discussed above (i.e.,
non-local terms of the Lagrange density that are reparameterization invariant)
is the reason that this dynamical mechanism works in the same way everywhere
in spacetime so that intensive parameters are keep at (universally) constant
(critical) values.

This is also the essence of the claim to fame of Baby Universe
theories: these use the effective breakdown of locality to render ``coupling
constants'' dynamical. In the context of quantum field theory, the picture
to have in mind is a universe completely permeated by a labyrinth of
quantum fluctuation worm-holes. Coupling constants only ``notice'' the
resulting
non-locality in the sense that these constants are fixed as (non-local)
averages over all space-time.

\vspace{.4cm}
\nin{\bf Back to simple model for multi-phase universe}
\vspace{.3cm}

Let us take as a prototype extensive variable the space-time integral

\begin{equation}
\frac{I_2}{V}\stackrel{def}{=}\int_Vd^4x|\phi^2(x)|\stackrel{def}{=}
\langle |\phi^2|\rangle. \end{equation}

There is one such integral for each Feynman path history of the universe.
Let us assume that the value of such an integral (i.e., an extensive
quantity) is fixed at some ``God-given'' value
$\langle |\phi^2| \rangle_{``God"}$. This requirement

\begin{equation} \langle |\phi^2| \rangle=\langle |\phi^2| \rangle_{``God"}
\label{criterion} \end{equation}
is a selection criterion that singles out a subset of all possible Feynman
paths - namely those for which (\ref{criterion}) is satisfied.

Inasmuch as this rule is a statement about the allowed values of Feynman
paths when integrated from time $t_{Big\;Bang}$ to $t_{Big\;Crunch}$, it is
appropriate to call this criterion (\ref{criterion}) a non-local Law of Nature
for times $t$ such that $t_{Big\;Bang}< t < t_{Big\;Crunch}$. The justification
for calling (\ref{criterion}) a non-local law is that it implements a criterion
for selecting a subset of all possible Feynman paths in configuration space
that is based on what will happen in the future (as well as the past of
course).

This is to be contrasted to the situation for a fixed quantity
$E_{TOT}/V$ in the 3-dimensional ice-water system. Here there is really no time
inasmuch as the system is in equilibrium. The fixed extensive quantity
$E_{TOT}/V$ is a remnant of the initial conditions of the system that
(due to conservation laws) survives as an easily measured quantity in
equilibrium. Fixed quantities such as $E_{TOT}/V$ in the 3-dimensional
ice-water system select a subset of all possible microstate configurations
in a way analogous to the manner in which the requirement (\ref{criterion})
selects an allowed subset of all possible Feynman paths in the 4-dimensional
model.

Returning now to the toy model of the universe that we want to consider,
let us assume that the fixed value $\langle |\phi^2(x)|\rangle_{God}$
cannot be realised  in all space-time (i.e., throughout the history of the
universe) but instead must be realised as a ``coexisting'' combination of
two vacua (``phases'') $\langle |\phi^2(x)|\rangle_A$ and
$\langle |\phi^2(x)|\rangle_{B}$ corresponding to the two relative
minima
in the double-well potential of Sher mentioned above. We can think of a
two-minima Standard Model effective Higgs field potential. For the sake of
later
discussion, let us assume that
$\langle |\phi^2(x)|\rangle_{A}\leq \langle |\phi^2(x)|\rangle_{B}$. We shall
sometimes use ``us'' and ``other'' (i.e., the vacuum we live in and some
other vacuum) instead of respectively the subscripts $A$ and $B$. In what
follows, spacetime averages (designated with the brackets
$\langle\cdots\rangle$)
that have a  subscript  ``$A$'', ``$B$'', ``$us$'' or ``$other$'' always
denote values of spacetime averages that correspond to relative minima
of $V_{eff}$. Hence $\langle\cdots\rangle_{A}$, $\langle\cdots\rangle_{B}$,
$\langle\cdots\rangle_{``us"}$, and $\langle\cdots\rangle_{``other"}$ always
denote
average values $\langle f(\phi)\rangle$ of the integrand of
$\frac{1}{V}\int_Vd^4x f(\phi(x))$ that correspond either to
true vacua or meta-stable vacua.

One possibility (the one to be considered here) would be for the fixed value
of $\langle |\phi^2(x)|\rangle_{God}$ to be realised as a
combination of the two vacua $\langle |\phi^2(x)|\rangle_{us}$ and
$\langle |\phi^2(x)|\rangle_{other}$ that exist in two different time periods
in the
life of the universe:

\beq \langle |\phi^2(x)|\rangle_{God}=
\langle |\phi^2(x)|\rangle_{us}(t_{ignition}-t_0)+
\langle |\phi^2(x)|\rangle_{other}(t_{crunch}-t_{ignition}) \label{cos} \eeq

\nin where $t_{ignition}$ stands for  ``time of ignition'' and denotes the
time at which the transition from one vacuum to the other occurs.
The  scenario in which the thermodynamics of the Big Bang
is such that the universe starts off in what becomes a ``false'' vacuum is
considered in
the literature\cite{sher}. This would
correspond to the relative minimum in $V_{eff}$ that has the smaller value of
the space-time
integral $\int d^4x\sqrt{g(x)}|\phi^2(x)|$; i.e., the value
$\langle |\phi^2(x)|\rangle_{us}$. At the time $t_{ignition}$, the transition
\footnote{This transition is the essence of the
``vacuum bomb'' described to me by H.B. Nielsen}
$\langle |\phi^2(x)|\rangle_{us}\rightarrow \langle |\phi^2(x)|\rangle_{other}$
occurs.

Note that $I_2$ and $I_4$ are in this simple case a function of $t_{ignition}$
which means that $m^2_{Higgs}\stackrel{def}{=}
\frac{\partial V_{nl}(\{I_2,I_4\})}{\partial I_2}$ is also a function of
$t_{ignit}$.

\begin{figure}
\centerline{\epsfxsize=\textwidth \epsfbox{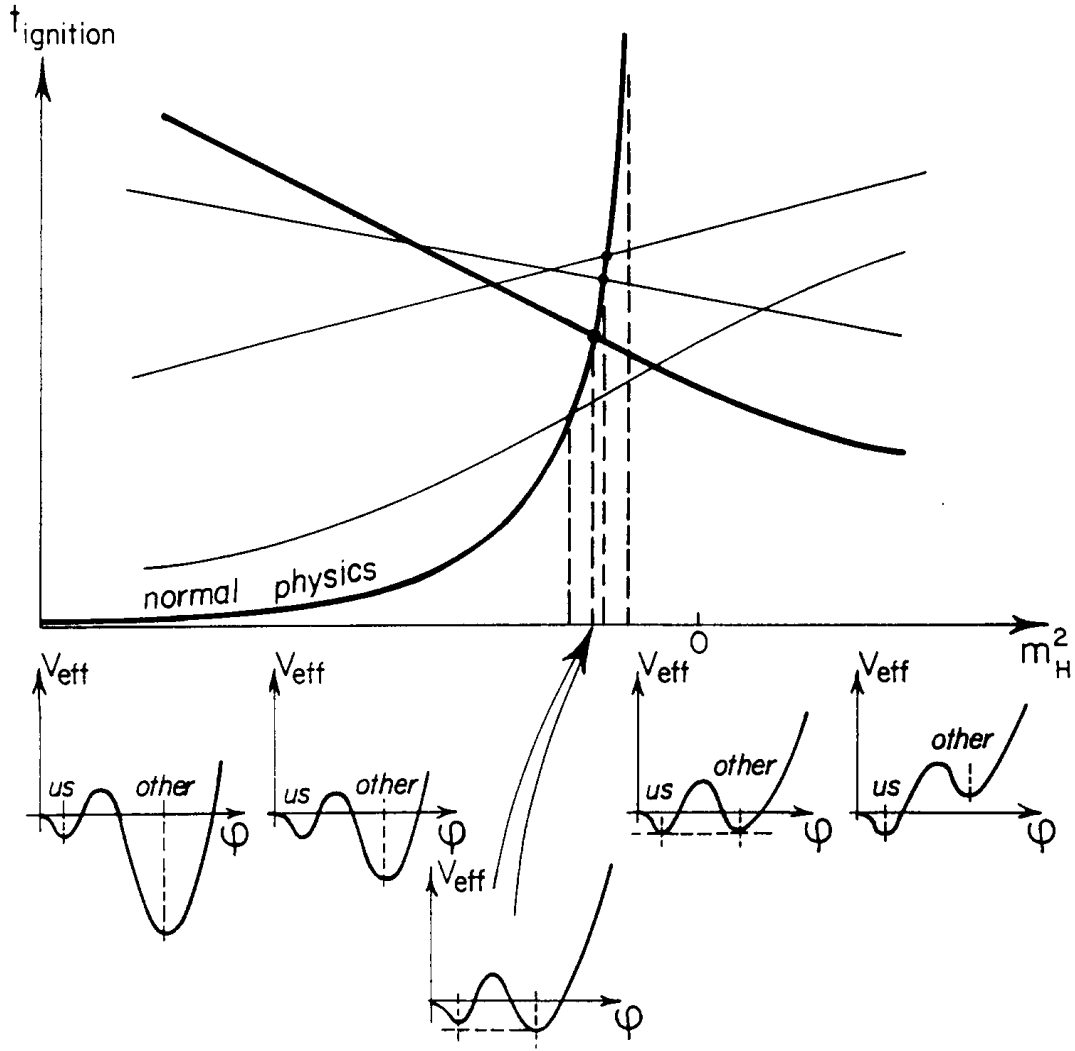}}
\caption[figurenonloc]{\label{fignonloc} The development of the double well
potential and $m_{Higgs}$ as a function of $t_{ignition}$. Note that
all the more or less randomly drawn non-locality curves intersect the
``normal physics'' curve
near where the vacua are degenerate (i.e., the MPCP solution).}
\end{figure}

Let us first use  ``normal physics'' to see how the relative depths of the
two minima of the double well are related to $m^2_{Higgs}$ and to
$t_{ignition}$. From the work of Sher, it can be deduced that a large
negative value of $m^2_{Higgs}$ corresponds to the minimum at
$\langle |\phi^2(x)|\rangle_{other}$ being deeper relative to the depth of the
minimum at $\langle |\phi^2(x)|\rangle_{us}$ than for less negative values
of $m^2_{Higgs}$ (see Figure \ref{fignonloc}). It can also be argued quite
plausibly that a minimum at $\langle |\phi^2(x)|\rangle_{other}$ much deeper
than that at $\langle |\phi^2(x)|\rangle_{us}$ would correspond to an
early (small) $t_{ignition}$ inasmuch as the ``false'' vacuum at
$\langle |\phi^2(x)|\rangle_{us}$ is very unstable. However, as the value
of the potential at $\langle |\phi^2(x)|\rangle_{other}$ approaches that at
$\langle |\phi^2(x)|\rangle_{us}$, $t_{ignition}$ becomes longer and longer
and approaches infinity as the depth of the wells at
$\langle |\phi^2(x)|\rangle_{us}$ and $\langle |\phi^2(x)|\rangle_{other}$
become
the same; it may be that $t_{ignition}$ becomes infinite before
the minimum at $\langle |\phi^2(x)|\rangle_{us}$ becomes as deep as the minimum
at $\langle |\phi^2(x)|\rangle_{other}$.
The development in the form of the double well potential
as $t_{ignition}$ becomes larger corresponds to less and less negative
values of $m^2_{Higgs}$. The development of the double well potential and
$m^2_{Higgs}$ as a function of $t_{ignition}$ is illustrated in Figure
\ref{fignonloc}.

In the simple cosmological model being considered,
it is readily seen that the ``coupling constant'' $m^2_{Higgs}$, regarded as
function of $t_{ignition}$, depends on the future as well as the past. Note
that the larger the difference
$|\langle |\phi^2(x)|\rangle_{other}-\langle |\phi^2(x)|\rangle_{us}|$
the greater the dependence of $I_{2\;``God"}\stackrel{def}{=}
\langle |\phi^2(x)|\rangle_{``God"}$ on $t_{ignition}$ (this corresponds
to a large heat of fusion in the ice-water analogy). If
$\langle |\phi^2(x)|\rangle_{us}=\langle |\phi^2(x)|\rangle_{other}$, the
dependence on $t_{ignition}$ disappears as does the non-locality in the theory.

\vspace{.4cm}
\nin{\bf Prediction of top quark and Higgs masses}
\vspace{.3cm}

In very recent work\cite{cdfhbn}, our multiple point criticality principle
has been used to predict the masses for the top quark and the Higgs mass.
This is done by making the the multiple
point criticality optimally effective. First the MPCP  is
assumed using the formulation that the two minima
- denoted $V_{us}$ and $V_{other}$ -
of the Standard Model Higgs field potential (with one-loop corrections)
 are degenerate:
$V_{us}=V_{other}$. Secondly, it is assumed that
$|\langle |\phi^2(x)|\rangle_{other}-\langle |\phi^2(x)|\rangle_{us}|$
is as large as ``possible'' subject to the constraint that we should stay
below the Planck scale.
This insures the largest possible
number of combinations of fixed extensive quantities that are not
realisable as a single phase (or, more generally, not realisable with a
smaller number of phases than are present at the multiple point). The analogy
to this assumption in the example above with water (in which ice, liquid,
and vapour coexist at the triple point) would be that Nature had
chosen large values for the heats fusion, vaporisation, and sublimation.

In the work\cite{cdfhbn} in which the top quark and Higgs masses are
predicted, the authors take
$|\langle |\phi^2(x)|\rangle_{other}-\langle |\phi^2(x)|\rangle_{us}|$
as being of order unity in Planck units. This amounts to taking the
vacuum expectation value of
$|\langle |\phi^2(x)|\rangle_{other}$ to be of order unity in Planck units
and
$\langle |\phi^2(x)|\rangle_{us}|$ (the vacuum corresponding to
physics as we humans know it) to be as determined by the Standard Model
(which turns out to be very small in terms of Planck scale units).
This amounts having the minimum
$V_{other}$ at the Planck scale; this assumption is consistent with
Planck scale units being regarded as important in fundamental physics
and at the same
time maximises the probability that the intensive quantities Higgs mass and
top quark mass acquire values dictated by the MPCP fin-tuning mechanism.
This follows because even randomly chosen values of
$\langle \phi^2(x)\rangle$ are not likely to fall outside the interval
$|\langle |\phi^2(x)|\rangle_{other}-\langle |\phi^2(x)|\rangle_{us}|$
that is of order unity in Planck scale units.

This assumption  of maximally effective MPCP fine-tuning
leads to
remarkably good predictions for the top quark and Higgs masses. The
prediction is $(M_t,M_H)=(173\pm 4,\;\; 135\pm 9)$ GeV. The authors of this
work\cite{cdfhbn} conclude that if this very good prediction of the top
quark mass and reasonable prediction for the expected Higgs mass are not
regarded as accidental, one is essentially forced to accept the validity
of the multiple point criticality principle fine-tuning mechanism
as well as the minimal
standard model (at least as far as the top quark and Higgs interactions
are concerned). The latter implies that super-symmetry would not be
allowed.

\vspace{.4cm}
\nin{\bf Avoiding paradoxes arising from non-locality}
\vspace{.3cm}

\begin{figure}
\centerline{\epsfxsize=\textwidth \epsfbox{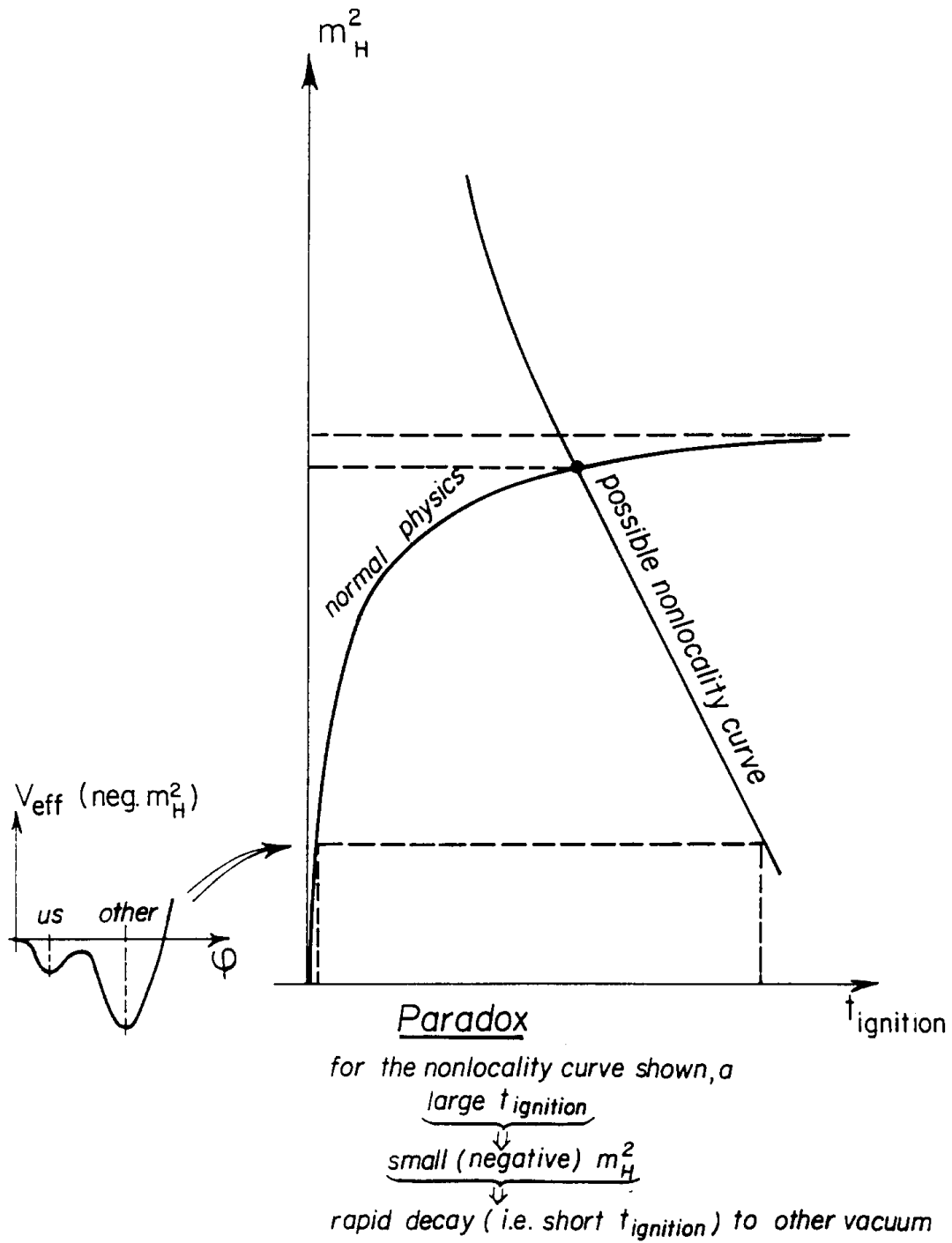}}
\caption[figparadox]{\label{paradox} Many non-locality curves could lead to
paradoxes similar to the ``matricide'' paradox. Such paradoxes are avoided if
the value of $m_{Higgs}$ is fine-tuned to the multiple point critical value.}
\end{figure}

In general the presence of non-locality leads to paradoxes. While
the form that the non-local action (or potential $V_{nl}$ in this discussion)
is, at present at least, unknown to us, we can make some guesses as to
its form. These guesses could correspond to the  non-locality curves
in Figure \ref{fignonloc}. 
In particular, non-locality curves having a negative slope as a function
of $t_{ignition}$ lead to paradoxes in the following manner.
Consider specifically the non-locality curve
in Figure \ref{paradox}. Now let us make the
assumption that  $t_{ignition}$ is large and see that this leads to a
contradiction. Assuming that $t_{ignition}$ is large, it is seen from
the non-locality function in Figure \ref{paradox} that this
implies that $m^2_{Higgs}$ has a large negative value.
But a large negative value of $m^2_{Higgs}$ corresponds to
a vacuum $\langle |\phi^2(x)|\rangle_{us}$ that is very
unstable and hence to a very short $t_{ignition}$ (corresponding to a rapid
decay to the stable vacuum $\langle |\phi^2(x)|\rangle_{other}$). So the
paradox
appears: the assumption of a {\em large} $t_{ignition}$ implies a
{\em small} $t_{ignition}$. This is very much akin to the ``matricide''
paradox encountered for example when dealing with ``time machines''.
It is well known\cite{nov1,nov2,nov3} that Nature avoids such paradoxes by
choosing
a very clever
solution in situations where these paradoxes lure.

In the case of the paradoxes that can come about due to non-locality of the
type considered here, the clever solution that Nature uses to avoid paradoxes
is that of obeying the Principle of Multiple Point Criticality\cite{bomb}.
This corresponds
to the intersection of the ``normal physics'' curve and the ``non-locality''
curve in Figure \ref{paradox}. This point corresponds to a value
of $m^2_{Higgs}$ for which the vacua at $\langle |\phi^2(x)|\rangle_{us}$ and
$\langle |\phi^2(x)|\rangle_{other}$ are (almost) degenerate. This is
equivalent to being at
the multiple point. The paradox is avoided by taking the multiple point
value of $m^2_{Higgs}$. But at the multiple point, an intensive
parameter has its value fine-tuned for a wide range of values for the quantity
conjugate  to this intensive quantity. Fine-tuning is to be understood as a
consequence of Nature's way of avoiding paradoxes that can come about due to
non-locality.

It should be pointed out that the paradox-free solution corresponding to the
intersection of the two curves in Figure \ref{paradox} occurs for a value of
$m^2_{Higgs}$ corresponding to ``our'' vacuum at $\langle
|\phi^2(x)|\rangle_{us}$
being very slightly unstable. The value of $m^2_{Higgs}$ corresponding to the
vacua at $\langle |\phi^2(x)|\rangle_{us}$ and $\langle
|\phi^2(x)|\rangle_{other}$
being (precisely) degenerate is slightly less negative than that corresponding
to the multiple point value of $m^2_{Higgs}$ at the intersection of the
curves.

It is very satisfying from the point of view of Random Dynamics to see that the
the multiple point value of $m^2_{Higgs}$ is very insensitive to which
``guess''
we use for the non-local action. Indeed all the ``non-locality'' curves in
Figure \ref{paradox} intersect the ``normal physics'' curve at values of
$m^2_{Higgs}$ that are tightly nested together. The reason for this is
that the $m^2_{Higgs}$ is a very slowly varying function of
$t_{ignition}$ as $m_{Higgs}^2(t_{ignition})$ approaches the value
corresponding to degenerate vacua. The more nearly parallel the ``normal
physics'' and the
``non-locality'' curves at the point of intersection, the less are the
(paradoxical) effects of non-locality.
For a point of intersection at values of $t_{ignition}$ sufficiently large
that $m^2_{Higgs}(t_{ignition})\approx m^2_{Higgs}(\infty)$),
the non-locality effects disappear as the curves become parallel since
both curves become independent of $t_{ignition}$.
If the curves were parallel, there would  also be the possibility  that they
don't
intersect in which there would be  no ``miraculous solution'' that could avoid
the paradoxes imbued in having non-locality.

If the interval $|\langle |\phi^2(x)|\rangle_{other}$-$\langle
|\phi^2(x)|\rangle_{us}|$
is large (e.g. of the order of the largest physically conceivable
scale (Planck?) if tuning is to be maximally effective)
and if $\langle |\phi^2(x)|
\rangle_{``God"}$ falls not too close to the ends of this interval, then
$t_{ignition}$ will be something of the order of half the life of the universe.
Actually, the approximate degeneracy of the vacua
$\langle |\phi^2(x)|\rangle_{other}$ and $\langle |\phi^2(x)|\rangle_{us}$
may be characteristic of the
temperature of the post-Big Bang universe in the present epoch and
{\em not} characteristic of the high temperature that prevailed immediately
following the Big Bang.
Such a much higher temperature universe might very plausibly
have favoured  the vacuum at $\langle |\phi^2(x)| \rangle_{us}$ very strongly
at the expense of the ``phase'' corresponding to the
$\langle |\phi^2(x)| \rangle_{other}>\langle |\phi^2(x)| \rangle_{us}$.
This could  be explained as being
due to the fact that  at high temperatures,
the free energy is considerably less than the total energy if the entropy
is large enough. A phase with a large number of light particles - for
example a Coulomb-like vacuum such as the ``us'' phase in which we live -
could very plausibly be so strongly favoured at high temperatures that other
phases - for example the ``other'' vacuum - simply disappeared at the high
temperature of the universe immediately following the Big Bang.

If this
were to have depleted the universe of the phase
$\langle |\phi^2(x)| \rangle_{other}$ at high temperatures,
it would indeed be difficult to
re-establish
it in a lower temperature universe even if the vacuum
$\langle |\phi^2(x)| \rangle_{us}$ were only meta-stable and
$\langle |\phi^2(x)| \rangle_{other}$ were
were the true vacuum at the lower temperature.
Such an exchange of the true vacuum is indeed a possibility in
going to lower temperatures inasmuch as the
difference between the total energy and the free energy decreases in going to
lower temperatures. Accordingly this difference becomes less effective in
favouring a Coulomb-like phase at the expense of a phase with heavy particles.

On the other hand, the value
of $\langle |\phi^2(x)| \rangle_{``God"}$ can easily
(i.e. as a generic possibility) assume a value that requires the
the universe to be in the ``phase'' $\langle |\phi^2(x)| \rangle_{other}$
 during a sizeable period of its life if the universe is to
have multiple point parameters in the course of its evolution (as required
for avoiding the paradoxes that accompany non-locality). How can Nature
overcome the energy barrier that must be surmounted in order to bring about
the decay of the slightly unstable (false) vacuum at
$\langle |\phi^2(x)| \rangle_{us}$ to the vacuum at
$\langle |\phi^2(x)| \rangle_{other}$? Even  producing just
a tiny ``seed'' of the vacuum $\langle |\phi^2(x)| \rangle_{other}$ would
be very difficult.
What miraculously clever means can Nature devise
so as to avoid deviations from a multiply critical evolution of the
universe?  One extraordinarily ingenious master plan that Nature may have
implemented has recently been proposed by Holger Bech Nielsen. Maybe life
was created with the express ``purpose'' of evolving some (super intelligent?)
physicists that could ignite a ``vacuum bomb'' by first creating in some very
expensive accelerator the required
``seed'' of the ``correct'' vacuum $\langle |\phi^2(x)| \rangle_{other}$ that
subsequently would en-gulf the universe in a (for us) cataclysmic transition
to the ``other'' phase thereby permitting the continued evolution of a
``paradox-free'' universe!

\subsubsection{Two-position fields/particles replicated at one position}
\label{twopos}

In Sections~\ref{retrieve}~and~\ref{nonlocmpc} we have suggested
that, {\em if fields are
defined the usual way as functions of space-time points} (and if for
simplicity
short distance non-localities are ignored),  we can interpret
long distance non-locality that is {\em independent of space-time}
as being incorporated into the laws of Nature rather
than being observable in some offensive way.
However, having once relinquished the principle of locality in this mild
ontological sense, there is no longer any compelling reason to assume that
fields depend on just one space-time point!
Rather it becomes quite natural to contemplate
the possibility of having, for example,  a field $\phi(x,y)$ that depends
on {\em two} space-time points $x=x_{\mu}$ and $y=y_{\mu}$. If we
for simplicity take $\phi(x,y)$ as a scalar field, it transforms under
reparameterization transformations (i.e., diffeomorphisms),
in both $x$ and $y$:
\begin{equation}
\phi \rightarrow \phi^{\prime} \;\;\mbox{where}\;\;\;
\phi^{\prime}(x^{\prime},y^{\prime})= \phi(x(x^{\prime}),y(y^{\prime})).
\end{equation}
Physically a field such as $\phi$ is just a function of
a couple of space-time points regarded as abstractly defined (i.e.
coordinate independent) events.

The integrals that can be used for constructing a nonlocal
but still reparameterization invariant action, depending on such
double-position fields $\phi(x,y)$, can hardly be imagined to  be anything but
double integrals of the form
\begin{equation}
\int d^4x \int d^4y {\cal F}(\phi(x,y), \phi(x), \phi(y), \partial \phi(x,y)/
\partial x^{\mu}, ...).
\end{equation}

Roughly such a model can be thought of as one in which space-time is
8-dimensional rather than just 4-dimensional but in which there are two
types of particles:

a) ``ordinary'' particles (or fields) having only one position
and really only depending on four out of the eight coordinates,
e.g. on $x$ but not on $y$.

\noindent and

b) ``double position particles (or fields)'' (e.g., $\phi(x,y)$)  that can
take values in the entire 8-space.

In practice we presumably have something
we may call ``vacuum'' for both sorts of fields: vacuum values in the
classical approximation are
constant (zero say)  over most combinations of $x$ and $y$. Assuming ``we''
humans are primarily composed of ``ordinary'' particles, we could not readily
interact with the Fourier components of the $\phi(x,y)$ field
unless these components  have zero momentum along either $x$
or $y$.
Genuine excitations of $\phi(x,y)$ locally {\em in the 8-dimensional space}
can only occur by interaction of {\em two} ordinary particles and therefore
are presumably rather suppressed.
This makes it very difficult in practice to
discover the non-locality related to the two-position fields,
{\em unless} there are some huge amounts of
matter in 8-space so to speak.

If reparameterization
invariance is not to be broken spontaneously, we must have
fields - for example $\phi(x,y)$ -that are constant in all 8-space points
$(x,y)$  except
along the diagonal $x=y$ (and, presumably, infinitesimally close to $x=y$).
However, the absence of non-locality that we experience phenomenologically
is probably insured if there is not spontaneous breakdown under
reparameterizations in a {\em local} region. So if the reparameterization
in one neighbourhood differed from that prescribed by reparameterization
invariance at another very
far removed region, this might well not be observed as a breaking
of locality.
So a priori it would not be forbidden phenomenologically if
the field $\phi(x,y)$ takes on  some other values (i.e., departing
from the almost everywhere dominant constant vacuum value) along a thin band
representing a graph of a function yielding $y$ as function of $x$.
We assume that the structure is the same everywhere along this  band
so that there is still reparameterization invariance under
the special type of transformation that transforms  the $x$ and $y$ at a point
$(x,y)$ in the band in the same way.
Field configurations corresponding to this band make up  a 4-dimensional
manifold in the
8-dimensional $(x,y)$ space, along which there can be a systematic
communication between a space-time point $x$ and its image $y$.

If we had  efficient communication by non-locality between say
$x$ and $y$ due to the above-outlined spontaneous breakdown of
reparameterization invariance, one can enquire as to whether
this effect would be {\em perceived} as a breaking of locality.
Probably not:  rather we would {\em interpret
the related space time points} - the ones with $(x,y)$ on the band -
as representing different degrees of freedom at a
single space-time point $x$ say. Because we would experience
the space-time point
$x$ and its image $y$  as the same space-time point, locality
is effectively restored. Concurrent with this, we would experience
a replication of the field degrees of freedom at one space-time point!
In Nature we seem to see a 3-fold replication with
respect to the fermions, in the sense that we observe
three generations. A tripling of the number of fields can
easily be achieved with
(just) two-position fields that are applied a couple of times:
the two-position field may have several ``bands'' as proposed, so that
one point - $x$ say - can be connected to several (e.g. two) far away points
$y$, by two different bands in the same $\phi(x,y)$-field.

Indeed, in  the experimentally supported Standard Model, we find a trinity
of similar (but not exactly replicated) field types:
the three generations of quarks and leptons!
But the three generations found
in Nature are not exact replicas of each other, as one at first might expect
if these truly represented
particles of the same sort just at different points in space and time.
However these generations may correspond to superpositions of
states at different related space-time points;
there is also the possibility of
some sort of (later) spontaneous breakdown of the symmetry
between the different related space time points. There
is therefore no necessity  for  perfect symmetry between the
different generations, in order to uphold the interpretation that these
are due to the non-locality with spontaneously broken reparameterization
symmetry.

Moreover, we would expect that a 3-fold (approximate) replication mechanism
due to non-locality would not only triple the quark and
lepton fields but also the boson fields! In fact, such a tripling of boson
fields
is an intrinsic feature of our long standing
$SMG^3$ ``anti-grand unification'' gauge group model. In this model,
we predict that  the values of the fine-structure constants at
the multiple point
of the phase diagram for the gauge group $SMG^3$ should agree with
experimental values.
For the moment let us content ourselves with the observation
that non-locality can easily lead to a picture in which
not only the fermions are tripled but also, essentially unavoidably, the
bosons. That is to say, we would predict,
roughly speaking, 3 photon-types, three $W^+$'s, three $W^-$ 's, three
$Z^0$'s, and 24 gluons.
There may also be some predicted but
not observed gauge bosons (and presumably even  more Higgses or replacements
for them) that would need to be given large masses.

\subsubsection{Concluding remarks on non-locality and MPC}

Relinquishing  strict adherence to the principle of locality has
several attractive features from a theoretical point of view.
In other work\cite{randyn1,randyn2,frogniel} we have in the spirit
of Random Dynamics also investigated
the possibility of removing this principle from the list of initial
assumptions necessary for constructing a fundamental theory.

Starting with the problem of the cosmological constant
being almost zero\footnote{Nowadays it seems from cosmological studies that it
may not be
exactly zero} - from a Planck scale point of view at least
- we have argued that there should be a  breakdown of the
principle of locality for field interactions from a
fundamental point of view. We have also argued that if fundamental physics
obeys the multiple point
criticality principle in 4-dimensions, non-locality is implied (at least in the
mild form in which reparameterization invariance remains intact) inasmuch as it
seems
difficult to find another explanation if one accepts the phenomenological
validity of the MPCP. Going the other
way, we have also argued that if non-locality is an inherent
feature of fundamental physics, the solution whereby  paradoxes due to
non-locality are avoided coincides with the choice of the multiple point values
of coupling constants.

As already mentioned, it appears that a  theory with non-locality
(that retains reparameterization invariance) or, equivalently, a theory for
which
the MPCP is valid  provides a
promising approach to explaining  a series of fine-tuning problems.
The principle of multiple point criticality - which
essentially asserts the coexistence of a number of phases -
leads to such an impressive number of good
predictions for fine-tuned quantities that one is almost
forced to take it seriously!

Essentially all of the well known fine-tuning mysteries in high energy physics
are potentially solved: 1) the vanishing of the dressed cosmological constant;
2) the small Higgs field expectation values and masses; 3) the hierarchy of
quark and
lepton masses; 4) strong CP conservation and 5) the multiple point values of
fine-structure constants.

The values obtained assuming the MPCP for the three Standard Model gauge
couplings are discussed at length elsewhere in this thesis. The manner in
which the MPCP might solve the mysteries of the first four
fine-tuning enigma are briefly mentioned now.

\begin{enumerate}

\item {\em Cosmological constant $\Lambda_{eff}=0$ at transition from
finite to infinite universe}

According to the multiple point criticality model for fine-tuning, one should
expect to find
the fine-tuned parameters observed in Nature at parameter values that
coincide with MPCP values. Since the cosmological constant value
$\Lambda_{eff}=0$ corresponds to the border between finite and infinite
space spheres (universes), it is
not surprising that computer simulations\cite{amb,ham}
of quantum gravity indicate singular
behaviour at the value $\Lambda_{eff}=0$ for the
cosmological constant.
We can therefore claim that the
phenomenologically indicated value $\Lambda_{eff}=0$ coincides with the
MPCP prediction for the value of the cosmological constant.

\item {\em Our ``unfortunate '' prediction of the Higgs mass}

A promising proposal for the explanation of the fine-tuning
problem of the expected small value of
the Weinberg-Salam Higgs mass (compared say to the
Planck scale) is at hand if the
multiple point  criticality idea for gauge couplings is assumed to be valid
in a slightly extended form
that requires not only
that a maximum number of  Coulomb-like and confinement-like  ``phases''
(i.e. \pcps) be accessible at the  multiple  point  (by
making infinitesimal changes in the multiple point coupling values) but
also that
(Planck scale) multiple point parameters should be at the border between
Higgsed and un-Higgsed phases.
At such boundaries there is a change of sign in $m_{Higgs}^2$ which, for
transitions that  are not too strongly first order, would imply that values of
$m_{Higgs}^2$  relative to the Planck scale are strongly suppressed.
So in a certain sense,
our fine-tuning model explains the physical
fine-tuning problem related to the so-called hierarchy problem.
The hierarchy problem  might  then  be  said  to  be
Nature's way of telling us that the Yang-Mills vacuum is very close to
a Higgs phase boundary!
It should be pointed out that the need for solving this problem
has often been used to argue for super-symmetry.

Let us elaborate briefly on this a priori promising but de facto
experimentally refuted approach.  Using one-loop corrections,
logarithms can get introduced into the condition for
the coexistence of phases in such a way that
the Higgs mass, or equivalently
the Higgs field expectation value appears only in
a logarithm in  the equation
imposing the coexistence of the Weinberg Salam Higgs phase and
the unbroken phase (in which $W^+$, $W^-$ and $Z^0$ would be
massless).
In this event,  we can argue that the fine-tuning mystery
of why the scale of the Higgs expectation value is so very low compared say
to the Planck scale is solved: once the ratio is determined
by its {\em logarithm}, it can easily become very large.

The calculation of what happens when we impose the requirement
of the equality of the Higgs field potential depths for the
two minima corresponding to the mentioned phases - the  Higgs and
the unbroken (= Coulomb) phase - has in fact already been performed;
this specially
adjusted Higgs potential corresponds to the Linde-Weinberg
bound\cite{wein,linde} which is slightly different that the better known
Coleman-Weinberg bound\cite{cole}. In the Linde-Weinberg bound,
the Higgs bare mass is fine-tuned to the requirement
that the two minima be equally deep. This is precisely
what our prediction using multiple point criticality
(or fundamental non-locality) would suggest:
we predict the Linde-Weinberg situation as the solution to the
problem of why the Higgs field expectation value is  so small.
Indeed, in the Linde-Weinberg situation,
the vacuum expectation value of the Higgs field is
obtained from a logarithm and in this way comes to deviate
exponentially from the
input mass. In the philosophy of a fundamental scale, we
would of course take the input mass to be at the Planck scale (or whatever
the fundamental
scale is taken to be). This looks wonderful at first:
we have solved the problem of the small Higgs expectation value
by postulating fundamental non-locality! The technical hierarchy problem,
that, in going to different orders in perturbative calculations,
includes  quadratic divergences (if you do not have super-symmetry
at least) which are expected to be of the order of the cut-off scale,
is now argued away. This is done by claiming that in going to different orders
in the perturbative calculations, we should for each order recalculate
the amount of space-time volume which is in the unbroken phase
(and the amount which is in the Higgs phase). After this recalculation - which
is not really performed inasmuch as we do not actually know the non-local
action -
we find that we have just to use the Linde-Weinberg situation in which case
the Higgs mass and expectation value are not
renormalised away.

However, the Linde-Weinberg
bound does not agree well with experiment. The prediction obtained from
the requirement of remaining within this bound
- this must also be taken as our prediction from non-locality - is a
Higgs mass
which, using the usual expectation value known from
experimental weak interactions, turns out to be 7.8 GeV
and, in the same connection, a top-quark mass of less
than about 90 GeV.
This is a failure of our model that might be resolvable if there were
several Higgs fields
in which case the simple Linde-Weinberg calculation would not hold true.
With this scenario, one could hope to retain the exponential
behaviour of the Higgs field expectation value and
thereby still
solve the hierarchy-related problem of why the Higgs particle is so light.

Another possible scenario would be that Nature ends up at a  point in the
multiply critical manifold (presumably of high dimension) at which phase
transitions are maximally {\em first order}. In this picture, $m^2_{Higgs}$
becomes small by using the well-known two-minima Higgs field potential
of the pure Standard Model with one-loop corrections\cite{sher} in
conjunction with the MPCP fine-tuning mechanism. This latter states that the
Higgs field minima should be {\em degenerate} (i.e., have identical energy
densities) and furthermore, that the generic probability for having a universe
that can only be realized with two or more coexisting phases
should be maximised (maximum ``heat of melting''). This latter is implemented
in\cite{cdfhbn}
by postulating that one of the minima is at the largest thinkable energy -
the Planck scale.
This is tantamount to taking the ``heat of melting'' to be of order unity in
Planck scale units; it is then generically unlikely that
a universe can be realised as a single phase.
The two minima of the degenerate vacua of the
Higgs field potential get in this way separated by energies
that differ maximally;
i.e., of order of unity in Planck units.
By assuming that this gap in energy is as large as possible,
it is almost insured that even randomly chosen values of
``extensive variables'' (e.g., energy)
will fall within this gap and that the conjugate intensive quantities (e.g.,
Higgs mass and top quark mass) will accordingly be fine-tuned.
This implementation of the multiple point criticality fine-tuning mechanism
yields extremely good
predictions\cite{cdfhbn}
for the top quark mass and the expected Higgs mass. This has been discussed
in Section~\ref{paradoxx}.

Furthermore, by postulating that Nature seeks out multiple point criticality
at maximally
first order transitions (on for example a multi-dimensional multiply critical
surface), we might also alleviate the universality problem inherent to first
order transitions\cite{hbnpc} inasmuch as the requirement of maximal
first-orderness might place us at a {\em unique} position in the multiply
critical
manifold in parameter space. This could lead to unique values of continuum
couplings.

\item {\em Mass Hierarchies of Quark-Lepton Generations}        \\
As one of the fine-tuning mysteries we also count the question of why
the masses for most quarks and leptons are so small compared to the
weak interaction scale, which is the mass you would expect
if the Yukawa couplings were simply of order unity \cite{nl1,nl2}.
The explanation suggested is that, in the physics at the
fundamental scale, there are some Higgs fields that get
relatively small expectation values, much in analogy to
the above speculation about the Weinberg-Salam Higgs field. They might
even easily be exponentially light, due to the fact that their
masses are determined via a logarithm. But
even just the presence of the phase transition makes the Higgs masses small.
After all, it is at a transition between positive and negative mass
squared that we have the boundary between the Higgsed  and the
Coulomb phases. This then means that the transitions between
left and right handed components of fermions, which
need such Higgses for their occurrence, are suppressed.
Recent work \cite{frog1,frog2,frog3} explores
 the mass hierarchy of fermions
(leptons and quarks)
as a consequence of Higgsed
gauge symmetries that are only weakly broken,
due to the Higgs fields having only ``small'' expectation
values.
Such a mechanism can rather naturally explain the
large gaps between the generations in the mass spectrum.

{}From the point of view of a very general picture with approximately conserved
quantum numbers, searches have been made for clues as to which approximately
conserved gauge quantum numbers should exist beyond the Standard Model.
Using the extra quantum numbers in the
$SMG^3$ model (that is used here to in connection with our MPCP predictions
of fine-structure constants)
leads to a rather natural explanation of
the generation mass gaps. However,
a problem is encountered in getting the top quark mass
sufficiently heavy
compared to the bottom quark and  $\tau$ lepton masses.
However adding a fourth
Abelian gauge group $U(1)_f$ helps\cite{frog1,frog2,frog3}.


\item {\em Strong CP-conservation also at a junction of phases}

Yet another fine-tuning problem is the strong CP-problem.
The $\Theta_{QCD}$ dependence of the
first order de-confining phase transition has recently been
studied\cite{schierholz}. It is found
that in the continuum limit there is a critical point at $\Theta_{QCD} = 0$,
where the confinement phase corresponding to $\Theta_{QCD} = 0$ meets the
``Higgs'' phase corresponding to $\Theta_{QCD} \neq 0$. It then follows,
assuming that QCD is in the confinement phase, that $\Theta_{QCD} = 0$ and
CP is conserved by the strong interactions.
                             However, from our point
of view, we look for a more ``ontological''
type of solution and do  not accept that Nature at the bare level
should be precisely renormalised to, for example, reveal the confining phase
of long distance QCD. We can nevertheless,
in the spirit of our multiple point principle above,
use the phase diagram of \cite{schierholz} to suggest
the possibility that $\Theta_{QCD}=0$ can be characterised
as a meeting point for phases. Therefore, even the strong
CP-problem of why  $\Theta_{QCD}$ is so small can find
an explanation derivable from the coexistence of phases
in the same spirit as our solution of
the other fine-tuning problems.

\end{enumerate}

\subsection{A speculative alternative proposal for the stability of the
multiple point}\label{specpro}

The most current and probably most eloquent explanation for Nature's
affinity for the multiple point stems from the apparent necessity of
having long distance  (diffeomorphism invariant) nonlocal interactions
if the fine-tuning of physical constants is to be dynamical (discussed
in Section~\ref{fintnnonloc}).
However, an earlier pursuit of a theoretical explanation for why
the multiple point should be realised in Nature has some merits. This
proposal of a  mechanism for the stability of the multiple point (outlined
in more detail in Appendix~\ref{specproapp}) assumes a
nonlocal lattice gauge glass with a random plaquette action  (e.g.;
an action with quenched  random  values  for  character
expansion parameters). Here ``non-locality'' is restricted to fundamental
scale distances:
action terms for Wilson loops of extent $A$ very large compared to the
lattice constant $a$ are effectively nonlocal as seen from  scales  in
the intermediate length range $[a,A]$. We define gauge couplings  that
run due to the inclusion of successively larger Wilson loops in  going
towards the infrared. The effect of these terms on the running of  the
couplings is describable by  an  extra  term  in  the  Callan-Symanzik
$\beta$-function. The inclusion of these glassy nonlocal action  terms
in such a generalised $\beta$-function (really a multicomponent vector
of generalised $\beta$-functions) is in addition to  but  opposite  in
sign to the normal Yang-Mills renormalization group  contribution. It
is argued  that rapid variations in the generalised  $\beta$-functions at
the multiple point can easily lead to zeros of these $\beta$-functions
close to the multiple point. It is estimated that already at  energies
near the Planck scale the running plaquette  action  parameter  values
are presumably very close to those of the  ``infrared  stable''  fixed
point zeros of the generalised $\beta$-functions which in turn are  close
to the multiple point (see Appendix~\ref{appstab}).


\begin{flushright}
\begin{tiny}
theph.tex 31 May 96 alf 
\end{tiny}
\end{flushright}

\section{The phases of a non-simple gauge group}\label{phasecon}

\subsection{Features of the phases that can convene at the multiple
point}\label{featuresofph}

Since it is  postulated  that  Nature  seeks  a  special  point  in
plaquette action parameter space - the multiple  point  -  where  many
``phases'' come together, it is necessary to clarify what is meant by these
``phases'' as well as how such different ``phases'' are distinguished.
This is the purpose of this Section.

First it should be  made  clear  that  when  referring  to  phases  of
groups at the Planck scale, it is phase transitions between
what are usually referred to as lattice artifact  ``phases''.
Distinguishing  quantitatively  different  physical  behaviours  -  here
referred    to     as     ``Higgs-like'',     ``Coulomb-like''     and
``confinement-like phases'' - at the lattice scale (here taken as  the
Planck scale) is motivated from results\cite{lau1,lau2,lau3,lau4,lau5}
 obtained using the mean  field
approximation $(MFA)$ which intrinsically distinguishes phases on  the
basis  of  the  qualitative  differences  in  the  physics  that   are
discernible  at  the  scale  of  the  lattice.   All
``phases'' involving non-Abelian \dofx will of course, for sufficiently
long  distances,  turn  out  to  be  confining  with  no  long   range
correlations (corresponding to finite glue-ball masses) when, as is the
case in this section, matter fields are ignored.

An important feature of the phases of interest here is  that  they  be
identifiable at the scale of the lattice. The following
discussion of some qualitative features of such  phases  motivates
the more or less well known fact that transitions between the  lattice
artifact phases of interest to us are first order transitions (at least
near the multiple point). Let us write the logarithm of the
partition function as $\log Z
\propto V_{tot}\cdot \log f(\beta)$ where $V_{tot}$ is the geometrical
volume of the system and $f(\beta)\stackrel{def}{=}\frac{1}{V_{tot}}\log Z$
is the free energy density. As $Z$
is essentially the product of the average height
$\langle e^{S_{action}}\rangle$
of the distribution of plaquette variables  times  the  average  width
$W(\beta)$  of   this   distribution, we can write
$f(\beta)=
\langle s_{action} \rangle +s_{entropy}$ where
$\langle s_{action}\rangle\stackrel{def}{=}
\frac{1}{V_{tot}}\langle S_{action} \rangle$ and
$s_{entropy}\stackrel{def}{=} \frac{1}{V_{tot}}\log W(\beta)$.

Let us clarify the properties required of two phases
- call them phases  $I$  and  $II$  - in order that they be
distinguishable using  a  small
geometric sampling volume of the order of a  few  lattice  cubes.  The
hope for being able to do this lies in not having much overlap in  the
regions of configuration space
\footnote{A point in configuration space
assigns an element of the gauge group to each degree of freedom  (link
variable in the case of a lattice).}
that are appreciably populated in
different phases
\footnote{Within each phase, the  configuration  space
appreciably populated by quantum fluctuations varies continuously as a
function of the plaquette action parameters.}.
This being the case, it
is  meaningful  to introduce  free   energy   density   functions
$f_I(\beta)$ and $f_{II}(\beta)$ that very much  dominate  respectively
in  the phases $I$ and $II$:

\beq  f(\beta)  \approx  \left  \{\begin{array}{l} f_I \mbox{  in phase I} \\
f_{II} \mbox{  in  phase  II}
\end{array} \right. \eeq

\nin which corresponds to a partition  function  $Z$ that approximately
factorizes with one factor for each of  the two phases $I$ and $II$:

\beq Z\approx Z_I(\beta)+ Z_{II}(\beta).\eeq

The disjointness of the regions of  configuration
space appreciably populated in phases $I$ and $II$  is  tantamount  to
requiring that the quantity

\beq                                                              \log
(\frac{Z_I(\beta)}{Z_{II}(\beta)})=V_{tot}(f_I(\beta)-f_{II}(\beta))
\label{quan} \eeq

\nin be very  large  or  very  small  depending  on
whether we are in phase $I$ or $II$ (assuming that we are sufficiently
far  removed   from   the   phase   boundary   where   by   definition
$\frac{Z_I(\beta)}{Z_{II}(\beta)}=1$).  The  relation  (\ref{quan})  is
equivalent to saying that if, in going from phase $I$  to  $II$,  this
quantity that contains $V_{tot}$  as  a  factor  changes  by  a  large
amount, a change could also be detected using a smaller volume  (e.g.,
a lattice scale volume $V_{lattice\;scale}$).

This scenario relies in  an  essential  way  on  the  assumption  that
the quantity  $f_I(\beta)-f_{II}(\beta)$,  which  vanishes  at  the  phase
boundary, rapidly becomes large or small in going to parameter  values
$\beta$ removed from the boundary. This in turn relies on having to  a
good approximation a discontinuous change at the phase boundary in the
configuration space appreciably populated  by  the  two  phases.  This
would be reflected as a discontinuity at the boundary in the entropy

\beq   \Delta   s_{entropy}=   s_{I,\;entropy}(\beta_{crit}+\epsilon)-
s_{II,\;entropy}(\beta_{crit}-\epsilon). \eeq

\nin Such a discontinuity can come about  at  the  phase  boundary  if
there is a compensating jump in the average plaquette action

\beq        \Delta         \langle         s_{action}\rangle         =
\langle                   s_{I,\;action}(\beta_{crit}+\epsilon)\rangle-
\langle s_{II,\;action}(\beta_{crit}-\epsilon)\rangle. \eeq

\nin in accord with having (by definition) $f_I=f_{II}$ at  the  phase
boundary; i.e., in the limit $\epsilon \rightarrow 0$, we must have

\beq      0= \log
f_I(\beta_{crit}+\epsilon)-\log f_{II}(\beta_{crit}-\epsilon)=
\Delta \langle s_{action}\rangle - \Delta s_{entropy} \eeq

\nin But the average plaquette action is essentially the derivative of
$\log Z$ w.r.t $\log \beta$ so a jump in $\langle s_{action}  \rangle$
is tantamount to  a  discontinuity  in  $\frac{\partial  \log  Z}{\log
\beta} $ which is of course the defining feature of first order  phase
transitions. In fact it is known from Monte Carlo  calculations  based
on lattice gauge theories that phase transitions at the multiple point
are first order. So the idea of distinguishing lattice artifact phases
at the scale of the lattice is not inconsistent with known results.

However,  determination  of  gauge
couplings based on first  order  phase  transition  are  intrinsically
plagued by problems of  non-universality.  From  this  point  of  view,
second order phase transitions would be  preferable.  However,  it  is
rather meaningless to talk about second order phase transitions  at  a
given scale. The phases separated by a second  order
phase transition are  defined relative  to
distinct (different) fixed points in  parameter  space;  a  particular
phase is identified by  which  fixed  point  the  running  coupling(s)
converge to in  the
long wavelength limit. That second order phase transitions  can  first
be identified in the limit  of  long  wavelengths  can  be  understood
intuitively when it is recalled that the
derivative of $\log Z$ w.r.t.  $\beta$  is  continuous  at  the  phase
boundary of a second order phase transition. It is therefore difficult
to avoid overlap in the regions  of  configuration  space  appreciably
populated  by   the   two   phases.   Hence   the   quantity
$f_I(\beta)-f_{II}(\beta)$, which can be expected to be large for first
order transition, become smaller (and a more poorly defined  quantity)
in the transition from a first order to second order phase  transition
- even for points in the phases $I$ and  $II$  far  removed  from  the
phase     boundary.     In     this      case,      the      condition
$\frac{Z_I(\beta)}{Z_{II}(\beta)}\neq 1$ (which really has no  meaning
for a fully second order phase transition) that is necessary for being
able to see that there are two phases requires  a  larger  and  larger
sample volume $\approx V_{tot}$. This is another way of saying that we
have to take a long wavelength limit in order to see  a  second  order
phase transition.

For small sampling volumes $V_{sample}$, first
order  phase  transitions  dominate  completely  over   second   order
transitions in determining the physically realized phase boundary.

At the multiple point of a phase diagram for a lattice  gauge  theory,
the different lattice artifact phases  are  presumably  separated  by
{\em   first   order}   phase   transitions   and   accordingly    are
distinguishable using a small sampling volume which is assumed  to be of
the order of the Planck  scale  lattice.  Such  phases  are  therefore
completely governed by which {\em micro} (e.g. Planck scale)  physical
fluctuation  patterns  yield  the   maximum   value   of   $\log   Z$.
Qualitatively  different  short  distance  physics  could  consist  of
different distributions of group elements along various  subgroups  or
invariant subgroups of the gauge group for different regions of (bare)
plaquette action parameter space. It is therefore the physics  at  the
scale of the lattice that is of interest because it is  lattice  scale
physics that dominates the different  $\log  Z$  ansatz  that  prevail
(i.e., are maximum) in different parameter space regions separated  by
first order transitions. However,  this  does  not  mean  that  longer
distance behaviour is  unchanged  in  passing  from  one  ``phase''  to
another. As an example, consider the string tension at the  transition
between two different {\em lattice  scale  phases}  that  both  really
correspond to  confining  phases  in  the  usual  sense:  in  what  we
designate as confining in the  mean  field  approximation  ($MFA$)  or
``confinement at the lattice scale'', the string tension has an  order
of magnitude given by dimensional arguments from the lattice.  On  the
other
hand, in what we call the ``Coulomb phase at the lattice  scale'',  or
the Coulomb phase in the $MFA$ approximation, the  string  tension  is
much smaller; i.e,. smaller by an exponential factor.

\subsection{Goal: the classification of the phases  of  the  vacuum
for a lattice gauge theory with non-simple gauge group}

Now we want to assign names to the different lattice artifact  phases,
i.e., qualitatively different physical behaviours of the  vacuum  of  a
lattice gauge theory at  the  lattice  scale.  The
different phases that are possible depend  on  the  gauge  group  $G$.
There is a possible phase for each combination of a subgroup $K_i\in G$
and an invariant
subgroup $H_j \triangleleft K_i$. Pairs such as $(K_i,H_j)$ can be used
as labels for the possible phases. The {\em criterion} as to
which phase is realized is
according  to  whether  or  not  there  is  spontaneous
breakdown of the gauge symmetry remaining after a partial fixing of the
gauge. It is necessary to use the freedom to choose a gauge  in  order
to put Elitzur's Theorem  out  of  commission.  Otherwise  spontaneous
symmetry breakdown is precluded insofar as  Elitzur's  Theorem  states
that any gauge variant quantity vanishes when averaged over  the  full
gauge symmetry.

For illustrative purposes, one can think of a lattice formulation  of  a
gauge theory with gauge group $G$. Let  the  dynamics  of  the  system
be described by a Lagrangian ${\cal  L}(A^{\mu},\phi)$  that
is invariant under (local) gauge transformations $\Lambda$  of
the  gauge  potential  $A^{\mu}$  and the (complex)  scalar
field  $\phi$.  In  the
continuum, the fields  $A^{\mu}$  and  $\phi$  transform  under  gauge
transformations as

\beq    gA^{\mu}(x)\rightarrow     \Lambda^{-1}(x)gA^{\mu}(x)\Lambda(x)+
i\Lambda^{-1}(x)\partial^{\mu}\Lambda(x)
\;(\mbox{$g=$ coupling constant})\label{gpot} \eeq

\beq \phi(x) \rightarrow \Lambda(x)\phi(x) \label{phi} \eeq

In the lattice  formulation,  each  of  the  four  components  of  the
$A^{\mu}$ field corresponds  to  a  group-valued  variable  $U(\link)$
defined on links $\link$ of the lattice; under a local gauge transformation,
the $U(\linkxy)$ transforms as

\beq U(\linkxy)\rightarrow \Lambda^{-1}(x)U(\linkxy)\Lambda(y)=
\Lambda^{-1}(x)U(\linkxy)\Lambda(x+a\delta^{\nu})\approx
\Lambda^{-1}(x)U(\linkxy)(\Lambda(x)+\partial^{\nu}\Lambda(x)a\delta_{\nu})
\label{lgt}\eeq

$$=\Lambda^{-1}(x)U(\linkxy)\Lambda(x)(1+\partial^{\nu}(\log
\Lambda(x))a\delta_{\nu})
\approx \Lambda^{-1}(x)U(\linkxy)\Lambda(x)\exp(i\partial^{\nu} (\log
\Lambda(x))a\delta_{\nu})
$$

\nin This is readily verified: write
$U(\linkxy)=\exp(iA_{\nu}(x)a\delta^{\nu})\approx
1+iA^{\nu}(x)a\delta_{\nu}$ in
which case the gauge transformation above is

$$\Lambda^{-1}(x)(1+iA^{\nu}(x)a\delta_{\nu})\Lambda(x)(1+
\partial^{\nu}(\log \Lambda(x))a\delta_{\nu})\approx $$

$$1+\Lambda^{-1}(x)(iA^{\nu}(x)a\delta_{\nu})\Lambda(x)+
\partial^{\nu}(\log \Lambda(x))a\delta_{\nu} $$

$$=1+\Lambda^{-1}(x)(iA^{\nu}(x)a\delta_{\nu})\Lambda(x)+
\Lambda^{-1}(x)\Lambda(x)
\frac{1}{\Lambda(x)}\partial^{\nu}(\Lambda(x))a\delta_{\nu} $$

\nin which corresponds to the transformation (\ref{gpot}) for the gauge
potential.

On the lattice, the group-valued field $\phi$ is defined on lattice sites;
the transformation rule is as in (\ref{phi}) above.

As the aim is to determine the transformation properties
of the  {\em vacuum}  under  gauge
transformations consistent with the gauge choice, it is  necessary  to
characterise the vacuum for the Yang-Mills field $U(\linkxy)$
as well as the scalar
field $\phi(\sitex)$ in some way. The Yang-Mills vacuum could be
characterised  by  a
probability density $P(U(\linklo))$:

\beq  P(U(\link^{\!\!\!\!l_0}))  =  \int  \prod_{\link \;\; \not   = \;\;
\linklo} d^{{\cal H}aar}U(\link) e^S. \eeq

Here $U(\linklo) \in SMG^3$ is the link variable  associated  with  an
arbitrary link $\linklo$ of the lattice A similar probability  density
function can be envisaged for the site-defined scalar field $\phi$:

\beq P(\phi(\sitex)) = \int \prod_{\site\; \neq \; \sitex}d\phi(\site)
e^S \eeq

\nin where $\sitex$ is an arbitrary site of the lattice. Assuming
that translational  and  rotational  symmetry  is  not  spontaneously
broken, the density functions $P(U(\linklo))$ and $P(\phi(\sitex))$ can
be taken as the same for respectively all links and sites.

\nin We assume that the action $S$  contains  a  term  that  partially
fixes the gauge in accord with the  choice  of  gauge.
                                  It it also assumed that  the  action
contains a parameter $\epsilon$ that multiplies a term that explicitly
breaks the symmetry under consideration; $\epsilon$ is  set  equal  to
zero at the end of the calculation.

By expanding the vacuum density functions  $P(U(\link^{\!\!\!\!l_0}))$
and    $P(\phi(\sitex))$    in    continuous    unitary    irreducible
representations  of  the  gauge  group,  it  can  be  seen  that   the
investigation of the transformation properties of these functions  is
reduced to an examination of  the  transformation  properties  of  the
expansion            coefficients            $\int            d^{{\cal
H}aar}U(\linklo)D_{kl}^{(\nu)*}(U(\linklo))              P(U(\linklo))
\stackrel{def.}{=} \langle  D_{kl}^{(\nu)*}(U(\linklo))\rangle$  where
$D_{kl}^{(\nu)}(U(\linklo))$  denotes  a   matrix   element\cite{hamermesh}
   of   the
continuous unitary irreducible representation  in  the  representation
$(\nu)$.

\subsection{Some motivation for vacuum phase classification scheme}

The scheme to be used for vacuum phase classification
utilises transformation  properties  of  the  vacuum  that  are
suggested by examining the requirements for getting a  massless  gauge
particle as the Nambu-Goldstone  boson  accompanying  the  spontaneous
breakdown  of  the   vacuum   $\langle   U(\linkxy)\rangle   $.   These
requirements should certainly be consistent  with  the  transformation
properties of the vacuum $\langle U(\linkxy)\rangle $ that we  use  to
define a Coulomb phase in the classification scheme to be given.

As already pointed out, a gauge choice must  be  made  in  order  that
spontaneous breakdown of gauge symmetry is at all possible.  Otherwise
Elitzur's Theorem insures that all  gauge  variant  quantities  vanish
identically.

Once a gauge choice is made - the Lorenz gauge is  strongly  suggested
inasmuch as  we want, in order to classify phases, to retain the freedom to
make  gauge  transformations
with constant and linear gauge functions  - the symmetry  under
the remaining gauge symmetry must somehow be broken in order to get  a
Nambu-Goldstone boson that, according to the Nambu-Goldstone  Theorem,
is present for each generator of  a  spontaneously  broken  continuous
gauge symmetry. However  the  proof  of  the  Nambu-Goldstone  Theorem
requires  the  assumption  of  translational   invariance.   This   is
tantamount to the requirement that the vacuum is invariant under gauge
transformations having gauge functions such that these  transformations
correspond to gauge transformations generated by
the
commutator  of  the  momentum  operator  with  the  generator  of  the
spontaneously broken symmetry.

Recalling from~(\ref{lgt}) that a link variable $U(\linkxy)$ transforms
under gauge
transformations as

\beq  U(\linkxy)  \rightarrow   \Lambda^{-1}(x)U(\linkxy)\Lambda(x)\cdot
\underbrace{\exp(        i\partial^{\nu}(\log         \Lambda(x))\cdot
a\delta_{\nu}}_{\mbox{\footnotesize gradient part of transf.}}
\label{utransf},\eeq

\nin it is seen that, for the special case of an  Abelian  gauge  group,  a
gauge function that is linear in the coordinates (or higher order  in
the coordinates) is required for  spontaneous  breakdown  because  the
only possibility for spontaneously breaking the  symmetry  comes  from
the ``gradient'' part of the transformation (\ref{utransf}). However,  we
need also to  take  into  account  the  requirement  of  translational
invariance.

We show now that these two requirements. i.e.,

\begin{enumerate}

\item spontaneous breakdown via the  gradient  in  (\ref{utransf})

\item translational  invariance

\end{enumerate}

\nin can only be satisfied for gauge transformations with linear gauge
functions. Let $Q_{\nu}$ denote the generator of such transformations.
For  such  transformations,  the  first   requirement   is   obviously
satisfied.  The  second  requirement  is   equivalent   to   requiring
that the vacuum $\langle U(\linkxy)\rangle  $  is  annihilated
by the commutator $[P_{\mu},Q_{\nu}]=ig_{\mu\nu}Q$ where  $Q$  denotes
the generator of gauge transformations with constant gauge  functions.
So the condition for having translational invariance  translates  into
the requirement  that  the  vacuum  $\langle  U(\linkxy)\rangle  $  be
invariant under gauge transformations with constant  gauge  functions.
An examination of (\ref{utransf}) verifies that this is always  true  for
Abelian gauge groups and also for  non-Abelian  groups  if  the  vacuum
expectation value $\langle U(\linkxy)\rangle $ lies in the  centre  of
the group (which just means that the vacuum is not ``Higgsed'').

Note that while gauge transformations with gauge  functions  quadratic
(and higher order) in  the  coordinates  would  suffice  for  giving
spontaneous symmetry breaking, such  gauge  functions  would  preclude
translational invariance.


In summary, the conditions to be  fulfilled  in  order
that the Nambu-Goldstone boson accompanying a spontaneous breakdown of
gauge symmetry can be identified with a massless gauge  particle  (the
existence of which is the  characteristic  feature  of  a  Coulomb-like
phase) suggest that  the  Coulomb  phase  vacuum  is  invariant  under
gauge   transformations   having  a  constant   gauge   function    but
spontaneously broken under gauge transformations  having  linear  gauge
functions. These transformation properties of the vacuum  will  emerge
in Section~\ref{lingf} as the defining  features  of  a  Coulomb-like  phase.

\subsection{``Phase'' classification according to symmetry properties
of vacuum}\label{phclass}

When the gauge field $U(\linkxy)$ takes values in  a  non-simple  gauge
group such as $SMG^3$ having many subgroups and  invariant  subgroups
(including discrete subgroups),
it is possible for \dofx corresponding say to  different  subgroups  to
take  group  values  according  to  distributions  that   characterise
qualitatively  different  physical  behaviours  along   the   different
subgroups. Some \dofx can have a fluctuation pattern  characteristic
of a Higgsed phase; some  of  the  \dofx  having  fluctuation  patterns
characteristic  of  an  un-Higgsed  phase  can  be  further  classified
according  to  whether  they  have  Coulomb-like  or
confinement-like  patterns  of  fluctuation.  The  point  is   that   a
``phase'', which of course corresponds to  a region  in  the  action
parameter space, can, for a non-simple gauge group, be described in terms of
characteristics that differ along different subgroups.

The fluctuation patterns for the various \dofx corresponding to these
subgroups  can be classified according to the
transformation properties of the vacuum under the two classes of gauge
transformations $\Lambda_{Const}$ and $\Lambda_{Linear}$; following
a partial fixing of the gauge, the (lattice artifact) phases
of  the  vacuum are to be classified\cite{frogniel,nonabel}
according  to whether or not there is spontaneous breakdown of gauge
symmetry under gauge
transformations corresponding to the sets of gauge functions
$\Lambda_{Const}$ and $\Lambda_{Linear}$
that are respectively constant and linear in the spacetime coordinates:

\beq \Lambda_{Const}\in
\{\Lambda:\br^4\rightarrow G|\exists\alpha[\forall x\in \br^4 [\Lambda(x)=
e^{i\alpha}]] \}\eeq

\nin   and

\beq\Lambda_{Linear}\in
\{\Lambda:\br^4\rightarrow G|\exists\alpha_{\mu}[\forall x\in \br^4[\Lambda(x)=
e^{i\alpha_{\mu}x^{\mu}}]]\}. \label{lin} \eeq

Here $\alpha=\alpha^at^a$ and $\alpha_{\mu}=\alpha^a_{\mu}t^a$ where
$a$ is a ``colour'' index in the case of non-Abelian subgroups.
The $t^a$ denote a basis of  the
Lie     algebra     satisfying     the      commutation      relations
$[t^a,t^b]=c^{ab}_ct^c$  where  the  $c^{ab}_c$  are   the   structure
constants.

Spontaneous  symmetry   breakdown   is   manifested   as
non-vanishing values for gauge variant quantities.  However,  according
to Elitzur's theorem, such quantities cannot survive  under  the  full
gauge symmetry. Hence a partial fixing of the gauge  is  necessary  before  it
makes sense to talk about the spontaneous breaking of symmetry.  We
choose the Lorentz gauge for the reason that  this  still  allows  the
freedom of making gauge transformations of the types $\Lambda_{Const}$
and $\Lambda_{Linear}$ to be used in classifying the lattice  artifact
``phases'' of the vacuum. On  the  lattice, the choice of the Lorentz gauge
amounts  to  the
condition  $\prod_{\;\linkx \;  \begin{tiny}  \mbox{emanating   from   }
\end{tiny}     \sitex     }     U(\link)=1$     for     all     sites
\begin{large}$\cdot$\end{large}.

It will  be
seen that the set of possible ``phases'' corresponds one-to-one to
the set of  all  possible  subgroup  pairs  $(K,H)$
\footnote{In this classification scheme it has been assumed that the action
energetically favours $U(\Box)\approx \bunit$; however, a vacuum also having
fluxes corresponding to nontrivial elements of the centre could be
favoured if for instance there are negative values for
coefficients of plaquette terms in the action. Such terms would lead to new
partially confining phases that were Coulomb-like
but for which fluctuations in the  \dofx are centred at a
nontrivial element of the centre instead of at the identity.}
consisting  of  a
subgroup $K\subseteq SMG^3$ and invariant subgroup $H\triangleleft K$.
Each ``phase'' $(K,H)$ in general corresponds to a partitioning of the
\dofx (these latter can be labelled by a Lie algebra basis) - some that
are Higgsed, others that are un-Higgsed; of the latter, some  \dofx  can
be confining, others Coulomb-like. It is therefore useful to think of a
group element $U$ of the gauge group as being parameterised  in  terms
of  three  sets  of  coordinates  corresponding  to  three   different
structures that are appropriate to the symmetry properties  used
to define a given phase $(K,H)$ of the vacuum.
These  three  sets  of  coordinates,
which are definable in terms of the gauge group $SMG^3$, the  subgroup
$K$, and the invariant subgroup $H\triangleleft K$,
are the {\em homogeneous space}  $SMG^3/K$,  the  {\em  factor  group}
$K/H$, and $H$ itself:


\beq U=U(g,k,h) \;\;\mbox{with }\;g\in SMG^3/K, k\in K/H, h\in H. \eeq

\nin The coordinates $g\in SMG^3/K$ will  be  seen  to  correspond  to
Higgsed \dof, the coordinates $k\in  K/H$  to  un-Higgsed,  Coulomb-like
\dofx and the coordinates $h\in H$ to un-Higgsed, confined \dof.

\subsubsection{Higgsed or un-Higgsed vacuum}\label{higgsedvac}

The  \dofx  belonging  to  the
subgroup $K$ are said to be un-Higgsed if, after fixing the gauge in accord
with say the Lorentz condition,  $K\subseteq G$ is the maximal
subgroup of gauge transformations belonging to the set $\Lambda_{Const}$
that
leaves the vacuum invariant\footnote{The vacuum invariance referred  to
really   means   the   invariance   of   the   coefficients   $\langle
D^{(\mu)}_{ij}(U(\linkxy)) \rangle_{vac}$ in an expansion  in  (matrix
elements   of)   continuous   unitary   irreducible    representations
$D^{(\mu)}_{ij}(U(\linkxy))$.  The  expansion  referred  to  is   that
corresponding to  some  link  variable  probability  density  function
$P(U(\link^{\!\!\!\!l_0}))  =  \int  \prod_{\link  \;\;  \not  =  \;\;
\linklo} d^{{\cal H}aar}U(\link)  e^S.$}.  For  the  vacuum  of  field
variables defined on sites (denoted by  $\langle  \phi(\sitex)\rangle
$), invariance under transformations $\Lambda_{Const.}$ is possible  only  if
$\langle\phi(\sitex)\rangle=0$. For the vacuum of field variables defined on
links
(denoted by $\langle U(\linkxy)\rangle$,  invariance  under  transformations
$\Lambda_{Const}$ requires that $\langle U(\linkxy)\rangle $ takes values in
the centre of the subgroup $K$. Conventionally, the idea of Higgsed \dofx
pertains to field variables defined on sites. With the above criterion using
$\Lambda_{Const}$, the notion of Higgsed \dofx is generalised to also
include link variables.

If $K\subseteq G$ is the maximal subgroup for which the transformations
$\Lambda_{Const}$ leave the vacuum invariant,
the field variables taking values in the homogeneous  space
$G/K$ (see for example \cite{dubrovin,nakahara}) are by definition Higgsed in
the vacuum.
For these degrees of freedom, gauge symmetry
is  spontaneously   broken   in   the   vacuum   under   gauge
transformations $\Lambda_{Const}$.

Before leaving the discussion of  the  Higgs  phase,  a  simple  model
allowing a Higgsed vacuum is presented. Taking by way of  example  the
gauge group $G=SO(3)$, we consider a model  for  which  there  is  the
possibility of Higgsed  \dof. To this end consider a field variable
defined on sites - denoted $\phi(\site)$ - and taking
values  in  the  cosets  of  the
homogeneous space  $G/K=SO(3)/SO(2)$ in such a way that the gauge symmetry
is spontaneously broken down to that of $SO(2)$.

Upon identifying the field \dofx $\phi(\site)$ with the cosets of
$SO(3)/SO(2)$,
the action of the gauge group on the variables $\phi(\site)$ corresponds to
moving these variables around on a $S_2$ inasmuch as $S_2$ is the orbit
of an element of $SO(3)/SO(2)$ (e.g., the ``north pole'') under the action
of representatives
of the cosets of $SO(3)/SO(2)$ (the same applies of course to the action
of the whole gauge group because every $g\in G$ is the representative of
some coset of $SO(3)/SO(2)$). Formally, the action of the group\cite{gilmore}
$SO(3)$ on the homogeneous space $SO(3)/SO(2)$ is given by

\beq g\cdot \phi(\sitex)= \phi^{\prime}(\sitex)\cdot R
\left \{ \begin{array}{c} g\in SO(3) \\ \phi, \phi^{\prime} \in SO(3)/SO(2) \\
R\in SO(2) \end{array} \right. \eeq

\nin As $g\cdot \phi(\sitex)\in SO(3)$, this can
element can always be expressed as the (unique)
coset decomposition $\phi^{\prime}(\sitex)\cdot R$.
The mapping of $SO(3)/SO(2)$ onto
itself under the action of the group $SO(3)$ is given by

\beq SO(3): \frac{SO(3)}{SO(2)}\stackrel{g}{\longrightarrow}\frac{SO(3)}{SO(2)}
\eeq

\[ \phi\stackrel{g}{\longrightarrow}\phi^{\prime} \;\;\; \phi, \phi^{\prime}
\in SO(3)/SO(2). \]

Think of having an $S_2$ at each site $\sitex$ of the lattice.
In this picture, the variable $\phi(\sitex)$ at each site $\sitex$
corresponds to a point on the $S_2$ at this site. A priori there is no
special point in this homogeneous space $SO(3)/SO(2)\stackrel{-}{\simeq}
S_2$ which implies\footnote{Even if one were to succeed in embedding a
homogeneous space in an affine space in a natural manner, such an
embedding would not in general be convex. Therefore it would generally
be necessary to construct the convex closure (e.g., in
a vector space) if we want to talk about the averages of
field variables.
As an example, think of the homogeneous space $SO(3)/SO(2)$
which is metrically equivalent with an $S_2$ sphere. In this case, one
could obtain the complex closure as a ball in the linear embedding space
$\br^3$. Alternatively, we can imagine supplementing the $SO(3)/SO(2)$ manifold
with the necessary (strictly speaking non-existent) points needed in order to
render averages on the $S_2$ meaningful. Either procedure eliminates the
problem that an average taken on a non-convex envelope is generally unstable;
e.g., for an $S_2$ the average of two points near a pole can jump
discontinuously when these two points are moved around slightly in the
vicinity of the pole. In particular, by including the points in the ball
enclosed by an $S_2$, it is possible for $\langle \phi \rangle$ to have
a value lying
in the symmetric point. This point, corresponding to $\langle \phi \rangle=0$,
is of course unique in not leading to spontaneous breakdown under rotations
of the $S_2$.}
$\langle \phi(\site) \rangle=0$.
The Higgs mechanism comes into play when, for all sites on the lattice,
the vacuum  distribution of the $\phi(\site)$ - modulo parallel transport
between sites by link variables - clusters about some point in
$SO(3)/SO(2)\stackrel{-}{\simeq} S_2$ and we conclude
\footnote{There are technical problems here. This conclusion presumably
requires the validity of the ``cluster decomposition principle''
\cite{streater}}
that
$\langle \phi(\site) \rangle\neq 0$. In the Higgsed situation, the point
of $S_2$ about which there is a clustering of
the values of $\phi(\site)$ for all sites of the lattice (modulo parallel
transport) defines the ``north pole'' of the rotations of the $SO(2)$
gauge symmetry surviving the spontaneous breakdown of the $SO(3)$ symmetry
by Higgsing.

The Higgs mechanism outlined above can be provoked if a term

\beq \kappa dist^2(\phi(\sitex), U(\linkxy)\phi(\sitey)) \eeq

\nin where $\kappa$ is a parameter and
$dist^2(\phi(\sitex), U(\linkxy)\phi(\sitey))$ is the suitably defined
squared distance
on the $S_2$ at the site $\sitex$  between the point $\phi(\sitex)$
and the point $\phi(\sitey)$ after the latter is ``parallel transported''
to $\sitex$ using the link variable $U(\linkxy)\in G$.

In terms of of elements $g\in SO(3)$,

\beq dist^2(\phi(\sitex), U(\linkxy)\phi(\sitey))\stackrel{def}{=} \eeq
\[ inf\{dist^2(g_x\cdot SO(2), U(\linkxy)g_y\cdot SO(2)|\]
\[ \mbox{
$g_x$ \& $g_y$ \begin{footnotesize} are reps. of respectively the cosets
\end{footnotesize}$\phi(\sitex)$ \& $\phi(\sitey)$}\} \]

In order to provoke the Higgs mechanism, not only must the parameter
$\kappa$ be sufficiently large to ensure that it doesn't pay not to
have clustered values of the variables $\phi(\cdot)$. It is also
necessary that ``parallel transport'' be well defined so that it makes sense
to talk about the values of $\phi(\site)$ being organised (i.e., clustered)
at some coset of $SO(3)/SO(2)$. This would obviously not be the case if the
theory were confined. In confinement, $\langle U(\linkxy) \rangle=0$ and
parallel transport is meaningless. In the continuum theory, this would
correspond to having large curvature (i.e., large $F_{\mu\nu}$) which
in turn would make parallel transport very path dependent
\footnote{Even when confinement is absent, there are technical difficulties
in defining parallel transport over large spacetime distances; presumably
it is necessary to average over a bundle of spacetime parallel paths.}.

\subsubsection{Confined and Coulomb-like \dofx in the vacuum}\label{lingf}

In the vacuum, the un-Higgsed \dofx -  taking  values  in  the  subgroup
$K$~- can be in a confining phase or a Coulomb-like phase according  to
the  way  these  \dofx  transform  under  gauge  transformations
$\Lambda_{Linear} \in K$ having linear gauge functions.

Degrees  of  freedom  taking  values   in   the   invariant   subgroup
$H\triangleleft K$ are by definition confined in the vacuum if $H$  is
the  maximal  invariant   subgroup   of   gauge
transformations $\Lambda_{Linear}$  that  leaves
the vacuum invariant; i.e.,  $h$  consists  of  the  set  of  elements
$h=\exp\{i\alpha^1_at_a\}$ such that the  gauge  transformations  with
linear gauge function  $\Lambda_{Linear}$  exemplified  by\footnote{In
the quantity $x^1/a$, $a$ denotes the lattice constant; modulo lattice
artifacts, rotational invariance  allows  the  (arbitrary)  choice  of
$x^1$ as the axis $x^{\mu}$  that  we  use.}$\Lambda_{Linear}\stackrel
{def.}{=}h^{x^1/a}$ leave the vacuum invariant.

If $H\triangleleft K$ is the maximal invariant  subgroup  of  \dofx  that  are
confined in the vacuum, the cosets belonging to the factor group $K/H$
are by definition in a Coulomb phase (again, in the Lorentz gauge).
For \dofx corresponding to this set of
cosets, there is invariance under coset representatives of the type
$\Lambda_{Const}$ while gauge  symmetry
is spontaneously broken in  the  vacuum  under  coset representatives
of the type $\Lambda_{Linear}$.

In conclusion, the phase classification  scheme  used  here  allows  a
unique  phase  for  each  subgroup  $K$  and  invariant  subgroup  $H$
satisfying the condition $H\triangleleft  K  \subseteq  G$  where  $G$
denotes the gauge group. For a non-simple group
such as the $SMG$, any given phase $(K\subseteq G, H\triangleleft K)$
generally has \dofx of  all  possible  types:  Higgsed,  confined,  and
Coulomb-like.


The different possible ``phases'' of the vacuum will be classified  on
the basis of the transformation properties of the vacuum  under  gauge
transformations of the types  $\Lambda_{Const.}(x)=  e^{i\alpha^at^a}$
and                $\Lambda_{Linear}(x)=e^{i\alpha^a_{\mu}t^ax^{\mu}}$
\cite{frogniel}.

For  the  \dofx
that have as values the cosets of the factor  group  $K/H$,  there  is
invariance of the vacuum expectation  value  under  the  corresponding
constant gauge transformations, while there is  spontaneous  breakdown
under the linear gauge transformations corresponding  to  these  \dof.
Such \dofx will be said to demonstrate ``Coulomb-like'' behaviour.


Having now formal criteria for distinguishing the different phases  of
the vacuum, it would be useful to elaborate a bit further on  what  is
meant by having  a  phase  associated  with  a  subgroup  -  invariant
subgroup  pair  $(K_i\subseteq  G,  H_j  \lhd  K_i)$.  A  phase  is  a
characteristic region of action parameter space. Where does an  action
parameter space come from and what makes a region of it characteristic
of a given phase $(K_i\subseteq G, H_j \lhd K_i)$? An action parameter
space comes about by choosing  a  functional  form  of  the  plaquette
action. Having an action  allows  the  calculation  of  the  partition
function  and  subsequently   the   free   energy.   As   each   phase
$(K_i\subseteq G,  H_j  \lhd  K_i)$  corresponds  to  different  micro
physical  patterns  of  fluctuations  along  the  subgroup  $K_i$  and
invariant subgroup $H_j$ (recall  that  we  are  dealing  with  phases
separated by first order phase transitions),  the  partition  function
and hence the free energy is a different  function  of  the  plaquette
action parameters for each phase $(K_i\subseteq G, H_j \lhd  K_i)$.  A
region of plaquette action parameter  space  is  characteristic  of  a
(i.e., in a) given phase $(K_i\subseteq G, H_j \lhd K_i)$ if the  free
energy $\log Z_{K_i\subseteq G,  H_j  \triangleleft  K_i}$  associated
with this phase has the largest value of  all  free  energy  functions
(i.e., one free energy function for each phase $(L\subseteq  G,  M\lhd
L)$) in this region of plaquette action parameter space.

In seeking the multiple point, we  seek  the
point or surface in parameter space where ``all'' (or a maximum number
of)  phases  $(K_i\subseteq  G,  H_j  \lhd  K_i)$  ``touch''  one
another.

\begin{flushright}
\begin{tiny}
theact.tex 31 May 96 alf 
\end{tiny}
\end{flushright}

\section{Generalised action is required}\label{secgenact}


\subsection{The general idea}\label{genidea}

In
implementing the multiple point criticality  principle,  the  goal,  a
priori at least, is to bring {\em all} phases  $(K_i\subseteq  G,  H_j
\lhd K_i)$  together  in  plaquette  action  parameter  space  at  the
multiple point/surface. In practice we settle for a point in parameter
space at which a chosen set of \pcps come together.  In  referring  to
the multiple point, it is, in practice at least, such a  point  that  is
meant.

A  condition  that  can  be used as a criterion for when a   \pcp
$(K_i\subseteq G,H_i  \triangleleft  K_i)$  is  in  contact  with  the
multiple  point is that  there must be a region  of  parameter   space
infinitesimally close to the multiple point in which the
free energy function $\log  Z_{K_i\subseteq  G,  H_j
\triangleleft K_i}$ corresponding to this phase has a  value  that  in
this region is greater than the values of the  free  energy  functions
corresponding to all other \pcps. This  obviously  requires  that  the
free energy function for each phase $(K_i\subseteq G, H_j  \lhd  K_i)$
depends specifically on an ansatz for the micro  physical  fluctuation
pattern along the subgroup $K_i\subseteq G$ and the invariant subgroup
$H_j \lhd K_i$ - at least at the multiple point. The picture to have  in
mind is that, for each ansatz (i.e., each partially confining phase)
that can be  realized,
the corresponding free energy function, which is defined everywhere in
parameter space, dominates over the free energy functions of all other
\pcps in {\em some} region of parameter space. A point in parameter  space
a neighbourhood of which contains a region in which each chosen  phase
has a dominant free energy function is the multiple point.

Note that  a free energy function $\log Z_{K_i\subseteq G, H_j
\triangleleft K_i}$ is determined, via  the  partition  function
$Z$, from the plaquette action. The latter has in turn a  functional
form that is determined by the choice of action parameters.  So  if  a
phase $(K_i\subseteq G, H_j \triangleleft K_i)$ is to  be  in  contact
with the multiple point,
the  choice  of  action  parameters  must  include   parameters   that
specifically govern the micro physical fluctuation patterns along this
subgroup  -  invariant  subgroup  pair  in  such  a   way   that   the
corresponding  free  energy  function  $\log  Z_{K_i\subseteq  G,
H_j\triangleleft  K_i}$  can  come  to  dominate  in  some  region  of
parameter space at the multiple point.

A further requirement on  the  action  parameters  is  that  they  are
sufficiently independent in the  sense  that  the  adjustment  of  the
parameters relevant for making the free energy function for one  phase
dominant in a neighbourhood of the  multiple  point  does  not,  in  so
doing, preclude the variation of the free  energy  function  for some
other phase that is necessary in order that it can  dominate  in  some
other region adjoining the multiple point.

\subsection{The $SMG$ (and $U(N)$) can have an infinity of
phases}\label{iftyofph}

What is needed is a way to seek the multiple point in a
phase diagram
for gauge  groups  such  as  $SMG$,  $SMG^3$  and
$U(N)$ (collectively referred to by the  symbol  ``$G$'');  $U(N)$  is
useful because it has many features  in  common  with  the  $SMG$  and
$SMG^3$ while, for expositive purposes, being simpler  to  deal  with.
All  these  groups  have  nontrivial  subgroups  $K\subseteq  G$   and
invariant subgroups $H\triangleleft K$.
In fact, for all  the  gauge
groups $G$ considered, the number of possible phases $(K_j\subseteq G,
H_i\triangleleft K_j)$ is infinite. For example,  for  all  these  gauge
groups $G$, there is an infinity of possible subgroups $K_j$ including
$U(1)$, the Cartesian product of $U(1)$ subgroups and  an
invariant $U(1)$ subgroup as  well  as  subgroups  of  such  Cartesian
products, and then also discrete subgroups of the centre of non-Abelian
subgroups. Possible invariant  subgroups  $H_i$  include  the  infinitely  many
discrete (invariant) subgroups $\bz_N$ of $U(1)$ the latter of  course
being an invariant subgroup as well as a subgroup of the gauge  groups
considered.

So the goal, in principle at least,
is to bring together the infinity of possible
phases $(K_j,H_i)$ at the multiple point. In general this would require an
infinite number of action parameters that could,
in principle at least,  be  taken  as
the coefficients of an expansion in group characters corresponding  to
each imaginable way of associating group characters with Wilson loops.
A priori then, the number $n$ of possible phases as well as the dimension
$d$ of action parameter space are infinite.

It is not clear whether the number $n$ of possible phases approaches infinity
more or less rapidly than the number of possible
action parameters that could be used to span a parameter space in which the
phase diagram for a system with $n\rightarrow \infty$ phases could be
constructed.
The suspicion is that the
number of phases $n$ increases less  rapidly  than  the  total
number of available plaquette parameters as  $n\rightarrow  \infty  $.
This suspicion is motivated by considering the  phases  involving  the
infinity of $\bz_N$ invariant subgroups of say $U(1)$. For a given  $N$,
the number of conjugacy classes is  $N$  (i,e.,  one  class  for  each
discrete element of the (invariant) subgroup $\bz_N$). To each of the  $N$
conjugacy classes there corresponds a coefficient that  could  be
used as the basis of the action parameter space.  On  the  other
hand, the number of subgroups of a $\bz_N$ subgroup (= the  number  of
possible phases w.r.t which there can be  confinement  for  a  $\bz_N$
subgroup) is limited to the number of possible {\em factors}  of  $N$.
As the number of factors of  $N$  is  generally  less  than  $N$,  the
indication is that when all subgroups $\bz_N$ are taken  into  account
(i.e., $N \rightarrow \infty  $)  the  potential  number  of  possible
plaquette action parameters increases more rapidly than the number  of
possible phases w.r.t. which there can be confinement.

\subsection{Practical considerations in implementing the
$MPCP$}\label{practical}

In  practice we shall
restrict consideration to a {\em finite} number $n$ of possible phases
$(K_j,H_i)$. In an infinite dimensional action parameter space this finite
number $n$ of phases could (using an appropriate action) presumably
be made to come together
along an  infinite  dimensional  manifold  ${\cal
M}_{crit,\;\infty}$ of co-dimension no greater than  $n-1$.
The  symbol  ${\cal  M}_{crit,\;\infty}$  denotes
the manifold along which  all  $n$  phases  under  consideration  have
critical parameter values; the  second  subscript  $\infty$  indicates
that this manifold is embedded in an  infinite  dimensional  plaquette
action parameter space.
The claim that the restriction to a finite  number  $n$
of (judiciously chosen) phases will lead  to  a  multiple  point  that
gives good values for gauge couplings is tantamount to the  assumption
that having criticality or not for the infinitely many  ``forgotten''
phases  is
not important for the values of the  action  parameters  used  to
calculate the continuum couplings;
i.e., the coordinates of the multiple point for  the  restricted  phase
diagram would not be changed much if the ``forgotten'' phases had been
appended to the multiple point of the restricted $n$ phase system.




Concurrent with the restriction to a finite number
$n$ of phases, it will be assumed from now on that
the action parameter space can be restricted
to  a  finite  $d$-dimensional  subspace  of  the in principle infinite
dimensional action parameter space.
The idea is to choose a combination of action terms
in such a way that all of the $n$ phases $(K\subseteq  G,  H
\triangleleft K)$ under consideration can, in the the  $d$-dimensional
parameter subspace, be brought together  along  a  sub-manifold  ${\cal
M}_{crit\;d}$ of the $codim$ $n-1$ manifold ${\cal M}_{crit\; \infty}$
embedded  in
the most general infinite dimensional  parameter  space  as  described
above. Denoting a point  in  this  restricted  parameter  space  by  a
$d$-dimensional vector $\vec{\beta}$,  the  points  lying  in   the
manifold ${\cal M}_{crit\;d}$ embedded in the $d$-dimensional parameter  space
can formally be defined as follows:

\beq \{\vec{\beta}|  \forall \;  H_i   \lhd K_j\subseteq  G \;  \exists   \;\;
\vec{\epsilon}_{infinitesimal}=\eeq

\[ =(\epsilon_1, \cdots , \epsilon_d)  \;\;
[\log   Z_{K_j,H_i}(\vec{\beta}+\vec{\epsilon})=\log Z=\mbox
{max}\{\log Z_{L,M}|M\lhd L\subseteq G\}]\}.\]

There can be a constraint on the number $n$ of phases
that can be provoked  by  varying
the action parameters chosen  to  span  a  $d$-dimensional  parameter
space. Consider a multiply critical  manifold  ${\cal  M}_{crit.\;d}$
embedded in the $d$-dimensional parameter space along  which  all  $n$
phases  meet.
If  $codim({\cal  M}_{crit.\;d})=n-1$,  then  the
manifold ${\cal M}_{crit.\;d}$ is referred to as  being  a  {\em
generic}  multiply  critical  manifold.
Multiply  critical  manifolds  that   are   generic   can   be   found
systematically for many parameter choices. Sometimes, however,  it  is
possible to have multiply critical manifolds along which more than the
generic number of phases come together; i.e., along such surfaces,  $n
> codim({\cal M}_{crit.\; d})+1$. Such manifolds  are  referred  to  as
being non-generic and can generally only be found for judicious choices
of  the  $d$-dimensional  parameterisation when, for example, there is some
degree of symmetry.   On   the   other   hand,
$codim({\cal M}_{crit.\;d})$ cannot exceed the dimension  $d$  of  the
action parameter space. In summary, it is seen that

\[ d \geq codim({\cal M}_{crit.\;d}) \left\{ \begin{array}{ll}
=n-1  &  \mbox{\footnotesize  in  the   generic   case}   \\   <   n-1
& \mbox{\footnotesize in the non-generic case} \end{array} \right. . \]

The case $d=n-1(=codim({\cal M}_{crit.\;d}))$, corresponds to the
largest possible number of phases than can come  together  generically
in a $d$-dimensional  parameter  space:  $n-1$  phases  meet  along  a
multiply critical manifold ${\cal  M}_{crit.\;d}$  of  dimension  zero
(generic multiple  point). If $d>n-1$, then we would generically expect to
have
a    multiple    point     critical     surface\footnote{Our     crude
approximation\cite{long} used to extract qualitative features  of  the
phase diagram implies that the continuum couplings at different points
on the multiple surface  have  the  same  values.  Variations  in  the
continuum couplings along the multiple point surface cannot be seen in
the approximation where we  use  the  truncated  Taylor  expansion  of
(\ref{tay2}) and (\ref{tay3}).}  of  co-dimension  $n-1$.
For  a  special  choice  of
parameterisation, utilising for example  the  symmetries  afforded  by
multiple occurrences of a given gauge subgroup/invariant  subgroup,  it
may be possible to have $n > codim({\cal M}_{crit.\;d})+1$; i.e.,  the
number of phases $n$ coming together at the multiple point exceeds the
number of phases that meet at a generic multiple point.

In order to see the difference between a  generic  and  a  non-generic
multiple point, consider a journey in  action  parameter  space
that starts at  a  random  point  and  subsequently  seeks  out  phase
boundaries for which a  successively  greater  number  of  phases  are
accessible by making infinitesimal changes in the action parameters at
the points along the journey. In the generic case, the co-dimension $codim$ of
the boundary  goes  up  by  one  each  time  an  additional  phase  is
encountered so that a phase boundary surface/curve is in contact  with
$codim+1$ phases. In  the  non-generic  case,  the  number  of  phases
accessible goes up faster than the co-dimension - at least once in  the
course of the journey in parameter space. So for the non-generic case,
it is possible to have points in phase space  (e.g.,  a  non-generic  multiple
point) forming a surface at which infinitesimal variations  in  action
parameters give access to a number of phases exceeding $codim+1$ where
$codim$ is the co-dimension of this surface.

The  choice
of plaquette action parameters that span this restricted $d$-dimensional
subspace
amounts to a choice of which  types  of
terms are to be included in the plaquette  action.  These
parameters are the coefficients  of  (gauge  invariant)  action  terms
defined on fields the vacua of which can transform in  characteristic
ways under the gauge symmetry remaining  after  a  partial  fixing  of
the gauge: in the phase classification scheme  outlined  in  the
previous section, the phases of the vacuum  were  in  fact  labelled
according to the  transformation  properties  of  the  vacuum  using  the
gauge symmetry remaining after fixing the gauge in accord with  the  Lorentz
condition.

Included in the gauge  symmetry  remaining  after  adopting  the  Lorentz
condition are the class of gauge  transformations $\Lambda_{Const}$
generated by elements of the Lie algebra that are constants.
The
maximal subgroup $K$ of gauge transformations of the class $\Lambda_{Const}$
that leave the vacuum  invariant  is  the  criterion
used to identify the subgroup $K$ as the field \dofx that in the vacuum
are not Higgsed. Accordingly, the field \dofx (in  the  vacuum)  taking
values in the set of cosets belonging to the homogeneous  space  $G/K$
are in a Higgsed phase. This is just  the  generalised
notion of a Higgsed phase \cite{frogniel} that was used in
Section~\ref{higgsedvac}.

Also included in the gauge transformations allowed  after  fixing  the
gauge in accord with the Lorentz condition are gauge
transformations $\Lambda_{Linear}$ generated by  elements of the
Lie algebra that depend linearly on spacetime. The maximal  invariant
subgroup $H\triangleleft K$  of  gauge  transformations belonging to the
class $\Lambda_{Linear}$ that  leave  the   vacuum
invariant is used as the defining feature of an invariant subgroup $H$
of field \dofx that are confined in the vacuum. The vacuum  expectation
values of field \dofx taking values in the factor group
$K/H$   are   accordingly   invariant   under   gauge    transformations
of the class
$\Lambda_{Const.}$  but  led   to   spontaneous breakdown
under  gauge  transformations  of  the class  $\Lambda_{Linear}$.
These  transformation  properties  are  the  defining  features  of
field \dofx that are in a Coulomb-like phase.

The  boundaries  between  the
various  phases  are   characterised   by   singularities   in the
derivatives of   $\log
Z=\mbox{max}\{\log  Z_{(L,M)}|L\subseteq  G,\;M\lhd  L\}$  where   $G$
denotes the gauge group. At  the
boundary delineating a phase for which an  subgroup  $K_j\subseteq  G$
and  an  invariant  subgroup  $H_i\lhd  K_j$  are  realized  as  being
respectively  un-Higgsed  and  confinement-like,  $\log   Z_{(K_j,H_i)}$
dominates in a part of the neighbourhood of such a boundary. Recall  that
the quantity $\log Z_{(K\subseteq G,H\triangleleft  K)}$ is by  definition
$\log Z$ when  the  latter  is  calculated  for  a  field
configuration distribution for which the lattice gauge theory is in  a
phase for which the homogeneous space $G/K$ is Higgsed, the  invariant
subgroup $H\lhd K $ is ``confining'' and the  factor  group  $K/H$  is
``Coulomb-like''\footnote{i.e., Bianchi  identities  are  ignored  for
\dofx along $H$ and $MFA$ is used w. r. t.  $G/H$  with  link  averages
$\langle U(\link)_{G/H}\rangle \neq 0$.}.


In implementing the multiple point criticality principle $MPCP$ in practice,
we  seek  a  multiple
point in some restriction to  a  finite  dimensional
subspace  of  the  in  principle  infinite  dimensional  action
parameter space. This just amounts to making an action ansatz.
Consider an action parameter space that has  been
chosen so that we can realize a given phase $(K   \subseteq   G,   H
\triangleleft K)$.
In this paper, we consider only the special case $K=G$ corresponding to
not having \dofx that are Higgsed. However, we want to include a suggestion of
the manner in which one - at least in a discretised gauge theory - could
also have convening phases at the
multiple point that are Higgsed w.r.t. to various \dofx even though we
shall not make use of Higgsed phases in the sequel.

In order to bring about a Higgsing of the gauge group $G$ down to the
subgroup $K$, one could use
action terms  defined  on  gauge  invariant
combinations  of  site-defined  fields  $\phi(\sitex)$  and  the  link
variables $U(\link)$. The fields $\phi(\sitex)$  take values on homogeneous
spaces $G/K$ of
the gauge group $G$ where $K\subseteq G$. Such action terms
\footnote{For example,         such         term         could         be
$dist(\phi(\sitex),U(\linkxy)\phi(\sitey))$  where  ``dist''  denotes  a
distance  function  appropriate  to  the  (convex)  manifold  of   the
homogeneous space.} are designed so that for sufficiently large values
of a coefficient $\kappa$,  the  field  $\phi(\sitex)$  acquires  a
non-vanishing vacuum expectation value: $\langle  \phi(\sitex)  \rangle
\neq 0$ with the result  that  the  gauge  symmetry  is  spontaneously
Higgsed from that of the gauge group $G$ down to that of the  subgroup
$K$. Then degrees of freedom corresponding to the cosets of  $G/K$  are
Higgsed and \dofx corresponding to elements of $K$ are unHiggsed.  We
have seen that the defining feature of the subgroup $K$ is that it is the
maximal  subgroup  of  gauge  transformations  having  constant  gauge
functions that leave the vacuum invariant.

Other coefficients - call them $\beta$ and  $\xi$  -  multiply  action
terms defined on factor groups $K/H$ of the un-Higgsed subgroup
$K$ where $H \triangleleft K$. Two types of coefficients $\beta$
and $\xi$ having to  do  with  respectively  continuous  and  discrete
invariant subgroups $H$ are distinguished.  For  sufficiently  large
values of the parameters $\beta$  and/or  $\xi$,  the  gauge  symmetry is
spontaneously broken from that of $K$ down to that of the  invariant
subgroup $H$
under  gauge  transformations of the class $\Lambda_{Linear}$.
The \dofx corresponding to cosets of the factor  group  $K/H$
are Coulomb-like while  elements  of  the  invariant  subgroup
$H$ correspond to confined \dof. The feature used to define
$H$ is that $H\triangleleft K$ is the maximal invariant subgroup of the
class of gauge transformations $\Lambda_{Linear}$
that leaves the vacuum invariant.

Actually, were we to include the
possibility of Higgsed phases, an extra interaction between the Higgs
field and the gauge field (in addition to the one implemented by the use of
covariant derivatives in the kinetic term for the Higgs field) would be
needed in order to make the various phases meet at the multiple point.
Otherwise there is the risk that the fine-structure constant changes (e.g.,
does not remain equal to $\alpha_{crit}$)
in going from
$\langle \phi(\sitex) \rangle=0$ to $\langle \phi(\sitex) \rangle\neq 0$.
A suitable interaction term might be of a rather explicit form; for
example, it could be implemented by replacing the parameters
$\beta$ and $\xi$ by functions of the Higgs fields so that the interaction
effectively (i.e., via the Higgs fields) will depend on the subgroup
$K \subseteq SMG^3$ of un-Higgsed \dof.

This could be accomplished using a term in the action of the form

\beq   c|\phi(\sitex)|^2    \mbox{Tr}(U(\Box)). \label{higgsint}\eeq

\nin A term such as (\ref{higgsint}) comes into play when the gauge symmetry
is spontaneously broken by Higgsing from $G$ down to $K\subseteq G$. It
could
compensate changes in the critical coupling that accompany such a
spontaneous breakdown inasmuch as it is obvious that


\beq \langle \phi(\sitex) \rangle \left \{ \begin{array}{l} = 0 \mbox{ in
phase  }(G,\bunit)  \\ \neq  0  \mbox{  in  phase  }  (K,\bunit),\;\;\;
\phi(\sitex)\in G/K  \end{array} \right. .\eeq

In other words, a term such as (\ref{higgsint}) vanishes in the phase
$(G,\bunit)$ where
$\langle \phi(\sitex) \rangle = 0$ but can,  in  going  into  the  phase
$(K,\bunit)$  where  $\langle   \phi(\sitex)   \rangle   \neq   0$,
make a contribution to the inverse  squared  coupling
for $K$.

\subsection{Seeking the multiple point in a phase diagram for
$SMG^3$}\label{seekingthemp}

Ideally it is at the multiple point of the gauge group $SMG^3$ that critical
values of action parameters are sought. This is technically rather difficult
however. Hence it is fortuitous that there are approximate ways of
constructing the phase diagram for $SMG^3$ such that
we do not have to deal with the whole $SMG^3$ at once.
Instead, it is possible to  piece together an approximation of the phase
diagram for
$SMG^3$ by
separately considering subgroups the treatment of
which can be found
in the literature
\cite{bhanot1,bhanot2,bachas,lautrup,flyvbjerg,creutz,drouffe}. This is
possible
because
$SMG^3$ is a Cartesian
product group and therefore allows
an action that is additive in
contributions
from each $SMG$ group factor of
$SMG^3$. That is, for the gauge group

$$SMG_{Peter}\times SMG_{Paul}\times SMG_{Maria}$$

\nin it is possible to have an action of the form

\beq S=S_{Peter}+ S_{Paul}+ S_{Maria}.\label{addact}\eeq

With such an action, we are restricted to  bringing  together  at  an
approximative multiple point the confining phases that correspond
to {\em factorizable} invariant subgroups which  means  invariant  subgroups
that are Cartesian products of  invariant  subgroup  factors  each  of
which can be identified as coming from  just  one  of  the  isomorphic
$SMG$ factors (labelled by ``$Peter$'', ``$Paul$'', $\cdots$)
of $SMG^{N_{gen}}$. So if  we  restrict  ourselves
to an additive action of the type (\ref{addact}),
the phase diagram for  the  ``fundamental''
gauge group $SMG^{N_{gen}}$ is completely determined from a  knowledge
of  the  phase  diagram  for  just  one  of  the  group   factors
(e.g. $SMG_{Peter}$)  of
$SMG^{N_{gen}}$.
The additive action approximation yields the same value of the coupling for
the  $U(1)$
subgroup of each of the $SMG$ factors (labelled by the $N_{gen}$ indices
``$Peter$'', ``$Paul$'', $\cdots$). The same applies for the  three
$SU(2)$'s and $SU(3)$'s. In going to the diagonal subgroup, all three
$SMG$ fine-structure constants (i.e. for $U(1)$, $SU(2)$ and $SU(3)$) are
each enhanced by the same factor $N_{gen}=3$:


\beq \frac{1}{\alpha_{diag}}(\mu_{Pl})=\frac{1}{\alpha_{Peter}}(\mu_{Pl})+
\frac{1}{\alpha_{Paul}}(\mu_{Pl})+\frac{1}{\alpha_{Maria}}(\mu_{Pl})=
\label{diagcoup} \eeq

\[ =\frac{1}{\alpha_{multicr.}}+\frac{1}{\alpha_{multicr.}}+
\frac{1}{\alpha_{multicr.}}=\frac{3}{\alpha_{multicr.}}. \]

For the non-Abelian subgroups, it turns out that the  approximate
multiple point found in this way lacks contact with relatively few  of
the possible partially confining  phases  whereas  such an approximate
multiple point
lacks contact with  an infinity of partially confining phases of
$U(1)^3$.
Accordingly, we have found that the approximate  multiple point
critical couplings obtained using an  additive  action  (\ref{addact})
yield excellent predictions for the non-Abelian fine-structure constants
whereas the analogous prediction for the $U(1)$ fine-structure constant
is off by about 100 \%. We shall see that the reason for this has to
do with the ``Abelian-ness'' of $U(1)$: interactions between the $N_{gen}=3$
replicated $SMG$ factors in the AGUT gauge group $SMG^3$ give rise to terms
of the type $F_{Peter}^{\mu\nu}F_{\mu\nu\;Paul}$
(here the indices $``Peter,\;Paul,\cdots''$ label the $N_{gen}=3$
replicated $SMG$ factors in the AGUT gauge group).
Having such terms (which would break gauge symmetry in the case of non-Abelian
groups) results in $\frac{1}{2}N_{gen}(N_{gen}+1)=6$
quadratic $F_{\mu\nu}$ contributions
to the Lagrangian instead of the $N_{gen}=3$
possible terms in the non-Abelian case.
For the $SU(2)$
and $SU(3)$
subgroups of $SMG^3$, the only possible non-factorizable
subgroups
are diagonal subgroups of Cartesian products of $\bz_2$ and $\bz_3$ subgroups
from the different $SMG$ factors of $SMG^3$. The multiple point obtained when
these few phases are neglected is therefore a good approximation. In this
approximation, the couplings for $SU(2)$ and $SU(3)$ can be found separately
and then multiplied by the factor $N_{gen}=3$ by which the inverse squared
couplings for the non-Abelian subgroups are enhanced in going to the
diagonal subgroup of $SMG^3$.

For the $U(1)$ couplings, the situation is more complicated because of the
``mixed'' terms in the Lagrangian
         and the related fact that, for $U(1)$, the phases involving
non-factorizable subgroups are important in the determination of the multiple
point. Therefore it is necessary to effectively treat $U(1)^3$ rather than a
single $U(1)$ when determining multiple point $U(1)$ couplings.

The problem with phases that are lacking when the action is restricted to
being additive - i.e., phases corresponding to confinement along
non-factorizable subgroups - is present unless all the
group factors of a Cartesian product group are
without common nontrivial isomorphic subgroups of the centre.
In the  case of the Cartesian product group $SMG^3$,  the centre
(which itself is a Cartesian product)
has nontrivial repeated subgroup factors that are in different $SMG$
factors of $SMG^3$. Diagonal subgroups of such repeated subgroup factors are
non-factorizable in the sense that they cannot be factorized into
parts that each
are unambiguously  associated with just one $SMG$ factor of $SMG^3$.
With an additive action, it is not possible to  have  confinement alone along
the diagonal subgroups of such repeated factors.


Getting the phases  that  are  confined  w.r.t.  non-factorizable
invariant subgroups to convene at the  multiple point  (together
with phases for factorizable invariant subgroups) requires interaction
terms in the action that obviously are incommensurate with having  an
additive action. Having such interaction terms means that in
in general     
it   is   necessary   to   seek   the
multiple point  for  the  whole  $SMG^3$.  For  simplicity,  we  might
approximate  the  problem  by  considering  $U(1)^3$,  $SU(2)^3$   and
$SU(3)^3$ separately - but even this may ignore  some  non-factorizable
subgroups that could confine by  having  appropriate  interaction
terms in the action.
As mentioned above, for  the  non-Abelian  groups,  an  even
rougher approximation  is  rather  good:  finding the multiple point
couplings for  $SU(2)$  and  $SU(3)$
instead of respectively for $SU(2)^3$ and $SU(3)^3$ corresponds to finding
the multiple point using the approximation of an  additive
action (\ref{addact}).

Having non-factorizable subgroups requires
having invariant (and therefore necessarily central) ``diagonal-like''
subgroups (i.e., diagonal subgroups or subgroups that are diagonal up to
automorphisms within subgroups of the centre).
The centre of $SMG^3$ is the Cartesian product

\beq [(U(1)\times \bz_2\times \bz_3)/``\bz_6"]_{Peter} \times
     [(U(1)\times \bz_2\times \bz_3)/``\bz_6"]_{Paul} \times
     [(U(1)\times \bz_2\times \bz_3)/``\bz_6"]_{Maria} \eeq

In the case of the non-Abelian subgroups $SU(2)^3$ and $SU(3)^3$,
the possibility
for non-factorizable subgroups is limited to the finite number of
``diagonal-like'' subgroups that can be
formed from $\bz_2^3$ and $\bz_3^3$ (i.e., the respective centres of
$SU(2)$ and $SU(3)$). An examples is

\beq \{(g,g)|g\in \bz_3 \} \stackrel{-}{\simeq} \bz_3 \eeq

\nin where the element $(g,g)$ is the special (diagonal) case of say
an element $(g_{Peter},g_{Paul})\subset SMG^3$ for which
$g_{Peter}=g_{Paul}\stackrel{def}{=}g$. Other examples are

\beq \{(g,g^{-1})|g\in \bz_3 \} \stackrel{-}{\simeq} \bz_3, \eeq

\beq \{(g,g,g)|g \in \bz_3 \} \stackrel{-}{\simeq} \bz_3, \eeq

\beq \{(h,h^{\prime},h^{\prime\prime})|h,h^{\prime},h^{\prime\prime}\in \bz_2,
\mbox{two out of three of the }h,h^{\prime},h^{\prime\prime} \mbox{odd}\}\eeq
\[ =\{(1,1,1), (1,-1,-1), (-1,1,-1), (-1,-1,1)\}\stackrel{-}{\simeq}\bz_2
\times \bz_2 \]
and
\beq \{(h,h,g,g)|h\in \bz_2, g\in \bz_3 \}\stackrel{-}{\simeq}
\bz_2\times \bz_3 \}. \eeq

In the case of $U(1)^3\subset SMG^3$, any subgroup is invariant
(because $U(1)^3$ lies entirely in the centre of $SMG^3$). In particular,
any diagonal-like  subgroup
is invariant and constitutes
therefore a
non-factorizable subgroup along which there separately can be confinement.
While the non-factorizable (invariant) subgroups for $SU(2)^3$ and $SU(3)^3$
are exclusively of dimension $0$, such
subgroups can occur for $U(1)^3$ with
dimension 0, 1, 2 and 3.
For $U(1)$, non-factorizable subgroups occur as diagonal-like subgroups of
all possible  Cartesian products  having two or three repeated subgroup
factors (with different labels ``$Peter$'', ``$Paul$'', $\cdots$). These
repeated factors can be
discrete $\bz_N$ subgroups (for all $N\in \bz$) and  also $U(1)$ subgroups.
The latter are of
importance as regards plaquette action terms that are bilinear
in gauge fields:
unlike the case for continuous non-Abelian subgroups, it is possible to have
{\em gauge invariant} quadratic action terms of, for example, the type
$F_{\mu\nu\;Peter}F^{\mu\nu}_{Paul}$ defined on, for example,
$U(1)_{Peter}\times U(1)_{Paul}\subset U(1)^3$. Because the subgroup
$U(1)_{Peter}\times U(1)_{Paul}$ lies
in the centre of $SMG^3$, diagonal-like subgroups are {\em invariant} and
it is therefore meaningful to consider the transition between phases that
are confining and Coulomb-like for such diagonal-like subgroups. By
introducing terms in the action  of the type
$F_{\mu\nu\;Peter}F^{\mu\nu}_{Paul}$, we can extend the space of parameters and
thereby find additional phases that we subsequently can try to make accessible
at the multiple point.
In fact,
such terms can explain the factor ``6'' enhancement of Abelian inverse
squared couplings in going to the diagonal subgroup of $U(1)^3$.
The analogous factor for the non-Abelian diagonal subgroup couplings
is recalled as being only three - i.e., $N_{gen}=3$.

Because the procedure for getting the Abelian and non-Abelian couplings
from the parameter values at the multiple point of $SMG^3$ utilise
different approximations of the full $SMG^3$ group, it is convenient to
treat the two case separately. The non-Abelian couplings, being in many ways
simpler than for $U(1)$, are treated first (in Section~\ref{mpcnonabel}
immediately following this section). After this, $U(1)$ is considered in
Sections~\ref{mpcabelmeth} and \ref{calculation}.

\begin{flushright}
\begin{tiny}
thenon.tex 31 May 96 alf 
\end{tiny}
\end{flushright}

\section{Implementing the $MPCP$ for the purpose of determining
the non-Abelian $SMG$ gauge couplings}\label{mpcnonabel}

\subsection{General remarks}\label{secgenrem}

For the purpose of determining the non-Abelian couplings from $MPC$, it is a
good approximation to seek the multiple point for a single $SMG$ rather
than the full gauge group $SMG^3$ to which the $MPCP$ in principle is to
be applied. The multiple point for the gauge group $SMG$ is the point/surface
where the infinity of phases $(K,H)$ (with $H\lhd K \subseteq SMG$) convene
in the (presumably infinite-dimensional) phase diagram for $SMG$. Also,
we shall deal only with a finite subset of the infinity
of possible phases $(K,H)$. First of all, consideration will be restricted to
the subset of phases for which $K$ is identical with the gauge group
\footnote{Sometimes $G=U(N)$ will be considered for the purpose of
illustration; $U(N)$, which is simpler to deal with than the $SMG$,
is a prototype for the $SMG$ insofar as both groups are factor groups
modulo a discrete subgroup common
to both a $U(1)$ subgroup and (the centre of) a $SU(N)$ subgroup.}
$G$
where of course we are interested in the case $G=SMG$
(but sometimes consider the similar but simpler case of $G=U(N)$).
Taking $K=G$ is
tantamount to the assumption that all degrees of freedom
of $G$ are ``un-Higgsed'' (i.e., all $\Lambda_{Const}\in G$ leave the vacuum
invariant). Secondly, only a finite number of the infinity of discrete
(invariant) subgroups $\bz_N \subseteq U(1)$ are considered: namely the ones
that are also invariant subgroups of (the centres of) the non-Abelian
subgroups of the $SMG$ (i.e., $\bz_2\subset SU(2)$ and $\bz_3\subset SU(3)$).
The analogous restriction in the case of $U(N)$ would be to the
$\bz_N\subset SU(N)$.

The restrictions described above amount to a specification of a finite
number $n$ of phases that we seek to bring together at the multiple point.
In the case of $U(N)$ this
is the set of phases that can be confined
w.r.t. the following 5 invariant subgroups $H$:

\begin{equation} H=\left\{ \begin{array}{l}  \mbox{\boldmath  $1$}  \\
U(1) \;\;\;\;\;\;\;\;\;\;\;\;\;\ \\ SU(N) \\
U(N) \\ Z_N
\end{array} \right.  \label{invun} \end{equation}

\nin In the case of the $SMG$, this set would be the 13 invariant
subgroups

\begin{equation} H=\left\{ \begin{array}{l}  \mbox{\boldmath  $1$}  \\
U(1) \\ SU(2) \\ SU(3) \\ (U(1) \times SU(2))/Z_2=U(2) \\ (U(1) \times
SU(3))/Z_3=U(3) \\ SU(2)  \times  SU(3)  \\  SMG=S(U(2)  \times  U(3)) \\
\bz_2 \\ \bz_3 \\ \bz_2 \times \bz_3 \\ SU(2)\times \bz_3 \\
SU(3)\times \bz_2 \\
\end{array} \right. .\label{invh} \end{equation}

Having now specified the finite set of phases that are to convene
at a multiple point, the question of choosing a plaquette action that can
do this must now be addressed. It is obviously necessary to
restrict the class of actions considered to those having a finite number of
parameters. One could  for  example  in
some way truncate the  most  general  group  character  expansion.  An
alternative approach corresponding to that used here is to consider
the restriction to the class  of  plaquette  actions  $S_{\Box}$  that
correspond to  distributions  $e^{S_{\Box}}$ of  plaquette  variables  that
consists of narrow ``peaks''\footnote{We are assuming  that  a  procedure
using a weak coupling approximation is valid even in a neighbourhood at
the critical $\beta$'s. The approximation in which  $\frac{1}{6}\beta$
is considered to be very small is discussed  below  and  in  reference
\cite{long}.} centred at elements $p$ belonging to discrete subgroups
of the centre of the gauge  group.  In  accord  with the  above-mentioned
restriction to a finite number  of the infinity of  discrete
subgroups  $\bz_N\subset
U(1)$, this amounts,  for  the gauge group  $G=SMG$,
to  sharply  peaked  maxima
centred  at  elements  $p\in  \mbox{span}\{\bz_2,\bz_3\}$. For $G=U(N)$,
$p\in \bz_N$.
This  restricted  class  of  plaquette  actions  can  be
formulated as low order Taylor expansions around such elements
$p$ in the span of the discrete subgroups considered.
The simplest of this restricted class of plaquette actions (neglecting
zero order terms) would then be of the form

\beq  \left.  \frac{\partial^2   S_{act.}}{\partial   k_a\partial
k_b}\right|_                           {\vec{k}^{(p)}}
(k_a-k^{(p)}_a)(k_b-k^{(p)}_b) \label{tay2}\eeq

Here the $k_a$    are  coordinates  on  the  group
manifold in a neighbourhood of just one of the
elements $p$ belonging to the span of
discrete subgroups considered. The coordinates at $p$ are denoted
$\vec{k}^{(p)}=(k_1^{(p)},k_2^{(p)},\cdots            ,
k_{dim(G))}^{(p)})$.

{}From the assumption  of  a
distribution $e^{S(\Box)}$ of {\em narrow} peaks centred at the
elements $p \in \bz_N$ (thinking for the moment of $G=U(N)$ for illustrative
purposes), a group element  not  close  to  some  $p  \in
\bz_N$ leads to a vanishing value of $e^{S(\Box)}$; i.e., a
non-vanishing value of $e^{S(\Box)}$ at  a  given  group  element
$\vec{k}=(k_1, k_2, \cdots ,k_{dim(G)})$
gets its value in our ansatz action solely from the
Taylor expansion centred at just one element $p \in \bz_N$ (i.e., the
one  for  which  the  quantity   $\sum_{a,b}g^{ab}(\vec{k}^{(p)})
(k_a-k^{(p)}_a)(k_b-k^{(p)}_b)$ is minimum).  Here
$g^{ab}$ is  the  metric  tensor defined by requiring
invariance under left and right group multiplication supplemented with
normalisation conventions.  We  define  the
quantities $\beta_i$ (at the point $p$) by

\beq  \left.  \frac{\partial^2   S_{act.}}{\partial   k_a\partial
k_b}\right|_
{\vec{k}^{(p)}}
\stackrel{def.\;of\;\beta_i}{=}\sum_i\beta_ig_i^{ab}
(p) \label{tay3} \eeq

\nin where the  index $i$ labels the Lie subgroups
of the gauge group invariant w.r.t. the algebra.

Let us define $dist(p,u)\stackrel{def}{=}\int_p^uds$ where $ds=g_{ab}dk^adk^b$
is the left-right invariant Riemann space metric. We sometimes write
(\ref{tay2}) as

\beq \sum_i\beta_i dist_i^2(\vec{k^{(p)}}, \vec{k}) \eeq
or
\beq \sum_i\beta_i dist_i^2(k^{(p)},k) \eeq
where $dist^2_i$ is the component of $dist^2$ along the $i$th invariant
subgroup. At least for small distances on the group manifold, this
decomposition is well defined.

Later we shall also need to define the volume of - for example - an invariant
subgroup. It turns out to be useful in dealing with distances along- and
volumes of invariant subgroups to let the quantities $\beta_i$ be absorbed in
the metric tensor defined on the various invariant subgroups.

With this
convention, we write the squared distance along  the  $i$th  invariant
subgroup as $Dist^2_i$ with uppercase ``$D$''. The relation to squared
distances  $dist^2_i$
in the non-absorbed notation is

\beq Dist^2_i=\beta_i dist^2_i. \label{abmet}\eeq

\nin Using the absorbed  notation,  we  can
define $Dist^2 \stackrel{def.}{=} \sum_i Dist^2_i.$ Formally we have

\beq   Dist(1,k)   =   \int_1^k   ds   \;\;   \makebox{   with    }\;\;
ds^2=g_{ab}dk^adk^b \eeq

\noindent where the $k^a$ are a set of coordinates (on the group)  for
the element $k\in G$. These coordinates are chosen so as to split into
separate sets along each of the three Lie groups associated  with  the
three  ``basic''  invariant  sub-Lie  algebras   $U(1),\;SU(2),$   and
$SU(3)$. This could be done by choosing a basis  in  the  Lie  algebra
with each  basis  vector  in  only  one  of  the  ``basic''  invariant
sub-algebras whereupon the group could be parameterised by means of the
exponential map.

The metric tensor $g_{ab}$ in the adjoint representation is given by

\begin{equation}           g_{ab}(k)=\left(            \begin{array}{rrr}
(g^{U(1)}\cdot\beta_1 )^{(1 \times  1)}&  0^{(1  \times  3)}  &  0^{(1
\times 8)} \\ 0^{(3 \times 1)} & (g^{SU(2)}(k)\cdot \beta_2)^{(3 \times
3)}& 0^{(3 \times 8)} \\  0^{(8  \times  1)}  &  0^{(8  \times  3)}  &
(g^{SU(3)}(k) \cdot \beta_3)^{(8 \times 8)} \end{array} \right) \end{equation}

\noindent    where    $g_{ab}^{U(1)}$,    $g_{ab}^{SU(2)}(k)$,     and
$g_{ab}^{SU(3)}(k)$  are  the  metrics  along  the  various  invariant
subgroups.

Using this metric with absorbed $\beta$'s, a volume can be defined for any
sub-manifold  -  for
instance, a subgroup $H$ - of the group $G=S(U(2) \times U(3))$.  Such
a volume, denoted by ``$Vol H$'' with uppercase ``$V$'', is given by

\beq                                               \frac{1}{(dim(H))!}
\int_Hd^{dim(H)}h\sqrt{g_{a_1b_1}(k(h))g_{a_2b_2}               \cdots
g_{a_{dim(H)}b_{dim(H)}}(k(h))} \cdot \eeq

$$ \cdot\left(\frac{\p  k^{a_1}}{\p  h^{l_1}}\frac{\p  k^{a_2}}{\p h^{l_2}}
\cdots \frac{\p  k^{a_{dim(H)}}}{\p  h^{l_{dim(H)}}}  \varepsilon^{l_1
l_2   \cdots    l_{dim(H)}}
\frac{\p    k^{b_1}}{\p
h^{m_1}}\frac{\p     k^{b_2}}{\p h^{m_2}}     \cdots      \frac{\p
k^{b_{dim(H)}}}{\p   h^{m_{dim(H)}}}   \varepsilon^{m_1   m_2   \cdots
m_{dim(H)}}\right)^{\frac{1}{2}} $$

\noindent the coordinates $h^1,h^2, \dots h^{dim(H)}$ parameterise the
manifold $H$ and $\varepsilon^{l_1, l_2, \dots l_{dim(H)}}$ is totally
antisymmetric in $dim(H)$ indices.

It is also possible by means of the metric  $g_{ab}(k)$  to  induce  a
metric on a factor group $G/H=\{gH \mid g\in G \}$ by  defining  the
distance between two cosets $g_1H$ and $g_2H$ as the distance
between representatives for these cosets that is shortest:

\beq Dist(g_1 H,g_2 H)\stackrel{def.}{=}inf \{ Dist(k_1 ,k_2)  \mid k_1   \in
g_1 H \mbox{ and } k_2 \in g_2 H \}.\eeq

\noindent With this metric on a factor group, a volume  for  a  factor
group can be defined:

\beq Vol(G/H)=\int d^{dim(G/H)}f \sqrt{det \{g_{pq}^{(factor\;gr.)}(f)
\}} \eeq

\noindent where $g_{pq}^{(factor\;gr.)}$  denotes  the  metric  induced
into the factor group, i.e.,

\beq ds_{factor\;gr.}^2=(inf\{Dist(k_1,k_2)  \mid k_1 \in  gH,  k_2  \in
(g+dg)H \})^2 \eeq

$$=g_{pq}^{(factor\;gr.)}(gH)df^p df^q .$$

\noindent Here $g+dg$ is a representative for  a  coset  $(g+dg)H$
infinitesimally close to $gH$. In the coordinatization of $G/H$,  $gH$
and $(g+dg)H$ take respectively the  coordinate  sets  $f^p$  and
$f^p+df^p$.

The need will arise for consideration of 4$th$ and 6$th$  order  terms
when applying the principle  of  multiple  point  criticality  to  the
subgroup $U(1)^3=U(1)\times U(1)  \times  U(1)  \subset  SMG^3$. A 4$th$ order
term would be of the form

\beq
\frac{\partial^4S}
{\partial k_a\partial k_b\partial k_c\partial k_d}|
_{k^{(p)}_a,k^{(p)}_b,k^{(p)}_c,k^{(p)}_d}
(k_a-k^{(p)}_a)(k_b-k^{(p)}_b)
(k_c-k^{(p)}_c)(k_d-k^{(p)}_d) \eeq

\nin where

\beq \frac{\partial^4S}
{\partial k_a\partial k_b\partial k_c\partial k_d}|
_{k^{(p)}_a,k^{(p)}_b,k^{(p)}_c,k^{(p)}_d}
\stackrel{def.}{=}\sum_{i,j}\gamma_{ij}g_i^{ab}(\vec{k}^{(p)})
g_j^{cd}(\vec{k}^{(p)}) \label{int}\eeq

\nin The parameter $\gamma_{ij}$ is the coefficient of a term of the  action
that is present if there is an interaction between the \dofx corresponding to
the $ith$ and $jth$ subgroups.
The effects of such terms are elaborated upon in Appendix~\ref{appint}.
A 6$th$ order term would describe the interaction between the \dofx
associated with three different subgroups.


\subsubsection{Problems with universality}\label{universality}

A priori, the model considered  here  lacks  universality  because  the
determination of multiple point critical surfaces  is
based on first-order phase transitions. However an action ansatz of the
type considered (i.e.,  narrow  maxima  centred  at  elements
$p_{i,\;N}=e^{i2\pi p_{i,\;N}/N}\in \bz_N \;\;(p_{i,\;N}\in \bz)$, has
the advantage that  it  intrinsically  has  at  least  an  approximate
universality. This can be argued as follows.

Consider  the  space  of  all  (analytic)  plaquette  action  mappings
$S_{\Box}:G\rightarrow \br$. Let this  space  be  spanned  by  a  basis
consisting of the set of coefficients of the terms of all orders
of a Taylor expansion of an analytic function.

If we restrict our considerations to actions sharply peaked at identity
(and at elements of discrete subgroups of centre), this is essentially
the same as to say that we are considering actions that are very  much
dominated by the second order term  of  a  Taylor  expansion.  Actions
dominated by  second  order  terms  include  all  actions  with  large
coefficients to second order terms regardless of higher order terms.

Physical quantities (e.g., continuum couplings) or  properties  (e.g.,
phase transitions) are generally calculated  as  functional  integrals
weighted with the exponentiated action. If the the exponent (i.e., the
action) of this  weight  function  depends  essentially  only  on  the
(large) coefficient of the second order term of  a  Taylor  expansion,
this is tantamount to saying that physical quantities essentially have
contour curves that are orthogonal to  the  axis  of  coefficients  of
second order terms of Taylor expansion provided the coefficient values
are sufficiently large.

Universality for our model means that we can move along  the  infinite
dimensional multiple point surface (i.e., surface of co-dimension $n-1$
embedded in our action function space) without  intersecting  contours
of continuum couplings. This will generally not be  the  case  because
the multiple point surface and  the  contours  of  constant  continuum
couplings will depend in different ways on the higher order  terms  in
Taylor expansion of action such that multiple point surface intersects
contours of constant continuum coupling  values  thereby  dashing  any
hope of universality.

However, if the location of the multiple point surface in action space
spanned by Taylor expansion parameters is at a position  corresponding
to sufficiently large values of coefficients of  second  order  Taylor
expansion term, it too (i.e., just like all other physical  quantities
such as continuum couplings)  will  essentially  only  depend  on  the
second order term of \underline{any} action $S$ with this large  value
of $S^{\prime\prime}$. This would tend  to  make the multiple point
surface parallel to surfaces of constant continuum  couplings  thereby
leading to approximate universality.

\subsection{The modified Manton action}\label{modmantonact}

Non-generic multiple points will be sought using the simplest possibility from
among the class of actions that can be expressed as low order Taylor
expansions around elements $p$ in the span of discrete subgroup(s)
of the centre of the gauge group. This simplest possible
action
can be characterised as the
projection of the class of all possible actions onto the  subspace  of
actions having the form of second order  Taylor  expansions  in  group
variables with one Taylor expansion  at  each  element  $p$
in the span of the relevant discrete subgroup(s).
For illustrative purposes, one can
think of the gauge group $U(N)$
with  $p\in \bz_N \subset U(N)$.
Assuming for simplicity that the peaks  expanded  around  each  element
$p\in \bz_N$ are symmetric (so that odd-order derivatives  vanish),
the simplest of this restricted class of plaquette actions (neglecting
zero order terms) would then be of the form (\ref{tay2}) and (\ref{tay3}).
When (\ref{tay3}) is used with $G=U(N)$,  $i\in  \{U(1),SU(N)\}$
and $p\in \bz_N$. So the action to be used
- from now on referred to as  the modified Manton  action  -
leads to Gaussian peaks at each $p$; $\beta_i$ is the action parameter
that determines the width of the Gaussian distribution along the $i$th
Lie  subgroup.  The  parameter  $\beta_i$  is  assumed   to   be   the
same\footnote{In  a  more  sophisticated  ansatz,  this  need  not  be
assumed.} at all elements $p\in \bz_N$. A full  specification  of the
modified Manton action  also  requires  parameters  that  specify  the
relative height of peaks at the different elements $p\in  \bz_N$.  For
$N=2$ or $3$, one parameter - denote it by $\xi_N$ - is sufficient.

Using the parameters of the modified Manton action (e.g., for  $U(N)$,
the parameters are $\beta_i$ and $\xi_N$ as described above),  non-generic
multiple points are readily found. The selection of subgroups to which
these parameters correspond make up what we call  the  ``constituent''
invariant subgroups\footnote{ By constituent  invariant  subgroups  we
refer essentially to the Cartesian product  factors  of  the  covering
group together with a selection  of  the  discrete  subgroups  of  the
centre - namely the ones of special importance in obtaining the  gauge
group as a factor group (e.g., $\bz_2$ is  of  special  importance  in
obtaining the factor group $U(2)$ from the covering group  $U(1)\times
SU(2)$   because   $U(2)\overline{\simeq}(U(1)\times   SU(2))/\bz_2$.}
(e.g., for $U(N)$, the ``constituent invariant subgroups  are  $U(1)$,
$SU(N)$, and $\bz_N$). The corresponding set  of  parameters  has  the
advantage that they are essentially independent: variation of  one  of
these parameters leads to a change in the width  of  the  distribution
$e^{S_{\Box}}$  along  the  corresponding  ``constituent''   invariant
subgroup that is approximately un-coupled from the distributions of group
elements along other ``constituent'' invariant  subgroups.  Also,  all
possible invariant subgroups $H$ of the gauge groups  considered  here
can be constructed as factor groups associated with a  subset  of  the
set of the ``constituent'' invariant subgroups.

For the $SMG$, the ``constituent''  invariant  subgroups  are  $U(1)$,
$SU(2)$,  $SU(3)$,  $\bz_2$,  and  $\bz_3$.   The   $\beta_i$   ($i\in
\{U(1),SU(2),SU(3)\}$)  constitute  three   of   the   five   required
parameters  for  the  modified  Manton  action.  The   remaining   two
parameters are designated  as  $\xi_2$  and  $\xi_3$.  These  are
associated with the discrete invariant subgroups $\bz_2$  and  $\bz_3$
and determine the  relative  heights  of  peaks  of  the  distribution
$e^{S_{\Box}}$ at various elements  $p\in  \mbox{span}\{\bz_2,\bz_3\}$.

The  modified
Manton action for the $SMG$ is

\beq   S_{\Box}(U(\Box))=    \left\{    \begin{array}{l}    \sum_{i\in
\{U(1),SU(2),SU(3)\}}\beta_idist_i^2(U(\Box),      p)+\log      \xi(p)
\;\;\;\mbox{for $U(\Box)$ near $p\in \mbox{span}\{\bz_2,\bz_3\}$ }  \\
\approx  -\infty\;\;\;  \mbox{for  $U(\Box)$  not   near   any   $p\in
\mbox{span}\{\bz_2,\bz_3\}$} \end{array} \right. \label{modman} \eeq

\nin where the  symbol  $dist_i^2(U(\Box),p)$  denotes  the  component
of the distance squared (defined in Section~\ref{genrem})
\footnote{Left-right invariance of a Riemann space metric
$ds^2=g_{ab}dk^a dk^b$ specifies $dist_i^2$ for each simple
invariant sub-algebra $i$ up to a normalisation factor. The decomposition
of $dist(p,u)=\int^u_p ds$ into components $dist_i$ along the $i$th invariant
subgroup is at least well defined
for small distances.} between the group element $U(\Box)$  and  the  nearest
element $p\in \mbox{span}\{\bz_2,\bz_3\}$ along  the  $i$th  invariant
Lie subgroup and

\beq  \log  \xi(p)=\left  \{  \begin{array}{ll}  0  &  \mbox{for   }
p\in  \bz_3  \\   \log   \xi_2   &   \mbox{for   }
p\not \in \bz_3 \end{array} \right \} + \left \{ \begin{array}{ll} 0 &
\mbox{for } p\in \bz_2 \\ \log \xi_3 &  \mbox{for
} p\not \in \bz_2 \end{array} \right \}. \eeq

\nin This action gives rise to  a
distribution  $e^{S_{\Box}}$  having   6   Gaussian   peaks   at   the
elements\footnote{The notation used here is that of  the  footnote  in
the beginning of the Introduction where the $SMG$ is embedded into $SU(5)$.}

\beq p_r\stackrel{def.}{=}
\left( \begin{array}{cc} e^{\frac{i2\pi r}{2}}\bunit^{2\times 2} &
\begin{array}{ccc} 0 & 0  &  0
\\ 0 & 0 & 0 \end{array} \\ \begin{array}{cc} 0 & 0 \\ 0 & 0 \\ 0 &  0
\end{array} & e^{\frac{-i2\pi r}{3}}\bunit^{3\times 3} \end{array} \right )
\in \mbox{span}\{\bz_2,\bz_3\}=\bz_6 \;\;\;\mbox{with  } r=0,1,\cdots,5.  \eeq

\nin all  having
   widths     $(2\beta_1)^{-1/2}$,     $(2\beta_2)^{-1/2}$,     and
$(2\beta_3)^{-1/2}$   in
respectively the $U(1)$, $SU(2)$, and $SU(3)$ subgroup directions. For
$r=0$ (i.e., at the group identity), the peak height is 1; for  $r=3$,
(i.e., at the nontrivial  element  of  $\bz_2$)  the  peak  height  is
$\xi_2$; for $r=2,4$ (i.e., the nontrivial elements of  $\bz_3$),  the
peak heights are $\xi_3$; for $r=1,5$, the  peak  heights  are  $\xi_2
\xi_3$.

Note that the assumption of very sharp peaks at the  elements  $p  \in
\mbox{span}\{\bz_2,\bz_3\}$  means  that   actions   of   this   class
are in essence completely specified by parameters corresponding to the
coefficients of second order terms in their  Taylor  expansions  about
these elements $p$. Higher order  terms  in  the  Taylor  expansion  are
irrelevant. Roughly speaking, this also means that physical quantities
such as continuum couplings and the  log$Z_H$  for  various  invariant
subgroups $H$ (and therefore the realizable phases) can only depend on
the coefficients of second order action  terms.  Hence,  the  multiple
point critical surface can be expected to be approximately parallel to
surfaces of constant continuum coupling  values  thereby  yielding  an
approximate universality.

\subsection{Factorising the free energy}\label{facfreeenergy}

The critical coupling values for the transitions on the  lattice
which we are interested in here happen to be so relatively weak that a
weak coupling approximation makes sense. We therefore make  use
of  such  an  approximation  in  conjunction   with   a   mean   field
approximation ($MFA$) in our very crude  exploratory  studies  of  the
qualitative features to be expected for the phase diagram of a lattice
gauge theory. A sensible weak  coupling  approximation  requires  that
$\beta^{-1}$ is small compared to the squared extension of  the  group
where of course $\beta$ is the coefficient of the  real  part  of  the
trace of a plaquette variable in the action. Because  we  assume  weak
coupling, we use the approximation of a  flat  space  measure  in  the
evaluation of functional integrals  with  the  limits  of  integration
extended to $\pm \infty$.

It is natural to enquire as to how such a weak coupling  approximation
can have a chance of being sensible in dealing with  the  onset  of  a
confinement-like phase at the lattice phase transition.  Recall  first
that we use as the defining feature of a confinement-like  phase  that
Bianchi identities can be ignored  to  a  good  approximation\footnote{This is
consistent with the definition in section 3.3: when fluctuations are so strong
that gauge symmetry is not broken by a gauge transformation with a linear
gauge function $\Lambda_{Linear}$ (leading to a translation of the
gauge potential $A_{\mu}$  by a constant), then the
fluctuations can also be assumed to be so strong
that the effect of Bianchi identities are washed out.}.  But  the
variables to which the Bianchi identities  apply\footnote{Recall  that
Bianchi identities  impose  a  constraint  (e.g.,  modulo  $2\pi$  for
$U(1)$) on the divergence of flux from a volume enclosed by
plaquettes} are not plaquette variables but rather the  variables
taken by  3-dimensional  volumes  enclosed  by  plaquettes  -  in  the
simplest case, just the cubes  bounded  by  six  plaquettes.  Now  the
distribution of such cube variables is the 6-fold convolution  of  the
distribution  of  plaquette  variables were it not
for Bianchi identities.  So  if  the  distribution  of
plaquette variables has a width proportional to  $\beta^{-1}$,
the width of the distribution of group elements taken by  cubes
enclosed  by  six  plaquettes  is  proportional  to  $(\beta/6)^{-1}$.
Therefore, the question of the  validity  of  using  a  weak  coupling
approximation at the  phase  transition  on  a  lattice  is  really  a
question of whether the number 6 can be regarded as large compared  to
unity. Accepting  this  as  true,  we  can  conclude  that  even  when
$\beta^{-1}$ is small compared to the squared extension of  the  group
as required for a meaningful weak coupling approximation, the quantity
$(\beta/6)^{-1}$ is large enough compared to the squared extension  of
the  group  so  as  to  justify  the  the  use  of  the  Haar  measure
distribution for the Bianchi-relevant cube variables obtained  as  the
convolution of six plaquette variable distributions. This  amounts  to
ignoring the Bianchi identities. A phase for which Bianchi  identities
can be ignored is, according to our ansatz, a confinement-like phase.

In this approximation we obtain an  expression  for  the  free  energy
$\log        Z$         in         terms         of         quantities
$Vol(L)\stackrel{def.}{=}\frac{vol(L)}{``fluctuation\;vol."}\cdot(\pi e)^
{\frac{dim(L)}{2}}$  where $L$  is
a factor group/invariant subgroup of the $SMG$
and   $vol(L)$
is the volume of $L$. The ``fluctuation volume'' is defined as the width of
the flat distribution that yields the same entropy $S_{ent}$ as the original
distribution; i.e.,

\[ \Delta S_{ent}= < -\log(\frac{e^{\beta dist^2}}{\sqrt{\pi/\beta}})-
\log(``flat\; distribution")>=0.
\]

\nin For large $\beta$ (weak coupling approximation), we have the
approximation $Vol(L)\approx \beta^{\frac{dim(L)}{2}}vol(L)$ which we shall
use in the sequel.

For the \pcp that is confined w. r. t.  the  invariant
subgroup $H$, the free energy per active\footnote{Active link means  a
link not fixed by a  gauge  choice - for example,  the  axial  gauge.}  link
is\cite{long}                  \begin{equation}                   \log
Z_{per\;active\;link}=\mbox{max}_H\{\log Z_H|H\lhd G\} \eeq

\nin where

\beq \log Z_H =                    \log
[\frac{(\pi/6)^{(dim(G/H))/2}}{Vol(G/H)}                       ]+2\log
[\frac{(\pi)^{(dim(H))/2}}{Vol(H)}] \label{fe} \end{equation}

\[= \log  [\frac{(\pi/6)^{(dim(G))/2}}{Vol(G)}
]+\log      [\frac{(6\pi)^{(dim(H))/2}}{Vol(H)}]
\]

\nin where it is understood that  $\log  Z_H$  is  calculated  using  an
ansatz that results in confinement for the invariant subgroup $H$  and
Coulomb-like behaviour for the  factor  group  $SMG/H$.

In our approximation (\ref{fe}), it can be shown that at the  multiple
point, any two invariant subgroups $H_1$ and $H_2$ of the gauge  group
must satisfy the condition

\beq \log(\sqrt{6\pi})^{dim(H_J)-dim(H_I)}=\log\frac{Vol(H_J)}{Vol(H_I)}
\label{piv}\eeq

\nin where it is recalled that the quantity $Vol(H_J)$ is  essentially
the volume of the subgroup $H_J$ measured in  units  of  the  critical
fluctuation volume.
In special case where $H_I=\bunit$, we get

\beq \frac{\log Vol(H_J)}{dim(H_J)}=\log(\sqrt{6\pi}); \label{facprop} \eeq

\nin i.e., for {\em any} invariant subgroup $H_J$  the  quantity  $Vol
(H_J)$ per Lie algebra dimension must be equal (in this approximation)
to  the  same  constant
(``$\sqrt{6\pi}\;$'') at the multiple point.

The condition expressed by (\ref{facprop}) is referred to as the free energy
factorisation property and must be fulfilled in order to have a multiple point
in the approximation we are considering. In general this property can only
be fulfilled if one uses an action that can also provoke phases confined
{\em solely} w.r.t. discrete subgroups. Consideration of this latter type
phase will be postponed until after the digression in the next section
(Section~(\ref{digression})).
in which we consider the approximation in which the confinement of discrete
subgroups occurs {\em only} when the continuous (sub)groups to which discrete
subgroups belong are also confined; i.e., discrete subgroups are not {\em
alone}
confined.

\subsection{Digression: developing techniques for constructing phase diagrams
for non-simple gauge groups (without confinement of discrete
subgroups)}\label{digression}

Before proceeding to show the factorisation property (\ref{facprop}) which
is necessary for having a multiple point, a digression is made at this point.
The formula (\ref{fe}) will now be derived in the special case where $p$ in
(\ref{modman}) always coincides with the identity element. This amounts to
the restriction to the first 8 of the 13
invariant subgroups $H$ in the list (\ref{invh}) - namely the ones that do
not involve the discrete subgroups $\bz_2$ and $\bz_3$. These 8
invariant subgroups correspond, in the approximation considered, to the
\pcps of the $SMG$ that can be realized using the set
$\{U(1),SU(2),SU(3)\}$ as the constituent invariant
subgroups.
The construction of crude phase diagrams using this
set will be considered in some detail. It will be seen that the absence of
action terms that can render the discrete subgroups $\bz_2$ and $\bz_3$
critical coincides with not being able to have the first 8 phases of
(\ref{invh}) meet at a multiple point. The crucial role of the discrete
subgroups $\bz_2$ and $\bz_3$ will be demonstrated explicitly in a
later section.

With the restriction to the set of invariant subgroups
$\{U(1),SU(2),SU(3)\}$, the action in (\ref{modman}) becomes

\begin{equation}
S_{action}=\sum_i         \beta_i \sum_{\Box}
dist_{i\in \{U(1),\;SU(2),\;SU(3)\}}^2(1,U(\Box)) \label{action}
\end{equation}

\noindent
The  symbol  $dist_i(g,k)$  denotes  the
component of the distance between the group elements $g$ and $k$ along
the  invariant  subgroup  $i$ (see Section~\ref{genrem}).
Here  $\{i\}=\{U(1),\;SU(2),\;SU(3)\}$.

The  symbol
$U(\Box)$ of course designates  the  group-valued  plaquette  variable
obtained from the group composition of  the  link  variables  for  the
links enclosing the plaquette in question:

\beq U(\Box)=\prod_{\link\; \in\; \partial \; \Box} U(\link) \eeq

With the convention in which  the  $\beta$'s  are  absorbed  in  the
metric (see Section~\ref{genrem}), the Manton action (\ref{action}) becomes:

\begin{equation}
S_{action}=-\sum_{\Box}Dist^2(1,U(\Box))=-\sum_i \sum_{\Box}
 Dist_i^2(1,U(\Box)).
\end{equation}

As alluded to in Section~\ref{genrem}, the  concept  of  projecting
a  distance  onto  invariant
subgroups is well defined only for the covering group (e.g.,
$ R \times SU(2) \times SU(3)$ for  the
$SMG$). For a group
obtained by identifying a discrete subgroup of the centre of the covering
group   the  decomposition  of  the
distance between two group elements $g$ and $k$ along basic  invariant
subgroups can be ambiguous due to the global structure of  the  group.
This could be a problem for the {\em group} $S(U(2)\times U(3))$ used here
as the $SMG$ (obtained from $R\times SU(2)\times SU(3)$ by identifying
elements of the discrete subgroup
$(2\pi, -\bunit^{2\times 2}, e^{i\frac{2\pi}{3}}
\bunit^{3\times 3})^p$ with $p\in \bz$).
However,  this ambiguity is absent when, for dominant configurations,
group elements are clustered together at the identity as is inherently
the  case  in the weak coupling approximation used here.

Given the gauge group $G$ and the phase that  is  partially  confining
with confinement along $H \subseteq  G$,  we  have  claimed  that  the
Coulomb-like Yang-Mills fields can be  regarded,  essentially  without
loss of information, as mappings of the set of links  of  the  lattice
into the factor group $G/H$ (instead of the whole  group  $G$)
\footnote{``Confinement'' w. r. t. the subgroup $H$ means  quantum
fluctuations  in
the corresponding degrees  of  freedom  that  are
sufficiently large so  that  Bianchi  identities can be ignored.
Defining  the  symbol  $P_{\link}(g)$
as the link variable probability density at the field value $g
\in G$, a field theory with confinement along a subgroup $H  \subseteq
G$, means that $P_{\link}(g)$ is  more  slowly  varying  along  the
cosets $gH=\{g \cdot h \mid h \in H \subseteq G; g\i G \}$ of $G/H$ than
along group orbits ``orthogonal'' to these cosets.  So
the  information  needed  to  specify  the
distribution $P_{\link}(g)$ over the entire  gauge  group  $G$  is,
roughly  speaking,   contained   in   a    specification    of    how
$P_{\link}(g)$ varies over the set of cosets of the factor group $G/H$.
The density function $P_{\link}(g)$ can essentially be  replaced by
function  defined  on
the cosets of the factor group $G/H$. Relative to  the
full gauge group $G$,  fluctuations along the  Coulomb-like
                   degrees  of  freedom  are  more  or  less  strongly
concentrated about the coset $\bunit \cdot H$ where  $\bunit$
is  the  identity
element of the gauge group $G$. The  phase
with confinement-like behaviour along the invariant subgroup $H\triangleleft G$
will sometimes be referred to as the phase  that  is  partially
confining  w.r.t. $H$. It is to be implicitly understood then that the
\dofx identified with the cosets of the factor group have
Coulomb-like behaviour.}.
Therefore
the  actions  that  regulate  the  distribution  of  Coulomb-like  and
confinement-like link variables are in a sense  defined  on  different
groups each of which  contribute  mutually  orthogonal  components  to
distances $Dist$ in $G$. It is useful in the calculations that  follow
to rewrite the plaquette action $S_{action}$ as the sum of contributions
from confinement-like and Coulomb-like degrees of freedom:
$S_{action}=S_{action,   \;conf}+S_{action,   \;Coul}$.   Using   this
decomposition for a given phase, action contributions from  components
of the distance along invariant Lie sub-algebras corresponding to gauge
field degrees of freedom that  are  Coulomb-like  are  separated  from
invariant sub-algebra components corresponding to confined gauge  field
degrees of freedom. We can write

\begin{equation}
S_{action}=-\sum_{\Box} Dist^2(1,U(\Box))
=-\sum_{\Box}   \sum_i   Dist_i^2(1,U(\Box))
\end{equation}

$$
=S_{action,    \;
Coul.}+S_{action,                      \;                      conf.}=
-\sum_{\Box}( Dist_{Coul.}^2(1,U(\Box))+ Dist_{conf.}^2(1,U(\Box)))
$$

\noindent where the indices ``$Coul$.'' and ``$conf$'' denote sets of the
three ``basic'' invariant subgroups that  correspond  respectively  to
Coulomb-like and confinement-like (physical) gauge  field  degrees  of
freedom in any given one of the $2^3=8$ general phases  obtainable  as
combinations of Coulomb or confining ``basic'' invariant sub-algebras:

$$         - Dist_{Coul.}^2(1,U(\Box))=-\sum_{i           \in
``Coul."}Dist_i^2(1,U(\Box) $$

$$-Dist_{conf.}^2(1,U(\Box)))=-\sum_{i           \in
``conf."}Dist_i^2(1,U(\Box). $$

In  limiting  our  considerations  to  these  $2^3=8$
phases, we  are  explicitly  ignoring  the  denumerable  infinity  of
distinct   phases   that   could   arise   if   one   considered   all
group-theoretically  possible  invariant  subgroups (e.g. the discrete
subgroups a few of which will be taken into account in a later section).

Here we shall consistently  use  the  terminology
``confining phase''  even  though,  in  many  cases,  the  designation
``strong coupling phase '' or ``short range correlation phase''  would
be more appropriate. Inasmuch as  we  are  unconcerned  as  to
whether a ``confining phase'' also has  matter  field  representations
(e.g. quarks) that actually suffer confinement, the  word  confinement
is  possibly  misleading.  Here the   essential   property   that
distinguishes Coulomb and confinement phases is whether the phase  has
long or short range correlations.

The mean  field approximation (MFA) calculation  in  the  weak  coupling
approximation begins with the addition and subtraction  of  an  ansatz
action $S_{ansa}$  to  the  plaquette  action  $S_{action}$.

For the ansatz action $S_{ansa}$ we take

\begin{equation}
S_{ansa}= -\sum_{j \in ``Coul"} \alpha_j \sum_{\link} dist_j^2(1,U(\link))
= -\sum_{j \in ``Coul"}  \tilde{\alpha_j} \sum_{\link} Dist_j^2(1,U(\link))
\end{equation}

\noindent where $\tilde{\alpha}_j=\alpha_j / \beta_j $. The parameters
$\tilde{\alpha_j}$  are  to  be  adjusted  in  accordance   with   the
requirement  that  the  distribution  defined  by  $S_{ansa}$   should
effectively approximate the distribution given by the original action.
In fact, we approximate by fitting $S_{ansa}$ to the  most  rapidly
varying (i.e., Coulomb-like) behaviour of the gauge  field  which  means
that the $\alpha$'s are to be chosen so as to  minimise  the  quantity
$S_{action,  \;Coul}-S_{ansa}$.  This  implies  that   the   necessary
variation of $S_{ansa}$ can be realized by defining it on the $de \;\;
facto$ domain of $S_{Coul}$ - namely the cosets of  the  factor  group
$G/H$ where again it is to be remembered that the symbol  $H$  denotes
one of the (invariant) subgroups $H$  as  enumerated  in  (\ref{invh})
along  which  (physical)  gauge  field  degrees  of  freedom  are   in
confinement. In practice the $\alpha$'s are chosen so as  to  maximise
the partition function in an approximate form to  be  elaborated  upon
below.

In light of the discussion above, the partition function

\begin{equation}       Z=\int_G       \cdots       \int_G        {\cal
D}U(\link)\exp(S_{action})    \stackrel{def.}{=}     \int_G     \cdots \int_G
\prod_{\link} d^{{\cal  H}aar}U(\link)  \exp(S_{action})  \label{part}
\end{equation} \noindent can now be written as

\begin{equation}     Z      =\int_G      \cdots      \int_G      {\cal
D}U(\link)\exp(S_{action,     \;     Coul.}-S_{ansa}+S_{action,     \;
conf.}+S_{ansa}). \end{equation}

\[
\stackrel{def}{=}\exp(``-\beta F")\]

\noindent The partition function $Z$ can be reformulated:

$$Z=\langle \exp(S_{action,  \;  Coul}-S_{ansa})  \rangle_{S_{ansa}+S_{action,
\;conf.}} \cdot Z_{S_{ansa}+S_{action, \; conf.}}
$$

\noindent where we define

\begin{equation} Z_{S_{ansa}+S_{action \;conf.}}=\int_G \cdots  \int_G
{\cal D}U(\link)\exp ( S_{action, \;conf.}+S_{ansa}) \end{equation}

\noindent    and    where     the     notation     $\langle     \cdots
\rangle_{S_{ansa}+S_{action, \; conf.}}$ denotes the average obtained under
functional integration over the  (above-mentioned  Cartesian  product)
Haar measure weighted with the  exponentiated  ``action''  $S_{action,
\;conf.}+S_{ansa}$; i.e.

\begin{equation}      \langle      A      \rangle_{S_{ansa}+S_{action,
\;conf}}=\frac{\int_G \cdots \int_G {\cal D}U(\link) \exp  (S_{action,
\;conf.}+S_{ansa}) \cdot A} {\int_G  \cdots  \int_G  {\cal  D}U(\link)
\exp (S_{action, \;conf.}+S_{ansa})}. \end{equation}

\begin{equation}      \langle      A      \rangle_{S_{ansa}+S_{action,
\;conf}}=\frac{\int_G \cdots \int_G {\cal D}U(\link) \exp  (S_{action,
\;conf.}+S_{ansa}) \cdot A} {\int_G  \cdots  \int_G  {\cal  D}U(\link)
\exp (S_{action, \;conf.}+S_{ansa})}. \end{equation}

\noindent Repeating this procedure, we rewrite

\begin{equation}
Z_{S_{ansa}+S_{action, \; conf.}}
\end{equation}

\noindent as

\begin{equation} \langle  \exp(S_{action,  \;conf.})\rangle_{S_{ansa}}
\cdot Z_{S_{ansa}}. \end{equation}

\noindent where the average $\langle \cdots\rangle_{S_{ansa}}$ is defined by

\begin{equation}  \langle  A  \rangle_{S_{ansa}}=\frac{\int_G   \cdots
\int_G {\cal D}U(\link) A\exp(S_{ansa})} {\int_G \cdots  \int_G  {\cal
D}U(\link) \exp(S_{ansa})}. \end{equation}

\noindent and

\begin{equation}    Z_{S_{ansa}}=\int_G    \cdots     \int_G     {\cal
D}U(\link)\exp ( S_{ansa}) \end{equation}

The original expression for the partition function in Eqn. (\ref{part})  has
been re-expressed as the product of three factors:

\begin{equation}
Z= \langle\exp(S_{action, \; Coul.}-S_{ansa}) \rangle_{S_{ansa}+S_{action,
\; conf.}} \cdot \langle \exp S_{action, \; conf.} \rangle_{S_{ansa}}
\cdot Z_{S_{ansa}}. \label{zansa}
\end{equation}

Using            the            inequality

\begin{equation}   \langle    \exp(S_{action,    \;    Coul}-S_{ansa})
\rangle_{S_{ansa}+S_{action, \;conf.}} \label{ine}  \end{equation}  $$
\geq  \exp(\langle  S_{action,  \;Coul}-S_{ansa}\rangle_{S_{ansa}+S_{action,
\;conf.}}), $$

\noindent we note that if the difference between the two distributions
given by $S_{ansa}$ and $S_{action, \; Coul}$ is small, the inequality
becomes  an  approximate  equality.  This  being  the  case,  we   can
hence-forward fit the parameters $\alpha_i$  by  maximising  the  right
hand side of this inequality.

\begin{equation}
Z\geq        Z_{MFA} \stackrel{def.}{=} \end{equation}

\beq        =
 \exp(\langle S_{action,  \;  Coul.}-S_{ansa}\rangle_{S_{ansa}+S_{action,
\;         conf.}}         )         \langle          \exp(S_{_{action,
\;conf.}})\rangle_{_{S_{ansa}}}Z_{S_{ansa}}. \label{zmfa} \eeq

Before proceeding with the calculation, it is useful to establish some
conventions. We designate  the  squared  distance  between  the  group
identity  $\bunit$  and  a  nearby  group  element  $U=\exp(\ba)$   as
$dist^2(\bunit,\exp(\ba))$ where  $\ba$  is  an  element  of  the  Lie
algebra. Throughout this paper,  the  representation  of  Lie  algebra
elements, denoted with bold type, is assumed to be anti-Hermitian.

In the weak coupling  approximation,  functional  integrals  on  group
manifolds can be  approximated  by  integrals  in  the  tangent  space
located at the group identity. In the tangent space, we can  take  the
distance squared as the Cartan-Killing form $CK(\ba,\ba)$  defined  as
$-\mbox{Tr}((\rho_{def.\;rep.}(\ba))^2)$. It is assumed that  the  12
basis generators of the Lie algebra have  been  chosen  so  that  each
generator is orthogonal to all but one basic sub-algebra. A  completely
unambiguous notation is a cumbersome combination of sums over the
sub-algebras of basis invariant subgroups as well as over  Lie  algebra
generators within a given sub-algebra. For example, a distance measured
along the Coulomb-like degrees of freedom of a partially confining phase
is written with all the details as

\beq       dist_{Coul.}^2(\bunit,        \exp(\ba))=        \sum_{i\in
``Coul."}dist^2_i(\bunit,\exp(\ba_i)) \label{carkill} \eeq

\[ =\sum_{i\in  ``Coul."}dist^2_i(\bunit,\exp(\sum_{a_i\in
i\, th \;basic\; inv.\;  subalg.}
A^{a_i}\genai)) \]

\[                                                  \stackrel{def.}{=}
-Tr_{def.\;rep.}\left[\sum_{i\in``Coul."}(\sum_{a_i\in     i\,      th
\;basic\; inv.\; subalg.}A^{a_i}\genai\cdot \sum_{b_i\in  i\,th\;basic
\; inv.\; subalg.}A^{b_i}\genbi)\right] \]

\beq = \sum_{i\in  ``Coul."}  \sum_{a_i\in  i\,  th  \;basic\;  inv.\;
subalg.}1/2(A^{a_i})^2_i     \mbox{     (with     normalisation      }
Tr_{def.\;rep.}(\genai,\;\genbi)=-\delta_{a_{i}b_{i}}/2) \label{gory}\eeq

\nin From now on we use  a  more  streamlined  notation  that  can  be
defined by rewriting Eqn. (\ref{gory}):

\beq \sum_{i\in  ``Coul."}  \sum_{a_i\in
i\,      th      \;basic\;       inv.\;       subalg.}\frac{1}{2}(A^{a_i})^2_i
\stackrel{def.}{=}\sum_{i\in ``Coul."}\frac{1}{2}A^2_i \stackrel{def.}{=}
\frac{A^2}{2} \label{stream} \eeq

\nin That is, when we write $A^2_i$, it is assumed that we have  summed
over  all  ``col-or  indices''  $(a,b,\cdots)$   labelling   generators
spanning the $i$th basic sub-algebra. If we write  $(A^a)_i$,  we  mean
the $a$th Lie algebra component of $\ba$ that, moreover, is assumed to
lie entirely within the $i$th  sub-algebra.  With  this  understanding,
sub-algebra indices $i,j,\cdots$ on the colour indices $a,b,\cdots$  can
be omitted.






\nin The Lie algebra basis (in the defining  representation)  is
normalised such that $Tr(\gena,\;\genb)=\delta_{ab}/2$  where  $a$
and $b$ label the relevant subset of the $12$ generators of the $SMG$.
The choice of representation for calculating the trace is of course  a
matter   of   convention   since,   for   simple   groups,   different
representations simply give different  overall  normalisation  factors
that can be absorbed into the coupling constant.

We deal first with the argument of the exponential in the first factor
in        (\ref{zmfa}):

         \[ \langle         S_{action,         \;
Coul.}-S_{ansa}\rangle_{S_{ansa}+S_{action, \; conf.}}. \]

\nin  Commence  by
using    the    observation    made     above:     $\langle     \cdots
\rangle_{S_{ansa}+S_{action,  \;  conf.}}   \approx   \langle   \cdots
\rangle_{S_{ansa}}$ because the  essential  information  contained  in
$S_{action, \; Coul.}$ and  $S_{ansa}$  is  retained  when  these  are
defined on the set of cosets of $H$ and, if considered as  a  function
on $G$, they have little variation along each coset of $H$.  Expressed
slightly differently, we recall that both $S_{action, \;  Coul.}$  and
$S_{ansa}$ are proportional to $Dist^2_{Coul}$ which is slowly varying
along the ``$H$-parallel" cosets.  Recall  that  along  these  cosets,
fluctuation are sufficiently large so that Bianchi identities are  not
important for these degrees of freedom.


\beq \langle S_{action, \; Coul.}-S_{ansa} \rangle_{
S_{action,  \;  conf.}+S_{\;ansa}} \eeq

\[ \approx  \langle   S_{action,   \;
Coul.}-S_{ansa} \rangle_{G/H, \; S_{ansa}} \]


\begin{small}

\[                          =                          \int_{G/H}{\cal
D}U_{coset}(\link)(-\sum_{\Box} Dist^2_{Coul.}(1,U_{coset}(\Box))
+\sum_{\link}\sum_{i                                                   \in
``Coul."}\tilde{\alpha}_i Dist^2_i(1,U_{coset}(\link))) \cdot \]

\beq \cdot                       \exp(-\sum_{i                        \in
``Coul."}\tilde{\alpha}_i Dist^2_i(1,U_{coset}(\link))) \eeq

\end{small}

Before calculating this integral approximatively in the weak  coupling
limit, it is useful to describe how we implement the gauge  constraint
necessary to eliminate unphysical gauge degrees of freedom. In say the
axial gauge, it is seen that, for a 4-dimensional cubic lattice, 6  of
the 8 links emanating from a site  are  ``active  links'';  i.e.,  not
gauge-fixed. We  therefore  take  the  gauge  condition  into  account
approximatively by using the statistically  correct  distribution:  on
the average, 3 of the 4 link variables  demarcating  a  plaquette  are
active even though any given (2-dimensional) plaquette has either 2 or
0 links in the gauge-fixed direction. Denoting by  $U_1(\link)$,  $U_2(\link)$
and $U_3(\link) \in G/H$ the three link variables that on the average  are
active for any plaquette, we conclude, again on the  average,  that  a
plaquette variable $U(\Box)$ is specified by the group product $U_1(\link)
\cdot U_2(\link) \cdot U_3(\link) \in G/H$ of the three active links.  We  can
now  write  $Dist^2(1,U(\Box))=Dist^2(1,U_1(\link)  \cdot   U_2(\link)   \cdot
U_3(\link)).$

For a partially confining phase for which the link degrees of  freedom
corresponding to the $i$th basic sub-algebra are Coulomb-like, we shall
assume  a  link  distribution  for
distances $\sqrt{A^2/2}$  that is  Gauss  distributed  with  normalised
density $F(\sqrt{A^2/2})$ given by

\begin{equation}                   F_{_{\link}}(\sqrt{A^2/2})=(\prod_i
(\frac{\tilde{\alpha}_i}{\pi})^{\frac{d_i}{2}}) \cdot  Vol(G/H)  \cdot
e^{-\sum_{i\in``Coul."}\frac{1}{2}     \tilde{\alpha}_iA_i^2}d^{{\cal
H}aar}U(\link)   \;\;   \mbox{with}    \;\;    \sum_i    d_i=dim(G/H).
\label{distrib} \end{equation}

\noindent   and   where   we   have   normalised   $\int_{G/H}d^{{\cal
H}aar}U(\link)=1.$ In this (tangent space) approximation we have

\begin{small}

\[\langle S_{ansa} \rangle_{S_{ansa}}  \approx  \sum_{\link}  (\prod_i
(\frac{\tilde{\alpha}_i}{\pi})^{\frac{d_i}{2}})    \cdot     Vol(G/H)\cdot\]

\[\cdot                                        \int_{-\infty}^{\infty}
\frac{d^{dim(G/H)}(A^a)_i/\sqrt{2}}{Vol(G/H)}(\sum_{i\in
``Coul."}\frac{1}{2}  \tilde{\alpha}_i  A_i^2  )  \exp(  -\sum_{i  \in
``Coul."}\frac{1}{2}\tilde{\alpha}_i A_i^2)\]

\end{small}

\begin{equation}
=\sum_{\link}  \sum_{i  \in  ``Coul."}      \langle   \tilde{\alpha}_i
A_i^2/2              \rangle_{F_{_{\link}}(A_i/\sqrt{2})}=\sum_{\link}\sum_i
\frac{d_i}{2}
\end{equation}
\noindent  where  the  approximation  of  extending  the  limits  of
integration to the interval $\; ] -\infty; \; \infty[\;\;$ is justified in
consequence of our using the weak coupling approximation.


For  a  plaquette  variable,  we  want  the  distribution  of
distances corresponding to the group product of the three active  link
variables. In the tangent space approximation, such a distribution  is
obtained as the convolution  of  the  distributions  for  the
distances for each of the three link variables.  For  each  link,  the
distribution is  given  by  Eqn.  (\ref{distrib}).  That  is,
distances   corresponding   to   plaquette   variables   (denoted   by
$\bp\approx\ba_1+\ba_2+\ba_3$  in  the   flat   space   approximation)   are
distributed with the density

\beq F_{\Box}(\sqrt{P^2/2})=F_{\link} \ast  F_{\link}
\ast F_{\link}(\sqrt{A^2/2}).\eeq

\nin Using the normalisation $Norm$ given by

\begin{footnotesize}

\[Norm=\]

\[=\sum_{\Box}  \int_{G/H}  \cdots  \int_{G/H}  d^{{\cal
H}aar}U_{1\;coset}(\link) d^{{\cal H}aar}U_{2\;coset}(\link)  d^{{\cal
H}aar}U_{3\;coset}(\link) \cdot \exp(-\sum_{i           \in
``Coul."}
\tilde{\alpha}_iDist^2_i(1,U(\link)))\]

\end{footnotesize}

\noindent we have

\begin{footnotesize}

\begin{equation}\langle  S_{action,   \;   Coul.}   \rangle_{S_{ansa}}
= \label{conv} \end{equation}

\[\frac{1}{Norm}\sum_{\Box}  \int_{G/H}  \cdots  \int_{G/H}  d^{{\cal
H}aar}U_{1\;coset}(\link) d^{{\cal H}aar}U_{2\;coset}(\link)  d^{{\cal
H}aar}U_{3\;coset}(\link) \cdot  \]

\[\cdot  (-   Dist^2_{Coul.}(1,   \;    U_{1,    \;
coset}(\link)  \cdot   U_{2,   \;   coset}(\link)   \cdot   U_{3,   \;
coset}(\link))) \cdot  \exp(-\sum_{i           \in            ``Coul."}
\tilde{\alpha}_iDist^2_i(1,U(\link))) \approx\]


\[\approx                                                  \sum_{\Box}
\prod_i((\frac{\tilde{\alpha}_i}{\pi})^{\frac{d_i}{2}}\cdot
Vol(G/H))^3               \cdot  \]

\[ \cdot             \int_{-\infty}^{\infty}
\frac{d^{dim(G/H)}(P^a)_i/\sqrt{2}}{Vol(G/H)}(-1/2\sum_{i\in
``Coul."}P^2_i)(\int_{-\infty}^{\infty}
\frac{d^{dim(G/H)}(A^a)_{1,\;i}/\sqrt{2}}{Vol(G/H)}
\int_{-\infty}^{\infty}
\frac{d^{dim(G/H)}(A^a)_{2,\;i}/\sqrt{2}}{Vol(G/H)}\cdot\]

\[        \cdot        \exp(-\sum_{i         \in
``Coul."}(\frac{1}{2}\tilde{\alpha}_iA^2_{i,\;1})) \exp  (-
\sum_{i   \in   ``Coul."}(1/2   \tilde{\alpha}_iA_{i,    \;    2}^2))
\exp(-\sum_{i    \in     ``Coul"}(1/2     \tilde{\alpha}_i(P_i
-A_{i, \; 1}-A_{i, \; 2})^2))\]

\[=\sum_{\Box}Vol(G/H)(\prod_i(\frac{\tilde{\alpha}_i}{3\pi})^{\frac{d_i}{2}})
\cdot                                          \int_{-\infty}^{\infty}
\frac{d^{dim(G/H)}(P^a)_i/\sqrt{2}}{Vol(G/H)}(-1/2\sum_{i\in    ``Coul."}P^2_i)
\exp(-\sum_{i \in ``Coul."}\frac{\tilde{\alpha}_i}{3}P_i^2/2)\]

\beq=-\sum_{\Box}\sum_i  \langle   P_i^2/2   \rangle_{F_{\link}   \ast
F_{\link}   \ast   F_{\link}(\bp_i)}=-\sum_{\Box}   \sum_id_i    \cdot
\frac{3}{2\tilde{\alpha}_i}. \label{convol} \eeq

\end{footnotesize}

The factor $Z_{S_{ansa}}$ in the expression (\ref{zmfa}) for $Z_{MFA}$  is
readily calculated:

\begin{equation}  Z_{S_{ansa}}=\prod_{\link}  \int_G  d^{{\cal  H}aar}U(\link)
\exp(-      \sum_{i      \in      ``Coul."}       \tilde{\alpha}_i
Dist^2_{i}(1,U(\link)) \end{equation}

\[   \approx   \prod_{\link}   \int_{G/H}   d^{{\cal    H}aar}U_{coset}(\link)
\exp(-\sum_{i                    \in                     ``Coul."}
\tilde{\alpha}_iDist_i^2(1,U_{coset}(\link)).\]

\noindent where in  the  last  step  we  have  used  that  $S_{ansa}$,
regarded as a function on $G$, is slowly varying  along  each  of  the
cosets of $H$. Assuming a normalised Haar measure  in  the  functional
integral over the cosets of  $G/H$,  the  tangent  space  approximation
yields

\begin{equation}
Z_{S_{ansa}}=\prod_{\link}\int_{-\infty}^{\infty}
\frac{d^{dim(G/H)}(A^a)_i/\sqrt{2}}{Vol(G/H)}
\exp(-\sum_{i   \in   ``Coul."}   \frac{1}{2}    \tilde{\alpha}_iA_i^2)
\end{equation}

$$=\prod_{-}\frac{1}{Vol(G/H)}                                   \cdot
\prod_{i\in ``Coul."}((\frac{\pi}{\tilde{\alpha}_i})^{d_i/2})
$$

Finally,   the   factor   $\langle   \exp   S_{action,    \;    conf.}
\rangle_{S_{ansa}}$  in  Eqn. (\ref{zmfa})  must  be  calculated.  For   the
confining degrees of freedom we shall use  the  approximation  in
which Bianchi identities are ignored:

$$
\langle   \exp   S_{action,   \; ``conf."}   \rangle_{S_{ansa}}=\langle
\prod_{\Box}    \exp(-\sum_{i    \in    ``conf."}Dist^2_i(1,U(\Box)))
\rangle_{S_{ansa}}\approx$$

\begin{equation}   \approx    \prod_{\Box}
\langle     \exp(-\sum_{i      \in      ``conf."}Dist^2_i(1,U(\Box)))
\rangle_{S_{ansa}} \label{sconf} \end{equation}

$$\approx \prod_{\Box} \int_Hd^{{\cal H} aar}U(\link) \exp(-\sum_{i  \in
``conf."}Dist^2_i(1,\bp_i))). $$

\noindent Restriction of the region of integration to $H$ in the  last
step reflects the fact that contributions from  cosets  far  from  the
region near the coset {\bf 1}$\cdot H$ are suppressed  by  $S_{ansa}$.
Assuming a normalised Haar measure on $H$, the approximation in  which
we use the tangent space at the group identity yields

\begin{footnotesize}

\begin{equation}     \langle      \exp(S_{action,      \;      conf.})
\rangle_{S_{ansa}}\approx \label{sconfansa} \end{equation}

\[\approx \prod_{\Box} \int_{-\infty}^{\infty}  \cdots
\int_{-\infty}^{\infty}
\frac{d^{dim(H)}(P^a)_i/\sqrt{2}}{Vol(H)}
\exp(-\sum_{i\;\in\;``conf."}P_i^2/2)=\prod_{\Box} \frac{
\prod_{\;\;\;j            \in
``conf."}(\pi)^{d_j/2}}{Vol(H)} \]

\end{footnotesize}

\noindent where again, in allowing the  limits  of  integration  to  be
infinite, we are assuming that the finite group volume  is  effectively
very large  in  the  weak  coupling  approximation.

Before proceeding, we should  comment  briefly  on  the  validity  and
accuracy of the approximation of ignoring Bianchi identities  in  Eqn.
(\ref{sconf}). The reliability of this approximation  depends  on  the
values of the $\beta_i$'s (and hereby, $Dist^2_i$ in the  metric  with
absorbed $\beta_i$'s) in such a  way  that  it  is  better  for  small
$\beta_i$'s than for larger ones. We  have  estimated  the  effect  of
taking the Bianchi-identities into account in a  crude  way  and  find
that this is accomplished by the replacement of a  number  of  factors
$\langle        \exp(-\sum_{i        \in        ``conf."}        \dot
Dist_i^2(1,U(\Box)))\rangle_{S_{ansa}}$ by $1/\sqrt{6}$. In fact,  one
should replace one such factor for each independent  Bianchi-identity.
For increasing $\beta$-values, a crude criterion for  determining  the
$\beta$-value at which the Bianchi identity correction should start to
be included  is  that  the  correction  should  increase  rather  than
decrease the partition function.

We have also looked into a refined approximation for this correction
and found that the correction is  expressible  by
means  of  a  complete   elliptic   integral   that   arises   via   a
$\theta_3$-function  that  in  turn  is  obtained  as   a   sum   over
exponentials of squares of integers times a  constant.  We  shall  not
elaborate further on the refined correction here; we consider only the
crude correction outlined above.

In the region of ``large'' $\beta_i$'s  where  the  Bianchi-identities
are to be taken into account, it turns  out  that  the  slope  of  the
$``-\beta F"_{per \; active \; link, \; conf.}$ for degrees of freedom
that are confined is the same  as  that  of
$``-\beta F"_{per\;active\;link,\;Coul.}$ for these  same  degrees  of
freedom in the Coulomb phase. As  a  consequence,  the  free  energies
$``-\beta F"_{per \; active \; link, \; conf.}$ and $``-\beta
F \mbox{''}_{per \; active \; link, \; Coul.}$  for  respectively  the
confinement and Coulomb phases (both as  functions  of  $\log  \beta$)
cannot intersect for $\beta$-values larger than those  for  which  the
Bianchi identity correction commences. In  fact,  the  two  free
energies are not only parallel but in fact coincident  in  the  lowest
order approximation we consider (see Figure~\ref{betaf}).
\begin{figure}
\centerline{\epsfxsize=\textwidth \epsfbox{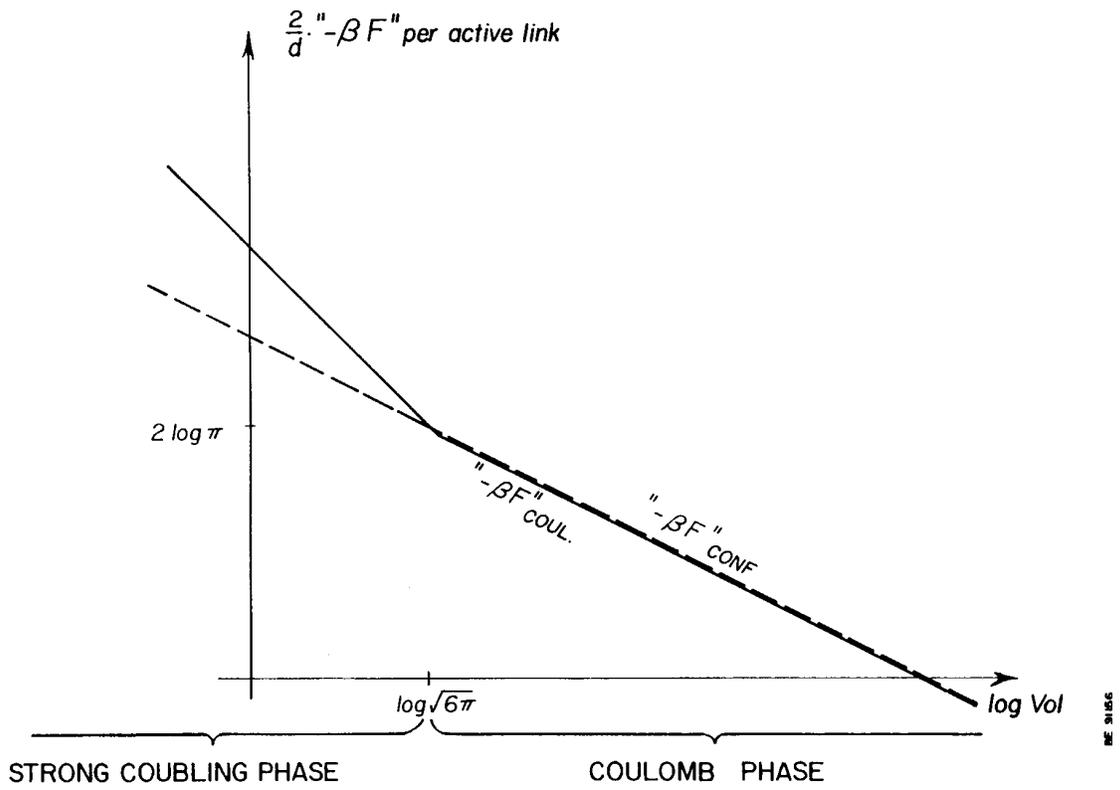}}
\caption[betaF]{\label{betaf} The  free  energies  $``-\beta
F"_{conf}$  and  $``-\beta  F"_{Coul.}$  as   a   function   of   $\log
Vol$.}
\end{figure}

In going to weaker couplings, the fact that the values of the  inverse
squared  couplings  $\beta_i$  for  which  Bianchi  identities  become
important coincides with the onset of the Coulomb  phase  should  $a\;
priori$ probably be regarded as a chance artifact of the lowest  order
approximation. However, we suspect that the transition to the  Coulomb
(organised) phase may be fundamentally  related  to  having  couplings
weak enough to allow the self-organising effects of  Bianchi  identity
constraints to be enforced in which case the coincidence  may  not  be
just a chance occurrence. In either  event,  the  fact  that  $``-\beta
F"_{per\;active\;link,\;conf.}$    is    parallel     to     $``-\beta
F"_{per\;active\;link,\;Coul.}$ for $\beta$ values exceeding that  for
which Bianchi identity corrections set in precludes an intersection of
$``-\beta      F"_{per\;active\;link,\;conf.}$      and      $``-\beta
F"_{per\;active\;link,\;Coul.}$ in  this  region  of  $\beta$  values.
Therefore ignoring Bianchi identity corrections is justified when  the
$\beta_{crit.}$ are determined by the phase transition in  the  lowest
order MFA.

For $\beta$ values corresponding to confinement,it is  justifiable  to
neglect Bianchi identity corrections in the lowest  order  calculation
of  $\langle  \exp  S_{action,\;conf.}  \rangle  _{S_{ansa}}$  as  the
corrections first become relevant at the $\beta_{i,\;crit.}$  -  i.e.,
upon leaving the  confinement  phase  when  the  critical  points  are
determined as the points of intersection for  $``-\beta  F"  _{per  \;
active \; link, \; conf.}$ and $``-\beta F" _{per \; active  \;  link,
\; Coul.}$.

Collecting  the  various  factors  in  the   expression   (\ref{zmfa})
for $Z_{MFA}$ that have been calculated yields

$$Z_{MFA}=\exp(\langle  S_{action,  \;
Coul.}-S_{ansa}\rangle_{S_{action,    \;    conf.}+S_{ansa}})    \langle
\exp(S_{action, \; conf.}) \rangle_{S_{ansa}}\cdot Z_{S_{ansa}}$$

\begin{equation}  Z=\exp(``-\beta  F")\geq Z_{MFA}   \approx
\exp(-\sum_{\Box}(\sum_{i        \in        ``Coul."}d_i        \cdot
\frac{3}{2\tilde{\alpha}_i})+  \sum_{\link}(\sum_{i  \in  ``Coul."}   d_i
\cdot\frac{1}{2})) \cdot\end{equation}

$$\cdot          \prod_{\Box}\prod_{j          \in          ``conf."}
\frac{(\pi)^{d_j/2}}{Vol(H)}   \cdot    \prod_{\link}    \frac{1}{Vol(G/H)}
\prod_{i \in ``Coul."}(\frac{\pi}{\tilde{\alpha}_i})^{d_i/2}.$$

\noindent that is,

\begin{equation} ``-\beta F" = \log Z \geq -\sum_{\Box}  \sum_{i  \in
``Coul."} d_i\frac{3}{2\tilde{\alpha}_i}+\sum_{-}\sum_{i \in ``Coul."}
\frac{d_i}{2}+\sum_{\Box}\log \frac{\prod_{j\in ``conf"}(\pi)^{d_j/2}}{Vol(H)}+
\end{equation}

$$+\sum_{-}  \log(\frac{1}{Vol(G/H)}  \prod_{i   \in
``Coul."}(\frac{\pi}{\tilde{\alpha}_i})^{d_i/2})               \label
{logz}$$

As we want to calculate a $``-\beta F"$ density rather than $``-\beta
F"$ for the entire lattice ($``-\beta F"$ for  an  infinite  lattice
would of course be divergent), we now choose to work with the quantity
$``-\beta F"= \log Z$ \underline{per active link.} For a 4-dimensional
hyper-cubic lattice, there are 3  active  links  per  site  (i.e.,  the
number of dimensions reduced by the  one  dimension  along  which  the
gauge is fixed) and 6 plaquettes per site. This  yields  2  plaquettes
per active link. The expression for the total  lattice  $``\beta  F"$
can now be rewritten as a ``free energy density per active link'':

\newcommand{\bfpal}{``-\beta F\mbox{''}_{per\;active\;link}}

\begin{equation}  \log  Z_{per  \;   active   \;   link}=\bfpal   \geq
\end{equation}

\begin{equation}
-2\sum_{i \in  ``Coul."}  d_i\frac{3}{2\tilde{\alpha}_i}+\sum_{i  \in
``Coul."}\frac{d_i}{2}+2  \log  \left  (  \prod_{j  \in  ``conf."}
\frac{\pi^{d_i/2}}{Vol(H)} \right )+ \label{bfpal} \end{equation}

$$+\log  \left  (  \frac{1}{Vol(G/H)}
\prod_{i  \in  ``Coul."}(\frac{\pi}{\tilde{\alpha}_i})^{d_i/2}  \right
).$$


\noindent The extremum for $\bfpal$ w. r. t. the $\tilde{\alpha}_i$ is
found for

\begin{equation} \tilde{\alpha}_i=6 \;\;\;(\forall i \in
\mbox{``Coul."})\end{equation}

\noindent Inserting  $\tilde{\alpha}_i=6$   in  (\ref{bfpal})   yields
$\bfpal$ to the approximation used here:

\begin{equation}  \bfpal=\sum_{i   \in   ``Coul."}(-2d_i\frac{3}{2\cdot
6}+\frac{d_i}{2}+\log((\frac{\pi}{6})^{d_i/2})))+ \end{equation}

$$+2\log           \left
(\frac{(\pi)^{\frac{\sum_{j     \in      ``conf."}d_j}{2}}}{Vol(H)}  \right
)+\log(\frac{1}{Vol(G/H)})$$

\begin{equation}
=\log  [\frac{(\pi/6)^{(dim(G/H))/2}}{Vol(G/H)}
]+2\log      [\frac{(\pi)^{(dim(H))/2}}{Vol(H)}]      \label{betafpal}
\end{equation}

\subsubsection{The phase diagram for $SMG$ in lowest order $MFA$}

The   expression   (\ref{betafpal})   for   $``-\beta
F_{per\;active\;link}$''
provides a  means  of   constructing  an approximate  phase
diagram depicting the phase boundaries separating the partially confining
phases of a non-simple gauge  group.
Using the weak coupling approximation,
the goal in this section is the construction of a phase  diagram for the
$SMG$ using an
action parameter  space spanned by the
logarithm of the volumes of the three
different    ``basic''     invariant     subgroups     $H_i$     where
$\{i\}=\{U(1),\;SU(2),\;  SU(3)\}$  (Figure\ref{smgfull}).  In   the
absorbed notation, increasing values along  the  $i$th  axis  of  this
space correspond to  increasing  values  of  the  ``scaling''  factor
$\beta_i^{dim(H_i)/2}$.  Recall  from  Eqn.   (\ref{abmet})   that   the
relationship between volumes with and without the  $\beta$'s  absorbed
into the metric is  $Vol(H_i)=\beta_i^{(dim(H_i))/2}\;vol(H_i)$  where
``$Vol$'' and ``$vol$'' designate respectively volumes in terms of the
metric with and without the $\beta_i$'s absorbed. In the ``$Vol$''

At the boundary between a  partially  confining  phase  that  confines
along an invariant subgroup  $H_I$  and  another  partially  confining
phase that is in confinement along an invariant  subgroup  $H_J$,  the
condition to be fulfilled for the critical volumes of the subgroups is
obtained by equating the ``free energy'', $``-\beta F$'', for these  two
phases. This leads to the condition

\beq                                                  \log
(6\pi)^{\frac{dim(H_J)-dim(H_I)}{2}}=\log \frac{Vol(H_J)}{Vol(H_I)}
\label{bc}. \eeq

Uppercase indices $I,\;J,\cdots$ label the invariant subgroups
corresponding  to
the eight possible partially confining phases  of the $SMG$ obtainable
when the discrete subgroups are not included among the basic invariant
subgroups (i.e., using just
the basic invariant subgroups $H_i$ with $i  \in
\{U(1),\;SU(2),\; SU(3)\}$). Of  course  the
invariant subgroups $H_J$ do  not  in  all  cases  coincide  with one of  the
three  ``basic''  invariant  subgroups, but each $H_J$ is spanned  by
some subset of the basic invariant subgroups $H_i$ (having perhaps a discrete
subgroup of the centre identified with the group identity).

The quantities ``$Vol$'' corresponding to the first eight invariant subgroups
in the list (\ref{invh}) w.r.t. which there can be confinement in the
approximation being considered in this Section are listed in the first column
of Table~\ref{tab8ph}. The second column contains the corresponding ``$Vol$''
quantities  for the Coulomb-like factor group \dof.

\begin{tiny}
\begin{table}\caption[table8ph]{\label{tab8ph}   The
first entry in a row is the quantity ``$Vol$'' for an invariant
subgroup $H$ w.r.t which there can be confinement.
The second entry in a row is the quantity ``$Vol$''for the corresponding
factor  group $G/H$ \dofx that behave in a Coulomb-like fashion.}

\begin{equation}   \begin{array}{ll}   Vol(\{1\})=1         &
Vol(SMG)=Vol(SMG/\{1\})=Vol(U(1)_{large})Vol(SU(2))Vol(SU(3))/6\\
Vol(U(1)(large))=Vol(U(1)_{large})   &   Vol(SMG/U(1))=Vol(So(3)\times
SU(3)/Z_3)=Vol(SU(2))Vol(SU(3))/6\\      Vol(SU(2))=Vol(SU(2))       &
Vol(SMG/SU(2))=Vol(U(3)_{factorgr.})=Vol(U(1)_{large})Vol(SU(3))/6\\
Vol(SU(3))=Vol(SU(3))                                                &
Vol(SMG/SU(3))=Vol(U(2)_{factor\;gr.})Vol(U(1)_{large})/6\\
Vol(U(2)_{subgroup})=Vol(U(1)_{large})Vol(SU(2))/2                   &
Vol(SMG/U(2))=Vol({SU(3)/Z_3}_{factorgr.})=Vol(SU(3))/3\\
Vol(U(3)_{subgroup})=Vol(U(1)_{large})Vol(SU(3))/3                   &
Vol(SMG/U(3))=Vol(So(3)_{factorgr.})=Vol(SU(2))/2\\    Vol(SU(2)\times
SU(3)_{subgroup})=Vol(SU(2))Vol(SU(3))     &      Vol(SMG/(SU(2)\times
SU(3)))=Vol(U(1)_{small})=Vol(U(1)_{large})/6\\
Vol(SMG)=Vol(U(1)_{large})Vol(SU(2))Vol(SU(3))/6                     &
Vol(SMG/SMG)=Vol(\{1\})=1\\ \end{array}. \end{equation}
\end{table}
\end{tiny}

In particular, setting one  of  the  invariant  subgroups  ($H_I$  for
example) in Eqn. (\ref{bc}) equal to  the  group  identity  ${\bunit}$
yields the value of the volume of the subgroup $H_J$ at the  interface
with          the           totally           Coulomb phase. That is,
$\log Vol(H_J)_{subgr.}=\log((6\pi)^{(dim(H_J))/2})$   at   the    boundary
between the totally Coulomb phase and the  partially  confining  phase
with confinement along $H_J$. Within this partially  confining  phase,
$\log Vol(H_J)_{subgr.} \leq \log((6\pi)^{(dim(H_J))/2})$.

Alternatively, setting one  of  the  invariant  subgroups  ($H_J$  for
example) in (\ref{bc}) equal to the entire gauge group $G$ yields
the value of the volume for the invariant  factor  group  $G/H_I$  at  the
interface      with      the      totally       confining       phase. That is
$\log Vol(G/H_I)_{factor\;gr.}=\log(6\pi)^{(dim(G/H_I))/2}$    at     the
boundary  between  the  totally  confining  phase  and  the  partially
confining phase that is Coulomb-like w.r.t.  $(G/H_I)_{factorgr.}$.
Within this partially confining  phase,  $\log Vol(G/H_I)_{factorgr.}  \geq
\log(6\pi)^{(dim(G/H_J))/2}$.

In general (but not considering Higgsed \dof), each partially
confining phase can be characterised by the
invariant subgroup $H_J$ along  which  the  corresponding  degrees  of
freedom are confined {\em or} the factor  group
$G/H_J$ the cosets of which constitute  the  Coulomb-like  degrees  of
freedom. Denoting by $L_{H_J}$ and $L_{G/H_J}$ the Lie algebras of
respectively the invariant subgroup $H_J$ and the factor group $G/H_J$,
we have of course that these Lie algebras
span the gauge  group  Lie  algebra
$L_G$:  $L_G=L_{H_J}  \oplus L_{G/H_J}$.

To understand the approximate phase diagram of the $SMG$ sought here,
it is  useful  to
have  as  a  reference  the  parallel  study  of   the   simpler   Lie
algebra-identical Cartesian product group $U(1)  \times  SU(2)  \times
SU(3)$. As stated previously, the group that we use
as the Standard Model Group ($SMG$) is consistently taken to be

\begin{equation} SMG = U(1)  \times  SU(2)  \times  SU(3)/``\bz_6"
\label{SMG} \end{equation}

\nin where

\begin{equation} ``\bz_6" \stackrel{def}{=}  \{
(2\pi,-{\bunit}^{(2 \times 2)}, \exp(i2\pi/3){\bunit}^{(3 \times  3)})^p
\mid p \in {\bf Z} \}. \label{z6} \end{equation}

\noindent This group has of course the same Lie  algebra  as  the
Cartesian product group $U(1) \times SU(2) \times SU(3)$ but has  quite
another global structure - namely that resulting when the
subgroup of centre elements generated by the  element  $(2\pi,\;  -{\bunit}
^{(2\times  2)},  \;  \exp(i2\pi/3){\bunit}^{(3\times   3)})$   are
identified with the group identity. In other words, the elements of the
Standard  Model
Group ($SMG$) are, in this  work,  the  set  of  cosets
$g\cdot(2\pi,\;-{\bunit}^{(2\times   2)},\;\exp(i2\pi/3){\bunit}
^{(3\times 3)})^p$ where $g\in U(1) \times SU(2) \times  SU(3)$  and
$p \in Z$. Each such coset consists of 6 elements of the group
$U(1)\times SU(2)\times SU(3)$.

A characteristic feature of the phase diagram for the $SMG$ is that there is
an extended phase boundary separating the totally confining and the totally
Coulomb-like phases in the action parameter space spanned  by just
the  three  variables  $\log
Vol(H_i)$ ($i \in \{U(1),\;SU(2),\; SU(3)\}$).  This  feature, which makes
it impossible for all of the 8 possible \pcps to convene
at a multiple point,   is  not
present in the phase diagram for the Cartesian product group $U(1) \times
SU(2) \times SU(3)$  where  in  fact  the  boundary  separating the  totally
confining phase from the totally  Coulomb-like phase is just one point in the
3-dimensional phase space where the 8 partially confining phases come
together.

This difference in the phase diagrams for $SMG$
and $U(1) \times SU(3) \times SU(3)$ stems from the different way in which
the volume of an invariant  subgroup  and  the  Lie  algebra-identical
factor group are related to each other and to the volume of the  whole
gauge group.

Generally, the volume of  a  confining  subgroup  $Vol(H_J)$  and  the
volume of the factor group with the  same  Lie  algebra
are simply related. For a Cartesian product group  $G=H_1
\times H_2 \times \cdots \times  H_J  \times  \cdots$,
                           the  relation  between  the  volume  of   a
subgroup factor $H_J$ and the volume of the factor  group
with      the      same      Lie       algebra       is particularly simple:

$$ Vol(H_J)=Vol(G/{\textstyle \prod_{_{cart.\;prod.,\;I\not = J}}}
\cdots \times H_I \cdots)$$

\noindent         where $G/{\textstyle \prod_{_{cart.\;prod.,\;I\not = J}}}
\cdots \times H_I \cdots $

\nin
is the factor group having the  same  Lie  algebra  as  the  subgroup  $H_J$.
Characteristic for a Cartesian product group is  that  the  centre  is
``disjoint'' in the sense that the centre of a Cartesian product group is
multiplicative in the centres of the subgroup factors. The centres of
$SU(N)$ subgroups and $U(1)$ are isomorphic to respectively $\bz_N$
(cyclic group of order $N$) and $S_1$ (a circle).

The situation is a little more complicated when various subgroups of the
centre of the Cartesian product group are identified with the identity
element of the group (as is the case for $SMG$).
In the case of $SMG$ \ref{SMG}, which is obtained from
the Cartesian product group $U(1)\times SU(2)\times SU(3)$ by identifying
the  discrete ``$\bz_6$'' subgroup \ref{z6} of centre elements with
the unit element
(\ref{SMG}), the volume is smaller by a factor 6 (the number of centre
elements identified with the group identity) than that for the
Cartesian product group $U(1)  \times  SU(2) \times SU(3)$.
The greater complexity of the  phase  diagram  for  the
$SMG$ arises because of the  different  ways  that  the  total  volume
reduction  factor  ``6''  can  be  shared  between  (confined)
invariant subgroups and corresponding (Coulomb-like) factor groups.

{\bf The group  $U(N)=(U(1)  \times  SU(N))/``Z_N"$  as  illustrative
analogy to $SMG=(U(1) \times SU(2) \times SU(3))/``Z_6"$}\\

In order to see how we get the phase diagram for the $SMG$ in the approximation
considered here, it is useful to see how the distinctive difference
in the  phase  diagrams
for the gauge group $SMG=S(U(2) \times U(3))$ of Eqn. (\ref{SMG})  and
the Lie algebra-identical Cartesian product gauge group  $U(1)  \times
SU(2) \times SU(3)$ comes about. To this end,
it is enlightening (because it is
easier)  to  consider  the
essentially analogous pair of groups $U(N)$ and the Cartesian  product
group $U(1) \times SU(N)$ (which of course has the same Lie algebra as
$U(N)$). The relation of the group $U(N)$ to the
Cartesian product group $U(1)  \times  SU(N)$  is  analogous  to  the
relation  of the  $SMG$ (Eqn. (\ref{SMG})) to  the  Cartesian  product
group $U(1) \times SU(2) \times SU(3)$ because $U(N)$, like the $SMG$,
is obtained by identifying a discrete subgroup of  the  centre  of  the
Cartesian product  group  with  the  group identity:

\begin{equation}
U(N)=(U(1) \times SU(N))/``\bz_N" \end{equation}

\nin where

\begin{equation}
``\bz_N" \stackrel{def}{=} \{(2\pi, \exp(i2\pi/N){\bunit}^{(N\times  N)})^p\}
\;\;\; (p \in \bz).
\end{equation}

For both $U(N)$ and $U(1) \times SU(N)$, the choices for confining
subgroups  are  the  four  invariant
subgroups $SU(N)$, $U(1)$, $G$ (i.e., the whole group $U(N)$ or  $U(1)
\times SU(N)$) and {\bf 1} (the group identity). The  Lie  algebra-identical
factor groups, listed in the same  order,  are  $G/U(1)$,  $G/SU(N)$,
$G/\bunit$, and $G/G$. The invariant $(subgroup,\;\; factor\;\; group)$
pairs  characterising (in a redundant way)  the  four  partially  confining
phases  are
$(U(1)_{subgroup},(G/U(1))_{factor\;gr.})$,
$(SU(N)_{subgroup},(G/SU(N))_{factor\;gr.})$,
$(G_{subgroup},(G/G)_{factor\;gr.})$,                         $(\bunit
_{subgroup},(G/\bunit)_{factor\;gr.})$.

The last two $(subgroup,\;\; factor\;\; group)$ pairs
designate the two  special
cases of partially confining phases in which all (physical) Yang-Mills
degrees of freedom are respectively confining  and  Coulomb-like.  The
two remaining $(subgroup,\;\; factor\;\; group)$ pairs
correspond to phases  for
which some of the (physical) Yang-Mills  degrees  of  freedom  are  in
confinement   while    others    are    in    the    Coulomb    phase.
For example,
$(U(1)_{subgroup},(G/U(1))_{factor\;gr.})$  designates  the  partially
confining phase with confining degrees of  freedom  along  $U(1)$  and
degrees of freedom that are Coulomb-like w.r.t.  the  factor  group
$G/U(1)$.

The phase diagram for $U(N)$ is different from that for the  Cartesian
product group $U(1) \times SU(N)$:  the  phase  in  which  the  $U(1)$
degrees of freedom are confining and the $SU(N)$  degrees  of  freedom
are Coulomb-like and the phase  in  which  the  roles  of  $U(1)$  and
$SU(N)$ are exchanged have phase  boundaries  that do not have any points
in common in the action parameter space (phase diagram) considered
(Figure~\ref{sun}). Rather there is an extended
boundary between the totally Coulomb and the totally confining phases
in contrast to the phase diagram for $U(1)\times SU(2)$ where the boundary
between the totally
Coulomb-like and totally confinement-like phases is just a point
(the multiple point) at which the other partially confining phases
also convene.

For the Cartesian product group $U(1)  \times  SU(N)$,  subgroups  and
corresponding (i.e., Lie algebra-identical) factor groups have equal
volumes. In the metric in which
the $\beta$'s are not absorbed into  the  metric  tensor,  this  means
simply            that             $vol(U(1))_{subgr.}=vol((U(1)\times
SU(N))/SU(N))_{factor\;gr.}$ and  $vol(SU(N))_{subgr.}=vol((U(1)\times
SU(N))/U(1))$ where we continue to use the notation in which ``$vol$''
with lower  case  ``$v$''  designates  group  volumes  without  absorbed
$\beta$'s. For $U(1)\times SU(N)$  the equality   $\log
Vol(U(1))_{subgr.}=\log
Vol((U(1)    \times    SU(N))/SU(N))_{factor    gr.}$ is fulfilled for the
same value of the absorbed quantity $(\beta_1)^{\frac{1}{2}}$ on the right
and left sides of the equation. The same applies for the absorbed quantity
$(\beta)^{\frac{N^2-1}{2}}_N$  in the equality  $\log
Vol(SU(N))_{subgr.}=\log Vol((U(1)  \times  SU(N))/U(1))_{fac.  gr.}$.
Therefore, the phase boundaries separating the confinement and Coulomb
phases - for either the $U(1)$ or the $SU(N)$ gauge degrees  of  freedom -
coincide and the phase diagram consists of  the two  straight  lines

$$\log   Vol(U(1))=\log(6\pi^{(dim   (U(1)))/2})=    \log
\sqrt{6\pi}$$

\noindent  and

$$log  Vol(SU(N))=\log(6\pi^{(dim(SU(N)))/2}).$$

\nin that intersect at the multiple point where all 4 \pcps convene.
(see Figure~\ref{cartprod}).
\begin{figure}
\centerline{\epsfxsize=\textwidth \epsfbox{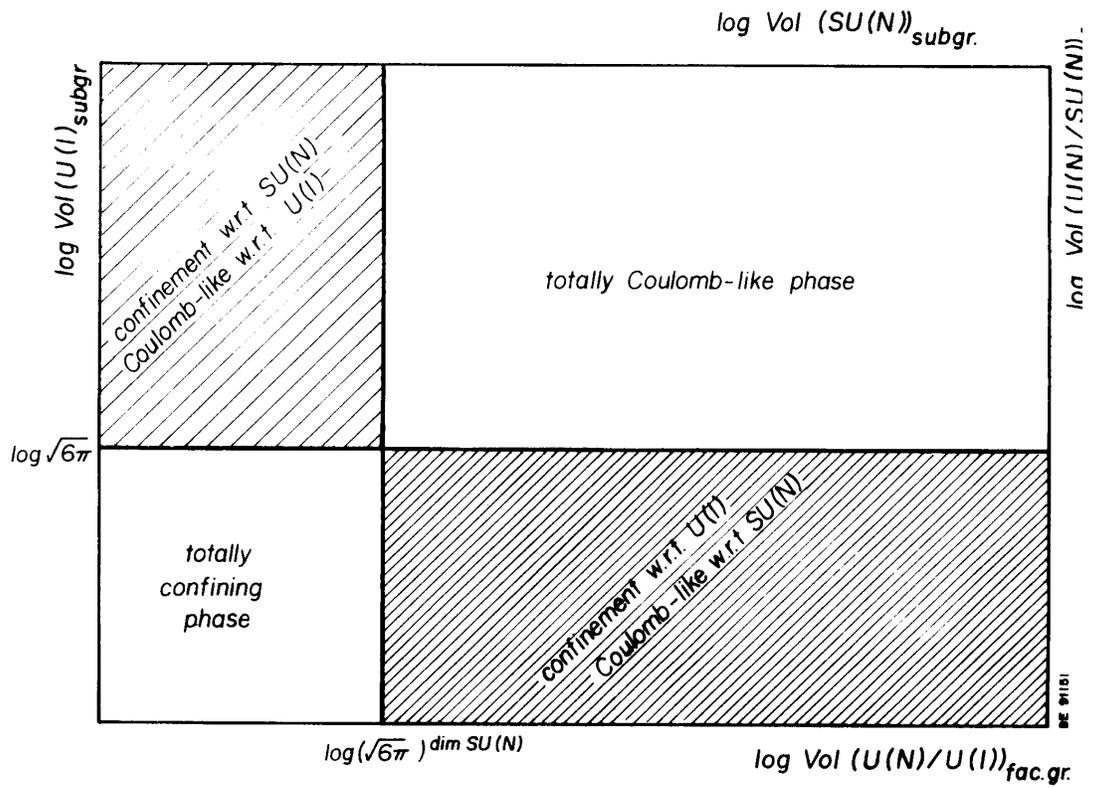}}
\caption[tabcartprod]{\label{cartprod} Phase  diagram  for
the Cartesian product group $U(1) \times SU(N)$.}
\end{figure}

{\bf The Phase Diagram for the Group $U(N)$}\\

For $U(N)$ this is  not  the  case: rather $vol(U(1))_{subgr.}
= N\cdot vol(U(N)/SU(N))_{factor\;gr.}$ and $vol(SU(N))_{subgr.}
= N\cdot vol(U(N)/U(1))_{factor\;gr.}$.
The reason is that the identification (with the group identity)
of  a $N$-element
discrete  subset  common to the
centre of the subgroups $U(1)$  and  $SU(N)$
 reduces  the  volume  of
$U(N)$ relative the  Lie  algebra-identical  Cartesian  product  group
$U(1) \times SU(N)$ by a factor  equal  to  the  $N$  centre  elements
identified with the identity in obtaining $U(N)$ from $U(1) \times SU(N)$.
That is,

\begin{equation}   Vol    U(N)=\frac{Vol(U(1)    \times    SU(N))}{N}=
\frac{VolU(1)_{subgr.}Vol(SU(N))_{subgr.}}{N}.             \label{vol1}
\end{equation}

\noindent But it is also true that

\begin{equation}  Vol(U(N))=  VolU(1)_{subgr.}  Vol(U(N)/U(1))_{fac.\;
gr.} \label{vol2} \eeq

$$= VolSU(N)_{subgr.}  Vol(U(N)/SU(N))_{fac.\;  gr.}.  $$

\noindent From Eqn. (\ref{vol1})  and  Eqn.  ({\ref{vol2})  it  can  be
concluded that for $U(N)$ one has

\beq Vol(U(N)/SU(N))_{fac.\;gr.}=\frac{VolU(1)_{subgr.}}{N} \eeq

\noindent and

\beq Vol(U(N)/U(1))_{fac.\;gr.}=\frac{VolSU(N)_{subgr.}}{N} \eeq

\noindent i.e., in going from  $U(1)  \times  SU(N)$  to  $U(N)$,  the
volume of a factor group with a given sub-Lie algebra is reduced by  a
factor $N$ relative to the subgroup with this  sub-Lie  algebra.  This
happens because  the  subgroup  corresponding  to  any  given  sub-Lie
algebra has the same volume in $U(N)$ as in $U(1) \times SU(N)$, so it
is the factor group that bears the entire $1/N$ group volume reduction
factor.

Inasmuch as critical values for the inverse squared  couplings  (i.e.,
the  $\beta_i$'s)  must  ultimately  be  extracted  from  the   volume
expressed in the metric into which the $\beta_i$'s are absorbed  (Eqn.
(\ref{abmet})), it is important to realize that a given $\beta_i$  can
be absorbed into: (1) a subgroup volume or  (2) a  factor  group
volume. This is true for $U(N)$ as well as the $SMG$:

\begin{enumerate}

\item Consider a subgroup $H_J$ that coincides with a the $i$th basic
invariant subgroup $H_i$ or a Cartesian product of basic invariant subgroups
that includes $H_i$. In these cases, $\beta_i$ is absorbed into
the quantity $Vol(H_J)$  that, in directions within the $i$th basic Lie
sub-algebra, is
calculated using
in the  metric  induced  by  the  group  along  the  invariant
\underline{subgroup} $H_i$.
Distances within this  subgroup  are   proportional   to
$\sqrt{\beta_i}$. This essentially means that distances along $H_i$ are
measured in units of the root mean square extent of quantum fluctuations.

For  the
partially confining phase confined along the Cartesian product of say the
$i$th and $j$th basic invariant subgroups: $H_J=H_i\times H_j$,
the interface with  the  totally  Coulomb  phase
coincides with the critical  value  of
$Vol(H_J$) as
determined by Eqn. (\ref{bc}) when the invariant  subgroup $H_I$ is set
equal to  the  gauge  group  identity:

\beq (6\pi)^{(dim(H_i)+dim(H_j))/2}=Vol(H_i)Vol(H_j)= \eeq

\[ =\beta^{dim(H_i)/2}\beta^{\beta(H_j)/2}vol(H_i)vol(H_j). \]

In the case of $U(N)$, taking $H_I=${\bf 1} in (\ref{bc})  and  the  other
invariant  subgroup  $H_J$  to  be
$H_J=U(1)$ or $SU(N)$ leads to the equations

\begin{equation} (6\pi)^{(dim(U(1))  )/2}=Vol_{crit}(U(1)_{subgr.})
=\sqrt{\beta_{1\;crit}}vol(U(1)_{subgr})\end{equation}

\nin and

\begin{equation}
(6\pi)^{(dim(SU(N)))/2}=Vol_{crit}(SU(N)_{subgr})
=(\sqrt{\beta_{N\;crit}})^{(N^2-1)}vol(SU(N)_{subgr})\eeq

\nin at the interface of the totally Coulomb-like phase and the phases
confined w.r.t. respectively $U(1)$ and $SU(N)$.


\item Another occurrence  of  $\beta_i$
is  in  the  Volume
expressed in  the  metric  induced  by  the  group along an
invariant  {\em factor  group}  the Lie subalgebra of which
includes that of the $i$th basic invariant subgroup $H_i$. Assuming
that the Lie algebra of the factor group coincides with that of the invariant
subgroup $H_I$, we can write this factor group as $G/H_J$ $(J\neq I)$.
where $G$ denotes the gauge group $U(N)$ or $SMG$ as appropriate to
the context.
By assumption we have  $H_i\subseteq H_I$ (i.e., the Lie algebra of the
factor group includes but does not necessarily coincide with the Lie
sub-algebra of the  i$th$ basic invariant subgroup).
The Volume  of  the factor group is the Volume of the
Lie sub-algebra-identical basic invariant subgroup
divided by a natural number $n$:

\beq Vol(G/H_J)=\frac{Vol(H_I)}{n} \eeq

This natural number $n$is the the total number
of identified\footnote{The term ``identified elements'' denotes group
elements identified with the group identity.} centre elements of the gauge
group divided by the number of these same identified
group elements that are identified in the invariant subgroup
$H_J$ if the latter is not simply a basic invariant subgroup (or a
Cartesian product of basic invariant
subgroups). As an example, consider the factor group $SMG/U(2)$ having the same
Lie algebra as $SU(3)$. For $Vol(SMG/U(2))$ (which contains an absorbed
$\beta_{SU(3)}$, we have that
$Vol(SMG/U(2))=Vol(SMG/((U(1)\times SU(2))/\bz_2)=$ $
Vol(SU(3)_{subgr}/(6\cdot\frac{1}{2})$.
The ``natural number''
$n=6\cdot \frac{1}{2}=3$ arises as the number of centre elements
of $U(1)\times SU(2)\times SU(3)$ that are identified in the
$SMG$ (i.e., 6) divided by the the number of these that are already
identified in
$U(2)$ (i.e., 2). For the critical value of Vol($SMG/U(2)$), the absorbed
quantity $\beta_{SU(3)}$ is a factor $(\frac{6}{2})^{2/dim(SU(3))}$ larger
that the $\beta_{SU(3)}$ value absorbed in the critical value of
Vol($SU(3)$).

For the  partially  confining
phase having a Coulomb-like distribution of the cosets
$g\cdot H_J$ with $g\in G$ (or, equivalently,
the \pcp that is confined along $H_J$)),
the interface with the totally confined phase coincides with
the critical value of $Vol(G/ H_J$ as determined
by  Eqn.
(\ref{bc}) upon setting one of the subgroups in this equation equal to
the whole gauge group. Thinking of $G=U(N)$, this yields for
the critical volume of the factor group with  the  Lie
algebra of $U(1)$ (i.e., $U(N)/SU(N)$)

\begin{equation}
(6\pi)^{(dim(U(N))-dim(SU(N)))/2}=(6\pi)^{dim(U(1))/2} \label{a1}
=Vol((U(N)/SU(N))_{factor\;group})= \end{equation}

$$
=(\beta_{U(1)})_{fac\;gr}^{(dim(U(1))/2}\cdot
vol((U(N)/SU(N))_{factor\;group})=
(\beta_{U(1)})_{fac\;gr}^{(dim(U(1))/2}\cdot \frac{vol(U(1))}{N}.$$

\noindent But we also have that (\ref{a1}) is the critical value
of $Vol(U(1))=(\beta_{U(1)})^{\frac{dim(U(1))}{2}}_{subgr}vol(U(1))$. Using
that $vol(U(1))=N\cdot vol(U(N)/SU(N))$, we see that the relation between
the $\beta_{U(1)}$ absorbed into $Vol(U(1))$ and that absorbed into
$Vol(U(N)/SU(N))$ is

\beq  (\beta_{U(1)})_{fac\;gr}=
N^{\frac{2}{dim(U(1))}}(\beta_{U(1)})_{sub\;gr} \eeq

\nin An analogous argument leads to the relation

\beq (\beta_{SU(N)})_{fac\;gr}=
N^{\frac{2}{dim(SU(N))}}(\beta_{SU(N)})_{sub\;gr} \eeq

\nin The subscripts ``$fac\;gr$'' and ``$subgr$'' on $\beta$, put
in above for clarity, are normally omitted inasmuch as these subscripts can
be figured out from the context in which a $\beta$ appears.


$$=Vol((U(N)/U(1))_{factor\;group})
=(\beta_N)^{(dim(SU(N))/2}vol((U(N)/U(1))_{factor\;group}).$$





\end{enumerate}

In a coordinate system spanned by the variables $\log Vol(H_i)$  ($H_i
\in \{U(1),\;SU(N) \}$), the phase diagram for  $U(N)$  is  completely
determined by locating two special points  ``1'' and ``2''  of  the
phase diagram: denote by ``1'' a point in the partially
confining
phase that is Coulomb-like solely along $U(1)$
but which is very close to the corner
where this phase is in contact with the totally Coulomb and totally
confining phases; denote by ``2'' a point in the \pcp that is confined
solely
along $U(1)$ but which again is very close to the corner where this phase is
in contact with the phases that are totally confining and totally Coulomb-like
(see Figure~\ref{sun}).
\begin{figure}
\centerline{\epsfxsize=\textwidth \epsfbox{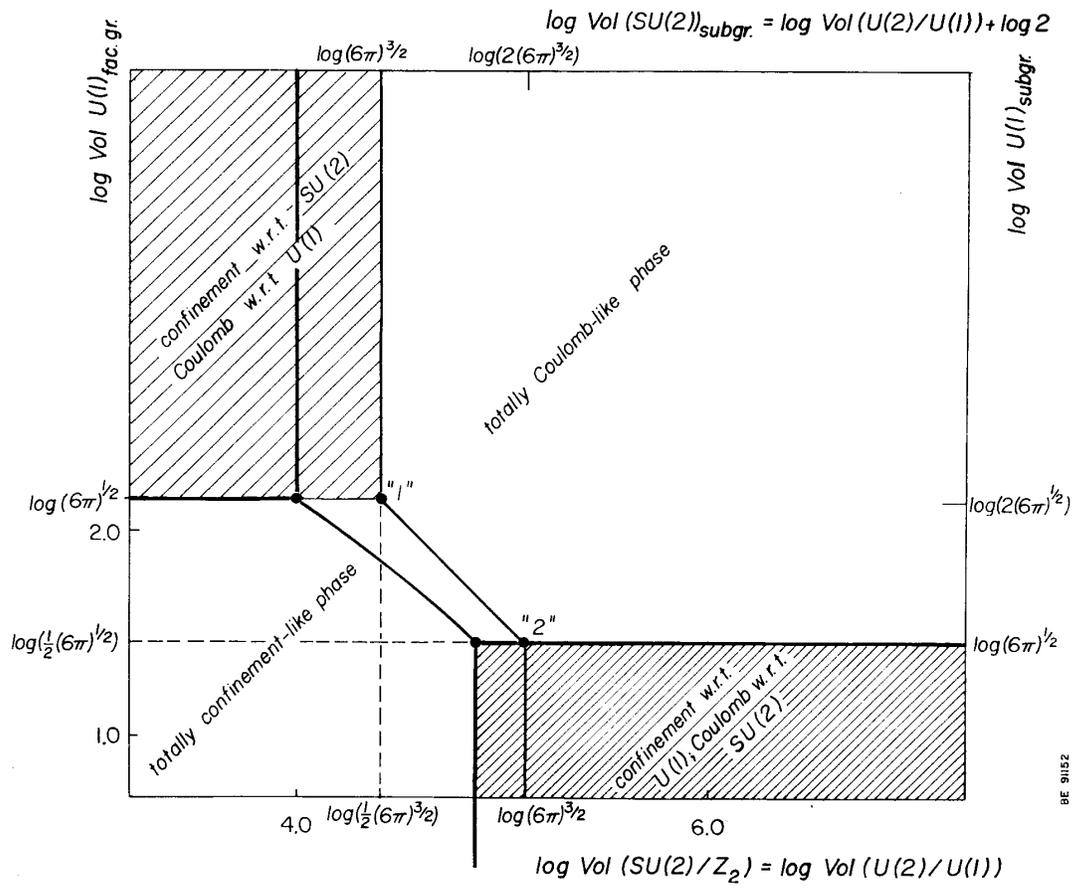}}
\caption[figsun]{\label{sun} Phase  diagram  for   $U(N)$.}
\end{figure}
As has been pointed out, the centre of $U(N)$ contains a $\bz_N$ that
is shared by both of the invariant subgroups $SU(N)$ and $U(1)$. Near
``1'' but in the phase with only $SU(N)$ confining, the (Coulomb-like)$U(1)$
\dofx are realized as the cosets $g\cdot SU(N)\in U(N)/SU(N)$ ($g\in U(N)$).
The picture one
can have in mind is that the distribution of $U(1)$ \dofx is tightly
clustered about the coset
$\bunit\cdot SU(N)$ whereas the confining $SU(N)$ \dofx
fluctuate (within the the cosets) somewhat more. Effectively, the large
fluctuations along these cosets\footnote{Recall that the defining feature
of a confining phase is that the
correlations between plaquette variables introduced by Bianchi identities
can be neglected for confinement-like \dofx (as opposed to the Coulomb-like
phase for which such correlations are assumed to be important).
But Bianchi identities show up as constraints on closed 3-volumes. The simplest
``Bianchi variable'' is therefore the cube enclosed by 6-plaquettes. If we
pretend that the Bianchi identity constraint is absent, then
distribution of such variables would essentially be the 6-fold convolution of
the distribution of plaquette variables. The criterion for whether or not
Bianchi identities are important is as follows: calculate the distribution
of the 6-fold convolution of the distribution of a plaquette variable
{\em without} regard to Bianchi identities; if this distribution is
essentially flat, we take this as the indication that the neglect of Bianchi
identities is justified.
The situation in which Bianchi
identities are effectively absent corresponds to distributions of plaquette
variables that are (at least approximately) independent of each other.
Note that  even a distribution
of plaquette variables for which a weak coupling approximation is not
meaningless can lead to a distribution of cube variables that
is essentially flat. In this
case, Bianchi identities are (by definition) not important and the
corresponding \dofx are (again by definition) confined.}
$G\cdot SU(N)$ make the elements within these cosets
equivalent to the group identity as far as the $U(1)$ \dofx are concerned.
This applies also to the elements of the $\bz_N$ subgroup of the centre
that is shared with the $U(1)$: at the point ``1'', large fluctuations
along $U(N)$ render the elements of the $Z_N\subset U(N)$  essentially
equivalent to the identity. But this same $Z_N$ is shared with $U(1)$:
$Z_N\subset U(1)$. Hence the Coulomb-like
$U(1)$ \dofx are realized on the manifold (consisting of the cosets
$u \cdot\bz_N \in U(1)/\bz_2\simeq U(N)/SU(N)$) $(u\in U(1)$) having
a volume that is
reduced by a factor $\# \bz_N\stackrel{def}{=}N$ relative to the volume
of the manifold of the $U(1)$ subgroup.

Near the point ``2'' (but in the phase with confinement along $U(1)$)
in the phase diagram for $U(N)$, the roles of
$U(1)$ and $SU(N)$ are reversed: the larger fluctuations
are within the cosets of the  confinement-like $U(1)$ \dofx and the
Coulomb-like $SU(2)$ \dofx are realized as the cosets
$g\cdot U(1)\in SU(N)/\bz_2 \simeq U(N)/U(1)$ $(g\in U(N))$.

The coordinates of corner ``1'' are

\beq       (\log Vol(SU(N))_{subgr.}),\log
Vol((U(N)/SU(N))_{fac.\;gr.}))=(\log(\sqrt{6\pi})^{N^2-1},\log\sqrt{6\pi})\eeq

\nin and the coordinates of corner ``2'' are

\beq           (\log           Vol((U(N)/U(1))_{fac.\;gr.}),\log
Vol((U(1))_{subgr.})=(\log(\sqrt{6\pi})^{N^2-1},\log\sqrt{6\pi}).\eeq

To plot the coordinates of the two corners ``1'' and ``2'' in  a  space
spanned by axes corresponding to the variables $\log Vol(H_i)$ $(i \in
\{U(1),SU(N) \})$, we shall work with two  sets  of  labels  for  each
axis: one set for subgroups  and  another  set  for  the  Lie  algebra
identical factor group. These are simply  related:  the  set  of  axis
labels for factor groups is shifted relative to the  axis  labels  for
subgroups by an amount $\log N$ in the direction of  increasing  $\log
Vol(H_i)$. This simply reflects the fact that for a factor group,  the
quantity $Vol_{fac.\;gr.}$ attains the same  numerical  value  as  the
quantity $Vol_{subgr.}$ for the Lie  algebra-identical  subgroup  only
after $\beta_i$ is  ``scaled  up''  by  a  factor  $N^{2/d_i}$  (where
$d_i=dim(H_i)$ with $H_i \in \{ U(1), SU(N) \})$.

To be in the partially confining phase of $U(N)$ that is
Coulomb-like solely w.r.t. $U(1)$, one of the requirements is that
$\log Vol((SU(N))_{subgr})\leq (6\pi)^{\frac{N^2-1}{2}}$. To be in the phase
that is confining solely w.r.t. $U(1)$, one of the requirements is that
$\log Vol((U(N)/U(1))_{fac\;gr})\geq (6\pi)^{\frac{N^2-1}{2}}$.
But it has been seen that when $Vol(SU(N))=Vol(U(N)/U(1))$, the $\beta_{SU(N)}$
absorbed into $Vol(U(N)/U(1))$ is a factor $N^{2/dimSU(N)}$ larger than
than the $\beta_{SU(N)}$ absorbed into $Vol(SU(N))$.
So there is an interval of length $\log  N$  extending  from  $\log
Vol(SU(N))_{subgr.}=\log(6\pi)^{dim\;SU(N)/2}$ to $\log
Vol(U(N)/U(1))_{factor\; gr.}=\log (6\pi)^{dim SU(N)/2}$ in which
it is only possible to realize the totally confining or totally Coulomb-like
phases of $U(N)$.  The same is seen to be true for
the   interval   of   length   $\log   N$   extending    from    $\log
Vol(U(1))_{subgr.}=\log(\sqrt{6\pi})$      to      $\log
Vol(U(N)/SU(N))_{factor gr.}=\log (\sqrt{6\pi})$.

The straight line connecting the points ``1'' and ``2'' in the  phase
diagram of Figure~\ref{sun} is the phase interface
separating the  total  confinement  and
total Coulomb phases of the theory. To see this,  write  the  equation
for this line:

\beq  \log   Vol((U(N)/SU(N))_{fac.\;gr.})=\log   \sqrt{6\pi}-   (\log
Vol(SU(N))_{subgr.}-\log ((\sqrt{6\pi})^{N^2-1})). \eeq

\nin Rearranging yields

\beq
\log(6\pi)^{N^2/2}=\log(Vol(SU(N))_{subgr.}Vol((U(N)/SU(N))_{fac.\;gr.})=
\log VolU(N) \eeq

\nin which from Eqn. (\ref{bc}) with  $H_I=U(N)$  and  $H_J={\bunit}$
is seen to be the condition  to  be  satisfied  at  the  interface
separating the totally Coulomb and totally confining phases.

In other words, starting at the point ``1'' in Figure~\ref{sun}
(where $U(1)$ alone is Coulomb), the other end  of  the  phase  boundary
separating the totally confining and the totally  Coulomb-like phases  (point
``2'' where  alone  $SU(N)$  is  Coulomb in  Figure~\ref{sun})  is
attained by going $-\log N$ along the $\log Vol(U(1))$ axis and $+\log
N$ along the $\log Vol(SU(N))$ axis.

{\bf Phase Diagram for the Standard Model Group}\\

With the relationship between $U(1) \times SU(N)$ and $U(N)$  and  the
corresponding phase diagrams in mind, the  slightly  more  complicated
phase diagram for the $SMG$ can essentially be constructed by  analogy
insofar as the $SMG$ and $U(1) \times SU(2)\times SU(3)$  are  related
in a way much like  the  relation  between  $U(N)$  and  $U(1)  \times
SU(N)$.

For the $SMG$, the corner of the partially confining $SMG$ phase where
$U(1)$ alone is Coulomb (call it  point  ``1''  as  it  is  the  point
analogous to point ``1'' in Figure~\ref{sun}) and  the  corner  of  the
partially confining phase where alone $U(1)$  is  confining  (call  it
point ``2'' as it is the point analogous to point ``2'' in  Figure~\ref{sun})
have coordinates in a 2-dimensional phase diagram spanned by
$\log Vol  ((SU(2)  \times  SU(3))$ and $ \log
Vol(U(1))$ (containing the shaded planes of Figure~\ref{smgu1}) that are given
by
respectively
\begin{figure}
\centerline{\epsfxsize=\textwidth \epsfbox{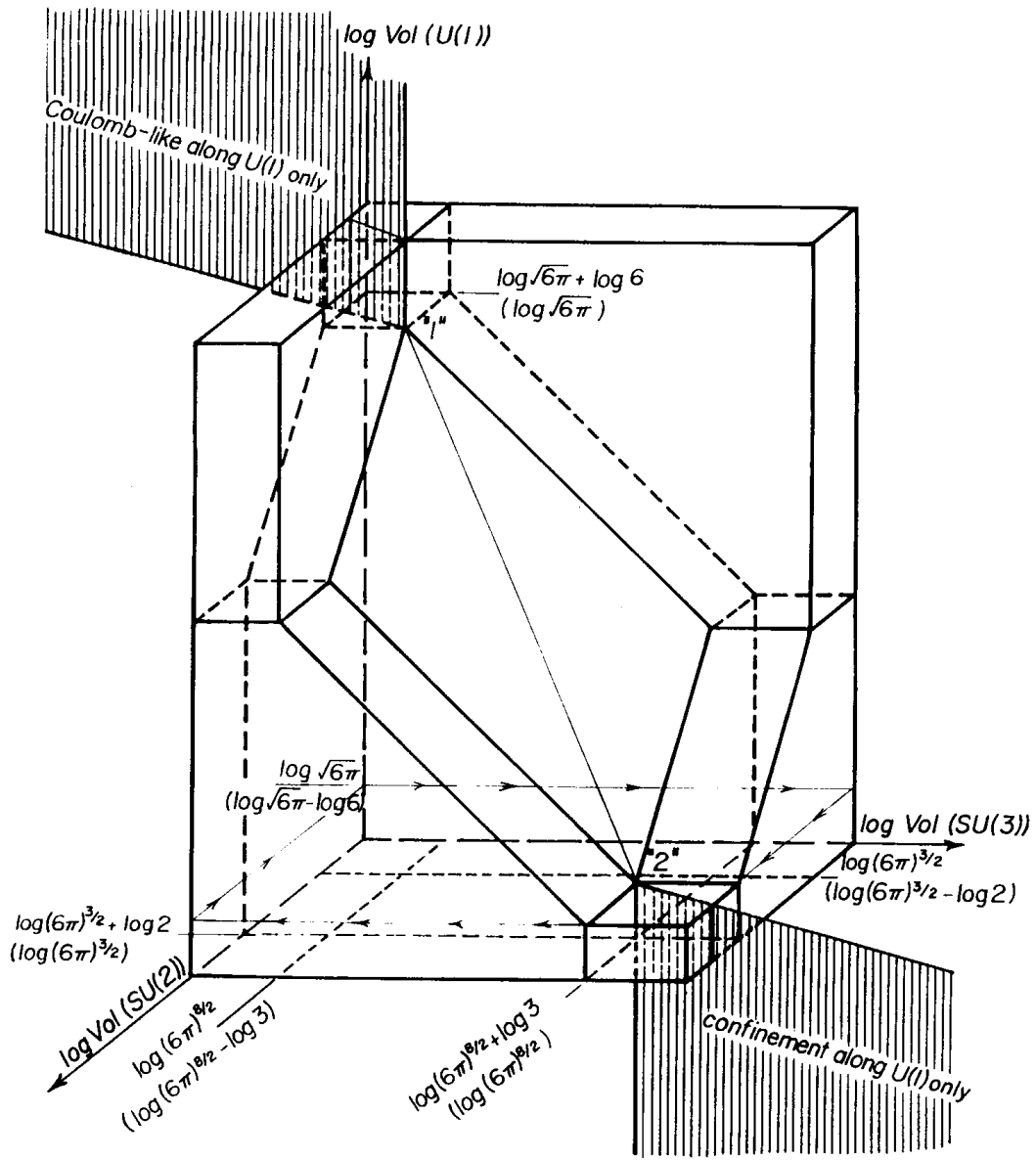}}
\caption[figsmgu1]{\label{smgu1}Construction of the phase  diagram  for  the
Standard
Model Group to lowest order.}
\end{figure}

\begin{equation}  (   \log   Vol(SU(2)\times   SU(3))_{subgr.},   \log
Vol(SMG/(SU(2)\times    SU(3))_{factor\;     gr.})=\end{equation}

$$(\log(\sqrt{6\pi})^{3+8}),\log( \sqrt{6\pi}))$$

\noindent at point ``1'' and

\begin{equation}  (   \log   Vol(SMG/U(1))_{factor   \;   gr.},   \log
Vol(U(1))_{subgr.})=\end{equation}

$$=(\log((\sqrt{6\pi})^{3+8}), \log( \sqrt{6\pi}))$$

\noindent at point ``2''. The (somewhat redundant) subscripts  $``subgr.$''
and  $``fac.\; gr.$'' indicate explicitly which set of
labels on  a  $\log  Vol$
axis  that  the
numbers on the right-hand side of the equalities refer to. Starting  at
point ``1'' in Figure~\ref{smgu1}, point ``2''  is  reached  by  decreasing
$\log Vol((SMG/(SU(2) \times SU(3)))_{factor\; gr.}$ by $\log  6$  and
increasing $\log Vol_{crit.}((SU(2) \times SU(3))_{subgr.}$  by  $\log
6$.

While point ``1'' and ``2'' are the termini of a line  separating  the
total confinement and total Coulomb phase, this is  not  the  complete
story for the $SMG$. Recall that
in the $U(N)$ phase
diagram, {\em all} the elements of the $\bz_N$ of centre elements
shared by $U(1)$ and $SU(N)$ are shared in such a way
that, in going from ``1'' to ``2'' (or vice versa)  there is a redistribution
of fluctuations along the cosets of just two possible coset structures
(recall that is this section possibility of having $\bz_N$ confined is
ignored).
At point ``1'' there
are large fluctuations along the cosets $g\cdot SU(N) \in U(N)/SU(N)$
($g\in U(N)$) while the coset-valued $U(1)$ \dofx are rather tightly
clustered about the coset $\bunit \cdot SU(N)$. At point ``2'', fluctuations
have become large along the cosets $g\cdot U(1)\in U(N)/U(1)$ $g\in U(N)$
while the Coulomb-like $SU(N)$ \dofx are tightly clustered about the coset
$\bunit \cdot U(1)$. For $U(N)$, there are only two sets of cosets;
fluctuations are large along one or the other of these sets at the points
``1'' or ``2''. The volume reduction factor $N=\#\bz_N$ for $U(N)$ relative to
$U(1)\times SU(N)$
is bourn solely by the Coulomb-like $U(1)$
\dofx at ``1'' and solely by the Coulomb-like $SU(N)$ \dofx at ``2''.

Unlike the case for $U(N)$, the $SMG$  has  among
the  possible  invariant  subgroups  two  that  have  centre  elements
identified relative to the corresponding  subgroups  in  $U(1)  \times
SU(2) \times SU(3)$:
these are $U(2) \overline{\simeq} (U(1)\times SU(2))/\bz_2$  and
$U(3)\isomorph (U(1)\times SU(3))/\bz_3$. We want now to think about
going from ``1'' to ``2'' by way of ``3'' in the $SMG$ phase diagram.
Now in our $SMG\stackrel{def}{=} S(U(2)\times U(3))$ we know that
the elements of ``$\bz_6$'' are identified (which means that
they are rendered indistinguishable because of fluctuation patterns of
one sort or another). Now at point ``1'', the fluctuation pattern
that causes the identification of the elements of $\bz_6$  is that of
large fluctuations along $SU(2)$ and $SU(3)$. In going from point ``1''
towards point ``3'', fluctuations along $SU(3)$ decrease and fluctuations
along $U(1)$ increase until, at ``3'', the $\bz_3 \in ``\bz_6"$ are
identified due to fluctuations along $U(1)$ {\em instead} of fluctuations
along $SU(3)$ while the increased fluctuations along $U(1)$ means that
the $\bz_2\in ``\bz_6"$ are now accessed by fluctuations {\em both}
along $U(1)$ and $SU(2)$. But the group for which the elements of a $\bz_2$
can be accessed either by going along a $U(1)$ subgroup or a $SU(2)$ subgroup
is just $U(2)$ which in fact coincides with the confined \dofx at point ``3''
(see Figure~\ref{smgfull}). In going from point ``3'' to point ``2'',
fluctuations along
$U(1)$ increase even more while now fluctuations along $SU(2)$ decrease (and
fluctuations along $SU(3)$ remain small). Upon reaching point ``2'', all
elements of ``$\bz_6$'' are accessed exclusively by fluctuations along
$U(1)$ corresponding to $U(1)$ alone confining at point ``2''.
\begin{figure}
\centerline{\epsfxsize=\textwidth \epsfbox{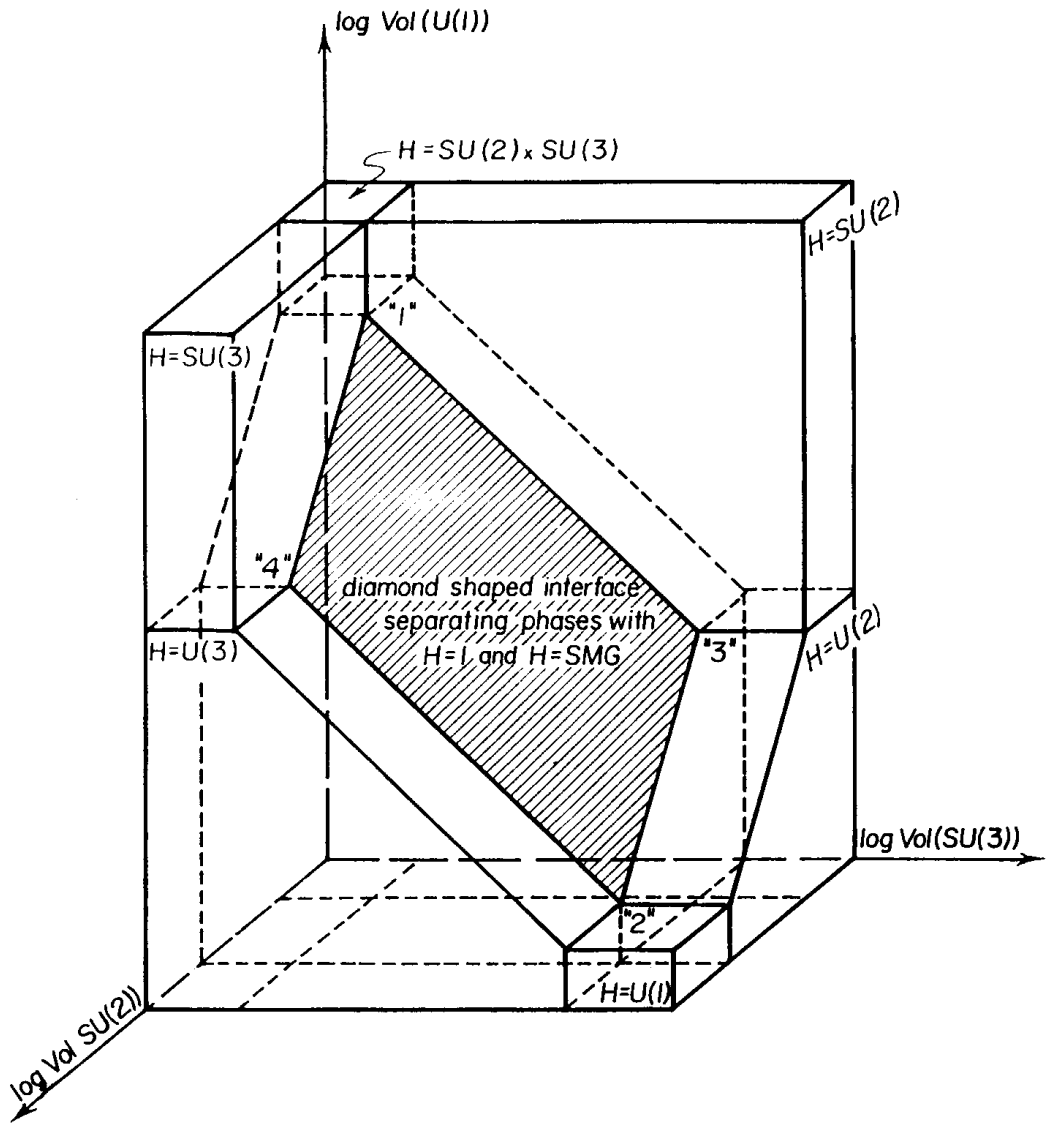}}
\caption[figsmgfull]{\label{smgfull}The lowest order phase
diagram for the $SMG$ showing the 8 partially confining phases and the
``diamond'' interface separating the  totally  confining  and  totally
Coulomb-like phase.}
\end{figure}
Having
$U(2)$ and $U(3)$ allows the volume factor of six bourn by the $U(1)$
Coulomb-like \dofx at ``1'' to be redistributed to Coulomb-like $SU(2)$ and
$SU(3)$
\dofx at respectively corner ``3'' and ``4'' (see Figure~\ref{smgfull}).
At the intermediate point ``3''
en route from ``1'' to ``2'', the volume reduction factor bourn by the
$U(1)$ \dofx is reduced from six at ``1'' to three at ``3'';
at point ``3'' the remaining fac two of the
total volume reduction factor of six is bourn by the
$U(2)=(U(1)\times SU(N))/\bz_N$ subgroup
of confined \dof. An analogous scenario involves a point ``4'' in
Figure~\ref{smgfull} where
the Coulomb-like $U(1)$ \dofx bear a volume reduction factor of 2 while the
remaining factor of 3 is bourn by the confining $U(3)$ \dof.

In the $MFA$ weak coupling approximation, having $U(2)$ and $U(3)$  as
invariant subgroups in the  $SMG$  has  the  effect  of  changing  the
{\em phase boundary line} separating the
total confinement and total Coulomb phases  into  a  {\em phase
boundary plane} in the  space  spanned  by  the
$\log Vol_i$. Here $i$ is any one of  the  three  ``basic''  invariant
subsets of the $SMG$. The reason as suggested above
is that the presence  of  the  $U(2)$
and $U(3)$ subgroups allows two alternatives to  the  straight
line route connecting the corner of the partially confining phase with
$U(1)$ alone Coulomb with the corner of the partially confining  phase
with $U(1)$ alone in confinement. These two alternative routes  define
the boundaries of the $SMG$ phase boundary plane separating the  total
confinement and total Coulomb  phases.  Starting  at  point  ``1''  in
Figure~\ref{smgu1}, one alternative route to point ``2''  is  that
for which $\log Vol(SU(2)_{subgr.})$ is held  at  the  constant  value
$(\sqrt{6\pi})^3$  until  point  ``3''  is   reached   by   decreasing
respectively     increasing     the     $\log     Vol((SMG/(SU(2)\times
SU(3))_{factor\; gr.})$ and $\log Vol(SU(3)_{subgr.})$ coordinates  of
point ``1'' by  the  amount  $\log  3$.  At  point  ``3'',  $\log  Vol
((SMG/U(2))_{factor  \;  gr.})$  is  held  at   the   constant   value
$(\sqrt{6\pi})^8+\log 3$ while decreasing respectively  increasing  the
coordinates $\log Vol((SMG/(SU(2)\times SU(3)))_{factor\;gr.})-\log
3$ and $\log Vol(SU(2)_{subgr.})$ by the amount $\log 2$. This  brings
us to point ``2''. Exchanging the roles of $SU(2)$ and $SU(3)$  yields
the second alternative route to point ``2'' via point  ``4''  .  These
two alternative routes via point ``3''  and  point  ``4''  define  the
boundaries  of  the  diamond-shaped  interface  separating  the  total
confinement and total Coulomb phases as shown in Figure~\ref{smgu1}.

\subsubsection{Next order in perturbation}

The weak coupling approximation used above is really the approximation
in which the fluctuations around the unit element are  so  small  that
one can approximate the group by the Lie algebra  (the  tangent  plane
approximation). For the  invariant  subgroup(s)  in
the confining phase, the calculations  above  utilise  only  that  the
fluctuation of the plaquette variable is small enough for the validity
of this approximation whereas it is used for both the  link  and  the
plaquette variables for the Coulomb phase degrees  of  freedom.
For the same values of the $\beta_i$ 's, the fluctuations  are  larger
for the confinement phase than  for  the  Coulomb  phase.
Therefore it is to be anticipated that the numerically most  important
correction will come from the next order correction estimate for  the
confining degrees of freedom.


Next order corrections come about by using a corrected
Haar measure that reflects the curvature of the group manifold and a corrected
rule of composition of Lie algebra vectors for non-Abelian groups when group
multiplication is referred to the Lie algebra. In  the  tangent  space
approximation for the group manifolds with curvature  that  were  used
in  calculations  to  leading  order,  the  rule  of
composition is approximated by simple vector addition of  Lie  algebra
elements. This approximation neglects the non-commutativity of the
non-Abelian group generators.

When both corrections - i.e., the correction for the Haar measure  and
the  correction   for   non-commutativity
                         - are included, the  defining  condition  for
the phase boundary separating the totally Coulomb and totally confined
phases becomes

\beq             \log             Vol(SMG)=\log              (6\pi)^6-
\underbrace{\frac{1}{2\beta_2}-\frac{2}{\beta_3}}_{\mbox{Haar
measure}}+
\underbrace{\frac{1}{24\beta_2}+\frac{1}{6\beta_3}}_{\mbox{non-commutativity}}
\eeq

The leading order terms of this equation, i.e.,
$Vol(SMG)=\log (6\pi)^6$ are identical with the condition
(\ref{bc}) in the special case where $H_J=SMG$ and $H_I=\bunit$.
The details of these  next to lowest order corrections  are dealt with in
Appendices~\ref{apphaar}, \ref{appcorr} and \ref{appnoncom}.

\subsection{Using discrete subgroups of the centre to get
a multiple point}\label{usingdiscrete}

In the $SMG$ phase diagram of Figure~\ref{smgfull}, it is seen that the 8 \pcps
that can be
realized in a parameter space spanned by parameters proportional to
$\beta_1$, $\beta_2$, and $\beta_3$ do not convene at a multiple
point/surface. When the set of constituent invariant subgroups is restricted
to the set $\{SU(3), SU(2), U(1)\}$, the factorisation property (\ref{facprop})
cannot be realized for all of the first 8 invariant subgroups of
(\ref{invh}).

This factorisation property can be realized for all the  13  invariant
subgroups $H \lhd SMG$ (or all the 5 invariant subgroups $H\lhd U(N)$)
using the previously defined set of constituent  invariant  subgroups.
That is, it is possible to factorise $Vol(H)$ into a product  of  some
subset of a common set of 5 factors  $(1/p_i)Vol(K_i)$  (with  $p_i\in
N^+$) (3 such factors for $U(N)$)  corresponding  to  the  constituent
invariant subgroups $K_i \in  \{\bz_2,\bz_3,U(1),SU(2),SU(3)  \}$  for
the $SMG$ (for $U(N)$, the constituent invariant subgroups are $K_i\in
\{SU(N),U(1),\bz_N \}$). Then it is  possible  by  adjustment  of  the
parameters   $\beta_1,\beta_2,\beta_3,\xi_2,\xi_3$   to    make    the
quantities $\log Z_{H\;per\;active\;link}$ equal  for  each  invariant
subgroup $H\lhd SMG$ (the same  applies  to  each  invariant  subgroup
$H\lhd U(N)$ using the parameters  $\beta_1,\beta_N,\xi_N$).  This  is
equivalent to finding  a  {\em  non-generic}  multiple  point  (because
$5<n_{SMG}-1=13-1$ for the $SMG$ and $3<n_{U(N)}-1=5-1$ for $U(N)$).

We explicitly demonstrate the factorisability of $Vol(H)$ in the sense
that we show that it is of  the  form  $Vol(H)=\mbox{product  of  some
factors} (1/p_i)Vol(K_i)$ for each  invariant  subgroup  $H$  of  both
$SMG$ and $U(N)$. To do this, we use a calculational trick in which we
replace each $H$ by a Cartesian product group  related  to  $H$  by  a
homomorphism  that  is  locally  bijective.  This  (to  $H$)   locally
isomorphic Cartesian  product  group  consists  of  the  covering  Lie
(sub)groups  corresponding  to  the  gauge  \dofx  that  the  invariant
subgroup  $H$  involves  supplemented  by  the  discrete   constituent
invariant subgroups contained in these  Lie  subgroups.  For  all  the
invariant subgroups $H$, the Cartesian product group  replacement  can
be obtained by simply omitting factors in the Cartesian product  group
replacement $\bz_2\times \bz_3\times U(1)\times SU(2)\times SU(3)$ for
the whole $SMG$. Of course such a Cartesian product group  in  general
differs in global structure from the invariant subgroup  $H$  that  it
replaces. However, as we are only interested in the quantity  $Vol(H)$
for invariant subgroups $H$, we can use a  correction  factor\footnote
{We define the quantity $p_{_{H}}=\#(H\cap D)$ where  $D$  (which  has
$\#(D)=36$) is the discrete  subgroup  of  the  centre  that  must  be
divided out of the Cartesian product  group  $\bz_2\times  \bz_3\times
U(1)\times SU(2)\times  SU(3)$  in  order  to  get  the  $SMG$;  i.e.,
$((\bz_2\times   \bz_3\times    U(1)\times    SU(2)\times    SU(3))/D)
\overline{\simeq}SMG  \stackrel{def.}{=}  S(U(2)\times   U(3))$.}
$1/p_{_{H}}$ to adjust the quantity $Vol$ of the Cartesian  product
group replacement for $H$ so as to make it equal to $Vol(H)$.

As an example, consider the  invariant  subgroup  $H=U(2)\subset  SMG$
which is locally isomorphic to the Cartesian product group $U(1)\times
SU(2)\times \bz_2\times \bz_3$. By this we  mean  that,  assuming  the
modified  Manton   action   (\ref{modman})   and   a   weak   coupling
approximation, the Cartesian  product  group  $U(1)\times  SU(2)\times
\bz_2 \times \bz_3$ simulates the subgroup $U(2) \subset SMG$  in  the
sense that the regions on the group manifold of $U(2)\subset  SMG$  in
which the probability distribution $e^{S_{\Box}}$ is concentrated  can
be  brought  into  a  one  to  one  correspondence  with  centres   of
fluctuation sharply peaked around  points  in  the  Cartesian  product
group.  In  other  words,  for  $U(2)\subset  SMG$,   the   region   of
correspondence with the Cartesian product group is the composite of  6
small      neighbourhoods      around      the      elements      $p\in
\mbox{span}\{\bz_2,\bz_3\}$. Even though the Cartesian  product  group
in this example contains $2\cdot 2\cdot 3$ elements for  each  element
in $U(2)$, the action on the Cartesian product group is defined so  as
to be $-\infty$ everywhere except at one  of  the  these  12  elements
where  this  action  then  has  the  same  value  as  the  action   at
corresponding element of $U(2)$.

In order to make the quantity $Vol(U(1)\times SU(2)\times  \bz_2\times
\bz_3)$ equal to $Vol(U(2))$ (for $U(2)\subset SMG$) , the former must
be reduced by a factor $p_{U(2)}$ obtained as follows:  Remember  that
the $U(1)$ embedded in the $SMG$ has a length $6\cdot  2\pi$  so  that
the  $U(2)$  subgroup  lying  in  the  $SMG$  is  $(U(1)_{12\pi}\times
SU(2))/\bz_2$.      Comparing      the      quantity       $Vol(U(2))=
Vol(U(1)_{12\pi})\cdot Vol(SU(2))/\bz_2$ and the  quantity  $Vol$  for
the locally isomorphic  Cartesian  product  group:  $Vol(U(1)_{12\pi})
\times Vol(SU(2))\times  Vol(\bz_2)\times  Vol(\bz_3))$,  it  is  seen
that, relative to $Vol$ for the Cartesian product group, the  quantity
$Vol(U(2))$ is down by $(\#\bz_2)\cdot (\#\bz_2)\cdot (\#\bz_3)=2\cdot
2\cdot 3=12\stackrel{def}{=}p_{U(2)}$.

\begin{table}
\caption[tabnon1]{\label{non1} The quantity $Vol(H)$ for  any one of the 13
invariant subgroups $H$  of
the $SMG$ (these are listed in left  column),  can  be  written  as  a
product of some subset of the set of five  quantities  $(1/p_i)Vol(K_i)$.
The common set of factors  $1/p_{_{\tbz_2}}=1$,  $1/p_{_{\tbz_3}}=1$,
$1/p_{_{U(1)}}=1/6$,
$1/p_{_{SU(2)}}=1/2$, and $1/p_{_{SU(3)}}=1/3$, some subset of which  make
possible the factorisation of all the $Vol(H)$  into  the  product  of
corresponding   subsets   of   the   quantities
$(1/p_{_{\tbz_2}})Vol(\bz_2)$,
$(1/p_{_{\tbz_3}})Vol(\bz_3)$,  $(1/p_{_{U(1)}})Vol(U(1))$,
$(1/p_{_{SU(2)}})Vol(SU(2))$,
and $(1/p_{_{SU(3)}})Vol(SU(3))$, are given in the last five columns.}

\begin{small}

\[ \begin{array}{llc||ccccc}
Vol(H) & locally\;isomorph.\;cart.\;prod.\;gr. & \frac{1}{p_{_H}} &
\frac{1}{p_{_{\tbz_2}}} & \frac{1}{p_{_{\tbz_3}}}
& \frac{1}{p_{_{U(1)}}} & \frac{1}{p_{_{SU(2)}}} & \frac{1}{p_{_{SU(3)}}} \\ \\
Vol(\bunit) & & 1 &  &   &  &  &  \\
Vol(\bz_2) & Vol(\bz_2) & 1  & 1 &  &  &  &  \\
Vol(\bz_3) & Vol(\bz_3) & 1 &  & 1  & & &  \\
Vol(\bz_2 \times \bz_3) & Vol(\bz_2 \times \bz_3) & 1 & 1 & 1 &  &  &  \\
Vol(SU(2)) & Vol(SU(2)\times \bz_2) & 1/2 & 1 & &  & 1/2  &  \\
Vol(SU(3)) & Vol(SU(3)\times \bz_3) & 1/3 & & 1 & &  & 1/3 \\
Vol(SU(2)\times  \bz_3) & Vol(SU(2)\times \bz_2\times  \bz_3) & 1/2 & 1 & 1 &
& 1/2 &  \\
Vol(SU(3)\times \bz_2) & Vol(SU(3)\times \bz_3 \times \bz_2) & 1/3 & 1 & 1 &
& & 1/3  \\
Vol(U(1)) & Vol(U(1)\times \bz_2\times \bz_3) & 1/6 & 1 & 1 & 1/6 &  & \\
Vol(SU(2)\times SU(3)) & Vol(SU(2)\times SU(3)\times \bz_2\times \bz_3) & 1/6 &
1 & 1 &  & 1/2 & 1/3  \\
Vol(U(3)) & Vol(U(1)\times SU(3)\times \bz_2\times \bz_3) & 1/18 & 1 & 1 & 1/6
& & 1/3 \\
Vol(U(2)) & Vol(U(1)\times SU(2)\times \bz_2\times \bz_3) & 1/12 & 1 & 1 & 1/6
& 1/2 &  \\
Vol(SMG) & Vol(U(1)\times SU(2)\times SU(3)\times \bz_2\times \bz_3) & 1/36  &
1 & 1 & 1/6 & 1/2 & 1/3
\end{array} \]
\end{small}
\end{table}
\vspace{.5cm}

\clearpage

\begin{table}
\caption[tabnon2]{\label{non2} In a  manner  analogous  to  that
of  Table~\ref{non1}, the quantities $Vol(H)$ for the  5
invariant subgroups $H$ of $U(N)$ (listed in left column)  factorise  into
products of subsets of the constituent  quantities  $(1/p_i)Vol(K_i)$.
The coefficient $1/p_i$ of any corresponding $Vol(K_i)$ is, as seen in
the last three columns, the same for all the $Vol(H)$ in which such  a
$Vol(K_i)$ contributes in the factorisation of $Vol(H)$.
Figures~\ref{mpphdiag1} and \ref{mpphdiag2}, which depict the phase
diagram for $U(2)$, illustrate
how the 5 \pcps of a $U(N)$ group meet at the multiple point in our
approximation.}
\[ \begin{array}{llc||ccc}
Vol(H)& locally\;isomorphic\; cart.\;prod.\;gr. & \frac{1}{p_{_H}} &
\frac{1}{p_{_{\tbz_N}}} & \frac{1}{p_{_{U(1)}}} & \frac{1}{p_{_{SU(N)}}} \\ \\
Vol(\bunit) & & 1 &  & & \\
Vol(\bz_N) & Vol(\bz_N) & 1 & 1 & & \\
Vol(U(1)) & Vol(U(1) \times \bz_N)  & 1/N & 1 & 1/N &  \\
Vol(SU(N)) & Vol(SU(N)\times \bz_N) & 1/N & 1 &  & 1/N \\
Vol(U(N)) & Vol(U(1)\times SU(N)\times \bz_N) & 1/N^2 & 1 & 1/N & 1/N
\end{array} \]
\end{table}

In Table~\ref{non1}, we  demonstrate  explicitly  that  the  volume  correction
factors $1/p_H$ for all the invariant subgroups  $H\lhd  SMG$  can  be
factored into a subset of 5 factors $1/p_i$ associated  with  each  of
the     ``constituent''     invariant     subgroups      $K_i      \in
\{\bz_2,\bz_3,U(1),SU(2),SU(3)\}$. For a given $i$,  $p_i$  is  always
the same in any $p_H$ in which $p_i$ is a factor. Listed in the  first
column of Table~\ref{non1} are the quantities $Vol(H)$  for  all  13  invariant
subgroups $H$ of the $SMG$;  listed  in  the  second  column  are  the
quantities $Vol$ for the corresponding, locally  isomorphic  Cartesian
product groups. The third column consists  of  the  volume  correction
factors $1/p_H$ by  which  the  quantities  $Vol$  for  the  Cartesian
product group in the second column must be multiplied in order
to get the corresponding quantity $Vol(H)$ in the first column. In the
next five columns, we give the factorisation of the correction factors
$1/p_H$ into subsets of five rational quantities $1/p_i$ with  $i  \in
\{\bz_2,\bz_3,U(1),SU(2),SU(3)\}$ that are associated  with  the  five
quantities $Vol(\bz_2)$, $Vol(\bz_3)$, $Vol(U(1))$, $Vol(SU(2))$,  and
$Vol(SU(3))$. Table~\ref{non2} is constructed in an analogous fashion
for the 5
invariant subgroups of $U(N)$ using $\bz_N,\; U(1)$,  and  $SU(N)$  as
the constituent invariant subgroups. For both the  $SMG$  and  $U(N)$,
the important point is that,  for  any  invariant  subgroup  $H$,  the
factorisation

\beq Vol(H)=\prod_i(\frac{Vol(K_i)}{p_i}) \;\;\;  (\mbox{\footnotesize
$i$ runs over a  subset of constituent  invariant  subgroups})
\label{volk} \eeq

\nin is such that the correction factor $1/p_i$
corresponding to a  given  constituent  invariant  subgroup  $K_i$  is
always  the  same  (unless  the  quantity  $Vol(K_i)$  is  absent   in
the product (\ref{volk}) in which  case  there  is  no  entry  in  the
column headed by $1/p_i$) in Tables~\ref{non1}~and~\ref{non2}.

The meeting of 13 \pcps at the (non-generic) multiple point in
the phase diagram for the $SMG$ in the 5-dimensional action  parameter
space is virtually impossible to depict clearly in a figure.  However,
the group $U(N)$, which has many features in common with the $SMG$,  has
a phase diagram with a non-generic multiple point in 3  dimensions
when we use an action ansatz analogous to that  used  for  the  $SMG$:
Gaussian peaks at elements of $\bz_N\subset U(N)$.
The  phase  diagram for $U(2)$ seen in
Figures~\ref{mpphdiag1}~and~\ref{mpphdiag2}  shows,
in our approximation, the
5 \pcps (corresponding to
the 5 invariant subgroups of a $U(N)$ group)
that meet at the multiple point.

\begin{figure}
\centerline{\epsfxsize=\textwidth \epsfbox{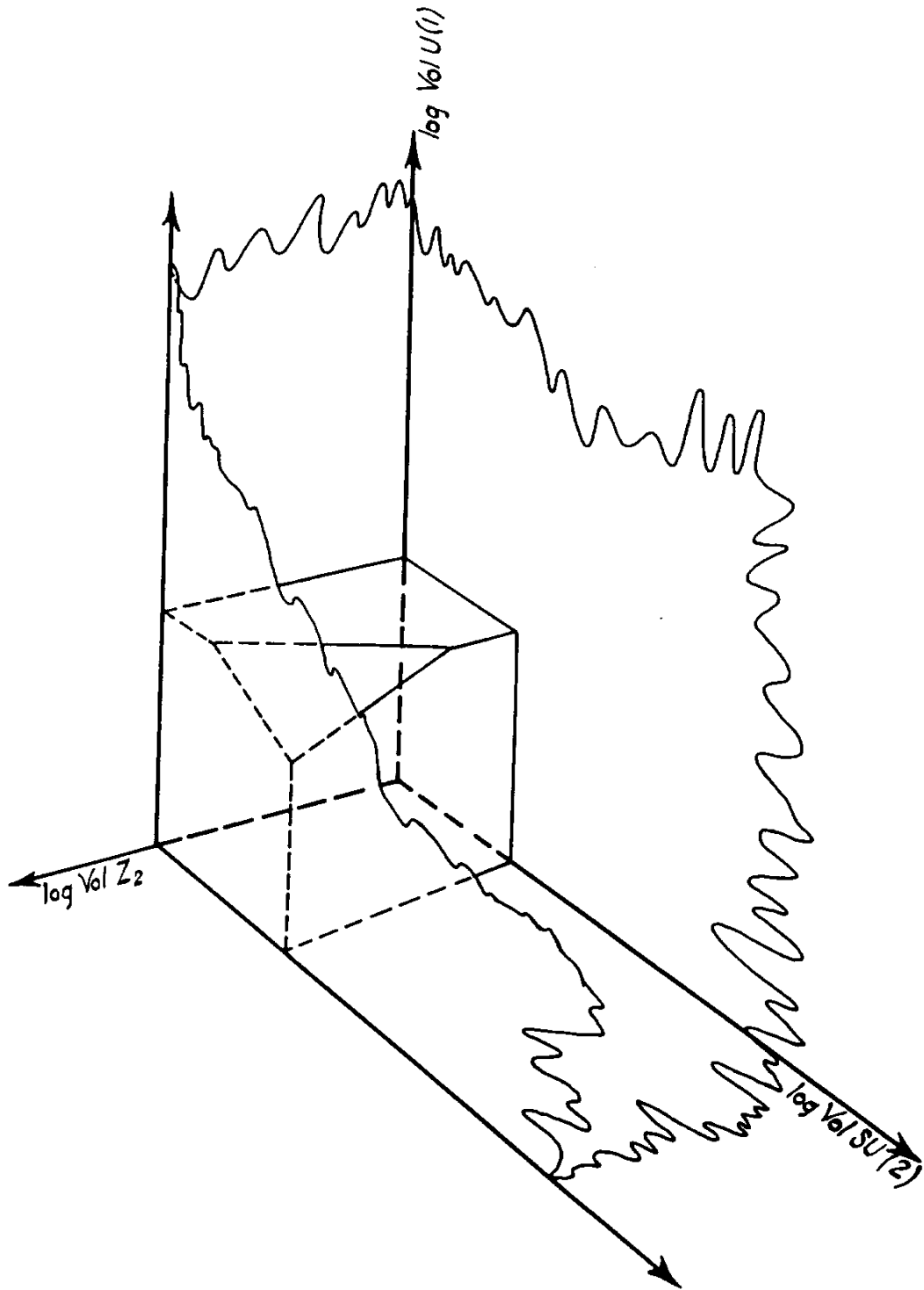}}
\caption[figmpphdiag1]{\label{mpphdiag1} For the gauge group $U(2)$,
this figure shows the region of allowed parameters
($\log Vol(SU(2)),\log Vol(U(1)),
\log Vol(\bz_2)$) for the modified Manton action: $\log(\pi e)^{3/2}\leq
\log Vol(SU(2)) \;(\approx \frac{3}{2}\log\beta_2+\log vol(SU(2)))< \infty,$
$\log (\pi e)^{1/2}\leq
\log Vol(U(1)) \;(\approx \frac{1}{2}\log\beta_1+\log vol(U(1)))< \infty,
0\leq \log Vol(\bz_2) \leq \log 2$. These intervals reflect our having
used $Vol H$ that, up to a factor $(\pi e)^{\frac{dim(H)}{2}}$,
are measured in units of the fluctuation volume. The cube with the chopped off
corner
represents the region of total confinement. Walls that extend to
$+\infty$ are terminated in the drawing with
irregular wavy boundaries.}

\end{figure}

\begin{figure}
\centerline{\epsfxsize=\textwidth \epsfbox{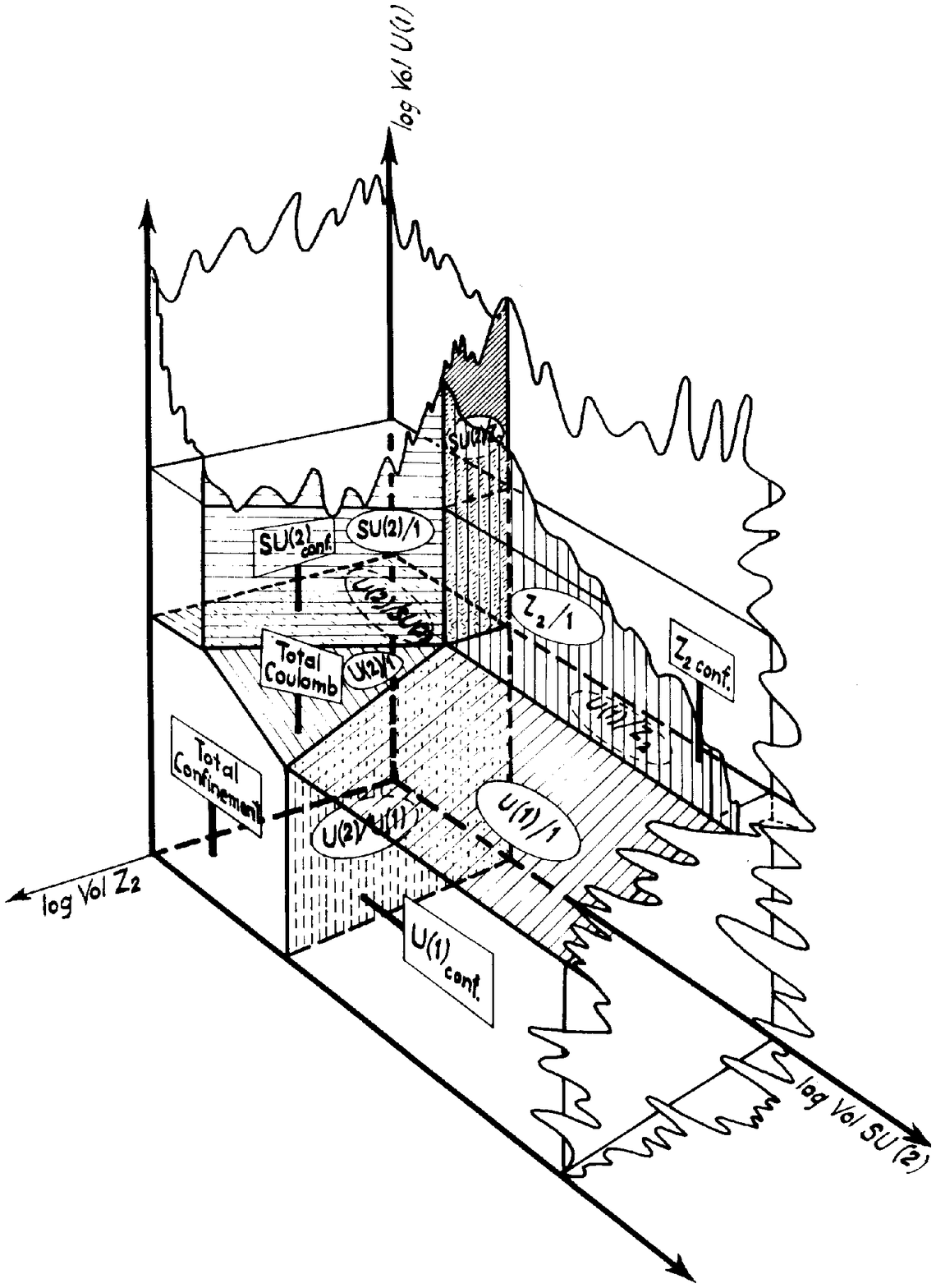}}
\caption[figmpphdiag2]{\label{mpphdiag2} Phase diagram for lattice gauge
theory with gauge group $U(2)$ in our weak
coupling approximation with modified Manton action. We have drawn the
figure with positive  effective
dimension for the discrete constituent invariant
subgroup $\bz_2$.
Rectangular signs on signposts are marked
with the confining
invariant subgroup $H$ and indicate the regions corresponding
to the 5 possible
phases; these 5 phases are seen to meet at the multiple point.
The oval signs lie in the phase boundaries and specify the factor group
$L=H_1/H_2$ formed from the two invariant subgroups $H_1$ and $H_2$
that are confined on the two sides of the boundary. It is these groups $L$
that change behaviour in crossing the phase boundary in question.
Unbroken shading lines indicate phase boundaries as seen from within the
totally Coulomb-like phase.}

\end{figure}

\subsubsection{Need for discrete subgroup parameters}\label{needfordisc}

It is instructive to answer the question: Why do we need the  discrete
group action parameters? Recall that in our  modified  Manton  action,
there can be sharp ``peaks'' in  the  distribution  $e^{S_{\Box}}$  of
plaquette variables centred not only at the group identity  but  also
at nontrivial elements of discrete subgroups. However, to motivate the
answer to our question, we revert for a moment to an action $S_{\Box}$
leading to a distribution $e^{S_{\Box}}$ of plaquette  variables  with
just one ``peak'' (at the identity) - this is just the  normal  Manton
action. Then the quantities $Vol$ corresponding to the {\em  the  same
Lie algebra ideals} in the Lie Algebra of the $SMG$  -  i.e.,  volumes
measured in units proportional to the fluctuation volume  -  obey  the
relations

\beq \begin{array}{lcl}
Vol(U(1)_{subgr.}) & = & 6Vol(SMG/(SU(2)\times SU(3))) \\
Vol(SU(2)_{subgr.}) & = & 2Vol(SMG/U(3)) \\
Vol(SU(3)_{subgr.}) & = & 3Vol(SMG/U(2)) \\
Vol(U(2)_{subgr.}) & = & 3Vol(SMG/SU(3)) \\
Vol(U(3)_{subgr.}) & = & 2Vol(SMG/SU(2)) \\
Vol(SU(2)\times SU(3))_{subgr.} & = & 6Vol(SMG/U(1)).
\end{array} \label{vols} \eeq

The important feature of this list is that each equality  relates  the
quantity $Vol$ for a subgroup and factor group that both correspond to
the same Lie algebra ideal (in the Lie algebra of the  $SMG$).  It  is
seen that $Vol$ for a subgroup with a given Lie algebra is larger than
a factor group with the same Lie algebra by an integer factor equal to
the number of centre elements of the subgroup that are  identified  in
the factor group. It is important not to lose sight of the  fact  that
quantity $Vol$ is, by definition, the volume of the group measured  in
the units proportional to  the  average  group  volume  accessed by
quantum  fluctuations.  This  fluctuation   volume   is   $\prod_{i\in
\{U(1),SU(2),SU(3)\}}(2\beta_i)^{-dim(i)/2}$ where  it  is  understood
that $i$ runs over the appropriate Lie sub-algebras. In the sequel,  it
is to be understood that when we refer to the volume of  a  group,  we
mean the quantity $Vol$.

Without the extra ``peaks'' of the modified Manton action, it  is  not
possible to vary the volume on the  left  hand  side  of  one  of  the
equations (\ref{vols}) {\em independently} of the volume on the  right
hand side. For example, because of  the  (nontrivial)  integer  factor
disparity  in  the  volumes  on  the  two  sides  of   the   equations
(\ref{vols}), we cannot have
$Vol(U(2)_{subgr.}) = Vol(SMG/SU(3))$ at the same point in the $\beta$
parameter space. In  particular,  two  such  volumes  can  never  have
critical values for the same values of the $\beta$'s which means  that
the two corresponding partially confining  phases  (i,e.,  confinement
w.r.t. $U(2)$ and $SU(3)$) cannot meet at a multiple point.

This feature is seen in Figure~\ref{mpphdiag2} which shows the phase
diagram  for  the
gauge group $U(2)$. In the plane  defined  by  log$Vol(\bz_2)=\log  2$
(the maximum value of log$Vol(\bz_2)$),  we  have  the  phase  diagram
corresponding to  the  normal  Manton  action  (one  ``peak''  in  the
distribution of plaquette variables, centred  at  the  identity;  the
fluctuation volume is accordingly also centred at the identity).  Due
to the fact that say  $Vol(SU(2)_{subgr.})=2Vol(U(2)/U(1))$  (measured
in  the  same  unit  of  volume  which  is proportional to
  $   \prod_{i\in
\{U(1),SU(2),SU(3)\}}(2\beta_i)^{-dim(i)/2}$),  it  is  impossible  to
have $Vol(SU(2)_{subgr.})=Vol(U(2)/U(1))$ for the same values  of  the
$\beta$ parameters; i.e., because the  volumes  of  the  subgroup  and
factor group corresponding to the same Lie algebra differ by a  factor
two, these two volumes cannot be critical for the same set of  $\beta$
parameters. This in turn precludes phases  partially  confined  w.r.t.
$SU(2)$   and   $U(1)$   from   coming   together.   In   the    plane
log$Vol(\bz_2)=\log 2$ of Figure~\ref{mpphdiag2}, it is indeed seen
that  these  \pcps
do not touch; the maximum number of phases that come together in  this
plane is three (i.e., not all four possible  phases)  where  three  is
generic number of phases that can meet in two dimensions.

In    order     to     succeed     in     having,     for     example,
$Vol(SU(2)_{subgr.})=Vol(U(2)/U(1))$ in the case of the group  $U(2)$,
it is necessary to introduce a parameter that allows us to change  the
volume $Vol(SU(2)_{subgr.})$ without changing the volume  of  the  Lie
algebra-identical factor group $Vol(U(2)/U(1))$. In this $U(2)$  case,
this is what is achieved by introducing the parameter log $Vol(\bz_2)$
which allows the variation of the relative heights  of  the  plaquette
distribution ``peaks'' centred at the two elements of $\bz_2$.

By  introducing  an  action  giving  rise  to   extra   ``peaks''   in
$e^{S_{\Box}}$ that are centred at elements that are (by  definition)
identified in a factor group but not in the subgroup having  the  same
Lie algebra as the factor group, we admit  the  possibility  of  extra
centres of fluctuation which increases the fluctuation  volume  (i.e.,
the unit in which $Vol$ is measured) for the subgroup but not for  the
factor group (because all centres of fluctuation are  identified  with
the group identity in the factor  group).  Hence  we  gain  a  way  of
varying the volume of the subgroup (measured in fluctuation-volume units)
{\em without} varying the volume of
the factor group. In the example of  $U(2)$  referred  to  above,  the
possibility of a peak in $e^{S_{\Box}}$ at the nontrivial  element  of
$\bz_2$ in addition to the peak at the identity means that  the  total
fluctuation volume for the  subgroup  $SU(2)$  can  be  made  up  of
contributions from both peaks whereas the fluctuation volume  of  the
Lie algebra-identical factor group $U(2)/U(1)$ can only come from  the
fluctuations centred at the identity since both elements  of  $\bz_2$
are identified in the factor group.

Relative to the approximate  $U(2)$  phase  diagram  of
Figure~\ref{mpphdiag2}, the
variation of the parameter log$Vol(\bz_2)$ can be described roughly as
follows.  Recall  from  above   that   in   the   plane   defined   by
log$Vol(\bz_2)=  \log  2$,  there  is  only  one   ``peak''   in   the
distribution $e^{S_{\Box}}$ (centred at the identity). In this plane,
the volume of the subgroup is, for given $\beta$  values,  identically
twice that of the factor group since there are by definition only half
as many elements (i.e., cosets) in the  factor  group  $U(2)/U(1)$  as
there are elements in the subgroup $SU(2)$ with the same Lie  algebra.
Now as the value of the  parameter  log$Vol(\bz_2)$  is  reduced,  the
pattern of fluctuations changes in such  a  way  that
the  nontrivial  element  of  $\bz_2$  becomes  a  centre  of  quantum
fluctuations with fluctuations that become progressively larger in the
sense that more and more probability is relocated  at  the  nontrivial
element of $\bz_2$. However, fluctuations about the nontrivial element
of $\bz_2$  are  not  ``noticed''  by  the  factor  group
because fluctuations about the identity  and  fluctuations  about  the
nontrivial element  correspond  to  fluctuations  about  the
same coset of the factor group - namely the  identity  of  the  factor
group. In our approximation, the parameter  log$Vol(\bz_2)$  decreases
until the two peaks of $e^{S_{\Box}}$ (one at each element of $\bz_2$)
have the same height and accordingly each contribute with half of  the
total fluctuation volume  (i.e.,  the  volume  accessible  to  quantum
fluctuations) of the subgroup $SU(2)$. This  coincides  with  reaching
the multiple point at which log$Vol(\bz_2)=0$.

So at the  multiple  point,  the  unit  of  volume  (i.e.,  the  total
fluctuation volume) used to measure the volume of the subgroup is just
twice the fluctuation  volume  of  the  Lie  algebra-identical  factor
group. This has the  consequence  that  the  volume  of  the  subgroup
$SU(2)$ is reduced by a factor two which is just the factor  by  which
$Vol SU(2)$ is larger than  the  Lie  algebra-identical  factor  group
$Vol(U(2)/U(1))$ in the absence of the extra parameter log$Vol(\bz_2)$
that at the multiple point leads  to  a  two-fold  increase  in
fluctuation volume.

More generally, having discrete group  action  parameters  allows  the
possibility of having a number of fluctuation centres  in  a  subgroup
that is just equal to the number of elements identified in going  from
a subgroup to the Lie algebra-identical factor group. At  the  multiple
point, these additional centres increase the total fluctuation  volume
and thereby the unit of volume measurement for the  subgroup  relative
to the fluctuation volume of the factor group by a factor equal to the
number of elements identified in the factor group.

In summary, we have developed a generalised action that deals with
the need for more than the usual number of parameters  in
the plaquette action if one wants to make the phases corresponding  to
confinement of the various invariant subgroups  -  including  discrete
(invariant) subgroups) share a common point (i.e., the multiple point)
in the phase diagram. With our plaquette action  parameterisation,  we
can show the existence of and also the  coincidence  in  one  multiple
point of phases  corresponding  to  all  invariant  subgroups  of  the
non-Abelian components of the $SMG$. The invariant subgroups that we do
not consider here correspond solely to additional discrete (invariant)
subgroups of $U(1)$. The defining feature of a confinement-like  phase
for an invariant subgroup is equivalent to the  assumption  that
Bianchi  identity
constraints can be neglected for such a phase in a crude weak coupling
approximation using a mean field approximation.

At the multiple point, we are dealing with first order phase transitions;
therefore, a priori at least, our multiple point principle suffers from
lack of universality. However, the fact that a weak coupling approximation
is at least approximately applicable - even for the determination of critical
couplings - leads to the irrelevance of terms greater than second order in
Taylor expansions of the action and consequently fosters the hope
of an approximative universality.

\subsection{Correction due to quantum fluctuations}\label{quantumcorr}

In our model, the $SMG$ gauge coupling constants are to be  identified
with the couplings for the diagonal subgroup  that  results  from  the
Planck scale breakdown of $SMG^3$. While in the naive continuum limit,
the diagonal subgroup field configurations consist (by definition)  of
excitations that are identical for the $N_{gen.}=3$ copies (labelled  by
names  $``Peter",\;``Paul",\cdots$)  of  any  $SMG$  gauge  degree  of
freedom $A^b_{\mu}$, a more realistic view must take into account  that
the $N_{gen.}$ copies of $A^b_{\mu}$ in $SMG^3$: \ppm  undergo  quantum
fluctuations relative to each other. In this section  this correction is first
estimated
for a confinement-like  phase (hereby justifying a disregard of Bianchi
identities) and subsequently corrected so as to be approximately
correct for a Coulomb-like phase.

Including the effect of fluctuations of a general quantum  field  $\theta$  in
the continuum  limit  is done using  the  effective  action
$\Gamma[\theta_{cl.}]$:

\beq
\Gamma[\theta_{cl.}]=S[\theta_{cl.}]-\frac{1}{2}\mbox{Tr}(log(S^{^{\prime
\prime}}[\theta_{cl.}])). \eeq

\nin The correction to the continuum couplings that we calculate below
consists in identifying the classical continuum action $\int d^4x
\frac{-1}{4g^2}(gF_{\mu\nu}^a)^2$ with the effective
action $\Gamma$ - instead of with the lattice  action  $S$  -  in  the
naive continuum limit approximation. In calculating  this  correction,
we  ignore   non-Abelian   effects   and   assume   that   the   action
$S_{_{Monte\;Carlo}}$             used             in              the
literature\cite{bachas,bhanot2,drouffe} deviates only  slightly  from
the Manton action  for  which  the  Tr$log$  correction  is  simply  a
constant. The $S_{Monte\;Carlo}$ could  for  example  be  the  popular
cosine action in the $U(1)$ case.  First,  however,  we  note  that  a
change  in  the   functional   form   of   the   action   by   $\delta
S(\theta_{cl.})$
leads to a  functional  change  in
the effective action $\Gamma(\theta_{cl.})$ that differs from  $\delta
S(\theta_{cl.})$ by  a  term  proportional  to  $\mbox{Tr}\frac{\delta
S^{^{\prime                        \prime}}[\theta_{cl.}]}{S^{^{\prime
\prime}}[\theta_{cl.}]}$:

\beq  \delta   \Gamma[\theta_{cl.}]=\delta   S[\theta_{cl.}]+\frac{1}{2}
\mbox{Tr}(\delta(\log (S^{^{\prime  \prime}}[\theta_{cl.}])))=  \delta
S[\theta_{cl.}]+\frac{1}{2}         \mbox{Tr}(\frac{\delta(S^{^{\prime
\prime}}[\theta_{cl.}])}{S^{^{\prime
\prime}}[\theta_{cl.}]}) \eeq

\nin  But  as   we   are   assuming   that   the   variation   $\delta
S[\theta_{cl.}]$ is done relative to the Manton action, we have

\beq \frac{1}{2}         \mbox{Tr}(\frac{\delta(S^{^{\prime
\prime}}[\theta_{cl.}])}{S^{^{\prime
\prime}}_{Manton}[\theta_{cl.}]}) = \frac{1}{2}         \mbox{Tr}(S^{^{\prime
\prime}}[\theta_{cl.}]\langle
(\theta-\theta_{cl.})^2\rangle) \eeq

\nin     where      we      have      used      that      $S^{^{\prime
\prime}}_{Manton}[\theta_{cl.}]    \propto     \langle
(\theta-\theta_{cl.})^2\rangle^{-1}=\;const.$     and     that,      up      to
a  constant,  $\delta(S^{^{\prime  \prime}}[\theta_{cl.}])=S^{^{\prime
\prime}}[\theta_{cl.}]$ (modulo a constant).

Neglecting
non-Abelian effects, we generalise  this  result  to  non-Abelian  gauge
groups and write it more concretely using $U=e^{i\theta^at^a}$ and
$S[U]=\sum_{\Box}S_{\Box}(U(\Box))$:

\begin{equation}   \Gamma    [U_{cl.}]=S[U_{cl.}]    +
\frac{1}{2}\Delta S_{\Box}(U_{cl.}(\Box))\langle (\theta_{\Box}^a(\Box)-\theta
_{\Box\;cl.}^a(\Box))^2\rangle
\;\;  (\mbox{summation
over}\;a) \label{gamma}\end{equation}

\nin We have for the Laplace-Beltrami operator

\begin{small}

\beq \frac{1}{2}\Delta    S(U(\Box))    \stackrel{def.}{=}
\lim_{\epsilon \rightarrow 0^+}\frac{\int  d^{N^2-1}\!f\;\;
\exp(-\frac{1}{2\epsilon}  f_a^2)(S(U  \cdot   e^{if_bt_{b}})-S(U))}{\int
d^{N^2-1}\!f\;\; \exp(-\frac{1}{2\epsilon} f_d^2)f_e^2} \;\; (\mbox{sum  over
}a,b,d,e). \eeq

\end{small}

\nin where $f_a$ and $t_{a}$  denote  respectively
the $a$th Lie algebra component and Lie algebra generator.
Upon expanding (in the representation $r$) the exponential
$\exp(if_b T_{b,\,r})$ representing $\exp(if_b t_{b})$
the argument  of  which  is  assumed  to  be  small  inasmuch  as  the
$f_b$ are assumed to be small,
there obtains

\begin{small}

\beq =  \lim_{\epsilon  \rightarrow  0^+} \sum_r \frac{\beta_r}{d_r}
\frac{\int   d^{N^2-1}\!f\;\;
\exp(-\frac{1}{2\epsilon} f_a^2)\mbox{Tr}(U \cdot (-\frac{1}{2}f_bf_cT_{b,\,r}
T_{c,\,r}))}{\int      d^{N^2-1}\!f\;\;      \exp(-\frac{1}{2\epsilon}
f_d^2)f_e^2} \;\;{\footnotesize (\mbox{sum  over  }a,b,c,d,e)}\eeq

\beq =\sum_r\frac{\frac{\beta_r}{d_r}Tr_r(U\cdot
(-\frac{1}{2}(T_{b,\,r})^2))}{N^2-1} \;\;(\mbox{sum over }b) \eeq

\end{small}

\nin where we have expanded the plaquette action  in  characters:  for
the representation  $r$  of  dimension  $d_r$  the  character  $\chi_r$
is given by $\chi_r=\mbox{Tr}_r(U(\Box))$ and

\beq S(U(\Box))=\sum_r \frac{\beta_r}{d_r}\mbox{Tr}_r(U(\Box)).     \eeq

\nin We have

\beq                        \frac{1}{2}                         \Delta
S(U(\Box))=\sum_r-\frac{1}{2}\frac{\beta_r}{d_r}
Tr_r(U(\Box))\frac{C^{(2)}_r}{N^2-1} \eeq


\nin where $C^{(2)}_r$ is the quadratic Casimir for the representation
$r$.         The         Casimir         is         defined         as
$C^{(2)}_r\bunit_r\stackrel{def.}{=}\sum_b(T_{b,\;r})^2    $.
For the groups  $SU(2)$  and  $SU(3)$,
the Lie algebra bases in the  fundamental  (defining)  representations
are      taken      respectively       as       $T_{b,\;r=f}=T_{b,\;r=
\underline{\sbtwo}}=\frac{\sigma^b}{2}$     and     $T_{b,\;     r=f}=
T_{b,\;r=\underline{\sbthree}}=\frac{\lambda^b}{2}$. The subscript $f$
denotes the fundamental representation; $\sigma^b$ and $\lambda^b$ are
the   Pauli   and   Gell-Mann   matrices   with   the    normalisation
$\mbox{Tr}(\frac{\sigma^a}{2}\frac{\sigma^b}{2})=\frac{\delta_a^b}{2}$
and                  $                   \mbox{Tr}(\frac{\lambda^a}{2}
\frac{\lambda^b}{2})=\frac{\delta_a^b}{2}$. With this basis convention,
and with the left-handed  quark  doublet  field  as  an  example,  the
covariant derivative is

\beq
D_{\mu i\;\;\;\alpha}^{\;\;\;\;j\;\;\,\beta}=
\partial_{\mu}\delta_i^{\;\;j} \delta_{\alpha}^{\;\;\beta}
-ig_2A_{\mu}^b\frac{(\sigma^b)_{ i}^{\;\;j}}{2}\delta_{i}^{\;\;j}
-ig_3A_{\mu}^b\frac{(\lambda^b)^{\;\;
\beta}_{\alpha}}{2}\delta_{\alpha}^{\;\;\beta}
-ig_1\frac{1}{6}A_{\mu}\delta_i^{\;\;j} \delta_{\alpha}^{\;\;\beta}.
\eeq

\nin where the index $b$ labels Lie algebra  components,  the  indices
$i,\;j$  label  matrix  elements  of   the   ($2$-dimensional)  fundamental
representation of $SU(2)$, and the indices $\alpha,\;\beta$ label  the
matrix elements of the ($3$-dimensional) fundamental representation of
$SU(3)$. The factor $\frac{1}{6}$ in  the  last  term  is  the  $U(1)$
quantum  number  $\frac{y}{2}$  where  $y$  is  weak  hyper-charge;  the
convention used is $Q=\frac{y}{2}+I_{W_{3}}$.

The above convention for the generators of $SU(2)$ and $SU(3)$ in  the
fundamental    representation    $f$    leads     to     a     Casimir
$C^{(2)}_f=\frac{N^2-1}{2N}$  for  an  $SU(N)$  group.  From  this  it
follows that, for the adjoint representation (denoted by $adj.$),  the
Casimir $C^{(2)}_{adj.}$ for an $SU(N)$ group is given by
$C^{(2)}_{adj.}=N$.

Ignoring Bianchi identities, we get for the deviations

\beq
\langle( \theta_{Peter}- \theta_{diag.})^2_a          \rangle=
\frac{N^2-1}{2}                                                          (
\frac{1}{2} \sum_r  \beta_r \frac{C^{(2)}_r}{N^2-1})^{-1}
(\mbox{sum over }a) \mbox{   (confinement phase)}\eeq

\[=-\{\frac{\beta_f}{d_f} Tr_f(U_{diag.}(\Box))C^{(2)}_f+
\frac{\beta_{adj.}}{d_{adj.}}
Tr_{adj.}(U_{diag.}(\Box))C^{(2)}_{adj.}\}\frac{1}{2}
\frac{N^2-1}{\beta_fC^{(2)}_f+ \beta_{adj.}C^{(2)}_{adj.}} \]

Letting the sum over representations run  only  over  the  fundamental
(=defining) and adjoint representations labelled  respectively  by  the
subscripts $f$ and $adj.$ (the only representations used in the Monte Carlo
runs of references \cite{bachas,bhanot2,drouffe}),  we  get  for  the
effective action (\ref{gamma})

\begin{equation} \Gamma (U_{diag.}(\Box))= \end{equation}
\[ = \frac{\beta_f}{d_f} Tr_f(U_{diag.}(\Box))
(1-\frac{C^{(2)}_f(N^2-1)}{2( \beta_fC^{(2)}_f+ \beta_{adj.}C^{(2)}_{adj.})})+
 \frac{\beta_{adj.}}{d_{adj.}} Tr_{adj.}(U_{diag.}(\Box))
(1- \frac{C^{(2)}_{adj.}(N^2-1)}{2(\beta_fC^{(2)}_f+
\beta_{adj.}C^{(2)}_{adj.})}) \]

\nin So with the continuum correction we have  to make the replacement

\beq \beta_r \rightarrow \beta_r(1-\frac{C^{(2)}_r(N^2-1)}
{2\sum_{\hat{r}} \beta_{\hat{r}}C^{(2)}_{\hat{r}}}) \;\;
\mbox{(for ``confinement phase'')} \label{eqn31}\eeq


This expression for the effective action has been obtained  using  the
approximation  that  all  plaquette  variables  can  be  regarded   as
independent (i.e., Bianchi identities have been disregarded).  This
approximation is appropriate for the confinement  phase.  However,  as we
are interested in criticality as approached  from  the  Coulomb  phase
(i.e., Coulomb phase in our scale dependent sense),   we want
the quantum fluctuation  correction  in  this  phase
where Bianchi identities must be respected.  These  identities  reduce
the number of degrees  of  freedom  per  plaquette  that  can
fluctuate independently by a factor  2.  In going to  the  Coulomb  phase,  the
continuum-corrected $\beta_r$ is modified  as
follows:

\beq \beta_r(1-\frac{C^{(2)}_r(N^2-1)}
{2\sum_{\hat{r}} \beta_{\hat{r}}C^{(2)}_{\hat{r}}})_{confinement}\rightarrow
\beta_r(1-\frac{C^{(2)}_r(N^2-1)}
{4\sum_{\hat{r}} \beta_{\hat{r}}C^{(2)}_{\hat{r}}})_{Coul.\;phase}\label{eqn32}
\eeq

\nin Without the continuum correction, we have for the fine  structure
constants at the multiple (i.e., triple ) point

\beq \frac{1}{ \alpha_{triple\;point,\,no\,cont.}}
=4\pi\sum_r \frac{C^{(2)}_r \beta_{r,\,triple\;point}}{N^2-1}
\;\;\mbox{(naive continuum limit)}. \eeq

\nin With the continuum-corrected $\beta_r$ in the  Coulomb  phase  we
have for the fine structure  constants  at  the  triple point

\begin{small}

\beq\frac{1}{\alpha_{_{triple\;point, \,cont.}}}=
4\pi\sum_r \frac{C^{(2)}_r}{N^2-1} \beta_{_{r\,triple\;point}}
(1-\frac{C^{(2)}_r(N^2-1)}
{4\sum_{\hat{r}}\beta_{_{ \hat{r},\,triple\;point}}C^{(2)}_{\hat{r}}})=
\label{corbian} \eeq
\[ 4\pi\sum_r\frac{C^{(2)}}{N^2-1} \beta_{r\;triple\;point}(1-\pi
C^{(2)}_r \alpha_{_{triple\;point,\;no\;cont.}})  \]

\end{small}

\subsection{Calculation of non-Abelian critical couplings at Planck
scale}\label{calculationnonabel}

It can be argued that at the multiple point of the phase  diagram  for
the whole $SMG^3=SMG\times SMG\times SMG$, the non-Abelian  (plaquette)
action parameters for each of the three Cartesian product factors take
the same values as at the multiple point  for  a  single  gauge
group $SMG$. This allows us to determine  the  multiple  point  action
parameters for the  gauge  group  $SMG^3$  from  a  knowledge  of  the
multiple point action parameters for just one of the $SMG$ factors  of
$SMG^3$.
Accordingly, we can calculate the multiple point critical couplings
from the couplings for the isolated $SU(2)$ and $SU(3)$ groups. To this end,
we have
used  figures  from  the
literature  \cite{bachas,bhanot2,drouffe1} to graphically extract the
coordinates $(\beta_f,\beta_{adj.})_{triple\;point}$ of the
triple point:

For $SU(2)$:  $(\beta_f,\beta_{adj.})_{triple\;point}= (0.54, 2.4)$

For $SU(3)$:  $(\beta_f,\beta_{adj.})_{triple\;point}= (0.8, 5.4)$


\begin{table}\caption[tabnon3]{\label{non3}}
\begin{small}
\begin{tabular}{|p{2.50 in}|l|}
\hline
\multicolumn{2}{|c|}{Table~\ref{non3}: SU(2) Gauge Coupling} \\
\hline \hline
\raggedright
Prediction~for~continuum~limit coupling~estimate,~$1/\alpha_
{2,\; triple\;point,\;cont.}$, using  & \\
\raggedright {\footnotesize 1.~not~exponentiated:} & $\overbrace{0.71\cdot   20
  }^{14.2}+\overbrace{0.89\cdot
1.7}^{1.5}=15._7    \pm    1$    \\     \raggedright     {\footnotesize
2.~exponentiated:}               &               $\overbrace{0.75\cdot
20}^{15._0}+\overbrace{0.89\cdot  1.7}^{1.5}=16._5  \pm  1$  \\  \hline
Experimental value\cite{kim,amaldi} for & \\ $1/\alpha_2$ reduced by a factor
3: & $\frac{1}{3}\cdot \alpha^{-1}_2(M_Z)=  \frac{1}{3}\cdot  (29.7\pm
0.2)=9.9\pm 0.07$ \\ ``desert extrapolation\cite{kim,amaldi}''  to  &
\\ Planck scale with one Higgs: &  $\stackrel{desert}{\longrightarrow}
\frac{1}{3}\cdot \alpha^{-1}_2 (\mu_{Pl.})=\frac{1}{3}\cdot 49._{5}=16.5$
\\ & \\ \hline \hline $\beta_{adj.,\;triple\;point}$ {\footnotesize  (i.e.,
at triple point)} & 2.4 (\mbox{ca. 5\% uncertainty from  MC})  \\
\hline  $\beta_{f\;triple\;point}$  {\footnotesize  (i.e.,  at  triple
point)} &  0.54  (\mbox{ca.  10\%  uncertainty  from  MC})  \\  \hline
\raggedright  $\beta_{adj.}$-contribution~to~$1/\alpha_{2,\;triple\;point}$
&  \\  {\footnotesize  (without   continuum   correction)}   &   $4\pi
\frac{C^{(2)}_{adj.}}{(2^2-1)}\beta_{adj,\;triple\;point}=4\pi \cdot  (2/3)
\cdot 2.4=20$ \\ & \\  \hline  \raggedright  $\beta_f$-contribution~to
{}~$1/\alpha_{2,\;triple\;point}$         &         \\         {\footnotesize
(without~continuum~correction):}                &                $4\pi
\frac{C^{(2)}_f}{(2^2-1)}\beta_{f\;triple\;point}=4\pi                \cdot
(\frac{3}{4}/3)\cdot  0.54=1.7$  \\  &  \\  \hline  \raggedright  Full
$1/\alpha_{2,\;triple\;point}$  {\footnotesize  (without  continuum}  &  \\
{\footnotesize                     correction):}                     &
$1/\alpha_{2,\;triple\;point,\;full,\;no\;cont.}=20+1.7=21.7$   \\   \hline
\raggedright       Continuum       correction        factor        for
$\beta_{adj.}$-contribution:  &  \\  \raggedright  {\footnotesize   1.
not~exponentiated (using (\ref{corbian})):}  &  $1-C^{(2)}_{adj.}  \pi
\alpha_{2,\;triple\;point,\;full,\;no.\;cont.}=$            \\            &
$1-2\pi/21.7=1-0.290=0.71$   \\   \raggedright    {\footnotesize    2.
exponentiated:}          &          $\exp(-C^{(2)}_{adj.}          \pi
\alpha_{2,\;triple\;point,\;full,\;no.\;cont.})$=           \\            &
$\exp(-2\pi/21.7)=\exp(-0.290)=0.75$  \\  &  \\  \hline   \raggedright
Continuum  correction  factor   for   $\beta_f$-contribution:   &   \\
\raggedright    {\footnotesize     1.     not~exponentiated     (using
(\ref{corbian})):}            &            $1-C^{(2)}_f            \pi
\alpha_{2,\;triple\;point,\;full,\;no.\;cont.}=$  \\   &   $1-(\frac{3}{4})
\pi/21.7=1-0.109=0.89$    \\    \raggedright     {\footnotesize     2.
exponentiated:}           &            $\exp(-C^{(2)}_f            \pi
\alpha_{2,\;triple\;point,\;full,\;no.\;cont.})=$ \\ & $\exp(-(\frac{3}{4})
\pi/21.7)=\exp(-0.109)=0.90$ \\ & \\ \hline \end{tabular} \end{small}
\end{table}


\begin{table}\caption[tabnon4]{\label{non4}}
\begin{small}
\begin{tabular}{|p{2.50 in}|l|}
\hline
\multicolumn{2}{|c|}{Table~\ref{non4}: SU(3) Gauge Coupling} \\
\hline \hline
\raggedright
Prediction~for~continuum~limit coupling~estimate,~$1/\alpha_{3,\;
triple\;point,\;cont.}$, using  & \\
\raggedright {\footnotesize 1.~not~exponentiated:} & $\overbrace{0.65\cdot
25}^{16._3}+\overbrace{0.84\cdot
1.7}^{1.4}=17._7    \pm    1$    \\     \raggedright     {\footnotesize
2.~exponentiated:}               &               $\overbrace{0.70\cdot
25}^{17._5}+\overbrace{0.85\cdot  1.7}^{1.4}=18._9  \pm  1$  \\  \hline
Experimental value\cite{kim,amaldi} for & \\ $1/\alpha_3$ reduced by a factor
3:   &    $\frac{1}{3}\cdot\alpha^{-1}_3(M_Z)=\frac{1}{3}\cdot(8.47\pm
0.5)=2.8\pm 0.2$ \\ ``desert extrapolation\cite{kim,amaldi}'' to & \\
Planck scale with  one  Higgs:  &  $\stackrel{desert}{\longrightarrow}
\frac{1}{3}\cdot \alpha^{-1}_3(\mu_{_{Pl.}})=\frac{1}{3}\cdot  53  \pm
0.7 =17.7\pm 0.3$ \\  &  \\  \hline  \hline  $\beta_{adj.,\;triple\;point}$
{\footnotesize  (i.e.,  at  triple  point)}  &   5.4   (ca.   5\%
uncertainty) \\ \hline $\beta_{f\;triple\;point}$ {\footnotesize (i.e.,  at
triple  point)}  &  0.8  (ca.   20\%   uncertainty)   \\   \hline
\raggedright  $\beta_{adj.}$-contribution~to~$1/\alpha_{3,\;triple\;point}$
&  \\  {\footnotesize  (without   continuum   correction)}   &   $4\pi
\frac{C^{(2)}_{adj.}}{(3^2-1)}\beta_{adj,\;triple\;point}=4\pi \cdot  (3/8)
\cdot 5.4=25$ \\ & \\  \hline  \raggedright  $\beta_f$-contribution~to
{}~$1/\alpha_{3,\;triple\;point}$         &         \\         {\footnotesize
(without~continuum~correction):}                &                $4\pi
\frac{C^{(2)}_f}{(3^2-1)}\beta_{f\;triple\;point}=4\pi                \cdot
(\frac{4}{3}/8)\cdot  0.8=1.7$  \\  &  \\  \hline  \raggedright   Full
$1/\alpha_{3,\;triple\;point}$  {\footnotesize  (without  continuum}  &  \\
{\footnotesize                     correction):}                     &
$1/\alpha_{3,\;triple\;point,\;full,\;no\;cont.}=25+1.7=26._7$  \\   \hline
\raggedright       Continuum       correction        factor        for
$\beta_{adj.}$-contribution:  &  \\  \raggedright  {\footnotesize   1.
not~exponentiated (using (\ref{corbian})):}  &  $1-C^{(2)}_{adj.}  \pi
\alpha_{3,\;triple\;point,\;full,\;no.\;cont.}=$            \\            &
$1-3\pi/26.7=1-0.35=0.65$   \\    \raggedright    {\footnotesize    2.
exponentiated:}          &          $\exp(-C^{(2)}_{adj.}          \pi
\alpha_{3,\;triple\;point,\;full,\;no.\;cont.})$=           \\            &
$\exp(-3\pi/26.7)=\exp(-0.35)=0.70$  \\  &  \\   \hline   \raggedright
Continuum  correction  factor   for   $\beta_f$-contribution:   &   \\
\raggedright    {\footnotesize     1.     not~exponentiated     (using
(\ref{corbian})):}            &            $1-C^{(2)}_f            \pi
\alpha_{3,\;triple\;point,\;full,\;no.\;cont.}=$  \\   &   $1-(\frac{4}{3})
\pi/26.7=1-0.16=0.84$    \\     \raggedright     {\footnotesize     2.
exponentiated:}           &            $\exp(-C^{(2)}_f            \pi
\alpha_{3,\;triple\;point,\;full,\;no.\;cont.})=$ \\ & $\exp(-(\frac{4}{3})
\pi/26.7)=\exp(-0.16)=0.85$ \\ & \\ \hline \end{tabular} \end{small}
\end{table}

\nin  The  calculation  of  $\alpha_2^{-1}$  and  $\alpha_3^{-1}$  are
presented in Tables~\ref{non3}~and~\ref{non4}  respectively.  In  these
tables,  the
subscripts ${adj.}$  and  $f$  denote  respectively  the  adjoint  and
fundamental representations of the groups considered.

In order to get an idea of the order of magnitude of the error involved
in estimating the average over the Laplace-Beltrami of the plaquette action
only to next to lowest order, we note that we can calculate
such an average to all orders in the case of a $\cos\theta$ action for a $U(1)$
gauge theory. In this case the
averaging is  readily  performed  and leads  to  an
exponential for which the first terms of a Taylor  expansion  coincide
with the terms we  calculated using (\ref{corbian}).
This  suggests  that  also  in  the
non-Abelian cases it might be quite reasonable to ``exponentiate''  our
``continuum'' corrections and subsequently use  the  change  made  by
such a procedure as a crude estimate of the error due to our omission
of the second order perturbative terms. By exponentiated continuum corrections
we mean by definition that, instead of the replacements (\ref{eqn31}) and
(\ref{eqn32}), we use respectively

\beq \beta_r\rightarrow \beta_r\exp(-\frac{C^{(2)}_r(N^2-1)}
{2\sum_{\hat{r}} \beta_{\hat{r}}C^{(2)}_{\hat{r}}}\;\;\;\mbox{(for
``confinement'')}\eeq

\nin and

\beq \beta_r\rightarrow \beta_r\exp(-\frac{C^{(2)}_r(N^2-1)}
{4\sum_{\hat{r}} \beta_{\hat{r}}C^{(2)}_{\hat{r}}} \;\;\;\mbox{(for ``Coulomb''
phase)} \eeq

\nin   As evidenced by Tables~\ref{non3}~and~\ref{non4},
this exponentiation yields a change of the  order  of  one  unit  in
$1/\alpha_{crit.,\;cont.}$ $\approx$ 20 from which we can estimate the
uncertainty due the neglect of higher order  terms  as  being  of  the
order of say 5 \%.

Since our deviations from the experimental couplings  extrapolated  to
the Planck scale\cite{kim,amaldi}  are of the same order of magnitude as
the uncertainty
in the Monte Carlo data and the
uncertainty due to chopping off the higher order continuum corrections,
a calculation of the next  order  corrections
and increased accuracy in the calculations
are called for in order to determine if our deviations are significant.

\begin{flushright}
\begin{tiny}
theu1b.tex 31 May 96 alf 
\end{tiny}
\end{flushright}

\section{Implementing the MPCP in determining the SMG U(1) coupling:
Methods for constructing phase diagrams}\label{mpcabelmeth}

The gauge group to which we ultimately  want  to  apply  the  {\bf M}ultiple
{\bf P}oint {\bf C}riticality {\bf P}rinciple (MPCP) is the
{\bf A}nti {\bf G}rand {\bf U}nified {\bf T}heory (AGUT) gauge group $SMG^3$
or some
group in which the latter is embedded in such a way that $SMG^3$ dominates
as the group to be considered. However
for the purpose of finding the multiple point
$U(1)$ coupling, it can be argued that we can approximately ignore the
interaction between  the
Abelian and non-Abelian subgroups  provided  we  identify  the  $U(1)_i$
factors in  $U(1)^3$  with  the  factor  groups  $SMG_i/(SU(2)  \times
SU(3))_i$ $ (i\in\{Peter, Paul, Maria\})$. In  this  approximation, we
essentially treat $SU(3)^3$, $SU(2)^3$ and $U(1)^3$ separately.
We shall now address the $U(1)$ \dofx by
endeavouring the construction of  some rather rough approximations
to  the  phase diagram for a lattice gauge theory with the gauge group
$U(1)^3$.  In
order to provoke the many possible phases $(K,H)$,
including in principle  the
denumerable infinity of  ``phases'' involving  the  discrete
subgroups of $U(1)^3$, it is necessary to use a  functional  form  for
the plaquette action that is quite general.

\subsection{Special problems with $U(1)^3$}

In the case of the non-Abelian subgroups of the $SMG$
that  we  have  dealt  with  in  earlier
work\cite{nonabel,albu}, the correction
factor in going from the multiple point couplings of $SMG^3$ to the
diagonal subgroup couplings is 3 corresponding to the value of the
number of generations $N_{gen}$. Recall that the diagonal subgroup couplings
are in our model predicted to coincide
with the experimental $U(1)$ coupling after extrapolation to the Planck scale.

However, the relation of the diagonal subgroup couplings to the multiple
point critical  couplings in the case of $U(1)^3$
turns out to be more complicated than for the
non-Abelian $SMG$  couplings.
The resolution of these  complications  helps  us  to
understand the  phenomenological  disagreement  found  when  a  naively
expected correction factor of $N_{gen}=3$ is used in going from the $U(1)$
couplings at  the multiple point  of $U(1)^3$ to the couplings
for the diagonal subgroup of $U(1)^3$.

For the fine-structure  constants  of  the  non-Abelian
groups $SU(2)$ and $SU(3)$, it  was  found  that  experimental  values
extrapolated to the Planck scale agree to within the uncertainties  of
our calculation with the predicted values
$1/\alpha_{diag\;multicr}=3/\alpha_{multicr.}$  (i.e.
the inverse fine-structure constants for the diagonal  subgroups  of the
non-Abelian subgroups of
$SMG^3$). While the factor 3 correction to the  multiple point  inverse
squared coupling values obtained for a  lattice  gauge  theory  yields
rather noteworthy agreement with the experimental values of  non-Abelian
fine-structure constants, the analogous relation does not  hold for
the $U(1)$ gauge algebra (weak hyper-charge). For $U(1)$  a  correction
factor  of  roughly  6 (or 7) is   indicated
phenomenologically. This would naively  suggest  that  at  the  Planck
scale we should postulate something like

\beq U(1)^{6\;or\;7}=\overbrace{U(1)\times \cdots \times
U(1)}^{6\;or\;7\;factors} \eeq

\nin rather than $U(1)^3$ as suggested by our preferred ``fundamental''
gauge group $SMG^{N_{gen}}$ with $N_{gen}=3$.

An explanation for this disparity when
we use $U(1)^3$ as the gauge group (rather than the naively indicated
$U(1)^6$ or $U(1)^7$)
can be sought by considering how
the ``Abelian-Ness'' of $U(1)$ distinguishes it from the non-Abelian subgroups.

\subsubsection{The normalisation problem for $U(1)$ \label{onlyz6}}

For $U(1)$, there is no natural unit of charge in contrast to the
non-Abelian groups $SU(2)$ and $SU(3)$. For these latter,  there is a
way to normalise the fine-structure constants by  means  of  the  commutators.
The commutation algebra  provides  a  means of unambiguously  fixing  a
convention for the gauge couplings that alone pertains to the Yang-Mills
fields without reference to the charge of, for example, a  matter  field;  the
Yang-Mills fields are themselves charged in  the  non-Abelian
case and can therefore be used to define a charge convention.
Essentially this is because the Lie algebra commutator relations
are non-linear and are therefore not invariant under re-scalings of the
gauge potential $g{\bf A}_{\mu}$. Such scalings,  if  not
forbidden, would  of  course  deprive  gauge  couplings  of  physical
significance.

Because such a rescaling is possible in the case of $U(1)$, the weak
hyper-charge fine-structure constant is only normalizable  by  reference
to some quantum of charge. This immediately raises the question of  which
particle should  be  declared  as  having  the  unit  quantum of charge
as its hyper-charge. An equivalent way to address this question is to ask
which $U(1)$-isomorphic factor group of $SMG$ should be identified with the
$U(1)$ on the lattice to give us the critical coupling.

It is only when - on the lattice - the group of real numbers $\br$
(in the covering group $\br\times SU(2)\times SU(3)$ of the $SMG$) is
compactified  to  a  $U(1)$
that a normalisation becomes possible and thereby that the idea  of  a
critical  coupling  acquires  a  meaning.  The  only  remnant  in  the
continuum of having chosen a specific group on the lattice is  the
quantisation rule of the charges (more generally, a constraint on  the
allowed representations) and the  lattice  artifact
monopoles. This suggests that we should take the length of the  $U(1)$
in such a way as to enforce empirical charge quantisation rules.  When
we state that the critical coupling for a $U(1)$ lattice gauge  theory
is given by

\beq  \alpha_{crit}\propto\frac{1}{4\pi\beta_{crit}}=\frac{1}{4\pi\cdot 1.01},
\eeq

\nin the meaning is that this $\alpha_{crit}$ is the fine-structure constant
at the phase transition {\em corresponding to the coupling to the smallest
charge quantum allowed on the lattice}. For the $SMG$ as we define it:

\beq
SMG   \stackrel{def}{=}   S(U(2)\times   U(3))\stackrel{def}{=}
(\br\times   SU(2)\times
SU(3))/\{(2\pi,\bunit^{2\times   2},e^{i\frac{2\pi}{3}}\bunit^{3\times
3})^n|n \in \bz\},  \eeq

\nin the charge quantisation rule
for weak hyper-charge is very sophisticated\cite{skew1,skew2,chi}:

\beq y/2+d/2+t/3=0\;\; (\mbox{mod }1). \label{rule}\eeq
This means that depending on whether the non-Abelian subgroups are
represented trivially or non-trivially, the smallest allowed quantum for the
weak hyper-charge is respectively $y/2=1$ and $y/2=1/6$. This complicated
quantisation rule can be regarded as a consequence of Nature having
chosen the gauge {\em group}\cite{michel,oraifear}
$S(U(2)\times U(3))$. In spite of the fact that the global structure of this
group imposes the severe restriction (\ref{rule}) on the possible
representations, it still allows all representations that are seen
phenomenologically.

The $U(1)$ centre of $SMG$
is embedded in the latter in a complicated way.
In order to determine the non-Abelian coupling of the $SMG$,
one must relate
the $U(1)$ centre of the $SMG$ and the simple $U(1)$ studied using Monte Carlo
methods on a lattice. Our earlier work
suggests that the disconnected $\bz_2$ and $\bz_3$ centres of respectively
the non-Abelian $SMG$ subgroups $SU(2)$ and $SU(3)$ should both
alone be confined in phases that convene at the multiple point. In
order to respect this requirement in the present work, it is necessary to
require that the class of $\bz_N$ discrete subgroups $\bz_N$
for which there can be phases convening
at the multiple point that are
solely confined along $\bz_N$ must be as follows: when $\bz_K$ is in this
class, then so are the groups $\bz_K+\bz_2=\bz_{K^{\prime}}$
(where $K^{\prime}$ is the smallest integer multi-plum of $K$ that is
divisible by 2) and the groups $\bz_K + \bz_3=\bz_{K^{\prime\prime}}$ (where
$K^{\prime\prime}$ is the smallest integer multi-plum of $K$ divisible by 3).
Hence, for the phases that convene at the multiple point, the greatest $N$
of a phase that is solely confined w.r.t a  subgroup $\bz_N$
must be such that $N$ is divisible by 2 and 3 and thus
also by 6.

A rule\footnote{In calculating the continuum coupling for a
continuous
Lie (sub)group, the effect on this continuum coupling due to having
discrete subgroups that convene at the multiple point can be
taken into account by calculating as if these discrete subgroups were
{\em totally} confined (instead of being critical as is
the case at the multiple point).}
from our earlier work\cite{long} states that the coupling for a continuous Lie
(sub)group $L$ at the multiple
point is given - to a good approximation - by the critical coupling
for a the factor group $L/\bz_{N_{max}}$ anywhere along the phase border where
the Coulomb-like \dofx corresponding to this factor group are critical. Here
$\bz_{n_{max}}$ denotes the largest discrete subgroup that alone confines in
a phase that convenes at the multiple point. We shall refer to this rule as
the $\bz_{N_{max}}$ factor group rule.


We shall argue below that the largest  discrete
subgroup of the $U(1)$ centre of $SMG$ that is solely confined in a phase
that convenes at the multiple point does not result in a  $U(1)$-isomorphic
factor group of length shorter than that corresponding to the
identification
of $SU(2)\times SU(3)$ with the identity. This corresponds to dividing  the
largest possible non-Abelian subgroup out of the $SMG$; the result
is a factor group isomorphic with $U(1)/\bz_6$:

\beq U(1)/\bz_6\stackrel{-}{\simeq}SMG/(SU(2)\times SU(3)).\label{facmax}\eeq


Consequently, we shall also argue that the
$U(1)$
critical coupling $\sqrt{4\pi\alpha_{crit}}$ obtained using Monte Carlo
simulations of a $U(1)$ lattice gauge
theory is to be identified with the charge quantum of the
factor group $SMG/(SU(2)\times SU(3))$. Subsequently we shall substantiate
that it is reasonable to take this charge quantum as the weak hyper-charge
of the left-handed positron (i.e., $y/2=1$).
The arguments for this choice
are indeed pivotal for the credibility of the proposed  model.  Had  we
for example taken the lattice critical coupling
$\sqrt{4\pi\alpha_{crit}}$ as the hyper-charge of
the left-handed quarks - which are assigned to the
$\underline{2}\otimes\underline{3}$ representation of $SU(2)\times SU(3)$:

\beq \left( \begin{array}{ccc} u_r & u_b & u_y \\ d_r^c & d_b^c & d_y^c
\end{array} \right), \eeq

\nin this would lead  to  an  $\alpha_{crit}(\mu_{Pl})$ that  was  a
factor $6^2=36$ times larger than that obtained
the left-handed positron.

\nin We return to these matters in Section~\ref{resolve}.

\subsubsection{The infinity of discrete subgroups of $U(1)^3$}\label{nor2}
Recall that at the multiple point, there are, in addition to phases confined
w.r.t. continuous subgroups,  also phases that are confined solely
w.r.t.  discrete subgroups. We use as the definition of  confinement that
Bianchi identities can be disregarded in the sense that plaquette variables
can be treated as independent variables. We define
Bianchi variables to be the group product of the plaquette variables
enclosing a 3-volume. The simplest Bianchi variable on a hyper-cubic lattice
are the 3-cubes enclosed by six plaquettes. Bianchi variables are identically
equal to the group identity.
This constraint introduces in general
correlations between the values taken by
plaquettes forming the boundary of a 3-volume. In the case of a first
order phase transition, there is a ``jump'' in the
width of the distribution of plaquette variables in going from a Coulomb to a
confining phase. Our claim is that this ``jump'' is explained
by a change in how effective Bianchi identities are in enforcing correlations
between plaquette variable distributions for different plaquettes forming
the closed surface of a 3-volume. In the Coulomb phase, Bianchi identities
can presume-ably only be satisfied by having the sum of phases (thinking now
of $U(1)$) of the plaquettes bounding a 3-volume add up to zero. At the
transition to a confining phase, the width of plaquette variable
distributions is large enough so that Bianchi identities are readily fulfilled
in any of a  large number of ways in which the values of plaquette variables
can sum to a non-zero multiple of $2\pi$. This greater ease (energetically)
with which
Bianchi identities can be satisfied for a variety of configurations
of values of boundary plaquette variables means that Bianchi
identities are less
effective in causing correlations between plaquette variables which in turn
allows even greater fluctuations in plaquette variables in a sort of chain
reaction that we claim is the explanation for the sudden decrease in the
Wilson loop operator at the Coulomb to confining phase transition.

Were it not for Bianchi identities, the distributions of values taken by
Bianchi variables would correspond (for a simple 6-sided cube) to the 6-fold
convolution of an independent plaquette variable distribution (i.e.,
uncorrelated with the distribution on other plaquettes). For such a
distribution, it turns out that the critical value of the inverse squared
coupling coincides with a change from a distribution centred at the group
identity to
an essentially ``flat'' (i.e.,
Haar measure) distribution. That the 6-fold convolution of independent
plaquette variable distributions becomes rather``flat'' at the critical value
of the coupling concurs nicely with our characterisation of
confinement as the condition that prevails when
the fulfilment of
Bianchi identities has become almost ``infinitely easy''
energetically and can therefore be neglected in the sense that plaquette
variable distributions for different plaquttes can be taken as
approximately independent.

If it is a discrete subgroup that is confined,
there will be subsidiary peaks in the exponentiated plaquette action
$e^{S_{\Box}}$ at nontrivial elements of this discrete subgroup. Confinement
occurs just when the subsidiary peaks are accessed with sufficient probability
so that the 6-fold convolution of the plaquette distribution over elements
of the discrete subgroup leads to comparable probabilities for accessing
all of these discrete subgroup elements
(i.e., when the 6-fold convolution
of a plaquette variable distribution takes values at all elements
of the discrete subgroup with roughly the same probability).

Having in the plaquette distribution
the presence of subsidiary  peaks (i.e., maxima of the distribution of
group elements) at nontrivial elements of discrete subgroups
affects the value of the critical coupling
of the {\em continuous} (i.e., Lie) group degrees of freedom
at the Coulomb to confinement phase transition.
However, once the discrete subgroup is in the confining phase, the dependence
of the Lie group critical coupling on the relative heights of the peaks has
essentially reached a plateau. This is so because fluctuations along the
discrete subgroup are by definition large enough so that
the transition-relevant distribution obtained as the 6-plaquette
convolution of the plaquette
distribution over the discrete group is essentially already flat
so that going deeper into confinement will hardly access more elements of the
Lie group.
So the Lie group coupling is essentially unchanged in going from
the multiple point to
where the discrete subgroup is deeply confined (meaning parameter values
for which the discrete peaks are equally high). Here the fluctuations
along the discrete subgroup and the cosets that are translations of it
are maximal
(i.e., equal probabilities for all the elements in a coset) and one therefore
needs effectively only to consider
the factor group obtained by dividing out the discrete
subgroup. This is the reasoning underlying the $\bz_{N_{max}}$ factor group
rule discussed above. The rule states that
to a good approximation, the multiple point continuous group
coupling equals the critical coupling for this factor group.

\subsubsection{Resolving the $U(1)$ normalisation problem}\label{resolve}

There is the problem with $U(1)$ that the principle of multiple point
criticality suggests that there should even be phases convening at the
multiple in which there is solely confinement of $\bz_N$  subgroups
of arbitrarily large $N$.
This would result in couplings that vanish. However,
if we also give the matter fields some arbitrarily large
number of the charge quanta of the $U(1)$ that corresponds to the lattice
compactification of $\br$, the coupling of these matter particles need not be
zero. But then our prediction would (only) be that the matter
coupling is a rational number times the multiple point critical coupling.

In order to suggest the manner in which this rational factor might arise,
let us speculate in terms of a model for how our universe came about.
First we describe the model; then we formulate two concise statements from
which the model follows. We end this Section by arguing for the validity of
the two statements.

Assume that at high temperatures (e.g.
immediately following the``Big Bang"), the phase that dominates is that having
the largest number of light particles. Recalling that the various phases
convening at the multiple point have the same vacuum energy density
(in Minkowski language), such a phase would constitute
the "highest pressure" phase that could be expected to expand at
the expense of
other phases.  We speculate that such a phase
has an optimal balance of
unconfined fermions and unconfined monopoles. However, unconfined monopoles are
present in phases
that are {\em confined} w.r.t. discrete subgroups (i.e., $\bz_N$ subgroups). So
in terms of our speculative
picture, we do not expect the high temperature dominant phase to be a totally
Coulomb-like phase but
rather a phase confined w.r.t. some discrete subgroups.
In this scenario, we would  claim that the phase in which we live
- ``our'' cold-universe phase - has the maximal number of
monopole charges consistent with having the phenomenologically known
electrically charged particles (quarks and leptons). This leads us to a
system of monopoles (in ``our'' cold-universe phase) causing confinement
for any fraction of the electric charges known to exist phenomenologically.
The picture to have in mind is that ``our'' cold-universe phase is but one
of many degenerate phases that can convene at the multiple point of a cold
universe. We speculate that the reason that only our phase is realized
is because ``our'' phase dominated so effectively at the high temperatures
following the ``Big Bang'' that all other phases disappeared with the result
that these phases are non-existent in the present low-temperature universe.
Had there existed ``seeds'' of these  phases in the present universe, they
could have competed more or less successfully with ``our'' phase.

Let us examine this proposal for ``our'' universe in the context of a
$U(1)$ lattice gauge theory. We denote by the symbol $U(1)_{fund}$
the $U(1)$ gauge group that is associated with the compactification
that establishes the Abelian degrees of freedom on the fundamental
lattice. Let us furthermore assume that there is some integer
$N_{max}$ such that $\bz_{N_{max}}$ is the largest discrete subgroup
of $U(1)_{fund}$ that can confine alone in one of the phases convening at
the cold-universe multiple point.
This corresponds to having Coulomb-like behaviour for the coset-\dofx
of the factor group $U(1)_{fund}/\bz_{N_{max}}$. This means that if a
$\bz_N$ with $N>N_{max}$ confines in a phase that convenes at the
multiple point, it does so not alone but because the continuous $U(1)$
\dofx also confine. Finally, let $N_{our}$ be defined such that $\bz_{N_{our}}$
is the largest
discrete subgroup that alone is confined in ``our'' phase (which is
assumed to be among the phases that meet at the multiple point).

With the assumption of an $N_{max}$, we can
immediately conclude that the $U(1)_{fund}/\bz_{N_{max}}$
representation of $U(1)_{fund}$ has the largest minimum allowed  charge
quantum. Let us denote this as $Q_{max}$. Furthermore, we can conclude
that the smallest allowed charge quantum - namely that of
$U(1)_{fund}$ - is $Q_{max}/N_{max}$.


In terms of monopoles, we have of course the dual situation: denoting
the smallest
allowed monopole charge for $U(1)_{fund}$ as  $m_{fund}$, the factor
group $U(1)_{fund}/\bz_{N_{max}}$ allows monopoles of fractional charge
the only restriction being that these must be multiples of $m_{fund}/N_{max}$.

The above proposal for ``our'' cold-universe phase as a vacuum that allows
monopoles causing confinement
for any fraction of the electric charges (measured in charge quanta of
$U(1)_{fund}$) known to exist phenomenologically
follows as a consequence of the validity of {\bf two statements:}

\begin{enumerate}

\item $N_{our}$ and $N_{max}$ are such that:

\[ N_{max}=6\cdot N_{our} \]

\[ N_{our} \mbox{ not divisible by 2 or 3.} \]


\item The critical coupling $e_{crit}=\sqrt{4\pi\alpha_{crit}}$ for a
$U(1)$ lattice gauge theory determined
using Monte Carlo methods should be identified with the charge quantum
$Q_{max}$ of the factor group $U(1)_{fund}/\bz_{N_{max}}$.

\end{enumerate}

Before substantiating these statements, we first discuss some conclusions
that that follow from assuming the validity of them.

As long as the conditions of {\bf  statement 1} are fulfilled,
$N_{max}$ can be arbitrarily
large without making the coupling at the multiple point vanish (see first
paragraph of (this) Section~\ref{resolve}).
The smallest allowed charge quantum in ``our'' phase is
$N_{our}(Q_{max}/N_{max})\stackrel{def}{=}Q_{our}$; the discrete subgroups
$\bz_2$ and $\bz_3$ are not confined in ``our'' phase. These
discrete subgroups
$\bz_2$ and $\bz_3$ - which are only found once as subgroups of
$\bz_{N_{max}}$ - are confined (alone) only in phases to which are associated
minimum allowed charge quanta larger than $Q_{our}$.
Using the {\bf statement 2}, we can fix the value of the smallest allowed
charge quantum in the phase with $/bz_{N_{max}}$ alone confined
as $\sqrt{4\pi\alpha_{crit}}$ and thus in ``our'' phase as
$Q_{our}=N_{our}\cdot(\sqrt{4\pi\alpha_{crit}}/N_{max})$.

It is now necessary to give an argument for which physical particles should
have $Q_{max}=\sqrt{4\pi\alpha_{crit}}$ as its charge quantum.
As stated above, earlier work leads us to  expect the $\bz_2$ and $\bz_3$
centres  of respectively
$SU(2)$ and $SU(3)$ to confine alone in phases convening at the multiple
point. The phase with $\bz_2\times \bz_3$ confined alone coincides
with the phase with Coulomb-like behaviour for the coset \dofx of the
factor group $SMG/(SU(2)\times SU(3))\stackrel{-}{\simeq} U(1)/\bz_6$
corresponding to the trivial representation of the $SU(2)\times SU(3)$
\dof. The left-handed positron $e^+_L$ is the singlet under
$SU(2)\times SU(3)$ that has the smallest charge.

At the end of this Section, we shall give a speculative argument for why
it is natural that the phase in which there  alone
is confinement of $SMG/(SU(2)\times SU(3))\stackrel{-}{\sim} U(1)/\bz_6$
should be identified with the phase in which there is confinement
solely of the discrete subgroup $\bz_{N_{max}}$ corresponding to
Coulomb-like \dofx for the cosets of
$\frac{U(1)_{fund}/\sbz_{N_{our}}}{\bz_6}\
=U(1)_{fund}/\bz_{N_{max}}$. This
identification puts the hyper-charge of the left-handed positron into
correspondence with the
factor group $U(1)_{fund}/\bz_{N_{max}}$ charge quantum
$\sqrt{4\pi\alpha_{crit}}$.

Use now the usual convention for hyper-charge: $y/2=Q/6Q_L$ (for particles
of hyper-charge $Q$) and associate $(y/2)_{e^+_L}=1$ with $Q=Q_{max}=
\sqrt{4\pi\alpha_{crit}}$ (the $U(1)$ lattice gauge critical coupling).
This determines the hyper-charge
quantum $Q_L$ of ``our'' phase (which has unconfined quarks and leptons
at the Planck scale) as $Q_L=\frac{\sqrt{4\pi\alpha_{crit}}}{6}$.
This is the charge quantum of the
$\underline{2}\otimes\underline{3}$ representation of $SU(2)\times SU(3)$.

The properties ascribed to ``our'' cold-universe phase are contingent
upon the validity of {\bf statements 1} and {\bf 2} above.  Let us now argue
for
the validity of these statements (in reverse order).

{\bf Statement 2} follows basically from  the $Z_{N_{max}}$
factor group
rule for the multiple point coupling of continuous \dofx as discussed on
page \ref{resolve}. This rule states that if the multiple
point for  $U(1)_{fund}$ has contact with a phase in which a discrete
subgroup $\bz_N\in U(1)_{fund}$ is alone confined, then to a very good
approximation,
the {\em multiple point} value of the coupling for the continuous \dofx (i.e.,
the coupling values that reflect the effect of also having a
phase confined alone w.r.t $\bz_N$ that convenes at the multiple point)
is obtained by assuming that this discrete
subgroup is {\em totally} confined (instead of having the multiple point
(i.e., critical) coupling value). This is tantamount to identifying
the multiple point value of the coupling of the continuous \dofx of
$U(1)_{fund}$ with the value of the critical coupling for the factor
group $U(1)_{fund}/\bz_N$. If there are more than one phase convening
at the multiple point that is confined solely w.r.t. some discrete
subgroup, then the best approximation to the multiple point coupling
of the continuous \dofx of $U(1)_{fund}$ is given by the critical value of
the coupling of the factor group with the largest discrete subgroup
$\bz_{N_{max}}$ divided out: i.e., the critical coupling value of
$U(1)_{fund}/\bz_{N_{max}}$. We referred to this approximation as the
$\bz_{N_{max}}$ factor group rule.

The approximate validity of
statement 2) follows using results from Monte Carlo simulations of
lattice gauge theories. From these results the critical value
$e_{crit}=\sqrt{4\pi\alpha_{crit}}$ of the coupling for factor groups
groups of the type $U(1)_{fund}/\bz_N$ with $N=2$ or 3 can be deduced.
As the identification of the critical coupling for
$U(1)_{fund}/\bz_{N_{max}}$ with the critical coupling for
$U(1)_{fund}/\bz_N$ ($N=2$ or 3) is good even for $N<<N_{max}$, the approximate
validity of statement 2) follows.

To establish the validity of {\bf statement 1},
write as above $N_{max}=pN_{our}$ where $p\in \bz$ and
$N_{our}$ is such that $\bz_{N_{our}}$ is the largest discrete
subgroup of $U(1)_{fund}$ that is confined in ``our'' phase. We
note first that $N_{our}$ cannot be divisible by 2 or 3.
Had this been
the case, we would have respectively the subgroups $\bz_2$ and $\bz_3$
confined in ``our'' phase. This would correspond to a restriction of
the possible Coulomb-like \dofx to those having the charge quantum of
a factor group isomorphic to $SMG/(SU(2)\times SU(3))$. The latter is
a singlet w.r.t $SU(2)\times SU(3)$ and accordingly has a charge
quantum too large to allow the
$\underline{2}\otimes\underline{3}$ representation of
$SU(2)\times SU(3)$ needed for having the phenomenologically observed
left-handed quarks and leptons. Phenomenologically at least, our phase
does not have confinement of quarks and leptons at the Planck scale.


However, in order to have the (unrealized) phases with $\bz_2$ and
$\bz_3$ alone confined among the degenerate cold-universe phases that
convene at the multiple point, it is necessary that $p$ be divisible
by 2 and 3: $p=q\cdot 6$. To establish statement 1) however,
we need to
argue that $q=1$. This somewhat speculative
argument goes as follows. Let us imagine that there are extra \dofx
that are hidden from us but which also tend to go into different
phases. Let us speculate that the extra hidden \dofx influence the
form of our ``fundamental'' Lagrangian. So really our ``fundamental''
Lagrangian is an effective Lagrangian; which effective Lagrangian is
realized as our ``fundamental'' Lagrangian can depend on which phases
that hidden \dofx are in. It is important for the argument that the
difference that these extra \dofx can make as to which effective
Lagrangian is realized as our ``fundamental'' Lagrangian can even be
manifested as different numbers of quanta of $U(1)_{fund}$ for quarks
and leptons for different effective Lagrangians. From this point of
view, figuring out which phase would have maximum pressure
immediately following
the
``Big Bang'' also requires looking at different
``possible'' effective Lagrangians (corresponding to
hidden \dofx being in
different phases and even perhaps having quarks and leptons
made up of
different numbers
of quanta of $U(1)_{fund}$)  before
``deciding'' on what our
``fundamental'' Lagrangian should be. These
different ``fundamental ''
Lagrangians (i.e., different effective
Lagrangians among which ours is found)
are different because the extra
to us hidden \dofx of other fundamental
theories can be in phases
having various
different minima. Using as input that observed quarks and
leptons must not be confined, this picture favours a choice for our
``effective''
Lagrangian that corresponds to quarks and
leptons having the largest possible
number
of the charge quanta of
$U(1)_{fund}$; i.e., the largest possible number of
the quanta
$Q_{max}/N_{max}$. This allows the largest possible discrete
subgroup
to be confined in ``our'' phase and accordingly the greatest
number of monopoles
consistent with having observed fermions.

Another
way of putting this is that
phenomenology tells us that $\bz_2$ and
$\bz_3$ cannot be confined in our
phase. So the corresponding
monopoles are not available for helping to have
a high pressure at the
high temperatures immediately following
the Big Bang. However, all
possible other monopoles can help create high
pressure at high
temperatures; the corresponding discrete subgroups are expected to be confined
in ``our''
phase.
The argument is that when the hidden \dofx can go into one or
another phase
that
lead to one or another ``effective'' Lagrangian
for us, the effective
Lagrangian that can be expected to become our
``fundamental'' Lagrangian
is one
that
doesn't ``waste'' monopoles in
the sense that the charge quanta of ``our''
phase (i.e., of the factor-group $U(1)_{fund}/\bz_{N_{our}}$)
do not consist of a smaller number of  fundamental
quanta $Q_{max}/N_{max}$
than absolutely necessary
in order to have
the phenomenologically forbidden $\bz_2$ and $\bz_3$
monopoles
convene in (unrealized) cold-universe degenerate phases
convening at
the multiple point\footnote{E.g., if there were two effective Lagrangians
${\cal L}_{eff\;1}$ and ${\cal L}_{eff\;2}$  - one leading to
$N_{max}=42\cdot N_{our_{1}}$ and the other to
$N_{max}=6\cdot N_{our_{2}}$ (assuming $N_{max}$ the same in both cases) -
we would expect ${\cal L}_{eff\;2}$ to be be realized as {\em the} ``our''
effective Lagrangian because
$Q_{our_{2}}= N_{our_{2}}\frac{Q_{max}}{N_{max}}=\frac{Q_{max}}{6}$ is
larger than $Q_{our_{1}}=\frac{Q_{max}}{42}$. Relative to the Lagrangian
${\cal L}_{eff\;2}$, the Lagrangian ${\cal L}_{eff\;1}$ lacks a
confined $\bz_7$ subgroup and therefore the pressure contribution from the
corresponding monopoles.}.
This dictates that $N_{max}$ is just a single
factor 6 larger than $N_{our}$ so that $q=1$ above as we set out to show.



\subsection{Portraying $U(1)^3$ and its subgroups}\label{sec-portray}

The  phase  diagram  for the  group   $U(1)^3\stackrel{def}{=}
U(1)_{Peter}\times
U(1)_{Paul}\times U(1)_{Maria} \subset SMG^3$ can be  expected  to  be
rather  complicated  because  of  its  many  subgroups.  There   is   a
denumerable infinity of compact subgroups of $U(1)^3$ (discrete  as  well  as
continuous subgroups ranging in dimension  from  zero  to  three).  We
shall seek an approximate $U(1)^3$ phase diagram in the context  of  a
Lattice gauge theory with a Manton action.

As mentioned above, even a continuum action term of  for  example  the
form $\int d^4x F_{\mu\nu}^{Peter}F^{\mu\nu\;Paul}$ is invariant under
gauge transformations in the case of Abelian groups such  as  $U(1)^3$
simply  because  $F_{\mu\nu}^{Peter}$   and   $F^{\mu\nu\;Paul}$   are
separately gauge invariant
\footnote{Under a gauge transformation, we have
\beq Tr[F_{\mu\nu}^{Peter}F^{\mu\nu\;Paul}]\rightarrow Tr[\Lambda^{-1\;Peter}
F_{\mu\nu}^{Peter}\Lambda^{Peter}\Lambda^{-1\;Paul}F^{\mu\nu\;Paul}
\Lambda^{Paul}] \neq Tr[F_{\mu\nu}^{Peter}F^{\mu\nu\;Paul}] \eeq
\nin unless gauge transformations commute with the $F_{\mu\nu}^{I}$'s
 $I \in \{``Peter", ``Paul",\cdots \}$.}.
In particular, a Manton action can have  a
term of this type and therefore a general Manton action can be written

\beq        S_{\Box;,Man}(\theta^{Peter},\theta^{Paul},\theta^{Maria})
=min\{\hat{\theta}^ig_{ik}\hat{\theta}^k|
\hat{\theta}^j=\theta^j   \mbox{   mod   }(2\pi)\}\eeq
\nin where $i,k\in
\{Peter, Paul,Maria\} \label{firstc}$ and $g_{ik}$ is the metric tensor.

We may choose more general coordinates  by  defining  new  coordinates
$\theta^i$ as linear combinations of the old  ones  $\tilde{\theta}^j$:
$\theta^i \rightarrow K^i_{\;k}\tilde{\theta}^k$. Under such a transformation,
an action term of for example the  type  $(F^{Peter}_{\mu\nu})^2$  may
transform into a linear combination involving also
terms of the type  $F^{\mu\nu\;Peter}F^{Paul}_{\mu\nu}$  and  vice
versa. Also, the identification $mod$ $2\pi$  is  transformed  into  a
more general identification modulo a lattice $L$ in the covering group
${\bf R}^3$:

\beq \vec{\theta}\;\;\;\;\;\widetilde{\mbox{\tiny   identified}}
\;\;\;\;\;\vec{\theta}+\vec{l} \;\;\mbox{where}\;\; \vec{l}\in L
\label{secondc}
\eeq

\nin The meaning of  (\ref{secondc}) is that $\vec{\theta}$ and
$\vec{\theta}+\vec{l}$  corresponds
to the same group element of $U(1)^3$.

Because the requirement of
gauge invariance for an action defined  on  the  Abelian  gauge  group
$U(1)_{Peter}\times U(1)_{Paul}\times U(1)_{Maria}$ does not  prohibit
linear combinations of  $F^{\mu\nu}_{Peter}$,  $F^{\mu\nu}_{Paul}$  and
$F^{\mu\nu}_{Maria}$ that can lead to bilinear terms of the type
$F^{\mu\nu}_{Peter}F_{\mu\nu\;Paul}$, there are many possible formulations
corresponding to the same physics (this assumes  of  course  that  the
functional form of the action and the quantisation rules  are  changed
appropriately in going from one formulation to  another).
So points in  the  phase  diagram  should  correspond  to  equivalence
classes of formulations having the same physics.

The gauge group $U(1)^3$ is a (compact) factor group of  the  covering
group ${\bf R}^3$ obtained by dividing out  a  discrete  subgroup  $L$
isomorphic to ${\bf Z}^3$ that we refer to as  the
identification lattice $L$. This is just  the  3-dimensional  lattice  of
elements of ${\bf R}^3$ that are identified with the unit  element  in
going to $U(1)^3$. If we assume that ${\bf R}^3$  is  provided
with an inner product, there will  be  a
recipe for constructing a unique Manton action

\begin{equation}
S_{\Box}(\vec{\theta})=min\{
                  \vec{\theta}^{\prime T}{\bf  g}\vec{\theta}^{\prime}
|\vec{\theta}^{\prime}\in     \vec{\theta}+L      \}.      \label{min}
\end{equation}

\noindent where ${\bf g}$ denotes the metric tensor. The point is that
we construct the metric ${\bf g}$ so that it describes the Manton action.
The  expression
(\ref{min}) is just the generalisation of (\ref{firstc}) to the case of
an arbitrary  choice  of  coordinates  instead  of  the  special  case
in (\ref{firstc}) where coordinates  are  referred  to  basis  vectors
$\vec{l}\in L$.

For ease of exposition, it is useful to consider $U(1)^2$ as a representative
prototype for $U(1)^3$.
Physically different Manton actions correspond to different classes of
isometric-ally  related  embeddings  of  the   identification   lattice
into the Euclidean plane (i.e., $\br^2$ provided with the action-related
metric).
A  pair  of
embeddings where  one  is  rotated  w.r.t.  the  other  correspond  to
physically the same Manton action. Such rotations could be implemented
by coordinate transformations that transfers the coordinate  set  from
one embedding into being the coordinate set of the rotated  embedding.
Obviously  the  two  lattice  constants  (call  them  $a_{Peter}$  and
$a_{Paul}$) and the angle (call it $\phi$)  between  the  two  lattice
directions  are  isometric-ally  invariant   (i.e.,   invariant   under
rotations).
Hence the specification of the properties of a  physically
distinct Manton action (for $U(1)^2$) requires three parameters. These
can be taken as the
three independent matrix elements of the metric  tensor.  We  re-obtain
the coordinate choice (\ref{firstc}) by  adopting  as  our  coordinate
choice  the  requirement  that  the  identification  lattice  has  the
coordinates\footnote{We require of  this  coordinate  system  that  it
allows the group composition  rule  (denoted  with  ``$+$'')  for  two
elements             $(\theta_{Peter},\theta_{Paul})$              and
$(\theta_{Peter}^{\prime},\theta_{Paul}         ^{\prime})$:
$(\theta_{Peter},\theta_{Paul})+(\theta_{Peter}^{\prime},\theta_{Paul}
^{\prime})         =          (\theta_{Peter}+\theta_{Peter}^{\prime},
\theta_{Paul}+\theta_{Paul}^{\prime})$.}

\beq 2\pi(n_{Peter},n_{Paul}) \;\;\;\mbox{with }\;\;\
n_{Peter}, n_{Paul} \in {\bf  Z}.
\label{coord2} \eeq

We give now a concrete example. Using the coordinates (\ref{coord2})
for the identification lattice,
the class of  embeddings  corresponding  to  a
given Manton  action  $S_{\Box}(\vec{\theta}$)  given  by  (\ref{min})  is
specified by the metric tensor

\begin{equation} {\bf g}=\left( \begin{array}{cc} g_{11} &  g_{12}  \\
g_{21}  &  g_{22}   \end{array}   \right)   =\left(   \begin{array}{cc}
\frac{\beta_{Peter}}{2}                                              &
\sqrt{\frac{\beta_{Peter}\beta_{Paul}}{4}}\cos\phi                \\
\sqrt{\frac{\beta_{Peter}\beta_{Paul}}{4}}\cos\phi &
\frac{\beta_{Paul}}{2}      \end{array}
\right). \end{equation}

\nin In particular, for $\vec{\theta}=(2\pi,0)$ it follows that

\beq      S_{\Box\;      Man}(\vec{\theta})=(2\pi,0){\bf       g}\left(
\begin{array}{c}    2\pi    \\     0     \end{array}     \right)     =
\frac{\beta_{Peter}}{2}(2\pi)^2. \label{metten}\eeq

\nin We define

\beq \frac{\beta_{Peter}}{2}(2\pi)^2 \stackrel{def}{=}a^2_{Peter}
\label{dist1dim}. \eeq

\[ \frac{\beta_{Paul}}{2}(2\pi)^2 \stackrel{def}{=}a^2_{Paul}. \]

\nin  where  $a_{Peter}$  and  $a_{Paul}$  denote   respectively   the
identification lattice constants in the respectively the  $Peter$  and
$Paul$ directions along the lattice.

Strictly speaking, two different metric tensors (\ref{metten})  may
correspond to the same physical action because there are different
ways of representing the same physics that are related by
(discrete)  isomorphic  mappings of  the  identification  lattice
into itself. But these discrete ambiguities do not affect  the  number
of (continuous) parameters needed - namely three for $U(1)^2$.

Using the covering group  $\br^3$  with  the  Manton-action  metric  and  the
embedded identification lattice, it is possible to depict, among other
things, the denumerable infinity of  compact  subgroups  of  $U(1)^3$.
Starting at the identity of the covering group  $\br^3$,  it  is  seen
that the identification lattice  induces  a  $U(1)$  subgroup  on  any
direction along which a lattice  point  is  encountered  at  a  finite
distance from the unit element of $\br^3$. Recall from above that  the
lattice constant $a_i$ is  inversely  proportional  to  the  coupling:
$a_i=2\pi \sqrt{\frac{\beta_i}{2}}$ $(i\in \{Peter, Paul, Maria \})$ .
So the larger the distance from the identity to the first  encountered
lattice point along some one-dimensional  subgroup  of  $U(1)^3$,  the
weaker is the coupling for this subgroup. In particular,  if  we  have
$a_i=a_{i\;crit}=2\pi   \sqrt{\frac{\beta_{i\;crit}}{2}}$   for    all
nearest neighbour lattice points, then all other  one-dimensional  subgroups
will be in a  Coulomb-like  phase  and  at  least
somewhat removed from the phase boundaries at which confinement  would
set in.

We want to let the $MPCP$ single out  the identification lattice $L$ -
which of course means a system of couplings - that
will bring the maximum number of phases together. We shall consider phases
corresponding to
subgroups of dimension ranging from 0 to 3  as  candidates  for  phases
that can meet at a multiple point.

If the $Peter$, $Paul$ and  $Maria$  directions  of  the  lattice  are
chosen to be mutually orthogonal (corresponding to a {\em cubic}
identification lattice), we have in this choice a proposal
for  a  multiple
point in the sense that, by  choosing  the  nearest  neighbour  lattice
constants to correspond to critical couplings, we have a Manton action
described by the geometry of this identification lattice such  that
various phases can be reached by  infinitesimal  changes  in
this lattice and  thereby in the  action  form.  By  such  infinitesimal
modifications, one can  reach  a  total  of  8  phases  with  confinement  of
8
subgroups. These subgroups are the ones  corresponding  to  directions
spanned by the 6 nearest neighbour
points  to,  for  example,  the  origin  (i.e.,  unit
element) of the orthogonal lattice: 1 zero-dimensional subgroup  (with
the Manton action, we do not  get  discrete  subgroups  confining),  3
one-dimensional  subgroups,  3  two-dimensional   subgroups   and   1
three-dimensional  ``subgroup''  (i.e.,  the  whole   $U(1)^3$).
For the choice of the orthogonal lattice, the action (\ref{firstc}) is
additive (i.e., without interactions)  in  the  $Peter$,  $Paul$,  and
$Maria$ terms and the diagonal coupling  is  multiplied  by  the  same
factor 3  as  for  the  non-Abelian  couplings  (see  (\ref{diagcoup})
above). However, as already mentioned, an additive action is without
interaction terms. These are important for
the $U(1)$ diagonal coupling.

It turns out that we can get a larger number of phases to convene at the
multiple point using a {\em hexagonal} lattice. Really this refers to
a special way of having interaction terms of the type $F_{\mu\nu}^{Peter}
F^{\mu\nu\;Paul}$ in such a way that there is an abstract symmetry similar
to that of a hexagonal lattice.
The hexagonal identification lattice results in a better implementation
of the $MPCP$. With the hexagonal  choice of lattice, it  is  possible
with infinitesimal departures from a lattice  with  critical  distance
to the nearest neighbours to provoke any one of
12  different  phases  in the ``volume'' approximation (after some slight
extra modifications; see Section (\ref{grvolapprox})
below) or 15 different phases in  the ``independent monopole'' approximation
(Section (\ref{moncon}) below):
{\em one} phase corresponding to confinement of the zero-dimensional subgroups,
{\em six} phases corresponding to confinement of one-dimensional subgroups,
{\em four} phases ({\em seven} in the ``independent monopole'' approximation)
corresponding to confinement of two-dimensional  subgroups
and {\em one}  phase  corresponding  to  confinement  of  the  whole
three-dimensional
$U(1)^3$.  The  choice  of  the  hexagonal  lattice  obviously  better
satisfies the $MPC$ principle. The fact  that  the  hexagonal  lattice
introduces  interactions  between  the  $Peter$,  $Paul$  and  $Maria$
degrees of freedom in the  Lagrangian  is  not  forbidden  for  $U(1)$
contrary to the situation for  the  non-Abelian  couplings  where  such
mixed terms in the Lagrangian would not be gauge invariant (unless they
were of fourth order or higher).

Originally the hexagonal identification lattice
was invented as a way of optimally realizing the multiple point criticality
idea for $U(1)^3$ and its {\em continuous} subgroups.
But we should also endeavour
to have phases confined alone w.r.t. {\em discrete} Abelian subgroups in
contact with the multiple point.
However, it is {\em a priori} not obvious that this hexagonal
identification lattice can be used for implementing the multiple
point criticality principle in the case of the discrete subgroups $\bz_N$
of $U(1)^3$
which,  according to the $MPCP$ should also be present at the multiple point.
For example, it seems unlikely that subgroups of $\bz_2^3$ can in analogy
to the $6+4+1+1=12$ continuous subgroups $U(1)^3$ (in the hexagonal scheme)
separately confine at the multiple point. The reason is that $\bz_2$
does not have sufficiently many
conjugacy classes so that the subgroups of $\bz_2^3$ can have a
generic multiple point at which 12 phases convene inasmuch as $\bz_2^3$
has only 8 elements and consequently only 8 conjugacy classes\footnote{
By including action terms involving several plaquettes it would in principle
be possible to have an action parameter space of dimension high enough to
have a generic confluence of the 12 phases each which is partially confined
w.r.t. a different discrete subgroup of $\bz_2^3$. However,
even assuming that our $MPCP$ were correct,
it might not be sufficiently favourable for Nature to implement it
to this extreme.}. Consequently, at most 8 phases can convene at a generic
multiple point if we restrict ourselves to  single
plaquette action terms and only allow confinement of $\bz_2^3$ and
subgroups thereof.

In general, having a phase for a gauge group $G$ that confines {\em alone}
along an (invariant) subgroup
$H$ requires that the distribution of elements along $H$ is rather
broad {\em and} that the cosets of the factor group $G/H$ {\em alone}
behave in a  Coulomb-like fashion which most often means that the distribution
of these cosets must
be more or less concentrated about the coset consisting of
elements identified with the identity.

Let us think of the hexagonal
identification lattice for $U(1)^2$ (the latter for the sake of
illustration instead of $U(1)^3$)
that is spanned by the variables $\theta_{Peter}$
and $\theta_{Paul}$ say. In the most general case, the action for a $U(1)^2$
gauge theory could be taken as an infinite sum of terms of the type

\beq  a_{nm}cos(n\theta_{Peter}+m\theta_{Paul}) \label{genact}  \eeq

Let us enquire as to what sort of terms could be used to attain criticality
for $\bz_{2\;Peter}\times \bz_{2\;Paul}\subset U(1)^2$ itself as well as for
subgroups
of $\bz_{2\;Peter}\times \bz_{2\;Paul}\subset U(1)^2$. Denote elements
of $U(1)^2$ as $(\theta_{Peter},\theta_{Paul})$ and use additivity in the
Lie algebra as the composition rule:

\beq (\theta_{1\;Peter},\theta_{1\;Paul})\circ
(\theta_{2\;Peter},\theta_{2\;Paul})
=(\theta_{1\; Peter}+\theta_{2\;Peter},\theta_{1\;Paul}+\theta_{2\;Paul}). \eeq

\nin Relative to the identity $(0,0)$, the elements of
$\bz_{2\;Peter}\times \bz_{2\;Paul}\subset U(1)^2$ (each of which
constitutes a conjugacy class) are  $(0,\pi)$, $(\pi,0)$, and $(\pi,\pi)$
(assuming a $2\pi$ normalisation).
Note that the terms in (\ref{genact}) having {\em even} values of both
$m$ and $n$ cannot be used to suppress the probability density at nontrivial
elements of $\bz_{2\;Peter}\times \bz_{2\;Paul}$ relative to the identity
element $(0,0)$; such {\em even} $n$ and {\em even} $m$ terms of (\ref{genact})
therefore
leave $\bz_{2\;Peter}\times \bz_{2\;Paul}$ and its subgroups totally confined.

Note however by way of example that {\em all} terms of (\ref{genact}) with {\em
odd} $n$
and {\em even} $m$ contribute to the suppression of the element
$(\pi,\theta_{Paul})\in \bz_{2\;Peter}\times \bz_{2\;Paul}$ relative to the
element
$(0,\theta_{Paul})\in \bz_{2\;Peter}\times \bz_{2\;Paul}$
(where $\theta_{Paul}\in \bz_{2\;Paul}$ can be anything) and can
therefore be used to render
the subgroup $\bz_{2\;Peter}$ critical (while the distribution over the
elements of the subgroup $\bz_{2\;Paul}$ is flat for any element of
$\bz_{2\;Peter}$ which means that $\bz_{2\;Paul}$ is left totally confined).
We observe that while all such odd-$n$ even-$m$ terms

\beq n=2p+1 \mbox{ for } p\in \bz \label{pee} \eeq

\nin suppress the probability density at
$(\pi,\theta_{Paul}\in \bz_{2\;Paul})$ relative to
$(0,\theta_{Paul}\in \bz_{2\;Paul})$, these odd-$n$ terms
also concentrate probability
density at $p$ different maxima along $U(1)_{Peter}
\setminus \bz_{2\;Peter}$; i.e., at
elements
$(0 <\theta_{Peter} < \pi, \theta_{Paul}\in \bz_{2\;Paul})$.
However these $p$ extra maxima in probability are not ``noticed''
by $\bz_{2\;Peter}\times \bz_{2\;Paul}$ and its subgroups because such
maxima are located at elements of
$U(1)_{Peter}\times U(1)_{Paul}$ that do not coincide with
elements of $\bz_{2\;Peter} \times \bz_{2\;Paul}$. The point to be gleaned
from this example is that for the purpose of rendering the $\bz_{2\;Peter}
\in U(1)_{Peter}\times U(1)_{Paul}$ \dofx critical, we can do the job with
any {\em one} representative from among the infinite number of terms of
(\ref{genact}) having coefficients $a_{nm}$ with $n$ odd and
$m$ anything.
We can therefore make the choice $n=1$ without loss of generality.
This choice will also be seen to be a convenient way to approximately
decouple the action parameters relevant to
\dofx corresponding to continuous subgroups of
$U(1)_{Peter}\times U(1)_{Paul}$
and the \dofx corresponding to discrete subgroups of
$U(1)_{Peter}\times U(1)_{Paul}$.

Generalising the above example, we can enumerate a choice for the
smallest set of parameters $a_{nm}$
in (\ref{genact}) that permits us maximal freedom in trying to get
partially confining phases w.r.t. subgroups of $U(1)_{Peter}\times U(1)_{Paul}$
(including $\bz_{2\;Peter} \times \bz_{2\;Paul}$ and subgroups thereof)
to convene at the multiple point. Such a choice is conveniently made as
follows:

\begin{itemize}
\item confinement alone along $\bz_{2\;Peter}$ and a peaked Coulomb-like
distribution of the cosets of the factor group
$(\bz_{2\;Peter}\times \bz_{2\;Paul})/\bz_{2\;Peter}$ is achieved using any
term $a_{nm}$ of (\ref{genact}) for which with $n$ is {\em even} and $m$ is
{\em odd}; we choose $a_{01}\stackrel{def}{=}\beta_{Paul}$ and set all other
$n$-even, $m$-odd terms equal to zero.

\item confinement alone along $\bz_{2\;Paul}$ and a peaked Coulomb-like
distribution of the cosets of the factor group
$(\bz_{2\;Peter}\times \bz_{2\;Paul})/\bz_{2\;Paul}$ is achieved using any
term $a_{nm}$ of (\ref{genact}) for which $m$ is {\em even} and $n$ is
{\em odd}; we choose $a_{10}\stackrel{def}{=}\beta_{Peter}$ and set all other
$m$-even, $n$-odd terms equal to zero.

\item confinement alone along\footnote{We want
the anti-diagonal subgroup if we want an analogy to the third direction
in the hexagonal identification lattice; however for $\bz_2$ the
anti-diagonal subgroup  coincides with the diagonal subgroup
$\{(1,1),(-1,-1)\}$. Here we have changed to a notation for the elements of
$U(1)_{Peter}\times U(1)_{Paul}$ corresponding to a multiplicative composition
of group elements.}
$\{(1,1),(-1,-1)\}
\subset \bz_{2\;Peter}\times \bz_{2\;Paul}$ and a peaked Coulomb-like
distribution of the cosets of the factor group
$(\bz_{2\;Peter}\times \bz_{2\;Paul})/\{(1,1),(-1,-1)\}$ is achieved using any
term $a_{nm}$ of (\ref{genact}) for which both with $n$ and $m$ is
{\em odd}; we choose $a_{11}\stackrel{def}{=}\beta_{interaction}$ and set all
other $n$-odd, $m$-odd terms equal to zero.

\end{itemize}

\nin This gives us effectively three free parameters with which we can try
to bring  {\em discrete} \pcps together at the multiple point. This choice
using

\beq a_{nm}=a_{10}\stackrel{def}{=}\beta_{Peter},\eeq
\[ a_{nm}=a_{01}\stackrel{def}{=}\beta_{Paul}\] and
\[ a_{nm}=a_{11}\stackrel{def}{=}\beta_{interaction}\] is the most
smooth choice. Other choices for action terms with
$n$ and/or $m$ odd could potentially result in additional maxima
in the probability density that are not centred at elements of
$\bz_{2\;Peter}\times \bz_{2\;Paul}\subset U(1)_{Peter}\times U(1)_{Paul}$
(e.g., for $p\neq 0$ in (\ref{pee})). But these additional maxima would
effectively not influence the distribution of continuum
\dofx as such additional maxima can easily  be suppressed by (dominant)
$n$-even, $m$-even action terms everywhere on $U(1)_{Peter}\times U(1)_{Paul}$
except at elements of $\bz_{2\;Peter}\times \bz_{2\;Paul}$ Representing these
dominant $n$-even, $m$-even action terms by the smoothest ones corresponds to
using just three non-vanishing
parameters to adjust the continuum \dofx along subsets of
$U(1)_{Peter}\times U(1)_{Paul}$:

\beq a_{20}\stackrel{def}{=}\gamma_{Peter}. \eeq
$$ a_{02}\stackrel{def}{=}\gamma_{Paul}$$ and
$$ a_{22}\stackrel{def}{=}\gamma_{interaction}.$$
So we end up with six parameters
where the three $n$-even, $m$-even ones can be used
to bring phases confined w.r.t.  continuous subgroups
of $U(1)_{Peter}\times U(1)_{Paul}$ together at the multiple point.
These parameters are approximately independent of
the parameters $\beta_{Peter}$, $\beta_{Paul}$ and $\beta_{interaction}$
than can be used to bring phases confined w.r.t.  discrete subgroups
of $U(1)_{Peter}\times U(1)_{Paul}$ together at the multiple point.
We end up with an action $S$

\beq S=\gamma_{Peter}\cos(2\theta_{Peter})+\beta_{Peter}\cos\theta_{Peter}+
       \gamma_{Paul}\cos(2\theta_{Paul})+\beta_{Paul}\cos\theta_{Paul} + \eeq
$$    +\beta_{interact}\cos(\theta_{Peter}+\theta_{Paul})
     +\gamma_{interaction}\cos (2(\theta_{Peter}+\theta_{Paul})).  $$

\nin Let us assume that $\gamma_{Peter}$, $\gamma_{Paul}$ and
$\gamma_{interaction}$ have been chosen so as to
bring $U(1)^2$ and the continuous subgroups
of $U(1)^2$ together at the multiple point.
This leaves three approximately independent parameters that can be used
as coefficients to plaquette action terms defined on
$\bz_2\times \bz_2$ and its subgroups. These parameters can be adjusted so as
to bring phases confined w.r.t. subgroups of $\bz_2\times \bz_2$
together at the multiple
point. That we have three (effectively) independent parameters up to a
constant action term
is in accord with $\bz_2\times \bz_2$ having just four elements (i.e., four
possible
conjugacy classes). With three parameters we can have a generic multiple point
at which four phases convene. However, the number of {\em possible} different
phases (regardless of whether they can all meet at the multiple point)
obtainable by varying the parameters
$\beta_{Peter}$, $\beta_{Paul}$, and $\beta_{interact}$ is five.
Two of the five possible phases
correspond to total confinement and
totally Coulomb-like behaviour for $\bz_{2\;Peter}\times \bz_{2\;Paul}$;
the remaining three possible phases correspond to confinement along
1-dimensional\footnote{Strictly speaking,
$\bz_{2\;Peter}\times \bz_{2\;Paul}$ and subgroups hereof are of course all
0-dimensional; when we talk about ``1-dimensional subgroups of discrete
groups'' we mean the (measure zero) sets that coincide with elements of,
e.g., the 1-dimensional subgroup $U(1)_{Peter}\in U(1)^3$.}
subgroups of $\bz_{2\;Peter}\times \bz_{2\;Paul}$ enumerated above in
connection with our procedure for choosing
$\beta_{Peter}$, $\beta_{Paul}$, and $\beta_{interact}$.
However, only two of these three phases with
confinement solely along 1-dimensional
subgroups can convene at a (generic) multiple point. This is different from
the
situation for $U(1)^2$ (i.e., for the continuum); it is
shown elsewhere that in this case, all three phases
that are confined solely along a
1-dimensional subgroups can convene at a single (generic) multiple point.


On the other hand,
for $\bz_N$ (with $N>3$) there are enough conjugacy
classes (and thereby potential action parameters) so that for any of the three
directions $\theta_{Peter}$, $\theta_{Paul}$ and
$\theta_{Paul}-\theta_{Peter}$ in $\bz_3$
we can independently choose to have a somewhat flat distribution
of group elements (corresponding to confinement-like behaviour)) along for
example the
$\theta_{Peter}$ direction
while at the same time having
a peaked distribution of the cosets of the factor group
$(\bz_{N\;Peter}\times \bz_{N\;Paul})/\bz_{N\;Peter}$ (corresponding to
Coulomb-like behaviour for these \dof).
This is of course just the \pcp confined w.r.t. $\bz_{N\;Peter}$. It turns
out that also for $\bz_3$, this is in principle at least just barely possible.

For $U(1)^3$,
an analogous difference between the subgroups $\bz_2^3$ and $\bz_N^3$
($N>3$) is found.
Of the six possible 1-dimensional subgroups of $\bz_2^3$, only three of the
corresponding \pcps
can convene at any (generic) multiple point
as compared to the situation for $U(1)^3$ where six such phases can convene
at the multiple point.

According to the multiple point criticality
principle, we should determine the critical $U(1)$ coupling corresponding to
the multiple point in a $U(1)^3$ phase diagram where a maximum number of
\pcps convene. This also
applies of course to the possible 1-dimensional
discrete subgroups. We deal with these latter subgroups by using an
appropriate correction to the continuum $U(1)$ coupling in a later Section.

Beforehand, it is not known whether it is even numerically possible
to have criticality for the discrete subgroups using the hexagonal symmetry
scheme for the couplings.
At least in the case of $\bz_2$, the subgroups in some directions are lacking
because there are not enough action parameters to bring them all to the
multiple point.
Hence the $\bz_2$ correction  should only have a weight
reflecting the contribution from the fraction of these 1-dimensional
discrete subgroups that (alone) can be confined at the
multiple point. For $\bz_2^3$,
it turns out that only one half (i.e., three out of six) of the hexagonal
nearest neighbour
1-dimensional  subgroups can convene at a (generic) multiple point.
In the
boundary case of $\bz_3^3$, it is not entirely clear as to whether
the contribution should also be
reduced by some factor.

On the other hand, for $\bz_N^3$ ($N>3$), it is not strictly excluded
to have the six 1-dimensional phases at a (generic) multiple point that
correspond to the six analogous phases of  $U(1)^3$.
This  reflects the fact that for $\bz_N^3$
with $N>3$,
there are sufficiently many conjugacy classes\footnote{Strictly speaking,
this is also true for $\bz_3$: there are eight conjugacy classes
corresponding to the eight elements of $\bz_3$. However, it can hardly be
useful to have
separate action terms for elements $g\in \bz_3^2$ and $-g\in \bz_3^2$. So for
the purpose of provoking different \pcps independently, there are effectively
only four conjugacy classes.  But four action parameters are in
principle at least just sufficient
to bring $1+3+1=5$ phases together at a generic multiple point.}
so that the hexagonal identification lattice that is so efficient in getting
phases corresponding to continuous
subgroups of $U(1)^3$ to convene at the multiple point can presumably also
bring the analogous phases of discrete subgroups  $\bz_N^3$ ($N>3$)
together at the multiple point.


When we talk about ``contributions'' of $\bz_N$ subgroups to $\frac{1}{g^2}$,
we are anticipating that in a later Section, we shall make approximate
corrections for our having initially
neglected that there should also be phases convening at the multiple point
for which the various discrete invariant subgroups are alone confining while
the corresponding continuous factor groups behave in a Coulomb-like fashion.
The correction
procedure that we use results in small corrections to the critical continuum
couplings that we loosely refer to as ``contributions'' to the inverse squared
couplings from $\bz_2$, $\bz_3$, etc.

In summary, it is possible for $\bz_N$ discrete subgroups of large enough $N$
to
realize all possible combinations of phases for the (nearest neighbour)
1~-~dimensional subgroups of the hexagonal identification lattice coupling
scheme. These \pcps should also convene at the multiple point; we deal with
this requirement in an approximate way in a later Section by making a
correction to $\frac{1}{g^2}$ for discrete subgroups $\bz_N$ with various
values of $N$.
The  result of the discussion above is that the approximate
correction that will be made to $(\frac{1}{g^2})_{mult\;point}$
coming from taking into account
that we also want to have \pcps w.r.t. $\bz_2^3$ at the multiple point is
reduced
by a factor $\frac{3}{6}=\frac{1}{2}$ relative to the analogous correction
for $\bz^3_N$ ($N> 3$). It may also well be that the contribution in the
marginal case of $\bz_3$ should also be reduced by some factor. These
considerations will be incorporated into the presentation
of our results.

\subsection{Mapping out the phase diagram for $U(1)^3$: approximative
techniques}

\subsubsection{Monopole condensate approximation - outline of procedure}
\label{moncon}
The philosophy of the first approximation to be used to estimate which
phase is obtained for given parameters is that the decisive factor  in
distinguishing the Coulomb-like phase (or Coulomb-like behaviour of some
of the degrees of freedom) from  the  confinement  phase  is  whether
quantum  fluctuations  are  such  that  the  Bianchi  identities   are
important or essentially irrelevant in introducing correlations between
plaquette variables.

That  is  to  say  we  imagine  that  the  phase  transition   between
a ``Coulomb'' and confining phase -  as  function  of  the
parameters $\beta$ - occur when the fluctuations of the  plaquette  variables
take such values that the fluctuation of the convolution of the number
of plaquette variable distributions (coinciding  with  the  number  of
plaquettes bounding a 3-cube - e.g., six  for  a  hyper-cubic  lattice)
become just large enough so as to be essentially spread out over the
whole  group
(or over the elements within the cosets of a factor group)
in question and thereby rendering Bianchi identities
essentially irrelevant.

The idea behind this philosophy is that when the fluctuations  are  so
large that a naive (i.e. neglecting Bianchi constraints) convolution
of  the  6  plaquettes  making  up  the
boundary  of  a  3-cube  fluctuates  over  the  whole  group  (leading
essentially to the Haar measure  distribution),  the  Bianchi-identity
is then assumed to be essentially  irrelevant  in  the  sense  that
each  plaquette
fluctuates approximately independently of the  other  plaquettes  that
form the boundary of a 3-cube. In this situation there is essentially
no (long range) correlations. This is  of  course  the  characteristic
feature of a confining phase.

If, however, fluctuations of the convolutions of  plaquettes  variable
distributions $e^{S_{\Box}}$ for the six plaquettes bounding a  3-cube
do not cover the whole group, the Bianchi identities are important in
the  sense  that  the  constraint  that  these  impose  leads   to   a
correlation  of  plaquette  variable   fluctuations   over   ``long''
distances  (i.e.,  at  length  scales   of at least   several   lattice
constants).  Such  ``long''  range  correlations  are  taken  as   the
characteristic feature of a Coulomb-like behaviour.

The idea of phase determination according to whether the  fluctuations
in plaquette variables are  small  enough  so  that  Bianchi  identity
constraints can introduce ``long'' range correlations or  not  can  be
translated into a  lattice  monopole  scenario:  a  Coulomb-like  phase
corresponds to a scarcity of monopoles while the vacuum of a confining
phase is copiously populated by monopoles. For a single  $U(1)$  gauge
group, a monopole (or rather the cross section in the time track of a
monopole)  is just  a  3-cube  for  which  the  values  of  the
bounding  plaquette  variables  -  defined say by  the  convention   that
Lie-algebra (angle) variables take values in the interval $[-\pi,\pi)$
- have fluctuations large enough so as to get back to the unit element
by first adding up to a circumnavigation  of  the  whole  group.
Such a traversal of the whole $2\pi$ length of the group  as  the  way
the Bianchi identity is realized is tantamount to  having  a  lattice  artifact
monopole. The  confinement  phase  is  characterised  by  the  copious
occurrence of such monopoles.

The  case  where  the  gauge  group  is  $U(1)^3$  is  slightly   more
complicated. As seen above, the group $U(1)^3$ can be  thought  of  as
the cosets of the group $\br^3$ modulo an identification lattice.  A
unique assignment of an  element  of  the  group  $\br^3$  to  each
$U(1)^3$-valued plaquette requires a convention which we  take  to  be
the choice of that element among the  coset  representatives having the
shortest distance to the  zero-element  of  $\br^3$.  With  such  a
convention, we can, for any 3-cube, now ask if the  sum  of  the  $\br^3$
representatives for the surrounding plaquette variables typically
add up to the unit element (as is characteristic  of  the  Coulomb-like
phase) or instead add up to one of  the  nontrivial  elements  of  the
identification-lattice (as is characteristic  of  a  confining  phase)
corresponding respectively to not having  a  monopole  or having  a  monopole
with  some   $N_{gen}$-tuple   of   magnetic    monopole    charges
$2\pi(n_{Peter},   n_{Paul},   n_{Maria})$    $(n_{Peter},    n_{Paul},
n_{Maria}\in \bz)$.

Monopoles come  about  when  the  Bianchi  identities  (one  for  each
of the  $N_{gen}$  $U(1)$  subgroups  labelled   by   names   $``Peter"$,
$``Paul"$ and $``Maria"$) are satisfied by having the values of  the
plaquette variables of a 3-cube add up to a lattice point  other  than
that corresponding to the  identity  element  of  $\br^3$.  In  other
words, a monopole is a  jump  from  the  origin  of  the  $\br^3$
identification lattice to another point of the identification  lattice
that takes place when values  of  the  variables  for  the  plaquettes
surrounding a 3-cube add up to a nonzero multiple  of  $2\pi$  for  at
least one of the $N_{gen}=3$  $U(1)$'s  of  $U(1)^3$  as  the  way  of
fulfilling the Bianchi identities.

Having a  phase  in  which  for  example a one-dimensional subgroup
 -  $U(1)_{Peter}$ say - is  confined
corresponds to having, statistically speaking, an abundance  of cubes
of  the  lattice  for  which   the  monopole   charge
$2\pi(n_{Peter},  n_{Paul},  n_{Maria})$  is  typically  $\pm 2\pi(1,0,0)$
but (depending on couplings) also with less frequent occurrences of the
monopole  charges  $\pm 2\pi(2,0,0)$,  $\pm 2\pi(3,0,0),\cdots$  as  well   as
only occasional monopoles with $n_{Paul}\neq 0$ and $n_{Maria}\neq 0$. Which
phase is realized is  determined  of  course  by  the  values  of  the
couplings. We recall that  the  information  about  the  couplings  is
``built  into''  the  distance   between   lattice   points   of   the
identification   lattice.   Confinement   along   for   example    the
$U(1)_{Peter}$ subgroup corresponds to having  a  less  than  critical
distance between nearest  neighbour  lattice  points  lying  along  the
$U(1)_{Peter}$ subgroup. It is also possible to have confinement along
two dimensional subgroups (including the orthogonal two-dimensional subgroups)
and the entire (three-dimensional) $U(1)^3$.

We want to use the monopole condensate  model  to  construct  a  phase
diagram for $U(1)^3$. A  confining
subgroup is generated in a direction along which the  spacing  between
nearest (identification lattice) neighbours is  smaller  than  that
corresponding  to  critical
coupling values. In general, the  critical
coupling for a given subgroup depends on which  phases  are
realized for the remaining $U(1)$ degrees of freedom.
For example, confinement for a given one dimensional subgroup of $U(1)^3$
occurs for
a  weaker  coupling  when
one or both of the other $U(1)$ \dofx are  confined than when both of these
other
\dofx are in Coulomb-like phases. In  the
roughest  monopole  approximation, these interactions
between phases is ignored. Accordingly, the critical distance in one
direction is taken to be independent  of
the distance between  neigh-boring  identification  lattice  points  in
other directions. This approximation is appropriate if we take the transition
as being second order
because the fluctuation pattern then goes smoothly through
the transition so that the transition for one subgroup does not abruptly
change the fluctuation pattern significantly for another subgroup.

In this approximation, seeking the multiple point  is  easy.  Multiple
point criticality is achieved simply by having the  critical  distance
between identification lattice points in all nearest  neighbour  directions.
In  this
approximation, the number of phases convening at the multiple point is
maximised by having the largest possible number  of  nearest  neighbour
directions (i.e., maximum number of one-dimensional  subgroups).  This
just  corresponds  to  having  the  tightest   possible   packing   of
identification lattice points. In three dimensions  (corresponding  to
$N_{gen}=3$) tightest packing is attained using a  hexagonal  lattice.
The  generalisation  to   $U(1)^3$   for   the   coordinate choice  of
(\ref{coord2}) is that the points  to  be  identified  with  the  unit
element are

\beq 2\pi(n_{Peter},n_{Paul},n_{Maria}) \;\;\;
(n_{Peter}, n_{Paul}, n_{Maria} \in {\bf Z}). \label{coord3} \eeq

\nin and with this coordinate choice the value of the Manton action at
the multiple point is given by

\beq S_{\Box\;Man}(\vec{\theta}(\Box))= \theta^i(\Box)g_{ik}\theta^k(\Box)
\;\; (i,k \in \{Peter, Paul, Maria\}) \label{manmult} \eeq

\nin where

\beq {\bf g}= \frac{\beta_{crit}}{2}\left( \begin{array}{ccc}
1 & \frac{1}{2} & \frac{1}{2} \\
\frac{1}{2} & 1 & \frac{1}{2} \\
\frac{1}{2} & \frac{1}{2} & 1 \end{array} \right). \label{manten} \eeq

Here we review briefly the symmetry properties
of the hexagonal lattice in the metric of (\ref{manten}).
A point of the lattice has 12
nearest neighbours that define a cub-octahedron. Under an isometric
transformation that leaves the identification lattice invariant (as a set),
one of the 12 nearest neighbours be transformed into another one
in 12 ways.
Moreover, the 4 points adjacent to
any one of the 12 nearest neighbour points must be transformed into each other
in 4 ways.
In this way we account for the $4\times 12$ operations that exhaust
the allowed symmetry operations of the point group characterising the symmetry
of the hexagonal lattice.

\begin{figure}
\centerline{\epsfxsize=\textwidth \epsfbox{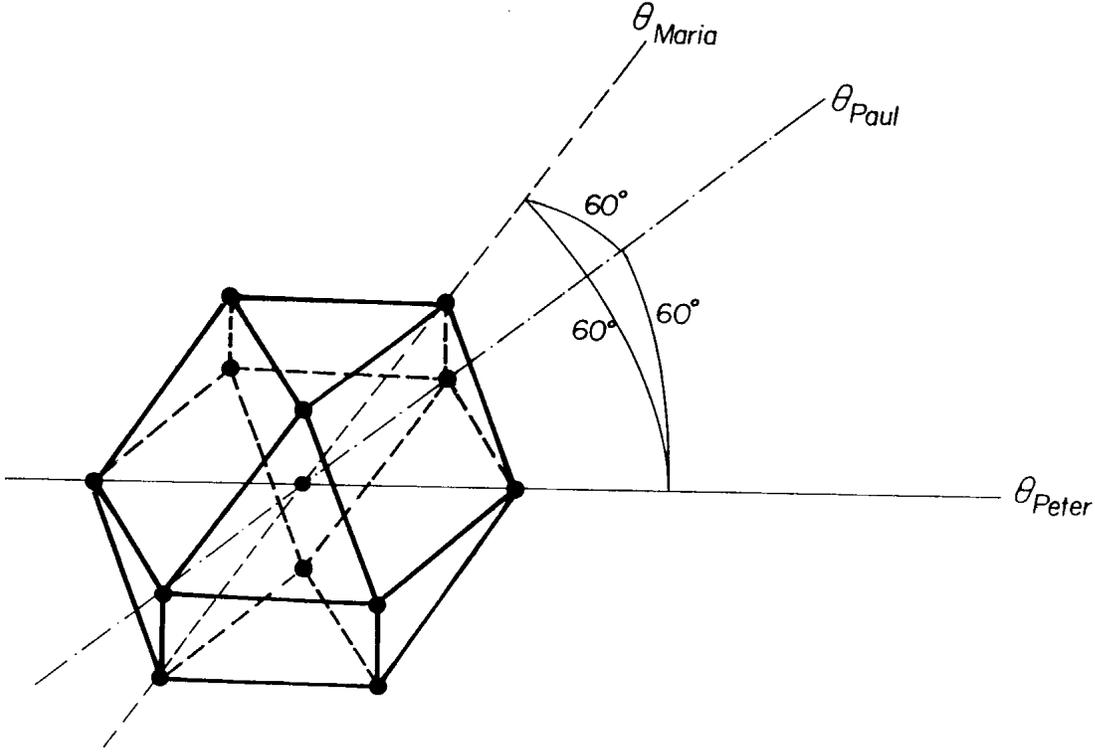}}
\caption[figcuboct]{\label{figcuboct} The nearest neighbours of a chosen point
in the identification lattice form a cub-octahedron. The metric used is that
which corresponds to taking the squared distance as the Manton action.}
\end{figure}

For the purpose of elucidating the symmetries of the hexagonal identification
lattice, it is useful to introduce an extra (superfluous) coordinate
$\theta_4$. First let us  rewrite $S_{\Box\;Manton}$
in (\ref{manmult})
as

\beq
\vec{\theta}^T{\bf g}\vec{\theta}=
\vec{\theta}^T\left\{\frac{\beta}{2}\left(\left(
\begin{array}{ccc} 1/2 & 0 & 0 \\ 0 & 1/2 & 0 \\ 0 & 0 & 1/2 \end{array}\right)
+\frac{1}{2} \left(\begin{array}{ccc} 1 & 1 & 1 \\ 1 & 1 & 1 \\ 1 & 1 & 1
\end{array}\right)\right)\right\}\vec{\theta} = \eeq

\[ = \frac{\beta}{4}(\theta_1^2+\theta_2^2+\theta_3^2 +
(\underbrace{-\theta_1-\theta_2-\theta_3}_{\stackrel{def}{=}\theta_4})^2)
=\frac{\beta}{4}\sum^4_{i=1} \theta_i^2
\]
 \nin where $\theta_1=\theta_{Peter},\cdots, \theta_3=\theta_{Maria};
\theta_4=-\sum_{i=1}^3\theta_i$.

In this coordinate system with the superfluous coordinate $\theta_4$, we
have the constraint

\beq \sum_{i=1}^4 \theta_i=0 \eeq

\nin and the hexagonal lattice is  characterised as the set of points with
coordinates

\beq (\theta_1,\theta_2,\theta_3,\theta_4)\in 2\pi{\bf Z}^4. \eeq

In this notation, it is apparent that the symmetry group
for the lattice and the
metric consists of the permutations combined with or without a
simultaneous sign shift of all four coordinates.

Each of the 12 nearest neighbours to any site of the identification lattice
(e.g. the group identity) have, in the 4-tuple coordinate notation, just
two non-vanishing coordinates (that sum to zero). The 1-dimensional subgroups
 correspond to the 6 co-linear pairs of these  12 nearest neighbours.

The
2-dimensional subgroups are of two types. One type, of which there are 4,
are spanned by the identity and any (non-co-linear) pair of the
12 nearest neighbour sites
that have a common non-vanishing coordinate.
A given subgroup of this type contains 6
nearest neighbour sites positioned at the corners of a hexagon; all 6 such
sites of a given 2-dimensional subgroup of this type have a vanishing
coordinate in common; e.g., the 6 nearest neighbours with a ``0''
for the first coordinate belong to the same 2-dimensional subgroup of this
type.
              That there are four such subgroups follows from the fact that
there are 4 possibilities for having a common vanishing coordinate in the
4-tuple notation. The other type of 2-dimensional subgroups - there
are 3 mutually orthogonal such subgroups - are each spanned by 2 pairs of
nearest neighbour sites where the two sites of each such pair have no common
non-vanishing coordinates. There are 3 such pairs:

\beq \begin{array}{c} (\pm 2\pi,0,\mp 2\pi,0) \\ (0,\pm 2\pi,0,\mp 2\pi)
\label{p1}\end{array}\eeq

\beq \begin{array}{c} (\pm 2\pi,\mp 2\pi,0,0) \\ (0,0,\pm 2\pi,\mp 2\pi)
\label{p2}\end{array}\eeq

\beq \begin{array}{c} (\pm 2\pi,0,0,\mp 2\pi) \\ (0,\pm 2\pi,\mp 2\pi,0).
\label{p3}\end{array}\eeq

\nin Any of the 3 pairs (\ref{p1}), (\ref{p2}), (\ref{p3}) span
one of the $\left(\begin{array}{c} 3 \\ 2 \end{array}\right)=3$ orthogonal
2-dimensional subgroups.

The 3-dimensional ``subgroup'' (which of course is the whole $\br^3$ space)
corresponds in the 4-tuple notation to the
(whole) hyper-plane specified by

\beq \{\vec{\theta}| \sum_{i=1}^4\theta_i=0 \}. \eeq

\nin The 0-dimensional subgroup corresponds simply to the identification
lattice site that is chosen as the group identity.

\subsubsection{Group volume approximation}\label{grvolapprox}

In this approximation, which is an alternative to the monopole
approximation, we calculate  the
free energy as a function of the couplings for  each  phase  ansatz  (i.e.
each partially confining phase). The criterion for having a phase in contact
with the multiple point is that there  is  some  region  of  plaquette
action parameter space infinitesimally close to  the  multiple  point
where the corresponding free energy function is the most stable  (i.e., larger
than the free energy functions of all the other phases  that  meet  at
the multiple point). In Section~\ref{facfreeenergy}, the  approximate
expression (\ref{fe}) for the free energy was derived\footnote
{In obtaining this relation, we used Gaussian integrals
in the Lie algebra to approximate group integrals, the approximation of
independent plaquettes for the confined subgroup $H$ (i.e.,
Bianchi  identities  are  neglected), and a weak  coupling  mean  field
description  for the Coulomb phase \dofx $G/H$.}
{\em per active link}\footnote{ For a 4-dimensional
hyper-cubic lattice, there are 3  active  links  per  site  (i.e.,  the
number of dimensions reduced by the  one  dimension  along  which  the
gauge is fixed) and 6 plaquettes per site. This  yields  2  plaquettes
per active link. So the quantity $\log Z$ per active site is the half of the
quantity $\log Z$ per plaquette.}. We use the notation
$\log Z_{H \triangleleft G}$  for the free energy function corresponding to
the phase for which $H$ is the largest confined invariant
subgroup of the gauge group $G$:

\beq (\log Z_{H\triangleleft G})_{\mbox{\tiny per active link}}=
\log\left [ \frac{(\pi/6)^{\frac{d_G}{2}}}{{\bf \beta}_G^{\frac{d_G}{2}}vol(G)}
\right ] +
\log \left [ \frac{(6\pi)^{\frac{d_H}{2}}}{{\bf \beta}_H^{\frac{d_H}{2}}vol(H)}
\right ]. \label{long1} \eeq

\nin where

\beq {\bf \beta}_{H}^{\frac{1}{2}dim\;H}\stackrel{def}{=}
\prod_i \beta_i^{\frac{1}{2}dim\;H_i} \eeq

\nin and the index $i$ runs over the Lie algebra ideals
\footnote{For example, for $H=SMG$, ${\bf \beta}_H^{\frac{1}{2}dim\;H}=
\beta_{U(1)}^{\frac{1}{2}}\beta_{SU(2)}^{\frac{3}{2}}
\beta_{SU(3)}^{\frac{8}{2}}$
and for $H=U(3)$, ${\bf \beta}_H^{\frac{1}{2}dim\;H}=
\beta_{U(1)}^{\frac{1}{2}}\beta_{SU(3)}^{\frac{8}{2}}$. Note that $vol(U(3))=
\frac{1}{3}vol(U(1))\cdot vol(SU(3))$ because $U(3)$ is obtained
by identifying the  3 elements of the
$\bz_3$ subgroup of the centre of $U(1)\times SU(3)$.}
of $H$.

Consider two partially confining phases in the case that  one  of
these is confined w.r.t to the invariant subgroup $H_I$ and the  other
is confined w.r.t. the invariant  subgroup  $H_J$.  At  any  point  in
parameter space  where  these  two  partially  confining  phases  meet
(including the multiple point of course) the condition to be satisfied
is $\log Z_{H_I\triangleleft G}= \log  Z_{H_J\triangleleft  G}$.  This
together with (\ref{long1}) leads to the following  condition  that  is
fulfilled at any point on the phase boundary separating these two phases:

\beq \log(6\pi)^{\frac{dim(H_J)-dim(H_I)}{2}}=\log
\frac{{\bf \beta}_{H_J}^{\frac{dim(H_J)}{2}}vol(H_J)}{{\bf
\beta}_{H_I}^{\frac{dim(H_I)}{2}}vol(H_I)}.
\label{long2} \eeq

\nin We  want  of  course  to  consider (\ref{long1}) in the   special
case   for   which
$G=U(1)^3$.


Using here a slightly different notation, designate by $\log Z_{H_n}$ the
free energy
per active link for  the
phase ansatz  for  which  one  of  the  above-mentioned  n-dimensional
subgroups $H_n$ of $U(1)^3$ ($dim(H_n)=n$; $n\in \{0,1,2,3\}$), is confining
and  the
factor group $U(1)^3/H_n$ behaves in a Coulomb-like way
($H_n$ could be one of the
1-dimensional      subgroups: e.g.,       $H_1=U(1)_{Peter}$
say). Let us denote by $a$ the lattice constant of the identification lattice.
Rewriting (\ref{long1}) and specialising to the case of  the  gauge
group $G=U(1)^3$ and $H_J=H_n$ reveals the dependence of  the
free energy per active link on the quantity log$a$:

\beq \log Z_{H_n}= C-(dim(U(1)^3)+dim(H_n))\log a \eeq

\nin where $dim(U(1)^3)=3$ and $dim(H_n)$ are respectively  the
dimension  of  the
gauge group (i.e., the $U(1)^3$ part of $SMG^3$) and the dimension of the
subgroup $H_n$  and  $C$  is  a
quantity that does not depend on the identification lattice constant $a$.
The slope of the various phase ans\"{a}tze is just

\beq \frac{d\log Z_{H_n}}{d\log a}=
-(dim(U(1)^3)+dim(H_n)). \label{deriv} \eeq

\nin Upon rewriting
(\ref{long2}), one  obtains  for  the  condition  defining  the  phase
boundary between the phase with confinement along the subgroup $H_n$
and the phase with confinement along $H_m$ the equation

\beq (6\pi)^{(dim(H_n)-dim(H_m))/2}=
\frac{(\frac{a^2}{2\pi^2})^{dim(H_n)/2}c_n(2\pi)^{dim(H_n)}}
{(\frac{a^2}{2\pi^2})^{dim(H_m)/2}c_m(2\pi)^{dim(H_m)}} \;\; (n,m\in
\{0,1,2,3\})
\label{cond} \eeq

\nin where the volume $vol(H_n)$ of the subgroup $H_n\subseteq U(1)^3$,
measured  in
the coordinate $\theta$, is 

\beq  vol(H_n) = c_n (2\pi)^{dim(H_n)}. \label{cn}\eeq
The quantity $c_n$ is  a
factor associated with the subgroup $H_n$ that depends on the geometry
of the identification lattice.

As an example, consider  first  a  cubic  identification
lattice (actually we shall end up using an hexagonal  lattice  as  this
better satisfies the principle of multiple point criticality). For the
cubic lattice  with
$a=a_{1\;crit}\stackrel{def}{=}2\pi\sqrt{\frac{\beta_{crit}}{2}}$,  it  is
possible  to  have  the
confluence  of three  phases of the type  corresponding  to   1-dimensional
subgroups of $U(1)^3$ at a multiple point - namely those corresponding
to the 1-dimensional subgroups along the $Peter$, $Paul$, and  $Maria$
directions     of     the     lattice     having     $a_{1\;crit}=2\pi
\sqrt{\frac{\beta_{crit}}{2}}$ (the subscript ``1'' on $a_{1\;crit}$
denotes that it is a one dimensional subgroup that is critical).
Furthermore, in the case of the  cubic
identification lattice, it will be seen that phases corresponding to
all subgroups of $G=U(1)^3$
are simultaneously  critical   when   the identification  lattice
constant   $a=a_{1\;crit}=2\pi \sqrt{\frac{\beta_{crit}}{2}}$.
This  follows by observing  that  the
free energy $\log Z_n$ ($n\in \{0,1,2,3\}$) for the different ans\"{a}tze
(i.e., confinement along the various possible subgroups)
are equal for the same value of the identification lattice constant $a$
(i.e.,  for
$a=a_{1\;crit}$) because the constants $c_n$ in (\ref{cond})  are
independent of the dimension $dim(H_n)$  of  the  subgroup  (and  therefore
equal). Hence the condition (\ref{cond})  that  defines  the  boundary
between two partially confining phases is independent of dimension. This then
means that  for  the
unit cell of the cubic  identification  lattice,  all  the  quantities
$\log Z_n$ ($n\in  \{0,1,2,3\}$)  intersect  for  $a=a_{1\;crit}=2\pi
\sqrt{\frac{\beta_{crit}}{2}}=2\pi \sqrt{\frac{1.01}{2}}=4.465$. So the use
of the cubic identification lattice with $a=a_{1\;crit}$ shows that it
is possible to have a multiple point at which  8  partially  confining
phases are in contact: there  is  one  totally  confining
phase (corresponding to $H_3$), three phases corresponding to  three
2-dimensional subgroups $H_2$, three phases corresponding  to  three
1-dimensional subgroups  $H_1$,  and  a  totally  Coulomb-like  phase
corresponding to $H_0$. In particular, the coupling corresponding to
the diagonal subgroup of $U(1)^3$ (in the first approximation, this is
the coupling that we  identify  with  the continuum  $U(1)$
coupling) is down by a factor  $\sqrt{3}$  relative  to  the  critical
coupling for a $U(1)$ lattice gauge theory. This  follows  because  the
inverse of the ratio of the length of  the  diagonal  to  the critical lattice
constant  is $\sqrt{3}$. Phenomenologically, a  factor
of roughly $\sqrt{6}$ rather than $\sqrt{3}$  is  needed  so  we  must
conclude that for the $U(1)$ continuum coupling, the prediction of the
multiple point criticality  principle  using  a  cubic  identification
lattice is at odds with experiment.

However,   the  multiple  point  criticality
principle states that we should  seek the values of  the  continuum
$U(1)$ coupling at a point in parameter space at which a {\em maximum}
number of phases come together. We have already seen that
         for       a hexagonal identification  lattice  in  the  covering
group $\br^3$ of the gauge group $U(1)^3$,
we can, in terms of the 12 nearest neighbours of a site in the
hexagonal identification lattice, identify a  total
of 15 subgroups corresponding to 15 partially confining  phases. Even though
we shall discover in the sequel that 3 of these 15 \pcps - the 2-dimensional
``orthogonal'' phases given by (\ref{p1}-\ref{p3}) - are not
realistically realizable in the volume approximation inasmuch as these phases
are ``pushed'' too far away from the multiple point in the volume
approximation, there
remains 12 partially confining phases that can be made to convene at the
multiple
point. This is, in view of the multiple point criticality principle,
an improvement upon  the total of 8 phases that can be realized at the
multiple point in the case of the cubic identification lattice.

It will be  seen  that  the  price  we  must  pay  for realizing these 12
remaining partially confining phases at the multiple point in the case of
the hexagonal identification lattice instead  of  the  8  partially
confining phases of the cubic identification lattice
is that these 12 phases no longer come together exactly at a common  value  of
the identification lattice constant $a$ if we use a pure Manton action
(\ref{min}).

For the hexagonal identification lattice, the problem is that when the
lattice constant $a$ is chosen so that $a=a_{1\;crit}
\stackrel{def}{=}2\pi\sqrt{\frac{\beta_{crit}}{2}}$ corresponding to
criticality for the 1-dimensional subgroups,  this  choice  fixes  the
values of the couplings for the  2-  and  3-dimensional  subgroups  at
sub-critical values. For example, for
$a=a_{1\;crit}$, the free energy functions $\log  Z_0$  and  $\log
Z_1$ are equal corresponding to the  coexistence  of  the  totally
Coulomb-like phase and the six phases that are
confined along 1-dimensional subgroups. However, if  for  example  one
wishes to have coexistence of the totally Coulomb-like phase and the four
phases that are confined along the four 2-dimensional subgroups, it will be
seen (Table~\ref{tab2})  that $\log a$ must be decreased by
$\frac{1}{4}\log  (4/3)$.
But this reduction in $\log a$ would put the phases  corresponding to
1-dimensional subgroups into confinement.

Information about the cubic and hexagonal lattices are tabulated in
Tables~\ref{tab1} and \ref{tab2}. Table ~\ref{tab1}  pertains  to  the
cubic lattice; Table ~\ref{tab2} to  the  hexagonal
identification lattice.  The entries in the  first  four (five)
rows and  columns  of  Table~\ref{tab1} (Table~\ref{tab2}) give  the
values  of  the identification  lattice
constant $a^2$ (in terms of $a^2_{1\;crit}$) at which pairs
(corresponding to a row
and column heading) of free  energy
phase ans\"{a}tze intersect; i.e., these entries are the quantities

\beq \frac{a^2}{a^2_{1\;crit}}=\left (\frac{c_n}{c_m}\right
)^{\frac{-2}{dim(H_n)-dim(H_m)}}
\;\;\;\;(n,m \in \{0,1,2,3\}) \label{vcf} \eeq

\nin obtained by rewriting (\ref{cond}) and using that $a_{1\;crit}^2=3\pi$
(obtained from (\ref{cond}) with
$n=1$ and  $m=0$. The quantities $a$ and
$a_{1,\;crit}=2\pi \sqrt{\frac{\beta_{crit}}{2}}$ are respectively  the
identification
lattice constant and the critical value  of  the (identification)  lattice
constant. The quantities $c_n$   are  the  volume
correction factors associated with the subgroup $H_n$ ($n \in  \{0,1,2,3\}$).
These are also tabulated in the tables below. All the
volume correction factors are unity for the cubic identification lattice.
For the hexagonal lattice, $c_0$ and $c_1$ are both unity whereas
$c_2=\sqrt{3/4}$ and $c_3=\sqrt{1/2}$ corresponding respectively to the
ratio of  area of a minimal parallelogram in the hexagonal lattice to the area
of a simple plaquette in the cubic lattice  and the ratio of the volume of a
(minimal) parallelpipidum of the hexagonal lattice to the volume of a simple
cube in the cubic lattice.

\begin{table}
\caption[table1]{\label{tab1}Parameters pertaining to the cubic identification
lattice. The
entries in the first four rows and columns are all unity because phases
corresponding to all subgroups convene at the multiple point for the
critical value of the coefficient $\frac{1}{e^2_{U(1)\;crit}}$
in the Manton action; i.e., the quantity
$\frac{a^2}{a^2_{1\;crit}}=\left (\frac{c_n}{c_m}\right )^{\frac{-2}
{dim(H_n)-dim(H_m)}}$ $(n,m \in \{0,1,2,3\})$  is unity for
all $m,n\in \{0,1,2,3 \}$. The quantities in the last three columns are
as explained in Table (\ref{tab2}).} \vspace{.5cm}
\begin{tabular}{|c||c|c|c|c||c|c|c|}
\hline
CUBIC & $\log Z_{H_0}$ & $\log Z_{H_1}$ & $\log Z_{H_2}$ & $\log Z_{H_3}$ &
$\frac{d\log Z_{H_n}}{d \log a}$ & \# phases & $c_n$ \\ \hline \hline
$\log Z_{H_0}$  &
1 & 1 & 1 & 1 & -3 & 1 & 1 \\  \hline
$\log Z_{H_1}$ &   & 1 & 1 & 1 & -4 & 3 & 1 \\ \hline
$\log Z_{H_2}$ &  &  & 1 & 1 & -5 & 3 & 1 \\ \hline
$\log Z_{H_3}$ &  &  &  & 1 & -6 & 1 & 1 \\ \hline
\end{tabular}
 \end{table}
\vspace{1cm}
\begin{table}
\caption[table2]{\label{tab2}Parameters pertaining to the hexagonal
identification lattice.
As regards the five rows and first five columns, the entry in the n$th$ column
and the m$th$ row is the coefficient
$\frac{a^2}{a^2_{1\;crit}}=\left (\frac{c_n}{c_m}\right )^{\frac{-2}
{dim(H_n)-dim(H_m)}}$ $(n,m \in \{0,1,2,3\})$ .
This is the quantity by which $\frac{1}{e^2_{U(1)\;crit}}$ must be multiplied
in order that
the phases confined w.r.t. the $n$-dimensional and $m$-dimensional subgroups
can come together at the multiple point. The slope of the
$\frac{d \log Z_{H_m}}
{d \log a}$, calculated from (\ref{deriv}), is given in the sixth column.
Column
seven gives the number of phases of dimension $m$. The entries in
column eight are the ``volume'' correction factors $c_n$ (see (\ref{cn}))
in the hexagonal lattice
relative to the corresponding (unit) ``volumes'' in the cubic lattice.}
\vspace{.5cm}
\begin{tabular}{|c||c|c|c|c|c||c|c|c|}
\hline
HEXAG. & $\log Z_{H_0}$  & $\log Z_{H_1}$ & $\log Z_{H_2 orthog}$ & $\log
Z_{H_2}$ & $\log Z_{H_3}$ &
$\frac{d\log Z_{H_n}}{d \log a}$ & \# phases & $c_n$ \\ \hline \hline
$\log Z_{H_0}$  & 1
& 1 & 1 & $\sqrt{\frac{4}{3}}$ & $\sqrt[3]{2}$ & -3 & 1 & 1 \\  \hline
$\log Z_{H_1}$ &  & 1  & 1 & $\frac{4}{3}$ & $\sqrt{2}$ & -4 & 6 & 1 \\ \hline
$\log Z_{H_2 orthog}$ & & &1 & & 2 & -5 & 3 & 1 \\ \hline
$\log Z_{H_2}$ &  &  &  & 1 &  $\frac{3}{2}$ & -5 & 4 & $\sqrt{\frac{3}{4}}$ \\
\hline
$\log Z_{H_3}$ &  &  &  & & 1 & -6 & 1 & $\sqrt{\frac{1}{2}}$ \\ \hline
\end{tabular}
 \end{table}
\vspace{1cm}

However, the amount by  which
the free energy functions for these different phases fail to intersect
at a common value of the identification lattice  is hopefully
small  enough  to
be dealt with meaningfully by perturbing  the  Manton
action (using 4$th$ and 6$th$ order terms) in such a way as to  allow
12 phases to convene at a multiple point.

We therefore
replace the Manton action (containing by definition  only  second
order terms) by a more complicated action:

\[ S_{\Box,\;Manton} \rightarrow S_{\Box,\;Manton}+S_{\Box,\;h.o.} \]

\nin where $S_{\Box,\;h.o.}$ designates higher than second order terms.
In choosing the higher order terms, we want to use the lowest possible
order terms that bring together the desired phases at the  multiple  point.

The number of additional terms needed depends on how  many  phases  we
want to bring together at the multiple point. As explained  above,
we have decided to settle for the 12 phases (corresponding to
one 0-dimensional, six 1-dimensional, four 2-dimensional, and one
3-dimensional subgroups)  that have the smallest possible volume on
the  hexagonal  lattice and which are not too far from being able to convene
at the multiple point with the Manton action alone.  These 12
phases seem to exhaust the  ones  for  which  a  modification  of  the
couplings using the procedure to be explained below can  be  regarded  as  a
small perturbation; for example, the diagonal subgroup
coupling (with pure Manton action) is so far removed from the
critical couplings of the 12 hexagonal lattice phase discussed above that we
{\em a priori} give up trying to have a phase confined along the
diagonal subgroup in contact with the multiple point. The same applies
presumably to the
2-dimensional ``orthogonal'' subgroups (\ref{p1}-\ref{p3}) as
already mentioned above.

Due to the  high  degree  of  symmetry of the hexagonal lattice,  the
conditions  for  the  criticality  are  identical   for  phases
corresponding to the four 2-dimensional subgroups
and the six 1-dimensional subgroups. So the  number
of parameters we need to get all 12 phases to convene is effectively  that
for four phases (corresponding to the four possible dimensionalities of
subgroups). This requires 4-1=3 parameters. This can be compared to  the
generic number of parameters necessary for the meeting  of  12  phases:
12-1=11 parameters. The point is that the symmetry  of  the  hexagonal
identification lattice allows a non-generic multiple point in an action
parameter space spanned by just three parameters. These can be chosen
as the Manton parameter (i.e., the coefficient  to  the  second  order
term in a Taylor expansion of the action) and two parameters that  are
coefficients to respectively a  4th  and  a  6th  order  term.
These 4th and 6th order terms are to be chosen so as to have
the same symmetry as the hexagonal lattice; otherwise we lose the symmetry
that allows a non-generic multiple point. Without the symmetry, we would
in general need 11 parameters instead of 3. It is also necessary  that
these two terms contribute differently to the  different  free  energy
functions for the different types of subgroup that we want  to
bring to the multiple point. Otherwise we  could  compensate  for  the
effect of these higher order terms for all subgroups by using a single
new effective coefficient to the Manton term. In other words, we  want
our high order terms to be such that these give  {\em  different}  new
effective coefficients to  the  second  order  action  term  for  {\em
different} subgroups. The effective second order coefficient is
defined  as the coefficient in the  Manton action that would give the
same fluctuation width inside the subgroup in question as there would be
with the higher order terms in place.
To this end we use linear combinations of spherical harmonics $Y_{lm}$
with $l=4$ and $l=6$ that have the same symmetry as the  cub-octahedron
(which can be taken as the ``unit cell'' of the hexagonal identification
lattice). These linear combinations, denoted $Y_{4\;comb}$ and
$Y_{6\;comb}$, are invariant under the symmetry of the
cub-octahedron.

In using the $Y_{4\;comb}$ and
$Y_{6\;comb}$
as perturbations to the Manton action, we obtain an effective Manton inverse
squared coupling strength  that varies with the
direction $\vec{\xi}$:

\beq \frac{1}{e^2_{eff}(\vec{\xi})}. \label{invcoupeff} \eeq

\nin Here $\vec{\xi}$ denotes a vector in $\br^3$ (the covering space of
$U(1)^3$).

The desired combinations $Y_{4\;comb}$ and $Y_{6\;comb}$ that have the symmetry
of the cub-octahedron turn out, after a rather strenuous calculation, to be

\beq Y_{4\;comb}=
\frac{2}{3}\sqrt{7}Y_{40}+\frac{4}{3}\sqrt{5}(Y_{43}+Y_{4,-3})/i\sqrt{2} \eeq

\nin and

\beq Y_{6\;comb}=
(-4\sqrt{\frac{3}{35}})Y_{60}+\sqrt{\frac{11}{10}}(Y_{66}+Y_{6,-6})/\sqrt{2}+
(Y_{63}+Y_{6,-3})/i\sqrt{2}. \eeq

\nin These have been calculated in a coordinate system in which the $z$-axis
coincides with a 3-axis of symmetry of the cub-octahedron. In  Table \ref{tab3}
these combinations $Y_{4\;comb}$ and $Y_{6\;comb}$ are averaged over the
1,2 and 3-dimensional subgroups of $U(1)^3$. The fact that both combinations
vanish for $U(1)^3$ (the 3-dimensional subgroup) reflects of course the
property that spherical harmonics vanish when integrated over the surface
of a sphere. Table \ref{tab3} also gives the values of $Y_{4\;comb}$ and
$Y_{6\;comb}$ along the diagonal subgroup of $U(1)^3$.

\begin{table}
\caption[table3]{\label{tab3}The $4th$ and $6th$ order action contributions
needed to realize
12 \pcps at the multiple point. The contributions have the symmetry of the
hexagonal identification lattice.}

\beq \begin{array}{|c|c|c|} \hline \mbox{subgroup} & Y_{4\;\;comb} &
Y_{6\;\;comb} \\ \hline
\langle Y_{l\in\{4,6\}}\rangle_{3-dim} & 0 & 0 \\ \hline
\langle Y_{l\in\{4,6\}}\rangle_{2-dim} & \frac{\sqrt{7}}{4} &
\frac{5}{4}\sqrt{\frac{3}{35}} \\ \hline
        Y_{l\in\{4,6\};\;1-dim} & \frac{\sqrt{7}}{4} &
\frac{117}{32}\sqrt{\frac{3}{35}} \\ \hline
        Y_{l\in\{4,6\};\;diag subgr} & \frac{2\sqrt{7}}{3} &
-4\sqrt{\frac{3}{35}} \\ \hline
\end{array} \eeq
 \end{table}

\begin{figure}
\centerline{\epsfxsize=\textwidth \epsfbox{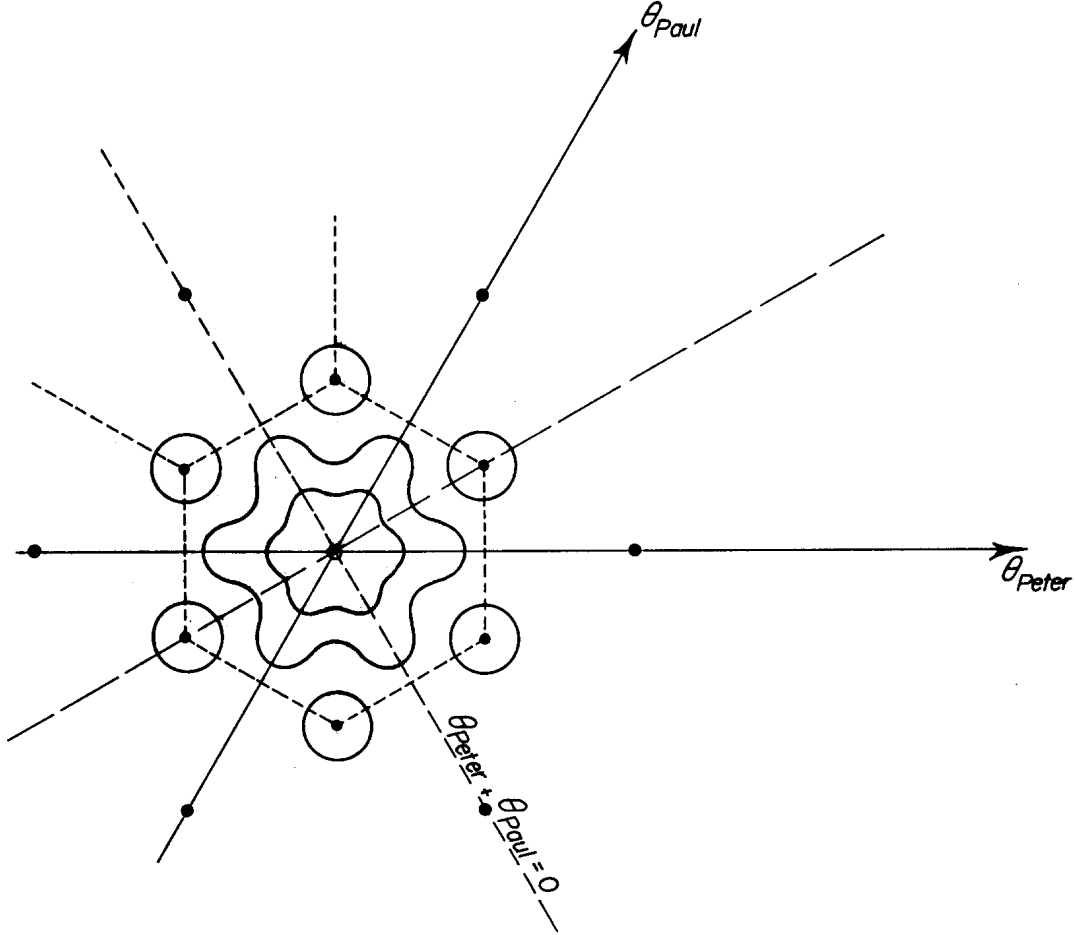}}
\caption[figwebbed]{\label{figwebbedfeet}Contours of constant perturbed Manton
action for $U(1)^2$ represented in the covering group $\br^2$ with the metric
in the plane of the paper that is identified with the Manton action metric.
The hexagonal  lattice of ``$\bullet$'' are points identified in compactifying
from $\br^2$ to $U(1)^2$.
The purpose of the correction - it is sixth order and gives  the contours
a ``webbed feet'' look - is to increase $\log Z$ for the phases with
confinement along one of the three 1-dimensional subgroups - i.e., along the
$\theta_{Peter}$ axis, the $\theta_{Paul}$ axis and along the line given by
$\theta_{Peter}+\theta_{Paul}=0$ - while disfavouring fluctuations along
directions that bisect the angles between these 1-dimensional subgroup
directions. This is accomplished by decreasing the
gradient of the action in these three subgroup directions while increasing
the gradient in
directions that bisect the above-mentioned three subgroups}
\end{figure}
Using the Tables \ref{tab1}, \ref{tab2}, and \ref{tab3}, let us now determine
the coefficients to the
2$nd$ order (i.e. Manton) as well as 4$th$ and 6$th$ order action terms
by using the requirement that averages over the
1, 2 and 3-dimensional subgroups of $U(1)^3$ are equal to the $U(1)$
critical inverse squared coupling $1/e^2_{U(1)\;crit}$ when the volume
correction factors for the hexagonal lattice are taken into consideration.
These latter are given by (\ref{vcf}). Using that $\beta=1/e^2=a^2/2\pi^2$
we can write the condition to be satisfied if the average over the subgroup
$H_n$ - i.e., $\langle 1/e^2(\vec{\xi}) \rangle_{H_n}$ -
is to have a value corresponding
to the boundary between a phase confined along $H_n$ and the totally
Coulomb phase:

\beq \langle \frac{1}{e^2(\vec{\xi})} \rangle_{H_n}=
\left(\frac{c_n}{c_0}\right)^{\frac{-2}{dim(H_n)-dim(H_0)}}
\frac{1}{e^2_{U(1)\;crit}} \label{avcoup} \eeq

\nin where $c_0$ and $H_0$ correspond to the totally Coulomb phase.
Eqn. (\ref{avcoup}) yields three equations -
one for each type of subgroup $H_n$
($n=dim(H_n)$).

For $n=3$ there are no contributions to
$\langle\frac{1}{e^2_{eff}(\vec{\xi})}\rangle_{3-dim\;subgr}$ from
$Y_{4\;comb}$ and $Y_{6\;comb}$. The second order coefficient
$\frac{1}{e_{Manton}^2}$ is therefore determined by the one equation

\beq \langle\frac{1}{e^2_{eff}(\vec{\xi})}\rangle_{3-dim\;subgr}=
\frac{1}{e_{Manton}^2}=\left(\frac{c_3}{c_0}\right)^{\frac{-2}{3-0}}
\frac{1}{e^2_{U(1)\;crit}}=2^{\frac{1}{3}}\frac{1}{e^2_{U(1)\;crit}}.\eeq

The coefficients to $Y_{4\;comb}$ and $Y_{6\;comb}$ - denoted respectively
as $B_4$ and $B_6$ - can be obtained from the equations for
$\langle\frac{1}{e^2_{eff}(\vec{\xi})}\rangle_{1-dim\;subgr}$ and
$\langle\frac{1}{e^2_{eff}(\vec{\xi})}\rangle_{2-dim\;subgr}$.
Assigning dimensionality to the strictly speaking dimensionless quantity
$1/e^2$, we use that
               [$B_4$]=[$\frac{1}{e^4}$] and [$B_6$]=[$\frac{1}{e^6}$].

For $n=1$ we have:

\beq \langle\frac{1}{e^2_{eff}(\vec{\xi})}\rangle_{1-dim\;subgr}^3=
B_6\langle Y_{6\;comb}\rangle_{1-dim\;subgr}+\left(\frac{1}{e^4_{Manton}}+
B_4\langle Y_{4\;comb} \rangle_{1-dim\;subgr}\right)^{\frac{3}{2}}= \eeq

\[
=B_6\frac{117}{32}\sqrt{\frac{3}{35}}+
\left(\frac{2^{\frac{2}{3}}}{e^4_{U(1)\;crit}}+B_4\frac{\sqrt{7}}{4}\right)^
{\frac{3}{2}}=
\left(\left(\frac{c_1}{c_0}\right)^{\frac{-2}{3-0}}
\frac{1}{e^2_{U(1)\;crit}}\right)^3=\left(1\frac{1}{e^2_{U(1)\;crit}}\right)^3
.\]

For $n=2$ we have:

\beq \langle\frac{1}{e^2_{eff}(\vec{\xi})}\rangle_{2-dim\;subgr}^3=
B_6\langle Y_{6\;comb}\rangle_{2-dim\;subgr}+\left(\frac{1}{e^4_{Manton}}+
B_4\langle Y_{4\;comb} \rangle_{2-dim\;subgr}\right)^{\frac{3}{2}}= \eeq

\[
=B_6\frac{5}{4}\sqrt{\frac{3}{35}}+
\left(\frac{2^{\frac{2}{3}}}{e^4_{U(1)\;crit}}+B_4\frac{\sqrt{7}}{4}\right)^
{\frac{3}{2}}=
\left(\left(\frac{c_2}{c_0}\right)^{\frac{-2}{2-0}}
\frac{1}{e^2_{U(1)\;crit}}\right)^3=
\left(\sqrt{\frac{4}{3}}\frac{1}{e^2_{U(1)\;crit}}\right)^3
.\]

\nin The values of the geometric factors $c_n$ are from Table \ref{tab2} and
the values of $\langle Y_{6\;comb}\rangle_{n-dim\;subgr\;H_n}$ and
$\langle Y_{4\;comb}\rangle_{n-dim\;subgr\;H_n}$ $(n=dim(H_n)\in\{0,1,2,3\})$
are taken from Table \ref{tab3}.

Solving these equations for the coefficients $B_4$ and $B_6$ yields

\beq B_4= -0.1463 \mbox{  and   } B_6=-0.7660 \eeq

We have now succeeded in fitting three coefficients of a modified Manton
(i.e. a plaquette action dominated by a second order ``Manton'' term
but having perturbative 4$th$ and 6$th$ order terms) in such a way
that 4 types of phases $H_n$ convene at a multiple point in the sense
that $\langle 1/e^2_{eff}(\vec{\xi}) \rangle_{H_n}$
($n=dim(H_n)\in\{0,1,2,3\}$) is equal to the $U(1)$
critical coupling up to a factor pertaining to the geometry of the hexagonal
identification lattice. Because the modified Manton action has the
symmetry of the hexagonal lattice, multiple point criticality for a phase
corresponding to a given dimension implies multiple point criticality for
all phases corresponding to a given dimension. For this reason we achieve
multiple point criticality for a total of 12 phases. The averaging
$\langle 1/e^2_{eff}(\vec{\xi})\rangle_{H_n}$ can be taken as an average over
all directions within the subgroup $H_n$ using a measure defined by being
invariant under rotations leaving the Manton metric invariant.

So we now have at our disposal a means of calculating a directionally
dependent effective inverse
squared coupling where the directional dependence comes from the
perturbative $4th$ and $6th$ order action terms. In a later section,
we shall want to calculate $1/e^2_{eff}$ in the direction corresponding to
the diagonal subgroup (in a chosen coordinate system).

Having now developed some tools for constructing approximate phase
diagrams for the gauge group $U(1)^3$ in which the (or some chosen)
multiple point can be sought out, we proceed to do calculations in
the next Section (Section~\ref{calculation}).

\begin{flushright}
\begin{tiny}
theu1d.tex 31 May 96 sent 

\end{tiny}
\end{flushright}

\section{Calculation of the numerical value of the continuum coupling}
\label{calculation}

\subsection{Outline of procedure}

The aim  now  is to calculate  the  continuum  $U(1)$  standard  model  weak
hyper-charge  coupling  corresponding  to  the   ``diagonal   subgroup''
coupling at  the  multiple  point  of  the  $AGUT$  gauge  group
$SMG^3$.
In principle, the multiple point should be sought
in a very high dimensional action parameter space that is
also in contact
with a multitude of phases that are alone confined w.r.t discrete
$\bz_N$ subgroups.
In an even  more  correct search for the multiple point involving
phases with confining discrete subgroups, we should really consider Abelian
and non-Abelian groups at the same time (i.e, the full $SMG^3$ or perhaps an
even larger group) because   discrete subgroups having the characteristic of
being non-factorizable could a priori simultaneously involve Abelian subgroups
and centres of semi-simple subgroups.

\begin{figure}
\centerline{\epsfxsize=\textwidth \epsfbox{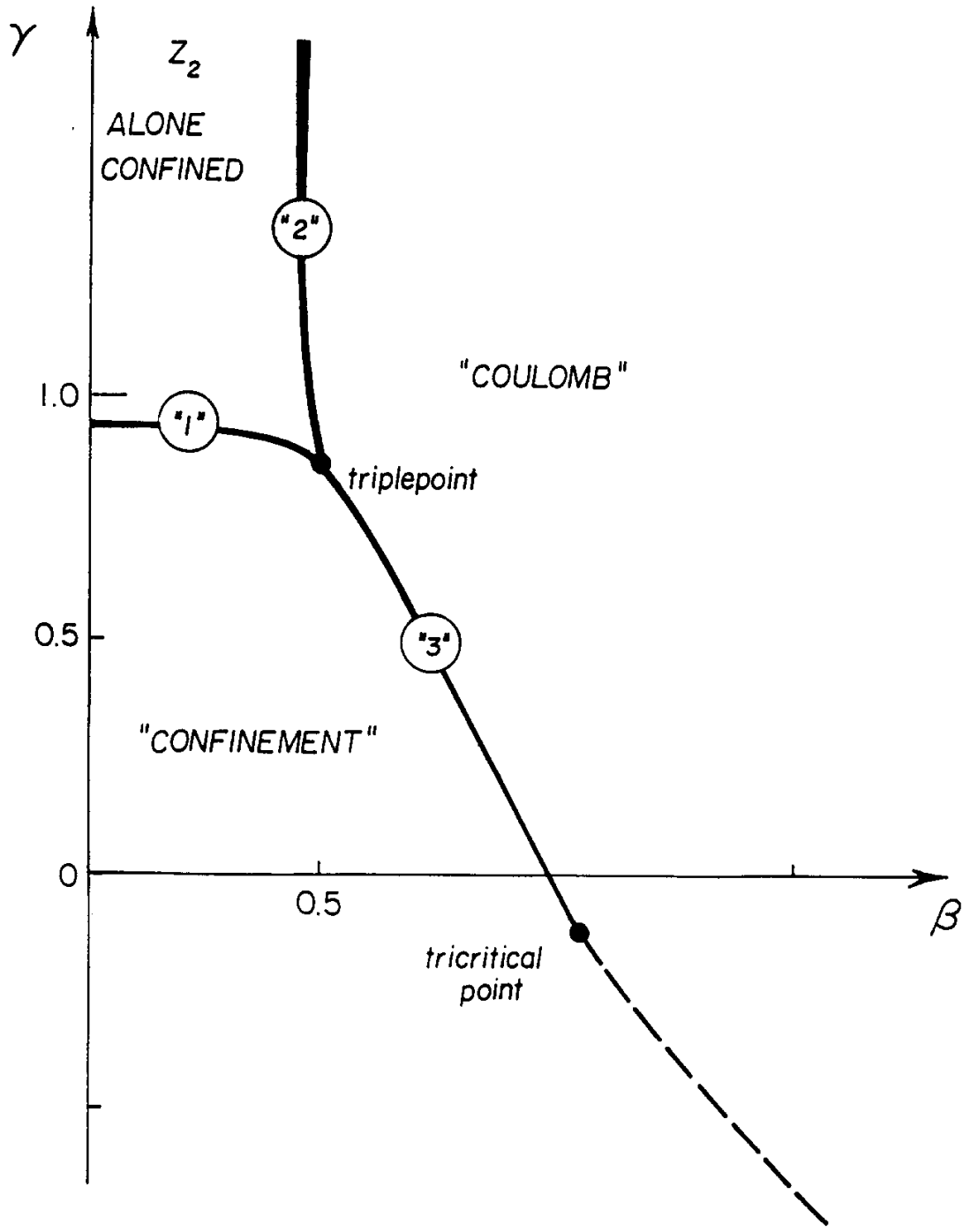}}
\caption[figurebhanot]{\label{figbhanot}The phase diagram for $U(1)$ when the
two-parameter action is used. This type of action makes it possible
to provoke the confinement of $\bz_2$ (or $\bz_3$) alone.}
\end{figure}

As a crude prototype to  a $U(1)^3$ phase diagram, we consider the (generic)
phase diagram spanned by the parameters of an action with $\cos \theta,
\cos \frac{\theta}{2}$ and $\cos \frac{\theta}{3}$ terms.
This action, which is one of the simplest generalisations of the
pure Wilson action, has been studied extensively\cite{bhanot1} and many
features of the
phase diagram (Figure~\ref{figbhanot}) are well understood. From the triple
point (TP) (which is the ``multiple point'' in this 2-dimensional phase
diagram)
emanate three characteristic phase borders: the phase border ``3''
separates the totally
confining and totally Coulomb-like phases;
the phase border ``1'' separates the totally confining phase from the phase
where only the discrete subgroup $\bz_2$ is confined;
this latter phase is separated
from the totally Coulomb-like phase by the phase border ``2''.

The calculational procedure to be used in determining the continuum $U(1)$
coupling is approximative and is done in two steps:

\begin{enumerate}

\item [A.] first  we  calculate  the
factor analogous to the factor $3=N_{gen}$  in  the  non-Abelian  case;
we call this  the enhancement factor and denote it as
$\frac{1/\alpha_{U(1)^3\;diag}}{1/\alpha_{U(1)_{crit\;TP}}}$. This factor
lies in the range 6.0 - 8.0 depending on the degree of ``first-orderness''
of the triple point (TP) transition at boundary ``2''.

\item [B.] In the second step, the continuum $U(1)$ coupling corresponding to
the multiple point
value for a single $U(1)$ is determined using an
analogy to a procedure proposed by Luck\cite{luck} and developed by
Jers\`{a}k\cite{jersak}.

\end{enumerate}

This two-step calculation can be done using more or less good approximations
as regards the extent to which the continuum $U(1)$ coupling value reflects
having phases solely confining w.r.t. discrete subgroups among the phases
that convene at the multiple point.
Let us outline the possible approximations in the order of increasing
goodness.

\begin{enumerate}

\item The roughest calculation would be to use a single parameter action with
hexagonal symmetry without regard to having phases at the triple point (TP)
that are confining solely w.r.t. the discrete subgroups $\bz_2$ and $\bz_3$
of $U(1)^3$. In this approximation, these discrete subgroups are treated as
though they were totally
confining inasmuch as it is a $U(1)$-isomorphic {\em factor group} obtained
essentially by dividing $\bz_2\times \bz_3$ out of the $U(1)$ centre of
$SMG$ that is identified with the lattice $U(1)$ critical coupling.

\item By using a two-parameter action (later a three parameter action)
leading to
the phase diagram of Figure~\ref{figbhanot}, the action
now acquires a (nontrivial) dependence on the elements within the
cosets of the factor group $U(1)/\bz_2$ (or the factor group
$U(1)/(\bz_2\times \bz_3)$ in the case of a three-parameter action)
that can reveal
how close the discrete subgroups are to being critical. However these
details are of little importance to the $U(1)$ continuum coupling; the
latter depends essentially only on a single yet to be defined parameter
$\gamma_{eff}$ the critical value of which is very nearly constant along
the phase boundary ``1''
of Figure~\ref{figbhanot}. Hence the $U(1)$ continuum coupling is also
approximately constant along this phase boundary in accord with the rule
described in the footnote on page~\pageref{rule}. The critical value
$\gamma_{eff\;crit}$ of
the parameter $\gamma_{eff}$ is expressible in terms of the critical
lattice parameters available from computer data for a lattice gauge
theory with a single $U(1)$.

\item The effect on the continuum coupling of having phases convening
at the multiple point  that are
confined solely w.r.t. $\bz_2$ and solely w.r.t. $\bz_3$
appears first when we take into account the discontinuity in $\Delta
\gamma_{eff}$ encountered in crossing the boundary ``2'' at the multiple
point. As we in both steps A. and B. above want to use the value of
$\gamma_{eff}$
corresponding to the {\em totally Coulomb-like phase} at the multiple point,
it is important for our calculation of the $U(1)$ continuum coupling to take
the ``jump'' $\Delta \gamma_{eff}$ into account. Inasmuch as the continuum
subgroup \dofx are in the same phase on both sides of boundary ``2'',
this discontinuity $\Delta \gamma_{eff}$ is entirely due to a phase
transition for the discrete subgroup(s). Moreover, the presence of a
discontinuity presumably reflects the degree of first-orderness
of the triple point transition at border ``2'' inherited from a pure $\bz_2$
and $\bz_3$ transition (i.e., for $\gamma >> 1$ in Figure~\ref{figbhanot}).

\item The discrete subgroups  $\bz_2$ and $\bz_3$ contribute
differently to the ``jump'' $\Delta \gamma_{eff}$ in crossing the boundary
``2'' due to the fact that
$\bz_2^3$ does not inherit the hexagonal symmetry of $U(1)^3$ while
$\bz_3^3$ is more likely to do so.
This is discussed at the end of Section~\ref{sec-portray}.

\end{enumerate}


It is important to recall that the normalisation of the $U(1)$
that we have argued for is implemented by the identification of the
$U(1)$ lattice critical coupling with the ($U(1)$-isomorphic)
{\em factor-group} $= SMG/(SU(2)\times SU(3))$ for  some  one  of
the Cartesian product factors say $SMG_{Peter}$. Since we have  argued
or assumed that phases with genuine discrete subgroups of this  $U(1)$
factor-group are not to be in contact with the multiple point chosen by
Nature, the only discrete subgroups that are to be taken  into  account
are     discrete     subgroups          of      the      $U(1)$
{\em subgroup} of $SMG$. The  relation  between  the  $U(1)$
subgroup  and  the  factor-group  $SMG/(SU(2)\times  SU(3))$   can   be
described  as  $U(1)_{factorgr}=U(1)_{subgr}/{\bf  Z}_6$.

In using $U(1)/\bz_6$ as the factor group to be identified with the lattice
critical $U(1)$, we identify the elements of $\bz_6$ and thereby ``hide''
any differences that there might be in the probabilities for being at
different elements of $\bz_6$ when a one-parameter action is used
(approximation 1 in list above). But the details of how the heights
of the peaks in probability at different elements of $\bz_6$ differ are
important if we want to arrange that the discrete subgroups of $\bz_6$ are
by themselves to be confined in phases convening at the triple point.
However such details become visible again if the (one-parameter)
Wilson action (roughest approximation {\bf 1} in the list above)
is replaced by the
(two-parameter) ``mixed'' fundamental-adjoint action
(approximation {\bf 2} and {\bf 3} in the list above).
By introducing an additional parameter in this way, we
render the group elements identified in the factor groups
$U(1)_{subgr}/\bz_2$ and  $U(1)_{subgr}/\bz_3$ inequivalent (i.e.,
the action acquires a dependence on the elements {\em within} the
cosets of these factor groups). So in
effect, by going from the Wilson action to the two-parameter action we
lift the factor group up into a kind of covering space. The result is that by
replacing the  $U(1)_{factorgr}$
critical coupling by the triple point
coupling  for the $U(1)_{subgr}$ of $SMG$,
we essentially arrange that the subgroups ${\bf Z}_2$ and ${\bf Z}_3$
can confine individually in phases that convene at the triple point (TP).

In both steps A. and B. of the calculation of the continuum $U(1)$ coupling,
we make use of the ``jump'' $\Delta \gamma_{eff}$
in the quantity $\gamma_{eff}$ that
in Section \ref{secgamef} below will be argued to be an effective coupling
in the sense that in the region of the phase
diagram near the phase border ``1'' in Figure~\ref{figbhanot}
(i.e., on both sides of ``1'') it
is to a good approximation valid that the phase realized (i.e.,
the totally confined or the phase with
only $\bz_2$ confined) is determined by the value of this one variable
$\gamma_{eff}$
($\gamma_{eff}$ is a certain combination of the parameters
$\gamma$ and $\beta$ of the two-parameter action
(see Figure~\ref{figbhanot})).
Consequently, the variable $\gamma_{eff}$ is necessarily constant along the
phase boundary ``1'' and we can also assume that the corresponding
continuum coupling has a constant
value along this boundary. The change in $\gamma_{eff}$ -
i.e. $\Delta \gamma_{eff}$ -
comes first at the boundary ``2'' in going into the totally Coulomb-like phase.
The value of  $\Delta \gamma_{eff}$ (calculated in Section~\ref{deltagam})
depends on the degree of ``first-orderness'' that at the
multiple point ($\gamma\approx 1$) is inherited from the pure $\bz_2$ and
$\bz_3$ transitions at $\gamma \rightarrow \infty$. Without
the correction for discrete subgroups embodied by $\Delta \gamma_{eff}$, the
multiple point coupling is obtained as if the discrete
subgroups were totally confining.

In the step A., the quantity  $\Delta \gamma_{eff}$, which
reflects
the degree of first-orderness inherited from the pure $\bz_2$ and $\bz_3$
transition in crossing boundary ``2'' at the multiple point, is
used to interpolate between the enhancement factor of about 8  obtained
with the
volume approximation and the enhancement factor of 6
obtained with the independent monopole approximation.
These approximations are most suitable for respectively first and second order
transformations. The calculation of the enhancement factor is done in
Section~\ref{enhance}.

In  step B. of the calculation (performed in Section~\ref{contcoup}),
the quantity $\Delta\gamma_{eff}$ is again used - this time in the
combination $\gamma_{eff}+\Delta \gamma_{eff}$ - to calculate the  $U(1)$
continuum coupling corresponding to
the triple point  values of a (single) $U(1)$ lattice gauge theory.
We seek the continuum coupling in the corner of the totally {\em Coulomb-like}
phase (necessary if Planck-scale confinement of observed fermions is to be
avoided)
that lies at the
triple point - that is, in the ``corner'' formed by the phase borders ``2''
and ``3''. According to the above argumentation, we know that the
continuum coupling at any position along the border ``1'': it is just equal
to the value at $\gamma=\gamma_{crit}$ and $\beta=0$. In particular, this is
true at the multiple point in the phase with only $\bz_2$ confining (i.e.,
in the ``corner'' formed by the phase boundaries ``1'' and ``2''). But as
argued above, we want the coupling corresponding to the Coulomb phase
``corner'' formed by borders ``2'' and ``3''. This requires a correction
that accounts for going from the multiple point corner formed by
``1'' and ``2'' to the multiple point corner formed by borders ``2''
and ``3'' (and in principle also a small correction from crossing
border ``1''). It  is this
transition, corresponding to the transition from a
phase with solely $\bz_2$
confining to a totally Coulomb-like phase that is accompanied by the ``jump''
denoted by $\Delta \gamma_{eff}$.



\subsection{The  approximation  of  an  effective   $\gamma$ \label{secgamef}}

In the literature (\cite{bhanot1,bhanot2}) we find the phase diagram for a
$U(1)$
group with a mixed lattice action having a $\gamma$  term  defined  on
the factor-group $U(1)/{\bf Z}_2$:

\beq
S_{\Box}(\theta)= \gamma cos(2\theta) + \beta cos(\theta). \label{mix}
\eeq

\nin With this action it is easy to provoke confinement of the whole
group as well as the totally Coulomb phase and phase confined solely w.r.t.
a discrete subgroup isomorphic to $\bz_2$ for a judicious choice of
the action parameters $\gamma$ and $\beta$ that span the phase diagram
in Figure~\ref{figbhanot}.
Indeed the phase diagram of Figure~\ref{figbhanot} clearly reveals a triple
point
common to three phases.
The interpretation of these phases as  the  three referred to above
is confirmed by  rough  mean  field
estimates for the phase borders.
In the case of the non-Abelian subgroups $SU(2)$ and $SU(3)$, two of the
phases in Figure~\ref{figbhanot}  are  actually  connected,  because one of
the  phase
borders ends at a tri-critical point.
However, this does not of course preclude the  existence
of a multiple (i.e., triple) point.

Before proceeding, it is useful to change  notation  by
scaling the variable $\theta$ down by a  factor  two
inasmuch as it is recalled (see
\ref{onlyz6}) that we want to normalise relative to the factor  group
\footnote{The motivation is that the  normalisation  that  we  use  to
define the coupling $\alpha_1$  (i.e.,  $\alpha_{1,\;Peter}$,  etc.)  is
relative  to  the  factor  group  rather  than  the  subgroup   (i.e.,
$U(1)_{factorgr}  =  SMG/(SU(2)\times  SU(3))  \sim  U(1)_{subgr}/{\bf
Z}_6$ rather than $U(1)_{subgr}$).  Instead  of  $U(1)/{\bf  Z}_6$,  we
consider for illustrative purposes  the  analogous  situation
$U(1)/{\bf  Z}_2$. This
case is also comparable with readily available results to be found   in   the
literature\cite{bhanot1}.} $U(1)/\bz_2$.

\beq
S_{\Box}(\theta)= \gamma \cos(\theta) + \beta \cos(\theta/2).
\label{newbhanot}\eeq

\nin Note that with this notational convention, Bianchi identities are
fulfilled modulo $4\pi$. In discussions of monopoles dealt with
in later sections, we shall have occasion to distinguish ``full''
$4\pi$-monopoles and ``minimal strength'' $\frac{2\pi}{4\pi}$-monopoles.
The latter
will be seen to correspond to the
``length'' of the factor group $U(1)/\bz_2$. These remarks first become
relevant and more transparent when, in a later section, we explain the idea
of ``minimal strength'' monopoles. In the case of $\bz_2$, such monopoles
are referred to as $\frac{2\pi}{4\pi}$-monopoles. These will be seen
to be the monopoles present relative
to a $\bz_2$ background field.

In order to obtain numerical results, the multiple
point coupling in this diagram will in a later Section be related to the point
at which
$\beta=0,\gamma=\gamma_{critical}$
inasmuch as we have a procedure for relating this point  to the continuum
coupling at the triple point (hereafter ``TP'') ``corner'' formed by
the phase boundaries ``1'' and
``2'' and subsequently at the for us interesting TP ``corner'' formed
 by the phase boundaries ``2'' and ``3'' (i.e., the totally Coulomb phase at
 the TP - see Figure~\ref{figbhanot}).
We shall actually argue that to
a very good approximation the continuum coupling does not vary  along
the phase border ``1'' (separating  the  ``total  confinement''  and  the
``phase with only ${\bf Z}_2$ confining'') in going from
$(\beta,\gamma)=(0,\gamma_{crit})$
to the corner at the TP formed by the phase boundaries ``1'' and ``2''.
It is first upon crossing the phase boundary ``2'' into the totally Coulomb
phase at the  TP that there is a change - a jump $\Delta \gamma_{eff}$ -
in the quantity $\gamma_{eff}$ that  immediately below will be seen to be an
effective coupling. This jump $\Delta \gamma_{eff}$ comes from a jump in
the  relative
probability of finding the plaquette variable at the(a)  non-trivial element of
$Z_2(\bz_3)$ upon making the transition at boundary ``2'' separating the
totally
Coulomb-like phase from the phase solely confined w.r.t. $\bz_2$ or $\bz_3$ in
Figure~\ref{figbhanot}. As the continuum \dofx are in the same phase on both
sides
of the boundary ``2'', the discontinuity $\Delta \gamma_{eff}$ must be entirely
due to the discrete subgroup transition which inherits a considerable
degree of the first order nature of the pure
(i.e., for $\gamma\rightarrow \infty$) discrete group transition.


In order to see how the effective coupling $\gamma_{eff}$ comes about, we
consider the
partition function for the action (\ref{newbhanot})

\beq Z=\int {\cal D}\theta(\link) \exp(\sum_{\Box}(\gamma cos(\theta)  +
\beta cos (\theta/2))). \eeq

\nin It can  be rewritten as

\beq  Z=  \int  {\cal  D}\hat{\theta}(\link)   \exp(\sum_{\Box}(\gamma
\cos(\hat{\theta})+\log(\cosh(\beta(\cos(\hat{\theta}/2)-1)))+\log(1+\langle
\sigma \rangle_{\hat{\theta}}(\tanh(\beta(\cos\hat{\theta}/2-1))))
\label{zexact}
\eeq

\nin  where

\beq \sigma= \mbox{sign}\cos(\theta/2) \eeq

\nin and where the variable  $\hat{\theta}$, which takes values on the interval
$0\leq \hat{\theta}\leq 2\pi$, is related to $\theta$ by

\beq \hat{\theta}=\left \{ \begin{array}{c} \theta \mbox{ for } \sigma=+1
\\     \theta\pm     2\pi     \mbox{ for }     \sigma=-1      \end{array}
\right \}\;(\mbox{mod } 4\pi) \eeq

\nin and

\beq    \langle    \sigma    \rangle_{\hat{\theta}}=\langle     \sigma
\rangle_{\mbox{\tiny with restriction }\theta (\Box_{\tiny A}\;) = \hat{\theta}
\;
(\mbox{\tiny mod }2\pi)}=
\eeq

\beq     =     \frac{\int     {\cal     D}\theta(\link)e^{S}      \delta
(\theta(\Box_{\tiny                   A})-\hat{\theta}\;(\mbox{mod}
2\pi))\sigma(\Box_{\tiny A}\;)}{\int {\cal  D}\theta(\link)e^{S}  \delta
(\theta(\Box_{\tiny A}\;)-\hat{\theta}\;(\mbox{mod} 2\pi))} \eeq

\nin where $\Box_{\tiny A}$ is some fixed plaquette (that due to
long distance translational invariance can be arbitrarily chosen).
Up to now, this (rather formal) treatment has been exact.

The effective coupling is defined by requiring equality of
averages of the second
derivatives of two expressions for the action: namely the action
$\gamma_{eff}\cos\theta$ and the action appearing as the exponent of
(\ref{zexact}); that is,

\beq \langle \frac{d^2}{d\theta^2}(\gamma_{eff}\cos\theta )\rangle=
\label{2act}\eeq

\[ \langle \frac{d^2}{d\theta^2}(\gamma
\cos(\hat{\theta})\;\;+\;\;\log(\cosh(\beta(\cos(\hat{\theta}/2)-1)))
\;\;+\;\;\log(1+\langle
\sigma \rangle_{\hat{\theta}}(\tanh(\beta(\cos\hat{\theta}/2-1)))
\rangle.  \]

\nin Before taking the derivative on the right-hand side of (\ref{2act}),
we expand the second and third terms of the action in the exponent of
(\ref{zexact}) in the small
quantity $\beta(\cos(\hat{\theta}/2)-1)$. To leading order, the second term
in the exponent of (\ref{zexact}) is

\beq                        \log(cosh(\beta(\cos(\hat{\theta}/2)-1)))=
\frac{1}{2}(\beta(\cos(\hat{\theta}/2)-1))^2          +          \dots
\approx \frac{1}{2}(\beta(-(\frac{\hat{\theta}^2}{8}))^2)
\cdots \label{secondterm}\eeq

\nin while the third term to leading order in
$\beta(\cos(\hat{\theta}/2)-1)$ is

\beq                       \log(1+\langle                       \sigma
\rangle_{\hat{\theta}}(tanh(\beta(\cos(\hat{\theta}/2)-1)))) \approx
\langle                       \sigma
\rangle_{\hat{\theta}}(\beta(\cos(\hat{\theta}/2)-1)).
\label{hop} \eeq


\nin Performing the derivatives in (\ref{2act}) yields

\beq \langle \gamma_{eff}\cos\hat{\theta}\rangle=\langle\gamma\cos\hat{\theta}+
\frac{\langle\sigma\rangle_{\hat{\theta}}\beta}{4}\cos(\hat{\theta}/2)
\rangle  \label{derivative}\eeq

\nin where on the right-hand side the term with $\cos\hat{\theta/2}$
arises as the second derivative of the leading term in (\ref{hop}):

\beq   \langle\sigma\rangle_{\hat{\theta}}\beta(\cos(\hat{\theta}/2)-
1))\eeq

\nin which is of degree one in $\beta(\cos(\hat{\theta}/2)-1)$.
In the approximation used, the contribution from the leading
term in (\ref{secondterm}) is neglected as this term is of second degree in
$\beta(\cos(\hat{\theta}/2)-1)$.

\nin Rewriting $\cos(\hat{\theta}/2)$ as $\frac{\cos(\hat{\theta}/2)}
{\cos\hat{\theta}}\cos\hat{\theta}$ on the right-hand side of
(\ref{derivative}), we can extract the effective coupling $\gamma_{eff}$
as

\beq \gamma_{eff}=
\gamma+\frac{\langle\sigma\rangle_{\hat{\theta}}\beta}{4}
\langle\frac{\cos(\hat{\theta}/2)}
{\cos\hat{\theta}}\rangle \approx
\gamma+\frac{\langle\sigma\rangle_{\hat{\theta}}\beta}{4}
\langle \cos^{-\frac{3}{4}}\hat{\theta}\rangle  \label{gammaeffect}
\eeq




In the roughest approximation, we take
$\langle \cos^{-\frac{3}{4}}\hat{\theta}\rangle=1$
in (\ref{gammaeffect}) and thereby obtain $\gamma_{eff}$ as

\beq \gamma_{eff}=\gamma +
\langle     \sigma      \rangle_{\hat{\theta}}\frac{\beta}{4}
\approx \gamma +
\langle     \sigma      \rangle\frac{\beta}{4}
\label{actz2}  \eeq

\nin where in the last step, $\langle     \sigma      \rangle_{\hat{\theta}}$
has been replaced by  $\langle \sigma \rangle$
inasmuch as $\langle     \sigma      \rangle_{\hat{\theta}}$ is
to a good approximation independent of $\hat{\theta}$. The reason is that the
region in $\hat{\theta}$ over which we shall average is not very large - even
for critical $\gamma$. This combined with the fact that
$\langle \sigma\rangle_
{\hat{\theta}}$ depends (for symmetry reasons) to lowest order on
$\hat{\theta}^2$ allows us to ignore the dependence of $\langle \sigma\rangle$
on $\hat{\theta}$.



Near the boundary ``1'' separating the totally confining phase from the phase
where $\bz_2$ alone is confined, it is claimed that
the physics is quite accurately described by a particular single
combination  of  the
two  lattice  action  parameters  $\beta$  and  $\gamma$  that can be
used as a replacement for the dependence on both parameters.
That this is a rather good approximation has to do
with the fact that  fluctuations
in the ${\bf Z}_2$ degrees of freedom are strong all the way along the
phase border ``1'' because $\bz_2$ is confined on both sides of this
boundary. This gives rise to a very effective averaging over the distribution
at
$\theta$ and $\theta+2\pi$; this combined with the argument
that the dependence of the distribution on $\hat{\theta}$ is small
means that
the  information content in both  $\gamma$  and
$\beta$ that is  relevant is manifested essentially as a  single  parameter
$\gamma_{eff}$.

In particular, both the continuum coupling and the question of
which phase is realized (i.e.
the position of the phase boundary ``1'')
should,  in  the  region  where this approximation is valid,
only depend the single  parameter  $\gamma_{eff}$.  Hence
the continuum coupling will not vary along  this  phase  border.  This
implies that $\gamma_{eff}$  will have the same value at the triple point
(TP)
as for $\beta=0$.
At the TP, there are three corners because three phases meet here;
each
has it own continuum  coupling provided the phase transitions are first order.
The  above  argument  leads  to  the
conclusion that the continuum coupling at the multiple point in
the corner of the phase with alone ${\bf  Z}_2$  confined
equals the value of this coupling in the same phase but where
$\beta=0$ and where  $\gamma$  is  infinitesimally  above  $\gamma_{ crit}$.
Analogously, the continuum coupling in the totally confining phase (to
the extent that this makes sense) is the same at the multiple point
corner and the
point in this phase where $\beta =0$  and  $\gamma$  is  infinitesimally
below the critical value.

If we want to be able to provoke confinement solely  along other
discrete subgroups
than $\bz_2$ (e.g., along $\bz_3$), an action more general than (\ref{mix})
is needed.
Such a more general action would be

\beq
S=\gamma\cos \theta +\beta_2\cos\frac{\theta}{2}+\beta_3\cos\frac{\theta}{3}
+\beta_6\cos\frac{\theta}{6} \label{fullact} \eeq

\nin Taking the second derivative of $S$:

\beq
-S^{\prime\prime} =\gamma\cos \theta +\frac{\beta_2}{4}\cos\frac{\theta}{2}
+\frac{\beta_3}{9}\cos\frac{\theta}{3}+\frac{\beta_6}{36}\cos\frac{\theta}{6}
\label{fullactder}  \eeq

\nin Assume that $\gamma$ is large compared to $\beta_2$, $\beta_3$, and
$\beta_6$. We can then write

\beq \gamma_{eff}=(-S^{\prime\prime}(0))P_0+(-S^{\prime\prime}(2\pi)P_2+
(-S^{\prime\prime}(4\pi))P_4+(-S^{\prime\prime}(6\pi))P_6
+(-S^{\prime\prime}(8\pi))P_8+(-S^{\prime\prime}(10\pi))P_{10} =\eeq

\beq
=\left(\gamma+\frac{\beta_2}{4}+\frac{\beta_3}{9}+\frac{\beta_6}{36}\right)P_0+
\left(\gamma-\frac{\beta_2}{4}-\frac{\beta_3}{18}+\frac{\beta_6}{72}\right)P_2+
\left(\gamma+\frac{\beta_2}{4}-\frac{\beta_3}{18}-\frac{\beta_6}{72}\right)P_4+
\eeq

\[+\left(\gamma-\frac{\beta_2}{4}+\frac{\beta_3}{9}-
\frac{\beta_6}{36}\right)P_6+
\left(\gamma+\frac{\beta_2}{4}-\frac{\beta_3}{18}-\frac{\beta_6}{72}\right)P_8
\left(\gamma-\frac{\beta_2}{4}-\frac{\beta_3}{18}+
\frac{\beta_6}{72}\right)P_{10}  \]

\nin where $P_0,P_2,P_4,P_6,P_8$ and $P_{10}$ are the probabilities
that a plaquette takes a value near (in the corresponding
sequence) $0,2\pi,4\pi,6\pi,8\pi$ and $10\pi$.
Regrouping, we have

\beq \gamma_{eff}=\underbrace{P_0+P_2+P_4+P_6+P_8+P_{10})}_{=1}\gamma + \eeq

\[+\frac{\beta_2}{4}\underbrace{(P_0(1)+P_2(-1)+P_4(1)+P_6(-1)+P_8(1)
+P_{10}(-1))}_{\langle \sigma_{\sbz_2} \rangle}+ \]
\[+
\frac{\beta_3}{9}\underbrace{(P_0(1)+P_2(-\frac{1}{2})+P_4(-\frac{1}{2})P_6(1)
+P_8(-\frac{1}{2})+P_{10}(-\frac{1}{2}))}_{\langle \sigma_{\sbz_3} \rangle}+ \]
\[
+\frac{\beta_6}{36}\underbrace{(P_0(1)+P_2(\frac{1}{2})+
P_4(-1)+P_6(-1)+P_8(-\frac{1}{2})+
P_{10}(\frac{1}{2}))}_{\langle \sigma_{\sbz_6} \rangle}
=\]

\beq =\gamma + \frac{\beta_2}{4}\langle\sigma_{\bz_{2}}\rangle+
               \frac{\beta_3}{9}\langle\sigma_{\bz_{3}}\rangle+
              \frac{\beta_6}{36}\langle\sigma_{\bz_{6}}\rangle
               \label{gamefz23}\eeq

\nin where

\beq \sigma_{\sbz_2}=\mbox{ sign }\cos(\theta/2) \eeq

\[ \sigma_{\sbz_3}=\mbox{ sign }\cos(\theta/3) \]

\[ \sigma_{\sbz_6}=\mbox{ sign }\cos(\theta/6) \]

Equation (\ref{gamefz23}) contains (\ref{actz2}) as a special case; the
more detailed derivation of (\ref{actz2}) was included for illustrative
purposes.

Note that with the action (\ref{fullact}), Bianchi identities are now
fulfilled modulo $12\pi$. The analogy to the remarks pertaining to monopoles
immediately following (\ref{newbhanot}) are for the action (\ref{fullact})
that ``full'' monopoles correspond to charge $12\pi$ and ``minimal strength''
monopoles - denoted $\frac{2\pi}{12\pi}$ - to the ``length'' of the
factor group $U(1)/\bz_6$. These ``minimal strength'' monopoles will be
described as monopoles relative to a $\bz_6$ background field or
alternatively as monopoles modulo a $\bz_6$ background. These remarks
become more relevant in the following section where we consider the
effect of including phases at the multiple point that are critical w.r.t.
$\bz_2$ and $\bz_3$.

\subsection{Estimating the degree of ``first -orderness'' in the
transition from the $\bz_2$ confining phase to the totally Coulomb
phase at the triple point} \label{deltagam}

In the limit
of very large $\gamma$ values, the phase transition at border ``2'' becomes
a pure $\bz_2$ transition inasmuch as all the probability is concentrated
at a $\bz_2$ subgroup of $U(1)$. We want to use known results for $\bz_2$
to estimate the degree of ``first-orderness'' of the transition in crossing
the boundary ``2'' at the multiple point.
A proper $\bz_2$ transition corresponds to infinite
$\gamma$ whereas $\gamma$ at the multiple point is of the order unity. However,
we expect the phase transition in crossing the border ``2'' at the triple
point to inherit to some extent the properties (i.e., a degree of
first-orderness) of a $\bz_2$ phase transition even though $\gamma$ at the
triple point is only of order unity. The reason is that, also at the
triple point, the transition at the border ``2'' (from the phase with
$\bz_2$ alone confining to the totally Coulomb phase) really only involves
the $\bz_2$ \dof. That the transition in crossing border ``2''
at the triple point presumably does not have the full degree of first-orderness
of a
pure $\bz_2$ transition is due to the importance of group elements of $U(1)$
that
depart
slightly (and continuously) from the elements of $\bz_2\subset U(1)$.
Having such
elements  make
possible ``$\frac{2\pi}{4\pi}$-monopoles'' the density of which increases
as $\gamma$ becomes smaller.
What are ``$\frac{2\pi}{4\pi}$-monopoles''? Here we make connection with
the remarks immediately following (\ref{newbhanot}) and, more generally,
the remarks in the last paragraph of the preceding section.
Think of the six plaquettes bounding a 3-cube.
In the phase with $\bz_2$ alone confining (and with $\gamma$ large but
not infinite), plaquette configurations of a 3-cube
can involve an odd number of
plaquettes that have plaquette variable
values near the nontrivial element of $\bz_2$
(in the notation of (\ref{newbhanot}) in which Bianchi identities are
fulfilled
modulo $4\pi$, the nontrivial element of $\bz_2$ corresponds to $2\pi$ so
$\bz_2=\{0,2\pi\}\subset U(1)$) in combination with small deviations from
$\bz_2$ (the deviations lie along $U(1)$ in which of course $\bz_2$ is
embedded)
such that together the six plaquette values of a 3-cube sum to zero
(mod $4\pi$ in the notation of (\ref{newbhanot})).
We can regard the flux through such a configuration as that coming from a
``$\frac{2\pi}{4\pi}$-monopole'' relative to a $2\pi$ ``background'' flux
coming from the general abundance of plaquettes having the value near
the (nontrivial) element $2\pi\in \bz_2\subset U(1)$.

If one considers an isolated $\bz_2$ theory (i.e., a $\bz_2$ that is not
embedded in a $U(1)$ as is the case for  infinite $\gamma$),
there can be no monopoles because there is for
$\bz_2$ no way to have 6 ``small'' elements that sum up to a circumnavigation
of the whole group.
However, for finite $\gamma$, the distribution of group elements
accessible due to quantum fluctuations spreads out slightly from $\bz_2$ to
$U(1)$
elements ``close to $\bz_2$'' with the result that
it is possible to have $\frac{2\pi}{4\pi}$-monopoles in the
sense introduced above. In other words,
in the phase with only $\bz_2$ confining, it is possible to have monopoles
modulo a $\bz_2$ background (i.e.,$\frac{2\pi}{4\pi}$-monopoles)
if $\gamma$ is not so large as to
preclude
continuous plaquette variable deviations from $\bz_2$ along $U(1)$ of
sufficient magnitude so that these deviations from $\bz_2$
for plaquette values of a 3-cube can
add up to the length of the factor group $U(1)/\bz_2$. When
Bianchi identities are satisfied modulo $4\pi$ by such configurations,
we can say that we get half (i.e., $\frac{2\pi}{4\pi}$) of the way to $0$
(mod $4\pi$)
using $\frac{2\pi}{4\pi}$-monopoles; the other half of the way to $0$
(mod $4\pi$)
is provided by the $2\pi$ background field having as the source an
odd number of plaquettes with values near the nontrivial element of
$\bz_2\subset U(1)$.

In the sequel, we shall restrict our attention to ``minimal strength''
monopoles\footnote{A ``minimal strength'' $\bz_N$ monopole is a
configuration of 6 plaquettes surrounding a 3-cube such that the sum of
continuous deviations from elements of $\bz_N$ add up to the length of the
factor group $U(1)/Z_N$.} (i.e.,
$\frac{2\pi}{4\pi}$-monopoles in the case of the action (\ref{mix}))
inasmuch as such ``minimal strength'' monopoles in the
dominant configuration in which a foursome of 3-cubes encircles a common
plaquette. This dominant configuration 
can be expected
to constitute the
vast majority of the monopoles present. In the case of the
action~(\ref{newbhanot}), the dominant monopoles
are the  $\frac{2\pi}{4\pi}$
monopoles (These are the only possible only less than full strength monopoles)
In the case of the action~(\ref{fullact}), minimal strength (and
presumably dominant)
monopoles are $\frac{2\pi}{12\pi}$ monopoles; in principle there could also be
monopoles of strength $4\frac{4\pi}{12\pi}$ and $\frac{6\pi}{12\pi}$.

\begin{figure}
\centerline{\epsfxsize=\textwidth \epsfbox{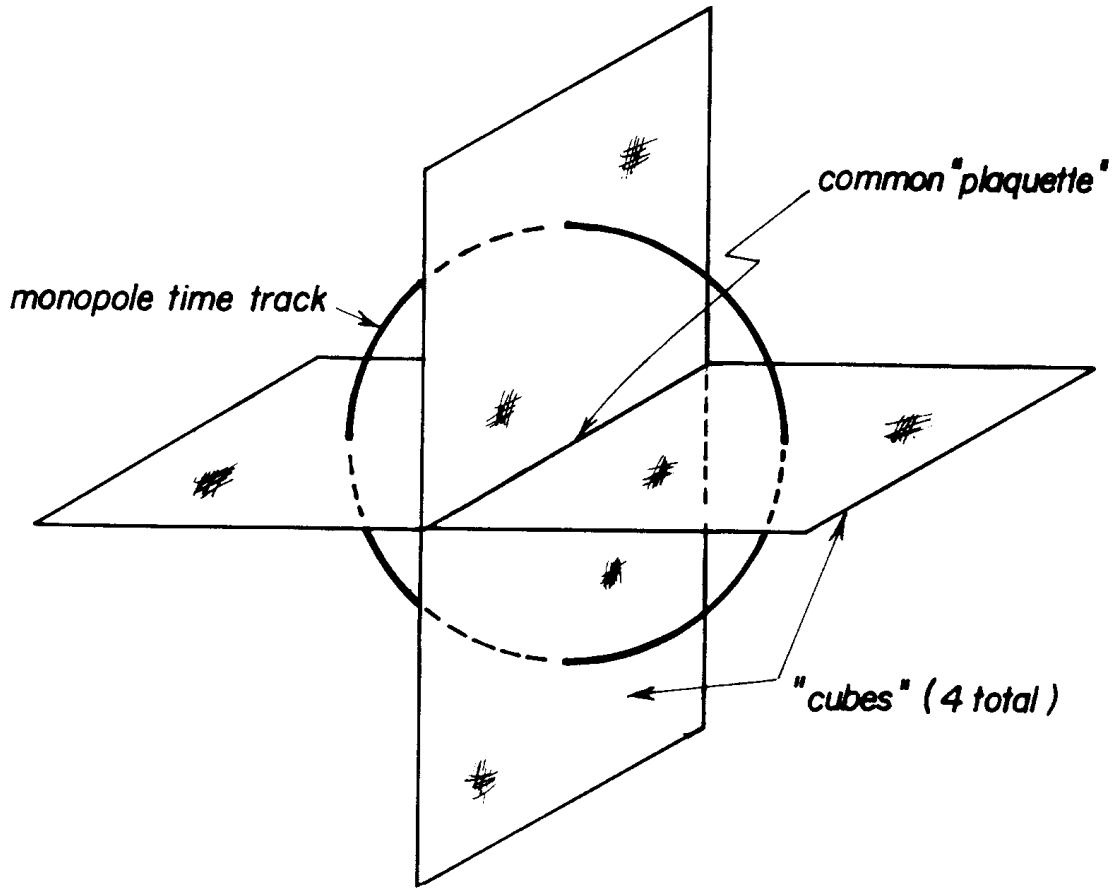}}
\caption[The dominant configuration for monopoles]{\label{domconfig}
The important monopoles are expected to be of minimal strength and to be
found essentially only in the dominant configuration of four cubes
surrounding a common plaquette. The dominant configuration is illustrated
above in a picture having one dimension less than the actual (4-dimensional)
dominant configuration. The actual dominant configuration - i.e.,
a plaquette
common to four 3-cubes has in the above dimensionally reduced picture
become a link common to four plaquettes.}
\end{figure}

We claim that as $\gamma \rightarrow \infty$, the
probability of having such a dominant configuration monopole decreases
exponentially;
accordingly
there is only a thin population of minimal ``strength monopoles'' (and an
even much thinner population of monopoles other than the ``minimal strength''
type). Hence it is
presumably a very good approximation to describe the presence of
monopoles as due solely to the dominant configuration of ``minimal strength''
monopoles.

In the case of the action (\ref{newbhanot}), this means
four 3-cube $\frac{2\pi}{4\pi}$-monopoles that encircle a common
plaquette having a value corresponding to the nontrivial element of $\bz_2$.
Consider by way of example the case where each of the four
3-cubes in this dominant configuration have
the value $2\pi/5$
on  five plaquettes (with the sixth ``encircled'' common plaquette
having the value $\pm 2\pi$). Such a 3-cube configuration would, relative to a
$\bz_2$ background flux (expected for large $\gamma$ and small
$\beta$'s), behave as a $\frac{2\pi}{4\pi}$-monopole with a flux of $2\pi/5$
emanating from each of five plaquettes.

The dominant-configuration $\frac{2\pi}{4\pi}$-monopoles
can be expected to occur with some
low but nonzero density in the lattice near the phase border ``2'' even
for large (but not too large) $\gamma$ values. Our suspicion, confirmed by
calculations below, is that the degree of ``first-orderness'' of the phase
transition at the boundary ``2'' is greater the smaller the chance that
small deviations from $\bz_2$ (lying in $U(1)$) can, for the six
plaquettes of a 3-cube, add up to a
$\frac{2\pi}{4\pi}$-monopole (or, stated equivalently, add up to
the length of the factor group $U(1)/\bz_2$).

As $\gamma$ decreases,
an increasing number of $\frac{2\pi}{4\pi}$-monopoles is encountered.
At the triple point (TP), where $\gamma\approx 1$,
the presence of a larger number of $\frac{2\pi}{4\pi}$-monopoles than for
very large $\gamma$
mitigates but does not eliminate the high degree of ``first-orderness''
characteristic of pure $\bz_2$ transitions (for which the deviations
from $\bz_2$ (along $U(1)$)
of six 3-cube
plaquette variable values cannot sum to the length of the whole $U(1)/\bz_2$
due to $\gamma$ being too large).

In order to deal quantitatively with the effect of
$\frac{2\pi}{4\pi}$-monopoles, and
thereby with the question of how much of the behaviour of a pure
$\bz_2$ transition is inherited by the phase transition at border ``2''
at the triple point, it is
useful to define {\em two} new variables $\on$ and $\B$:

\beq \on\stackrel{def}{=} \left \{ \begin{array}{l}
+1 \mbox{ if $U(\Box)$ closest to $e^{i0}\in \bz_2$} \\
-1 \mbox{ if $U(\Box)$ closest to $e^{i\pi}\in \bz_2$} \end{array} \right.
\eeq

\nin The other new variable $\B$ (the subscript ``$BIO$''
is an acronym for \underline{B}ianchi \underline{I}dentity
\underline{O}beying) is defined as follows:

\beq \B\stackrel{def}{=} \on \cdot \eeq

\[\cdot\left \{ \begin{array}{l}
+1 \mbox{ \footnotesize for ordinary $\Box$
(i.e. {\em not} the encircled $\Box$ in the dominant config.)} \\
-1 \mbox{ \footnotesize  for $\Box$ encircled by four 3-cube
$\frac{2\pi}{4\pi}$-monopoles in the dominant config.} \end{array} \right.
\]

\nin The variable $\B$ differs from
the variable $\on$ only by a sign change of $\on$ in the case where
the plaquette $\Box$ coincides with the ``encircled'' plaquette. The
``encircled''
plaquette is
always present in the four 3-cube
$\frac{2\pi}{4\pi}$-monopoles of the dominant monopole
configuration.

Let us make the observation that the values assigned by the variable
$\on$ to the plaquettes of a 3-cube satisfy the $\bz_2$-Bianchi identity if the
3-cube is not a $\frac{2\pi}{4\pi}$-monopole;
i.e., in our approximation, not one of the four 3-cube
$\frac{2\pi}{4\pi}$-monopoles encircling
a common plaquette
in the dominant $\frac{2\pi}{4\pi}$-monopole configuration. Note, however,
that the Bianchi
identity {\em is} violated by the values assigned by the variable $\on$ to the
plaquettes of a 3-cube when there is a $\frac{2\pi}{4\pi}$-monopole.
For instance, it is readily seen that
for the very special $\frac{2\pi}{4\pi}$-monopole example given above
($U(\Box)=e^{i\frac{2\pi}{4\pi}\cdot 2\pi}=-1$ on the ``encircled''
plaquette; $U(\Box)=e^{i2\pi/5}$ on the remaining 5 plaquettes of the
$\frac{2\pi}{4\pi}$-monopole
3-cube), the Bianchi identity is violated:

\beq \prod_{\Box\in \partial(\frac{2\pi}{4\pi}-monopole\;3-cube)}\on=
(-1)\cdot 1^5=-1  \neq 1 \eeq

\nin inasmuch as $\on=-1$ for $U(\Box)=-1$ and $\on=1$ for
$U(\Box)=e^{i2\pi/5}$.



More generally, a $\frac{2\pi}{4\pi}$-monopole (which really just means a
monopole
modulo a $\bz_2$ background) consists of a configuration of plaquette
variable values of a 3-cube that deviate continuously from elements of
$\bz_2$ in such a
way that the total sum of continuous deviations (lying in $U(1)$)
from $\bz_2$ equals, modulo
$4\pi$, $2\pi$ multiplied by the number of plaquettes  for which the continuous
deviations are centred at the nontrivial element of $\bz_2$. Note that in
order to have a monopole, an odd number of the six plaquettes of a three
cube must be near the nontrivial element (i.e., $2\pi$) of $\bz_2$

Even more generally, we have for a monopole modulo a $\bz_N$ background (i.e.,
a monopole for which the continuous $U(1)$ deviations from $Z_N\subset U(1)$
add up to a multiple of the length of the factor group $U(1)/\bz_N$):

\beq
\prod_{\mbox{\tiny $\Box \in$ 3-cube}}( U(\Box)g_{_{\tiny
nearest}}(U(\Box))^{-1}) =
\prod_{\mbox{\tiny $\Box \in$ 3-cube}} g_{\tiny nearest}(U(\Box))
\;\;\;\;\;\;\;
(g_{\tiny nearest}(U(\Box))\in \bz_N) \eeq

\nin where $g_{\tiny nearest}(U(\Box))$ is defined as that element of $\bz_N$
which is nearest to $U(\Box)$:

\beq dist^2(U(\Box),g_{\tiny\; nearest}(U(\Box)))\stackrel{def}{=}
inf\{ dist^2(U(\Box),g^{\prime})\} \;\;\;\;(g^{\prime}\in \bz_N) \eeq

\nin where $dist^2(U(\Box),g^{\prime})$ denotes the squared distance from a
plaquette variable value $U(\Box)$ and an element
$g^{\prime}\in \bz_N$.
We are really interested in $\bz_6=\bz_2\times \bz_3$ inasmuch
as we are also interested in the modification of first-orderness due to
an increasing number of monopoles modulo $\bz_3$ in going from large $\gamma$
to the triple point.
However, for the purpose of exposition, we continue to use the
example of monopoles modulo $\bz_2$.

With the modification of the variable $\on$ that defines the variable
$\B$, we have in $\B$ a variable that, for sufficiently large $\gamma$,
assigns values to
configurations of plaquettes that respects the $\bz_2$
Bianchi identities - also for $\frac{2\pi}{4\pi}$-monopole configurations
(when the monopoles are of the dominant configuration type which is the only
type for which the variable $\B$ is defined).

Note that the variable $\B$ differs from the variable $\on$ only if there
are $\frac{2\pi}{4\pi}$-monopoles. For $\gamma \rightarrow \infty$ such
monopoles disappear and $\B=\on\in \bz_2=\{+1,-1\}$.
In going to smaller values of $\gamma$
in the phase with only $Z_2$ confining, an increasing range of
fluctuations along $U(1)$ centred at the elements of
$\{0,2\pi\} =\bz_2 \subset U(1)$ provide  alternative
(Bianchi identity-obeying)
configurations that supplement the essentially discrete
group-valued plaquettes characteristic of large $\gamma$ configurations.

We want now to determine approximately the ($\gamma$ dependent) relation
between
the distributions of the two variables $\on$ and $\B$. The average value of
$\on$ is estimated
using the identity

\beq \langle \on \rangle = P(\B=+1)\langle\on \rangle_{(\B=+1)}+ \eeq
                       \[    +P(\B=-1)\langle\on \rangle_{(\B=-1)} \]

\nin where $P(\B=+1)$ and $P(\B=-1)$ denote respectively the probabilities
that $\B=+1$ and $\B=-1$ while $\langle\on \rangle_{\B=+1}$ and
$\langle\on \rangle_{\B=-1}$ denote averages of $\on$ subject respectively
to the constraints that $\B=+1$ and $\B=-1$.

Denoting by $\xi=\xi(\gamma)$ the
($\gamma$ dependent) probability that a plaquette coincides with the
``encircled plaquette'' of the dominant $\frac{2\pi}{4\pi}$-monopole
configuration,
there obtains

\beq \langle\on \rangle_{\B=+1}
=\frac{e^{\beta}\cdot 1 +\xi e^{-\beta}\cdot(-1)}{e^{\beta}+\xi e^{-\beta}}
\eeq

\nin and

\beq \langle\on \rangle_{\B=-1}
=\frac{e^{-\beta}\cdot (-1) +\xi e^{\beta}\cdot(+1)}{e^{-\beta}+\xi e^{\beta}}
\eeq


Using

\beq P(\B=+1)=\frac{1}{2}+\frac{1}{2}\langle \B \rangle \eeq

\nin and

\beq P(\B=-1)=\frac{1}{2}-\frac{1}{2}\langle \B \rangle     \eeq

\nin we have

\beq \langle \on \rangle=\frac{1}{2}\left(\frac{e^{\beta}-\xi e^{-\beta}}
{e^{\beta}+\xi e^{-\beta}}+
\frac{\xi e^{\beta}-e^{-\beta}}{\xi e^{\beta}+e^{-\beta}}\right)+
\frac{1}{2}\left(\frac{e^{\beta}-\xi e^{-\beta}}{e^{\beta}+\xi e^{-\beta}}-
\frac{\xi e^{\beta}-e^{-\beta}}{\xi e^{\beta}+e^{-\beta}}\right)
\langle \B \rangle \eeq

\beq \approx \langle \B \rangle(1-2\xi\cosh 2\beta)+2\xi\sinh 2\beta
\label{on} \eeq

\nin where in the last step we have used that $\xi$ is assumed to be small.

We want now to calculate the jump in (\ref{actz2}) in going from the
phase with only $\bz_2$ confining to the totally Coulomb phase at the
the triple point. That is, we want $\Delta \gamma_{eff}$ along the boundary
``2'' in Figure~\ref{figbhanot} as a function of $\gamma$:

\beq \Delta \gamma_{eff}=\Delta(\gamma+\langle \sigma \rangle
\frac{\beta_{crit}(\gamma)}{4})=
\Delta\langle \sigma \rangle \frac{\beta_{crit}(\gamma)}{4}=
\frac{\beta_{crit}(\gamma)}{4}\Delta\langle \on \rangle.
\label{hm}\eeq

\nin where we have made the identification
$\langle \sigma \rangle=\langle \on \rangle$. Substituting (\ref{on})
into (\ref{hm}) we get

\beq  \frac{\beta_{crit}(\gamma)}{4}\Delta\langle \on \rangle=
\frac{\beta_{crit}(\gamma)}{4}\Delta\langle \B \rangle
(1-2\xi\cosh 2\beta(\gamma)) \label{hmm} \eeq

\nin In our approximative procedure we identify $\Delta\langle \B \rangle$
with the jump in $\Delta \langle S_{\Box} \rangle$ for a $\bz_2$ gauge theory
since the phase transition ``2'' at the triple point is determined from
the phase of $\bz_2$.

Let us define a parameter $\beta_{BIO}$ as the action parameter $\beta$ in
a $Z_2$ gauge theory which optimally reproduces the distribution of the
variables $\B$ in the $U(1)$ theory (with the mixed action (\ref{newbhanot}))
by using an action of the form

\beq S=\beta_{BIO}\sum_{\Box}\B. \eeq

\nin In other words, $\beta_{BIO}$ is defined such that

\beq \langle\B  \rangle_{\mbox{\tiny in }U(1) \mbox{\tiny theory with }
S=S(\beta,\gamma)}=
\langle\B  \rangle_{\sbz_2 \mbox{ \tiny theory with}
S=\beta_{BIO}\sum_{\Box}\B } \eeq

We now want to obtain $\beta_{BIO}$ as a function of $\beta$ and $\xi$ (and
hereby $\gamma$ inasmuch as $\xi=\xi(\gamma)$) by equating the ratio
of the probabilities

\beq \frac{P(\B=1)}{P(\B=-1)} \label{probratio}\eeq

\nin for the two actions $S=S(\beta,\gamma)$ and $S_{BIO}=
\beta_{BIO}\sum_{\Box}\B$:

\beq
\frac{\overbrace{e^{\beta}}^{\on=+1}+\overbrace{\xi e^{-\beta}}^{\on=-1}}
{\underbrace{e^{-\beta}}_{\on=-1}+\underbrace{\xi e^{\beta}}_{\on=+1}}=
\frac{e^{\beta_{BIO}}}
{e^{-\beta_{BIO}}}.
\label{onto}\eeq

\nin This procedure for estimating $\beta_{BIO}$ is somewhat errant in that
Bianchi identities are ignored on both sides of equation (\ref{onto}) in
various ways: first in the calculation of the ratio (\ref{probratio}) and
second, and presumably less importantly, in the simulation-by a-$\bz_2$
theory that defines $\beta_{BIO}$. The hope is that these error roughly
cancel inasmuch as the same error is present on both sides of the equation.

Taking the logarithm of both
sides of (\ref{onto}) and solving for $\beta_{BIO}$ yields

\beq \beta_{BIO}=\beta+
\frac{1}{2}\log\frac{1+\xi e^{-2\beta}}{1+\xi e^{2\beta}} \label{betabio} \eeq

\nin We want to use (\ref{betabio}) to relate $\beta_{BIO}$ and
$\beta_{crit}(\gamma)$ along the boundary ``2'' in Figure~\ref{figbhanot}.
Using that  $\xi << e^{\beta},\;1$ there obtains

\beq \beta_{crit\;BIO}\approx
\beta_{crit}(\gamma)+\frac{1}{2}\xi(e^{-2\beta}-e^{2\beta})
=\beta_{crit}(\gamma)-\xi \sinh 2\beta. \label{betabioa} \eeq

\nin Substituting (\ref{betabioa}) for $\beta_{crit}(\gamma)$ on the right-hand
side
of (\ref{hmm}) yields

\beq
\Delta \gamma_{eff}=\frac{\beta_{crit}(\gamma)}{4}\Delta\langle \on \rangle=
\frac{1}{4}\beta_{crit\;BIO}\left(1+\frac{\xi\sinh 2\beta}{\beta_{BIO}}\right)
\Delta \langle \B \rangle(1-2\xi \cosh 2\beta) = \eeq

\beq = \frac{1}{4}\beta_{crit\;BIO}\Delta\langle \B \rangle
\left(1+\xi\left(\frac{\sinh 2\beta}{\beta_{crit\;BIO}}-2\cosh
2\beta\right)\right).
\label{just} \eeq

\nin Solving (\ref{betabioa}) for $\xi$ and substituting into (\ref{just})
yields

\beq  \Delta \gamma_{eff}=
\frac{1}{4}\beta_{crit\;BIO}\Delta\langle \B \rangle
\left(1+(\beta_{crit}(\gamma)-
\beta_{crit\;BIO})\left(\frac{1}{\beta_{crit\;BIO}}-
\frac{2}{\tanh 2\beta}\right)\right). \eeq

{}From the literature \cite{creutz1}
we have values for $\langle S_{\Box} \rangle_
{\sbz_2}=
\Delta\langle \B \rangle$ and $\beta_{crit\;BIO}$. The quantity
$\beta_{crit}(\gamma)-\beta_{crit\;BIO}$ is estimated graphically using
a $U(1)$ phase diagram found in the literature\cite{bhanot1}  corresponding
to the action (\ref{newbhanot}). It is
now finally possible to calculate $\Delta\gamma_{eff}$ at the triple point for
the transition from the phase with only $\bz_2$ confining to the totally
Coulomb-like phase.

It is indeed fortunate that the subtraction $\frac{1}{\beta_{BIO}}-
\frac{2}{\tanh 2\beta}$ almost cancels thereby rendering our calculation
of $\Delta \gamma_{eff}$ rather insensitive to the large uncertainty in the
graphical estimate of $\beta_{crit}(\gamma)-\beta_{BIO}$. This means that the
major contribution  to the change in first-orderness in going from very large
$\gamma$ to $\gamma \approx 1$ at the triple point is achieved simply by
determining  $\beta_{crit}(\gamma)$ by the condition that
$\beta_{crit\;BIO}=\beta_{crit}(\gamma=\infty)$. This makes it possible to
perform an analogous correction to the first-orderness in going from a
pure $\bz_3$ theory to the triple point for an action $\gamma\cos \theta +
\beta_3 \cos \frac{\theta}{3}$ without having access to the phase diagram
for the $U(1)$ theory with an action of this form
                  (that we need for the graphical estimate of
$\beta_{TP\;crit}(\gamma)-\beta_{\sbz_3\;BIO}$).

In subsequent calculations, we shall make use of the fact
that the probability $\xi$
of having a
$\bz_2$ and a $\bz_3$ monopole must be roughly equal.
The argument goes as follows: we can assume
that essentially all monopoles present will be of the ``minimal strength''
type. In the case of the $SMG$, this means  monopoles
modulo a $\bz_6$ background: i.e., $\frac{2\pi}{12\pi}$-monopoles. These
are built up of $U(1)$ elements close to $\bz_6$ such that the deviations
from $\bz_6$ of six 3-cube plaquette variables add up to the full extent of
$U(1)/\bz_6$.
Of course it is still assumed that these ``minimal strength'' monopoles
essentially
only are found in
dominant configuration of four 3-cubes that encircle a common plaquette
But such a
``minimal strength'' monopole is a superposition of a $\bz_2$ and a $\bz_3$
monopole:

\beq \frac{2\pi}{12\pi}=\frac{6\pi}{12\pi}-\frac{4\pi}{12\pi} \eeq

Assuming a rarity of
$\pm \frac{2\pi}{4\pi}$-monopoles (i.e.,
$\pm \frac{6\pi}{12\pi}$-monopoles in the $12\pi$ normalisation)
as well as $\frac{\pm 4\pi}{12\pi}$-monopoles  (i.e.,
monopoles corresponding to the strength of a nontrivial element
of the $\bz_3$ subgroup),
monopoles are for all practical purposes exclusively of
the $\frac{2\pi}{12\pi}$ type. And each of these ``minimal strength''
monopoles is formally a linear combination
of exactly one $\bz_2$-monopole and one $\bz_3$-monopole. Hence these latter
monopole types are ``present'' in essentially equal numbers.

As we would like to include not only the degree of first orderness
inherited from $\bz_2$ at the triple point, but also that inherited from
$\bz_3$, we need to generalise (\ref{hm})
and (\ref{just})  which were
derived for $\bz_2$ alone. The generalisation of
(\ref{hm}) is obtained by varying (\ref{gamefz23}):

\beq \Delta \gamma_{eff}=\frac{\beta_2}{4}\Delta(\langle
\sigma_{\sbz_2}\rangle)+
                         \frac{\beta_3}{9}\Delta(\langle
\sigma_{\sbz_3}\rangle)+
                         \frac{\beta_6}{36}\Delta(\langle
\sigma_{\sbz_6}\rangle).
\label{delgamun}                         \eeq

\nin where the notation has been changed such that $\langle \sigma \rangle
\stackrel{def}{=}\langle \on \rangle$
in (\ref{hm}) is in (\ref{delgamun}) denoted by
$\langle \sigma_{\sbz_2} \rangle$. For $\bz_3$ the analogous
quantity is denoted by  $\langle \sigma_{\sbz_3} \rangle$
                                      in (\ref{delgamun}).
Moreover, we have the notational change
$\beta_{crit\;BIO}\rightarrow \beta_{crit\;\sbz_2}$ in going from (\ref{hm}) to
(\ref{delgamun}). In (\ref{delgamun}) the analogous
quantities for $\bz_3$ and $\bz_6$ are  denoted respectively as
$\beta_{crit\;\sbz_3}$ and $\beta_{crit\;\sbz_6}$.
We have taken the $\beta_6$ term in (\ref{delgamun}) as being zero. This is
presumably justified by the smallness of the $\bz_6$ ``jump'' contribution
when treated (incorrectly) as being independent of $\bz_2$ and
$\bz_3$.

In going to the  new notation,  (\ref{just}) becomes (for $\bz_2$)

\beq \Delta \gamma_{eff}=\frac{1}{4}\beta_{crit\;\sbz_2}\Delta\langle S_{\Box}
\rangle_{\sbz_2}
\left(1+\xi\left(\frac{\sinh 2\beta}{\beta_{crit\;\sbz_2}}-2\cosh
2\beta\right)\right)
\eeq

\nin The generalisation that also includes the discontinuity
inherited from $\bz_3$ that contributes to
$\Delta\gamma_{eff}$
at the triple point transition from the
phase with just the discrete subgroups $\bz_2$ and $\bz_3$
confining to the totally Coulomb-like phase is

\beq
\Delta \gamma_{eff}=\sum_{N\in\{2,3\}}
\frac{1}{N^2}\beta_{crit\;\sbz_N}\Delta\langle S_{\Box}\rangle_{\sbz_N}
\left(1+\xi\left(\frac{\sinh 2\beta}{\beta_{crit\;\sbz_N}}-2\cosh
2\beta\right)\right)
 \label{final}. \eeq

\nin From the argumentation above, we know  that $\xi$ is expected to have the
same value in both terms of (\ref{final}).

In (\ref{final}) it is seen that the subgroups $\bz_2$ and $\bz_3$
both contribute a term to $\Delta\gamma_{eff}$ at the triple point.
Presumably it is a good
approximation to calculate $\Delta\gamma_{eff}$ as if contributions from
$\bz_2$ and $\bz_3$ are mutually independent inasmuch as these subgroups
factorize at the multiple point. However, even in this approximation there
will still be an indirect interaction between these subgroups via the continuum
\dofx in $U(1)$ and via the encircled plaquette in the dominant monopole
configuration.
Using (\ref{final}), the contributions from $\bz_2$ and  $\bz_3$
to $\Delta\gamma_{eff}$ are calculated and tabulated in
Table \ref{tab4}.

\begin{table}
\caption[table4]{\label{tab4}The quantity $\Delta\gamma_{eff}$ calculated
using the appropriate terms in (\ref{final}).
In the last row, the quantities for $\bz_6$ are calculated (incorrectly)
in a manner analogous to that used for $\bz_2$ and $\bz_3$. This procedure
presumably overestimates the effect of $\bz_6$ contributions.}
\beq
\begin{array}{|l|l|l|l|l|}\hline
  &\beta_{crit\;\sbz_N} & \Delta\langle S_{\Box}\rangle_{\sbz_N} & \xi &
\Delta\gamma_{eff} \\
  \hline
\bz_2 & 0.44 & 0.44 & 0.04 & 0.047_3 \\
   \hline
\bz_3 & 0.67 & 0.56 & 0.04 & 0.039_3 \\
   \hline
(\bz_6) & (1.00) & (0.13) & (0.0437) & (0.0033) \\
\hline\end{array}
\eeq
 \end{table}

\subsection{Calculating the enhancement factor for $1/\alpha_{U(1)}$
corresponding to the Planck scale breakdown of $U(1)^3$ to the diagonal
subgroup} \label{enhance}

The two approximations that we have developed in order to gain an insight
into the phase diagram for the group $U(1)^3$ - the independent monopole
approximation and the group volume approximation - are more or less
suitable according to whether the phase transitions are second or
first order.

To determine the correct enhancement factor, we interpolate between the
independent monopole approximation that gives this factor as 6 and the
volume approximation that puts this factor at about 8. This interpolation
is done by calculating the jump $\Delta W_{\Box\;``3"}$ in the Wilson
operator at the  boundary ``3'' transition at the TP
(see Figure~\ref{figbhanot}) that reflects the degree of first-orderness
inherited
at this transition from pure $\bz_2$ and $\bz_3$ transitions. As
$\Delta \gamma_{eff}$ expresses the degree of
first-orderness at the TP in going into the totally Coulomb-like phase,
$\Delta W_{\Box\;``3"}$ is calculated using the assumption that it
depends essentially on $\Delta \gamma_{eff}$.

The first approximation is the monopole  condensate  approximation  in
which the relevant quantity for which phase is realized is
the amount of fluctuation  in  the  convolution  of  the  6  plaquette
variables enclosing a 3-cube.

In the second approximation - based on the group volume approximation -
it turns out that to attain the multiple point in the hexagonal symmetry
scheme, it is necessary to introduce
additional  parameters  in  the  form  of
coefficients to 4$th$ and 6$th$ order perturbations to the Manton  action.
These additional parameters  are  used  to   get  the  free  energy
functions (corresponding to different phases) to coincide in parameter
space at a point - ``the'' multiple point. This point is shared by what
we expect is a maximum number of phases.

If, for example,  the Coulomb to confining phase transition for a
$Peter$-$U(1)$
subgroup is purely
second order,  this phase transition would not be expected to
cause any change in at what value of the distance along another subgroup axis
(e.g., the $Paul$-axis)  the
first identification-lattice point is encountered.
The reason is that there is no discontinuous change in the degree of
fluctuation in the $Peter$-plaquette variable in making the transition.
In this case we expect the independent monopole approximation
to work well.

On the other hand, if the phase transition is
very strongly first order so that the fluctuations along the $Peter$-subgroup
become discontinuously larger upon
passing into the $Peter$ confinement phase, this can be expected to affect
the threshold at which other subgroups go into confinement in a sort of
``interaction effect''. In this situation the volume-approximation
can be useful because it can take into account (and actually overestimates)
the influence that fluctuations along different directions in the group
can have on each other. The independent monopole approximation
tends to ignore
this effect.

Because the group volume approximation accounts for the interaction effect
between
fluctuations along different subgroups, it was necessary to use
$4th$ and $6th$ order action terms in order to
get a multiple point at which 12 phases convene (corresponding to continuous
invariant subgroups; we neglect an infinity of discrete subgroups in this
approximation). The effect of
the higher order terms is the preferential enhancement of quantum
fluctuations along the one dimensional (nearest neighbour) subgroup
directions of
the identification lattice thereby effectively eliminating the influence
that fluctuations along one subgroup have on the fluctuations along
another subgroup and vice versa.

In fact, the volume approximation effectively replaces the gauge group $G$
by its factor group $G/H$ when $H$ has confinement-like behaviour.
This amounts to treating the fluctuations along the component of the group
lying within the cosets $gH\;\;(g\in G)$
as being so large that, as far as Bianchi identities are concerned,
we can regard the distribution of elements within the cosets of $H$ as
essentially being that of the Haar measure
\footnote{We are interested in whether or not Bianchi identities introduce
correlations between plaquette variables that are sufficiently coherent
so as to lead to spontaneous breakdown of gauge symmetry under
transformations of the type (\ref{lin}). If the distribution along cosets
of $H$ is effectively the Haar measure,
all elements within a coset are accessed with equal probability and there
is not spontaneous breakdown under transformations of the type (\ref{lin})
as far the \dofx corresponding to the invariant subgroup $H$ are concerned.
Hence, the fulfilment of Bianchi identities in the case of the \dofx for which
we may not forget about them (i.e., when these identities can introduce
coherent correlations between plaquettes) is insured
by the more lenient requirement that Bianchi identities only need be fulfilled
after mapping the $U(\Box)\in G$ into the factor group $G/H$.
This is consistent with our
definition of confinement, which is that correlations between
values of different plaquette variables that are imposed by
Bianchi identities effectively disappear when a subgroup goes into the
confining phase. In the volume approximation, we can for calculational
purposes therefore assume the Haar
measure for the distribution of plaquette variables. Recall from earlier
sections that this is really not the case. Rather, going into confinement
at a first order phase
transition is accompanied by a discontinuous broadening of the  width of
the distribution
of elements within the cosets of the confined subgroup. But this is
sufficient to
suddenly allow the fulfilment of Bianchi identities by having  the sum
of plaquette variables add up to a nonzero multiples of $2\pi$ which
in turn reduces the effectiveness of Bianchi identities in introducing
coherent correlations between plaquettes which again allows larger plaquette
variable fluctuations
which again makes it even easier to avoid correlations from Bianchi
identities in a sort of self-perpetuating chain of events.}.

\subsubsection{The   independent   monopole
condensate approximation - the calculation}

In the  independent  monopole  approximation,  we  can
reach the multiple point using the  Manton  action alone (i.e.,  no
higher order terms). The diagonal $U(1)$  subgroup  to  be  identified
with the $U(1)$ of the $SMG$ is that given by $\theta(1,1,1)$  in  the
coordinate choice (\ref{coord3}).

The first identification lattice point  met by this diagonal
subgroup occurs for $\theta=2\pi$; i.e., the point $2\pi(1,1,1)$.
Hence the quantisation rule
$y/2 \in{ \bf Z}$ (for  particles  not  carrying  non-Abelian  gauge  coupling)
is achieved by the
naive        continuum        limit         identification         \beq
\exp(i\theta(-))=\exp(iag_1A_{\mu}y/2)\;\;\;\mbox{for}\;\;\;     y/2=1.
\eeq

\nin For $y/2=1$ (corresponding to $e^+_L$), the  covariant  derivative
is

\beq
D_{\mu}=
\partial_{\mu}
-ig_1A_{\mu}.
\eeq

The equation analogues to (\ref{dist1dim}) for the diagonal subgroup
(on the 3-dimensional identification lattice) is
\beq
\frac{\beta_{diag}}{2}(2\pi)^2=length(2\pi(1,1,1))= \label{diag6} \eeq

\[ =(2\pi)^2(1,1,1)
\frac{\beta_{crit}}{2}\left( \begin{array}{ccc} 1 & \frac{1}{2} & \frac{1}{2}
\\ \frac{1}{2}
& 1 & \frac{1}{2} \\ \frac{1}{2} & \frac{1}{2} & 1 \end{array} \right)
\left(   \begin{array}{c} 1 \\ 1 \\ 1 \end{array} \right)=(2\pi)^2\frac{
\beta_{crit}}{2} \cdot 6 \]

\nin for the multiple point. Contrary to the case of  the
non-Abelian couplings that are weakened  by  a  factor  $N_{gen}=3$  in
going to the diagonal subgroup of $U(1)^3$, the $U(1)$ coupling at  the
multiple
point is weakened by a factor 6 in going to the diagonal  subgroup  of
$U(1)^3$. In general, the weakening factor in the  hexagonal  case  in
going from $U(1)^{N_{gen}}$ to the diagonal subgroup $U(1)$ along  the
direction $(1,1,\cdots,1)$ is $N_{gen}+ \left(\begin{array}{c} N_{gen}
\\ 2 \end{array}\right)=N_{gen}(N_{gen}+1)/2$:

\nin so that
\beq g^2_{diag}=\frac{g^2_{crit}}{N_{gen}(N_{gen}+1)/2}.  \eeq

\subsubsection{The volume of groups scheme}

In the earlier section~\ref{grvolapprox}, we have developed a  means for
calculating
an effective inverse squared coupling having a directional dependence on
$4th$ and $6th$ order action terms. We now calculate
the effective inverse squared coupling (\ref{invcoupeff}) along the diagonal
subgroup:

\beq \frac{1}{e^2_{eff}(\vec{\xi})} \mbox{  (for $\xi=(1,1,1)$)  } = \eeq

\[
\left(B_6Y_{6\;\;comb}(diag)+\left(\frac{1}{e^2_{Manton}}+
B_4Y_{4\;\;comb}(diag)\right)^{\frac{3}{2}}\right)^{\frac{1}{3}} =\]

\[ =\left(\frac{-0.766}{e^6_{U(1)\;\;crit}}(-4\sqrt{\frac{3}{35}})+
\left(\frac{2^{\frac{2}{3}}}{e^4_{U(1)\;\;crit}}+
\frac{-0.146}{e^4_{U(1)\;\;crit}}\frac{2}{3}\sqrt{7}\right)^{\frac{3}{2}}
\right)^{\frac{1}{3}}= \]

\[ =1.34\cdot\frac{1}{e^2_{U(1)\;\;crit}} \;\;\;\mbox{  (in vol. approx.)}\]

\nin From (\ref{diag6}) we have that the inverse squared coupling
corresponding to the diagonal subgroup of $U(1)^3$ is
                                       a factor 6 larger than
$\frac{1}{e^2_{U(1)\;\;crit}}|_{\xi = (1,1,1)}\;$:

\beq \frac{1}{e^2(diag)}=6\cdot\frac{1}{e^2_{eff}(\vec{\xi})}\left |_{
\vec{\xi}=(1,1,1)}\right.=6\cdot 1.34=8.04 \eeq

\subsubsection{The calculation of the enhancement factor}

We have seen that the enhancement factor
$\frac{1/\alpha_{U(1)^3}}{1/\alpha_{U(1)}}$
has respectively the values 6.0 and $1.34\cdot 6.0$ according to whether the
``independent monopole'' or the ``volume'' approximation is used. These
approximations tend respectively to ignore and to overestimate the
dependence
that fluctuations in one subgroup can have on which phase is
realized along other subgroups or factor-groups.
This interaction effect depends on the
degree of first-orderness of the phase transition; this degree of
first-orderness is used in
our procedure
to determine to what extent the pure ``monopole  approximation'' should
be ``pushed'' towards the ``volume approximation''.
We seek a combination of these two approximations - with the relative weight
determined by the degree of first-orderness -
that is to be embodied in the value of $\Delta \gamma_{eff}$ that subsequently
is used in both steps of the calculation of the $U(1)$ continuum coupling.
In this section, we use $\Delta \gamma_{eff}$ to determine the $U(1)$ coupling
at the multiple point of a phase diagram for a $U(1)^3$ gauge group.

The correction
for the degree of first-orderness will be implemented by choosing the ``hop''
$\Delta W=\Delta\langle \cos \theta\rangle$ in the Wilson operator at the
TP transition to the totally Coulomb-like phase in such a way that it
reflects the residual first-orderness.
This transition
obviously has to separate confinement-like and Coulomb-like phases for the
{\em continuum} degrees of freedom.
There are two possibilities -
namely the TP transition at border ``1'' and the TP transition at border
``3'' corresponding let us say to respectively the jumps
$\Delta W_{\Box\;TP\;``1"}$
and $\Delta W_{\Box\;TP\;``3"}$ in the Wilson operator.

But we now
argue that $\Delta W_{\Box\;TP\;``1"}$ is not what we want because it
doesn't
reflect the degree of ``residual'' first-orderness (at the TP) that is due
to the $Z_N$ transition\footnote{In fact by using a trick of changing variables
(more on this below), we
can actually show that $\Delta W_{\Box\;TP\;``1"}\approx
\Delta W_{\Box\;\gamma=0\;``3"}$.
The latter reflects (less pronounced) residual first-orderness of the $Z_N$
transition quite far removed from the TP - namely that for $\gamma=0$
which is not so far from the tri-critical point at $\gamma=-0.11$ where
all remnants of the $Z_N$ transition disappear and the transition at
boundary ``3''continues for $\gamma<-0.11$ as a pure second order transition.}
The reason has to do with  $Z_N$ ($N=2,3$) being in the same
phase on both sides of the
border ``1'' at the TP. Accordingly, $\Delta W_{\Box\;TP\;``1"}$ cannot
reflect the discrete group transition.

So it is the discontinuity  $\Delta W_{\Box\;TP \;``3"}$ that we want to
use to interpolate between the  ``independent monopole'' and the ``volume''
approximation so as to obtain the
enhancement factor $\frac{\alpha_{U(1)^3\;diag}}{\alpha_{U(1)\;crit}}$ that
reflects the appropriate degree of first-orderness for the TP transition
in going from confinement to Coulomb-like behaviour for the continuum \dof.

In order to estimate the residual ``first-orderness'' present at the multiple
point in making the transition to the totally Coulomb-like phase from the
phase(s)
with confinement solely w.r.t to discrete subgroup(s), we shall
use the already proposed scenario in which we speculate that the increased
frequency of minimal strength monopoles (i.e., $\frac{2\pi}{4\pi}$
monopoles in the $4\pi$ normalisation implicit in
(\ref{newbhanot})) is
related
to the fact
that the phase transition along the border ``2'' in Figure~\ref{figbhanot}
becomes less and
less strongly
first order as $\gamma$ decreases. That is, we speculate that the increasing
role of minimal strength monopoles (in the $4\pi$ normalisation, the minimal
strength $\frac{2\pi}{4\pi}$ monopoles are the only monopoles; in the $12\pi$
normalisation, there are, in addition to minimal and most abundant
$\frac{2\pi}{12\pi}$ monopoles, also (less common) $\frac{6\pi}{12\pi}$-
and $\frac{4\pi}{12\pi}$-monopoles)
in typical plaquette
configurations
is the reason that the transitions to the totally Coulomb-like phase at border
``2'' and subsequently, also at border ``3'' in Figure~\ref{figbhanot}
becomes less and less first order as $\gamma$ is diminished.

As mentioned just above, it is well
known that, for $U(1)$, the phase transition at border ``3'' becomes  second
order at the tri-critical point (at a slightly negative value of $\gamma$)
and continues as a second order phase transition for $\gamma$ values less
than the tri-critical value $\gamma_{TCP}$.
The above picture is not inconsistent with the results of numerical studies
that clearly reveal even a pure $U(1)$  gauge theory with a Wilson action
(i.e., a theory with $\gamma=0$) as having a
weakly first order phase transition as evidenced by a ``jump''
$\Delta W_{\Box}$ in the Wilson operator $W_{\Box}$.
Indeed one finds in the work of Jers\`{a}k\cite{tricrit} {\em et al} fits that
relate the ``jump'' $\Delta W_{\Box}$ in the Wilson operator $W_{\Box}
\stackrel{def}{=} \langle \cos (\theta_{\Box}) \rangle$ to $\Delta \gamma
\stackrel{def}{=}\gamma-\gamma_{TCP}$ where $\gamma_{TCP}$ denotes the
value of $\gamma$ in the tri-critical point:

\beq  \Delta W_{\Box}=A(\gamma-\gamma_{TCP})^{\beta_{\mu}} \label{gamnul}\eeq

\nin The values for $\gamma_{TCP}$ and $\beta_{\mu}$ are given respectively as
$\gamma_{TCP}=-0.11\pm 0.05$ and $\beta_{\mu}=1.7\pm 0.2$ while the constant
$A$ is deduced to be $A=0.68_{35}$. For $\gamma=0$ (corresponding to a
Wilson action), there obtains $\Delta W_{\Box}=0.68(0.11)^{1.7}=0.016$.

Actually this latter discontinuity will be seen to be of interest to us
because it can be shown
that this jump is to a good approximation the jump $\Delta W_{\Box,\;``1"}$
encountered in crossing border ``1'' near the multiple point. The reasoning
is as follows: the jump $\Delta W_{\Box,\;``1"}$ is to a good approximation
constant
along the phase border ``1''; consequently, $\Delta W_{\Box,\;``1"}$ near the
multiple point is essentially the same as that at $\gamma=1.01$ and $\beta=0$
which, in turn, is,
by a simple change of  notation, identical with the discontinuity
$\Delta W_{\Box}$ at $\gamma=0$, $\beta=1.01$ that using (\ref{gamnul}) was
found to have the value  $\Delta W_{\Box}=0.016$.

So what is wanted for the purpose of calculating the enhancement factor is the
jump $\Delta W_{\Box,\;``3"}$ encountered at the multiple point in
traversing border ``3" separating the totally Coulomb-like and totally
confinement-like phases. What we have is a way to calculate
$\Delta W_{\Box;\;``2"}$: this procedure
relates $\Delta W_{\Box,\;``2"}$ to the cubic root\cite{lautrup,moriarty}
of the quantity
$\Delta \gamma_{eff}$ (see Section~\ref{deltagam}) encountered in crossing
the border ``2'' at the multiple point. Were
it not that the transition at border ``1'' is (weakly) first order but
instead second order, then we would have had
$\Delta W_{\Box,\;``1"}=0 $
and

\beq  \Delta W_{\Box,\;``2"}=\Delta W_{\Box,\;``3"}=
A(\Delta \gamma_{eff})^{\frac{1}{3}}\;\;\;\mbox{ (when
$\Delta W_{\Box,\;``1"}=0$)}
\label{2ndord}\eeq

\nin where $A=0.252$.
However, having argued that $\Delta W_{\Box,\;``1"}=0.16\neq 0 $
corresponding to a weakly first order transition in crossing border ``1''
 in the
vicinity of the multiple point, we conclude  on the grounds of
continuity that this jump must be the difference in the ``jumps''
$\Delta W_{\Box,\;``2"}$ and $\Delta W_{\Box,\;``3"}$ in crossing
respectively
the borders ``2'' and ``3'' at the multiple point (see Figure~\ref{figbhanot}).
Recall that
these jumps, observed in crossing the borders ``2'' and ``3'' near the
multiple point
are essentially assumed to be the residual effects of first-order pure
discrete subgroup transitions at
large $\gamma$.
So in principle at least, the ``jump'' $\Delta W_{\Box;\;``3"}$ is obtained by
correcting  \footnote{The reason that we do the calculation in this
circuitous way - instead of trying to directly estimate
$\Delta W_{\Box\;``3"}$
by first calculating the ``$\Delta \gamma_{eff}$'' at boundary ``3'' - is
that it is not clear what this latter $\Delta \gamma_{eff}$ means. The reason
that we calculate $\Delta \gamma_{eff}$ at boundary ``2'' is that the
phases on both sides of this boundary are very similar w.r.t the
continuum \dof. This allows us to conclude that our $\Delta \gamma_{eff}$ at
boundary ``2'' can be associated essentially alone with the discrete
subgroup transition.} $\Delta W_{\Box\;``2"}$ (calculated by using
(\ref{2ndord})) by the
amount of the ``jump'' $\Delta W_{\Box,\;``1"}$
in crossing border ``1''.
Using that $\Delta W_{\Box,\;``1"}$ is small, we
make this correction in an approximate way by increasing $\Delta \gamma_
{eff}$ in (\ref{2ndord}) by the corrective quantity

\beq \Delta \gamma_{corr\;``1"}\stackrel {def}{=}
(\frac{\Delta W_{\Box,\;``1"}}{A})^3  \label{corr} \eeq

\nin obtained by inverting (\ref{2ndord}). In this approximation, we obtain

\beq \Delta W_{\Box,\;``3"}\approx A(\Delta \gamma_{eff}+
\Delta \gamma_{corr\;``1"})^{\frac{1}{3}}= \label{jump3}\eeq

\[ = A(\Delta \gamma_{eff}+ (\frac{0.016}{0.252})^3)^{\frac{1}{3}}  \]

\nin where we have used that $\Delta W_{\Box,\;``1"}=0.016$ in (\ref{corr})
which in turn has been used in (\ref{jump3}). Strictly speaking, it is
inconsistent to assume additivity in the ``jumps'' $\Delta W_{\Box,\;``1"}$,
$\Delta W_{\Box,\;``2"}$, and $\Delta W_{\Box,\;``3"}$ (essential because
of continuity requirements) {\em and} at the same time that both
$\Delta W_{\Box,\;``2"}$ and $\Delta W_{\Box,\;``3"}$ are related to an
appropriate $\gamma_{eff}$ by a cubic root law. Consistency requires
$\Delta W_{\Box,\;``1"}=0$ corresponding to a second order transition. For
small $\Delta W_{\Box,\;``1"}$, this inconsistency is not bothersome
and the approximation (\ref{jump3}) is good. In fact the corrective term
$\Delta \gamma_{corr\;``1"}$ is  so small so as not to yield a difference
in $\Delta W_{\Box,\;``2"}$ and $\Delta W_{\Box,\;``3"}$ that is discernible
to within the calculational accuracy.

Equation (\ref{jump3}) provides a way of calculating
the for us interesting $\Delta W_{\Box,\;``3"}$ at the multiple point.
Various values of $\Delta W_{\Box,\;``3"}$ are tabulated in Table \ref{tab5}.
These are calculated for different values of $\Delta \gamma_{eff}$ that in
turn are obtained as combinations of the $\Delta \gamma_{eff}$ in
Table \ref{tab4} calculated
for the $\bz_2$, $\bz_3$, and $\bz_6$ discrete subgroups of $U(1)$.

Before we use these various $\Delta W_{\Box,\;``3"}$ values to calculate the
enhancement factor
$\frac{\alpha_{U(1)^3\;diag}}{\alpha_{U(1)\;crit}}$, we need to
develop a way of using the $\Delta W_{\Box,\;``3"}$ to
interpolate between the ``pure monopole''
and the ``volume'' approximation. We now do this for the general case of any
discontinuity $\Delta W_{\Box}$.
In general, when there is a ``jump'' $\Delta W_{\Box}$,
we estimate that we get the most
correct enhancement factor
$\frac{1/\alpha_{U(1)^3\;diag}}{1/\alpha_{U(1)}\;crit}$
by linearly interpolating between the enhancement factor ``6'' corresponding
to the independent monopole approximation and the enhancement factor
$1.34\cdot 6=8.04$ corresponding to the volume approximation. That is,
the enhancement factor is calculated as

\[ \left(\frac{1/\alpha_{U(1)^3\;diag}}{1/\alpha_{U(1)}}\right)_{actual}=
\left(\frac{1/\alpha_{U(1)^3\;diag}}{1/\alpha_{U(1)}}\right)_{ind\;mono}+ \]
\[ +\frac{\eta}{\tau}
\left[\left(\frac{1/\alpha_{U(1)^3\;diag}}{1/\alpha_{U(1)}}\right)_{vol}-
\left(\frac{1/\alpha_{U(1)^3\;diag}}{1/\alpha_{U(1)}}\right)_{ind\;mono}\right]
\]

\beq= 6+\frac{\eta}{\tau}[6(1.34-1)] \label{intpol}\eeq

\nin where $\frac{\eta}{\tau}$ is given by

\beq
\frac{\eta}{\tau}=
\frac{\left(\frac{Coul\;fluc}{conf\;fluc}\right)^2_{ind\;\;mono}-
     \left(\frac{Coul\;fluc}{conf\;fluc}\right)^2_{actual}}
     {\left(\frac{Coul\;fluc}{conf\;fluc}\right)^2_{ind\;\;mono}-
     \left(\frac{Coul\;fluc}{conf\;fluc}\right)^2_{vol}} \label{etatau}\eeq

\nin and $\eta$ is defined as the numerator while $\tau$ the denominator on
the right hand side of (\ref{etatau}).  Write

\beq \left(\frac{Coul\;fluc}{conf\;fluc}\right)^2\stackrel{def}{=}
     \frac{1-\langle \cos \theta\rangle_{Coul}}
     {1-\langle \cos \theta\rangle_{conf}} \label{coulconf}
=1-\frac{\Delta W_{\Box}}{1-\langle \cos \theta \rangle_{conf}} \label{yes}
\eeq

\nin where in the last step we have used that
$\langle \cos\theta \rangle_{Coul}=\langle \cos\theta \rangle_{conf}+
\Delta W_{\Box}$.

Using that
$\left(\frac{Coul\;fluc}{conf\;fluc}\right)^2_{ind\;mono}=1$ essentially
by definition, we have using (\ref{etatau}) and (\ref{yes}) that

\beq \eta=\frac{\Delta W_{\Box}}
{1-\langle \cos \theta \rangle_{conf}}= \Delta W_{\Box}/0.377 \label{eta} \eeq

\nin where in (\ref{eta}) we have used $\langle \cos\theta \rangle_{conf\;ph}
=0.623$.

Various values of $\eta$ are tabulated in Table \ref{tab5} corresponding
to the values of $\Delta W_{\Box,\;``3''}$ that are also tabulated
in the same Table.

The quantity $\frac{\eta}{\tau}$ are used to obtain the values for the
enhancement factors $\frac{1/\alpha_{U(1)^3\;diag}}{1/\alpha_{U(1)}\;crit}$
tabulated in the final two columns of Table \ref{tab5}.
The two columns correspond to
$\frac{1/\alpha_{U(1)^3\;diag}}{1/\alpha_{U(1)}\;crit}$ for
two different values of $\tau$. The first, corresponding
to the roughest approximation, is for $\tau=1$ inasmuch as we make the
approximation
$\left(\frac{Coul\;fluc}{conf\;fluc}\right)^2_{vol}
\approx 0$. The enhancement factors in the column at the extreme right hand
side are obtained using a  better estimate\footnote{$\tau=
1-\left(\frac{Coul\;fluc}{conf\;fluc}\right)^2_{vol}=
1-\frac{2(1-\langle \cos \theta \rangle)_{Coul/\; crit}}
{\langle \theta^2\rangle_{conf}}=1-\frac{2(1-0.65)}{\pi^2/3}\approx 1-0.21
=0.79.$
Here $\langle \theta^2 \rangle_{conf}$ is calculated as though one had the
ideal Haar measure distribution which is the distribution used in effect in
our volume approximation.}
of $\tau$:
\beq \tau=1-\left(\frac{Coul\;fluc}{conf\;fluc}\right)^2_{vol}=1-0.21=0.79.\eeq

The values of $\frac{1/\alpha_{U(1)^3\;diag}}{1/\alpha_{U(1)\;crit}}$ in the
last column of Table \ref{tab5} will appear in Table \ref{tab6} in
conjunction with the calculation of the Planck scale value of the continuum
$U(1)$ fine-structure constant $1/\alpha_{U(1)\;Pl.\;scale}$.





\begin{footnotesize}

\begin{table}
\caption[table5]{\label{tab5}Enhancement factors given in the
last four columns on the right are given for two ways of calculating
$\tau$ as well as with and without $\Delta\gamma_{corr``1"}$ included
in the calculation of $\Delta W_{\Box}$ in (\ref{jump3}).
For the quantity $\tau\stackrel{def}{=}
1-\left(\frac{Coul\;fluc}{conf\;fluc}\right)^2$
we have $\tau=1$ when confinement fluctuations are taken as infinite and
$\tau=0.79$  when confinement fluctuations are taken as finite.
The second and third columns contain
$\Delta W_{\Box}$ calculated respectively with and without the quantity
$\Delta\gamma_{corr``1"}$ in (\ref{jump3}).
The values for $\Delta\gamma_{eff}$ in the first column are taken from
Table~\ref{tab4}. The
quantity $\eta$ in the fourth and fifth columns is defined in (\ref{etatau})
and calculated according
to (\ref{eta})
with and without the quantity
$\Delta\gamma_{corr``1"}$ in the expression (\ref{jump3}) for
$\Delta W_{\Box}$. }
\vspace{.7cm}
\[\!\!\!\!\!\!\!\!\!\!\!\!\!\!\!\!\!\!\!\!\mbox{
\begin{tabular}{|c||c|c|c|c|c|c|c|c|c|}\hline
Procedure & $\Delta \gamma_{eff}$ & \mc{2}{c|}{$\Delta W_{\Box ,``3"}$ from
(\ref{jump3})} &
    \multicolumn{2}{c|}{$\eta$ from (\ref{eta})} &
    \multicolumn{4}{c|}{$\frac{1/\alpha_{{U(1)}^3_{``diag"}}}
{1/\alpha_{{U(1)}_{``crit"}}}$ from (\ref{intpol})} \\
    \cline{3-10} & & with  & $\Delta \gamma_{corr``1"}$  &
    with & $\Delta \gamma_{corr``1"}$ &
 \mc{2}{c|}{with $\Delta \gamma_{corr``1"}$} & \mc{2}{c|}{$\Delta
\gamma_{corr``1"}=0$}\\
\cline{7-10} & & $\Delta \gamma_{corr``1"}$ & $=0$ & $\Delta \gamma_{corr``1"}$
& $=0$ &  $\tau=0.79$ & $\tau=1$ & $\tau=0.79$ & $\tau=1$ \\ \hline \hline
Vol Approx: &&\mc{2}{c|}{}&\mc{2}{c|}{}&\mc{2}{c|}{}&& \\
Haar (ideal)& &\mc{2}{c|}{{\tiny ($\langle \cos \theta \rangle_{Coul}=0.65$)}}
&\mc{2}{c|}{1} &\mc{2}{c|}{} &  & 8.04 \\
Haar, compact & &\mc{2}{c|}{{\tiny ($\langle \cos \theta \rangle_{Coul}=0.65$)}
} & \mc{2}{c|}{0.79}  & \mc{2}{c|}{} & 8.04 &  \\
Mean field & &\mc{2}{c|}{}&\mc{2}{c|}{$\frac{1}{2}$} & \mc{2}{c|}{} &
7.29 & 7.02 \\ \hline
Ideal ind mono &0&\mc{2}{c|}{0} & \mc{2}{c|}{1} &\mc{2}{c|}{6} &\mc{2}{c|}{6}
\\ \hline
No discrete &0 & 0.016 &0 & 0.042$_4$  & 0  & 6.11$_0$ & 6.08$_7$  & 6 & 6 \\
subgroups  &&&&&&&&& \\ \hline
Using  & & & & & & & & &  \\
$\bz_2$ only: & 0.047$_3$ & 0.091$_3$ & 0.091$_1$ & 0.24$_2$ & 0.24$_2$ &
6.62$_5$ & 6.49$_4$ & 6.62$_4$  & 6.49$_3$ \\  \hline
Using & 0.047$_3$+ & & & & & & & & \\
$\bz_2+\bz_3$: & 0.039$_3$ & 0.11$_2$ & 0.11$_1$ & 0.29$_6$ & 0.29$_6$ &
6.76$_4$ &6.60$_4$ & 6.76$_4$  & 6.60$_3$ \\ \hline
Using & $\frac{0.047_3}{2}+$ & & & & & & & & \\
$\frac{1}{2}(\bz_2+\bz_3)$: & $+\frac{0.039_3}{2}$ & 0.088$_7$ & 0.088$_5$ &
0.23$_5$ & 0.23$_5$ &  6.60$_7$ & 6.48$_0$ & 6.60$_6$  & 6.47$_9$ \\ \hline
Using & $\frac{0.047_3}{2}$ & & & & & & & & \\
$\frac{1}{2}\bz_2+\bz_3$: & $+0.039_3$ & 0.10$_0$ & 0.10$_0$ & 0.26$_6$ &
0.26$_6$ & 6.68$_8$ & 6.54$_3$ & 6.68$_7$  & 6.54$_2$  \\ \hline
\end{tabular}}\]
\end{table}

\end{footnotesize}


\subsection{Continuum  critical
coupling from critical $U(1)$ lattice coupling}\label{contcoup}

The Planck scale prediction for the $U(1)$ fine-structure constant is to be
obtained as the product of the enhancement factor and the continuum critical
coupling that corresponds to the lattice critical coupling.

We have the enhancement factor in Table~\ref{tab5} (calculated using different
approximations) but we have yet to translate
the lattice $U(1)$ critical coupling
into a continuum one. This is the purpose of this section.


We use a procedure analogous to that used by Jers\`{a}k et al\cite{jersak}.
In this work the continuum coupling is calculated
numerically. Using Monte Carlo methods on the lattice, the Coulomb potential
is computed  and fitted to the formula proposed by Luck (\cite{luck}).
In the Coulomb phase with  the Wilson action, the fit yields

\beq
Wilson:\;\;\;   \alpha(\beta) =0.20 - 0.24 (\frac{\beta
-\beta_{crit}}{\beta})^{0.39}\;= \;
0.20 - 0.24 (1-\frac{1.0106}{\beta})^{0.39}. \label{contfor}
\eeq
For the Villain action (in the Coulomb phase) the analogous result is
\beq
Villain:\;\;\;   \alpha(\beta) = 0.20 - 0.33 (1-\frac{0.643 }{\beta})^{0.52}.
\label{vil} \eeq

It is of course our intention to substitute $\gamma_{eff}$ for what
Jersak et al. designates as $\beta$. This is justified in as much as
(\ref{contfor}) is valid for $\beta \geq\beta_{crit}$; i.e., for $\beta$ lying
{\em within the  Coulomb phase}. The replacement
of
\beq\beta -\beta_{crit}\eeq

\nin  by

\beq \gamma_{eff\;tot\;Coul\;ph}-\gamma_{eff.\;only\;\sbz_2\;conf}
\stackrel{def}{=} \Delta \gamma_{eff} \eeq

\nin is valid
inasmuch as the phases separated by the phase boundary ``2'' in
Figure~\ref{figbhanot} are both in the Coulomb phase as far as the
{\em continuum}
\dofx are concerned. Values obtained for $\alpha$ using (\ref{contfor})
with $\beta$ replaced by $\gamma_{eff}$ and $\beta-\beta_{crit}$ by
$\Delta \gamma_{eff}$ are tabulated for various values of the latter
in the third column of Table~\ref{tab6}.


\begin{footnotesize}
\begin{table}
\caption[table6]{\label{tab6}Our Planck scale prediction for the $U(1)$
fine-structure constant
is obtained as the product of the enhancement factor (from the last four
columns of Table~(\ref{tab5})) and the value of
$1/\alpha_{cont}$ obtained from the critical value of the lattice parameter
$\gamma_{eff}$ and the ``jump'' $\Delta \gamma_{eff}$ in this same quantity
in crossing the phase border ``3'' (see Figure~\ref{figbhanot}) at the multiple
point.
We list a number of
combinations that differ according to how the discrete subgroups are treated
w.r.t. whether the discrete subgroups are large enough to have the symmetry
of the hexagonal identification lattice
and how $\tau$ (see Table~\ref{tab5}) is calculated as an indication of the
sensitivity of our
prediction to such details. The prediction marked with ``$\bullet$'' indicates
the predicted value calculated in what we regard as the most correct manner.
Also included are results for the Villian action where (\ref{vil}) has been
used to calculate $\alpha(\beta)$. In this table, we use the same
$\Delta\gamma_{eff}$
in both (\ref{contfor}) and (\ref{vil}) (this is incorrect; see next table).
In the Villian case, the $\Delta W_{\Box\;``3"}$  used in calculating the
enhancement factor is calculated as $\Delta W_{\Box\;``3"}=
0.16(\Delta\gamma_{eff})^{0.29}$ (this is the counterpart of (\ref{jump3})
for the Wilson action)
with $\Delta_{``1"\;corr}=0$). The coefficient ``0.16'' is estimated
from
Monte Carlo data in (\cite{jersak}) and is rather uncertain\footnotemark.}
\beq \!\!\!\!\!\!\!\!\!\!\!\!\!
\begin{array}{|l|l|l|l|l|l|l|l|}\hline
\multicolumn{8}{|c|}{ \mbox{single $U(1)$}}  \\ \hline
\mbox{Procedure} &\Delta\gamma_{eff} & \alpha_{cont.} & 1/\alpha_{cont.}
&\mc{2}{c|}{\mbox{enh. fac Haar}}&
\mc{2}{c|}{\mbox{prediction $1/\alpha_{Pl. scale}$}} \\
\cline{5-8}  & & & & \tau=0.79 & \tau=1 & \tau=.79 & \tau=1 \\ \hline
\mbox{Wilson action (using (\ref{contfor}))}&&&&&&& \\ \hline
\mbox{$\bz_2$ only with $\Delta\gamma_{corr``1"}$ in (\ref{jump3})}
& 0.047_3 & 0.128_6 & 7.77_8 & 6.62_5 & 6.49_4 & 51.5 & 50.5 \\ \hline
\mbox{$\bz_2$ only without $\Delta\gamma_{corr``1"}$ in (\ref{jump3})}
& 0.047_3 & 0.128_6 & 7.77_8 & 6.62_4 & 6.49_3 & 51.5 & 50.5 \\ \hline
\mbox{$\bz_2+\bz_3$ with $\Delta\gamma_{corr``1"}$ in (\ref{jump3})}
& 0.086_6 & 0.110_8 & 9.02_1 &  6.76_4 & 6.60_4 & 61.0 & 59.6 \\ \hline
\mbox{$\bz_2+\bz_3$ without $\Delta\gamma_{corr``1"}$ in (\ref{jump3})}
& 0.086_6 & 0.110_8 & 9.02_1 & 6.76_4 & 6.60_3 & 61.0 & 59.6 \\ \hline
\mbox{$\frac{1}{2}(\bz_2+\bz_3)$ with $\Delta\gamma_{corr``1"}$ in (\ref{jump3}
)} & 0.043_3 & 0.130_9 & 7.64_0 & 6.60_7 & 6.48_0 & 50.5 & 49.5 \\ \hline
\mbox{$\frac{1}{2}(\bz_2+\bz_3)$ without $\Delta\gamma_{corr``1"}$ in
(\ref{jump3})} & 0.043_3 & 0.130_9 & 7.64_0 & 6.60_6 & 6.47_9 & 50.5 & 49.5 \\
\hline
\mbox{$\frac{1}{2}\bz_2+\bz_3$ with $\Delta\gamma_{corr``1"}$ in (\ref{jump3})}
 & 0.063_0 & 0.120_6 & 8.29_2 & 6.68_9 & 6.54_3 & 55.5\;\;\bullet & 54.3 \\
\hline
\mbox{$\frac{1}{2}\bz_2+\bz_3$ without $\Delta\gamma_{corr``1"}$ in
(\ref{jump3})} & 0.063_0 & 0.120_6 & 8.29_2 & 6.68_7 & 6.54_2 & 55.5\;\;\bullet
& 54.3 \\ \hline\hline
\mbox{Villian action (using (\ref{vil}))}&&&&&&& \\ \hline
\mbox{$\bz_2$ only}   &0.047_3 &0.11_8 &8.46_5 & 6.45_2  & 6.35_7 & 54.6 & 53.8
 \\ \hline
\mbox{$\bz_2+\bz_3$} &0.086_6 &0.091_1 &10.9_8 & 6.53_9 & 6.42_6 & 71.8 & 70.6
\\ \hline
\mbox{$\frac{1}{2}(\bz_2+\bz_3)$}&0.043_3&0.12_2 &8.22_6 & 6.44_1  & 6.34_8
& 53.0\;\;\bullet & 52.2 \\ \hline
\mbox{$\frac{1}{2}\bz_2+\bz_3$} & 0.063_0 & 0.10_6 & 9.42_4 &6.49_1 & 6.38_8
& 61.2\;\;\bullet & 60.2 \\ \hline
\end{array}
\eeq
\end{table}
\end{footnotesize}
\footnotetext{
Making that the assumption that the phase transitions for both the Wilson
and Villian actions are second order, we take the  difference
$\langle \theta^2 \rangle -\langle \theta^2 \rangle_{crit.}$
as being the same when the string tension is the same for both action
types. Using figure 4a in Jersak et al: Nucl. Phys. B251, 1985, 299,
we obtain the coefficient 0.23 as the coefficient of $(\Delta \gamma_{eff})
^{.29}$. Allowing for the fact that the transitions are not strictly
second order gives rise to a correction that results in a coefficient
of 0.16 instead of the 0.23.}
The $\gamma_{eff}$ used in Table~\ref{tab6} is that  calculated in
(\ref{actz2})
to lowest order in $\hat{\theta}$. We want now to go to next order in
$\hat{\theta}$. For the Wilson action suitable for having a phase confined
solely
w.r.t. $\bz_2$, the appropriate effective coupling is
given by (\ref{gammaeffect}). We denote this improved effective
coupling by $\gamma_{eff\;corr}$:

\beq \gamma_{eff\;corr}=
\gamma+\frac{\langle\sigma_{\bz_{2}}\rangle_{\hat{\theta}}\beta}{4}
\langle\frac{\cos(\hat{\theta}/2)}
{\cos\hat{\theta}}\rangle \approx
\gamma+\frac{\langle\sigma_{\bz_{2}}\rangle_{\hat{\theta}}\beta}{4}
\langle \cos^{-\frac{3}{4}}\hat{\theta}\rangle \mbox{      (Wilson action)}.
\label{gameffectwil} \eeq

\nin The improved $\gamma_{eff\;corr}$ for the
Villian action case has the $\cos\hat{\theta}$ in the denominator in the
average on the left-hand side
of (\ref{gameffectwil}) removed
corresponding to the Villian action being approximately a Manton action
(having a second derivative that is $\hat{\theta}$-independent) instead
of being
equal to $\cos\hat{\theta}$ as in the Wilson case. So for the Villian action
we have


\beq \gamma_{eff\;corr}= \gamma+\frac{\beta}{4}\langle\sigma_{\bz_{2}}\rangle
\cdot\langle\cos(\hat{\theta}/2) \rangle \approx
\gamma+\frac{\beta}{4}\langle\sigma_{\bz_{2}}\rangle
\cdot\langle\cos^{\frac{1}{4}}\hat{\theta} \rangle
\;\;\;\mbox{(Villian action)}. \label{gameffectvil} \eeq

The effective couplings (\ref{gameffectwil}) and (\ref{gameffectvil})
are for respectively the Wilson and
Villian actions. In both cases there can be a phase confined solely w.r.t.
$\bz_2$. The analogous couplings for the Wilson and Villian actions in
the case where there is a  phase confined solely w.r.t. $\bz_3$ are given
respectively by (\ref{gameffwilz3}) and (\ref{gameffvilz3}) below; i.e.,
by

\beq \gamma_{eff\;corr}=
\gamma+\frac{\beta}{9}\langle\sigma_{\bz_{3}}\rangle
\langle \cos^{-\frac{8}{9}}\hat{\theta}\rangle \mbox{      (Wilson action)}
\label{gameffwilz3} \eeq

\nin and

\beq \gamma_{eff\;corr}=
\gamma+\frac{\beta}{9}\langle\sigma_{\bz_{3}}\rangle
\cdot\langle\cos^{\frac{1}{9}}\hat{\theta} \rangle
\;\;\;\mbox{(Villian action)}.
\label{gameffvilz3} \eeq

Table~\ref{tab7} lists the $U(1)$ fine-structure constant at the Planck  scale
that is
calculated  for different combinations of contributions from
$\bz_2$ and $\bz_3$ in the Wilson and Villian action cases
using the improved effective couplings
$\gamma_{eff\;corr}$ in (\ref{gameffectwil}), (\ref{gameffectvil})
(\ref{gameffwilz3}) and (\ref{gameffvilz3}).

\begin{footnotesize}
\begin{table}
\caption[table6a]{\label{tab7}Here we use the improved effective couplings
$\gamma_{eff\;corr}$ in (\ref{gameffectwil}) and (\ref{gameffectvil})
corresponding respectively to Wilson and
Villian actions for which there is a phase confined solely w.r.t. $\bz_2$.
The analogous improved effective couplings (\ref{gameffwilz3}) and
(\ref{gameffvilz3}) are used respectively
for the Wilson and
Villian actions that can provoke phases confined solely w.r.t. $\bz_3$.
Strictly speaking, for the improved calculation of $\Delta \gamma_{eff}$ -
i.e., $\Delta \gamma_{eff\;corr}$ - we should (for say the Wilson action
in the case where we have a phase confined solely w.r.t $\bz_2$)
calculate as follows:
$ \Delta\gamma_{eff\;corr}= \frac{\beta}{4}\left(
\langle \sigma \rangle_{Coul}\left\langle\frac{\cos\frac{\hat{\theta}}{2}}
{\cos\hat{\theta}}\right\rangle_{Coul}-
\langle \sigma \rangle_{conf}\left\langle\frac{\cos\frac{\hat{\theta}}{2}}
{\cos\hat{\theta}}\right\rangle_{conf}\right)$
but because
$\langle \sigma \rangle_{conf} << \langle \sigma \rangle_{Coul}$ we have
$\Delta\gamma_{eff\;corr}
\approx
\frac{\beta}{4}\left(
\langle \sigma \rangle_{Coul}\left\langle\frac{\cos\frac{\hat{\theta}}{2}}
{\cos\hat{\theta}}\right\rangle_{Coul}\right).$
We calculate $\Delta \gamma_{eff\;corr}$ iteratively inasmuch as the latter
is needed to get $\Delta W$ which is needed to get $\langle \cos\hat{\theta}
\rangle$ which in turn is needed to calculate $\Delta \gamma_{eff\;corr}$.
The $\Delta W$ obtained iteratively using $\Delta \gamma_{eff\;corr}$ is also
used in calculating the enhancement factor in Table~\ref{tab7}.
In the case of the Villian action, the cos$\hat{\theta}$ in the denominator of
$\left\langle\frac{\cos\frac{\hat{\theta}}{2}}
{\cos\hat{\theta}}\right\rangle$ is removed. The case having a
phase confined solely w.r.t. $\bz_3$ is calculated in a way analogous to
that for $\bz_2$ for respectively
the Wilson and Villian action cases.}
\beq \!\!\!\!\!\!\!\!\!\!\!\!\!\!\!\!\!
\begin{array}{|l|l|l|l|l|l|l|l|}\hline
\multicolumn{8}{|c|}{ \mbox{corrected $U(1)$ }}  \\ \hline
\mbox{Procedure} &\Delta\gamma_{eff,\;corr} & \alpha_{cont.} & 1/\alpha_{cont.}
 &\mc{2}{c|}{\mbox{enh. factor Haar}}&
\mc{2}{c|}{\mbox{prediction $1/\alpha_{Pl. scale}$}} \\
\cline{5-8}  & & & & \tau=0.79 & \tau=1 & \tau=0.79 & \tau=1 \\ \hline
\mbox{Wilson action (using (\ref{contfor}))}&&&&&&& \\ \hline
\mbox{$\bz_2$ only with $\Delta\gamma_{corr``1"}$ in (\ref{jump3})} &
0.060_0 & 0.122_0 & 8.19_6 & 6.67_7 & 6.53_5 & 54.7 & 53.6 \\ \hline
\mbox{$\bz_2+\bz_3$ with $\Delta\gamma_{corr``1"}$ in (\ref{jump3})} &
0.109_4 & 0.103_1 & 9.69_7 & 6.82_6 & 6.65_3 & 66.2 & 64.5 \\ \hline
\mbox{$\frac{1}{2}(\bz_2+\bz_3)$ with $\Delta\gamma_{corr``1"}$ in (\ref{jump3}
)}&0.0561_5 & 0.123_9 & 8.07_2 & 6.66_2 & 6.52_3 & 53.8\;\;\bullet & 52.7 \\
\hline
\mbox{$\frac{1}{2}\bz_2+\bz_3$ with $\Delta\gamma_{corr``1"}$ in (\ref{jump3})}
&0.0810_8 & 0.112_9 & 8.85_4 & 6.74_8 & 6.59_1 & 59.7\;\;\bullet & 58.4 \\
\hline \hline
\mbox{Villian  action (using (\ref{vil}))}&&&&&&& \\ \hline
\mbox{$\bz_2$ only} &0.0431_8 &0.121_7&8.21_9 & 6.44_1 & 6.34_8 &52.9 & 52.2
\\
\hline
\mbox{$\bz_2+\bz_3$} &0.0811_9&0.0942_4&10.6_1&6.52_9&6.41_8&69.3&68.1 \\
\hline
\mbox{$\frac{1}{2}(\bz_2+\bz_3)$}&0.0404_4&0.124_1 &8.05_5&6.43_2&6.34_1&
51.8\;\;\bullet&51.1 \\ \hline
\mbox{$\frac{1}{2}\bz_2+\bz_3$}
&0.0594_1&0.108_7&9.20_4&6.48_3&6.38_2&59.7\;\;\bullet&58.7
\\ \hline
\end{array}
\eeq
\end{table}
\end{footnotesize}


\newpage

\section{Results of MPCP predictions compared with experimental
values of fine-structure constants and conclusion}\label{conclusion}

We use the principle of multiple point criticality to calculate the values
of the three standard model gauge couplings. These agree with experiment
to well within the calculational
accuracy of 5 to 10\%. In the context used here, the principle states that
Nature seeks out the action parameter values in the phase diagram of a lattice
gauge theory that correspond to the multiple point. At this point, a maximum
number of phases convene.
The gauge group is taken as
the $N_{gen}$-fold Cartesian product of the standard model group:
$SMG^{N_{gen}}$ where $N_{gen}=3$ is the number of fermion generations. So
there is a $SMG$ factor for each family of quarks and leptons.
This gauge group is referred to as the
\underline{A}nti \underline{G}rand
\underline{U}nified \underline{T}heory $(AGUT)$ gauge group.
 At the Planck scale,
the gauge couplings are predicted to have the multiple point values
corresponding to
the diagonal subgroup of $SMG^{N_{gen}}$. The diagonal subgroup, which is
isomorphic to the usual standard model group, arises as that surviving the
Planck scale breakdown of the more fundamental $SMG^{N_{gen}}$ under
automorphic symmetry operations.

In order to provoke the many phase that should convene at the multiple
point -
including those corresponding to confinement solely of discrete
subgroups of the gauge group - we need a rather general action
the parameters of which span a multi-dimensional phase-diagram space.
In many cases, such phases
would be called lattice artifacts because the boundary between such
lattice-scale phases disappears in going to long wavelengths and what
is distinguishable as a Coulomb-like phase at lattice scales becomes
indistinguishable from a confining phase at large distances. Such
phases are usually regarded as not being of physical significance
because they depend on the presence of a lattice which has been
introduced only as a calculational regulator that must leave no
trace of its presence upon taking a continuum limit.

Our point of view is that a Planck scale lattice is one way of implementing
the fundamental necessity of having a truly existing regulator at roughly
the Planck scale. We would claim that field theories are intrinsicly
inconsistent without the assumption of a fundamental regulator. While the
lattice seems to play a fundamental role in our model, it is really only
a way of manifesting the necessity of a fundamental regulator. We would
of course hope that critical behaviour for any field theory
formulated using other regulators (e.g., strings) would lead to approximately
the same critical values for the coupling constants so that $MPC$ predictions
based on the assumption that Nature had chosen a different regulator would not
yield very different values for couplings than those obtained using a lattice
regulator. Obtaining the same values of couplings when using different
regulators would suggest that the principle of multiple point
criticality has a validity that transcends the particulars of the regulator.

Our claim is then that even the presence of phases that are only
distinguishable on a Planck scale lattice can have profound consequences
for physics. And this is so despite the fact that such phases can
- even though quantitatively distinguishable at the lattice
scale (e.g., two phases with different finite correlation lengths) -
become qualitatively indistinguishable at long distances.
This situation is not unfamiliar in other situations. For
example, at the triple point of water, three different phases can be accessed
by suitable changes in intensive parameters by just a small amount. However,
two of the three phases are not qualitatively distinct: at the tri-critical
point, the distinction between liquid and vapour disappears. This however does
not change the fact that all three phases are important in defining the
triple point values of temperature and pressure.




Our Planck scale predictions for the gauge coupling constants come about as
the product of the continuum value of
the lattice critical coupling {\em and} the appropriate
enhancement factor in going from the multiple
point of $SMG^3$ to the diagonal subgroup.

The main difference between the Abelian and non-Abelian case is
a factor two  in the squared coupling weakening factor that comes
from going from $SMG^3$ to the diagonal subgroup of $SMG^3$:
the diagonal subgroup squared coupling for $U(1)$ is a factor

\beq N_{gen}+
\left (\begin{array}{c} N_{gen}\\ 2 \end{array} \right )=(N_{gen}+1)
N_{gen}/2=6 \eeq
weaker than the critical values obtained from
Monte Carlo data for a single $U(1)$. This is to be compared to
the naively expected weakening factor $N_{gen}=3$
that is found for the non-Abelian couplings in going to the diagonal
subgroup.
The reason for the difference in the weakening factor in going to the
diagonal subgroup of $SMG^3$  is that in the case of $U(1)$ there is the
possibility of interaction terms $F_{\mu\nu\; Peter}F^{\mu\nu}_{Paul}$ in
the Lagrangian (in the non-Abelian case, such terms are not gauge invariant).
The indices
$Peter, Paul,\cdots$ label the various $SMG$ factors of $SMG^{N_{gen}}$
(of which there are $N_{gen}=3$).
Had the phase transition at the multiple point been purely second order,
we would expect the enhancement factor for the inverse squared $U(1)$
coupling to be {\em exactly}
$\frac{1}{2}N_{gen}(N_{gen}-1)=6.$
However, the fact that transitions between
phases solely confined w.r.t. discrete subgroups and the totally
Coulomb-like phase inherit a residual first-orderness from the pure discrete
subgroup transitions leads to an enhancement factor larger than
$\frac{1}{2}N_{gen}(N_{gen}-1)=6$.
The enhancement factor
for $U(1)$ is calculated
using different approximations the result of which are tabulated in
Table~\ref{tab5}

The values we have calculated for the $U(1)$
gauge coupling (i.e., the values for the diagonal subgroup of $SMG^3$ at
the multiple point of $SMG^3$) and the values calculated for
the non-Abelian couplings
are predicted to coincide with experimental values that have been extrapolated
to the Planck scale using the assumption of a minimal standard model.
In the renormalization group extrapolation procedure\cite{amaldi} used,
we accordingly
assume a desert with just a single Higgs ($N_{Higgs}=1$). The number
of generations (families) is of course taken to be 3.

In doing the
renormalization group extrapolation of experimental values to Planck
scale, we start the running
at the scale of $M_Z=91.176\pm 0.023$ using
values from LEP experiments\cite{kim}. We also extrapolate the other way:
we extrapolate our Planck scale predictions down to the scale of $M_Z$
so as these can be directly compared with experimental values of
fine-structure constants. Predicted and experimental\cite{amaldi} values
of the three
fines-truce constants are compared at both the Planck scale and the scale
of $M_Z$ are compared in Table~\ref{tab8}. We have included predicted
values obtained using several different variations in some details of our
model. For the non-Abelian fine-structure constants, the naive continuum
limit and the continuum-corrected continuum limit values are taken
from our earlier work\cite{nonabel}.

\begin{table}
\caption[Table 8]{\label{tab8} Our predictions using slightly different
calculational methods (approximations) and assumptions; these are
compared with experimental  values  (Delphi
results) extrapolated using the renormalization group  to  the  Planck
scale. The minimal Standard Model has been assumed in doing the
extrapolation. The predicted values for $U(1)$ in the last eight rows
are taken from Table~\ref{tab7} (with $\tau=0.79$).}

\vspace{.4cm}

\begin{tabular}{||c|c|c||}
\hline\hline {\bf SU(3)} & $\alpha^{-1}(\mu_{Pl.})$  & $\alpha^{-1}(M_Z)$   \\
\hline
{\bf Experimental values} & {\bf 53.6} & {\bf 9.25}$\pm 0.43$ \\ \hline
Continuum corrected continuum limit& $56.7\;\bullet$  &  12.$_8 \;\bullet$
      \\ \hline
Monopole correction & $56\pm 6\;\bullet$    & $12._1\pm 6 \;\bullet$ \\
\hline
Naive continuum limit & $80._1$    & $36._2$    \\ \hline \hline
{\bf SU(2)} & $\alpha^{-1}(\mu_{Pl.})$  & $\alpha^{-1}(M_Z)$   \\ \hline
{\bf Experimental values} & {\bf 49.2} & {\bf 30.10}$\pm 0.23$ \\ \hline
Continuum corrected continuum limit & $49.5\;\bullet$  &  29.$_8\; \bullet$
       \\ \hline
Monopole correction     & $48._3\pm 6\;\bullet$  & $28._5\pm 6\; \bullet $   \\
  \hline
Naive continuum limit & $65._1$   &  $45._3$     \\ \hline \hline
{\bf U(1)} & $\alpha^{-1}(\mu_{Pl.})$  & $\alpha^{-1}(M_Z)$   \\
& {\tiny ($SU(5)$ norm. in parenthesis)} & {\tiny ($SU(5)$ normalisation in
parenthesis)} \\ \hline
{\bf Experimental values}: & {\bf 54.}$_8$ (32.9) & {\bf 98.70}$\pm 0.21$
($59.22\pm 0.13$) \\ \hline
Continuum corrected continuum limit & 66 ($39._6$):  &  109.1 (65.5)       \\
\hline
Naive continuum limit (w. enh. 6.8): & 84.6(50.8)    &127.7    (76.6)  \\
\hline
Independent monopole approx.  &     30 (18) & 73 (44)   \\ \hline
$\bz_2$ (Wilson action): & 54.7 (32.8) & 97.8 (58.7)\\ \hline
$\bz_2$ (Villian action): & 52.9 (31.7) & 96.0 (57.6) \\ \hline
$\bz_2+\bz_3$ (Wilson action): & 66.2 (39.7) & 109.3 (65.6)\\ \hline
$\bz_2+\bz_3$ (Villian action): & 69.3 (41.6) & 112.4 (67.5)\\ \hline
$\frac{1}{2}(\bz_2+\bz_3)$ (Wilson action): & 53.8 (32.3)$\;\; \bullet$ & 96.9
(58.2)$\;\;\bullet$ \\ \hline
$\frac{1}{2}(\bz_2+\bz_3)$ (Villian action): & 51.8 (31.1)$\;\;\bullet$ & 94.9
(57.0)$\;\;\bullet$\\ \hline
$\frac{1}{2}\bz_2+\bz_3$ (Wilson action):& 59.7 (35.8)$\;\;\bullet$ & 102.8
(61.7)$\;\;\bullet$ \\ \hline
$\frac{1}{2}\bz_2+\bz_3$ (Villian action): & 59.7 (35.8)$\;\;\bullet$ & 102.8
(61.7)$\;\;\bullet$ \\ \hline
\end{tabular}
\end{table}

In trying to estimate the uncertainty in our calculation of the
$U(1)$ gauge coupling, two points of
view can be taken:

a) we could take the viewpoint that we do not really know which of the phases
characterised by being solely
confined w.r.t. discrete subgroups should also convene at the multiple point
in certain  cases. In particular, we could claim that we do not
know to what extent that $\bz_2-$ and $\bz_3$-like subgroups, in analogy
to the U(1)-continuum,
give rise to a hexagonal phase system at the multiple point. If this is the
case, we
have to let our lack of knowledge about such details of the phase diagram (and
the multiple point chosen by Nature) be included in the uncertainty in our
prediction.

b) we could take the standpoint that our choice of procedure for
including the effects of having solely confining
$\bz_2-$ and $\bz_3$-like subgroups at the multiple point is correct
and that we accordingly can do our calculations based on a correct
picture of the pattern of phases that convene
at the multiple point,
also w.r.t. solely confining  discrete
subgroups. In this case,  uncertainties in our results are assumed to be
due only to uncertainties in the Monte Carlo procedures used and in the
approximations we
use in our corrections of Monte Carlo data in order to get our
predictions.

In the case a) we must regard the differences in predictions arising when
$\bz_2-$ and $\bz_3$-like subgroups are taken into account
in different ways as being a measure of the uncertainty. For the predicted
$U(1)$
coupling at the Planck scale, this viewpoint leads to an estimated uncertainty
of about 5\%. We implement this point of view in Table~\ref{tab9} by averaging
all combinations in which there is a $\frac{1}{2}\bz_2$ contribution. This
results in an average of the combinations having $\bz_3$ and those having
$\frac{1}{2}\bz_3$ as the contributions from $\bz_3$. This reflects our
lack of certainty as to how the $\bz_3$ contribution should be treated.

In addition to this uncertainty, there will of course be the uncertainties
in the Monte Carlo results which we have used which
may be taken as 5\%. Also, our corrections are presumably not performed to
better than some 4\%, so it is unlikely that the
uncertainty in our prediction in case b) is less than 6.4\%.
In case a) we should rather
take the  uncertainty as being 8\%. These percent-wise uncertainties
concern the squared couplings  referred to the Planck scale. These correspond
to absolute Planck scale uncertainties of $4._5$ and $ 3._5$ in
the inverse fine-structure constant in
respectively the cases a) and b).
But since the renormalization group correction
consists basically of adding a rather well-determined constant to the inverse
fine-structure constants, the absolute uncertainty
in the $1/\alpha$'s is the same at all scales.

It is remarkable that in spite of these uncertainties being rather modest
we have agreement with experiment within them!

It is interesting to formulate our predictions as a number that can be
compared with the famous $\alpha^{-1}=137.036\dots$. From Table~\ref{tab9}
we deduce that the phenomenologically observed value of
$\alpha^{-1}$ decreases by $8.2\pm 0.5$ in going from low
energies to that of $M_Z$:

\beq 137.036-(\alpha_1^{-1}(M_Z)+\alpha_2^{-1}(M_Z))=
137.036-(98.70\pm 0.23 + 30.10\pm 0.23)=8.2\pm 0.3
 \eeq

\nin Our theoretical prediction for the famous $\alpha^{-1}=137.036$
is in the case a)

\beq \alpha^{-1}_1(M_Z)\;+\;\alpha^{-1}_2(M_Z)\;+\;8.2\pm 0.5=   \eeq

\[ = 99.4\pm 5 \;+\;  29.2 \pm 6 \pm 3.5\; +\; 8.2\pm 0.3= 136._8\pm 9 \]

\nin and in the case b)

\beq \alpha^{-1}_1(M_Z)\;+\;\alpha^{-1}_2(M_Z)\;+\;8.2\pm 0.5= \eeq
\[ =102.8\pm 3.5\;+\;29.2\pm 6 \pm 3.5\; +\; 8.2\pm 0.3= 140._2 \pm 8. \]


\begin{table}
\caption[Table9]{\label{tab9}The predicted values of
$\alpha^{-1}(M_Z)$ for $SU(3)$ and $SU(2)$,
are obtained as the average of several calculational procedures. The first
set of uncertainties comes from Monte Carlo data and from
the approximation procedure that we used to get our predictions from the
Monte Carlo critical couplings. The second set of uncertainties are the RMS
deviations from the average value of $\alpha^{-1}(M_Z)$ using the several
different calculational procedures. The predicted $\alpha^{-1}(M_Z)$ values
for $U(1)$ and uncertainties arise as the result of the implementing
the viewpoints a) and b) elaborated upon immediately above.}

\beq
\begin{array}{|l|l|l|} \hline
  & \alpha^{-1}(M_Z) & \alpha^{-1}(M_Z)\\
  & predicted        & experimental     \\ \hline
SU(3) & 12.4\pm 6 \pm 6 & 9.25\pm 0.43     \\ \hline
SU(2) & 29.2 \pm 6 \pm 3.5 & 30.10\pm 0.23  \\ \hline
U(1)&\begin{array}{cc} a) & 99.4 \pm 5 \\ b) & 102.8 \pm 3._5 \end{array}
& 98.70\pm 0.23       \\ \hline
\end{array}
\eeq
\end{table}

Since $\alpha_s^{-1}$ is rather small at experimental scales,
the absolute uncertainty  is percent-wise large at these scales.
But really
it is probably best to see our $\alpha_s$-prediction (at Planck scale)
as a prediction of the logarithm of the ratio
of the strong interaction scale to the Planck scale which then
allows only a crude prediction of $\alpha_s(M_Z)$.
Note that the strong scale to Planck scale ratio
is actually one of Dirac's surprising $10^{20}$ factors! So this
``large number'' is found here as an exponential of an order one number
that is proportional to the number of generations ($\pi^2$ in
the denominator of the $\beta$-functions leads to couplings that walk
slowly with scale).

Assuming the coexistence of more than one phase separated by
transitions that are first order
is roughly equivalent to assuming
the principle of multiple point criticality. 
This principle offers the hope of a general explanation for the occurrence
of fine-tuned intensive quantities in
Nature. Indeed, the conspicuous values taken by a number of physical constants
 - e.g., the vanishing effective
cosmological constant, the fine-structure constants, $\Theta_{QCD}$ -
have values that coincide
with values obtained if it is assumed that Nature seeks out
multiple point values for intensive parameters\footnote{The
smallness of the Higgs mass relative to (say) the Planck scale is also
a conspicuous quantity
that could have been expected to be explainable as a multiple point value.
It is interesting that recent work\cite{cdfhbn} indicates that the high
value of
the top quark mass
precludes an explanation of the lightness of the  Weinberg-Salam Higgs as
a multiple point. However, the assumption that Nature has multiple point(s)
 together with the requirement that the phase
transition
between degenerate phases at the multiple point is {\em maximally} first order
leads to strikingly impressive predictions for the mass of the top
quark and the expected mass of the Weinberg-Salam Higgs.}.

As mentioned above, multiple point values of intensive parameters
occur in the presence of coexisting phases separated by first order
transitions.
Such coexistence could be enforced by having fixed but not fine-tuned amounts
of extensive quantities. We have shown in recent work\cite{nl1,nl2} that the
enforced coexistence of extensive quantities in spacetime is tantamount to
having long range nonlocal interactions of a special type: namely
interactions that are identical between fields at all spacetime points
regardless
of the spacetime distance between them. Such omnipresent nonlocal interactions,
which can be described by a very general form of a reparameterization invariant
action, would not be perceived as non-locality but rather most likely
absorbed into physical constants. Even still, the presence of nonlocal
interactions opens the possibility for having contradictions of a type
reminiscent of the ``grandfather paradox'' naively encountered in ``time
machines''. However, we can show\cite{bomb} that generically there is a
``compromise''
that averts paradoxes. It is interesting that this solution coincides with
multiple point values of intensive quantities such as fine-structure
constants and the cosmological constant.
Hence one can speculate that it is a mild form of non-locality,
intrinsic to fundamental physics, that is the underlying explanation
of Nature's affinity for the multiple point.

In a sense, the MPCP can also be said to have some
predictive power as to the form assumed by the ``the true action of Nature''.
For example, if two proposed actions differ in the number of ``phases'' that
can be brought together at a multiple point, the action that can bring
together the larger number of phases would, according to the MPCP, come
closer to being the ``true action of Nature''. The same sort of argument
may be applicable to proposed candidates for gauge groups:
having a MPCP would favour a non-simple gauge group over a
simple gauge group. While the implementation of the MPCP can be be said
to be rather complicated technically, the underlying idea is extremely
eloquent in its simplicity. This combined with the noteworthy
accuracy with which a number of constants of Nature are predicted
makes the MPCP a serious contender as an important link in our understanding
of fundamental physics.

\subsection{Loose speculations as to the relevance of the MPCP in the
evolution of living organisms}

In concluding this thesis, it is interesting to speculate as to the possibe
wider range of applicability of the principle of multiple point criticality.
In 1992, while I held an associate professorship at the Royal Danish School of
Pharmacy, I put forth the idea\cite{dfh1,dfh2} that complicated biological
regulatory mechanisms may achieve maximum stability by seeking out the
multiple point in the parameter space of what is
probably the intractably complicated action of a biological system.
By seeking out the multiple
point, Nature can utilize easy access to the maximum number of phases
as a means of maximizing the stability of the complicated forms of regulation
required for the dynamical stability of a biological system in the presence
of an ever-changing enviornment. Being at the multiple point is tantamount to
being simultaneously on the verge of undergoing a phase transition to any
one of many more or less ordered phases. A system in the most ordered phase
(a ``Coulomb-like'' phase characterised by \dofx all of which are correllated
over long distances)
that at the same time is infinitesimally close to the multiple point has
access to plethora of phases that are more or less chaotic (depending on the
degree of
``first-orderness'' of the transition) w.r.t. any of the various \dofx
(analogous for example to $U(1)$ \dof, $SU(3)/\bz_3$ \dof, $\bz_2$
\dof, etc., etc. in the lattice gauge theory implementation of the MPCP
with the gauge group $SMG^3$) simply by effecting infinitesimal changes in
intensive parameters. In being at the multiple point, a system is at what
can be said to be the
ultimative ``edge of chaos''.

The idea that the multiple point criticality principle might apply
in some sense to biological systems occurred to me in connection with
work I have been doing together with John Ipsen
(The Technical University of Denmark) and Holger Bech Nielsen (The Niels
Bohr Institute). First we were interested in the
use of a lattice gauge theory formulation as a way of implementing the
constraint of self-avoiding surfaces in computer simulations of
lipid membranes. Most recently we have been involved in designing
ways of doing computer simulations
using Kalb-Ramond lattice gauge theories as a way of introducing
a ``volume fugacity''.
Together with the usual area and topological fugacities as control
parameters, having this additional parameter
will allow simulation of membrane models that are considerably more
complicated that those presently tractable.
This ongoing work on membranes springs from earlier work with Ole Mouritsen
and also Peter Leth Christiansen (both from The
Technical University of Denmark) the focus of which was a Cand. pharm.
Ph.D. student who wanted to do serious studies of the way in which
pharmaceutical agents affect the very complicated interdependence of the
internal state of a lipid membrane and the activity of membrane-bound
proteins.
Phase transitional  regions
turn out to play an extremely important role in the mechanism by which
pharmaceutical agents exert physiologically observable effects as we
showed in a series\cite{mem1,mem2,mem3,mem4,mem5} of work.
It is really this work that suggested the validity of the principle of
multiple point criticality for biological systems inasmuch as it is
well-known that biological membranes are rather meticulously maintained
at values of intensive parameters (i.e. temperature) that lie very
close to a phase transition.

It is interesting that my suggestion that life processes seek out the
stability and regulatory flexibility afforded by being at the multiple point
is an idea that, superficially at least, bears a strong resemblence to a
school of
thought that has emerged at the Santa Fe Institute in recent years.
The first essential element in this thinking,
 which to a large extent has been put forth by
Stuart Kauffman and Christopher Langton (both at the Santa Fe Institute),
is that order, and in particular self-replicability, tend inherently to
emerge in systems having sufficient complexity. Kauffman's view is that
the dynamics of any sufficiently complicated system of agents
(e.g., cells, genes, protein fragments, business corporations,
members of a democracy, etc.) that
registers input and
generates output that subsequently is interpreted as input by other perhaps
neighbouring
agents will
evolve basins of attraction that entrap the system into persistent patterns
possessing a high degree of diffrentiation that interact in a highly
organized, stable and  self-sustaining way. This is sometimes referred to as
``catalytic closure''. In Kauffman's words, one has the unavoidable
emergence of ``order for free''.

Kauffman and Langton go even further: having made the case for ``order
for free'', they inquire as to the conditions under which the evolution of
self-organized systems is maximally robust when faced with the perpetual
changes in the enviornment that must be accommodated by a self-organized
system if it is to survive. In considering this problem, Kauffman is lead
to propose that the {\em control} of self-organized order also emerges
spontaneously. How does a systems evolves a
system of contol that assures maximum adaptability? Here adaptability refers
to the ability of a system to adjust the system of ``links''\footnote{
A system of ``links'' can be thought of as the (evolving) set of rules
that express the interdependence of the agents.} between its
agents so that the system fits its enviornment over time. Obviously a system
with a
deficiency of such links will not be well-suited to a cooperative effort
on the part of its agents in devising an optimal survival strategy in the face
of external changes. The uncoordinated performance of non-percollating
structures
cannot contribute optimally to the solution of a problem
(e.g., a change in the enviornment) faced by the system
as a whole. What is perhaps more surprising is that  an over-abundance
of links between agents is also debilitating for a system
challenged by changes in the enviornment. Such systems are frozen into
inactivity by the constraints of too many links and hence are unable to
be optimally innovative in solving problems. It seems that self-organized
systems
need, for optimal adaptability, to have a mechanism for dynamically
``self-tuning'' the system of links between its agents in such a way as to
be able to deal optimally with coevolutionary external changes.

Langton may have found a clue as to how such such a dynamical fine-tuning
mechanism comes about. Using
systems of cellular automata as a simplified form of a parallel-processing
(Boolean) network, Langton finds static and propagating structures at the
phase transition separating chaotic and ordered regimes that can support
information storage and tranmission as well as phenomena that can be
interpreted as a form of information processing. Kauffman and Langton
propose that at levels in systems which must coordinate complicated tasks,
selection attains a near-universal ``poised'' state
that hovers at the phase transition between spontaneous order and
chaos. In the words of C. Langton and N. Packard, it may be that life seeks
out and is sustained at the edge of chaos.

The principle of multiple point criticality,
in asserting that the universe cannot avoid seeking  out multiple point
values of intensive
parameters, is essentially asserting that the fundamental
laws of physics\footnote{This is to be understood in the sense that
the principle of multiple point
criticality determines that parameters of the action of the universe.}
reside at the ``edge of chaos''.

The point I wish to make is that
the multiple point (i.e., ``the
ultimative edge of chaos'') functions as an attractor that can provide
a useful stabilisation mechanism
that quite plausibly could
be inherited by biological systems
as the tendency of evolving life processes
to be maintained  at the edge of chaos. This would be consistent
with the idea of Kauffman and Langton that selection results in a
chain of biological systems in the course of evolution that
perpetually seek out
and  ``ride along'' on the ever-changing ``edge of chaos'' as the way of
achieving
the most sustainable evolution of  ever more complicated organisms.
Recent discussions\cite{stu1} and the exchange of notes\cite{stu2}
with Stuart Kauffman in Santa Fe
reveal that he fully shares my sentiments as to the importance of
exploring the idea that the apparent tendency of biological systems to
cling to the edge of chaos in the course of evolution may be a manifestation
of a fundamental physical principle of multiple point criticality.
We have made plans for continuing this avenue of investigation in the
months to come.

\section{Acknowledgements}

My heart-felt thanks and admiration go to Holger Bech Nielsen - great friend,
inspiring teacher and
enthusiastic colleague without whom the writing of a Ph.D. thesis
(and physics) undoubtedly wouldn't
have been nearly so much fun (and any  thesis would have turned out much
differently!). To Erling Jensen I am grateful for having learned the pleasures
of not always taking established thinking too seriously. To S{\o}ren
Ventegodt go my deep-felt thanks because his (untamed) enthusiasm is huge
enough also to bring me in out of the cold. I am also very grateful for
useful and inspiring interactions in various periods
 with Dino Butorovic, Ivica Picek, Niels
Brene, Leah Mizrachi, Kent J{\o}rgensen, Ole Mouritsen, John Ipsen,
Igor Volovich, Christian Surlykke, Sven Erik Rugh, Larisa  Laperashvili,
Collin Froggatt and
Kasper Olsen.
My thanks also go to many other kind and supportive people at
The Niels Bohr Institute.

\begin{flushright}
\begin{tiny}
30 May 96 alf 
\end{tiny}
\end{flushright}

\section{Appendices}
\subsection{Appendix:  Haar measure to next to leading order}\label{apphaar}

For the contribution to $``-\beta F"$
from the confining degrees  of  freedom,  the  most  important  next  order
correction to the tangent space or weak coupling  approximation  comes
from taking into account the influence on  the  Haar  measure  arising
from the curvature of the group manifold for a non-Abelian Lie group  $G$.
In the neighbourhood of the unit element the exponential map  $\;\;\;  exp  :
\mbox{Lie algebra} \rightarrow \mbox{Lie group} \;\;\;\;\;$ is one to  one,
and
it may be used to induce in the Lie algebra a  measure  from  the  Lie
group. However, for a  group  manifold  with  curvature,  it  is  only
infinitesimally close to  the  unit  element  that  the  correct  Haar
measure coincides with the ``flat'' measure:

\beq                     d^{{\cal                     H}aar}U=d^{{\cal
H}aar}(exp(A))=\rho_{flat}\;\;d^dA/\sqrt{2}=\frac{d^dA/\sqrt{2}}{vol(G)}.\eeq

\nin Here, as  in  the  preceding  text,  the  uniform  Haar  measure
density $\rho_{flat}$ is taken as $\rho_{flat}=\frac{1}{vol(G)}.$

Because a left translation in the Lie algebra and in the Lie group  do
not coincide exactly for a non-Abelian group (due  to  curvature),  the
Haar measure that is to be invariant under  left  translation  on  the
group will deviate  from  being  constant  relative  to  that  on  the
algebra. Consider all the group  elements  corresponding  to  the  Lie
algebra elements in a small volume element at the origin  of  the  Lie
algebra. These  group  elements  correspond,  after  left  (or  right)
translation by say the group element $exp(\ba)$, to a  volume  at  the
translated position in the Lie algebra that is expanded by some factor
relative to the small volume at the origin.  Invariance  of  the  Haar
measure  on  the  group  manifold  under   left   (or   right)   group
multiplication by $exp(\ba)$ requires therefore a compensating  factor
in the Haar  measure  density  $\rho^{{\cal H}aar}(\ba)$  at  the  left
translated position $exp(\ba)$ on the group manifold relative  to  the
density at the unit element of the group.

Let $\bb^1,\bb^2, \cdots \bb^{j=dim(G)}$ be infinitesimal  displacements
in the $dim(G)$ independent directions of the Lie algebra at the origin
of the Lie algebra.

Designate by $L_{exp(\sba)}$ the operation of left multiplication in  the
group: $L_{exp(\sba)}g=\exp(\ba)g\;\;(\exp(\ba),\;g  \in  G)$.

Denote by $\tla\bb^j$ the Lie  algebra  vector  that  is  vectorially
added  to  the  Lie  algebra  vector  $\ba$  when  the  group  element
$exp(\bb^j)$     is     translated     by     $L_{exp(\sba)}$;      i.e.,
$L_{exp(\sba)}exp(\bb^j)=exp(\ba+\tla \bb^j)$.


In the ``flat'' or tangent space approximation we  have  $\ba+\tla
\bb^j=\ba+\bb^j$ corresponding to the Lie algebra  composition
rule  of   simple   vector   addition   when   the   group   operation
$L_{exp(\sba)}exp(\bb^j)$ is referred to the Lie algebra.

For a group manifold with curvature, the Lie algebra composition  rule
- we denote it with the symbol ``$+^{\prime}$''  -  is  more  complicated.
Under  the   action   of   $L_{exp(\ba)}$   on   the   group   we   have
$\ba+\tla \bb^j =\ba+^{\prime}\bb^j$.

Denote by $\Delta V$ the volume element spanned by the $\bb^j$'s at the
origin of the Lie algebra: $\Delta V=\bb^1 \wedge \bb^2 \wedge  \cdots
\wedge \bb^{dim(G)}$. Let $\bc$ be an infinitesimal vector  specifying
a point within the volume $\Delta V$.

For  one   of   the   infinitesimal
displacements $\bb^j$ at the origin of the
Lie algebra, the effect of the group operation  $L_{exp(\sba)}$  referred
to the Lie algebra can be expressed as

\beq(\ba+\tla                               \bb)^j=(\ba+^{\prime}
\bb)^j=(\ba+^{\prime}\bnul)^j+(d(\ba+^{\prime}\;\cdot
)_{\sbnul})^j_k\bb^k=\ba^j+d((\ba+^{\prime}\;\cdot
)_{\sbnul})^j_k\bb^k \eeq

\nin where

\beq              d((\ba+^{\prime}               \cdot)_{\sbnul})^j_k=
\frac{\partial(\ba+^{\prime}\bc)^j}{\partial                \bc^k}
\mid_{\sbc=\sbnul}. \eeq

\nin It follows that the $\bb^j$ at the origin of the Lie algebra  has
become

\beq (\tla\bb)^j=\sum_k \frac{\p  (\ba+^{\prime}
\bc)^j}{\p \bc^k} \mid_{\sbc=\sbnul}\bb^k \eeq

\nin under the group operation $L_{exp(\sba)}$.

The volume element $\Delta V=\bb^1 \wedge  \bb^2  \wedge  \cdots
\wedge \bb^{dim(G)}$ at the origin of the Lie algebra expands into the  volume
$\tla \Delta V$ under the group operation $L_{exp(\sba)}$.  We  can
write

\beq \tla \Delta V=\tla\bb^1 \wedge \tla \bb^2 \wedge \cdots  \wedge
\tla \bb^{dim(G)} \eeq

\[ =\frac{\p(\ba+^{\prime}  \bc)^1}{\p \bc^k}\mid_{\sbc=\sbnul}\bb^k
\wedge   \cdots   \wedge   \frac{\p(\ba+^{\prime} \bc)^{dim(G)}}{\p
\bc^m}\mid_{\sbc=\sbnul}\bb^m \]

\[=  \left   \|\frac{\p(\ba+^{\prime}   \bc)^j}{\p \bc^k}   \left
|_{\sbc=\sbnul} \right . \right  \| (\bb^1  \wedge  \bb^2  \wedge
\cdots      \wedge      \bb^{dim(G)})      =det(d(\ba+^{\prime}
\cdot)_{\sbnul})\Delta V \]

Really, the map $(\ba+^{\prime} \cdot)_{\sbnul}$  is  the
composed map consisting of

\begin{enumerate}

\item a map from very near the  origin  of  the
Lie  algebra  to  very  near  the  identity  of  the  group;

\item left translation to very near the element  $exp(\ba)$  in  the
group and

\item a log mapping back to the Lie algebra  from  very  near  the
group element $exp(\ba)$.

\end{enumerate}

Linearising this composed map we get

\beq d(\ba+^{\prime}\cdot)_{\sbnul}=d(exp^{-1}  \circ  L_{exp(\sba)}\circ
exp)_{\sbnul} \eeq

\[   =d(exp^{-1})_{exp(\sba)}\circ   d(L_{exp(\sba)})_1    \circ
d(exp)_{\sbnul}=d(exp^{-1})_{exp(\sba)}\circ
d(L_{exp(\sba)})_{\sbunit} \]

\nin  where  in  the  last  step  we   have   essentially   identified
$exp_{\sbnul}$ with the unit operator by identifying  the  neighbourhood
of the unit element with the Lie algebra.

In terms of the composed map, we have

\beq       \tla       \Delta       V=det(d(exp^{-1}        \circ
L_{exp(\sba)})_{\sbnul})\Delta V; \eeq

\nin  i.e.,  under   the   group   operation   $L_{exp(\sba)}$,   an
infinitesimal volume element $\Delta  V$  at  the  origin  of  the  Lie
algebra is expanded by a factor equal to the Jacobian  determinant  of
the map $d(exp^{-1} \circ L_{exp(\sba)})_{\sbnul}$.

Invariance of the Haar measure under left translations requires:

\beq
{\cal H}aar(\Delta
V)=\rho^{{\cal H}aar}(\bnul)\Delta V=\rho^{{\cal H}aar}(\ba)  \tla  \Delta
V. \eeq

\nin The Haar measure, correct to next order,  for  a  group  manifold  with
curvature is given by

\beq    d^{{\cal H}aar}U=\rho^{{\cal    H}aar}d^dA(\ba)=\frac{det(d(exp^{-1}
 \circ
L_{exp(\sba)})_{\sbnul})^{-1}}{vol(G)}d^dA(\ba) \label{hbd} \eeq

We now make use of the theorem \cite{helgason1}

\begin{equation} d \;
exp_A = d(L_{exp \; A})|_{\bunit}\circ \frac{1-\exp(-ad\;\ba)}{ad  \;
A}  \label{helgasont},  \end{equation}

\nin which yields

\beq                                                    d(\ba+^{\prime}
\cdot)=d(exp^{-1}\circ
L_{exp(\sba)})_{\sbunit}=(\frac{1-exp(-ad\ba)}{ad\ba})^{-1}
\eeq

\nin where $ad\; \ba$ denotes the linear transformation associated  with
each element $A$ of the Lie algebra such that the operation  $ad\;\ba$
on any element $\bb$ of the Lie algebra yields the Lie algebra  vector
$[\ba,\bb]$. We have

\begin{equation}  ad  \; \ba  (\bb)  \stackrel{def.}{=}  [\ba,\bb].
\end{equation}


\nin Combining equations (\ref{helgasont}) and (\ref{hbd}) we find

\begin{equation}
d^{Haar}U=d^dA/vol(G)\cdot det(\frac{1-\exp(-ad\;\ba)}{ad \; A}).
\end{equation}

\nin Up to second order in $\ba$ we have

\begin{equation}
\frac{1-\exp(-ad\; \ba)}{ad \; \ba} = 1-\frac{1}{2}ad\; \ba +\frac{1}{6}(ad \;
\ba)^2
\end{equation}

\nin and this leads to

\begin{equation}
det(\frac{1-\exp(-ad\;  \ba)}{ad  \;  \ba})  =det(   1-\frac{1}{2}ad\;   \ba
+\frac{1}{6}(ad \; \ba)^2) \end{equation}

\[=\exp(Tr(\log(1-\frac{1}{2}ad\;   \ba    +\frac{1}{6}(ad    \;    \ba)^2)))
\]

\[=\exp(Tr(-\frac{1}{2}ad \; \ba +\frac{1}{6}(ad \;  \ba)^2-\frac{1}{8}(ad
\; \ba)^2)) \]

\[=\exp(Tr((ad \; \ba)^2)/24)=1+\frac{1}{24}Tr((ad \; \ba )^2).\]

\nin where we have used that $ad\;\ba$, as a commutator, is traceless.

\nin So the left (or right) invariant Haar measure to next to  leading
order is given by

\beq
d^{{\cal H}aar}U(\link)=\frac{1+\frac{1}{24}Tr(ad\ba)^2}{vol(G)}\tla\Delta
V \eeq

\nin As $ad\;\ba(\bb)=[\ba,\bb]$ is a commutator,  it  is  anti-Hermitian
and   its   square   is   negative   definite.    Hence    the    term
$+(1/24)Tr((ad\ba)^2)$ is actually negative and reflects the fact that
the group manifold of compact groups have positive Gauss curvature  in
the invariant metric. This means that the surface area of a sphere  in
the group manifold of a compact group is smaller than  in  flat  space
when measured in the invariant metric.  For  a  compact  group  it  is
intuitively reasonable that the Gauss curvature is positive insofar as
one would expect that the area of a sphere  on  a  group  manifold  of
finite size must, for large enough radius, return to zero.

For say $SU(N)$, we would like to  express  $Tr((ad\;\ba)^2)$  in  the
defining representation. To this end, start by writing the Lie algebra
vectors    $\ba$    and    $\bb$ as
$\ba=A^a\bx_a$   and   $\bb=B^a\bx_a$    where    the
$\bx_a$  is  a  basis  for  the  Lie  algebra   in   the   adjoint
representation. We have for the linear operator $ad\;\ba$

\beq
(ad\;\ba(\bb))^a\bx_a
=c^a_{bc}A^bB^c\bx_a=
(ad\ba)^a_cB^c\bx_a. \eeq

\nin We can write

\beq     ((ad(\ba)))^a_c      \stackrel{def.}{=}      ([\ba,
\cdot])^a_c=c^a_{bc}A^b \eeq

\nin In particular

\beq (ad(\bx_b))^a_c=c^a_{bc}. \label{adxmu} \eeq

\nin  Furthermore, for

\beq      (ad\;\ba\;ad      \;\bb)\bc\stackrel{def.}{=}[\ba,[\bb,\bc]] \eeq

\nin     we      get

\beq(ad\;\ba\;ad\;\bb)^i_j
=c^i_{ba}c^a_{cj}A^bB^c. \label{adab} \eeq

\nin Taking the  trace  yields  the  symmetric  bilinear  form  (inner
product)

\beq
Tr(ad\;\ba\;ad\;\bb)=
c^j_{ba}c^a_{cj}A^bB^c=g_{bc}A^bB^c.\eeq

\nin For a simple Lie algebra such as $SU(N)$, there exists, up  to  a
multiplicative constant, only one symmetric bilinear form.  Therefore,
the two bilinear forms (inner products)  $Tr(ad\;\ba  ad\;\bb)$
and $Tr(\ba_{defining} \bb_{defining})$  (where  $\ba_{defining}$  and
$\bb_{defining}$ denote the $N-$dimensional  defining  representations)
differ at most by a constant that can be determined by evaluating  the
same inner product in both representations. Assuming for the  operator
$ad\;\ba$ the existence of a set  of  eigenvectors  and  corresponding
eigenvalues, we can  take  $\ba_{defining}$  as  the  diagonal  matrix
$diag(\lambda_1,\lambda_2, \cdots, \lambda_{N-1}, \lambda_N)$ in which
case there obtains

\beq
Tr((ad\;\ba)^2)=\sum_{ij}(\lambda_i-\lambda_j)^2=2N\sum_i\lambda^2_i=2N\cdot
Tr(\ba^2_{defining}). \eeq

\nin In general we have

\beq        Tr(ad\;\ba        \circ         ad\;         \bb)=2N\cdot
Tr(\ba_{defining},\bb_{defining}). \label{adjdef} \eeq

For the Haar measure correct to next to leading order, we have arrived
at the expression

\beq d^{{\cal H}aar}U(\link)=\frac{1+(N/12)Tr_{def.}(\ba^2)}{Vol(G)}\tla\Delta
V \eeq

\nin in the defining representation. In a more usual notation, we  can
write the Haar measure as

\beq d^{{\cal H}aar}U(\link)=\frac{1+(N/12)Tr_{def.}(\ba^2)}{Vol(G)}d^dA. \eeq

\subsection{Appendix: Correction for curvature of group
manifold}\label{appcorr}

We want to reexamine our calculation of $``-\beta F"$
in light of the corrected Haar measure that,  to  next  to
leading order, takes  the  curvature  of  the  group  manifold  for  a
non-Abelian group into account.

To   this   end,    recall    the    expression    (\ref{zmfa})    for
$Z\geq Z_{MFA}\stackrel{def.}{=}exp(``-\beta F_{MFA}")$:

\[ Z_{MFA}       =
 \exp(\langle S_{action,  \;  Coul.}-S_{ansa}\rangle_{S_{ansa}+S_{action,
\;         conf.}}         )         \langle          \exp(S_{_{action,
\;conf.}})\rangle_{_{S_{ansa}}}Z_{S_{ansa}}.
\]

Consider first the calculation of $\langle          \exp(S_{_{action,
\;conf.}})\rangle_{_{S_{ansa}}}$. The leading order  approximation  of
Eqn. (\ref{sconfansa})  using  the  flat  Haar  measure  involves  the
integral

\[\langle      \exp(S_{action,      \;      conf.})
\rangle_{S_{ansa}}\approx \prod_{\Box} \int_{-\infty}^{\infty}  \cdots
\int_{-\infty}^{\infty}     \frac{d^{dim(H)}(P^a)_i/\sqrt{2}}{vol(H)}
\exp(-\sum_{i \in ``conf"}\beta_i P_i^2/2).\]

\nin where for later convenience we have rewritten  the  integrals  in
terms of the metric  without  absorbed  $\beta_i$'s.  Insofar  as  the
correction to the Haar measure coming from the curvature of the  group
manifold is  only  relevant  for  non-Abelian  groups,  one  need  only
consider the correction  for  partially  confining  phases  containing
non-Abelian basic invariant subgroups. We can write

\[\prod_{\Box}\int_Hd^{{\cal H}aar}U(\link)exp(-Dist_{U(1)
\Leftrightarrow U(1)\in  ``conf"}^2(\bunit,\bp_{U(1)})-\sum_{\!\!i\in
``conf"\wedge i\neq U(1)}Dist^2_i(\bunit,\bp_i)) \]

\nin where the next to leading order correction is contained in

\beq  \int_Hd^{{\cal H}aar}U(\link)exp(-\sum_{\!\!i\in  ``conf"\wedge   i
\neq U(1)}Dist^2_i(\bunit,\bp_i)) \eeq

\beq \approx  \int_{-\infty}^{\infty}  \cdots  \int_{-\infty}^{\infty}
\frac{d^{dim(H)}(P^a)_i/\sqrt{2}}{vol(H)}(\bunit-\sum_{\!\!i\in ``conf"\wedge i
 \neq
U(1)}\frac{N_i}{12}P^2_i/2)exp(-\sum_{\!\!i\in   ``conf"\wedge   i
\neq U(1)}\beta_iP^2_i/2) \eeq

\beq \approx  \int_{-\infty}^{\infty}  \cdots  \int_{-\infty}^{\infty}
\frac{d^{dim(H)}(P^a)_i/\sqrt{2}}{vol(H)}exp(-\sum_{\!\!i\in ``conf"\wedge i
\neq
U(1)}\frac{N_i}{12}P^2_i/2)exp(-\sum_{\!\!i\in   ``conf"\wedge   i
\neq U(1)}\beta_iP^2_i/2) \eeq

\beq \approx  \int_{-\infty}^{\infty}  \cdots  \int_{-\infty}^{\infty}
\frac{d^{dim(H)}(P^a)_i/\sqrt{2}}{vol(H)}exp(-\sum_{\!\!i\in ``conf"\wedge i
\neq
U(1)}(\beta_i+\frac{N_i}{12})P^2_i/2) \eeq

\nin The result is that the $\beta_{i}$'s for  $SU(2)$  and  $SU(3)$
are   effectively    increased    by    respectively    the    factors
$(1+\frac{2}{12\beta_{2}})$                                and
$(1+\frac{3}{12\beta_{3}})$ when the  next  to  leading  order
corrections to the Haar measure are included. As  a  consequence,  the
confinement free energy $``-\beta F_{conf.}"$

\beq 2 log(\frac{\pi^{(dim(H))/2}}{Vol(H)}) \eeq

\nin  for  the  group  $SU(N)$   is   shifted   by
$-2\frac{(N^2-1)N}{2\cdot12\beta_N}$  corresponding  to  a   shift   in
$logVol$ by $\frac{(N^2-1)N}{2\cdot12\beta_N}$ when  next  to  leading
order corrections are taken into consideration.

The effect of next to leading order corrections to the Haar measure on
the  Coulomb  phase  contribution  to  $``-\beta  F"$  must  also   be
considered. To this end, one should note that the integrals  performed
in   obtaining   $\langle   S_{ansa}   \rangle_{S_{ansa}}$,   $\langle
S_{action,\;Coul.} \rangle_{S_{ansa}}$ and $Z_{ansa}$ all involve  the
parameter $\tilde{\alpha}$ that, at the end of the calculation, is  to
be  adjusted  so  as  to  yield  a  link  variable  distribution  that
approximates  the  plaquette  action  optimally. The parameter $\tilde{\alpha}$
can be introduced in two ways:  the  fit  can  be  done  by
determining parameters $\tilde{\alpha}_{flat}$ that optimise the fit
in  the  flat   (tangent)   space   or   by   determining   parameters
$\tilde{\alpha}_{group}$  that  optimise  the  fit  on   the   group
manifold. The relation between these choices is

\beq
d^{{\cal H}aar}U(\link)(1+\frac{NP^2/2
}{12})exp(S_{ansa}(\tilde{\alpha}_{flat}))=
d^{{\cal H}aar}U(\link)exp(S_{ansa}(\tilde{\alpha}_{group})) \eeq

\nin   for    an    $SU(N)$    group.    Note    that    the    factor
$(1+\frac{NP^2/2}{12})$ converts (to next to leading order)
the  correct  Haar  measure  $d^{{\cal H}aar}U(\link)$  for  the  group
manifold into the flat (tangent)  space  measure.  If  the  parameters
$\tilde{\alpha}_{flat}$ are used  for  fitting,  the  results  of  the
calculations  for  $\langle  S_{ansa}  \rangle_{S_{ansa}}$,   $\langle
S_{action,\;Coul.} \rangle_{S_{ansa}}$ and $Z_{ansa}$  are  such  that
the Coulomb phase free energy $``-\beta F "_{Coul. \;  per\;  active\;
link}= 2 log(\frac{\pi^{(dim(H))/2}}{Vol(H)})$  is
unchanged in going to the Haar measure that  is  correct  to  next  to
leading order provided we ignore the effect of non-commutativity on the
convolution leading to the plaquette distribution from  that  for  the
link variables.

The choice of fitting parameters is not relevant in the calculation of
$\langle exp(S_{action,\;conf.} \rangle_{ansa}$ insofar as  the  later
is not included in the quantity to be fitted: recall that it  was  the
quantity  $S_{action,\;  Coul.}-S_{ansa}$  that   was   minimised   by
adjusting the link  distribution  parameters  $\tilde{\alpha_i}$.  The
calculation of  $\langle  exp(S_{action,\;conf.})  \rangle_{ansa}$ is
performed by integrating along  the  single  coset  $\bunit
\cdot H$ and is therefore independent of $S_{ansa}$.

Including
the next to leading order correction to  the  Haar  measure  causes  a
deformation of the ``diamond-shaped'' phase boundary separating the totally
confining and totally Coulomb-like phases.
To see this, note  that,  to  lowest  order,  the  determining  condition
$log(6\pi)^6=logVol(SMG)$  for  the  phase  diamond-shaped   interface
separating  the  totally  Coulomb  and  totally  confining  phases  is
obtained  by  equating  the  Coulomb  phase  and   confinement   phase
free energies:

\beq                ``-\beta_{crit.}                 F"\left|_{total\;
conf.\;(H=SMG)}\right.=``-\beta_{crit.}       F"\left       |_{totally
\;Coulomb-like\; (H=\bunit)} \right. \eeq

\nin The expression $log(6\pi)^6=logVol(SMG)$ is of course also just a
special case of Eqn. (\ref{bc}) with $H_J=SMG$ and $H_I=\bunit$.

Going to next  to  leading  order,  we  have  seen  that  the  Coulomb
contribution to $``-\beta F"$ is not affected if again the
effect of non-commutativity on the convolution leading to the plaquette
distribution from that for the link variables is disregarded:

\beq  ``-\beta_{crit.}   F"   \left|_{Coul.,\;0.\;order}   \right.   =
``-\beta_{crit.} F" \left|_{Coul.,\;next \;order} \right.  \label{BF1}
\eeq

\nin whereas  the  confinement  phase  free  energy effectively
acquires larger inverse squared couplings and consequently is displaced
in such manner that a given confinement free energy value is  attained
at smaller $logVol(SU(2))$ and $logVol(SU(3))$ values:

\beq ``-\beta_{crit.} F"  \left|_{conf.,\;next  \;order}  \right.\approx  2log
\frac{\pi^{dim(SMG)/2}}{(1+\frac{N_{SU(2)}}{12\beta_2})^{dim(SU(2))/2}
(1+\frac{N_{SU(3)}}{12\beta_3})^{dim(SU(3))/2}Vol(SMG)}    \label{BF2}
\eeq

\beq \approx ``-\beta_{crit.} F" \left|_{conf.,\;0.\;order}\right.
-\frac{2}{4\beta_2}-
\frac{2}{\beta_3} \eeq

\nin This results in the deformation of the original flat  ``diamond''
shaped interface separating the totally Coulomb and totally  confining
phases. The defining condition for the deformed diamond is

\beq                ``-\beta_{crit.}                F"\left|_{Coul.,\;
next\;order}\right.=``-\beta_{crit.}  F"\left|_{conf.,\;  next\;order}
\right.\eeq

\nin or

\beq log \left(\frac{(\pi/6)^{dim(SMG)/2}}{Vol(SMG)}\right)=2log\left(
\frac{\pi^{dim(SMG)/2}}{(1+\frac{2}{12\beta_2})^{3/2}
(1+\frac{3}{12\beta_3})^{8/2}
Vol(SMG)}\right) \label{haarcorr} \eeq

\nin This leads to the defining condition to next to leading order for
the phase boundary separating the totally Coulomb and totally confined
phases:

\beq logVol(SMG)=log(6\pi)^6-\frac{1}{2\beta_2}-\frac{2}{\beta_3}.  \eeq

\nin The latter replaces the lowest order condition

\beq logVol(SMG)=log(6\pi)^6 \eeq that determines  the  flat  diamond  phase
boundary of Figure \ref{smgfull}.  Figure  \ref{projection}  shows
the projection of the deformed  ``diamond-shaped''  interface  onto  a
plane parallel to the  $logVol(SU(3))$  and  $logVol(SU(2))$  plan  as
compared to the same projection  of  the  lowest  order  flat  diamond
interface.

\begin{figure}
\centerline{\epsfxsize=\textwidth \epsfbox{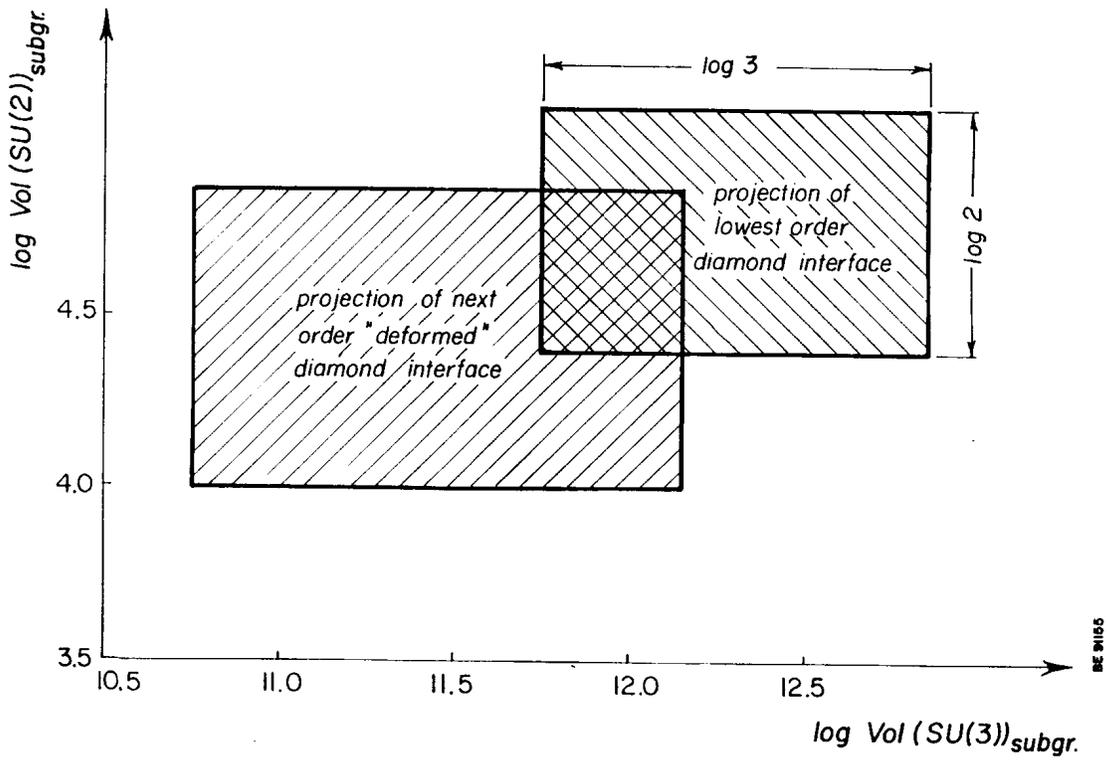}}
\caption{\label{projection} The projection
of the ``diamond-shaped'' interface corrected for the curvature of the
group    manifold     onto     a     plane     parallel     to     the
$logVol(SU(3))$-$logVol(SU(2))$  plane  as  compared   to   the   same
projection of the lowest order flat diamond interface. }\end{figure}

%
%
%

\subsection{Appendix: Correction    for    non-commutativity    in    $\langle
S_{act.,\;Coul}\rangle $ }\label{appnoncom}

In the lowest order approximation we used simple vector addition as the rule
of composition for Lie algebra vectors for non-Abelian groups. The correct
rule of composition (to next to lowest order) for Lie algebra  vectors
corresponding to non-Abelian group multiplication yields a deviation in
the  distribution  of  plaquette  variables  that  was   obtained   by
convoluting link variable  distributions.  For  a  plaquette  variable
composed of a number $m$ (typically $m=3$) of active  link  variables,  we
need the distribution of the convolution  of  $m$  link  variables  in
terms of Lie algebra elements corresponding to non-Abelian addition;
this kind of addition, hereafter designated with the symbol +' , leads to
vector sums denoted with the notation
$\bpprime=\sum'_r \ba_r = \baone +' \batwo +' \cdots +' \ba_m$.
In particular, we are interested in the width of the  distribution  of
plaquette  variables  corresponding  to   the   Lie   algebra   element
$\bpprime$.

For $m=2$ we have the  expansion  (Baker-Campbell-Hausdorff  formula):

\beq    \baone+'\batwo    =     \baone+\batwo+1/2[\baone,\batwo]     +
1/12[\baone-\batwo,[\baone,\batwo]] \eeq

\noindent The analogous expression for  $m>2$  is

         \beq \bpprime= \sum_r^m\ba_r + 1/2\sum_{r,s;r<s}^m [\ba_r,\ba_s]
            + 1/12\sum_{r,s}^m[\ba_r,[\ba_r,\ba_s]] \eeq

\nin (modulo a term with average zero).

\nin where higher order terms  and  terms
that vanish upon averaging over symmetric fluctuations are omitted.

For an arbitrary Cartan Killing form CK(.,.), there obtains

\beq   \langle   CK(\bpprime,\bpprime)\rangle   =   \sum_r^m   \langle
CK(\ba_r,\ba_r)\rangle   +   (1/4   -   2/12)\sum_{r,s;r<s}^m    \langle
CK([\ba_r,\ba_s], [\ba_r,\ba_s])\rangle \eeq

\[=  \sum_r^m  \langle  CK(\ba_r,\ba_r)\rangle  +   1/12\sum_{r,s;r<s}^m
\langle CK([\ba_r,\ba_s],[\ba_r,\ba_s])\rangle \]

Assuming the same  Gaussian  distribution  for  all  the  Lie  algebra
elements $\ba_r$, the sum $\sum_{r,s;r<s}
\langle CK([\ba_r,\ba_s],[\ba_r,\ba_s])\rangle$ in the last equation can  be
replaced by $\frac{m(m-1)}{2}  \langle  CK([\ba_r,\ba_s],[\ba_r,\ba_s])\rangle$
($r$ and $s$ now just two arbitrary, different  active  links  in  the
plaquette) in which case

\beq    \langle    CK(\bpprime,\bpprime)\rangle=    \sum_r^m    \langle
CK(\ba_r,\ba_r)\rangle      +      \frac{m(m-1)}{24}\langle CK([\ba_r,\ba_s],
[\ba_r,\ba_s])\rangle . \label{noncom3} \eeq

Let  $\bpone$  and  $\bptwo$  stand  for  a  couple  of  independently
distributed stochastic Lie algebra-valued variables  (assume  $\langle
\bpone^2\rangle  \approx  \langle  \bptwo^2  \rangle$   and   $\langle
\bpone\bptwo \rangle \approx 0$) distributed as $\sum_{r=1}^m  \ba_r$.
We  have  that  $m^2\langle  CK([\ba_r,\ba_s],   [\ba_r,\ba_s])\rangle
=\langle   CK([\bptwo,\bpone],[\bptwo,\bpone])\rangle.$    and    Eqn.
(\ref{noncom3}) can be written

\beq <CK(\bpprime,\bpprime)>= <CK(\bpone,\bpone)> + \frac{(m-1)}{(24m)}  \cdot
<CK([\bptwo,\bpone],[\bptwo,\bpone])> \label{ckpp}.\eeq

\nin Rewrite the last factor in (\ref{ckpp}) as

\beq  \langle  CK([\bptwo,\bpone],[\bptwo,\bpone])\rangle  =  -\langle
CK(\bpone,[\bptwo,[\bptwo,\bpone]])\rangle                   =-\langle
CK(\bpone,(\mbox{ad}\bptwo)^2 \bpone)\rangle \eeq

\nin where we have used the  invariance  of  the  Cartan-Killing  form
under cyclic permutations of  the  arguments  and  the  definition  of
$(\mbox{ad}\bptwo)^2$.    Note    that    the    operator     $\langle
(\mbox{ad}\bptwo)^2   \rangle$   is   invariant    under    similarity
transformations by a group element $U$ in the adjoint representation:

\beq    \rho_{adj.    \;rep.}(U)    (\mbox{ad}\bptwo)^2     \rho_{adj.
\;rep}(U)^{-1}=(\mbox{ad}\bptwo)^2 \eeq

\nin To see this, combine Eqns. (\ref{adxmu}) and (\ref{adab}) to get

\beq                                  (((\mbox{ad}\bptwo)^2)_{ij}=
(\mbox{ad}(\bx_b)\mbox{ad}(\bx_c))_{ij}P_{2,\;b}P_{2,\;c}. \eeq

\nin
Recall               however               that               $\langle
CK([\bptwo,\bpone],[\bptwo,\bpone])\rangle$ involves an  average
that    insures    that    $\langle     P_{2,\;b}P_{2,\;c}\rangle     =
\frac{\langle P^2_{2,\;a}\rangle}{d}\delta_{bc} $  in  which  case  we
have

\beq \langle CK(\bpone,(\mbox{ad}\bptwo)^2  \bpone)\rangle  =  \langle
CK(\bpone,P_{2,\;a}^2(\mbox{ad}(\bx_a))^2\bpone)\rangle \eeq

\nin                                                               But
$\frac{1}{d}P_{2,\;a}^2((\mbox{ad}(\bx_a))^2)$
is just the  Casimir  operator  for  the  adjoint  representation  and
therefore,  by  Schur's  Lemma,  just  proportional  to  the  identity
$\bunit$. This in turn means that we can write

\beq                                                           \langle
CK([\bptwo,\bpone],[\bptwo,\bpone])\rangle_{{\sbp}_1,\;{\sbp}_2}=-\frac{
\langle   (\mbox{ad}\bptwo)^2   \rangle_{{\sbp}_2}}{\bunit}    \langle
CK(\bpone,\bpone) \rangle_{{\sbp}_1} \eeq

\nin  where  $\langle  \cdots  \rangle_{{\sbp}_1,\;{\sbp}_2}$   denote
averages w. r. t. both $\bpone$ and $\bptwo$ whereas  $\langle  \cdots
\rangle_{{\sbp}_1}$ and  $\langle  \cdots  \rangle_{{\sbp}_2}$  denote
respectively averages w. r. t. just $\bpone$ and just  $\bptwo$.  Eqn.
(\ref{ckpp}) becomes

\beq                                          <CK(\bpprime,\bpprime)>=
<CK(\bpone,\bpone)>(1-\frac{(m-1)}{24m}\frac{\langle
(\mbox{ad}\bptwo)^2 \rangle_{{\sbp}_2}}{\bunit}) \label{noncom1} \eeq

\[                                            <CK(\bpprime,\bpprime)>=
<CK(\bpone,\bpone)>(1-\frac{(m-1)}{24m}\langle              Tr_{adj.\;
rep.}((\mbox{ad}\bptwo)^2)/d \rangle_{{\sbp}_2}) \]

\nin where $d$ is the dimension of the representation.

As we have defined squared distances  as  the  Cartan-Killing  form
defined in terms of the defining representation:

\beq     dist^2(\bunit,exp(\bptwo))= -Tr_{def.\; rep.}(\bptwo,\bptwo) \eeq

\nin we can rewrite Eqn.(\ref{noncom1}) as

\beq  \langle  Tr_{def.\;  rep.}(\bpprime,\bpprime)  \rangle  =\langle
Tr_{def.\;rep.}(\bpone,\bpone)\rangle
(1-\frac{(m-1)}{24m}\cdot           \frac{2N}{d}                \langle
Tr_{def.\;rep.}(\bptwo,\bptwo) \rangle) \label{noncom2} \eeq

\nin where we have used the result (\ref{adjdef}) for $SU(N)$ groups that

\beq
Tr_{adj.\;rep.}(\bptwo,\bptwo)=2N\mbox{Tr}_{def.\;rep.}(\bptwo,\bptwo)\eeq

\nin Assuming (for Coulomb-like degrees of freedom) the distribution of
Eqn. (\ref{distrib}) for the link variables:

\beq                                            F(\sqrt{A^2/2})\propto
\exp(-\tilde{\alpha}dist^2(1,\exp(\ba))                   =
\exp(\tilde{\alpha}    \mbox{Tr}_{def.\;     rep.}(\ba^2))    =
\exp(-\frac{1}{2}\tilde{\alpha}A^2), \eeq

\nin                we                 have

\beq                                                         -\langle
\mbox{Tr}_{def.\;rep.}(\ba_r,\ba_s)\rangle=\delta_{rs}\frac{d}{2 \alpha}.\eeq

\nin For a plaquette consisting of $m$ active links, the corresponding
result is

\beq          -          \langle          \mbox{Tr}_{def.          \;
rep.}(\bptwo,\bptwo)\rangle =\frac{md}{2 \alpha} \eeq

\nin From Eqn. (\ref{noncom2}) obtains

\beq \frac{\langle \mbox{Tr}_{def.\;rep.}(\bpprime,\bpprime)\rangle}
{\langle
\mbox{Tr}_{def.\;rep.}(\bp,\bp)\rangle}=1+\frac{(m-1)}{24m}\cdot
2N\frac{m}{2 \tilde{\alpha}\beta} \eeq

\nin         where         we         have          used          that
$\tilde{\alpha}\stackrel{def.}{=}\frac{\alpha}{\beta}$

Implementing the correction due to non-commutativity in the  expression
(\ref{convol}) for the convolution of $m=3$ link variables  requires  making
the replacement

\beq               \frac{1}{\tilde{\alpha}}                \rightarrow
\frac{1}{\tilde{\alpha}}(1+\frac{m-1}{24m}\cdot
2N\frac{m}{2 \tilde{\alpha}\beta}) \eeq

\nin   This   in    turn    means    that    the    contribution    to
$``-\beta F"_{per\;active\;link}$ coming from $\langle S_{act.,\;Coul.}
\rangle$ (Eqn.(\ref{bfpal})) is modified as follows:

\beq -2\sum_{i \in ``Coul."}d_i\frac{3}{2\tilde{\alpha}_i} \rightarrow
-2\sum_{i                                                          \in
``Coul."}d_i\frac{3(1+\frac{m-1}{24m}\cdot2N_i\cdot\frac
{m}{2 \tilde{\alpha}\beta})}{2\tilde{\alpha}_i} \eeq

\nin It is sufficiently accurate to assume that the extremum w. r. t.
$\tilde{\alpha}$ is the same as  for  the  lowest  order  calculation:
$\tilde{\alpha}_i=2m$ for all $i$.

For  $m=3$,  the  correction  for  non-commutativity  in  $``-\beta  F"
\mid_{Coul.}$ is:

\beq             -2\sum_{i\in              ``Coul."}d_i\frac{3}{2\cdot
2m}\frac{(m-1)}{24m}\cdot2N_i\frac{m}{2m\beta_i}=-\sum_{i\in
``Coul."}\frac{d_iN_i}{144\beta_i}\eeq

Taking this correction into account in Eqn. (\ref{haarcorr}) leads to

\beq log \left(\frac{(\pi/6)^{dim(SMG)/2}}{Vol(SMG)}\right)-\sum_{i\in
``Coul."}\frac{d_iN_i}{144\beta_i}=2log\left(
\frac{\pi^{dim(SMG)/2}}{(1+\frac{2}{12\beta_2})^{3/2}(1+
\frac{3}{12\beta_32})^{8/2}
Vol(SMG)}\right) \eeq

When both corrections - i.e., the correction for the Haar measure  and
the  correction   for   non-commutativity   in   calculating   $\langle
S_{act.,\;Coul.}\rangle$ - are included, the  defining  condition  for
the phase boundary separating the totally Coulomb and totally confined
phases becomes

\beq             \log             Vol(SMG)=\log              (6\pi)^6-
\underbrace{\frac{1}{2\beta_2}-\frac{2}{\beta_3}}_{\mbox{Haar
measure}}+
\underbrace{\frac{1}{24\beta_2}+\frac{1}{6\beta_3}}_{\mbox{non-commutativity}}
\eeq

While the sign of the correction to $\langle  S_{act.,\;Coul.}\rangle$
for non-commutativity tends to compensate the correction for  the  Haar
measure correction, the latter dominates in magnitude.


\subsection{Appendix: Interaction terms in the action}\label{appint}

It is illustrative to point  out  that  the  non-generic  multi-critical
point we get using the three parameters $\beta_1,\beta_N, \xi$  is  an
approximation that presupposes that there are no interaction terms  of
the type (\ref{int}) in the physical action we  seek  to  approximate.
If such terms were present in the action of Nature, we could introduce
a fourth
parameter $\gamma$ as a coefficient to  an  action  contribution  that
could compensate for interaction terms of  the  type  (\ref{int}).  It
is  interesting  to  briefly  consider  how  the  presence   of   such
interaction  terms  preclude  a  multi-critical  point   by   explicitly
introducing such a term. To this end, consider  a  term  of  the  form

\beq \gamma      dist^2(\bunit,U_{U(1)}(\Box))dist^2(\bunit,U_{SU(N)}(\Box))
\;\;\;\;(U_{U(1)}(\Box)       \in U(1),\;U_{SU(N)}(\Box) \in SU(N)).
\label{intterm}. \eeq

\nin which is a slightly special case of (\ref{int}).

If, for example,  we  consider  the  $SU(2)$
\dof, it is seen that the presence of this $\gamma$-term leads  to  an
increase in the effective inverse squared  coupling  for  the  $SU(N)$
\dof by a quantity depending on $dist^2(\bunit,U_{U(1)}(\Box))$:

\beq \beta_{SU(N),\;eff.}=\beta_{SU(N)}+dist^2(\bunit,U_{U(1)}(\Box)).
\label{inttermn} \eeq

An
analogous argument applies to the U(1) degrees of freedom:

\beq \beta_{U(1),\;eff.}=\beta_{U(1)}+dist^2(\bunit,U_{SU(N)}(\Box)).
\label{intterm1} \eeq

The presence of such terms  precludes a factorisation of the free
energy. Equivalently, (127) is invalidated by such interaction terms.
We can also reformulate the factorisation property (127) as an entropy
additivity property. In this completely equivalent but perhaps more
intuitive formulation, the condition for having the type of (non-generic)
multiple points that we consider is that the total entropy is additive
in entropy contributions from the ``constituent'' invariant subgroups.
The additivity requirement for the entropy essentially means that the
distribution of gauge degrees of freedom \underline{within} the cosets
of a factor group is the same for all the cosets of such a factor group.

\subsubsection*{The Entropy Formulation}

Assuming the validity of the weak coupling approximation, we  consider
first distributions  obtained  from  the  exponentiated  Manton  action:
$e^{\beta  dist^2(\sbunit,U_l)}/Z$ ($l\in L \subseteq G$) where
$dist^2(\bunit,l)$ is the
squared distance from the identity $\bunit$ to  the
group element $l$. If $l\in L$ is a coset of a factor group - i.e.,
if $L= G/H$ where $G$ is the gauge group and $H\triangleleft G$ -
the function
$dist^2$ is rendered unambiguous (for weak coupling) by requiring that
the element $g\in G$ such that $gH=l\in G/H$ is generated alone by the Lie
algebra of $G/H$.
Such a distribution is a {\em single}  normalised  Gauss  of
width $(2\beta)^{-(dim(L))/2}$ centred  at  the  group  identity.  We
introduce the symbol $Gauss_{\beta}(l)$ as the designation of  such
a  single-peaked  distribution of plaquette variables $l$ that is centred at
the identity. Bianchi identities are ignored.
For  the  distribution $Gauss_{\beta}(l)$,
the quantity $Vol(L)$  is  the  subgroup (or factor group) volume  with  the
inverse  squared couplings absorbed:
$Vol(L)\stackrel{def.}{=}\beta^{\frac{dim(L)}{2}}vol(L)$         where
$vol(L)$ is the volume of $L$. This  coincides
with the definition of $Vol(L)$ as the group volume measured in  units
proportional to the fluctuation volume:

$$ Vol(L)=\frac{vol(L)}{``fluctuation\;volume"}(\pi e)^{dim(L)/2} $$

The $Vol(L)$  are  well  defined  in  terms  of  the  entropy  if  the
distribution is additive in entropy contributions from the
``constituent'' invariant subgroups. Using the
normalised\footnote{The normalisation is such that integration over the
Haar measure $\int \frac{d^{{\cal H}aar}U}{vol(L)}$ ($U\in G$)
yields unity.} plaquette  distribution for a single plaquette
(ignoring  Bianchi  identities)

\beq    Gauss_{\beta}(l)\stackrel{def}{=}\frac{e^{-\beta
dist^2(\sbunit,l)}}{(\frac{\pi}{\beta})^{\frac{dim(L)}{2}}
\cdot\frac{1}{vol(L)}}  \;\;\;\;(l  \in
L),\eeq

\nin the entropy can then be calculated as:

\[ S_{entropy,\;Gauss_{\beta}}=
\langle-\log Gauss_{\beta}(l)\rangle_
{Gauss_{\beta}}
=\langle-\log(\frac{\exp                          (-\beta
dist^2(\bunit,l))}{(\frac{\pi}{\beta})^{(dim(L))/2}
\cdot\frac{1}{vol(L)}})\rangle \]

\[=-\log((\frac{\beta}{\pi})^{(dim(L))/2}vol(L))+\langle\beta
dist^2(\bunit,l)
\rangle \]

\[=-\frac{dim(L)}{2}(\log (\frac{\beta}{\pi})-1)-\log vol(L) \]

\[=\frac{dim(L)}{2}\log(\frac{e\pi}{\beta})-\log vol(L) \]

\[ S_{ent}=\frac{dim(L)}{2}\log(e\pi)-\log Vol(L). \]

\noindent The quantity $Vol(L)$ can  be  expressed  in  terms  of  the
entropy $S_{entropy}$ as

\beq             Vol(L)=e^{-S_{entropy}}\cdot(e\pi)^{\frac{dim(L)}{2}}
\label{volent1}.\eeq

As the quantity $Vol(L)$ is the group volume measured in units of  the
fluctuation  volume  determined  on  the  basis  of  a  single   Gauss
distributed peak $Gauss_{\beta}(l)$ centred at  the  identity,  we
append $Gauss_{\beta}$ as a subscript to the quantity  $Vol(L)$.
Hence, (\ref{volent1}) is rewritten

\beq Vol_{Gauss_{\beta}}(L)=e^{-S_{entropy,\;Gauss_{\beta}}}
\cdot(e\pi)^{\frac{dim(L)}{2}} \label{volent2}.\eeq

For illustrative purposes, let us write down the entropy in the case of
the  $U(N)$ group when the distribution is given by the {\em modified} Manton
action. By this is meant a distribution that, in addition to having a narrow
maximum at the unit element $\bunit$, also can have narrow maxima at nontrivial
elements of the discrete $\bz_N$ subgroups $\bz_2$ and $\bz_3$ inasmuch as
we only consider $U(2)$ and $U(3)$ (these are the $U(N)$ groups relevant for
the $SMG$). Having
now the possibility of these subsidiary peaks, we supplement the parameters
$\beta_{U(1)}$ and $\beta_{SU(N)}$ with a parameter $\xi^{(p)}$.
The parameter $\xi^{(p)}$ specifies the height of the subsidiary maxima
centred at nontrivial elements $p\in \bz_N$ relative to the maximum in the
distribution that is centred at the group identity $\bunit$. In the
approximation that there are no interaction terms in the action, it
is then possible to have non-generic multiple points in
a $3$-dimensional action  parameter space spanned by the variables

\beq (\beta_{U(1)}^{(p)},\beta_{SU(N)}^{(p)}, \xi^{(p)}). \eeq

\nin where the superscript $p$ labels the element $p \in \bz_N$ to which the
parameters pertain. This set of three parameters determines a  distribution
$\frac{e^{S_{mod.\;Manton}}}{Z}$ on  the
group       $U(N)$        ($N=2$        or        $3$) in the neighbourhood
of the element $p\in \bz_N\subset U(N)$.

As we  only  need  to  consider  the
local structure  of  the  gauge  group,  we  can  decompose  a  $U(N)$
plaquette variable $U_{U(N)}(\Box)$ into a $U(1)$ part and  a  $SU(N)$
part:    $U_{U(N)}(\Box)=(U_{U(1)}(\Box),U_{SU(N)}(\Box))$.

For $U(\Box)$ near the element $p\in \bz_N$ ($N=2$ or $3$), the
distribution $\frac{e^{S_{mod.\;Manton}}}{Z}$ is dominated by the Gauss
function

\beq Gauss_{(\xi^{(p)},\beta^{(p)}_{U(1)},\beta^{(p)}_{SU(N)})}(U(\Box))
\stackrel{def}{=}\eeq

$$ \frac{\exp(\beta^{(p)}_{U(1)}\;
dist^2(p,U_{U(1)}(\Box)))\cdot
\exp(\beta^{(p)}_{SU(N)}\;dist^2(p,U_{SU(N)}(\Box)))}{Z}$$

\nin where $(2\beta_{U(1)}^{(p)})^{\frac{1}{2}}$ respectively
$(2\beta_{SU(N)}^{(p)})^{\frac{1}{2}}$ is the width of
the maximum along the $U(1)$ and the $SU(N)$ direction(s) of the maximum
centred at $p\in \bz_N$. The are $N$ such Gauss-shaped maxima in the
distribution $\exp(S_{mod. Man.}(U(\Box)))/Z$ when $U(\Box)\in U(N)$;
$\xi^{(p)}$ is the height of the maximum at $p\in \bz_N$
relative to the height of the
Gauss-shaped maximum centred at the identity. The latter maximum
is denoted $Gauss_{(1,\beta_{U(1)}^{(\sbunit)},\beta_{SU(N)}^{(\sbunit)})}
U(\Box)$.
The partition function $Z$ that
normalises the $N$-maxima ($N=2$ or~$3$) modified Manton distribution
$\exp(S_{mod. Man.})/Z$ is

\beq Z=\frac{\pi^{N^2/2}}{vol(U(1))vol(SU(N))}\left (
\frac{1}{(\beta_{U(1)}^{(\sbunit)})^{\frac{1}{2}}
(\beta_{SU(N)}^{(\sbunit)})^{\frac{N^2-1}{2}}}+
\frac{N-1}{(\beta_{U(1)}^{(p)})^{\frac{1}{2}}
(\beta_{SU(N)}^{(p)})^{\frac{N^2-1}{2}}} \right ) \eeq

\nin It is easy to show that $\beta^{(p)}=\beta^{(\sbunit)}/\xi^{(p)}$
if we let $\beta^{(p)}$ be defined by

\beq e^{\beta^{(p)}dist^2(p,U(\Box))}=\xi^{(p)}e^{\beta^{(\sbunit)}dist^2
(\sbunit,U(\Box))}.\eeq

\nin The partition function can now be written

\beq Z=\frac{\pi^{N^2/2}}{(\beta_{U(1)}^{(\sbunit)})^{\frac{1}{2}}
(\beta_{SU(N)}^{(\sbunit)})^{\frac{N^2-1}{2}} vol(U(1))vol(SU(N))}
(1+\xi^{N^2}(N-1)) \eeq

For completeness we write down the distribution function for the
modified Manton action:

\beq \frac{e^S_{mod.\;Manton}}{Z}= \label{modman1} \eeq

\[ \left\{
\begin{array}{ll}
Gauss_{(1,\beta_{U(1)}^{(\sbunit)},\beta_{SU(N)}^{(\sbunit)})}U(\Box)  &
\mbox{if $U_{U(1)}(\Box),U_{SU(N)}(\Box)$ near identity of $U(N)$}  \\
Gauss_{(\xi^{(p)},\beta^{(\sbunit)}_{U(1)}/(\xi^{(p)})^2,
\beta_{SU(N)}^{(\sbunit)}/(\xi^{(p)})^2)}(U(\Box)) &
\mbox{if $U_{U(1)}(\Box),U_{SU(2)}(\Box)$ near nontrivial element of
$\bz_N$}  \\
0  & \mbox{if $U_{U(1)}(\Box),U_{SU(2)}(\Box)$ not near any element of $\bz_N$}
\end{array} \right. \]

\nin It is of course to be understood the $p\in \bz_N$ is chosen so as to
minimise $dist^2(p,U(\Box))$ ($U(\Box)\in U(N)$).
We have in  mind  a
weak  coupling  approximation  which  implies  that  the  distribution
$e^S_{mod.\;Manton}/Z$ is only appreciable near the elements of
$Z_N$  so  at  any
place in the group, at most  one Gaussian peak contributes to a non-vanishing
value of the distribution.  With
this restriction there is no ambiguity in talking  about  the  squared
distance between group elements.

The distribution (\ref{modman1}) is normalised w.r.t the Haar measure.
Recall that the quantity $Vol(L)$ is the volume of the subgroup
(or factor group $L$) measured in units of the {\em fluctuation volume}.
Note however that in the case of the modified Manton action,
the unit of volume for $Vol(L)$ is the sum of the fluctuation  volumes  of
several Gauss distributions that,  in  addition  to a distribution centred
at the identity, can
also  include  distributions
centred at nontrivial elements $p=e^{\pm i2\pi/N}\bunit \;\;(N=2,3)$ of
$Z_N$  subgroups having  height  $\xi$ and width
$(2\beta^{(p)})^{\frac{1}{2}}=(\frac{2\beta^{(\sbunit)}}{(\xi^{(p)})^2})
^{\frac{1}{2}}$.
This is  sometimes indicated explicitly by writing $Vol(L)$ using the
notation $Vol_{Gauss_{(\xi^{(p)},\beta^{(p)}_{U(1)},
\beta^{(p)}_{SU(N)}})}$.

The entropy of the distribution (\ref{modman1}) is

\beq S_{entropy}\stackrel{def}{=}-\langle \frac{e^S_{mod.\;Manton}}{Z} \rangle
=-\frac{1}{1+(N-1)\xi}\langle \log \frac{
Gauss_{(1,\beta^{(\sbunit)}_{U(1)},\beta^{(\sbunit)}_{SU(N)})}}
{1+(N-1)\xi}\rangle - \eeq
$$ -\frac{(N-1)\xi}{1+(N-1)\xi}\langle \log \frac{
\xi(Gauss_{(p,\beta_{U(1)}^{(p)},\beta_{SU(N)}^{(p)})}}{1+(N-1)\xi}\rangle -$$

$$ =-\frac{1}{1+(N-1)\xi}\log(\frac{1}{1+(N-1)\xi})
    -\frac{(N-1)\xi}{1+(N-1)\xi}\log(\frac{\xi^{(N^2+1)}}{1+(N-1)\xi})- $$
$$    -\langle \log Gauss_{(1,\beta_{U(1)}^{\sbunit},\beta_{SU(N)}^{\sbunit})}
\rangle $$

\nin where we have used that $Gauss_{(p,\beta_{U(1)}^{(p)},
\beta_{SU(N)}^{(p)})}=\xi^{N^2}Gauss_{1,\beta_{U(1)}^{(\sbunit)},
\beta_{SU(N)}^{(\sbunit)})}$.

\nin Rearranging yields \begin{small}

\beq S_{entropy}= \label{entfac}
\eeq

$$  \underbrace{-\log(\frac{1}{1+(N-1)\xi}\cdot
(\xi)^{\frac{(N^2+1)(N-1)\xi}{1+(N-1)\xi}})}_{S_{entropy,\;\sbz_N}}- $$

$$ -\underbrace{\langle \log Gauss_{(1,\beta_{U(1)}^{(\sbunit)})} \rangle}
_{S_{entropy,\;Gauss_{(1,\beta^{(\sbunit)}_{U(1)})}}}- $$

$$ -\underbrace{\langle \log Gauss_{(1,\beta_{SU(N)}^{(\sbunit)})} \rangle}
_{S_{entropy,\;Gauss_{(1,\beta^{(\sbunit)}_{SU(N)})}}}- $$

\end{small}

\nin  The  quantity  $S_{entropy,\,\sbz_N}$  is  the  entropy  due  to
presence of the discrete  Gaussian peaks at the  elements  of
$\bz_N$.  The  quantities  $S_{entropy,\;Gauss(1,\beta_{U(1)})}$  and
$S_{entropy,\;Gauss(1,\beta_{SU(N)})}$   are   the    entropies    due
respectively to fluctuations of (average)  widths  $(2\beta_1)^{-1/2}$
and $(2\beta_N)^{-1/2}$ within the Gaussian peaks situated  at  the
identity.






We have shown that  the  distribution  (\ref{modman1})  defined  on  the
non-simple  group  $U(N)\;\;(N=2$  or  $3)$  has  the  property  that

\beq
S_{entropy}=S_{entropy,\;\sbz_N}+S_{entropy,\;Gauss_{(1,\beta_{U(1)})}}+
S_{entropy,\;Gauss_{(1,\beta_{SU(N)})}}. \eeq

\nin We refer to this property as ``entropy additivity''. In the absence of
interactions between the \dof of the constituent invariant subgroups,
the  distribution  (\ref{modman1})  defined  on  $U(N)$  by  the   three
parameters $\beta^{(p)}_{U(1)},\beta^{(p)}_{SU(N)},$ and $\xi^{(p)}$
has  this property and
the $U(N)$ phase  diagram  spanned
by these three parameters has a non-generic  multiple  point  at
which $5$ phases convene.

In order to construct the phase diagram for $U(N)$ we equate the  free
energies (\ref{fe}) using the  ansatz  corresponding  respectively  to
confinement w. r. t. $H_1$ and $H_2$. This yields

\beq \frac{Vol(H_1)}{Vol(H_2)}=
(6\pi)^{\frac{dim(H_1)-dim(H_2)}{2}}  \eeq

\nin as the condition to be fulfilled at the phase boundary separating
a phase confined w. r. t. $H_1$ from a  phase confined w. r. t. $H_2$.

For distributions  such
as (\ref{modman1}) that, in addition to peaks at the group identity, also
have peaks at nontrivial elements of discrete  invariant  subgroups, we can
now write down an expression for $Vol(L)$ in terms of the entropy.
By analogy to (\ref{volent2}), we  have,  for  example,  for
$L=SU(N)/\bunit$

\beq Vol(SU(N)/\bunit)=e^{-S_{entropy,\;Gauss_{(1,\beta^{(\sbunit)}_{SU(N)})}}
-S_{entropy,\;\sbz_N}}\cdot(e\pi)^{\frac{N^2-1}{2}}     \label{volsun}
\eeq

\[ =Vol_{Gauss_{(1,\beta_{SU(N)})}}(SU(N))\cdot e^{-S_{entropy,\;\sbz_N}} \]

\[ =  \beta_N^{\frac{N^2-1}{2}}vol(SU(N)\cdot\frac{1}{1+(N-1)\xi}\cdot
(\xi)^{\frac{(N^2+1)(N-1)\xi}{1+(N-1)\xi}} \]

\nin where we have used  (\ref{volent2})  and  (\ref{entfac})  in  the
final step.
\vspace{.3cm}

\begin{figure}
\centerline{\epsfxsize=\textwidth \epsfbox{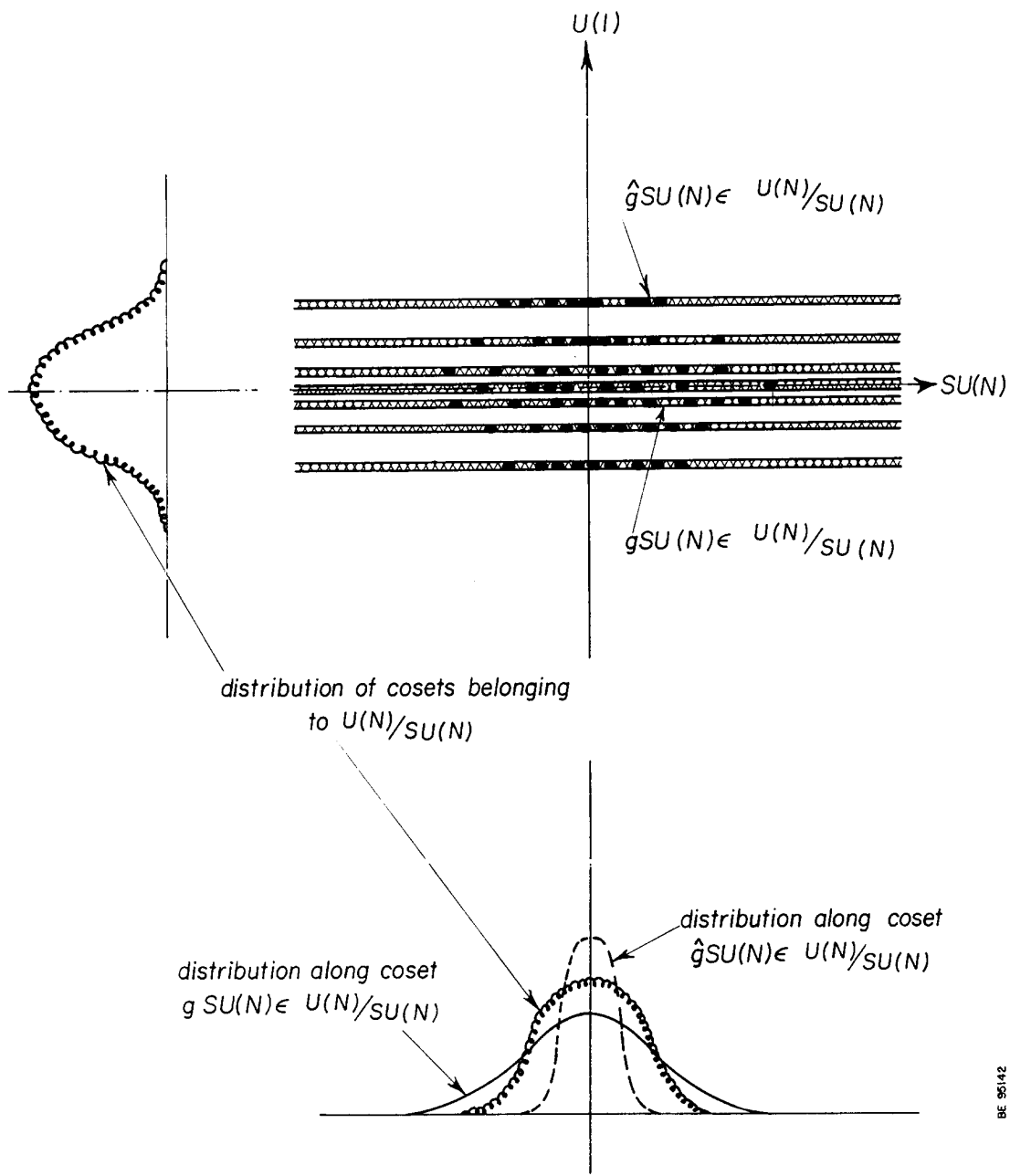}}
\caption[figintterm]{Action terms of the type~(\ref{intterm})
give rise to interactions in the sense that the distribution
of group elements within cosets varies along the distribution
of cosets.}
\end{figure}

Returning now to the discussion of interaction terms in the action of the
type (\ref{intterm}) that give rise to effective inverse squared couplings
of the types (\ref{inttermn}) and (\ref{intterm1}), we briefly argue that
if the coefficient $\gamma$ of such an interaction term (\ref{intterm})
is large, the entropy is not (approximately) additive in contributions from
the constituent invariant subgroups.
Moreover, a consequence of this is that there is no approximate
multi-critical point in the three-parameter space spanned by
$\beta^{(p)}_{U(1)},\;\beta^{(p)}_{SU(N)},$ and $\xi^{(p)}$. To illustrate
this, we consider again the
phase boundaries involving \dof isomorphic with the $SU(N)$.

One of these phase boundaries,
defined by the  condition $Vol(U(N)/U(1))=(6\pi)^{(N^2-1)/2)}$, separates
the totally confining phase from the phase with confinement
w. r. t. $U(1)$. The quantity $Vol(U(N)/U(1)$ is well defined because the
the distribution of cosets belonging to the factor group $U(N)/U(1)$ has a
well defined width and thereby a well defined entropy.

The other phase boundary defined by a condition on the $SU(N)$ \dof is that
separating the phase with confinement w. r. t. $SU(N)$ from the phase
having confinement only w. r. t. $\bz_N$. At this boundary, the condition
to be fulfilled  is $Vol(SU(N)/\bz_N)=(6\pi)^{(N^2-1)/2}$.
The quantity $Vol(SU(N)/\bz_N)$ is determined as an average over the
distributions of $SU(N)$ elements \underline{within} cosets
$gSU(N) \in U(N)/SU(N)$. We can use the restriction $g \in U(1)$ as we only
want to use $g$ as a label for the cosets of $U(N)/SU(N)$.
Unless  $\gamma  = 0$,
a distribution of $SU(N)$
elements within a coset $gSU(N)$ will depend on $g \in U(1)$.
Hence the quantity $Vol(SU(N)/\bz_N)$ is calculated as an average (with
weights given by the distribution of cosets $gSU(N) \in U(N)/SU(N)
\isomorph U(1)$) of
$SU(N)$ distributions that, for $\gamma \not= 0$, are different for each coset
$gSU(N)$. The calculation of $Vol(SU(N)/\bz_N)$ depends therefore on the
distribution of the cosets $gSU(N) \in U(N)/SU(n)$. Only if $\gamma = 0$ is
the distribution of $SU(N)$ \dof within each coset $gSU(N) \in U(N)/SU(N)$
the same. In this case one obtains the same result for $Vol(U(N)/\bz_N)$ for
all distributions of the cosets of the factor group $U(N)/SU(N)$ that
coincides with the value of $Vol(U(N)/U(1))$ at a non-generic
multi-critical point.

A similar line of reasoning applies to the $U(1)$ degrees of freedom.
For both the $U(1)$
and the $SU(N)$ \dof, we can say that an action with $\gamma=0$ leads to a
distribution of the group that is ``entropy-additive'' because the widths of
the distributions of $U(1)$ and $SU(N)$ elements within respectively the cosets
 and $gSU(N) \in U(N)/SU(N)\;\;(g\in U(1))$
$hU(1)\in U(N)/U(1)\;\; (h \in SU(N))$ are the same for all displacements $g$
and $h$.

\subsection{Appendix: Proposed model for  the  stability  of  the  multiple
point}\label{appstab}

We propose a  mechanism
for the stability of the multiple point that is based on a  model
which could be called a ``nonlocal gauge glass model''\footnote{The term
``gauge
glass'' was appropriately coined by Jeff Greensite  by  analogy  to  a
spin glass which is so named because the ``frozen  in''  structure  is
reminiscent of that of glass.}.
which is very  much  inspired  by
the project of ``random dynamics\cite{roysoc,randyn1,randyn2}''.ese two
assumptions - multiple point criticality and degenerate vacua
differing by a quantity of order unity in Planck units - leads to
The
essential feature is the influence of a bias effect that can occur  in
the presence of a plaquette (or multi-plaquette) action the functional form
of which is taken to be quenched random.
This could  mean  that  for  each
Wilson  loop $\Gamma$,  the  coefficients  (called  the  ``$\beta$'s'')  in say
 a
character expansion of the Wilson loop action are fixed at the  outset
as random values and remain fixed during the evaluation of the functional
integral. While translational invariance is broken at least at  small
scales because a different set of random $\beta$'s  is  associated  with
each Wilson  loop,  it  is  presumably  regained  at least
approximately in  going  to  large
distances inasmuch as it is assumed that the statistical  {\em distribution}
of quenched random variables is translation-ally invariant.

Randomly weighted terms in the  action  from  the  different  Wilson
loops would on the average contribute nothing to the  inverse  squared
coupling were it not for the {\em bias}:
the vacuum dominant value of a  Wilson
loop variable (a point in the gauge  group)  is  correlated  with  the
values of the quenched random coefficients for the Wilson  loop  under
consideration. This correlation comes about because the
vacuum field  configuration\footnote{Note that we envision
a relatively complicated vacuum state in which the link or rather plaquette
variables fluctuate around other elements than the unit element. However,
these ``other elements'' must necessarily be elements of the centre
if ``collapse'' ($\approx$ Higgs-like behaviour) is to be avoided; this may
require a connected centre\cite{center1,center2,gg} for the group that extends
almost
densely over the group.} adjustments  resulting
from the tendency to approximately  maximise  the  exponential  of  the
action $\exp(S)$ as a function of link variables  will  concurrently  tend  to
make the second derivative  w. r.  t.
Wilson loop variables of $\exp(S)$ more negative.

In the simplest model, the gauge glass we use is rather strongly  {\em
nonlocal} because we assume that the quenched random contributions  to
the  action  are  not  restricted  to  contributions  from   elementary
plaquettes, but in principle include {\em all} Wilson loops.  If  this
should lead to problems with locality,  we  can  postulate  that  only
loops up to some finite size are present in the action since the
most crucial prerequisite for the bias mechanism is the  inclusion  of
many Wilson loops with the size distribution being of  only  secondary
importance.

The bias effect can be formulated as
an additional term in the Callan-Symanzik $\beta$-functions (in
addition to the normal renormalization group contribution). To see this,
envision a series of calculations
of the effective  couplings $g(\mu)$ for successively larger
inverse energies $\mu^{-1}$.
For each value of $\mu^{-1}$, Wilson loops of size up $\mu^{-1}$ are
included in computing $g(\mu)$; therefore a calculation of $g(\mu)$
includes more and more Wilson loops in going towards the infrared.
The inclusion of progressively longer
and longer  loops  takes  place  in  a  background  field  made  up  of
contributions  from  the  already  included  smaller  loops  that  are
approximately described as a  background  continuum  Lagrange  density
$-\frac{1}{4g^2(\mu)}F_{\mu\nu}^2$. This process, in which the  coupling
$g(\mu)$ becomes smaller and smaller the more loops it  accounts  for,
culminates in $g(\mu)$ attaining  the  critical  value  whereupon  the
influence  of  additional  loops  on  the  vacuum   configuration   is
drastically  diminished  because  the   transition   to   a   $g(\mu)$
corresponding to the Coulomb phase leads  to  a  vacuum  configuration
that is much less readily influenced than in  the  confinement  phase.
Contributions from larger Wilson loops are no longer  correlated  with
the vacuum dominant field configuration that is almost solely  determined
by  the  Wilson  loops  of  smaller  spatial   extent.   Without   the
``protection'' of the bias effect, the contributions from these larger
loops cancel out on the average because of the assumed  randomness  in
the signs of action terms with the result that the effective couplings
will  no longer be modified much
by the inclusion of  larger  Wilson  loops  that
show up in going to larger length scales.

The  variation  of  the
effective coupling due to the bias effect might formally  be  included
in a generalised Callan-Symanzik $\beta$-function. (actually we mean a
multicomponent  vector  of  generalised  $\beta$-functions  with   one
component for each  parameter  of  a  single  plaquette  action  of  a
course-grained  lattice  at  the  scale  $\mu$).   These   generalised
$\beta$-functions (i.e., the components of the vector  of  generalised
$\beta$-functions) contain contributions that take into  account  that
the part of the Lagrangian of the theory that is used to define  gauge
couplings $g(\mu)$ is changing as we go to larger length scales.  That
this change has a non-vanishing average effect on the couplings is  due
to   the   bias   effect.   These   extra   contributions    to    the
$\beta$-functions, which are in addition to the normal renormalization
group effects, make the generalised $\beta$-functions explicitly scale
dependent.  Specifically,  we  envision  rapid   variations   of   the
$\beta$-functions as the bias effect is drastically  weakened  at  the
transition to a Coulomb-like phase. If  the  $\beta$-functions  become
zero, this would result in an infrared attractive fixed point near the
phase transitions at the multiple point.

An important point is that  multiple point
criticality is implied by  almost  any
mechanism that drives a gauge coupling to a critical value  because  a
mechanism that seeks out the critical coupling for  some  gauge  group
will probably function in the same way for all invariant subgroups  of
a gauge group. But this is tantamount to seeking out the multiple
point which by definition is the point or surface in the phase diagram
at the borderline between confining  and  Coulomb  phases  for  all
invariant subgroups. In particular, our model as outlined above would,
imply that  Wilson  loop  contributions  $\prod_{\;\link  \;  \in
\Gamma}U(\link)$ depending only on the cosets  in  $G/H$  w.r.t.  some
invariant subgroup $H$ would become very ineffective  in bringing about  a
further increase in  the  inverse  squared  couplings  (for  the  \dof
corresponding to the factor group)  once  it  is  only  the  invariant
subgroup $H$ of the group $G$ that  remains  ``confining'';  in  other
words,  the  couplings  for  $G/H$  stop  falling  (in   the   crudest
approximation) once $G/H$ ``reaches the Coulomb phase''.

Several alternatives to the nonlocal gauge glass explanation  for  the
stability of  the  multiple  point  (assuming  it  exists)  have  been
considered. We have for example in previous work, prior to the  advent
of the principle of multiple point criticality, used  the  entropy  as
the quantity to be maximised in the  predictions  of  gauge  couplings
from criticality. In this earlier work, we have for the non-Abelian Lie
subgroups of the $SMG^3$ considered criticality only  w.r.t.  the  Lie
subgroups $SU(2)$ and $SU(3)$ and not for criticality  w.  r.  t.  the
$\bz_2$ and $\bz_3$ discrete  invariant  subgroups  of,  respectively,
$SU(2)$ and $SU(3)$. We found that the entropy, calculated  to  lowest
order, was to first approximation constant on an interface  of  finite
extent that separated the totally Coulomb-like and totally confinement
phases in the parameter space of the $SMG$ phase diagram.  In  effect,
this interface prohibited other  \pcps  from  meeting  at  a  multiple
point. We now find that the addition of action  parameters  that  also
allow the discrete invariant subgroups to become critical  results  in
the shrinking of this interface into a point that coincides  with  the
multiple point. This can be expected to affect the entropy because, at
the multiple point, we are also on the verge of  confinement  for  the
groups $\bz_2$ and $\bz_3$. This means that  a  small  change  in  the
appropriate action parameters can bring  about  a  transition  from  a
Coulomb-like phase to a  confinement-like  phase  with  the  difference
between  the  two  being,  for  example,  defined  by  the  respective
perimeter and area law decay of Wilson loops (for charges 1/2  or  1/3
in the case of $\bz_2$ and $\bz_3$ respectively).

In the action  parameter  space  that  includes  parameters  that  can
be adjusted so as to have criticality w. r. t. the discrete  invariant
subgroups $\bz_2$ and $\bz_3$, the entropy is constant to lowest order
along a (hyper)surface separating the  totally  Coulomb  from  totally
confined phases. But a calculation to next order appears  to  lead  to
the conclusion that the entropy is not maximum  at  the  multiple
point thereby obviating the idea of maximum entropy as an  explanation
for the multiple point.

However, it can be claimed that  the  multiple point  is  such  a
characteristic ``corner'' of the phase diagram that  it  is  extremely
likely that there is some relevant physical quantity  or  property  that
is extremised  at this point. A possible scenario that  might  in  part
rescue the maximum entropy idea is that, at the  multiple point,
there are strong fluctuations along the discrete  subgroup  directions
of the gauge group that, for given entropy, might be very  effective  in
preventing potential Higgs fields from  bringing  the  model  into  a
Higgs phase.  In  other  words,  the  entropy  that  comes  from  the
``discretised'' lack of knowledge (as to which element of the discrete
invariant  subgroups  in  the  neighbourhood  of  which  the  plaquette
variable takes  a  value)  may  function  better  in  suppressing  the
tendency for ``Higgsing'' than the same  amount  of  entropy  arising
from  fluctuations  within  the  individual   Gaussian   distributions
$e^{S_{Manton}}$ centred at the elements of $\bz_2$  and  $\bz_3$.  If
this were true, one might use Higgs suppression as the property  to
be optimised at the multiple point.

Yet another admittedly rather speculative approach to  explaining  the
multiple point suggest  that  the  functional  integral  for  the
partition function in baby universe theory should have a maximal value
at the multiple point\cite{baby4}.

Meaningful continuum couplings for lattice gauge theories do not exist
for   couplings   that   exceed   the    critical
values\cite{surpc}. This is  corroborated  by  the  observation\cite{surpc}
that Mitrushkin\cite{mitr} only formally  obtains  a  strong  continuum
coupling in the Coulomb phase.

In summary, we have  in  this  appendix  supplemented  the
postulate  of  the  principle  of  multiple  point  criticality   with
proposals as to how a stable Planck  scale  multiple  point  might  be
realized. To this end, we described a gauge glass model inspired
by random dynamics. This model, which uses a quenched random action, has a
bias causing weaker couplings that is discontinuously
diminished at the multiple point. This leads to a zero of a generalised
Callan-Symanzik $\beta$ function  thereby establishing the multiple point
as an approximate ``infrared stable'' fixed point.

The speculative nature of these arguments in no way detracts
from the most important justification for the principle which  is  the
noteworthy phenomenological success.

\bibliography{rpred}
\end{document}